\documentstyle[12pt,epsf]{article}

\makeatletter
\def\mybibitem{\@ifnextchar[\@mlbibitem\@mbibitem}
\def\@mlbibitem[#1]#2{\item[\@biblabel{#1}\hfill]\if@filesw
      {\let\protect\noexpand
       \immediate
       \write\@auxout{\string\bibcite{#2}{#1}}}\fi\ignorespaces}
\def\@mbibitem#1{\item\if@filesw \immediate\write\@auxout
       {\string\bibcite{#1}{#1}}\fi\ignorespaces}

\newcommand{\msection}{%
\setcounter{equation}{0}%
\@startsection {section}{1}{\z@}%
{-3.5ex \@plus -1ex \@minus -.2ex}%
{2.3ex \@plus.2ex}%
{\normalfont\Large\bfseries}}

\newcommand{\mmsection}[1]{%
\immediate\write\@auxout
{\string\@writefile{toc}{\string\contentsline {section}{#1}{\the\value{page}}}}}

\newcommand{\mmsectionb}[1]{%
\section*{#1}
\immediate\write\@auxout
{\string\@writefile{toc}{\string\contentsline {section}{#1}{\the\value{page}}}}}

\newcommand{\mmmsection}[1]{%
\immediate\write\@auxout
{\string\@writefile{toc}{\string\contentsline {section}{\string\numberline {}#1}{\the\value{page}}}}}

\newcount\@indentflag \global\@indentflag=1 %
\newdimen\@eqtoeqnum \@eqtoeqnum=6pt %
\def\@indentamount{%
\ifcase\@indentflag 0pt\or\@centering\or0pt plus1fil\fi\relax
}
\def\FL{\global\@indentflag=0 }
\def\FR{\global\@indentflag=2 }

\def\@eqnnum{\hbox{\reset@font\rm(\theequation)}}
\let\make@eqnnum=\@eqnnum %
\def\eqnum#1{\dec@eqnnum \global\def\make@eqnnum{\reset@font\rm(#1)}%
\def\@currentlabel{#1}%
}
\def\inc@eqnnum{\addtocounter{equation}{1}}
\def\dec@eqnnum{\addtocounter{equation}{-1}}

\newbox\@testboxa
\newbox\@testboxb

\def\equation{\par\vskip-\lastskip\vskip\abovedisplayskip
\inc@eqnnum\let\@currentlabel=\theequation
\setbox\@testboxa=\hbox\bgroup\hskip\@totalleftmargin\hskip\@indentamount
\hbox\bgroup$\displaystyle
}

\def\endequation{$\egroup\hskip\@centering\egroup %
\setbox\@testboxb=\hbox{\make@eqnnum}%
\bgroup
\@tempdima\wd\@testboxa \advance\@tempdima by\wd\@testboxb
\ifcase\@indentflag
\advance\@tempdima by\@eqtoeqnum
\ifdim\@tempdima<\hsize %
\def\@tempa{0}%
\else
\def\@tempa{1}%
\fi
\or
\advance\@tempdima by2\@eqtoeqnum
\ifdim\@tempdima<\hsize %
\def\@tempa{0}%
\else %
\@tempdima\wd\@testboxa \advance\@tempdima by\wd\@testboxb
\advance\@tempdima by\@eqtoeqnum
\ifdim\@tempdima<\hsize %
\def\@tempa{0}%
\setbox\@testboxa\hbox{\hfill\box\@testboxa\kern\@eqtoeqnum}%
\else
\def\@tempa{1}%
\fi
\fi
\or
\advance\@tempdima by2\@eqtoeqnum
\ifdim\@tempdima<\hsize %
\def\@tempa{0}%
\setbox\@testboxb=\hbox{\kern\@eqtoeqnum\make@eqnnum}%
\else
\def\@tempa{1}%
\fi
\fi
\ifnum\@tempa=0 %
\hbox to\hsize{\unhbox\@testboxa\box\@testboxb}%
\else %
\vbox{\hbox to\hsize{\unhbox\@testboxa}%
\vskip6pt %
\hbox to\hsize{\hfil\box\@testboxb}}%
\fi
\egroup
\global\let\make@eqnnum\@eqnnum %
\vskip\belowdisplayskip\noindent\global\@indentflag=1 \global\@ignoretrue
}

\def\eqnarray{\par\vskip-\lastskip\vskip\abovedisplayskip
\inc@eqnnum\let\@currentlabel=\theequation
\global\@eqnswtrue\m@th
\global\@eqcnt\z@
\tabskip\@totalleftmargin\advance\tabskip by\@indentamount\let\\\@eqncr
\halign to\hsize\bgroup\hskip\@centering
$\displaystyle\tabskip\z@{##{}}$&\global\@eqcnt\@ne
\hfil${{}##{}}$\hfil
&\global\@eqcnt\tw@ $\displaystyle\tabskip\z@{##}$\hfil
\tabskip\@centering \if@eqnsw\phantom{\make@eqnnum\kern\@eqtoeqnum}\fi
&\llap{##}\tabskip\z@\cr}

\def\endeqnarray{%
\@@eqncr\egroup
\vskip\belowdisplayskip\noindent
\dec@eqnnum\global\@indentflag=1
\global\let\make@eqnnum\@eqnnum %
\global\@ignoretrue
}

\newcommand{\vereq}[2]{\lower3pt\vbox{\baselineskip1.5pt \lineskip1.5pt
\ialign{$\m@th#1\hfill##\hfil$\crcr#2\crcr\sim\crcr}}}

\newbox\tempboxa
\newdimen\captionboxsubcount 
\def\capsize#1{\captionboxsubcount=#1pt}
\newdimen\captionboxsub
\captionboxsub=\hsize \advance\captionboxsub by -\captionboxsubcount
\advance\captionboxsub by -\captionboxsubcount
\long\def\@makecaption#1#2{
 \setbox\@tempboxa\hbox{\footnotesize #1: #2}
 \ifdim \wd\@tempboxa >\captionboxsub 
\rightskip=\captionboxsubcount \leftskip=\captionboxsubcount 
  \footnotesize #1: #2 
\else \hbox to\hsize{\hfil\box\@tempboxa\hfil} 
 \fi}

\makeatother

\capsize{30}

\renewcommand{\section}{\msection}

\begin{document}

\renewcommand{\theequation}{\arabic{section}.\arabic{equation}}
\thispagestyle{empty}

\hfill{\normalsize\vbox{\hbox{February 2003} \hbox{DPNU-03-02}  }}
\vspace{0.5cm}
\begin{center}
{\LARGE\bf
Hidden Local Symmetry at Loop}\\
\Large
-- A New Perspective of Composite Gauge Boson \\
and 
Chiral Phase Transition --
\end{center}
\vspace{1cm plus 0.5cm minus 0.5cm}
\begin{center}
\Large
Masayasu {\Large\sc Harada} and
Koichi {\Large\sc Yamawaki}
\end{center}
\vspace{0.5cm plus 0.5cm minus 0.5cm}
\begin{center}
{\it
Department of Physics, Nagoya University,\\
Nagoya, 464-8602, Japan.}
\end{center}

\vspace{1.0cm plus 0.5cm minus 0.5cm}

\begin{abstract}
We develop an effective field theory of QCD and QCD-like theories
beyond the Standard Model, 
based on the hidden local symmetry (HLS) model
for the pseudoscalar mesons ($\pi$) as Nambu-Goldstone 
bosons and the vector mesons ($\rho$) as gauge bosons. 
The presence of gauge symmetry of HLS is vital
to the systematic low energy expansion or the chiral perturbation
theory (ChPT) with loops of $\rho$ as well as $\pi$.
We first formulate the ChPT
with HLS in details and further 
include quadratic divergences 
which are 
crucial to
the chiral phase transition.
Detailed calculations of the one-loop renormalization-group equation
of the parameters of the HLS model are given, based on which
we show the phase diagram of the full 
parameter space.
The bare parameters (defined at cutoff $\Lambda$) of the
HLS model 
are 
determined by the matching (``Wilsonian matching'') with
the underlying  
QCD at $\Lambda$ through the operator-product expansion of current
correlators. Amazingly,
the Wilsonian matching provides the effective field theory
with the otherwise
unknown information of the underlying QCD such as the explicit
$N_c$ dependence 
and predicts low energy phenomenology for
the three-flavored QCD
in remarkable agreement with the experiments. Furthermore, 
when the chiral symmetry restoration
takes place in the underlying QCD, the Wilsonian matching
uniquely leads to the Vector Manifestation (VM)  as a new pattern of
Wigner realization of chiral symmetry, with the $\rho$ 
becoming degenerate with the massless
$\pi$
as the chiral partner.
In the VM the vector dominance is badly violated. 
The VM is in fact realized in the large $N_f$ QCD when 
$N_f \rightarrow N_f^{\rm crit}-0$,
with the chiral symmetry restoration point 
$N_f^{\rm crit} \simeq 5\frac{N_c}{3}$ being
in rough agreement with the lattice simulation for $N_c=3$.
The large $N_f$ QCD near the critical point 
provides a concrete example of a strong
coupling gauge theory that generates a theory of weakly coupled
light composite gauge bosons. 
Similarly to the Seiberg duality in the SUSY QCD,
the $\mbox{SU}(N_f)$ HLS 
plays a role of a ``magnetic theory'' dual to the $\mbox{SU}(N_c)$ QCD
as an 
``electric theory''.  
The proof of the 
low energy theorem of the HLS at any loop
order 
is intact even including quadratic divergences.
The VM
can be realized also in hot and/or dense QCD.
\end{abstract}

\newpage

\tableofcontents

\listoffigures
 
\listoftables

\newpage

\section{Introduction}
\label{sec:intro}

As is well known, the
vector mesons are the very physical objects that the non-Abelian gauge
theory was first applied to in the history~\cite{Yang-Mills,Sakurai}. 
Before the advent of QCD the notion of
``massive gauge bosons'' was in fact very successful in the vector
meson phenomenology~\cite{Sakurai}.  
Nevertheless, little attention was paid to the idea
that the vector meson are literally gauge bosons, partly because of
their  non-vanishing mass. 
It is rather ironical that the idea of the vector mesons being  gauge
bosons was forgotten for long time, even after the Higgs mechanism was
established for the  electroweak gauge theory. 
Actually it was long considered that the vector meson mass cannot be  
formulated as the spontaneously generated gauge boson mass via Higgs
mechanism in a way consistent with the gauge symmetry and the chiral
symmetry.

It was only in 1984 that Hidden Local Symmetry (HLS) was proposed 
by collaborations including one of the present authors 
(K.Y.)~\cite{BKUYY,BKY:NPB,BKY:PTP,FKTUY}
to describe the vector mesons as genuine gauge bosons with the mass
being 
generated via Higgs mechanism in the framework of 
the nonlinear chiral Lagrangian. 

The approach is based on the general observation 
(see Ref.~\cite{BKY}) that 
the nonlinear sigma model 
on the manifold $G/H$ is gauge equivalent to 
another model having a larger symmetry 
$G_{\rm global} \times H_{\rm local}$, 
$H_{\rm local}$ being the HLS whose gauge fields are auxiliary fields
and can 
be eliminated when the kinetic terms are ignored.
As usual in the gauge theories, the HLS $H_{\rm local}$ is broken by
the gauge-fixing which then  
breaks also the $G_{\rm global}$.  
As a result, in the absence of the kinetic term of the HLS gauge bosons
we get back precisely the original nonlinear sigma model based  on 
$G/H$, with $G$ being a residual global symmetry under  
combined transformation of  $H_{\rm local}$ and $G_{\rm global}$  
and $H$ the diagonal sum of these two. 

In the case at hand, the relevant nonlinear sigma model is the
nonlinear chiral Lagrangian based on  
$G/H =\mbox{SU}(N_f)_{\rm L} \times \mbox{SU}(N_f)_{\rm R}
/\mbox{SU}(N_f)_{\rm V}$ for the QCD with massless $N_f$ flavors,
where $N_f^2-1$ massless Nambu-Goldstone (NG) bosons are identified
with 
the pseudoscalar mesons including the $\pi$ meson
in such an idealized limit of
massless flavors. The
underlying QCD dynamics 
generate the kinetic term of the vector 
mesons, which can be ignored for the energy region
much lower than the vector meson mass. Then the 
HLS model is reduced to the nonlinear chiral Lagrangian in the low energy limit
in accord with the 
low energy theorem of the chiral symmetry. 
The corresponding HLS model has the symmetry 
$[\mbox{SU}(N_f)_{\rm L} \times \mbox{SU}(N_f)_{\rm R}]_{\rm global} 
\times [\mbox{SU}(N_f)_{\rm V}]_{\rm local}$, with the gauge bosons 
of $[\mbox{SU}(N_f)_{\rm V}]_{\rm local}$ being identified with the
vector  
mesons ($\rho$ 
meson and its flavor partners). 

Now, a crucial step made 
for the vector mesons~\cite{BKUYY,BKY:NPB,BKY:PTP,FKTUY} was that 
the vector meson mass terms were introduced in a gauge invariant
manner, namely, in a way  
invariant under 
$[\mbox{SU}(N_f)_{\rm L} \times \mbox{SU}(N_f)_{\rm R}]_{\rm global} 
\times [\mbox{SU}(N_f)_{\rm V}]_{\rm local}$ and hence this
mass is regarded as generated via the Higgs mechanism after
gauge-fixing (unitary gauge) of HLS  
$[\mbox{SU}(N_f)_{\rm V}]_{\rm local}$.\footnote{
  There was a pre-historical 
  work~\cite{Balachandran-Stern-Trahern} discussing a
  concept similar to the HLS, which however did not consider a mass
  term  
  of vector meson and hence is somewhat remote from the physics of
  vector mesons. 
}
In writing the $G_{\rm global} \times H_{\rm local}$, 
we had actually introduced would-be Nambu-Goldstone (NG) bosons with
$J^{PC}=0^{+-}$ 
(denoted by $\sigma$, not to be confused with the scalar (so-called
``sigma'') mesons having  
$J^{PC}=0^{++}$) which are to be
absorbed into the vector mesons via Higgs mechanism in the unitary
gauge.  
Note that the usual quark flavor symmetry 
$SU(N_f)_{\rm V}$ of QCD corresponds to $H$ of $G/H$ which is a
residual unbroken diagonal symmetry  
after the spontaneous breaking of both $H_{\rm local}$ and  
$G_{\rm global}$ as mentioned above. 

The first successful phenomenology was established for the $\rho$ and
$\pi$ mesons in 
the two-flavors QCD~\cite{BKUYY}: 
\begin{eqnarray}
&&
  g_{\rho\pi\pi} =g \quad {\rm (Universality)}
\ ,
\\
&&
  m^2_\rho =2 g_{\rho\pi\pi}^2 F_\pi^2 \quad {\rm (KSRF(II))}
\ ,
\\
&&
  g_{\gamma\pi\pi}=0 
 \quad \mbox{\rm (Vector Dominance)} 
\ ,
\end{eqnarray}
for a particular choice of the parameter of the HLS Lagrangian $a=2$,
where 
$g_{\rho\pi\pi}$, $g$, $m_{\rho}$, $F_\pi$ and 
$g_{\gamma\pi\pi}$ are the $\rho$-$\pi$-$\pi$
coupling, the gauge coupling of HLS, the $\rho$ meson mass, the decay
constant of pion 
and the direct $\gamma$-$\pi$-$\pi$ coupling, respectively.
Most remarkably, we find a relation independent of the Lagrangian
parameters $a$ 
and $g$~\cite{BKY:NPB}:
\begin{equation}
g_\rho = 2 g_{\rho\pi\pi} F_\pi^2 \quad {\rm (KSRF (I))}
\ ,
\label{KSRF I; intro}
\end{equation}
which was conjectured to be a low energy theorem of HLS~\cite{BKY:NPB}
and then was argued to hold at general tree-level~\cite{BKY:PTP}.

Such a tree-level phenomenology including further developments (by the
end of 1987)
was reviewed in the previous Physics Reports by Bando, Kugo and one of 
the present authors (K.Y.)~\cite{BKY}. 
The volume included extension to the general group 
$G$ and $H$~\cite{BKY:PTP}, the case of Generalized HLS (GHLS) 
$G_{\rm local}$, i.e., the model
having the symmetry 
$G_{\rm global}\times G_{\rm local}$ which can accommodate
axialvector mesons ($a_1$ meson and its flavor 
partners)~\cite{BKY:NPB,Bando-Fujiwara-Yamawaki}, 
and the anomalous
processes~\cite{FKTUY}.
The success of the tree-level phenomenology is already convincing for
the HLS model 
to be a good candidate for the Effective Field Theory (EFT) of the
underlying QCD. It may also be useful for the QCD-like theories beyond
the  
SM such as the technicolor~\cite{Weinberg:TC1,Weinberg:TC2,Susskind}: 
the HLS model applied to the electroweak
theory, sometimes called a BESS 
model~\cite{Casalbuoni-DeCurtis-Dominici-Gatto:85,%
Casalbuoni-DeCurtis-Dominici-Gatto:87}, 
would be an EFT of
a viable technicolor such as the walking 
technicolor~\cite{Holdom85,Yamawaki-Bando-Matumoto,Akiba-Yanagida,
Appelquist-Karabali-Wijewardhana,BMSY} 
(See Ref.~\cite{Y,Hill-Simmons} for reviews)
which contains the techni-rho meson.

Thus the old idea of the vector meson being gauge bosons has been
revived by  
the HLS in a precise manner:
The vector meson mass is now gauge-invariant under HLS as well as
invariant under the chiral symmetry of  
the underlying QCD. It should be mentioned that the gauge invariance
of HLS does not exist 
in the underlying QCD and  is rather generated at the composite level
dynamically.  
This is no mystery, since the gauge symmetry is not a symmetry 
but simply redundancy of the description as was emphasized by
Seiberg~\cite{Seiberg} 
in the context of duality in the SUSY QCD.
Nevertheless, existence of the gauge invariance greatly simplifies
the physics as  
is the case in the SM. 
This is true even though the HLS model, based on the nonlinear sigma
model, is not renormalizable 
in contrast to the SM.
Actually, loop corrections are crucial issues for any theory of vector
mesons to become  
an EFT and this is precisely the place where the gauge invariance
comes into play.  
 
To study such loop effects of the HLS model as the EFT of QCD
extensively is 
the purpose of the present Physics Reports 
which may be regarded as a loop version to the previous one~\cite{BKY}.
We shall review, to the technical details, the physics of the 
loop calculations of HLS model developed
so far within a decade  in order to make the subject accessible to
a wider audience. Our results may also be applicable for the QCD-like
theories 
beyond the SM such as the technicolor and the composite $W/Z$ models.

Actually, in order that the vector meson theory be an EFT as a quantum
theory including loop corrections,
the gauge invariance in fact plays a vital role. 
It was first pointed out by 
Georgi~\cite{Georgi:1,Georgi:2} that the HLS makes possible   
the systematic loop expansion including the vector meson loops, 
particularly when the vector meson mass is light.
(Light vector mesons are actually realized in the Vector Manifestation
which will be fully discussed in this paper.)
The first one-loop calculation of HLS model was made by the present
authors in the Landau gauge~\cite{HY}
where the low energy theorem of HLS, 
the KSRF (I) relation, conjectured by
the tree-level arguments~\cite{BKY:NPB,BKY:PTP}, 
was confirmed at loop level.
Here we should mention~\cite{BKY:NPB} 
that being a gauge field the vector meson has a
definite off-shell extrapolation, which is crucial to discuss the low
energy theorem for the 
off-shell vector mesons at vanishing momentum.
Furthermore, a systematic loop expansion was precisely  
formulated in the same way as the usual chiral 
perturbation theory (ChPT)~\cite{Wei:79,Gas:84,Gas:85b}
by Tanabashi~\cite{Tanabashi} who then
gave an extensive analysis of the one-loop calculations in the
background field gauge. 
The low energy theorem of HLS was further proved
at any loop order in arbitrary covariant gauge by Kugo and the present 
authors~\cite{HKY:PRL,HKY:PTP}. 
Also finite temperature one-loop calculations of the HLS
was made in Landau gauge by Shibata and one of
the present authors
(M.H.)~\cite{Harada-Shibata}.

Here we note that there are actually many vector meson theories
consistent with the 
chiral symmetry such as the CCWZ matter field~\cite{CWZ,CCWZ}, the
Massive Yang-Mills field  
\cite{Schwinger:67,Schwinger:69,Wess-Zumino:67,Gas:69,Mei,Kay:85},
the tensor field method~\cite{Gas:84}:
They are {\it all 
 equivalent as far as the tree-level results are concerned} 
(see Sec.~\ref{ssec:ROMVM}).
However, as far as we know, the HLS model is the only theory which
makes the systematic derivative expansion possible. Since
these alternative models have no gauge symmetry at all, loop
calculations would run into 
trouble particularly in the limit of vanishing mass of the vector
mesons. 
    
More recently, new developments in the study of loop effects of the
HLS were made  
by the present authors~\cite{HY:letter,HY:matching,HY:VM,HY:VD}:
The key point was to include the quadratic divergence in the
Renormalization-Group Equation  
(RGE) analysis in the sense of Wilsonian RGE~\cite{Wilson-Kogut},
which was vital to the chiral phase transition triggered by
the HLS dynamics~\cite{HY:letter}: 
due to the quadratic running of $F_\pi^2$, 
the physical decay constant $F_\pi(0)$ (pole residue of the NG bosons) 
can be zero, even if 
the bare $F_\pi(\Lambda)$  defined at 
the cutoff $\Lambda$ (just a Lagrangian parameter)
is non-zero.
This phenomenon supports a view~\cite{HY:letter} that HLS is an
$\mbox{SU}(N_f)-$
``magnetic gauge theory'' 
dual (in the sense of Seiberg~\cite{Seiberg}) 
to the QCD as an $\mbox{SU}(N_c)$-``electric gauge theory'',
i.e., vector mesons 
are  ``Higgsed magnetic gluons'' dual to the 
``confined electric gluons'' of QCD:
The chiral restoration  
takes place independently
in both theories by their respective own dynamics for a
certain large number of {\it massless}
flavors $N_f$ ($N_c <N_f <11 N_c/2$), when both
$N_c$ and $N_f$ are regarded as large~\cite{HY:letter}. 
Actually, it was argued in various approaches that
the chiral restoration indeed takes place for the ``large 
$N_f$ QCD''~\cite{Banks-Zaks,%
Kogut-Sinclair,BCCDMSV:92,%
IKSY:92,IKSY:92b,IKSY:93,IKSY:94,IKKSY:96,IKKSY:98,%
Damgaard-Heller-Krasnitz-Olesen,%
Appelquist-Terning-Wijewardhana,%
Appelquist-Ratnaweera-Terning-Wijewardhana,%
Miransky-Yamawaki,%
Oehme-Zimmerman:1,Oehme-Zimmerman:2,Velkovsky-Shuryak}.
The chiral restoration implies that the QCD coupling
becomes not so strong as to give a chiral condensate 
and almost flat in the infrared region, 
reflecting the existence of an infrared fixed point (similarly to the one
explicitly observed in the two-loop perturbation) and thus 
the large $N_f$ QCD may be a dynamical model for the walking
technicolor~\cite{Holdom85,Yamawaki-Bando-Matumoto,Akiba-Yanagida,
Appelquist-Karabali-Wijewardhana,BMSY}.

One might wonder why the quadratic divergences are so vital to the
physics of the EFT,
since as far as we do not refer to the bare parameters as in the usual
renormalization where they are treated as free parameters, the
quadratic divergences are simply absorbed (renormalized) into the
redefinition 
(rescaling) of the $F_\pi^2$ 
no matter whatever value the bare $F_\pi^2$ may take.
However, the bare parameters of the EFT
are actually not free parameters but 
should be determined by matching with the underlying theory
at the cutoff scale where the EFT breaks down. This is precisely how 
the modern EFT based on 
the Wilsonian RGE/effective action~\cite{Wilson-Kogut}, obtained by 
integrating out the higher energy modes, necessarily contains
quadratic  
divergences as physical effects. 
In such a case
the quadratic divergence does exist as a physical effect as a matter
of principle, no matter whether it is a big or small effect.  
In fact, even in the SM, which is of course a renormalizable
theory and is usually analyzed without quadratic divergence for the
Higgs  
mass squared or $F_\pi$ (vacuum expectation value of the Higgs field)
renormalized into the observed value $\simeq 250 \, {\rm GeV}$, 
the quadratic divergence is actually  physical
when we regard the SM as an EFT of some more fundamental theory. In
the usual treatment without quadratic divergence, the bare
$F_\pi^2(\Lambda)$ 
is regarded as a free parameter and is freely tuned to be canceled
with the quadratic divergence of order $\Lambda^2$ to result in an
observed 
value $(250\, {\rm GeV})^2$, which is however 
an enormous fine-tuning if the cutoff
is physical (i.e., the SM is regarded as an EFT) and very big, 
say the Planck scale $10^{19}\, {\rm GeV}$, with the bare 
$F_\pi^2(\Lambda)$ 
tuned to an accuracy of order
$(250 \, {\rm GeV})^2/(10^{19}\, {\rm GeV})^2 \sim 10^{-33} \ll 1$. 
This is a famous naturalness problem, which, however, would not be a
problem  
at all if we simply ``renormalized out'' the quadratic divergence in
the SM. 
Actually, in the physics of phase transition such as in the lattice
calculation, 
Nambu-Jona-Lasinio (NJL) model, $CP^{N-1}$, etc., as well as the SM, 
bare parameters are precisely the   
parameters relevant to the phase transition and do have  a critical
value due to the quadratic/power divergence,  
which we shall explain in details in the text.   
In fact, even the usual nonlinear chiral Lagrangian can give rise to
the 
chiral symmetry restoration by the quadratic divergence of the $\pi$
loop~\cite{HY:letter,HY:VM}.
This is actually in accord with the lattice analysis that 
$O(4)$ nonlinear sigma model 
(equivalent to 
$\mbox{SU}(2)_{\rm L} \times \mbox{SU}(2)_{\rm R}$ nonlinear 
sigma model) give rise to the symmetry restoration for the hopping 
parameter (corresponding to our bare $F_\pi^2$) larger than a certain
critical value. 

The inclusion of the quadratic divergence is even more important for
the phenomenological analyses when the bare HLS theory defined at the
cutoff scale $\Lambda$ 
is matched with the
underlying QCD for the 
Operator Product Expansion (OPE) of the current correlators
(``Wilsonian matching'')~\cite{HY:matching}. 
Most notable feature of the Wilsonian matching is to
provide the HLS theory with the otherwise unknown information of the
underlying QCD such as the precise $N_c$-dependence which is explicitly
given through the OPE.
By this matching we actually determine the bare parameters of 
the HLS model,  and hence the quadratic divergences become really physical.
Most notably the bare $F_\pi(\Lambda)$ is given by
\begin{equation}
F_\pi^2(\Lambda)\simeq 2(1+\delta_A)\, \left(\frac{N_c}{3}\right)
\left(\frac{\Lambda}{4\pi}\right)^2 \, ,
\label{Fpilambdaint}
\end{equation}
where $\delta_A$ ($\sim 0.5$ for $N_f=3$) stands for the OPE corrections to the term $1$ (free quark 
loop).
For $N_c=N_f=3$ we choose 
\begin{equation}
\Lambda \simeq 1.1 \, \mbox{GeV} \ ,
\end{equation} 
an optimal value for the descriptions of both the QCD and the HLS to
be valid 
and the Wilsonian matching to make sense, which coincides with the
naive  
dimensional analysis (NDA)~\cite{Man:84}\footnote{
  The NDA does not hold for other
  than $N_c=N_f=3$, in particular, 
  near the chiral restoration point $N_f \sim N_f^{\rm crit}$ 
  with $F_\pi(0) \rightarrow 0$ while $\Lambda$ 
  remaining almost unchanged.
  For the general case other than $N_c=N_f=3$ we actually fix
  $\Lambda$  
  as $\frac{N_c}{3} \alpha_s (\Lambda_{N_c,N_f})= 
  \alpha_s (\Lambda_{3,3})|_{N_c=N_f=3}\sim 0.7$, 
  with $\Lambda_{3,3}=1.1 \, \mbox{GeV}$,
  where $\alpha_s(\mu)$ is the one-loop QCD running coupling.
  See Sec.~\ref{sssec:NDP}.
},
$\Lambda \sim 4\pi F_\pi(0)$, 
where 
\begin{equation}
F_\pi(0) =86.4 \pm 9.7\, \mbox{MeV}
\end{equation}
(the ``physical value'' in the chiral limit 
$m_u=m_d=m_s=0$)\footnote{%
  This value is determined from the ratio 
  $F_{\pi,{\rm phys}}/F_\pi(0)=1.07\pm 0.12$
  given in
  Ref.~\cite{Gas:85b}, where  $F_{\pi,{\rm phys}}$ is 
  the physical pion decay constant,
  $F_{\pi,{\rm phys}} = 92.42\pm0.26\,\mbox{MeV}$~\cite{PDG:02},
  and $F_\pi(0)$ the one at the chiral limit $m_u=m_d=m_s=0$.
  This should be distinguished from the popular ``chiral limit value''
  $88 \, \mbox{MeV}$~\cite{Gas:84} 
  which was obtained for $m_\pi^2=0$ while $m_K^2\ne 0$ kept to
  be the physical value. 
}.
Then we have $F_\pi^2(\Lambda) \sim 3\, 
(\frac{\Lambda}{4\pi})^2 \sim 3 \, (86.4 \, {\rm MeV})^2$. 
Were it not for quadratic divergence, we would have predicted 
$F_\pi^2(0) \sim F_\pi^2(\Lambda) \sim 3\, (86.4 \, {\rm MeV})^2$, three times 
larger than the reality.
It is essentially the quadratic divergence that pulls
$F_\pi^2$ down to the physical value $F_\pi^2(0) 
\sim \frac{1}{3}F_\pi^2(\Lambda) \sim (86.4 \, {\rm MeV})^2$.
As to other physical quantities, the predicted values 
through the RGEs in the case of $N_c=N_f=3$ are
in remarkable agreement with the experiments~\cite{HY:matching}.
It should be noted that without quadratic divergence the matching
between HLS and QCD would simply break down 
and without 
vector mesons
even the Wilsonian matching including the quadratic
divergences would break down.

When the chiral symmetry is restored in the underlying QCD with
$\langle \bar q q \rangle =0$, this Wilsonian matching determines the
bare  
parameters as 
$a(\Lambda)=1$, $g(\Lambda)=0$ and $F_\pi^2(\Lambda)\simeq
2.5 \frac{N_c}{3} (\frac{\Lambda}{4\pi})^2 \ne 0$
($\delta_A \simeq 0.25$ for $\langle  \bar q q \rangle =0$),
which we call ``VM conditions'' after the ``Vector Manifestation
(VM)'' to be followed by these conditions. The VM conditions
coincide with the Georgi's vector limit~\cite{Georgi:1,Georgi:2},
which, however, in contrast to the ``vector realization'' proposed in  
Ref.~\cite{Georgi:1,Georgi:2} with $F_\pi^2(0) \ne 0$,  
lead us to a novel pattern of the chiral symmetry restoration,
the VM~\cite{HY:VM} with $F_\pi^2(0) \rightarrow 0$. The VM 
is a Wigner realization accompanying massless degenerate 
(longitudinal component of) $\rho$
meson (and its flavor partners), 
generically denoted as $\rho$, and the pion (and its flavor partners),
generically denoted as $\pi$, 
as the chiral partners~\cite{HY:VM}:
\begin{equation}
m_\rho^2 \rightarrow 0=m_\pi^2 \ , 
\quad F_\pi^2(0) \rightarrow 0 \ , 
\end{equation}
with $m_\rho^2/F_\pi^2(0) \rightarrow 0$ near the critical point.
The chiral restoration in the
large $N_f$ QCD can  actually be identified  
with the VM.
An estimate of the critical $N_f$ of the chiral restoration 
is given by a precise cancellation between the bare $F_\pi^2(\Lambda)$ and the
quadratic divergence $\frac{N_f}{2}\frac{\Lambda^2}{(4\pi)^2}$:  
\begin{equation}
0=F_\pi^2(0) = 
 F_\pi^2(\Lambda) - \frac{N_f}{2} \frac{\Lambda^2}{(4\pi)^2}
             \simeq \left(2.5 \frac{N_c}{3} - \frac{N_f}{2}\right) 
             \left(\frac{\Lambda}{4 \pi}\right)^2
             \quad,
\end{equation}
which yields 
\begin{equation}
N_f^{\rm crit} \sim  5 \frac{N_c}{3}
\end{equation}
in rough agreement with the recent lattice 
 simulation~\cite{IKSY:92,IKSY:92b,IKSY:93,IKSY:94,IKKSY:96,IKKSY:98},
$6 < N_f^{\rm crit} < 7 \, (N_c=3)$ 
but in disagreement with
that predicted by the 
(improved) ladder Schwinger-Dyson equation with the two-loop
running coupling~\cite{Appelquist-Terning-Wijewardhana}, 
$N_f^{\rm crit}
\sim 12\frac{N_c}{3}$.
Further investigation of the phase structure
of the HLS model in a full parameter space
leads to an amazing fact that Vector Dominance (VD) is no longer a
sacred discipline of the 
hadron physics but rather an accidental phenomenon realized only for
the realistic world of 
the $N_c=N_f=3$ QCD~\cite{HY:VD}: In particular, at the VM critical point
the VD is badly violated.

Quite recently,
it was found by Sasaki and one of the present authors
(M.H.)~\cite{Harada-Sasaki} 
that the VM can really take place for the chiral symmetry restoration
for the finite temperature QCD. 
Namely, the vector meson mass vanishes near the chiral restoration
temperature in accord with the 
picture of Brown and Rho~\cite{Brown-Rho:91,Brown-Rho:96,%
Brown-Rho:01a,Brown-Rho:01b}, which is in sharp contrast
to the conventional chiral restoration \`a la
linear sigma model where the
scalar meson mass vanishes near the critical temperature.

In view of these we do believe that the HLS at loop level opened a
window to a new era of 
the effective field theory of QCD and QCD-like theories beyond the SM.

Some technical comments are in order:

In this report we confine ourselves to the chiral symmetric limit
unless otherwise mentioned, 
so that pseudoscalar mesons are all precisely  massless NG bosons.

Throughout this report
we do not include the axialvector meson ($a_1$ meson and their 
flavor partners), denoted generically by $A_1$,
since our cutoff scale $\Lambda$ is taken as 
$\Lambda 
\simeq 1.1 \, {\rm GeV}$ 
for the case $N_f=3$,
an optimal value where both the derivative expansion in HLS and the
OPE in the underlying QCD make sense. 
Such a cutoff is lower than the $a_1$ meson mass and hence the
axialvector mesons are decoupled 
at least for $N_f=3$. If, by any chance, 
the axialvector mesons are to become
lighter than the cutoff near the phase transition point, our 
effective theory analysis should be modified, based on the generalized
HLS Lagrangian 
having $G_{\rm global}\times G_{\rm local}$ 
symmetry~\cite{BKY:NPB,Bando-Fujiwara-Yamawaki}.

We also omit the scalar mesons which may be lighter than the 
cutoff scale~\cite{Harada-Sannino-Schechter:PRD,
Harada-Sannino-Schechter:PRL,Tornqvist-Roos,IITITT:96,%
Morgan-Pennington,Janssen-Pearce-Holinde-Speth},
since it does not contribute to the two-point functions (current
correlators) which we are studying 
and hence irrelevant to our analysis in this report.

In this respect we note that in the HLS perturbation theory there are
many counter terms  
(actually 35 for $N_f \ge 4$)~\cite{Tanabashi} compared with the usual
ChPT  
($10 +2+1=13$)~\cite{Gas:84,Gas:85b}
but only few of them are relevant to the two point function (current
correlators) and hence our 
loop calculations are reasonably tractable.

It is believed according to the NDA~\cite{Man:84}  
that  the usual ChPT (without quadratic divergence) breaks down at the
scale 
$\Lambda$ such that the loop correction is small: 
\begin{equation}
\frac{p^2}{(4 \pi F_\pi(0))^2}<
\frac{\Lambda^2}{(4 \pi F_\pi(0))^2} \sim 1 \quad ({\rm NDA}) \, .
\end{equation}
However the loop corrections 
generally have
an additional factor $N_f$, i.e., $N_f p^2 /(4\pi F_\pi(0))^2$
and hence when $N_f$ is crucial, we cannot ignore the factor $N_f$.
Then we should change the NDA to:~\cite{SS,CDGS}
\begin{equation}
\Lambda \sim \frac{4 \pi F_\pi(0)}{\sqrt{N_f}}\, , 
\end{equation}
which yields even for $N_f=3$ case a somewhat smaller value 
$\Lambda \sim 4 \pi F_\pi(0)/\sqrt{3} \sim m_\rho< 1.1 \, \mbox{GeV}$. 
This is reasonable since
the appearance of $\rho$ pole invalidates the ChPT anyway. 
This is another reason why we should 
include $\rho$ in
order to extend the theory to the higher scale 
$\Lambda \sim 1.1\,  {\rm GeV}$
where both the QCD (OPE) and the EFT (derivative expansion) make sense
and 
so does the matching between them. 
Now, the inclusion of quadratic divergence implies that the loop
corrections are given in terms of $F_\pi(\Lambda)$ instead of 
$F_\pi(0)$ and 
hence we further change the NDA to:
\begin{equation}
\Lambda \sim \frac{4 \pi F_\pi(\Lambda)}{\sqrt{N_f}} 
\, , 
\end{equation}
which is now consistent with the setting $\Lambda \sim 1.1 {\rm GeV}$, 
since $F_\pi(\Lambda) \sim \sqrt{3} F_\pi(0)$ for $N_f=3$
as we mentioned earlier.
As to the quadratic divergence for $F_\pi^2$ in the HLS model, 
the loop contributions get an
extra factor $1/2$ due to the additional $\rho$ loop, 
$\frac{N_f}{2} p^2 /(4\pi F_\pi(\Lambda))^2$, and hence 
the loop expansion would be valid up till 
\begin{equation}
\Lambda \sim \frac{4 \pi F_\pi(\Lambda)}{\sqrt{\frac{N_f}{2}}} 
\, ,
\end{equation}
which is actually the scale (or $N_f$ when $\Lambda$ is fixed) 
where the bare $F_\pi^2(\Lambda)$ is
completely balanced by the quadratic divergence 
to yield the chiral restoration $F_\pi^2(0)=0$.
Hence the region of the validity of the expansion is 
\begin{equation}
\frac{\frac{N_f}{2} \Lambda^2}{(4\pi F_\pi(\Lambda))^2} 
\sim \frac{N_f}{2N_c}
 <1 \, ,
\end{equation}
where $F_\pi^2(\Lambda)$ was estimated by Eq.~(\ref{Fpilambdaint})
with $\delta_A \sim 0.5$. This is satified in the large $N_c$ limit
$N_f/N_c \ll 1$, which then can be extrapolated over to the critical
region $N_f \sim 2 N_c$.  
Details will be given in the text.

This paper is organized as follows:

In Sec.~\ref{sec:BRCPT}
we briefly review the (usual) chiral perturbation  
theory (ChPT)~\cite{Wei:79,Gas:84,Gas:85b} (without vector mesons),
which gives the systematic low energy expansion of Green functions
of QCD related to light pseudoscalar mesons.

In Sec.~\ref{sec:HLS}
we give an up-to-date
review of the model based on the HLS~\cite{BKUYY,BKY}
at tree level.
Following Ref.~\cite{BKY}
we briefly explain some essential ingredients of the HLS
in Secs.~\ref{ssec:GHM}--\ref{ssec:PPLO}.
In Sec.~\ref{ssec:VMSLEC} we give a relation of the HLS to the
ChPT at tree level.
Section~\ref{ssec:ROMVM} is devoted to study the relation of the
HLS to other models of vector mesons:
the vector meson is introduced as the 
matter field in the 
CCWZ Lagrangian~\cite{CWZ,CCWZ} (the matter field method);
the massive Yang-Mills field
method~\cite{Schwinger:67,Schwinger:69,
Wess-Zumino:67,Gas:69,KRS,Mei};
and the anti-symmetric tensor field method~\cite{Gas:84,Eck:89a}.
There we show the equivalence of these models to the HLS model.
In Sec.~\ref{ssec:AP}, following 
Refs.~\cite{FKTUY} and \cite{BKY}, we briefly review
the way of incorporating vector mesons
into anomalous processes,
and then perform analyses on several physical processes using
up-to-date experimental data.

In Sec.~\ref{sec:CPHLS}
we review the chiral perturbation theory with HLS.
First we show that, {\it thanks to the gauge invariance} of the HLS,
we can perform the 
{\it systematic derivative expansion with including
vector mesons} in addition to the pseudoscalar Nambu-Goldstone 
bosons in Sec.~\ref{ssec:DEHLS}.
The Lagrangians of ${\cal O}(p^2)$ and ${\cal O}(p^4)$ are given in
Secs.~\ref{ssec:OP2L} and \ref{ssec:OP4L}.
In Sec.~\ref{ssec:BGFM} we introduce 
the background field gauge to calculate the one-loop corrections.
Since the effect of quadratic divergences are important in this
report, we explain the meaning of the quadratic
divergence in our approach in Sec.~\ref{ssec:QD}.
The explicit calculations of the two-point functions in the background
field gauge are performed in Sec.~\ref{ssec:TPFOL}.
The low energy theorem (KSRF (I)) at one-loop level is studied in
Sec.~\ref{ssec:LETOL} in the framework of
the background field
gauge, and the renormalization group equations for the relevant
parameters are given in Sec.~\ref{ssec:RGEWS}.
In Sec.~\ref{ssec:MHC} we show
some examples of the relations 
between the
parameters of the HLS and the ${\cal O}(p^4)$ ChPT parameters
following Ref.~\cite{Tanabashi}.
Finally in Sec.~\ref{ssec:PSH} we study the phase structure of
the HLS following Ref.~\cite{HY:VD}.

Section~\ref{sec:WM}
is devoted to review the ``Wilsonian matching'' proposed in
Ref.~\cite{HY:matching}.
First, we introduce the ``Wilsonian matching conditions''
in Sec.~\ref{ssec:MHUQ}.
Then, we determine the bare parameters of the HLS using those
conditions in Sec.~\ref{ssec:DBPHL} and
make several physical predictions in Sec.~\ref{ssec:RWM}.
In Sec.~\ref{ssec:PQNf2}
we consider QCD with $N_f=2$
to show how the $N_f$-dependences of the physical quantities
appear.
Finally, in Sec.~\ref{ssec:SR},
we study the spectral function sum rules 
related to the vector and axialvector current
correlators.

In Sec.~\ref{sec:VM}
we review ``Vector Manifestation'' (VM) of the
chiral symmetry proposed in Ref.~\cite{HY:VM}.
We first explain the VM and show that it is needed when
we match the HLS with QCD at the chiral restoration point
in Sec.~\ref{ssec:VMCSR}. Detailed characterization is
also given there.
Then, in Sec.~\ref{ssec:CPTLNQ} we review
the chiral restoration in the large $N_f$ QCD and discuss 
in Sec.~\ref{ssec:VMLNQ} that VM is
in fact be realized in the 
the chiral restoration of the large $N_f$ QCD. Seiberg-type duality is
discussed in Sec.~\ref{ssec:STD}.

In Sec.~\ref{sec:RALOLET}
we give a brief review of the proof of the low energy theorem 
in Eq.~(\ref{KSRF I; intro}) at any loop order,
following Refs.~\cite{HKY:PRL,HKY:PTP}. We also show that the proof is 
intact even when including the quadratic divergences.

In Sec.~\ref{sec:THDMC}
we discuss the application of the chiral perturbation with HLS to the
hot and/or dense matter calculations.
Following Ref.~\cite{Harada-Shibata}
we first review the calculation of the hadronic thermal corrections
from $\pi$- and $\rho$-loops in 
Sec.~\ref{ssec:HTE}.
In Sec.~\ref{ssec:VMNT}
following Ref.~\cite{Harada-Sasaki}
we review the application of the present approach to the hot matter
calculation,
and in Sec.~\ref{ssec:Admc} we briefly review the application
to the dense matter calculation following Ref.~\cite{Harada-Kim-Rho}.

Finally, in Sec.~\ref{sec:SD}
we give summary and discussions.

We summarize
convenient formulae and Feynman rules used in this
paper in Appendices~\ref{sec:CF}, \ref{app:FR} and \ref{app:FRLG}.
A complete list of the divergent corrections to the ${\cal O}(p^4)$
terms is shown in Appendix~\ref{app:RHKE}.

\newpage

\section{A Brief Review of the Chiral Perturbation Theory}
\label{sec:BRCPT}

In this section we briefly review the Chiral Perturbation 
Theory (ChPT)~\cite{Wei:79,Gas:84,Gas:85b},
which gives the systematic low-energy expansion of Green functions
of QCD related to light pseudoscalar mesons.
The Lagrangian is constructed via non-linear realization of
the chiral symmetry based on the manifold
$\mbox{SU}(N_f)_{\rm L}
\times\mbox{SU}(N_f)_{\rm R} /\mbox{SU}(N_f)_{\rm V}$,
with $N_f$ being the number of light flavors.
Here we generically use $\pi$ for the pseudoscalar NG bosons
(pions and their flavor partners)
even for $N_f\neq2$.
For physical pions, on the other hand, we write their charges
explicitly as $\pi^{\pm}$ and $\pi^0$.

In Sec.~\ref{ssec:GFQ}
we give a conceptual relation between the generating functional of QCD
and that of the ChPT following Ref.~\cite{Gas:84,Gas:85b}.
Then, after introducing the derivative expansion in 
Sec.~\ref{ssec:DE}, we review how to perform the order counting
systematically in the ChPT in Sec.~\ref{ssec:OC}.
The Lagrangian of the ChPT up until ${\cal O}(p^4)$ is given in
Sec.~\ref{ssec:L}.
We review the renormalization and the values
of the coefficients of the 
${\cal O}(p^4)$ terms in Secs.~\ref{ssec:R} and
\ref{ssec:VLEC}.
The particle assignment in the realistic case of $N_f=3$ is shown in
Sec.~\ref{ssec:PA2}.
Finally, we review the applications of the ChPT to 
physical quantities such as the vector form factors of the pseudoscalar
mesons (Sec.~\ref{ssec:l9}) and 
$\pi\rightarrow e\nu \gamma$ amplitude (Sec.~\ref{ssec:l10}).

\subsection{Generating functional of QCD}
\label{ssec:GFQ}

Let us start with the QCD Lagrangian with external source fields:
\begin{equation}
{\cal L}_{\rm QCD} = {\cal L}_{\rm QCD}^0 + 
\overline{q}_L \gamma^\mu {\cal L}_\mu q_L
+ \overline{q}_R \gamma^\mu {\cal R}_\mu q_R
+ \overline{q}_L \left[ {\cal S} + i {\cal P} \right] q_R
+ \overline{q}_R \left[ {\cal S} - i {\cal P} \right] q_L
\ ,
\label{QCD Lag}
\end{equation}
where ${\cal L}_\mu$ and ${\cal R}_\mu$ are external gauge fields
corresponding to $\mbox{SU}(N_f)_{\rm L}$ and 
$\mbox{SU}(N_f)_{\rm R}$, and
${\cal S}$ and ${\cal P}$ are external scalar and pseudoscalar source
fields. 
${\cal L}_{\rm QCD}^0$ is the ordinary QCD Lagrangian with $N_f$
massless quarks:
\begin{equation}
{\cal L}_{\rm QCD}^0 =
\bar{q} i { D \kern-9pt\mbox{\it/} } {\kern 4pt} q
- \frac{1}{2} \mbox{tr} 
\left[ G_{\mu\nu} G^{\mu\nu} \right]
\ ,
\end{equation}
where
\begin{eqnarray}
&&
D_\mu q = \left( \partial_\mu - i g_s G_\mu \right) q \ ,
\nonumber\\
&&
G_{\mu\nu} = \partial_\mu G_\mu - \partial_\nu G_\mu
- i g_s \left[ G_\mu \,,\, G_\nu \right]
\ ,
\end{eqnarray}
with $G_\mu$ and $g_s$ being the gluon field matrix and the QCD gauge
coupling constant.

Transformation properties of the external gauge fields ${\cal L}$
and ${\cal R}$ are given by
\begin{eqnarray}
&&
{\cal L}_\mu \rightarrow 
g_{\rm L} {\cal L}_\mu g_{\rm L}^\dag 
- i \partial_\mu g_{\rm L} \cdot g_{\rm L}^\dag
\ , 
\nonumber\\
&&
{\cal R}_\mu \rightarrow 
g_{\rm R} {\cal R}_\mu g_{\rm R}^\dag 
- i \partial_\mu g_{\rm R} \cdot g_{\rm R}^\dag
\ , 
\label{external gauges}
\end{eqnarray}
where
$g_{\rm L}$ and $g_{\rm R}$ are the elements of the
left- and right-chiral
transformations: $g_{\rm L,R} \in 
\mbox{SU}(N_f)_{\rm L,R}$.
Scalar and pseudoscalar external source fields
${\cal S}$ and ${\cal P}$ transform as
\begin{equation}
\left( {\cal S} + i {\cal P} \right) \rightarrow
g_{\rm L} \left( {\cal S} + i {\cal P} \right) g_{\rm R}^\dag \ .
\end{equation}
If there is an explicit chiral symmetry breaking due to the current
quark mass, it is introduced as the vacuum expectation value (VEV) of
the external scalar source field:
\begin{equation}
\langle {\cal S} \rangle = {\cal M} =
\left( \begin{array}{ccc}
m_1 & & \\
 & \ddots & \\
 & & m_{N_f} \\
\end{array} \right) \ .
\label{QMM:Nf}
\end{equation}
In the realistic case $N_f=3$ this reads
\begin{equation}
{\cal M} = 
\left( \begin{array}{ccc}
m_u & & \\
& m_d & \\
& & m_s \\
\end{array} \right)
\ .
\label{QMM:3}
\end{equation}

Green functions associated with vector and axialvector currents,
and scalar and pseudoscalar densities are generated by the functional
of the above source fields ${\cal L}_\mu$, ${\cal R}_\mu$, ${\cal S}$
and ${\cal P}$: 
\begin{equation}
\exp \left( i W[{\cal L}_\mu, {\cal R}_\mu, {\cal S}, {\cal P}] 
\right)
= 
\int [d q][d \overline{q}] [d G] \exp \left( i \int d^4x
{\cal L}_{\rm QCD} \right) \ .
\end{equation}
The basic concept of the ChPT is
that the most
general Lagrangian of NG bosons and external sources,
which is consistent with the chiral symmetry, can reproduce
this generating functional in the low energy region:
\begin{equation}
\exp \left( i W[{\cal L}_\mu, {\cal R}_\mu, {\cal S}, {\cal P}] 
\right)
= 
\int [d U] \exp \left( i \int d^4x {\cal L}_{\rm eff}
\left[ U, {\cal L}_\mu,
{\cal R}_\mu, {\cal S}, {\cal P} \right] \right)
\ ,
\label{Gfun}
\end{equation}
where $N_f \times N_f$ special-unitary matrix
$U$ includes the $N_f^2-1$ NG-boson fields.
In this report, for definiteness, we use
\begin{equation}
U = e^{2i\pi/F_\pi} \ , \quad
\pi = \pi_a T_a \ ,
\end{equation}
where $F_\pi$ is the decay constant of the NG bosons $\pi$.
Transformation property of this $U$ under the chiral symmetry is given
by
\begin{equation}
U \rightarrow g_{\rm L} \, U \, g_{\rm R}^\dag \ .
\end{equation}

It should be noticed that the above effective Lagrangian 
generally includes
infinite number of terms with unknown coefficients.
Then, strictly speaking, we cannot say that the above generating
functional agrees with that of QCD before those coefficients are
determined.
Since the above generating functional is the most general one
consistent with the chiral symmetry,
it includes that of QCD.
As one can see easily, the above generating functional has no
practical use if there is no way to control the infinite number of
terms.
This can be done in the low energy region based on the
derivative expansion.

\subsection{Derivative expansion}
\label{ssec:DE}

We are now interested in the phenomenology of pseudoscalar mesons in
the energy region around the mass of $\pi$, 
$p \sim m_\pi$.
On the other hand,
the chiral symmetry breaking scale $\Lambda_\chi$
is estimated as~\cite{Man:84}
\begin{equation}
\Lambda_{\chi} \, \sim \, 
4 \pi F_\pi \, \sim \, 1.1 \, \mbox{GeV} ,
\label{lam chi}
\end{equation}
where we used $F_\pi = 88$\,MeV estimated in the chiral
limit~\cite{Gas:84}.
Since $\Lambda_{\chi}$ is much
larger than $\pi$ mass scale, $m_\pi \ll \Lambda_\chi$,
we can expand the generating functional in Eq.~(\ref{Gfun})
in terms of 
\begin{equation}
\frac{p}{\Lambda_\chi} \qquad \mbox{or} \qquad
\frac{m_\pi}{\Lambda_\chi} 
\ .
\end{equation}
As is well known as Gell-Mann--Oakes--Renner relation~\cite{Gel:68},
existence of the approximate chiral symmetry implies 
\begin{equation}
m_\pi^2 \sim {\cal M} \Lambda_\chi \ .
\label{GMOR rel}
\end{equation}
So one can expand the effective Lagrangian in terms of the derivative
and quark masses by assigning 
\begin{eqnarray}
{\cal M} &\sim& {\cal O}(p^2) \ , \nonumber\\
\partial &\sim& {\cal O}(p) \ . \nonumber
\end{eqnarray}

\subsection{Order Counting}
\label{ssec:OC}

One can show that the low energy expansion discussed in the previous
subsection corresponds to the loop expansion based on the effective
Lagrangian.
Following Ref.~\cite{Wei:79},
we here demonstrate this correspondence by using the scattering
matrix elements of $\pi$.

Let us consider the matrix element with $N_e$ external $\pi$ lines.
The dimension of the matrix element is given by
\begin{equation}
D_1 \equiv \mbox{dim} ( M ) = 4 - N_e \ .
\end{equation}
The form of an interaction with $d$ derivatives, $k$ $\pi$ fields
and $j$ quark mass matrices
is symbolically expressed as
\begin{equation}
g_{d,j,k} (m_\pi^2)^j (\partial)^d (\pi)^k  \ ,
\end{equation}
where
\begin{equation}
\mbox{dim} ( g_{d,j,k} ) = 4 - d - 2 j - k \ .
\end{equation}
Let $\bar{N}_{d,j,k}$ denote the number of the above interaction
included in a diagram for $M$.  
Then the total dimension carried by coupling constants
is given by
\begin{equation}
D_2 = \sum_d \sum_j \sum_k \bar{N}_{d,j,k} ( 4 - d - 2 j - k) \ .
\end{equation}
One can easily show
\begin{equation}
\sum_k \bar{N}_{d,j,k} k  = 2 N_i  + N_e \ ,
\end{equation}
where $N_i$ is the total number of internal $\pi$ lines.
By writing
\begin{equation}
N_{d,j} \equiv \sum_k \bar{N}_{d,j,k} \ ,
\end{equation}
$D_2$ becomes
\begin{equation}
D_2 = \sum_d \sum_j N_{d,j} ( 4 - d - 2 j ) - 2 N_i - N_e \ .
\end{equation}
By noting that the number of loops, $N_L$, is related to $N_i$ and
$N_{d,j}$ by
\begin{equation}
N_L = N_i - \sum_d \sum_j N_{d,j} + 1 \ ,
\end{equation}
$D_2$ becomes
\begin{equation}
D_2 = 2 - 2 N_L + N_e + \sum_d \sum_j N_{d,j} ( 2 - d - 2 j ) \ .
\end{equation}
The matrix element can be generally
expressed as
\begin{equation}
M = E^D m_\pi^{D_3} f\left( E/\mu ,\, M_\pi/\mu \,\right) \ ,
\label{M D D3}
\end{equation}
where $\mu$ is a common renormalization scale and $E$ is a common
energy scale.
The value of $D_3$ is determined by counting the number of vertices
with $m_\pi$;
\begin{equation}
D_3 = \sum_{d,j} N_{d,j} (2 j) \ .
\end{equation}
$D$ is given by subtracting the dimensions carried by the coupling
constants and $m_\pi$ from the total dimension of the
matrix element $M$:
\begin{equation}
D = D_1 - D_2 - D_3 = 
2 + \sum_{d,j} N_{d,j} ( d - 2 ) + 2 N_L \ .
\end{equation}
As we explained in the previous subsection,
the derivative expansion is performed in the low energy region around
the $\pi$ mass scale: The common energy scale is on the order of
the $\pi$ mass, $E \sim m_\pi$, and both $E$ and $m_\pi$ are much
smaller than the chiral symmetry breaking scale $\Lambda_\chi$,
i.e.,
$E$, $m_\pi \ll \Lambda_\chi$.
Then. the order of the matrix element $M$ in the derivative expansion,
denoted by $\bar{D}$,
is determined by counting the dimension
of $E$ and $m_\pi$ appearing in $M$:
\begin{equation}
\bar{D} = D + D_3 = 2 + \sum_{d,j} N_{d,j} ( d + 2 j - 2 )
+ 2 N_L \ .
\end{equation}
Note that $N_{2,0}$ and $N_{0,1}$ can be any number: these do not
contribute to $\bar{D}$ at all.

We can classify the diagrams contributing to the matrix element $M$
according to the value of the above $\bar{D}$.
Let us list examples for $\bar{D} = 2$ and $4$.
\begin{enumerate}
\item $\bar{D} = 2$ \\
This is the lowest order.
In this case, $N_L=0$: There are no loop contributions.
The leading order diagrams are tree diagrams in which the vertices are
described by the two types of terms: $(d,\,j)=(2,0)$ or 
$(d,\,j)=(0,1)$.
Note that $(d,\,j)=(2,0)$ term includes $\pi$ kinetic term, and
$(d,\,j)=(0,1)$ term includes $\pi$ mass term.

\item $\bar{D} = 4$

\begin{enumerate}
\item $N_L=1$ and $N_{d,j}=0$ [$(d,j) \neq (2,0)$, $(0,1)$]\\
One loop diagrams in which all the vertices are of leading order.

\item $N_L=0$

\renewcommand{\theenumiii}{(\roman{enumiii})}
\begin{enumerate}
\item $N_{4,0}=1$, $N_{d,j}=0$ 
[$(d,j) \neq (4,0)$, $(2,0)$, $(0,1)$]

\item $N_{2,1}=1$, other $N_{d,j}=0$ 
[$(d,j) \neq (2,1)$, $(2,0)$, $(0,1)$]

\item $N_{0,2}=1$, other $N_{d,j}=0$ 
[$(d,j) \neq (0,2)$, $(2,0)$, $(0,1)$]
\end{enumerate}

Tree diagrams in which only one next order vertex is included.
The next order vertices are described by
$(d,\,j)=(4,0)$, $(2,1)$ and $(0,2)$.

\end{enumerate}

\end{enumerate}

It should be noticed that we 
included only logarithmic divergences in the above arguments.
When we include quadratic divergences using,
e.g., a method in Sec.~\ref{ssec:QD},
loop integrals 
generate the terms proportional to the cutoff which
are renormalized by the dimensionful coupling constants.

\subsection{Lagrangian}
\label{ssec:L}

One can construct the most general form of the Lagrangian order by
order in the derivative expansion consistently with the chiral
symmetry.
Below we summarize
the building blocks together with the orders in the derivative
expansion and the transformation properties under the chiral symmetry:
\begin{eqnarray}
U \ , \quad  & {\cal O}(1) \ , 
\quad& U \rightarrow g_{\rm L} U g_{\rm R}^\dag \ ,
\nonumber\\ 
\chi   \ , \quad &
{\cal O}(p^2)  \ , \quad 
& \chi \rightarrow g_{\rm L} \chi g_{\rm R}^\dag \ , \nonumber\\
&&
\chi \equiv 2 B ( {\cal S} + i {\cal P} ) \ , 
\nonumber\\
{\cal L}_\mu  \ , \quad & {\cal O}(p)  \ , \quad & 
{\cal L}_\mu \rightarrow 
g_{\rm L} {\cal L}_\mu g_{\rm L}^\dag 
- i \partial_\mu g_{\rm L} \cdot g_{\rm L}^\dag
\ ,
\nonumber\\
{\cal R}_\mu  \ , \quad & {\cal O}(p)  \ , \quad & 
{\cal R}_\mu \rightarrow 
g_{\rm R} {\cal R}_\mu g_{\rm R}^\dag  
- i \partial_\mu g_{\rm R} \cdot g_{\rm R}^\dag
\ ,
\label{trans prop ChPT}
\end{eqnarray}
where $B$ is a quantity of order $\Lambda_\chi$.
Here orders of ${\cal L}_\mu$ and ${\cal R}_\mu$ are
determined by requiring that
all terms of the covariant derivative of $U$
have the same chiral order:
\begin{equation}
\nabla_\mu U  = \partial_\mu U - i {\cal L}_\mu U
+ i U {\cal R}_\mu \ .
\label{covdel ChPT}
\end{equation}
To construct the effective Lagrangian we need to
use the fact that QCD
does not break the parity as well as the charge conjugation, and
require that
the effective Lagrangian is invariant under the 
transformations under the parity ($\hbox{\boldmath$P$}$) 
and the charge conjugation ($\hbox{\boldmath$C$}$): 
\begin{eqnarray}
U \ \ 
  &\displaystyle\mathop{\longleftrightarrow}%
    _{\hbox{\boldmath$\scriptstyle P$}}& \ \ 
  U^\dag \ , \nonumber\\
\chi \ \ 
  &\displaystyle\mathop{\longleftrightarrow}%
    _{\hbox{\boldmath$\scriptstyle P$}}& \ \ 
  \chi^\dag \ , \nonumber\\
{\cal L}_\mu \ \ 
  &\displaystyle\mathop{\longleftrightarrow}%
    _{\hbox{\boldmath$\scriptstyle P$}}&
  \ \ {\cal R}_\mu \ ,
  \nonumber\\
U \ \ &\displaystyle\mathop{\longrightarrow}%
    _{\hbox{\boldmath$\scriptstyle C$}}&
  \ \ U^T \ , \nonumber\\
\chi \ \ 
  &\displaystyle\mathop{\longleftrightarrow}%
    _{\hbox{\boldmath$\scriptstyle C$}}& \ \ 
  \chi^T \ , \nonumber\\
{\cal L}_\mu \ \ 
  &\displaystyle\mathop{\longleftrightarrow}%
    _{\hbox{\boldmath$\scriptstyle C$}}& \ \ 
  - \left( {\cal R}_\mu \right)^T \ ,
\end{eqnarray}
where the superscript $T$ implies the transposition of the matrix.

The leading order Lagrangian is constructed from the
terms of ${\cal O}(p^2)$ ($\bar{D}=2$ in the previous subsection)
which have the structures of $(d,j) = (2,0)$ or 
$(0,1)$:~\cite{Gas:84,Gas:85a}
\begin{equation}
{\cal L}_{(2)}^{\rm ChPT} =
\frac{F_\pi^2}{4} \mbox{tr}
\left[ \nabla_\mu U^\dag \nabla^\mu U \right]
+ \frac{F_\pi^2}{4} \mbox{tr}
\left[ \chi U^\dag + \chi^\dag U \right ]
\ .
\label{leading ChPT}
\end{equation}
This leading order Lagrangian leads to the equation of motion for 
$U$ up to ${\cal O}(p^4)$:
\begin{equation}
\nabla^\mu \nabla_\mu U^\dag \cdot U
- U^\dag \nabla^\mu \nabla_\mu U + 
U^\dag \chi - \chi^\dag U - \frac{1}{N_f}
\mbox{tr} \left[ U^\dag \chi - \chi^\dag U \right] 
= {\cal O}(p^4) \ .
\label{EOM:ChPT}
\end{equation}

The next order is counted as ${\cal O}(p^4)$
($\bar{D}=4$ in the previous subsection),
the terms in which 
are described by $(d,j) = (4,0)$, $(2,1)$ or $(0,2)$.
To write down possible terms we should note the following identities:
\begin{eqnarray}
&&
U^\dag \nabla_\mu U + \nabla_\mu U^\dag \cdot U = 0 \ ,
\nonumber\\
&&
\nabla_\mu U^\dag \cdot \nabla_\nu U + 
\nabla_\nu U^\dag \cdot \nabla_\mu U +
U^\dag \nabla_\mu \nabla_\nu U + \
\nabla_\mu \nabla_\nu U^\dag \cdot U = 0 \ .
\label{identities:ChPT}
\end{eqnarray}
Now, let us list all the possible terms below:

Generally, there are four terms for $(d,j)=(4,0)$:
\begin{eqnarray}
P_0 &\equiv& 
\mbox{tr}
  \left[ \nabla_\mu U \nabla_\nu U^\dag
  \nabla^\mu U \nabla^\nu U^\dag \right]
\ , \nonumber\\
P_1 &\equiv& \left( \mbox{tr}
  \left[ \nabla_\mu U^\dag \nabla^\mu U \right]
  \right)^2
\ , \nonumber\\
P_2 &\equiv& 
\mbox{tr} \left[ \nabla_\mu U^\dag \nabla_\nu U \right]
  \mbox{tr}\, \left[ \nabla^\mu U^\dag \nabla^\nu U \right]
\ , \nonumber\\
P_3 &\equiv& 
\mbox{tr}
  \left[ \nabla_\mu U^\dag \nabla^\mu U 
  \nabla_\nu U^\dag \nabla^\nu U \right]
\ . 
\label{P0 - P3}
\end{eqnarray}
In the case of $N_f=3$ we can easily show that the relation
\begin{equation}
P_0 = - 2 P_3 + \frac{1}{2} P_1 + P_2 
\label{rel:3:p0}
\end{equation}
is satisfied.  Then only three terms are independent.  
On the other hand, in the case of $N_f=2$ 
the relations
\begin{equation}
P_0 = P_2 - \frac{1}{2} P_1 \ , \qquad
P_3 = \frac{1}{2} P_1 \ ,
\label{rel:2:p0 p3}
\end{equation}
are satisfied, and only two terms are independent.

There are two terms for $(d,j)=(2,1)$:
\begin{eqnarray}
P_4 &\equiv& 
\mbox{tr} \left[ \nabla_\mu U^\dag \nabla^\mu U \right]
  \mbox{tr}\, \left[ \chi^\dag U + \chi U^\dag \right ]
\ , \nonumber\\
P_5 &\equiv& 
\mbox{tr} \left[ \nabla_\mu U^\dag \nabla^\mu U 
  \left(\chi^\dag U + U^\dag \chi \right) \right ]
\ .
\label{P4 P5}
\end{eqnarray}
In the case of $N_f=2$, we can show
\begin{equation}
P_5 = \frac{1}{2} P_4 \ .
\label{rel:2:p5}
\end{equation}

There are three terms for $(d,j)=(0,2)$:
\begin{eqnarray}
P_6 &\equiv& 
  \left( \mbox{tr} \left[ \chi^\dag U + \chi U^\dag \right ] \right)^2
\ , \nonumber\\
P_7 &\equiv& 
  \left( \mbox{tr} \left[ \chi^\dag U - \chi U^\dag \right ] \right)^2
\ , \nonumber\\
P_8 &\equiv& 
\mbox{tr} \left[ 
  \chi^\dag U \chi^\dag U + \chi U^\dag \chi U^\dag \right ]
\ .
\label{P6 - P8}
\end{eqnarray}

In the present case there are other terms which include the field
strength of the external gauge fields ${\cal L}_\mu$ and 
${\cal R}_\mu$:
\begin{eqnarray}
P_9 &\equiv& - i \, 
\mbox{tr} \left[
  {\cal L}_{\mu\nu} \nabla^\mu U \nabla^\nu U^\dag
  + {\cal R}_{\mu\nu} \nabla^\mu U^\dag \nabla^\nu U
\right]
\ , \nonumber\\
P_{10} &\equiv& 
\mbox{tr} \left[
  U^\dag {\cal L}_{\mu\nu} U {\cal R}_{\mu\nu} \right]
\ .
\label{P9 P10}
\end{eqnarray}
In addition there are terms which include the external sources only:
\begin{eqnarray}
Q_1 &\equiv& 
\mbox{tr} \left[ 
  {\cal L}_{\mu\nu} {\cal L}^{\mu\nu} + 
  {\cal R}_{\mu\nu} {\cal R}^{\mu\nu}
\right]
\ , \nonumber\\
Q_2 &\equiv& 
\mbox{tr} \left[ \chi^\dag \chi \right]
\ . \nonumber
\end{eqnarray}

One might think that there are other terms such as
\begin{equation}
\widetilde{P}_1 \equiv
\mbox{tr} \left[
\nabla_\mu \nabla^\mu U^\dag \cdot \nabla_\nu \nabla^\nu U 
\right]
\ .
\end{equation}
However, when we want to obtain Green functions up until
${\cal O}(p^4)$, this term is absorbed into the terms listed above by
the equation of motion in Eq.~(\ref{EOM:ChPT}) and the identity in 
Eq.~(\ref{identities:ChPT}):
\begin{equation}
\widetilde{P}_1 = P_3 + \frac{1}{4N_f} P_7 - \frac{1}{4} P_8 + 
\frac{1}{2} Q_2 + {\cal O}(p^6) \ .
\end{equation}
Namely, difference between the Lagrangians with and without
$\widetilde{P}_1$ term is counted as ${\cal O}(p^6)$ which is higher
order.

By combining the above terms the ${\cal O}(p^4)$ Lagrangian for
$N_f=3$ is given by
\begin{eqnarray}
{\cal L}_{(4)}^{\rm ChPT} &=&
\sum_{i=1}^{10} L_i P_i + \sum_{i=1}^2 H_i Q_i 
\nonumber\\
&=&
L_1\, \left( \mbox{tr}
  \left[ \nabla_\mu U^\dag \nabla^\mu U \right]
  \right)^2
\nonumber\\
&& 
{}+ L_2\,
\mbox{tr} \left[ \nabla_\mu U^\dag \nabla_\nu U \right]
  \mbox{tr}\, \left[ \nabla^\mu U^\dag \nabla^\nu U \right]
\nonumber\\
&& 
{}+ L_3\,
\mbox{tr}
  \left[ \nabla_\mu U^\dag \nabla^\mu U 
  \nabla_\nu U^\dag \nabla^\nu U \right]
\nonumber\\
&& 
{}+ L_4\,
\mbox{tr} \left[ \nabla_\mu U^\dag \nabla^\mu U \right]
  \mbox{tr}\, \left[ \chi^\dag U + \chi U^\dag \right ]
\nonumber\\
&& 
{}+ L_5\,
\mbox{tr} \left[ \nabla_\mu U^\dag \nabla^\mu U 
  \left(\chi^\dag U + U^\dag \chi \right) \right ]
\nonumber\\
&& 
{}+ L_6\,
\left( \mbox{tr} \left[ \chi^\dag U + \chi U^\dag \right ] \right)^2
\nonumber\\
&& 
{}+ L_7\,
\left( \mbox{tr} \left[ \chi^\dag U - \chi U^\dag \right ] \right)^2
\nonumber\\
&& 
{}+ L_8\,
\mbox{tr} \left[ 
  \chi^\dag U \chi^\dag U + \chi U^\dag \chi U^\dag \right ]
\nonumber\\
&& 
{}- i\, L_9\,
\mbox{tr} \left[
{\cal L}_{\mu\nu} \nabla^\mu U \nabla^\nu U^\dag
+ {\cal R}_{\mu\nu} \nabla^\mu U^\dag \nabla^\nu U
\right]
\nonumber\\
&& 
{}+ L_{10}\,
\mbox{tr} \left[
U^\dag {\cal L}_{\mu\nu} U {\cal R}_{\mu\nu} \right]
\nonumber\\
&& 
{}+ H_1\,
\mbox{tr} \left[ 
{\cal L}_{\mu\nu} {\cal L}^{\mu\nu} +
{\cal R}_{\mu\nu} {\cal R}^{\mu\nu}
\right]
\nonumber\\
&& 
{}+ H_2\,
\mbox{tr} \left[ \chi^\dag \chi \right]
\ ,
\label{p4:ChPT}
\end{eqnarray}
where $L_i$ and $H_i$ are dimensionless parameters.
$L_i$ is important for studying
low energy phenomenology of the pseudoscalar mesons.
For $N_f=2$ case we have
\begin{equation}
{\cal L}_{(4)}^{\rm ChPT} = \sum_{i=1,2,4,6,7,8,9,10} L_i P_i 
+ \sum_{i=1}^2 H_i Q_i \ .
\label{ChPT:Lag:2}
\end{equation}
For $N_f\geq4$ we need all the terms:
\begin{equation}
{\cal L}_{(4)}^{\rm ChPT} = \sum_{i=0}^{10} L_i P_i 
+ \sum_{i=1}^2 H_i Q_i \ .
\label{ChPT:Lag:4}
\end{equation}

\subsection{Renormalization}
\label{ssec:R}

The parameters $L_i$ and $H_i$ are renormalized at one-loop level.
Note that all the vertices in one-loop diagrams are from
${\cal O}(p^2)$ terms.
We use the dimensional regularization, and perform 
the renormalizations of the parameters by
\begin{equation}
L_i = L_i^r(\mu) + \Gamma_i \lambda(\mu) \ , \qquad
H_i = H_i^r(\mu) + \Delta_i \lambda(\mu) \ ,
\end{equation}
where $\mu$ is the renormalization point,
and $\Gamma_i$ and $\Delta_i$ are certain numbers given later.
$\lambda(\mu)$ is the divergent part given by
\begin{equation}
\lambda(\mu) = - \frac{1}{2\left(4\pi\right)^2}
\left[   \frac{1}{\bar{\epsilon}} - \ln \mu^2 + 1 \right]
\ ,
\end{equation}
where
\begin{equation}
\frac{1}{\bar{\epsilon}} = \frac{2}{4-n}
- \gamma_E + \ln 4\pi 
\ .
\end{equation}
The constants $\Gamma_i$ and $\Delta_i$ for $N_f=3$
are given by~\cite{Gas:84,Gas:85b}
\begin{equation}
\begin{array}{lllll}
\Gamma_1 = \frac{3}{32} \ ,
& \Gamma_2 = \frac{3}{16} \ ,
& \Gamma_3 = 0 \ , 
& \Gamma_4 = \frac{1}{8} \ ,
& \Gamma_5 = \frac{3}{8} \ ,
\\
\Gamma_6 = \frac{11}{144} \ ,
& \Gamma_7 = 0 \ ,
& \Gamma_8 = \frac{5}{48} \ ,
& \Gamma_9 = \frac{1}{4} \ , 
& \Gamma_{10} = - \frac{1}{4} \ ,
\\
& \Delta_1 = - \frac{1}{8} \ ,
& \Delta_2 = \frac{5}{24} \ .
& &
\end{array}
\end{equation}
Those for $N_f=2$ are given by
\begin{equation}
\begin{array}{llll}
\Gamma_1 = \frac{1}{12} \ ,
& \Gamma_2 = \frac{1}{6} \ ,
&\Gamma_4 = \frac{1}{4} \ ,
& \Gamma_6 = \frac{3}{32} \ ,
\\
\Gamma_7 = 0 \ ,
& \Gamma_8 = 0 \ ,
& \Gamma_9 = \frac{1}{6} \ ,
& \Gamma_{10} = - \frac{1}{6} \ ,
\\
& \Delta_1 = - \frac{1}{12} \ ,
& \Delta_2 = 0 \ . &
\end{array}
\end{equation}

\subsection{Values of low energy constants}
\label{ssec:VLEC}

In this subsection we 
estimate the order of the low energy constants.

By using the renormalization done just before, 
there is a relation
between a low energy constant at a scale $\mu$ and the same constant
at the different scale $\mu'$:
\begin{equation}
L_i^r(\mu') - L_i^r(\mu) = \frac{\Gamma_i}{2\left(4\pi\right)^2}
\, \ln \frac{{\mu'}^2}{\mu^2} \ .
\end{equation}
If there is no accidental fine-tuning of parameters, we would expect
the low energy constants to be at least as large as the coefficient
induced by a rescaling of order 1 in the renormalization point $\mu$.
Then,
\begin{equation}
L_i^r(\mu) \sim {\cal O}\left( 10^{-3} \right) \mbox{---}
{\cal O}\left( 10^{-2} \right) \ .
\label{est:Li}
\end{equation}
The above estimation can be compared with the values of 
the low energy constants
derived by fitting to several experimental data.
We show in Table~\ref{tab:lec} the values for the $N_f=3$ 
case at $\mu=m_\eta$~\cite{Gas:85b} and 
$\mu=m_\rho$~\cite{Eck:89a}.
\begin{table}[htbp]
\begin{center}
\begin{tabular}{|l|c|c|l|}
\hline
 & $L_i^r(\mu=m_\eta)$\cite{Gas:85b} 
 & $L_i^r(\mu=m_\rho)$\cite{Eck:89a} 
 & source \\
\hline
$L_1^r(\mu)$ & $(0.9\pm0.3)\times10^{-3}$ 
  & $(0.7\pm0.3)\times10^{-3}$ & $\pi\pi$ $D$-waves, Zweig rule \\
$L_2^r(\mu)$ & $(1.7\pm0.7)\times10^{-3}$ 
  & $(1.3\pm0.7)\times10^{-3}$ & $\pi\pi$ $D$-waves \\
$L_3^r(\mu)$ & $(-4.4\pm2.5)\times10^{-3}$ 
  & $(-4.4\pm2.5)\times10^{-3}$ & $\pi\pi$ $D$-waves, Zweig rule \\
$L_4^r(\mu)$ & $(0\pm0.5)\times10^{-3}$ 
  & $(-0.3\pm0.5)\times10^{-3}$ & Zweig rule \\
$L_5^r(\mu)$ & $(2.2\pm0.5)\times10^{-3}$ 
  & $(1.4\pm0.5)\times10^{-3}$ & $F_K$:$F_\pi$ \\
$L_6^r(\mu)$ & $(0\pm0.3)\times10^{-3}$ 
  & $(-0.2\pm0.3)\times10^{-3}$ & Zweig rule \\
$L_7^r(\mu)$ & $(-0.4\pm0.15)\times10^{-3}$ 
  & $(-0.4\pm0.15)\times10^{-3}$ & Gell-Man--Okubo, $L_5$, $L_8$ \\
$L_8^r(\mu)$ & $(1.1\pm0.3)\times10^{-3}$ 
  & $(0.9\pm0.3)\times10^{-3}$ & $K^0$-$K^+$, $R$, $L_5$ \\
$L_9^r(\mu)$ & $(7.4\pm0.7)\times10^{-3}$ 
  & $(6.9\pm0.7)\times10^{-3}$ & $\langle r^2 \rangle^\pi_{\rm e.m.}$
\\ 
$L_{10}^r(\mu)$ & $(-6.0\pm0.7)\times10^{-3}$ 
  & $(-5.2\pm0.7)\times10^{-3}$ & $\pi\rightarrow e\nu\gamma$ \\
\hline
\end{tabular}
\end{center}
\caption[Low energy constants in the ChPT]{%
Values of the low energy constants for $N_f=3$.
Values at $\mu=m_\eta$ is taken from Ref.~\cite{Gas:85a} and those at
$\mu=m_\rho$ is taken from Ref.~\cite{Eck:89a}.}
\label{tab:lec}
\end{table}
This shows that the above estimation in Eq.~(\ref{est:Li})
reasonably agrees with the
phenomenological values of the low energy constants.

\subsection{Particle assignment}
\label{ssec:PA2}

To perform phenomenological analyses we need a particle
assignment.
In a realistic case $N_f=3$ there are eight NG bosons which are
identified with $\pi^{\pm}$, $\pi^0$, $K^{\pm}$, $K^0$,
$\overline{K}^0$ and $\eta$.
[Strictly speaking, the octet component $\eta_8$ of $\eta$ is
identified with the NG boson.]
These eight pseudoscalar mesons
are embedded into $3\times3$ matrix $\pi$ as
\begin{equation}
\pi = \frac{1}{\sqrt{2}}
\left(
\begin{array}{ccc}
\frac{1}{\sqrt{2}} \pi^0 + \frac{1}{\sqrt{6}} \eta_8
  & \pi^+ & K^+ \\
\pi^- & - \frac{1}{\sqrt{2}} \pi^0 + \frac{1}{\sqrt{6}} \eta_8 
  & K^0 \\
K^- & \bar{K}^0 & - \frac{2}{\sqrt{6}} \eta_8
\end{array}
\right)
\ .
\end{equation}
The external gauge fields ${\cal L}_\mu$ and ${\cal R}_\mu$ include
$W_\mu$, $Z_\mu$ and $A_\mu$ (photon) as
\begin{eqnarray}
{\cal L}_\mu &=& e Q A_\mu + 
\frac{g_2}{\cos\theta_W} \left( T_z - \sin^2 \theta_W \right) Z_\mu
+ \frac{g_2}{\sqrt{2}} \left( W^+_\mu T_+ + W^-_\mu T_- \right) \ ,
\nonumber\\
{\cal R}_\mu &=& e Q A_\mu - 
\frac{g_2}{\cos\theta_W} \sin^2 \theta_W Z_\mu \ ,
\label{W Z Q}
\end{eqnarray}
where $e$, $g_2$ and $\theta_W$ are the electromagnetic coupling
constant, the gauge coupling constant of SU(2)$_{\rm L}$ 
and the weak mixing angle,
respectively.
The electric charge matrix $Q$ is given by
\begin{equation}
Q = \frac{1}{3}
\left(\begin{array}{ccc}
2 & 0 & 0 \\
0 & -1 & 0 \\
0 & 0 & -1 
\end{array}\right)
\ .
\label{charge matrix}
\end{equation}
$T_z$ and $T_+ = \left( T_- \right)^\dag$ are given by
\begin{equation}
T_z = \frac{1}{2} \left( \begin{array}{ccc}
1 & 0 & 0 \\
0 & -1 & 0 \\
0 & 0 & -1 
\end{array} \right) \ , \qquad
T_+ = \left( \begin{array}{ccc}
0 & V_{ud} & V_{us} \\
0 & 0 & 0 \\
0 & 0 & 0 
\end{array} \right) \ ,
\end{equation}
where $V_{ij}$ are elements of Kobayashi-Maskawa matrix.

\subsection{Example 1: Vector form factors and $L_9$}
\label{ssec:l9}

In this subsection, as an example, we illustrate 
the determination of the value of the low energy constant $L_9$
through the analysis
on the  vector form factors (the electromagnetic
form factors of the pion and kaon and the $K_{l3}$ form factor).
We note that in the analysis of this and
succeeding subsections we neglect effects of the isospin breaking.

In the low energy region the electromagnetic form factor of the
charged particle is given by
\begin{equation}
F_V^{\phi^{\pm}} (q^2) =
1 + \frac{1}{6} \langle r^2 \rangle_V^{\phi^{\pm}} q^2 + \cdots
\ ,
\end{equation}
where $\langle r^2 \rangle_V^{\phi^{\pm}}$ is the charge radius of the
particle $\phi^{\pm}$ and $q^2$ is the square of the photon momentum.
The electromagnetic form factor for the neutral particle is given by
\begin{equation}
F_V^{\phi^0} (q^2) =
\frac{1}{6} \langle r^2 \rangle_V^{\phi^0} q^2 + \cdots
\ .
\end{equation}
Similarly, one of the $K_{l3}$ form factors is given by
\begin{equation}
f_{+}^{K\pi}(q^2) = f_{+}^{K\pi}(0) 
\left[ 
  1 + \frac{1}{6} \langle r^2 \rangle^{K\pi} q^2 + \cdots
\right]
\ ,
\end{equation}
where $\langle r^2 \rangle^{K\pi}$ is related to the 
linear energy dependence $\lambda_+$ by
\begin{equation}
\langle r^2 \rangle^{K\pi} = \frac{6\lambda_+}{m_{\pi^\pm}^2}
\ .
\end{equation}

In the ChPT 
$\langle r^2 \rangle_V^{\pi^{\pm}}$,
$\langle r^2 \rangle_V^{K^{\pm}}$,
$\langle r^2 \rangle_V^{K^0}$ and $\langle r^2 \rangle^{K\pi}$
are calculated as~\cite{Gas:85b}
\begin{eqnarray}
\langle r^2 \rangle_V^{\pi^{\pm}}
&=&
\frac{12L^r_9(\mu)}{F_\pi^2} - \frac{1}{32\pi^2F_\pi^2}
\left[  2 \ln \frac{m_\pi^2}{\mu^2} + \ln \frac{m_K^2}{\mu^2} + 3
\right]
\label{charge radius pi}
\\
\langle r^2 \rangle_V^{K^0}
&=&
- \frac{1}{16\pi^2 F_\pi^2} \ln \frac{m_K}{m_\pi}
\nonumber\\
\langle r^2 \rangle_V^{K^{\pm}}
&=&
\langle r^2 \rangle_V^{\pi^{\pm}} +
\langle r^2 \rangle_V^{K^0}
\ ,
\label{charge radius K0}
\\
\langle r^2 \rangle^{K\pi}
&=&
\langle r^2 \rangle_V^{\pi^{\pm}} - \frac{1}{64\pi^2 F_\pi^2}
\left[
  3 h_1 \left( \frac{m_\pi^2}{m_K^2} \right)
  + 3 h_1 \left( \frac{m_\eta^2}{m_K^2} \right)
  + \frac{5}{2} \ln \frac{m_K^2}{m_\pi^2}
  + 3 \ln \frac{m_\eta^2}{m_K^2} - 6
\right]
\ ,
\label{charge radius Kp}
\end{eqnarray}
where 
\begin{equation}
h_1 (x) \equiv
  \frac{1}{2} \frac{ x^3 - 3x^2 - 3x + 1}{(x-1)^3} \, \ln x
  + \frac{1}{2} \left( \frac{x+1}{x-1} \right)^2 - \frac{1}{3}
\ .
\end{equation}
In Ref.~\cite{Gas:85a} the value of $L_9^r(m_\eta)$ is determined by
using the experimental data of $\langle r^2 \rangle_V^{\pi^{\pm}}$
given in \cite{Dal:82}.
There are several other experimental data after Ref.~\cite{Gas:85a}
as listed in
Table~\ref{tab:radii}, and they are not fully consistent.
Therefore, following Ref.~\cite{Gas:85b} we determine the value of
$L_9^r$ from the the linear energy dependence
$\lambda_+$ of the $K^0_{e3}$ form factor.
By using the experimental value of $\lambda_+$ given in
PDG~\cite{PDG:02}
\begin{equation}
\lambda_+ = 0.0282 \pm 0.0027 \ ,
\end{equation}
the value of $L_9^r(m_\rho)$ is estimated as
\begin{equation}
L_9^r(m_\rho) = \left( 6.5 \pm 0.6 \right) \times 10^{-3} \ .
\label{L9:val}
\end{equation}
Using this value, we obtain the following predictions for the charge
radii:
\begin{eqnarray}
\langle r^2 \rangle_V^{\pi^{\pm}}
&=& 0.400 \pm 0.034 \ \left(\mbox{fm}\right)^2
\ ,
\nonumber\\
\langle r^2 \rangle_V^{K^{\pm}}
&=& 0.39 \pm 0.03 \ \left(\mbox{fm}\right)^2
\ ,
\nonumber\\
\langle r^2 \rangle_V^{K^0}
&=& -0.04 \pm 0.03\ \left(\mbox{fm}\right)^2
\ ,
\end{eqnarray}
where the error bars are estimated by~\cite{Gas:85b}
$\delta \langle r^2 \rangle_V^{\pi^{\pm}} =
(\epsilon/2) \langle r^2 \rangle^{K\pi}$,
$\delta \langle r^2 \rangle_V^{K^{\pm}} =
(\epsilon/3)\langle r^2 \rangle_V^{\pi^{\pm}}$
and $\delta \langle r^2 \rangle_V^{K^0} =
(\epsilon/3)\langle r^2 \rangle_V^{\pi^{\pm}}$ with
$\epsilon=\pm0.2$.
It should be noticed that the resultant charge radius of $K^0$ does
not include any low energy constants.
We show in Table~\ref{tab:radii} the comparison of the above
predictions with several experimental data for the charge radii.
\begin{table}[htbp]
\begin{center}
\begin{tabular}{|l||c|c|c|}
\hline
& 
$\langle r^2 \rangle_V^{\pi^{\pm}}$ $(\mbox{fm})^2$ & 
$\langle r^2 \rangle_V^{K^{\pm}}$ $(\mbox{fm})^2$ & 
$\langle r^2 \rangle_V^{K^0}$ $(\mbox{fm})^2$ \\
\hline
ChPT & $0.400\pm0.034$ & $0.39\pm0.03$ & $-0.04\pm0.03$ \\
\hline
Dally(77) \cite{Dal:77} 
 & $0.31\pm0.04$   &               &                  \\
Molzon(78) \cite{Mol:78}
 &                 &               & $-0.054\pm0.026$ \\
Dally(80) \cite{Dal:80}
 &                 & $0.28\pm0.05$ &                  \\
Dally(82) \cite{Dal:82}
 & $0.439\pm0.030$ &               &                  \\
Amendolia(84) \cite{Ame:84}
 & $0.432\pm0.016$ &               &                  \\
Barkov(85) \cite{Bar:85}
 & $0.422\pm0.013$ &               &                  \\
Amendolia(86) \cite{Ame:86b}
 & $0.439\pm0.008$ &               &                  \\
Amendolia(86) \cite{Ame:86a}
 &                 & $0.34\pm0.05$ &                  \\
Erkal(87) \cite{Erk:87}
 & $0.455\pm0.005$ & $0.29\pm0.04$ &                  \\
\hline
\end{tabular}
\end{center}
\caption[Charge radii in the ChPT]{%
Predictions for the charge radii of $\pi^{\pm}$, $K^{\pm}$
$K^0$ in the ChPT with the existing experimental data.
}\label{tab:radii}
\end{table}

\subsection{Example 2: $\pi\rightarrow e\nu \gamma$ and $L_{10}$}
\label{ssec:l10}

In this subsection, we study the $\pi\rightarrow e\nu \gamma$ decay,
and then estimate the value of the low energy constant $L_{10}$.
The hadronic part is evaluated by one-pion matrix element of the
vector current $J_\mu^a(x)$ and the axialvector current
$J_{5\nu}^b(y)$ ($a,b=1,2,3$) as~\cite{Gas:84}
\begin{eqnarray}
&&
i \int d^4x d^4y \, e^{ik\cdot x} e^{ip\cdot y}
\left\langle 0 \left\vert 
  \mbox{T}\, J_\mu^a(x) J_{5\nu}^b(y)
\right\vert \pi^c (q) \right\rangle
\cdot \varepsilon^{\ast\mu}(k)
\nonumber\\
&&
=
- \epsilon_{abc} F_\pi \, \varepsilon^{\ast\mu}(k)
\left[
  g_{\mu\nu} + \frac{q_\mu p_\nu}{q\cdot k}
  + \frac{4\left(L_9^r(\mu)+L_{10}^r(\mu)\right)}{F_\pi^2}
    \left( q\cdot k \, g_{\mu\nu} - q_\mu k_\nu \right)
\right]
\ ,
\end{eqnarray}
where $\varepsilon^{\ast\mu}(k)$ is the polarization vector of the
photon, $\varepsilon^{\ast}(k) \cdot k =0$.
It should be noticed that the sum
$L_9^r(\mu)+L_{10}^r(\mu)$ is independent of the renormalization scale
although each of $L_9^r(\mu)$ and $L_{10}^r(\mu)$ does depend on it.
The coefficient of the third term is related to the 
axialvector form factor of 
$\pi\rightarrow \ell \nu \gamma$~\cite{Bry:82,Ker:86}
as
\begin{equation}
\frac{F_A}{\sqrt{2} m_{\pi^\pm}} = 
  \frac{4\left(L_9^r(\mu)+L_{10}^r(\mu)\right)}{F_\pi}
\ .
\label{FA l9 l10}
\end{equation}
By using the experimental value given by PDG~\cite{PDG:02}
\begin{equation}
\left. F_A \right\vert_{\rm exp} = 0.0116 \pm 0.0016 \ ,
\label{exp FA}
\end{equation}
the sum $L_9^r(\mu)+L_{10}^r(\mu)$ is estimated as~\footnote{%
For 2-loop estimation see Ref.~\cite{Bij:97}.
}
\begin{equation}
L_9^r(\mu)+L_{10}^r(\mu)
= \left( 1.4 \pm 0.2 \right) \times 10^{-3} \ .
\end{equation}
By using the value of $L_9^r(m_\rho)$ in Eq.~(\ref{L9:val}),
$L_{10}^r(m_\rho)$ is estimated as
\begin{equation}
L_{10}^r(m_\rho) = \left( -5.1 \pm 0.7 \right) \times 10^{-3} \ .
\label{L10:val}
\end{equation}

\newpage

\section{Hidden Local Symmetry}
\label{sec:HLS}

In this section we give an up-to-date
review of the model based on the hidden local
symmetry (HLS)~\cite{BKUYY,BKY},
in which
the vector mesons are introduced as the gauge bosons of the
HLS.
Here we generically use $\pi$ for the pseudoscalar NG bosons (pions
and their flavor partners) and $\rho$ for the HLS gauge bosons ($\rho$
mesons and their flavor partners).

We first discuss the necessity for introducing the vector mesons in
the effective field theory showing 
a schematic view of the $P$-wave $\pi\pi$ scattering
amplitude in Sec.~\ref{ssec:NVM}.
Then, following Ref.~\cite{BKY}
we briefly review the model possessing 
the $G_{\rm global} \times H_{\rm local}$ symmetry,
where $G = \mbox{SU($N_f$)}_{\rm L} \times 
\mbox{SU($N_f$)}_{\rm R}$  is the 
global chiral symmetry and 
$H = \mbox{SU($N_f$)}_{\rm V}$ is the HLS,
in Sec.~\ref{ssec:GHM}.
The Lagrangian of the HLS with lowest derivative terms is shown in
Sec.~\ref{ssec:LOL} with including the external gauge fields.
After making the particle assignment in Sec.~\ref{ssec:PA},
we perform the physical analysis in Sec.~\ref{ssec:PPLO}.
There the parameters of the HLS are determined and
several physical predictions such as the 
$\rho^0\rightarrow e^+e^-$ decay width and
the charge radius of pion are made.

By integrating out the vector meson field in the low-energy region,
the HLS Lagrangian generates the chiral Lagrangian for the
pseudoscalar mesons.
The resultant Lagrangian is a particular form of the most general 
chiral perturbation theory (ChPT)
Lagrangian, in which the low energy parameters $L_i$ are specified.
In Sec.~\ref{ssec:VMSLEC}
we briefly review how to integrate out the vector mesons.
Then we give predicted values of the low energy constant of the ChPT.

There are models to describe the vector mesons other than the
HLS.
In Sec.~\ref{ssec:ROMVM}
we review three models:
The vector meson is introduced as the 
matter field in the 
CCWZ Lagrangian~\cite{CWZ,CCWZ} (the matter field method);
the massive Yang-Mills field
method~\cite{Schwinger:67,Schwinger:69,
Wess-Zumino:67,Gas:69,KRS,Mei};
and the anti-symmetric tensor field method~\cite{Gas:84,Eck:89a}.
There we show the equivalence of these models to the HLS model.

In QCD with $N_f=3$ there exists a non-Abelian anomaly
which breaks the chiral symmetry explicitly.
In the effective chiral Lagrangian this anomaly is appropriately
reproduced by introducing the Wess-Zumino 
action~\cite{WZ,Witten}.
This can be generalized so as to incorporate vector mesons as
the gauge bosons of the HLS~\cite{FKTUY}.
We note that
the low energy theorems for anomalous processes such as
$\pi^0\rightarrow 2\gamma$ and $\gamma\rightarrow 3\pi$ are fulfilled
automatically in the HLS model.
In Sec.~\ref{ssec:AP}, following 
Refs.~\cite{FKTUY} and \cite{BKY}, we briefly review
the way of incorporating vector mesons,
and then perform analyses on several physical processes.

\subsection{Necessity for vector mesons}
\label{ssec:NVM}

\begin{figure}[htbp]
\begin{center}
\epsfxsize = 14cm
\  \epsfbox{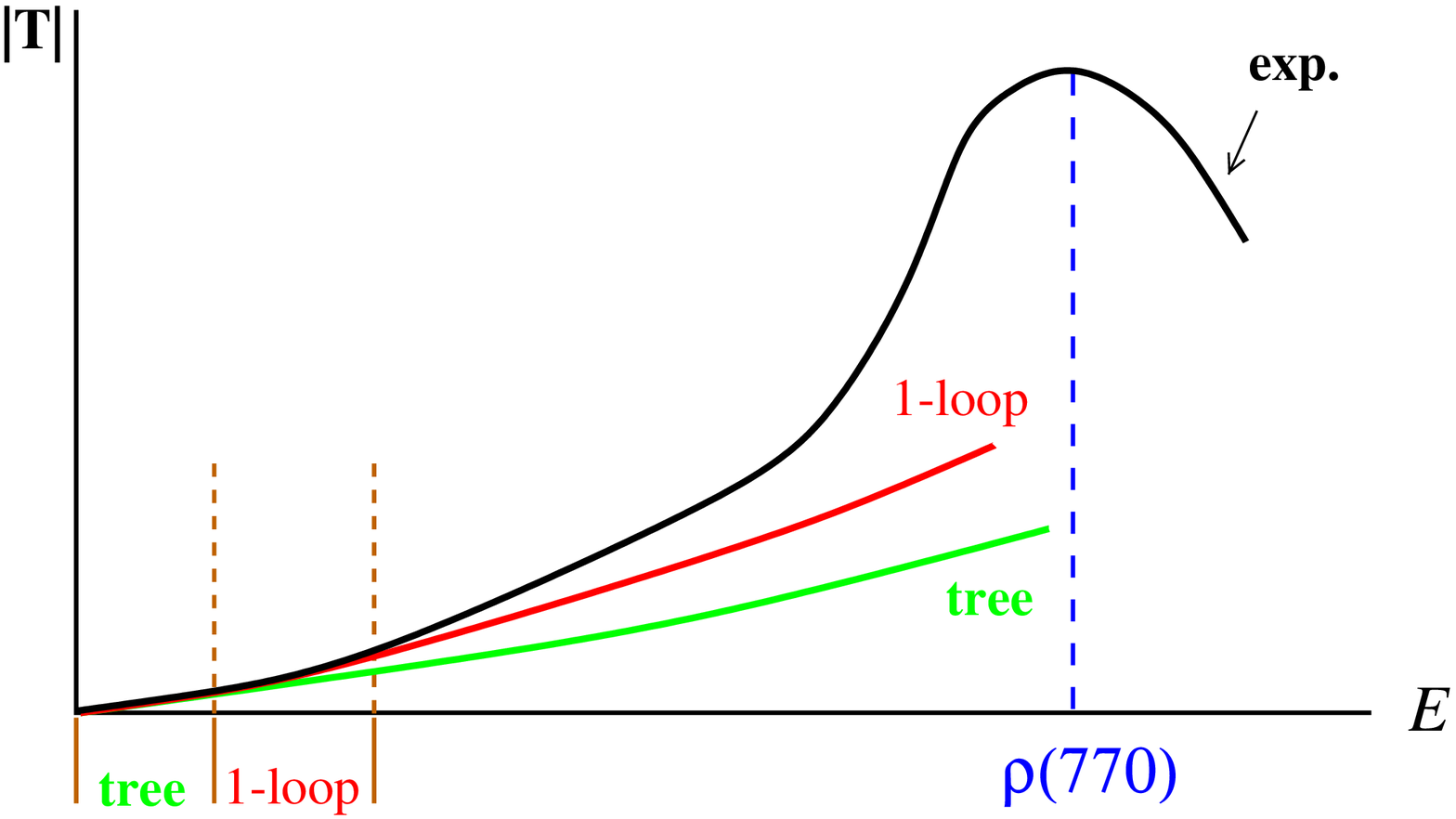}
\end{center}
\caption[$P$-wave $\pi\pi$ scattering amplitude (schematic view)]{%
Schematic view of $P$-wave $\pi\pi$ scattering amplitude.
}
\label{fig:pipi}
\end{figure}

Let us show a schematic view of the $P$-wave $\pi\pi$ scattering
amplitude in Fig.~\ref{fig:pipi}~\cite{DRV}.
As is well known, the ChPT reviewed in
section~\ref{sec:BRCPT}
explains the
experimental data in the low energy region around $\pi\pi$ threshold.
Tree prediction of the ChPT explains the experiment in the threshold
region.
If we include one-loop corrections, the applicable energy region is
enlarged.
In the higher energy region
we know the existence of $\rho$ meson, and the ChPT may not be
applicable.
So the ChPT is not so useful to explain all the data below
the chiral symmetry breaking scale estimated in Eq.~(\ref{lam chi}):
$\Lambda_\chi \sim 1.1$\,GeV.
One simple way is to include $\rho$ meson in the energy region.
A consistent way to include the vector mesons is the HLS.
Further, we can perform the similar systematic low energy expansion in
the HLS as we will explain in Sec.~\ref{sec:CPHLS}.

\subsection{$G_{\rm global} \times H_{\rm local}$ model}
\label{ssec:GHM}

Let us first describe the model based on the
$G_{\rm global} \times H_{\rm local}$ symmetry, where
$G = \mbox{SU($N_f$)}_{\rm L} \times 
\mbox{SU($N_f$)}_{\rm R}$  is the 
global chiral symmetry and 
$H = \mbox{SU($N_f$)}_{\rm V}$ is the HLS.
The entire symmetry $G_{\rm global} \times H_{\rm local}$ is
spontaneously broken down to a diagonal sum $H$ which is 
nothing but the $H$ of $G/H$ of the non-linear sigma model.
This $H$ is then the flavor symmetry.
It is well known that
this model is gauge equivalent to the non-linear sigma model
corresponding to the coset space 
$G/H$~\cite{Cremmer-Julia:78,Cremmer-Julia:79,%
D'Adda-Luscher-Vecchia:78,D'Adda-Vecchia-Luscher:79,%
Gates-Grisaru-Rocek-Siegel,%
Golo-Perelomov}.

The basic quantities of 
$G_{\rm global} \times H_{\rm local}$ linear model
are 
SU($N_f$)-matrix valued variables $\xi_{\rm L}$ and 
$\xi_{\rm R}$ which are introduced by dividing $U$ in the ChPT as
\begin{equation}
U = \xi_{\rm L}^\dag \xi_{\rm R} \ .
\label{div1}
\end{equation}
There is an ambiguity in this division.
It can be identified with the local gauge transformation which is
nothing but the HLS, $H_{\rm local}$.
These two variables transform under the full symmetry as
\begin{equation}
\xi_{\rm L,R}(x) \rightarrow \xi_{\rm L,R}^{\prime}(x) =
h(x) \cdot \xi_{\rm L,R}(x) \cdot g^{\dag}_{\rm L,R}
\ ,
\label{xi:trans}
\end{equation}
where 
\begin{equation}
h(x) \in H_{\rm local} \ , \quad
g_{\rm L,R} \in G_{\rm global} \ .
\end{equation}
These variables are parameterized as
\begin{eqnarray}
&&
\xi_{\rm L,R} = e^{i\sigma/F_\sigma} e^{\mp i\pi/F_\pi} \ ,
\quad
\left[ \, \pi = \pi^a T_a \,,\, \sigma = \sigma^a T_a \right] \ ,
\label{def:xiLR}
\end{eqnarray}
where $\pi$ denote the Nambu-Goldstone (NG) 
bosons associated with 
the spontaneous breaking of $G$ chiral symmetry and 
$\sigma$ 
denote the NG bosons absorbed into the gauge bosons.
$F_\pi$ and $F_\sigma$ are relevant decay constants, and
the parameter $a$ is defined as
\begin{equation}
a \equiv \frac{F_\sigma^2}{F_\pi^2} \ .
\end{equation}

{}From the above $\xi_{\rm L}$ and $\xi_{\rm R}$ we can construct two
Maurer-Cartan 1-forms:
\begin{eqnarray}
\alpha_{\perp\mu} &=&
\left( 
  \partial_\mu \xi_{\rm R} \cdot \xi_{\rm R}^\dag -
  \partial_\mu \xi_{\rm L} \cdot \xi_{\rm L}^\dag 
\right)
/ (2i) \ ,
\label{def:al perp}
\\
\alpha_{\parallel\mu} &=&
\left( 
  \partial_\mu \xi_{\rm R} \cdot \xi_{\rm R}^\dag +
  \partial_\mu \xi_{\rm L} \cdot \xi_{\rm L}^\dag
\right)
/ (2i) \ ,
\label{def:al parallel}
\end{eqnarray}
which transform as
\begin{eqnarray}
&& \alpha_{\perp\mu} \rightarrow 
  h(x) \cdot \alpha_{\perp\mu} \cdot h^\dag (x) \ ,
\\
&& \alpha_{\parallel\mu} \rightarrow 
  h(x) \cdot \alpha_{\perp\mu} \cdot h^\dag (x)
  -i  \partial_\mu h(x) \cdot h^\dag(x) \ .
\end{eqnarray}
The covariant derivatives of $\xi_{\rm L}$ and $\xi_{\rm R}$ are read
from the transformation properties in Eq.~(\ref{xi:trans}) as
\begin{equation}
D_\mu \xi_{\rm L,R} = \partial_\mu \xi_{\rm L,R} - i V_\mu 
\xi_{\rm L,R} \ ,
\end{equation}
where
\begin{equation}
V_\mu = V_\mu^a T_a 
\end{equation}
are the gauge fields corresponding to $H_{\rm local}$.  
These transform as
\begin{equation}
V_\mu \rightarrow
 h(x) \cdot V_\mu \cdot h^\dag (x)
  - i \partial_\mu h(x) \cdot h^\dag(x) 
\ .
\end{equation}
Then the covariantized 1-forms are given by
\begin{eqnarray}
\widehat{\alpha}_{\perp\mu} &=&
\frac{1}{2i}
\left( 
  D_\mu \xi_{\rm R} \cdot \xi_{\rm R}^\dag -
  D_\mu \xi_{\rm L} \cdot \xi_{\rm L}^\dag 
\right)
\ ,
\label{def:al hat perp 0}
\\
\widehat{\alpha}_{\parallel\mu} &=&
\frac{1}{2i}
\left( 
  D_\mu \xi_{\rm R} \cdot \xi_{\rm R}^\dag +
  D_\mu \xi_{\rm L} \cdot \xi_{\rm L}^\dag
\right)
\ .
\label{def:al hat parallel 0}
\end{eqnarray}
The relations of these covariantized 1-forms to $\alpha_{\perp\mu}$
and $\alpha_{\parallel\mu}$ in Eqs.~(\ref{def:al perp}) and
(\ref{def:al parallel}) are given by
\begin{eqnarray}
&& \widehat{\alpha}_{\perp\mu} = \alpha_{\perp\mu} \ ,
\nonumber\\
&& \widehat{\alpha}_{\parallel\mu} = \alpha_{\parallel\mu} 
- V_\mu \ .
\end{eqnarray}
The covariantized 1-forms 
$\widehat{\alpha}_{\perp\mu}$ and $\widehat{\alpha}_{\parallel\mu}$
in Eqs.~(\ref{def:al hat perp 0}) and (\ref{def:al hat parallel 0})
now transform homogeneously:
\begin{equation}
\alpha_{\perp,\parallel}^\mu \rightarrow 
  h(x) \cdot \alpha_{\perp,\parallel}^\mu \cdot h^\dag (x) \ .
\end{equation}
Thus we have the following two invariants:
\begin{eqnarray}
&& {\cal L}_{\rm A} \equiv
F_\pi^2 \, \mbox{tr} 
\left[ \widehat{\alpha}_{\perp\mu} \widehat{\alpha}_{\perp}^\mu
\right]
\ ,
\label{def:LA 0}
\\
&& a {\cal L}_{\rm V} \equiv
F_\sigma^2 \, \mbox{tr}
\left[ 
  \widehat{\alpha}_{\parallel\mu} \widehat{\alpha}_{\parallel}^\mu
\right]
=
F_\sigma^2 \, \mbox{tr}
\left[
  \left(
    V_\mu - \alpha_{\parallel\mu}
  \right)^2
\right]
\ .
\label{def:LV 0}
\end{eqnarray}
The most general Lagrangian
made out of $\xi_{\rm L,R}$ and $D_\mu\xi_{\rm L,R}$ with the lowest
derivatives is thus given by
\begin{equation}
{\cal L} = {\cal L}_{\rm A} + a {\cal L}_{\rm V} \ .
\label{Lag:LA LV}
\end{equation}

We here show that the system with the Lagrangian in 
Eq.~(\ref{Lag:LA LV}) is equivalent to the chiral Lagrangian 
constructed via non-linear realization of the chiral symmetry based
on the manifold $\mbox{SU}(N_f)_{\rm L} \times \mbox{SU}(N_f)_{\rm R}
/\mbox{SU}(N_f)_{\rm V}$,
which is given by 
the first term of Eq.~(\ref{leading ChPT}) with dropping the external
gauge fields.
First, ${\cal L}_V$ vanishes when we substitute the equation of motion
for $V_\mu$:~\footnote{%
This relation is valid since we here do not include the kinetic term
of the HLS gauge boson.
When we include the kinetic term, this is valid only in the low energy
region [see Eq.~(\ref{EOM V})].
}
\begin{equation}
V_\mu = \alpha_{\parallel\mu} \ .
\end{equation}
Further, with the relation
\begin{equation}
\widehat{\alpha}_{\perp\mu} 
= \frac{1}{2i}\,\xi_{\rm L} \cdot \partial_\mu U \cdot 
  \xi_{\rm R}^\dag 
= \frac{i}{2} \,\xi_{\rm R} \cdot \partial_\mu 
  U^\dag \cdot \xi_{\rm L}^\dag
\label{perp U0}
\end{equation}
substituted
${\cal L}_{\rm A}$ becomes identical to the first term of the chiral
Lagrangian in Eq.~(\ref{leading ChPT}):
\begin{equation}
{\cal L} = {\cal L}_{\rm A} =
\frac{F_\pi^2}{4} \mbox{tr}
\left[ \partial_\mu U^\dag \partial^\mu U \right]
\ .
\end{equation}

Let us show that the HLS gauge boson $V_\mu$ agrees with 
Weinberg's ``$\rho$-meson''~\cite{Wei:68} when we take the
unitary gauge of the HLS.
In the unitary gauge, $\sigma=0$, two 
SU($N_f$)-matrix valued variables $\xi_{\rm L}$ and 
$\xi_{\rm R}$ are related with each other by
\begin{equation}
\xi_{\rm L}^\dag = \xi_{\rm R} \equiv \xi = e^{i\pi/F_\pi}
\ .
\end{equation}
This unitary gauge is not preserved under the $G_{\rm global}$
transformation, which in general has the following form
\begin{eqnarray}
G_{\rm global} \ : \ 
\xi \rightarrow \xi^{\prime} 
&=& \xi \cdot g_{\rm R}^\dag = g_{\rm L} \cdot \xi 
\nonumber\\
&=&
\exp\left[i\sigma^\prime( \pi, g_{\rm R}, g_{\rm L} )/F_\sigma \right]
\exp\left[i\pi^\prime/F_\pi \right]
\nonumber\\
&=&
\exp\left[i\pi^\prime/F_\pi\right]
\exp\left[- i\sigma^\prime(\pi, g_{\rm R}, g_{\rm L})/F_\sigma\right]
\ .
\end{eqnarray}
The unwanted factor 
$\exp\left[i\sigma^\prime( \pi, g_{\rm R}, g_{\rm L} )/F_\sigma
\right]$
can be eliminated if we simultaneously perform the $H_{\rm local}$
gauge transformation with
\begin{eqnarray}
H_{\rm local} \ : \ 
h = 
\exp\left[i\sigma^\prime( \pi, g_{\rm R}, g_{\rm L} )/F_\sigma
\right]
\equiv
h \left( \pi, g_{\rm R}, g_{\rm L}\right) \ .
\end{eqnarray}
Then the system has a global symmetry $G =
\mbox{SU($N_f$)}_{\rm L} \times 
\mbox{SU($N_f$)}_{\rm R}$ under the following combined transformation:
\begin{equation}
G \ : \ \xi \rightarrow h \left( \pi, g_{\rm R}, g_{\rm L}\right)
\cdot \xi \cdot g_{\rm R}^\dag
= g_{\rm L} \cdot \xi \cdot 
h^\dag \left( \pi, g_{\rm R}, g_{\rm L}\right)
\ .
\label{com trans}
\end{equation}
Under this transformation the HLS gauge boson $V_\mu$ in the unitary
gauge transforms as
\begin{eqnarray}
G \ : \ 
V_\mu \rightarrow
 h \left( \pi, g_{\rm R}, g_{\rm L}\right) \cdot V_\mu \cdot
h^\dag \left( \pi, g_{\rm R}, g_{\rm L}\right)
-i \partial_\mu h \left( \pi, g_{\rm R}, g_{\rm L}\right)
\cdot h^\dag \left( \pi, g_{\rm R}, g_{\rm L}\right)
\ ,
\label{trans W rho}
\end{eqnarray}
which is precisely the same as Weinberg's 
``$\rho$-meson''~\cite{Wei:68}.

\subsection{Lagrangian with lowest derivatives}
\label{ssec:LOL}

Let us now construct the Lagrangian of the HLS 
with lowest derivative terms.

First, we introduce the external gauge fields ${\cal L}_\mu$
and ${\cal R}_\mu$ which include $W$ boson, $Z$-boson and photon
fields as shown in Eq.~(\ref{W Z Q}).
This is done by gauging the $G_{\rm global}$ symmetry.
The transformation properties of ${\cal L}_\mu$
and ${\cal R}_\mu$ are given in Eq.~(\ref{external gauges}).
Then, the covariant derivatives of $\xi_{\rm L,R}$ are now 
given by
\begin{eqnarray}
D_\mu \xi_{\rm L} &=&
\partial_\mu \xi_{\rm L} - i V_\mu \xi_{\rm L}
+ i \xi_{\rm L} {\cal L}_\mu \ ,
\nonumber\\
D_\mu \xi_{\rm R} &=&
\partial_\mu \xi_{\rm R} - i V_\mu \xi_{\rm R}
+ i \xi_{\rm R} {\cal R}_\mu \ .
\label{covder}
\end{eqnarray}
It should be noticed that in the HLS these external gauge fields are
included without assuming the vector dominance.
It is outstanding feature of the HLS model that $\xi_{\rm L,R}$ have
two independent source charges and hence two independent gauge bosons
are automatically introduced in the HLS model.
Both the vector meson fields and external gauge fields are
simultaneously incorporated into the Lagrangian fully consistent with
the chiral symmetry.
By using the above covariant derivatives two 
Maurer-Cartan 1-forms are constructed as
\begin{eqnarray}
\widehat{\alpha}_{\perp\mu} &=&
\left( 
  D_\mu \xi_{\rm R} \cdot \xi_{\rm R}^\dag -
  D_\mu \xi_{\rm L} \cdot \xi_{\rm L}^\dag 
\right)
/ (2i) \ ,
\nonumber\\
\widehat{\alpha}_{\parallel\mu} &=&
\left( 
  D_\mu \xi_{\rm R} \cdot \xi_{\rm R}^\dag +
  D_\mu \xi_{\rm L} \cdot \xi_{\rm L}^\dag 
\right)
/ (2i) \ .
\label{def:al hat}
\end{eqnarray}
These 1-forms are expanded as
\begin{eqnarray}
\widehat{\alpha}_{\perp\mu} &=&
\frac{1}{F_\pi} \partial_\mu \pi 
+ {\cal A}_\mu 
- \frac{i}{F_\pi} \left[ {\cal V}_\mu \,,\, \pi \right]
- \frac{1}{6F_\pi^3} 
  \Bigl[ \bigl[ \partial_\mu \pi \,,\, \pi \bigr] \,,\, \pi \Bigr]
+ \cdots
\ ,
\label{al perp exp}
\\
\widehat{\alpha}_{\parallel\mu} &=&
\frac{1}{F_\sigma} \partial_\mu \sigma - V_\mu
+ {\cal V}_\mu 
- \frac{i}{2F_\pi^2} \bigl[ \partial_\mu \pi \,,\, \pi \bigr]
- \frac{i}{F_\pi} \left[ {\cal A}_\mu \,,\, \pi \right]
+ \cdots
\ ,
\label{al para exp}
\end{eqnarray}
where
${\cal V}_\mu = \left({\cal R}_\mu + {\cal L}_\mu\right)/2$ and
${\cal A}_\mu = \left({\cal R}_\mu - {\cal L}_\mu\right)/2$.

The covariantized 1-forms in Eqs.~(\ref{def:al hat})
transform homogeneously:
\begin{equation}
\widehat{\alpha}_{\parallel,\perp}^\mu \rightarrow
h(x) \cdot \widehat{\alpha}_{\parallel,\perp}^\mu \cdot h^\dag(x) \ .
\label{hat al:trans}
\end{equation}
Then we can construct two independent terms with lowest derivatives
which are invariant under the full 
$G_{\rm global}\times H_{\rm local}$ symmetry as
\begin{eqnarray}
&& {\cal L}_{\rm A} \equiv
F_\pi^2 \, \mbox{tr} 
\left[ \hat{\alpha}_{\perp\mu} \hat{\alpha}_{\perp}^\mu \right]
=
\mbox{tr} \left[ \partial_\mu \pi \partial^\mu \pi \right]
+ \cdot 
\ ,
\label{def:LA}
\\
&& a {\cal L}_{\rm V} \equiv
F_\sigma^2 \, \mbox{tr}
\left[ 
  \hat{\alpha}_{\parallel\mu} \hat{\alpha}_{\parallel}^\mu
\right]
=
\mbox{tr} \biggl[
\left( \partial_\mu \sigma - F_\sigma V_\mu \right)
\left( \partial^\mu \sigma - F_\sigma V^\mu \right)
\biggr]
+ \cdots
\ ,
\label{def:LV}
\end{eqnarray}
where the expansions of the covariantized 1-forms in 
Eq.~(\ref{al perp exp}) and (\ref{al para exp}) were substituted to
obtain the second expressions.
These expansions imply that ${\cal L}_A$ generates the kinetic term
of pseudoscalar meson, while
${\cal L}_V$ generates the kinetic term of the would-be NG boson
$\sigma$ in addition to the mass term of the vector meson.

Another building block is the gauge
field strength of the HLS gauge boson defined by
\begin{equation}
V_{\mu\nu} \equiv \partial_\mu V_\nu - \partial_\nu
V_\mu - i [ V_\mu , V_\nu ] \ ,
\end{equation}
which also transforms homogeneously:
\begin{equation}
V_{\mu\nu} \rightarrow h(x) \cdot V_{\mu\nu} \cdot h^\dag(x) \ .
\end{equation}
Then a simplest term with $V_{\mu\nu}$ is the kinetic term of the
gauge boson:
\begin{equation}
{\cal L}_{\rm kin}(V_\mu) = - \frac{1}{2g^2} \, \mbox{tr} 
\left[ V_{\mu\nu} V^{\mu\nu} \right] \ ,
\label{HLS gauge kinetic}
\end{equation}
where
$g$ is the HLS gauge coupling constant.

Now the Lagrangian with lowest derivatives
is given
by~\cite{BKUYY,BKY}
\begin{eqnarray}
{\cal L} &=& 
{\cal L}_{\rm A} + a {\cal L}_{\rm V} + 
{\cal L}_{\rm kin}(V_\mu)
\nonumber\\
&=&
F_\pi^2 \, \mbox{tr} 
\left[ \hat{\alpha}_{\perp\mu} \hat{\alpha}_{\perp}^\mu \right]
+ F_\sigma^2 \, \mbox{tr}
\left[ 
  \hat{\alpha}_{\parallel\mu} \hat{\alpha}_{\parallel}^\mu
\right]
- \frac{1}{2g^2} \, \mbox{tr} 
\left[ V_{\mu\nu} V^{\mu\nu} \right] 
\ .
\label{leading Lagrangian 0}
\end{eqnarray}

\subsection{Particle assignment}
\label{ssec:PA}

Phenomenological analyses are performed with setting 
$N_f=3$ and extending the HLS to
$H_{\rm local} = \left[\mbox{U($3$)}_{\rm V}\right]_{\rm local}$.
Accordingly, the chiral symmetry is extended to
$G_{\rm global} = 
\left[\mbox{U($3$)}_{\rm L} \times
\mbox{U($3$)}_{\rm R}\right]_{\rm global}$.
Then 
the pseudoscalar meson field matrix becomes
\begin{eqnarray}
\pi 
&=& \sum_{a=0}^8 T_a \pi^a
\nonumber\\
&=& \frac{1}{\sqrt{2}}
\left(
\begin{array}{ccc}
\frac{1}{\sqrt{2}} \pi^0 
    + \frac{1}{\sqrt{6}} \eta_8 + \frac{1}{\sqrt{3}} \eta_0
  & \pi^+ & K^+ \\
\pi^- 
  & - \frac{1}{\sqrt{2}} \pi^0 
    + \frac{1}{\sqrt{6}} \eta_8 + \frac{1}{\sqrt{3}} \eta_0
  & K^0 \\
K^- & \bar{K}^0 
  & - \frac{2}{\sqrt{6}} \eta_8 + \frac{1}{\sqrt{3}} \eta_0
\end{array}
\right)
\ ,
\end{eqnarray}
where appropriate combinations of 
$\eta_8$ and $\eta_0$ become $\eta$ and $\eta^{\prime}$.

The HLS gauge boson field matrix is expressed as
\begin{equation}
V_\mu = \sum_{a=0}^8 T_a V_\mu^a \ ,
\quad T_0 = \frac{1}{\sqrt{6}} \ .
\end{equation}
Strictly speaking,
we need to introduce the effect of the violation
of Okubo-Zweig-Iizuka 
(OZI) rule~\cite{Okubo,Zweig:1,Zweig:2,Iizuka}
when we perform the systematic low-energy expansion.
That effect is expressed by the following Lagrangian:
\begin{eqnarray}
{\cal L}_{{\rm OZIB},(2)} &=&
\frac{F_{\pi,B}^2}{N_f} 
\mbox{tr} \left[ \hat{\alpha}_{\perp\mu} \right]
\mbox{tr} \left[ \hat{\alpha}_{\perp}^\mu \right]
+
\frac{F_{\sigma,B}^2}{N_f} 
\mbox{tr} \left[ \hat{\alpha}_{\parallel\mu} \right]
\mbox{tr} \left[ \hat{\alpha}_{\parallel}^\mu \right]
- \frac{1}{2N_f g_B^2} \, 
\mbox{tr} \left[ V_{\mu\nu} \right]
\mbox{tr} \left[ V^{\mu\nu} \right] 
\ .
\label{HLS OZIB}
\end{eqnarray}
However, it is well known that the OZI rule
works very well for vector meson nonet.
Then it is natural to take~\footnote{%
Note that OZI violating effect to the pseudoscalar meson decay
constant is needed for phenomenological analysis
(see, e.g., Ref.~\cite{SSW}).}
\begin{equation}
F_{\sigma,B} = 0 \ , \quad
\frac{1}{g_B} = 0 \ .
\end{equation}
In such a case, it is 
convenient to introduce the following particle assignment for the
vector meson nonet:
\begin{eqnarray}
\rho_\mu=
V_\mu/g
= \frac{1}{\sqrt{2}}
\left(
\begin{array}{ccc}
\frac{1}{\sqrt{2}} \left( \rho_\mu^0 + \omega_\mu \right)
  & \rho_\mu^+ & K_\mu^{\ast,+} \\
\rho_\mu^- 
  & - \frac{1}{\sqrt{2}} \left( \rho_\mu^0 + \omega_\mu \right)
  & K_\mu^{\ast,0} \\
K_\mu^{\ast,-} & \bar{K}_\mu^{\ast,0} & \phi_\mu
\end{array}
\right)
\ ,
\end{eqnarray}
where we used the ideal mixing scheme:
\begin{equation}
\left( \begin{array}{c}
\omega_\mu \\
\phi_\mu
\end{array} \right)
=
\left( \begin{array}{cc}
\sqrt{\frac{1}{3}} & \sqrt{\frac{2}{3}} \\
- \sqrt{\frac{2}{3}} & \sqrt{\frac{1}{3}}
\end{array} \right)
\left( \begin{array}{c}
V_\mu^8/g \\
V_\mu^0/g
\end{array} \right)
\ .
\end{equation}

The embedding of 
$W_\mu$, $Z_\mu$ and $A_\mu$ (photon) in 
the external gauge fields ${\cal L}_\mu$ and ${\cal R}_\mu$
were done in 
Sec.~\ref{ssec:PA2}.
Here, just for convenience, we list it again:
\begin{eqnarray}
{\cal L}_\mu &=& e Q A_\mu + 
\frac{g_2}{\cos\theta_W} \left( T_z - \sin^2 \theta_W \right) Z_\mu
+ \frac{g_2}{\sqrt{2}} \left( W^+_\mu T_+ + W^-_\mu T_- \right) \ ,
\nonumber\\
{\cal R}_\mu &=& e Q A_\mu - 
\frac{g_2}{\cos\theta_W} \sin^2 \theta_W Z_\mu \ ,
\label{external gauge fields}
\end{eqnarray}
where $e$, $g_2$ and $\theta_W$ are the electromagnetic coupling
constant, the gauge coupling constant of SU(2)$_{\rm L}$ 
and the weak mixing angle,
respectively.
The electric charge matrix $Q$ is given by
\begin{equation}
Q = \frac{1}{3}
\left(\begin{array}{ccc}
2 & 0 & 0 \\
0 & -1 & 0 \\
0 & 0 & -1 
\end{array}\right)
\ .
\label{charge matrix 2}
\end{equation}
$T_z$ and $T_+ = \left( T_- \right)^\dag$ are given by
\begin{equation}
T_z = \frac{1}{2} \left( \begin{array}{ccc}
1 & 0 & 0 \\
0 & -1 & 0 \\
0 & 0 & -1 
\end{array} \right) \ , \qquad
T_+ = \left( \begin{array}{ccc}
0 & V_{ud} & V_{us} \\
0 & 0 & 0 \\
0 & 0 & 0 
\end{array} \right) \ ,
\end{equation}
where $V_{ij}$ are elements of Kobayashi-Maskawa matrix.

\subsection{Physical predictions at tree level}
\label{ssec:PPLO}

Let us study some phenomena using the Lagrangian with lowest
derivatives
given in Eq.~(\ref{leading Lagrangian 0}).
In this Lagrangian all the vector mesons are degenerate even when
we apply the HLS to the case of $N_f=3$.
The mass splitting among the vector-meson nonet (or octet) is
introduced when we include the higher derivative terms
(see Sec.~\ref{sec:CPHLS}).
So we study some phenomenology related to the $\rho$ meson.
By taking the unitary gauge of the HLS ($\sigma=0$)
and substituting the expansions of 
$\widehat{\alpha}_{\perp\mu}$ and 
$\widehat{\alpha}_{\parallel\mu}$ given in 
Eqs.~(\ref{al perp exp}) and (\ref{al para exp}) into
the Lagrangian in Eq.~(\ref{leading Lagrangian 0}),
we obtain
\begin{eqnarray}
{\cal L} &=& 
\mbox{tr} \left[
  \left(
    \partial_\mu \pi - i \left[ A_\mu Q \,,\, \pi \right]
    + \cdots
  \right)^2
\right]
\nonumber\\
&& 
{}+ a F_\pi^2 \, \mbox{tr} \left[
  \left(
    g \rho_\mu - e A_\mu Q 
    + \frac{i}{2F_\pi^2} \left[ \partial_\mu \pi \,,\, \pi \right]
    + \cdots
  \right)^2
\right]
\label{Lag unexp}
\\
&=&
\mbox{tr} 
\left[ \partial_\mu \pi \partial^\mu \pi \right] +
a g^2 F_\pi^2 \, \mbox{tr} \left[ \rho_\mu \rho^\mu \right]
+ 2 i \left( \frac{1}{2} a g \right) \, \mbox{tr} 
\left[ \rho^\mu \left[ \partial_\mu \pi \,,\, \pi \right] \right]
\nonumber\\
&&
{}- 2 e a g F_\pi^2 A^\mu  
\,\mbox{tr} \left[ \rho_\mu Q \right] 
{}+ 2i e \left( 1 - \frac{a}{2} \right) A^\mu \, \mbox{tr} 
\left[ Q \left[ \partial_\mu \pi \,,\, \pi \right] \right]
\nonumber\\
&&
{}+ a e^2 F_\pi^2 A_\mu A^\mu \,\mbox{tr} \left[ Q Q \right]
+ \frac{4-3a}{12F_\pi^2} \,\mbox{tr} 
  \biggl[
    \left[ \partial_\mu \pi \,,\, \pi \right]
    \left[ \partial^\mu \pi \,,\, \pi \right]
  \biggr]
+ \cdots \ ,
\label{Lag exp}
\end{eqnarray}
where we have gauged only a subgroup of $G_{\rm global}$,
$I_{\rm global} = \mbox{U}(1)_{\rm Q} \subset H_{\rm global}
\subset G_{\rm global} = \mbox{SU}(3)_{\rm L} \times 
\mbox{SU}(3)_{\rm R}$, with the photon field $A_\mu$ in 
Eq.~(\ref{external gauge fields}), and the vector meson field
$\rho_\mu$ related to $V_\mu$ by rescaling the kinetic term 
in Eq.~(\ref{HLS gauge kinetic}):
\begin{equation}
V_\mu = g \rho_\mu \ .
\end{equation}
{}From this we can easily read the $\rho$ meson mass $m_\rho$, 
the $\rho\pi\pi$ coupling constant $g_{\rho\pi\pi}$,
the $\rho$--$\gamma$ mixing strength $g_\rho$ and 
the direct $\gamma\pi\pi$ coupling constant $g_{\gamma\pi\pi}$:
\begin{eqnarray}
&& m_\rho^2 = a g^2 F_\pi^2 \ , \label{mrho}\\
&& g_{\rho\pi\pi} = \frac{1}{2} a g \ , \label{grpp}\\
&& g_\rho = a g F_\pi^2 \ , \label{grho}\\
&& g_{\gamma\pi\pi} = \left( 1 - \frac{a}{2} \right) e
\ .
\label{ggampp}
\end{eqnarray}
We should note that {\it the $\rho$ acquires a mass through the Higgs
mechanism}
associated with spontaneous breaking of the HLS $H_{\rm local}$.
We also note that
the photon denoted by $A_\mu$ in Eq.~(\ref{Lag exp}) also acquire the
mass through the Higgs mechanism since the photon is introduced by
gauging the subgroup
$I_{\rm global} = \mbox{U}(1)_{\rm Q} \subset G_{\rm global}$
which is spontaneously broken together with the HLS.
Thus $H_{\rm local} \times (\mbox{gauged-})I_{\rm global}
\rightarrow \mbox{U}(1)_{\rm em}$.

When we add the kinetic term of the photon field $A_\mu$ in the
Lagrangian in Eq.~(\ref{Lag exp}),
the photon field mixes with the neutral vector meson
($\rho^0$ for $N_f=2$).
For $N_f =2$ the mass matrix of the photon and $\rho^0$ are given by
\begin{equation}
a F_\pi^2 \,
\left( \rho^0_\mu \,,\, A_\mu \right)
\left(\begin{array}{cc}
  g^2 & e g \\ e g & e^2 
\end{array}\right)
\left(\begin{array}{cc}
  \rho^{0\mu} \\ A^\mu
\end{array}\right)
\ ,
\end{equation}
which is diagonalized by introducing new fields defined by
\begin{eqnarray}
&& 
  \widetilde{\rho}^0_\mu \equiv 
    \frac{1}{\sqrt{g^2+e^2}} \left( g \rho^0_\mu - e A_\mu \right)
\ ,
\nonumber\\
&& 
  \widetilde{A}_\mu \equiv 
    \frac{1}{\sqrt{g^2+e^2}} \left( g \rho^0_\mu + e A_\mu \right)
\ .
\label{rho A mix}
\end{eqnarray}
The mass eigenvalues are given by
\begin{eqnarray}
&&
  m_{\widetilde{\rho}^0}^2 = a F_\pi^2 \left(g^2 + e^2\right) 
\ ,
\label{mass rho0}
\\
&&
  m_{\widetilde{A}}^2 = 0 
\ .
\label{mass photon}
\end{eqnarray}
The charged vector mesons $\rho^{\pm}$ of course do not
mix with the photon, and the masses of them are given by
\begin{equation}
  m_{\rho^{\pm}}^2 = a F_\pi^2 g^2 \ .
\label{mass rho pm}
\end{equation}
For $N_f=2$ the above situation implies that the 
$[\mbox{SU}(2)_{\rm V}]_{\rm HLS} \times \mbox{U}(1)_{\rm Q}$
symmetry is
spontaneously broken down to $\mbox{U}(1)_{\rm em}$.
The massless gauge boson of the remaining $\mbox{U}(1)_{\rm em}$
is nothing but the physical photon field
$\widetilde{A}_\mu$ in 
Eq.~(\ref{rho A mix}).
This situation is precisely the same as that occurring in the
Glashow-Salam-Weinberg model.
Comparing the mass of neutral $\rho$ in Eq.~(\ref{mass rho0}) with
the mass of charged $\rho$ in Eq.~(\ref{mass rho pm}),
we immediately conclude that
the neutral $\rho$ is heavier than the charged $\rho$:
$m_{\widetilde{\rho}^0} > m_{\rho^{\pm}}$.
Furthermore, we have the following prediction for the mass difference
between the neutral $\rho$ and the charged $\rho$:
\begin{equation}
m_{\widetilde{\rho}^0} - m_{\rho^{\pm}} \simeq 
\frac{e^2}{2 g} \sqrt{a} F_\pi \sim 1\, \mbox{MeV}\ ,
\label{mrho0 - mrhopm}
\end{equation}
where we used $e^2 = 4\pi /137 \simeq 0.092$,
$F_\pi \simeq 92$\,MeV [see Eq.~(\ref{exp Fp})],
$g \simeq 5.8$ [see Eq.~(\ref{gval:tree})]
and $a \simeq 2.1$ [see Eq.~(\ref{aval:tree})].
For $N_f=2$ the above mass difference in Eq.~(\ref{mrho0 - mrhopm})
is consistent with the experimental value of the $\rho^0$-$\rho^{\pm}$
mass difference~\cite{PDG:02}:
\begin{equation}
\left. m_{\rho^0} - m_{\rho^{\pm}} 
\right\vert_{\rm exp} = 0.5 \pm 0.7 \, \mbox{MeV} \ .
\end{equation}
Future experiment is desirable for checking the prediction
(\ref{mrho0 - mrhopm}) of the HLS.

Now we turn to a discussion of the implication of the relations among
the masses and coupling constants in
Eqs.~(\ref{mrho})--(\ref{ggampp}).
For a parameter choice $a=2$, the above results reproduce the
following outstanding phenomenological facts~\cite{BKUYY}:
\begin{enumerate}
\renewcommand{\labelenumi}{(\theenumi)}
\item $g_{\rho\pi\pi} = g$ 
  (universality of the $\rho$-coupling)~\cite{Sakurai}
\item $m_\rho^2 = 2 g_{\rho\pi\pi}^2 F_\pi^2$
  (KSRF II)~\cite{KSRF:KS,KSRF:RF}
\item $g_{\gamma\pi\pi}=0$ ($\rho$ meson dominance of the
electromagnetic form factor of the pion)~\cite{Sakurai}
\end{enumerate}
Moreover, independently of the parameter $a$,
Eqs.~(\ref{grpp}) and (\ref{grho}) lead to the KSRF
relation~\cite{KSRF:KS,KSRF:RF} (version I)
\begin{equation}
g_\rho = 2 F_\pi^2 g_{\rho\pi\pi} \ .
\label{KSRF I}
\end{equation}
This parameter independent relation comes from the ratio of the two
cross terms $\rho_\mu A^\mu$ and 
$\rho_\mu \left[ \partial^\mu \pi \,,\, \pi \right]$ in 
$a {\cal L}_{\rm V}$ term [second term in Eq.~(\ref{Lag unexp})],
so that it is obviously independent of $a$ which is an overall factor.
Moreover, the ratio is precisely fixed by the symmetry 
$G_{\rm global} \times H_{\rm local}$ of our Lagrangian with the
subgroup $I_{\rm global} \subset G_{\rm global}$ being gauged, and
hence is a 
{\it direct consequence of the HLS independently of dynamical
details}. 
Since 
the off-shell extrapolation of the vector meson fields are
well defined in the HLS,
the KSRF (I) relation also makes sense
for the off-shell $\rho$ at soft
momentum limit:
\begin{equation}
g_\rho(p_\rho^2 = 0) = 2 
g_{\rho\pi\pi}( p_\rho^2 =0; q_{\pi 1}^2 = 0, q_{\pi 2}^2 = 0 )
\, F_\pi^2 \ ,
\label{LET tree}
\end{equation}
where $p_\rho$ is the $\rho$ momentum and
$q_{1}$ and $q_{2}$ are the pion momenta.
This relation is actually 
a low-energy theorem of the HLS~\cite{BKY:NPB} to be valid
independently of any higher derivative terms which are irrelevant to
the low-energy limit.
This low-energy theorem was first
proved at the tree level~\cite{BKY:PTP}, then at one-loop
level~\cite{HY} and any loop order~\cite{HKY:PRL,HKY:PTP}.

Importance of this low-energy theorem is that although it is proved
only at the low-energy limit, the KSRF (I) relation actually holds
even at the physical point on the mass-shell.
$g_{\rho\pi\pi}$ and $g_\rho$
in Eq.~(\ref{KSRF I})
are related to 
the $\rho\rightarrow\pi\pi$ decay width and
the $\rho\rightarrow e^+e^-$ decay width as
\begin{eqnarray}
&&
\Gamma\left( \rho \rightarrow \pi\pi \right) =
\frac{\left\vert \vec{p}_\pi \right\vert^3}{6\pi m_\rho^2 }
\left\vert g_{\rho\pi\pi} \right\vert^2
\ , \quad
\left\vert \vec{p}_\pi \right\vert
= \sqrt{\frac{ m_\rho^2 - 4 m_\pi^2}{4}} 
\ ,
\label{rho width}
\\
&&
\Gamma\left( \rho \rightarrow e^+ e^- \right) =
\frac{4\pi \alpha^2}{3} \left\vert \frac{g_\rho}{m_\rho^2}
\right\vert^2
\frac{m_\rho^2 + 2 m_e^2}{m_\rho^2} \sqrt{ m_\rho^2 - 4 m_e^2 }
\ .
\label{rho ee width}
\end{eqnarray}
By using 
the experimental values~\cite{PDG:02}
\begin{eqnarray}
&&
  F_\pi = 92.42 \pm 0.26 \, \mbox{MeV} \ ,
\label{exp Fp}
\\
&&
  m_\rho = 771.1 \pm 0.9 \, \mbox{MeV} \ , 
\label{exp mr}
\\
&&
  m_\pi = 139.57018 \pm 0.00035 \, \mbox{MeV} \ , 
\label{exp mp}
\\
&&
  \Gamma\left( \rho \rightarrow \pi \pi \right)_{\rm exp}
  = 149.2 \pm 0.7 \, \mbox{MeV} \ ,
\label{exp rho pp}
\\
&&
  \Gamma( \rho \rightarrow e^+ e^- )_{\rm exp}
  = 6.85 \pm 0.11 \, \mbox{keV} \ ,
\label{exp rho ee}
\end{eqnarray}
the values of $g_{\rho\pi\pi}$ and $g_\rho$ are estimated as
\begin{eqnarray}
&&
  \left. g_{\rho\pi\pi} \right\vert_{\rm exp} = 6.00 \pm 0.01 \ ,
\label{grpp exp val}
\\
&&
  \left. g_\rho \right\vert_{\rm exp} = 0.119 \pm 0.001 \, 
  \mbox{GeV}^2 \ .
\label{grho exp val}
\end{eqnarray}
{}From these experimental values we obtain
\begin{equation}
\left. \frac{g_\rho}{2 g_{\rho\pi\pi} F_\pi^2} \right\vert_{\rm exp}
=1.15 \pm 0.01 \ .
\label{LET val}
\end{equation}
This implies that the KSRF (I) relation in Eq.~(\ref{KSRF I})
is well satisfied, which may be regarded as a decisive test of 
the HLS.~\footnote{%
  When we use $\Gamma\left( \rho \rightarrow \pi\pi \right)$ as an
  input and predict the $\rho\rightarrow e^+e^-$ decay width from the
  low-energy theorem,
  we obtain
  $ \Gamma\left( \rho \rightarrow e^+ e^- \right)
  = 5.11 \pm 0.23 \,$keV.
}
The above small deviation of the experimental values from the KSRF (I)
relation is on-shell corrections
due to the non-zero $\rho$ mass.
Actually, 
as we shall show in Sec.~\ref{sec:WM},
the difference of the value in Eq.~(\ref{LET val}) from one
is explained by the corrections from the higher derivative terms.

Now, let us 
determine three parameters $F_\pi$, $a$ and $g$ from the
experimental data.
The value of $F_\pi$ is just taken from the experimental value in 
Eq.~(\ref{exp Fp}).
We determine the values of $a$ and $g$ from 
$\left. g_{\rho\pi\pi} \right\vert_{\rm exp}$ 
in Eq.~(\ref{grpp exp val})
and $m_\rho$ in Eq.~(\ref{exp mr}) through Eqs.~(\ref{mrho}) and
(\ref{grpp}).
Then
the values of the parameters $a$ and $g$ are determined as~\footnote{%
  If we determine $g$ and $a$ from $g_\rho$ 
  of Eq.~(\ref{grho exp val}) and $m_\rho$ of Eq.~(\ref{exp mr}),
  then we have
  \begin{eqnarray}
  &&
    g = \frac{m_\rho^2}{g_\rho} = 5.01 \pm 0.79 \ , \quad
    \left(
      \frac{g^2}{4\pi} = 2.00 \pm 0.63 
    \right)
  \ ,
  \nonumber\\
  &&
    a = \frac{g_\rho}{g F_\pi^2} = 2.77 \pm 0.44 \ .
  \nonumber
  \end{eqnarray}
  There is about 15\% difference between the above value of $g$ and
  that 
  in Eq.~(\ref{gval:tree}), as implied by Eq.~(\ref{LET val}).
  Since Eqs.~(\ref{mrho}), (\ref{grpp}) and (\ref{grho}) lead to
  \begin{eqnarray}
  a = \frac{4 g_{\rho\pi\pi}^2 F_\pi^2}{m_\rho^2}
  = \frac{g_\rho^2}{m_\rho^2 F_\pi^2}
  \ ,
  \nonumber
  \end{eqnarray}
  there is about 30\% difference between the above value of $a$
  and that in Eq.~(\ref{aval:tree}).
  One might think that we could use the above values of $g$ and $a$ for
  phenomenological analysis.
  However, as we will show in Sec.~\ref{sec:WM}, the deviation of
  the 
  prediction of $g_\rho$ in Eq.~(\ref{grhoval:tree}) 
  from the experimental value in Eq.~(\ref{grho exp val})
  is explained by
  including the higher derivative term ($z_3$ term).
  Thus, we think that it is better to use 
  the values in Eqs.~(\ref{gval:tree})
  and (\ref{aval:tree}) for the phenomenological analysis at tree
  level. 
  Actually, the values of $g$ and $a$ in Eqs.~(\ref{gval:tree})
  and (\ref{aval:tree}) are consistent with those 
  obtained by the analysis based on the Wilsonian matching
  as shown in section~\ref{sec:WM}.
  [See $g(m_\rho)$ in Table~\ref{tab:WM mrho} and
  $a(0)$ in Table~\ref{tab:res}.]
}
\begin{eqnarray}
&& 
  g = \frac{m_\rho^2}{2 g_{\rho\pi\pi} F_\pi^2} 
  = 5.80 \pm 0.91 \ , \quad
  \left(
    \frac{g^2}{4\pi} = 2.67 \pm 0.84
  \right) 
\ , \label{gval:tree} \\
&&
  a = \frac{2 g_{\rho\pi\pi}}{g}
  = 2.07 \pm 0.33 \ , \label{aval:tree}
\end{eqnarray}
where we add 15\% error for each parameter, which is expected from the
deviation of the low-energy theorem in 
Eq.~(\ref{LET val}).
{}From these values the predicted value of $g_\rho$ is given by
\begin{eqnarray}
&& g_\rho = 0.103 \pm 0.023 \, \mbox{GeV}^2 \ ,
\label{grhoval:tree}
\end{eqnarray}
which is compared with the value in Eq.~(\ref{grho exp val}) obtained
from $\Gamma( \rho \rightarrow e^+ e^- )$.

\begin{figure}[htbp]
\begin{center}
\epsfxsize = 14cm
\ \epsfbox{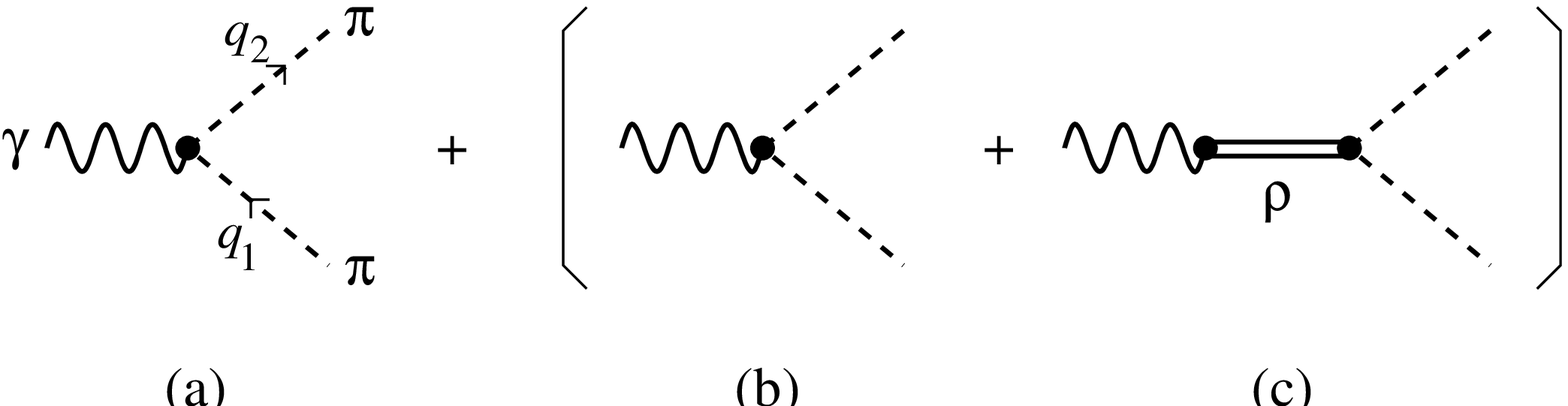}
\end{center}
\caption[Electromagnetic form factor in the HLS]{%
Electromagnetic form factor in the HLS:
(a) the direct $\gamma\pi\pi$ interaction from ${\cal L}_{\rm A}$ term
in the Lagrangian (\ref{leading Lagrangian 0});
(b) the direct $\gamma\pi\pi$ interaction from $a {\cal L}_{\rm V}$
term;
(c) the $\gamma\pi\pi$ interaction mediated by $\rho$ exchange.
}\label{fig:EM form}
\end{figure}

Before making physical predictions,
let us see 
the electromagnetic form factor of the pion.
If one sees only the direct
$\gamma\pi\pi$ coupling in Eq.~(\ref{ggampp}),
one might think that the electric charge of $\pi$ would not be
normalized to be unity, 
and thus the gauge invariance of the photon would be violated.
This is obviously not the case,
since the Lagrangian~(\ref{leading Lagrangian 0}) or 
(\ref{Lag unexp}) is manifestly gauge
invariant under $\mbox{U}(1)_{\rm em}$ by construction.
This can also be seen diagrammatically as follows.
The term proportional to $a$ in $g_{\gamma\pi\pi}$ of
Eq.~(\ref{ggampp}) comes from the vertex derived
from $a {\cal L}_{\rm V}$ term in the Lagrangian
(\ref{leading Lagrangian 0}),
and then it is exactly canceled with the $\rho$-exchange contribution
coming from the same $a {\cal L}_{\rm V}$ term
in the low energy limit.
Thus, the electric charge of $\pi$ is properly normalized.
To visualize this, we show the diagrams contributing to the
electromagnetic form factor of $\pi^{\pm}$ in Fig.~\ref{fig:EM form}.
The contributions from the diagrams in Fig.~\ref{fig:EM form} are
summarized as
\begin{eqnarray}
&&
  \Gamma^{\rm(a)}_\mu(q_2,q_1) = e (q_1+q_2)_\mu
\ ,
\nonumber\\
&&
  \Gamma^{\rm(b)}_\mu(q_2,q_1) = e (q_1+q_2)_\mu 
  \left( - \frac{a}{2} \right)
\ ,
\nonumber\\
&&
  \Gamma^{\rm(c)}_\mu(q_2,q_1) = e (q_1+q_2)_\mu 
  \frac{ g_\rho g_{\rho\pi\pi} }{m_\rho^2 - p^2}
\ ,
\end{eqnarray}
where $p^\mu = q_2^\mu - q_1^\mu$.
By summing these contributions with noting 
the relation $g_\rho g_{\rho\pi\pi} = a m_\rho^2/2$, 
the electromagnetic form factor of
$\pi^{\pm}$ is given by
\begin{equation}
F_{V}^{\pi^{\pm}}(p^2) =
1 - \frac{a}{2} + \frac{a}{2} \frac{ m_\rho^2}{m_\rho^2 - p^2}
\ .
\label{EM form HLS 0}
\end{equation}
In this form we can easily see that the contributions from the
diagrams (b) and (c) in Fig.~\ref{fig:EM form} are exactly canceled
in the low energy limit $p^2=0$,
and thus the electromagnetic form factor of pion is properly
normalized:
\begin{equation}
F_{V}^{\pi^{\pm}}(p^2=0) = 1 \ .
\end{equation}

Now, we make physical predictions using the values of the
parameters in Eqs.~(\ref{exp Fp}), (\ref{gval:tree}) and
(\ref{aval:tree}).
An interesting physical quantity is the 
charge radius of pion 
$\langle r^2 \rangle_V^{\pi^{\pm}}$, which is defined through the
electromagnetic form factor of $\pi^{\pm}$ in the low energy region as
\begin{equation}
F_{V}^{\pi^{\pm}}(p^2) = 1 + \frac{p^2}{6} 
\langle r^2 \rangle_V^{\pi^{\pm}} + \cdots \ .
\end{equation}
{}From the electromagnetic form factor in Eq.~(\ref{EM form HLS 0}),
which is derived from the Lagrangian with lowest derivatives in 
Eq.~(\ref{leading Lagrangian 0}),
the charge radius of $\pi^{\pm}$
is expressed as
\begin{equation}
\langle r^2 \rangle_V^{\pi^{\pm}} = 6 
\frac{g_\rho g_{\rho\pi\pi}}{m_\rho^4} 
= \frac{3a}{m_\rho^2}
\ .
\end{equation}
By using the value of $a$ in Eq.~(\ref{aval:tree}) and the
experimental value of $\rho$ meson mass in Eq.~(\ref{exp mr})
this is evaluated as
\begin{equation}
\langle r^2 \rangle_V^{\pi^{\pm}} = 
0.407 \pm 0.064 \ (\mbox{fm})^2 \ .
\end{equation}
Comparing this with the experimental values shown in 
Table~\ref{tab:radii} in Sec.~\ref{sec:BRCPT},
we conclude that the HLS model with lowest derivatives
reproduces the
experimental data of the charge radius of pion very well.

Another interesting physical quantity is the axialvector 
form factor $F_A$ of $\pi \rightarrow \ell\nu \gamma$ studied in
Sec.~\ref{ssec:l10}.
In the HLS with lowest dirivatives
there is no contribution
to this axialvector form factor, and thus
$F_A = 0$.
This, of course, does not agree with the experimental data in
Eq.~(\ref{exp FA}).
However, as we shall show in Sec.~\ref{sec:WM},
the prediction of the HLS reasonably agree with the experiment
when we go to the next order, ${\cal O}(p^4)$.

Finally in this subsection,
we consider the low-energy theorem on the $\pi\pi$ scattering
amplitude, which is a direct consequence of the chiral symmetry.
If one sees the contact 4$\pi$-interaction in Eq.~(\ref{Lag exp}),
one might think that the HLS violated the low-energy theorem of the
$\pi\pi$ scattering amplitude.
However, this is of course not true since the 
Lagrangian~(\ref{leading Lagrangian 0}) is chiral-invariant and hence
must respect the low-energy theorem trivially.
This can be also seen diagrammatically as follows:
The term proportional to $a$ in the contact 4$\pi$-interaction is
derived from $a {\cal L}_{\rm V}$ term in the 
Lagrangian~(\ref{leading Lagrangian 0}),
which is exactly canceled by the $\rho$-exchange contribution in the
low-energy limit.
\begin{figure}[htbp]
\begin{center}
\epsfxsize = 14cm
\ \epsfbox{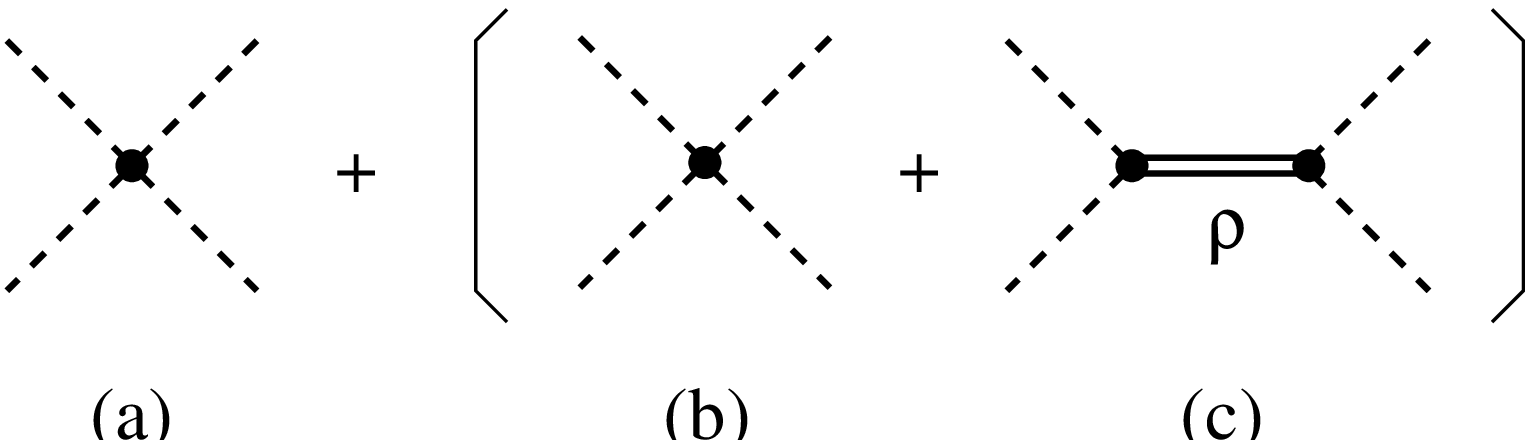}
\end{center}
\caption[$\pi\pi$ scattering in the HLS]{%
Diagrams contributing to the $\pi\pi$ scattering in the HLS:
(a) contribution from the contact 4$\pi$-interaction from 
${\cal L}_{\rm A}$ term in the Lagrangian 
(\ref{leading Lagrangian 0});
(b) contribution from the contact 4$\pi$-interaction from 
$a {\cal L}_{\rm V}$ term;
(c) contribution from the $\rho$-exchange.
The diagram (c) implicitly includes three diagrams:
$s$-channel, $t$-channel and $u$-channel
$\rho$-exchange diagrams.
}\label{fig:pi pi scat}
\end{figure}
To visualize this, we show 
the diagrams contributing to the $\pi\pi$ scattering
in Fig.~\ref{fig:pi pi scat}.
Contributions to the $\pi\pi$ scattering amplitude $A(s,t,u)$
are given by~\footnote{%
The invariant amplitude for 
$\pi_i(p_1) + \pi_j(p_2) \rightarrow \pi_k(p_3) + \pi_l(p_4)$
is decomposed as
$ \delta_{ij} \delta_{kl} A(s,t,u)
+ \delta_{ik} \delta_{jl} A(t,s,u)
+ \delta_{il} \delta_{jk} A(u,t,s)$,
where $s$, $t$ and $u$ are the usual Mandelstam variables:
$s=(p_1+p_2)^2$, $t=(p_1+p_3)^2$ and
$u=(p_1+p_4)^2$.
}
\begin{eqnarray}
&&
  A^{\rm(a)}(s,t,u) = \frac{s}{F_\pi^2}
\ ,
\label{pipi amplitude}
\\
&&
  A^{\rm(b)}(s,t,u) = - \frac{3 a s}{4F_\pi^2}
\ ,
\\
&&
  A^{\rm(c)}(s,t,u) = - g_{\rho\pi\pi}^2
  \left[
    \frac{u-s}{m_\rho^2-t} + \frac{t-s}{m_\rho^2-u} 
  \right]
\ .
\end{eqnarray}
Noting that $a/(4F_\pi^2) = g_{\rho\pi\pi}^2/m_\rho^2$,
we obtain
\begin{eqnarray}
&&
  A^{\rm(b+c)}(s,t,u) = - \frac{g_{\rho\pi\pi}^2}{m_\rho^2}
  \left[
    \frac{t(u-s)}{m_\rho^2-t} + \frac{u(t-s)}{m_\rho^2-u} 
  \right]
\ ,
\end{eqnarray}
where we used $s+t+u=0$.
Thus, the sum of the contributions from (b) and (c) does not
contribute in the low-energy limit and only the diagram (a)
contributes, which is perfectly consistent with the low-energy theorem
of the $\pi\pi$-scattering amplitude.
This can be easily seen as follows:
In the low-energy region we can neglect the kinetic term of $\rho$,
i.e., ${\cal L}_{\rm kin}(V_\mu) = 0$ in 
Eq.~(\ref{leading Lagrangian 0}),
and then the field $V_\mu$ becomes just an auxiliary field.
Integrating out the auxiliary field $V_\mu$ leads to
$a {\cal L}_{\rm V} = 0$ in Eq.~(\ref{leading Lagrangian 0}).
There remains only ${\cal L}_{\rm A}$ 
term which is nothing but the chiral
Lagrangian with the least derivative term.
Then the result precisely reproduces the low-energy theorem.

\subsection{Vector meson saturation of the low energy constants\\
(Relation to the ChPT)}
\label{ssec:VMSLEC}

Integrating out the vector mesons in the Lagrangian of the
HLS given in Eq.~(\ref{leading Lagrangian 0})
we obtain the Lagrangian for pseudoscalar mesons.
The resultant Lagrangian includes ${\cal O}(p^4)$ terms of the ChPT
in addition to 
${\cal O}(p^2)$ terms.
To perform this it is convenient to introduce the following quantities:
\begin{eqnarray}
\alpha_{\perp\mu} &=&
\left( 
  {\cal D}_\mu \xi_{\rm R} \cdot \xi_{\rm R}^\dag -
  {\cal D}_\mu \xi_{\rm L} \cdot \xi_{\rm L}^\dag 
\right)
/ (2i) \ ,
\nonumber\\
\alpha_{\parallel\mu} &=&
\left( 
  {\cal D}_\mu \xi_{\rm R} \cdot \xi_{\rm R}^\dag +
  {\cal D}_\mu \xi_{\rm L} \cdot \xi_{\rm L}^\dag 
\right)
/ (2i) \ ,
\end{eqnarray}
where ${\cal D}_\mu\xi_{\rm L}$ and ${\cal D}_\mu\xi_{\rm L}$ are
defined by
\begin{eqnarray}
&& {\cal D}_\mu\xi_{\rm L} =
  \partial_\mu \xi_{\rm L} + i \xi_{\rm L} {\cal L}_\mu \ ,
\nonumber\\
&& {\cal D}_\mu\xi_{\rm R} =
  \partial_\mu \xi_{\rm R} + i \xi_{\rm R} {\cal R}_\mu 
\ .
\label{covder:2}
\end{eqnarray}
The relations of these $\alpha_{\perp\mu}$ and $\alpha_{\parallel\mu}$
with $\widehat{\alpha}_{\perp\mu}$ and
$\widehat{\alpha}_{\parallel\mu}$ in Eq.~(\ref{def:al hat})
are given by
\begin{eqnarray}
&& \widehat{\alpha}_{\perp\mu} = \alpha_{\perp\mu} \ , \nonumber\\
&& \widehat{\alpha}_{\parallel\mu} = \alpha_{\parallel\mu} - V_\mu \ .
\label{rel: al perp para}
\end{eqnarray}

{}From the Lagrangian in Eq.~(\ref{leading Lagrangian 0})
the equation of motion for the vector meson is given by
\begin{equation}
F_\sigma^2 \left( V_\mu - \alpha_{\parallel\mu} \right)
- \frac{1}{g^2} \left( \partial^\nu V_{\mu\nu}
- i \left[ V^\nu \,,\, V_{\mu\nu} \right] \right) = 0 \ .
\label{para0}
\end{equation}
In the leading order of the derivative expansion the solution of
Eq.~(\ref{para0}) is given by
\begin{equation}
V_\mu = \alpha_{\parallel\mu} 
+ \frac{1}{m_\rho^2} {\cal O}(p^3) \ .
\label{EOM V}
\end{equation}
Substituting this into the field strength of the HLS gauge boson and
performing the derivative expansion we obtain
\begin{eqnarray}
V_{\mu\nu} &=& \hat{\cal V}_{\mu\nu} + i 
\left[ \hat{\alpha}_{\perp\mu} \,,\, \hat{\alpha}_{\perp\nu} \right]
+ \frac{1}{m_\rho^2} {\cal O}(p^4) 
\nonumber\\
&=&
\xi_{\rm L} \left( U {\cal R}_{\mu\nu} U^\dag + {\cal L}_{\mu\nu}
+ \frac{i}{4} \nabla_\mu U \cdot \nabla_\nu U^\dag
- \frac{i}{4} \nabla_\nu U \cdot \nabla_\mu U^\dag
\right) \xi_{\rm L}^\dag
+ \frac{1}{m_\rho^2} {\cal O}(p^4) 
\nonumber\\
&=&
\xi_{\rm R} \left( {\cal R}_{\mu\nu} + U^\dag {\cal L}_{\mu\nu} U
+ \frac{i}{4} \nabla_\mu U^\dag \cdot \nabla_\nu U
- \frac{i}{4} \nabla_\nu U^\dag \cdot \nabla_\mu U
\right) \xi_{\rm R}^\dag
+ \frac{1}{m_\rho^2} {\cal O}(p^4) 
\ ,
\label{V0}
\end{eqnarray}
where we used
\begin{equation}
\hat{\alpha}_{\perp\mu} 
= \frac{i}{2} \xi_{\rm L} \cdot \nabla_\mu U \cdot \xi_{\rm R}^\dag
= \frac{1}{2i} \xi_{\rm R} \cdot \nabla_\mu 
  U^\dag \cdot \xi_{\rm L}^\dag
\ .
\label{perpU}
\end{equation}
By substituting Eq.~(\ref{perpU}) 
into the HLS Lagrangian,
the first term in the HLS Lagrangian 
(\ref{leading Lagrangian 0}) becomes
the first term in
the leading order ChPT Lagrangian
in Eq.~(\ref{leading ChPT}):
\begin{equation}
\left. {\cal L}_{(2)}^{\rm ChPT} \right\vert_{\chi=0}
=
\frac{F_\pi^2}{4} \mbox{tr}
\left[ \nabla_\mu U^\dag \nabla^\mu U \right]
\ .
\label{leading ChPT:2}
\end{equation}
In addition, 
the second term in Eq.~(\ref{leading Lagrangian 0})
with Eq.~(\ref{para0}) substituted
becomes of ${\cal O}(p^6)$ in the ChPT
and the third term (the kinetic term of the HLS gauge boson) 
with Eq.~(\ref{V0}) becomes
of ${\cal O}(p^4)$ in the ChPT:
\begin{eqnarray}
{\cal L}_4^V &=&
\frac{1}{32g^2} 
\left( 
\mbox{tr}\left[ \nabla_\mu U \nabla^\mu U^\dag \right] \right)^2
+ \frac{1}{16g^2} 
\,\mbox{tr}\left[ \nabla_\mu U \nabla_\nu U^\dag \right]
\mbox{tr} \left[ \nabla^\mu U \nabla^\nu U^\dag \right]
\nonumber\\
&&
{} -  \frac{3}{16g^2} \, \mbox{tr} \left[
  \nabla_\mu U \nabla^\mu U^\dag \nabla_\nu U \nabla^\nu U^\dag
\right]
\nonumber\\
&&
- i \frac{1}{4g^2} \, \mbox{tr}\left[
  {\cal L}_{\mu\nu} \nabla^\mu U \nabla^\nu U^\dag
  + {\cal R}_{\mu\nu} \nabla^\mu U^\dag \nabla^\nu U
\right]
\nonumber\\
&&
- \frac{1}{4g^2} \,\mbox{tr}
\left[ {\cal L}_{\mu\nu} U {\cal R}^{\mu\nu} U^\dag \right]
\nonumber\\
&&
- \frac{1}{8g^2} \left[ {\cal L}_{\mu\nu} {\cal L}^{\mu\nu}
 + {\cal R}_{\mu\nu} {\cal R}^{\mu\nu} \right]
\ ,
\end{eqnarray}
where we fixed $N_f=3$.
Comparing this with the ${\cal O}(p^4)$ terms of the ChPT Lagrangian
given in Eq.~(\ref{p4:ChPT}),
we obtain the contributions of vector mesons to the low-energy
parameters of the ChPT:
\begin{equation}
\begin{array}{lll}
\displaystyle L_1^V = \frac{1}{32g^2} \ , \quad
& \displaystyle L_2^V = \frac{1}{16g^2} \ , \quad
& \displaystyle L_3^V = - \frac{3}{16g^2} \ ,
\\
& & \\
\displaystyle L_9^V = \frac{1}{4g^2} \ , \quad
& \displaystyle L_{10}^V = - \frac{1}{4g^2} \ . \quad
& 
\end{array}
\end{equation}
In Table~\ref{tab:L from V}
we show the values of $L_i^V$ obtained 
by using the value of $g$ determined in the previous subsection,
$g= 5.80 \pm 0.91$
[Eq.~(\ref{gval:tree})], with the values of $L_i^r(m_\rho)$
in Ref.~\cite{Eck:89a}.
\begin{table}[htbp]
\begin{center}
\begin{tabular}{|l|c|c|}
\hline
 & $L_i^r(m_\rho)\times 10^3$ & $L_i^V\times 10^3$ \\
\hline
$L_1$ & $0.7\pm0.3$ & $0.93\pm0.29$ \\
$L_2$ & $1.3\pm0.7$ & $1.86\pm0.58$ \\
$L_3$ & $-4.4\pm2.5$ & $-5.6\pm1.8$  \\
$L_4$ & $-0.3\pm0.5$ & \\
$L_5$ & $1.4\pm0.5$ & \\
$L_6$ & $-0.2\pm0.3$ & \\
$L_7$ & $-0.4\pm0.15$ & \\
$L_8$ & $0.9\pm0.3$ & \\
$L_9$ & $6.9\pm0.7$ & $7.4\pm2.3$ \\
$L_{10}$ & $-5.2\pm0.7$ & $-7.4\pm2.3$ \\
\hline
\end{tabular}
\end{center}
\caption[Vector meson contribution to the low-energy constants of the
ChPT]{%
Values of low-energy constants derived from the HLS Lagrangian 
with lowest derivatives.}
\label{tab:L from V}
\end{table}
This shows that the low-energy constants $L_1$, $L_2$, $L_3$ and $L_9$
are almost saturated by the contributions from vector mesons at the
leading order~\cite{Eck:89a,Eck:89b,DRV}.
$L_{10}$ will be saturated by including the next order correction
[see section~\ref{sec:WM}].

\subsection{Relation to other models of vector mesons}
\label{ssec:ROMVM}

There are models to describe the vector mesons other than the
HLS.
In this subsection, we introduce several models of the vector
mesons, and show the equivalence between those and the HLS.

In Ref.~\cite{Eck:89b} it was shown that
the vector meson field can be introduced as the matter field in the 
CCWZ Lagrangian~\cite{CWZ,CCWZ}.
Hereafter we call this model the matter field method.
The equivalence of the model to the HLS was studied in 
Refs.~\cite{Eck:89b,Bir:96}.
However, the higher order terms of the HLS, which we will show in 
Sec.~\ref{sec:CPHLS}, were not considered.
Here we show the equivalence including the higher order terms in
the HLS after briefly reviewing the matter field method.

Another popular model is the so-called ``Massive Yang-Mills'' field
method~\cite{Schwinger:67,Schwinger:69,%
Wess-Zumino:67,Gas:69,KRS,Mei}.
Although the notion of the ``Massive Yang-Mills'' itself
does not literally make sense due to the mass term introduced by hand, 
the real meaning of ``Massive Yang-Mills'' approach was 
revealed~\cite{Yamawaki:87,BKY}
in terms of the generalized HLS (GHLS) including the axialvector 
mesons~\cite{BKY:NPB,Bando-Fujiwara-Yamawaki}: The ``Massive-Yang Mills''
Lagrangian is nothing but a special gauge of the GHLS with a
particular 
parameter choice and hence equivalent to the HLS model after
eliminating the 
axialvector 
mesons~\cite{Sch:86,Yamawaki:87,Golterman-HariDass,Meissner-Zahed}.
(For reviews, see Refs.~\cite{BKY,Mei}.)
We here briefly review the equivalence to 
the  ``Massive Yang-Mills'' model in view of GHLS.

In Refs.~\cite{Gas:84,Eck:89a}
the vector mesons are introduced as anti-symmetric tensor fields.
The equivalence was studied in Refs.~\cite{Eck:89b,Tan:96}.
Especially in Ref.~\cite{Tan:96},
the equivalence was shown with including
the higher order terms of the HLS.
Here we briefly review the model and equivalence mostly following
Ref.~\cite{Tan:96}.

In the following discussions we restrict ourselves to the chiral
limit.
The extensions to the case with the explicit chiral symmetry breaking
by the current quark masses are automatic.
As we will show below, there are differences in the off-shell
amplitude since the definitions of the off-shell fields are different
in the models.
Moreover, 
we can make the systematic derivative expansion in the HLS 
as we will show in Sec.~\ref{sec:CPHLS},
while we know no such systematic expansions in other models.
Thus, 
{\it the equivalence is valid only for the tree level on-shell
amplitude}.

\subsubsection{Matter field method}
\label{sssec:MFM}

Let us show the equivalence between the matter field
method and the HLS.

We first briefly describe the nonlinear sigma model based on the
manifold $G/H$~\cite{CWZ,CCWZ} with restricting ourselves to the case
for $G = \mbox{SU}(N_f)_{\rm L} \times \mbox{SU}(N_f)_{\rm R}$ and
$H = \mbox{SU}(N_f)_{\rm V}$, following Ref.~\cite{BKY}.~\footnote{%
An explanation in the present way for general $G$ and $H$ was given in
Ref.~\cite{BKY}.%
}

Let $\xi(\pi)$ be ``representatives'' of the (left) coset space $G/H$,
taking the value of the unitary matrix representation of $G$, which
are conveniently parametrized in terms of the NG bosons $\pi(x)$ as
\begin{equation}
\xi(\pi) = e^{i\pi(x)/F_\pi} \ , \quad
\pi(x) = \pi^a(x) T_a \ ,
\label{xi def}
\end{equation}
where we omit the summation symbol over $a$.
The transformation property of $\xi(\pi)$ under the chiral symmetry
is given by
\begin{equation}
G \ : \  \xi(\pi) \rightarrow 
\xi(\pi^{\prime}) 
= h(\pi,g_{\rm R},g_{\rm L}) \cdot \xi(\pi) \cdot g_{\rm R}^\dag
= g_{\rm L} \cdot \xi(\pi) \cdot h^\dag(\pi,g_{\rm R},g_{\rm L}) 
\ .
\end{equation}
The fundamental objects are the following Maurer-Cartan 1-forms
constructed from $\xi(\pi) \in G/H$:~\footnote{%
In Refs.~\cite{Eck:89a,Eck:89b} $u_\mu$ and $\Gamma_\mu$ were used
instead of $\alpha_{\perp\mu}$ and $\alpha_{\parallel\mu}$.
The relations between them are given by
$u_\mu = 2 \alpha_{\perp\mu}$ and
$\Gamma_\mu = -i \alpha_{\parallel\mu}$.
}
\begin{eqnarray}
&&
  \alpha_{\perp}^\mu = \frac{1}{2i}
  \left[
    {\cal D}^\mu \xi \cdot \xi^\dag -
    {\cal D}^\mu \xi^\dag \cdot \xi
  \right]
\ ,
\nonumber\\
&&
  \alpha_{\parallel}^\mu = \frac{1}{2i}
  \left[
    {\cal D}^\mu \xi \cdot \xi^\dag +
    {\cal D}^\mu \xi^\dag \cdot \xi
  \right]
\ ,
\label{def: al para}
\end{eqnarray}
where ${\cal D}^\mu \xi$ and ${\cal D}^\mu \xi^\dag$ are defined by
\begin{eqnarray}
&&
  {\cal D}^\mu \xi^\dag \equiv \partial^\mu \xi^\dag
  + i \xi^\dag {\cal L}^\mu
\ ,
\nonumber\\
&&
  {\cal D}^\mu \xi \equiv \partial^\mu \xi
  + i \xi {\cal R}^\mu
\ .
\end{eqnarray}
The transformation properties of these 1-forms are given
by
\begin{eqnarray}
&&
  \alpha_{\perp}^\mu \rightarrow
  h(\pi,g_{\rm R},g_{\rm L}) \cdot \alpha_{\perp}^\mu
  \cdot h^\dag (\pi,g_{\rm R},g_{\rm L}) 
\ ,
\nonumber\\
&&
  \alpha_{\parallel}^\mu \rightarrow
  h(\pi,g_{\rm R},g_{\rm L}) \cdot \alpha_{\parallel}^\mu
  \cdot h^\dag (\pi,g_{\rm R},g_{\rm L}) 
 -i\partial^\mu h(\pi,g_{\rm R},g_{\rm L}) \cdot
  h^\dag (\pi,g_{\rm R},g_{\rm L}) 
\ .
\end{eqnarray}
Only the perpendicular part $\alpha_\perp^\mu$ transforms
homogeneously, so that we can construct $G$-invariant from 
$\alpha_\perp^\mu$ alone:
\begin{equation}
{\cal L}_{\rm CCWZ} = 
F_\pi^2 \mbox{tr} \left[ \alpha_\perp^\mu \alpha_{\perp\mu} \right]
\ ,
\end{equation}
where the factor $F_\pi^2$ is added so as to normalize the kinetic
terms of the $\pi(x)$ fields.
It should be noticed that
with the relation
\begin{equation}
\alpha_{\perp\mu} 
= \frac{1}{2i}\,\xi^\dag \cdot \nabla_\mu U \cdot \xi^\dag
= \frac{i}{2} \,\xi \cdot \nabla_\mu U^\dag \cdot \xi
\ ,
\label{perp U0 2}
\end{equation}
substituted
${\cal L}_{\rm CCWZ}$ 
becomes identical to the first term of the chiral
Lagrangian in Eq.~(\ref{leading ChPT}).

Following Ref.~\cite{Eck:89b},
we include the vector meson as the matter field in the adjoint
representation,
\begin{equation}
\rho^{\rm(C)}_\mu = \sum_a \rho^{{\rm(C)} a}_\mu \, T_a \ .
\end{equation}
This transforms
homogeneously under the chiral symmetry:
\begin{equation}
\rho^{\rm(C)}_\mu \rightarrow
  h(\pi,g_{\rm R},g_{\rm L}) \cdot \rho^{\rm(C)}_\mu 
  \cdot h^\dag (\pi,g_{\rm R},g_{\rm L}) 
\ .
\label{trans rho C}
\end{equation}
The covariant derivative acting on the vector meson field is defined
by
\begin{equation}
D^{\rm(C)}_\mu \, \rho^{\rm(C)}_\nu \equiv
  \partial_\mu \rho^{\rm(C)}_\nu - i 
  \left[ \alpha_{\parallel\mu} \,,\, \rho^{\rm(C)}_\nu \right]
\ .
\end{equation}
It is convenient to define the following anti-symmetric combination of
the above covariant derivative:
\begin{equation}
\rho^{\rm(C)}_{\mu\nu} \equiv
D^{\rm(C)}_\mu \, \rho^{\rm(C)}_\nu - 
D^{\rm(C)}_\nu \, \rho^{\rm(C)}_\mu 
\ .
\label{def:rhomn:E}
\end{equation}
In addition we need 
the
field strengths of the external source fields ${\cal L}_\mu$ and
${\cal R}_\mu$.
These are given by
\begin{eqnarray}
&&
\widehat{\cal V}_{\mu\nu} \equiv \frac{1}{2}
  \left[ 
    \xi {\cal R}_{\mu\nu} \xi^\dag + \xi^\dag {\cal L}_{\mu\nu} \xi
  \right]
\ ,
\nonumber\\
&&
\widehat{\cal A}_{\mu\nu} \equiv \frac{1}{2}
  \left[ 
    \xi {\cal R}_{\mu\nu} \xi^\dag - \xi^\dag {\cal L}_{\mu\nu} \xi
  \right]
\ ,
\label{A V def 2}
\end{eqnarray}
which transform homogeneously:
\begin{eqnarray}
&&
\widehat{\cal V}_{\mu\nu} \rightarrow 
h(\pi,g_{\rm R},g_{\rm L}) \cdot \widehat{\cal V}_{\mu\nu}
  \cdot h^\dag (\pi,g_{\rm R},g_{\rm L}) 
\ ,
\nonumber\\
&&
\widehat{\cal A}_{\mu\nu} \rightarrow 
h(\pi,g_{\rm R},g_{\rm L}) \cdot \widehat{\cal A}_{\mu\nu}
  \cdot h^\dag (\pi,g_{\rm R},g_{\rm L}) 
\ .
\end{eqnarray}
Note that these expressions of $\widehat{\cal V}_{\mu\nu}$
and $\widehat{\cal A}_{\mu\nu}$ agree with those
in Eq.~(\ref{A V def}) when the unitary gauge of the HLS 
is taken.
The above $\alpha_{\perp\mu}$,
$\rho^{\rm(C)}_\mu$, $\rho^{\rm(C)}_{\mu\nu}$,
$\widehat{\cal V}_{\mu\nu}$ and $\widehat{\cal A}_{\mu\nu}$
together with the covariant derivative acting these fields defined by
\begin{equation}
D^{\rm(C)}_\mu \equiv \partial_\mu - i 
  \left[ \alpha_{\parallel\mu} \,,\, \ \ \right]
\ ,
\end{equation}
are the building blocks of the Lagrangian of the matter
field method.

The Lagrangian of the matter field method is constructed from the
building blocks given above.
An example of the Lagrangian
including the vector meson is given by~\cite{Eck:89b}
\begin{eqnarray}
{\cal L}_{C} &=&
  - \frac{1}{2} \mbox{tr} 
  \left[
    \rho^{\rm(C)}_{\mu\nu} \rho^{{\rm(C)}\mu\nu} 
  \right]
  + M_\rho^2 \mbox{tr} 
    \left[ \rho^{\rm(C)}_\mu \rho^{{\rm(C)}\mu} \right]
\nonumber\\
&&
  {}- f_V \mbox{tr} 
    \left[ \rho^{\rm(C)}_{\mu\nu} \, \widehat{\cal V}^{\mu\nu} \right]
  - 4 i g_V \mbox{tr}
    \left[ 
      \rho^{\rm(C)}_{\mu\nu} \alpha_\perp^\mu \alpha_\perp^\nu
    \right]
\ .
\label{Lag:E0}
\end{eqnarray}
In order to make the procedure more systematic,
the terms including the pseudoscalar meson are added to the Lagrangian 
in Eq.(\ref{Lag:E0}) in Ref.~\cite{Eck:89b}.
The entire Lagrangian is given by
\begin{eqnarray}
{\cal L}_{\bar{C}} &=&
{\cal L}_{\rm CCWZ} + {\cal L}_{C} 
+ \sum_{i=1,2,3,9,10} \gamma_i^{\rm(C)} \, P_i 
\ ,
\label{Lag:E}
\end{eqnarray}
where $P_i$ is the ${\cal O}(p^4)$ terms in the ChPT defined 
in Eqs.~(\ref{P0 - P3}), (\ref{P4 P5}), (\ref{P6 - P8})
and (\ref{P9 P10}).
By using $\alpha_{\perp\mu}$ these $P_i$ ($i=1,2,3,9,10$) are
expressed as
\begin{eqnarray}
&&
  P_1 = 16\, \mbox{tr} \left( \left[
    \alpha_{\perp\mu} \alpha_\perp^\mu
  \right] \right)^2
\ ,
\nonumber\\
&&
  P_2 = 16\, \mbox{tr} \left[
    \alpha_{\perp\mu} \alpha_{\perp\nu}
  \right] 
  \mbox{tr} \left[
     \alpha^\mu_\perp \alpha^\nu_\perp
  \right] 
\ ,
\nonumber\\
&&
  P_3 = 16\, \mbox{tr} \left[ 
    \alpha_{\perp\mu} \alpha_\perp^\mu
    \alpha_{\perp\nu} \alpha_\perp^\nu
  \right]
\ ,
\nonumber\\
&&
  P_9 = - 8 i \, \mbox{tr}\left[ 
    \widehat{\cal V}_{\mu\nu} \alpha_\perp^\mu \alpha_\perp^\nu 
  \right]
\ ,
\nonumber\\
&&
  P_{10} = \mbox{tr}
  \left[ \widehat{\cal V}_{\mu\nu} \hat{\cal V}^{\mu\nu} \right]
  - \mbox{tr}
  \left[ \widehat{\cal A}_{\mu\nu} \hat{\cal A}^{\mu\nu} \right]
\ .
\end{eqnarray}

Now that we have specified the Lagrangian for the matter field method,
we compare this with the HLS Lagrangian.
This is done by rewriting the above vector meson field
$\rho^{\rm(C)}_\mu$ into $\hat{\alpha}_{\parallel\mu}$ of the HLS
as
\begin{equation}
\rho^{\rm(C)}_\mu = \zeta \hat{\alpha}_{\parallel\mu}
= \zeta \left( \alpha_{\parallel\mu} - V_\mu \right)
\ ,
\label{rel: rho E}
\end{equation}
where $\zeta$ is a parameter related to the redefinition of the vector
meson field $V_\mu$ in the HLS.
It should be noticed that this relation is valid only when we take the
unitary gauge of the HLS.
The covariant derivative $D_\mu^{\rm(C)}$ is related to that in the HLS
$D_\mu$ as
\begin{equation}
D^{\rm(C)}_\mu = 
\partial_\mu 
- i \left[ V_\mu \,,\, \ \ \right]
- i \left[ \hat{\alpha}_{\parallel\mu} \,,\, \ \ \right]
= D_\mu 
- i \left[ \hat{\alpha}_{\parallel\mu} \,,\, \ \ \right]
\ .
\label{rel: covdel E}
\end{equation}
Then $\rho^{\rm(C)}_{\mu\nu}$ in Eq.~(\ref{def:rhomn:E}) is rewritten
as
\begin{eqnarray}
  \rho^{\rm(C)}_{\mu\nu} 
&=&
  \zeta
  \left(
    D_\mu \hat{\alpha}_{\parallel\nu}
    - D_\nu \hat{\alpha}_{\parallel\mu}
    - 2 i \left[ 
      \hat{\alpha}_{\parallel\mu} \,,\, \hat{\alpha}_{\parallel\nu}
   \right]
  \right)
\nonumber\\
&=&
- i \zeta 
\left[ 
  \hat{\alpha}_{\parallel\mu} \,,\, \hat{\alpha}_{\parallel\nu}
\right]
+
i \zeta
\left[ 
  \hat{\alpha}_{\perp\mu} \,,\, \hat{\alpha}_{\perp\nu}
\right]
+ \zeta \widehat{\cal V}_{\mu\nu}
- \zeta V_{\mu\nu}
\ ,
\label{rel: rhomn E}
\end{eqnarray}
where
\begin{equation}
D_\mu \hat{\alpha}_{\parallel\nu} \equiv
  \partial_\mu \hat{\alpha}_{\parallel\nu} - i 
  \left[ V_\mu \,,\, \hat{\alpha}_{\parallel\nu} \right]
\ ,
\end{equation}
and to obtain the second expression we used the following identity
[see Eq.~(\ref{rel:parallel})]:
\begin{eqnarray}
D_\mu \hat{\alpha}_{\parallel\nu}
- D_\nu \hat{\alpha}_{\parallel\mu}
&=&
i \left[ 
  \hat{\alpha}_{\parallel\mu} \,,\, \hat{\alpha}_{\parallel\nu}
\right]
+
i \left[ 
  \hat{\alpha}_{\perp\mu} \,,\, \hat{\alpha}_{\perp\nu}
\right]
+ \widehat{\cal V}_{\mu\nu}
- V_{\mu\nu}
\ .
\label{rel:parallel 0}
\end{eqnarray}
In addition, as shown in Eq.~(\ref{rel: al perp para})
$\alpha_{\perp\mu}$ agrees with $\widehat{\alpha}_{\perp\mu}$ in the
unitary gauge of the HLS:
\begin{equation}
\alpha_{\perp\mu} = \widehat{\alpha}_{\perp\mu} \ .
\label{rel: al perp}
\end{equation}
Here we should note that
the expressions of $\widehat{\cal V}_{\mu\nu}$
and $\widehat{\cal A}_{\mu\nu}$ in Eq.~(\ref{A V def 2}) are
equivalent to 
those in the HLS with unitary gauge.
Then, together with this fact,
Eqs.~(\ref{rel: rho E}), (\ref{rel: covdel E}),
(\ref{rel: rhomn E}) and (\ref{rel: al perp}) 
show that 
{\it all the building blocks of the
Lagrangian of the matter field method are expressed by the building
blocks of the HLS Lagrangian}.
Therefore,  
{\it
for any Lagrangian of the matter field method consisting of such
building blocks, whatever the form it takes, we can construct the
equivalent Lagrangian of the HLS}.

Let us express the Lagrangian in Eq.~(\ref{Lag:E}) using the building
blocks of the HLS, and obtain the relations between the parameters in
the matter field method and those in the HLS.
The first and the third term in Eq.~(\ref{Lag:E}) 
are already expressed by
$\alpha_{\perp\mu}$, $\widehat{\cal V}_{\mu\nu}$
and $\widehat{\cal A}_{\mu\nu}$, so we concentrate on the second term,
${\cal L}_{C}$.
This is expressed as
\begin{eqnarray}
{\cal L}_C &=&
\zeta^2 M_\rho^2 \, \mbox{tr} \left[ 
  \hat{\alpha}_{\parallel\mu} \hat{\alpha}_{\parallel}^{\mu} 
\right]
- \frac{\zeta^2}{2} \, \mbox{tr} \left[ V_{\mu\nu} V^{\mu\nu} \right]
\nonumber\\
&&
{} + \left( - 3 \zeta^2 - 12 \zeta g_V \right)
\mbox{tr} \left[ 
  \hat{\alpha}_{\perp\mu} \hat{\alpha}_\perp^\mu
  \hat{\alpha}_{\perp\nu} \hat{\alpha}_\perp^\nu
\right]
+ \left( - 3 \zeta^2 \right)
\mbox{tr} \left[
  \hat{\alpha}_{\parallel\mu} \hat{\alpha}_\parallel^\mu
  \hat{\alpha}_{\parallel\nu} \hat{\alpha}_\parallel^\nu
\right]
\nonumber\\
&&
{} + \left( - 2\zeta^2 - 4 \zeta g_V \right)
\mbox{tr} \left[
  \hat{\alpha}_{\perp\mu} \hat{\alpha}_{\perp\nu}
  \hat{\alpha}^\mu_\parallel \hat{\alpha}^\nu_\parallel
\right]
+ \left( 2 \zeta^2 + 4 \zeta g_V \right)
\mbox{tr} \left[
  \hat{\alpha}_{\perp\mu} \hat{\alpha}_{\perp\nu}
  \hat{\alpha}^\nu_\parallel \hat{\alpha}^\mu_\parallel
\right]
\nonumber\\
&&
{} + \left( \frac{\zeta^2}{2} + 2 \zeta g_V \right)
\left(
\mbox{tr} \left[
  \hat{\alpha}_{\perp\mu} \hat{\alpha}_\perp^\mu
\right] \right)^2
+ \left( \zeta^2 + 4 \zeta g_V \right)
\mbox{tr} \left[
  \hat{\alpha}_{\perp\mu} \hat{\alpha}_{\perp\nu}
\right]
\mbox{tr} \left[
  \hat{\alpha}^\mu_\perp \hat{\alpha}^\nu_\perp
\right] 
\nonumber\\
&&
{} + \left( \frac{\zeta^2}{2} \right)
\left( \mbox{tr} \left[ 
  \hat{\alpha}_{\parallel\mu} \hat{\alpha}_\parallel^\mu
\right] \right)^2
+ \left( \zeta^2 \right)
\mbox{tr} \left[
  \hat{\alpha}_{\parallel\mu} \hat{\alpha}_{\parallel\nu}
\right]
\mbox{tr} \left[
  \hat{\alpha}^\mu_\parallel \hat{\alpha}^\nu_\parallel
\right]
\nonumber\\
&&
{} + \left( - \frac{\zeta^2}{2} + \zeta f_V \right)
\mbox{tr}
 \left[ \widehat{\cal V}_{\mu\nu} \widehat{\cal V}^{\mu\nu} \right]
+ \left( \zeta^2 + \zeta f_V \right)
\mbox{tr}\left[ \widehat{\cal V}_{\mu\nu} V^{\mu\nu} \right]
\nonumber\\
&&
{} + i \left( 2 \zeta^2 + 4 \zeta g_V \right)
\mbox{tr}\left[ 
  V_{\mu\nu} \hat{\alpha}_\perp^\mu \hat{\alpha}_\perp^\nu 
\right]
{} + i \left( 2 \zeta^2 \right)
\mbox{tr}\left[ 
  V_{\mu\nu} \hat{\alpha}_\parallel^\mu \hat{\alpha}_\parallel^\nu 
\right]
\nonumber\\
&&
{} + i \left( - 2 \zeta^2 - 2 \zeta f_V - 4 \zeta g_V \right)
\mbox{tr}\left[ 
  \widehat{\cal V}_{\mu\nu} \hat{\alpha}_\perp^\mu 
  \hat{\alpha}_\perp^\nu 
\right]
{} + i \left( 2 \zeta^2 + 2 \zeta f_V \right)
\mbox{tr}\left[ 
  \widehat{\cal V}_{\mu\nu} \hat{\alpha}_\parallel^\mu
  \hat{\alpha}_\parallel^\nu
\right]
\ ,
\label{Lag:E HLS}
\end{eqnarray}
where we used the following relation valid for $N_f=3$:
\begin{eqnarray}
&&
\mbox{tr} \left[
  \hat{\alpha}_{\perp\mu} \hat{\alpha}_{\perp\nu}
  \hat{\alpha}^\mu_\perp \hat{\alpha}^\nu_\perp
\right]
=
\mbox{tr} \left[
  \hat{\alpha}_{\perp\mu} \hat{\alpha}_{\perp\nu}
\right] 
\mbox{tr} \left[
  \hat{\alpha}^\mu_\perp \hat{\alpha}^\nu_\perp
\right]
+ \frac{1}{2}
\left(
\mbox{tr} \left[
  \hat{\alpha}_{\perp\mu} \hat{\alpha}_\perp^\mu
\right] \right)^2
- 2 
\mbox{tr} \left[ 
  \hat{\alpha}_{\perp\mu} \hat{\alpha}_\perp^\mu
  \hat{\alpha}_{\perp\nu} \hat{\alpha}_\perp^\nu
\right]
\ ,
\nonumber\\
&&
\mbox{tr} \left[
  \hat{\alpha}_{\parallel\mu} \hat{\alpha}_{\parallel\nu}
  \hat{\alpha}^\mu_\parallel \hat{\alpha}^\nu_\parallel
\right]
=
\mbox{tr} \left[
  \hat{\alpha}_{\parallel\mu} \hat{\alpha}_{\parallel\nu}
\right]
\mbox{tr} \left[
  \hat{\alpha}^\mu_\parallel \hat{\alpha}^\nu_\parallel
\right]
+ \frac{1}{2}
\left( \mbox{tr} \left[ 
  \hat{\alpha}_{\parallel\mu} \hat{\alpha}_\parallel^\mu
\right] \right)^2
- 2
\mbox{tr} \left[
  \hat{\alpha}_{\parallel\mu} \hat{\alpha}_\parallel^\mu
  \hat{\alpha}_{\parallel\nu} \hat{\alpha}_\parallel^\nu
\right]
\ .
\end{eqnarray}
The combination of the above ${\cal L}_C$ with ${\cal L}_{\rm CCWZ}$
and $P_i$ terms as in Eq.~(\ref{Lag:E}) gives the Lagrangian of the
matter field method written by using the HLS fields.
To compare this with the HLS Lagrangian we need to include the higher
order terms in addition to the terms given in 
Eq.~(\ref{leading Lagrangian 0}).
As we will show in Sec.~\ref{sec:CPHLS}, 
we can perform the systematic low-energy
derivative expansion in the HLS.
The ${\cal O}(p^4)$ terms in the counting scheme 
are listed in Eqs.~(\ref{Lag: y terms}),
(\ref{Lag: w terms}) and (\ref{Lag: z terms}) in 
Sec.~\ref{ssec:OP4L}.
Thus, the comparison of 
the ${\cal L}_{\bar{C}}$ written in terms of the HLS
field with the HLS Lagrangian including ${\cal O}(p^4)$ terms
leads to the relations between the parameters in two methods.
First, comparing 
the second and third terms in 
Eq.~(\ref{leading Lagrangian 0})
with 
the first and second terms 
in Eq.~(\ref{Lag:E HLS}), we obtain
\begin{equation}
\frac{1}{g^2} = \zeta^2 \ , \quad
F_\sigma^2 = \zeta^2 M_\rho^2 \ .
\end{equation}
Second, comparing the $y_i$ terms of the HLS in (\ref{Lag: y terms})
with the third to tenth terms in 
Eq.~(\ref{Lag:E HLS}) combined with $P_i$ ($i=1,2,3$) terms,
we obtain
{\arraycolsep = 10pt
\renewcommand{\arraystretch}{1.8}
\begin{eqnarray}
\begin{array}{ll}
\displaystyle
y_1 = - 3 \zeta^2 - 12 \zeta g_V + 16 \gamma_3^{\rm(C)} \ , &
\displaystyle
y_3 = - 3 \zeta^2 \ , 
\\
\displaystyle
y_6 = - 2\zeta^2  - 4 \zeta g_V \ , &
\displaystyle
y_7 = \zeta^2 + 4 \zeta g_V \ , 
\\
\displaystyle
y_{10} = \frac{\zeta^2}{2} + 2 \zeta g_V + 16 \gamma_1^{\rm(C)} \ , &
\displaystyle
y_{11} = \zeta^2 + 4 \zeta g_V + 16 \gamma_2^{\rm(C)} \ , 
\\
\displaystyle
y_{12} = \frac{\zeta^2}{2} \ , &
\displaystyle
y_{13} = \zeta^2 \ .
\end{array}
\end{eqnarray}%
}
\noindent%
Finally, 
comparing the $z_i$ terms of the HLS in (\ref{Lag: z terms})
with the eleventh to sixteenth terms in 
Eq.~(\ref{Lag:E HLS}) combined with $P_i$ ($i=9,10$) terms,
we obtain
{\arraycolsep = 10pt
\renewcommand{\arraystretch}{1.8}
\begin{eqnarray}
\begin{array}{ll}
\displaystyle
z_1 = - \frac{\zeta^2}{2} - \zeta f_V + \gamma_{10}^{\rm(C)} \ , &
\displaystyle
z_2 = - \gamma_{10}^{\rm(C)} \ , 
\\
\displaystyle
z_3 = \zeta^2 + \zeta f_V \ , &
\displaystyle
z_4 = 2 \zeta^2 + 4 \zeta g_V \ , 
\\
\displaystyle
z_5 = 2 \zeta^2 \ , &
\displaystyle
z_6 = - 2 \zeta^2 - 2 \zeta f_V - 4 \zeta g_V - 8 \gamma_9^{\rm(C)} \ , 
\\
\displaystyle
z_7 = 2 \zeta^2 + 2 \zeta f_V \ . & 
\end{array}
\label{cor: z E}
\end{eqnarray}%
}

Now let us discuss the number of the parameters in two methods.
The Lagrangian of the HLS is given by the sum of the ${\cal O}(p^2)$
terms in Eq.~(\ref{leading Lagrangian}) and ${\cal O}(p^4)$ terms in
Eqs.~(\ref{Lag: y terms}), (\ref{Lag: w terms}) and 
(\ref{Lag: z terms}), which we call ${\cal L}_{\rm HLS(2+4)}$.
The Lagrangian of the matter field method include
the mass and kinetic terms of the vector meson and the interaction
terms with one vector meson field
in addition to the ${\cal O}(p^2) + {\cal O}(p^4)$ terms of the ChPT
Lagrangian.
Then we consider $y_i$ ($i=1$, $10$, $11$) 
and $z_i$ ($i=1,\ldots,5$) terms
in addition to the leading order terms in ${\cal L}_{\rm HLS(2+4)}$.
First of all,
${\cal L}_{\rm CCWZ}$ in Eq.~(\ref{Lag:E}) exactly agrees with
${\cal L}_{\rm A}$ in 
Eq.~(\ref{leading Lagrangian}) or
Eq.~(\ref{leading Lagrangian 0}), 
so that
we consider other terms.
For the four-point interaction of the pseudoscalar mesons,
${\cal L}_{\bar{C}}$ as well as ${\cal L}_{\rm HLS(2+4)}$
include three independent terms:
There are correspondences between
$\gamma_i^{\rm(C)}$ ($i=1,2,3$) in ${\cal L}_{\bar{C}}$ and
$y_i$ ($i=10,11,1$) in ${\cal L}_{\rm HLS(2+4)}$.
Similarly, comparing the terms with the external gauge fields,
we see that
$\gamma_{10}^{\rm(C)}$ and $\gamma_{9}^{\rm(C)}$ correspond to
$z_1 - z_2$ and $z_6$, respectively.

The remaining parameters in ${\cal L}_{\bar{C}}$
are $M_\rho$, $f_V$ and $g_V$, while those in 
${\cal L}_{\rm HLS(2+4)}$ are
$F_\sigma$, $g$, $z_3$ and $z_4$.
One might think that the HLS Lagrangian contains more parameters than
the matter field Lagrangian does.
However, one of $F_\sigma$, $g$, $z_3$ and $z_4$ can be absorbed into
redefinition of the vector meson field~\cite{Tan:96}
as far as 
we disregard the counting scheme in the HLS and
take ${\cal L}_{\rm HLS(2+4)}$ as just a model Lagrangian.
Then the numbers of the parameters in two methods exactly agree with
each other as far as the on-shell amplitude is concerned.

Here we show how one of $F_\sigma$, $g$, $z_3$ and $z_4$ can be
absorbed into redefinition of the vector meson field
in the HLS~\cite{Tan:96}:
\begin{equation}
V_\mu \rightarrow V_\mu + (1 - K) \hat{\alpha}_{\parallel\mu}
\ .
\label{redef:V}
\end{equation}
This redefinition leads to~\cite{Tan:96}
\begin{eqnarray}
&&
V_{\mu\nu} \rightarrow 
K V_{\mu\nu} + (1-K) \widehat{\cal V}_{\mu\nu}
+ K (1-K) 
i \left[ 
  \hat{\alpha}_{\parallel\mu} \,,\, \hat{\alpha}_{\parallel\nu}
\right]
+ (1-K)
i \left[ 
  \hat{\alpha}_{\perp\mu} \,,\, \hat{\alpha}_{\perp\nu}
\right]
\ ,
\nonumber\\
&&
\hat{\alpha}_{\parallel\mu} \rightarrow 
K \hat{\alpha}_{\parallel\mu}
\ .
\end{eqnarray}
Then the Lagrangian ${\cal L}_{\rm HLS(2+4)}$ is changed as
\begin{eqnarray}
{\cal L}_{\rm HLS(2+4)} &\rightarrow&
K^2 F_\sigma^2 \, \mbox{tr} \left[ 
  \hat{\alpha}_{\parallel\mu} \hat{\alpha}_{\parallel}^{\mu} 
\right]
- \frac{K^2}{2g^2} \, \mbox{tr} \left[ V_{\mu\nu} V^{\mu\nu} \right]
\nonumber\\
&&
{}+ \left( - \frac{K(1-K)}{g^2} + K z_3 \right)
\mbox{tr}\left[ \widehat{\cal V}_{\mu\nu} V^{\mu\nu} \right]
\nonumber\\
&&
{} + i \left( - \frac{2K(1-K)}{g^2} + K z_4 \right)
\mbox{tr}\left[ 
  V_{\mu\nu} \hat{\alpha}_\perp^\mu \hat{\alpha}_\perp^\nu 
\right]
+ \cdots
\ ,
\label{lag:HLS redef}
\end{eqnarray}
where dots stand for the terms irrelevant to the present
discussion.
Since $K$ is an arbitrary parameter, we choose
\begin{equation}
K = 1 - \frac{g^2}{2} z_4 \ ,
\label{choice K E}
\end{equation}
so that the fourth term in Eq.~(\ref{lag:HLS redef}) disappear.
The redefinitions of the other parameters such as
\begin{eqnarray}
&& 
  F_\sigma \rightarrow F_\sigma/K \ , \quad
  g \rightarrow g K \ ,
\nonumber\\
&&
  z_3 \rightarrow \frac{z_3}{K} + \frac{1-K}{g^2} \ ,
  \quad \cdots \ ,
\label{redef:para E}
\end{eqnarray}
give the HLS Lagrangian ${\cal L}_{\rm HLS(2+4)}$ without $z_4$ 
term.~\footnote{%
  In Ref.~\cite{Tan:96} instead of $z_4$ term
  $z_3$ term is eliminated.  
  Here we think that eliminating $z_4$ term is more convenient
  since $z_3$ term is needed to explain the deviation of the on-shell
  KSRF I relation from one. [See Eq.~(\ref{LET val}) and analysis in
  Sec.~\ref{sec:WM}.]%
}

In rewriting ${\cal L}_{\bar{C}}$ into the HLS form there is an
arbitrary parameter $\zeta$ as in Eq.~(\ref{rel: rho E}).
This $\zeta$ corresponds to the above parameter $K$ for the
redefinition of the vector meson field in the HLS.
We fix $\zeta$ to eliminate
$z_4$ in Eq.~(\ref{cor: z E}):
\begin{equation}
\zeta = - 2 g_V \ .
\label{fix:zeta}
\end{equation}
Then we have the following correspondences between the parameters in
the HLS and those in the matter field method:
\begin{eqnarray}
\frac{1}{g} = 2 g_V \ ,\quad 
F_\sigma = 2 g_V M_\rho \ ,
\quad
z_3 = 2 g_V ( 2 g_V - f_V ) \ .
\end{eqnarray}

In the above discussions,
we have shown that ${\cal L}_{\bar{C}}$ is rewritten into
${\cal L}_{\rm HLS(2+4)}$ and the number of parameters are exactly
same in both Lagrangian.
Although the on-shell amplitudes are equivalent in two methods,
off-shell structures are different with each other.
This is seen in the $\rho\pi\pi$ coupling $g_{\rho\pi\pi}$
and the $\rho$-$\gamma$ mixing strength $g_\rho$.
The on-shell $g_{\rho\pi\pi}$ and $g_\rho$ are given by
\begin{eqnarray}
\left. g_{\rho\pi\pi} (p_\rho^2=m_\rho^2)\right\vert_{\rm HLS}
= g\, \frac{F_\sigma^2}{2F_\pi^2}
&=&
\left. g_{\rho\pi\pi} (p_\rho^2=m_\rho^2) \right\vert_{\rm C}
= \frac{g_V M_\rho^2}{F_\pi^2} \ ,
\nonumber\\
\left. g_{\rho} (p_\rho^2=m_\rho^2)\right\vert_{\rm HLS}
= g F_\sigma^2 \left[ 1 - g^2 z_3 \right]
&=&
\left. g_{\rho} (p_\rho^2=m_\rho^2)\right\vert_{\rm C}
= M_\rho^2 f_V \ ,
\end{eqnarray}
where we add $(p_\rho^2=m_\rho^2)$ to express the on-shell quantities.
In the low energy limit ($p_\rho^2=0$),
on the other hand, they are given by
\begin{eqnarray}
\left. g_{\rho\pi\pi} (p_\rho^2=0)\right\vert_{\rm HLS}
= g\, \frac{F_\sigma^2}{2F_\pi^2}
&\neq&
\left. g_{\rho\pi\pi} (p_\rho^2=0) \right\vert_{\rm C}
= 0 \ ,
\nonumber\\
\left. g_{\rho} (p_\rho^2=0)\right\vert_{\rm HLS}
= g F_\sigma^2 
&\neq&
\left. g_{\rho} (p_\rho^2=0)\right\vert_{\rm C}
= 0
\ .
\end{eqnarray}
These implies that {\it
two methods give different results for the off-shell amplitude
although they are completely equivalent as far as the
on-shell tree-level amplitudes are concerned}.

We should stress here that the redefinition in Eq.~(\ref{redef:V})
is possible only
when we omit the counting scheme in the HLS and
regard ${\cal L}_{\rm HLS(2+4)}$ as the model Lagrangian.
When we introduce the systematic derivative expansion in the HLS as we
will show in Sec.~\ref{sec:CPHLS},
the HLS gauge coupling constant $g$ is counted as ${\cal O}(p)$ while
other parameters are counted as ${\cal O}(1)$.
Since 
the redefinition in Eq.~(\ref{redef:V}) mixes ${\cal O}(p^2)$ terms
with ${\cal O}(p^4)$ terms,
we cannot make such a redefinition.
Actually, the redefinition of the parameters in 
Eq.~(\ref{redef:para E})
is inconsistent with the counting rule.
As a result of the systematic derivative expansion,
all the parameters in the HLS are viable.
Thus the complete equivalence is lost in such a case.
Of course,
we have not known the systematic derivative expansion including the
vector meson in the matter field method~\footnote{%
  See discussions in Sec.~\ref{ssec:DEHLS}.
}, 
so that the discussion of the equivalence itself does not make 
sense.

\subsubsection{Massive Yang-Mills method}
\label{sssec:MYMM}

The ``Massive Yang-Mills''
fields~\cite{Schwinger:67,Schwinger:69,Wess-Zumino:67}
(for reviews, Ref.~\cite{Gas:69,Mei})
for vector mesons $\rho$ ($\rho$ meson and its flavor partners) and
axialvector mesons $A_1$ ($a_1$ mesons and 
its flavor partners) were introduced by gauging the chiral symmetry in
the nonlinear  
chiral Lagrangian in the same manner
as the external gauge fields ($\gamma, W^{\pm}, Z^0$ bosons) in
Eqs.~(\ref{external gauges}) and (\ref{W Z Q}) but  
were interpreted as vector and axialvector mesons instead of the
external gauge bosons.
Although axialvector mesons as well as vector mesons
must be simultaneously introduced  in
order that 
the chiral symmetry is preserved by this gauging, the gauged chiral
symmetry is explicitly 
broken anyway in this approach by the mass of these mesons introduced
by hand. 
Hence the ``Massive Yang-Mills'' field method as it stands does not
make sense as a gauge theory.
However, it was shown~\cite{Yamawaki:87} 
that the same Lagrangian can be regarded as a gauge-fixed
form of the
generalized HLS (GHLS) Lagrangian~\cite{BKY,Bando-Fujiwara-Yamawaki}
which is manifestly gauge-invariant under 
GHLS. In this sense the GHLS and the Massive Yang-Mills field method
are equivalent~\cite{Yamawaki:87,Meissner-Zahed,Golterman-HariDass}.

The GHLS is a natural extension of the HLS  
from $H_{\rm local}$ to $G_{\rm local}$ (``generalized HLS'') such
that the symmetry 
$G_{\rm global} \times H_{\rm local}$ is extended to  
to $G_{\rm global} \times 
G_{\rm local}$~\cite{BKY,Bando-Fujiwara-Yamawaki}. 
By this the axialvector mesons are incorporated together with the
vector mesons as the gauge bosons 
of the GHLS.
 
Let us introduce dynamical variables by extending Eq.~(\ref{div1}): 
\begin{equation}
U= \xi_{\rm L}^\dag \xi_{\rm M} \xi_{\rm R}, 
\end{equation}
where these dynamical variables transform as 
\begin{eqnarray}
&&
\xi_{\rm L,R} \rightarrow {\tilde g}_{\rm L,R}(x) \cdot 
\xi_{\rm L,R} \cdot g^{\dag}_{\rm L,R}
\ ,
\\
&&
\xi_{\rm M} \rightarrow {\tilde g}_{\rm L}(x) \cdot 
\xi_{\rm M} \cdot {\tilde g}_{\rm R}^{\dag}(x)
\ ,
\end{eqnarray} 
with ${\tilde g}_{\rm L,R} \in G_{\rm local}=
\left[
  \mbox{SU}(N_f)_{\rm L} \times \mbox{SU}(N_f)_{\rm R}
\right]_{\rm local}$ and
$g_{\rm L,R} \in
G_{\rm global}
=
\left[
  \mbox{SU}(N_f)_{\rm L} \times \mbox{SU}(N_f)_{\rm R}
\right]_{\rm global}$.
The covariant derivatives read:
\begin{eqnarray}
&&
D_\mu \xi_{\rm L} = \partial_\mu \xi_{\rm L} - i L_\mu \xi_{\rm L} 
+i \xi_{\rm L} {\cal L}_\mu
\ , \quad (L \leftrightarrow R)
\ ,
\\
&&
D_\mu \xi_{\rm M} =
\partial_\mu \xi_{\rm M} -i L_\mu \xi_{\rm M} + i \xi_{\rm M} R_\mu,
\end{eqnarray}
where we also have introduced the external
gauge  fields,
${\cal L}_\mu/{\cal R}_\mu={\cal V}_\mu \mp {\cal A}_\mu$
for gauging the
$G_{\rm global}$ in addition to the GHLS gauge bosons 
$L_\mu/R_\mu=V_\mu \mp  A_\mu$ for 
$G_{\rm local}$ as in Eq. (\ref{covder}).

There are four lowest derivative terms invariant under
($\mbox{\rm gauged-}G_{\rm global}) \times G_{\rm local}$:
\begin{eqnarray}
&&
{\cal L} = a {\cal L}_{\rm V}
    +b {\cal L}_{\rm A} + c {\cal L}_{\rm M} +
    d {\cal L}_\pi
\ ,
\nonumber\\
&& \quad 
{\cal L}_{\rm V} = F_\pi^2 \,
\mbox{tr}\left[
  \left(
  \frac{
      D_\mu \xi_{\rm L} \cdot \xi_{\rm L}^\dag
   + \xi_{\rm M} D_\mu \xi_{\rm R} \cdot \xi_{\rm R}^\dag
   \xi_{\rm M}^\dag
  }{2 i}
  \right)^2
\right]
\nonumber\\
&& \qquad\ 
=\frac{F_\pi^2}{4} \,
\mbox{tr}\left[
  \left(
    L_\mu- \widehat{\cal L}_\mu +
    \xi_{\rm M} (R_\mu- \widehat{\cal R}_\mu) \xi_{\rm M}^\dag
  \right)^2
\right]
\ ,
\nonumber\\
&& \quad
{\cal L}_{\rm A} = F_\pi^2 \,
\mbox{tr}\left[
  \left(
  \frac{
     D_\mu \xi_{\rm L}\cdot\xi_{\rm L}^\dag 
   - \xi_{\rm M} D_\mu \xi_{\rm R} \cdot 
     \xi_{\rm R}^\dag \xi_{\rm M}^\dag
  }{2 i} 
  \right)^2
\right]
\nonumber\\
&& \qquad\ 
=\frac{F_\pi^2}{4} \,
\mbox{tr}\left[
  \left(
    L_\mu- \widehat{\cal L}_\mu -
    \xi_{\rm M} (R_\mu- \widehat{\cal R}_\mu) \xi_{\rm M}^\dag
  \right)^2
\right]
\ ,
\nonumber\\
&& \quad
{\cal L}_{\rm M} = F_\pi^2 \, 
\mbox{tr}\left[
  \left( \frac{D_\mu \xi_{\rm M}\cdot \xi_{\rm M}^\dag}{2 i} \right)^2
\right]
=
F_\pi^2\mbox{tr}\left[A_\mu A^\mu\right]
\ ,
\nonumber\\
&& \quad
{\cal L}_\pi = F_\pi^2 \,
\mbox{tr}\left[
\left(
  \frac{
      D_\mu \xi_{\rm L}\cdot\xi_{\rm L}^\dag 
    - \xi_{\rm M} D_\mu \xi_{\rm R} \cdot \xi_{\rm R}^\dag
      \xi_{\rm M}^{\dag}-D_\mu \xi_{\rm M}\cdot \xi_{\rm M}^\dag
  }{2 i}
  \right)^2
\right]
\nonumber\\
&& \qquad\ 
=
F_\pi^2\mbox{tr}\left[ \widehat{\cal A}_\mu\widehat{\cal A}^\mu \right]
=
\frac{F_\pi^2}{4}
\mbox{tr} \left[ \nabla_\mu U \nabla^\mu U^\dag \right]
\ , 
\label{GHLS}
\end{eqnarray}
in addition to the kinetic terms of the HLS and the external gauge
bosons, where
we defined ``converted'' external fields:
\begin{eqnarray}
\widehat{\cal L}_\mu=\xi_{\rm L} {\cal L}_\mu \xi_{\rm L}^\dag
-i \partial_\mu \xi_{\rm L} \cdot \xi_{\rm L}^\dag
= \widehat{\cal V}_\mu -\widehat{\cal A}_\mu
\ , \\
\widehat{\cal R}_\mu=\xi_{\rm R} {\cal R}_\mu \xi_{\rm R}^\dag
-i \partial_\mu \xi_{\rm R} \cdot \xi_{\rm R}^\dag 
= \widehat{\cal V}_\mu +\widehat{\cal A}_\mu
\ ,
\end{eqnarray}
which transform exactly in the same way as the GHLS gauge fields
$L_\mu$ and $R_\mu$, respectively:
$\widehat{\cal L}_\mu \rightarrow {\tilde g}_{\rm L} 
\widehat{\cal L}_\mu
{\tilde g}_{\rm L}^\dag - i \partial_\mu  {\tilde g}_{\rm L}\cdot
{\tilde g}_{\rm L}^\dag$ (similarly for $L \leftrightarrow R$).
Note that ${\cal L}_\pi$ in Eq.~(\ref{GHLS}) is actually the gauged
nonlinear chiral 
Lagrangian, the first term of Eq.~(\ref{leading ChPT}). 

In this GHLS Lagrangian we have two kinds of independent gauge fields, 
one 
for $G_{\rm local}$ ($L_\mu/R_\mu$) including the vector ($\rho$)
and axialvector ($A_1$)
mesons 
and the other for (\mbox{gauged-})$G_{\rm global}$
(${\cal L}_\mu/{\cal R}_\mu$) including the external gauge fields
$\gamma$, $W^{\pm}$, $Z^0$.  
This is an outstanding  feature of the whole HLS approach, since the
basic dynamical variables 
$\xi_{\rm L}$ and $\xi_{\rm R}$ have two 
independent source charges,  
$G_{\rm local}$ for GHLS ($H_{\rm local}$ for HLS) and 
$G_{\rm global}$.
These two kinds of independent gauge fields are automatically
introduced through the  
covariant derivative.

Now, a particularly interesting parameter choice in the 
Lagrangian Eq.~(\ref{GHLS}) is $a=b=c=1$ ($d=0$), which actually
yields a successful phenomenology for axialvector mesons as well as
the vector mesons~\cite{BKY:NPB,Bando-Fujiwara-Yamawaki}:
By taking a special gauge $\xi_{\rm M}=1$,
the Lagrangian reads
\begin{eqnarray}
&&
F_\pi^2 \mbox{tr} 
  \left[ \left( V_\mu - \widehat{\cal V}_\mu \right)^2 \right]
+ F_\pi^2 \mbox{tr} 
  \left[ \left( A_\mu - \widehat{\cal A}_\mu \right)^2 \right]
+F_\pi^2  \mbox{tr} 
  \left[ A_\mu A^\mu \right]
\nonumber\\
&&
=
F_\pi^2 \mbox{tr} 
\left[ \left( V_\mu - \widehat{\cal V}_\mu \right)^2 \right]
+2 F_\pi^2 \mbox{tr} 
\left[ 
  \left( A_\mu - \frac{1}{2}\widehat{\cal A}_\mu \right)^2
\right]
+\frac{1}{2}  \frac{F_\pi^2}{4}
\mbox{tr} \left[
\nabla_\mu U \nabla^\mu U^\dagger
\right]\ .
\end{eqnarray}
The kinetic term of $\pi$ should be normalized by the rescaling
$\pi(x) \rightarrow \sqrt{2} \pi(x)$, 
$ F_\pi \rightarrow \sqrt{2} F_\pi$.
Then the Lagrangian finally takes the form:
\begin{eqnarray}
2 F_\pi^2 \mbox{tr} 
  \left[ \left( V_\mu - \widehat{\cal V}_\mu \right)^2 \right]
+4 F_\pi^2 \mbox{tr} 
  \left[
    \left( A_\mu - \frac{1}{2}\widehat{\cal A}_\mu \right)^2
  \right]
+ \frac{F_\pi^2}{4}
\mbox{tr} \left[
\nabla_\mu U \nabla^\mu U^\dagger
\right].
\label{AVLag}
\end{eqnarray}
This is the basis for the successful phenomenology including
the axialvector mesons in addition to the vector 
mesons~\cite{BKY:NPB,Bando-Fujiwara-Yamawaki}.

{}From this we can reproduce the HLS Lagrangian with
$G_{\rm global}\times H_{\rm local}$
in the energy region lower than the axialvector meson
mass $m_\rho < p < m_{A_1}$. 
In this region we can ignore the kinetic term of $A_\mu$
and hence the equation of motion for $A_\mu$ reads:
$A_\mu - \frac{1}{2}\widehat{\cal A}_\mu=0$, by which we can
solve away the $A_\mu$ field in such a way that the
second  term of Eq.~(\ref{AVLag}) simply yields zero.
Since the first and the third terms of Eq.~(\ref{AVLag}) are the same as 
$2 {\cal L}_V$ and ${\cal L}_A$ terms in Eq.~(\ref{leading Lagrangian 0}),  
we indeed get back
the HLS Lagrangian Eq.~(\ref{leading Lagrangian 0}) with $a=2$.
(The same argument can apply to the arbitrary choice of the 
parameters $a$, $b$, $c$, $d$ in Eq. (\ref{GHLS}), 
which by solving away $A_1$
reproduces
the HLS Lagrangian (\ref{leading Lagrangian 0}) with arbitrary
$a$.)

On the other hand, by taking another special gauge
$\xi_{\rm M}=U$, $\xi_{\rm L}=\xi_{\rm R}=1$, 
Eq.~(\ref{GHLS}) with $a=b=c=1$ ($d=0$)
is shown to coincide with the otherwise unjustified
Massive Yang-Mills Lagrangian:~\cite{Yamawaki:87}
\begin{equation}
{\cal L} = F_\pi^2 \, \mbox{tr} 
  \left[ (V_\mu - {\cal V}_\mu)^2 \right]
+ F_\pi^2 \, \mbox{tr} 
  \left[ (A_\mu - {\cal A}_\mu)^2 \right]
+ \frac{F_\pi^2}{4} \, \mbox{tr} 
  \left[ D_\mu U D^\mu U^\dag \right]
\ ,
\label{EMYM}
\end{equation}
with $D_\mu U\equiv \partial_\mu U - i L_\mu U + i U R_\mu$.
This takes the same form as the Massive Yang-Mills Lagrangian when the
external fields ${\cal L}_\mu$ and ${\cal R}_\mu$ are 
switched off, and hence the GHLS and the Massive Yang-Mills are
equivalent to each  
other~\cite{Yamawaki:87,Meissner-Zahed,Golterman-HariDass}.
(Reverse arguments were also made, starting with the Massive Yang-Mills
Lagrangian and arriving at the GHLS Lagrangian by a ``gauge 
transformation''~\cite{Kay:85,Sch:86}, although in the Massive
Yang-Mills 
notion there is no gauge symmetry in the literal sense.)

In spite of the same form
of the Lagrangian, however, the meaning of the fields is quite
different: In the absence of the 
external fields, the GHLS fields in this gauge-fixing no longer transform as
the gauge fields  
in sharp contrast to the Massive Yang-Mills notion. Namely, the GHLS
gauge bosons $L_\mu$ and $R_\mu$ actually   
transform as matter fields under global 
$G (\subset G_{\rm global}\times 
G_{\rm local})$: $L_\mu \rightarrow g_L L_\mu g_L^\dagger$,
$R_\mu \rightarrow g_R R_\mu g_R^\dagger$, and hence the mass term
does not contradict the gauge invariance 
in the GHLS case (This is because the mass term in the GHLS model is
from the Higgs mechanism  
$G_{\rm local} \times G_{\rm global} \rightarrow G$ in much the same
way as that in the HLS model.) 
In the presence of the external gauge fields, on the other hand, both
the external fields and 
the HLS fields do transform as gauge bosons under 
the same (\mbox{gauged-})$G$
symmetry which is a diagonal 
sum of the $G_{\rm local}$ and (\mbox{gauged-})$G_{\rm global}$. The
existence 
of the two kinds of gauge bosons 
transforming under the same group are due to the two independent
source charges of the GHLS model.  
[Equation~(\ref{EMYM}) 
was also derived within the notion of the Massive
Yang-Mills~\cite{Sch:86}, 
without clear conceptual origin of such two independent gauge
fields.]

To conclude the Massive Yang-Mills approach can be regarded as a 
gauge-fixed form of
the GHLS model and hence equivalent to the HLS model for the 
energy region $ m_\rho < p < m_{A_1}$, after solving away the axialvector
meson field.

\subsubsection{Anti-symmetric tensor field method}
\label{sssec:ATFM}

Let us show the equivalence between the 
anti-symmetric tensor field method (ATFM) and the HLS.

In Refs.~\cite{Gas:84,Eck:89a}
the vector meson field is introduced as an anti-symmetric tensor 
field $V^{\rm(T)}_{\mu\nu} = - V^{\rm(T)}_{\nu\mu}$, 
which transforms homogeneously under the
chiral symmetry:
\begin{equation}
V^{\rm(T)}_{\mu\nu} \rightarrow 
h(\pi,g_{\rm R},g_{\rm L}) \cdot V^{\rm(T)}_{\mu\nu}
\cdot h^\dag(\pi,g_{\rm R},g_{\rm L}) 
\ .
\end{equation}
The transformation property of the field is same as that of the matter
field method.
Then the covariant derivative acting on the field is defined in the
same way as in the matter field method:
\begin{equation}
D^{\rm(T)}_{\mu} \equiv
\partial_\mu - i \left[ \alpha_{\parallel\mu} \,,\, \ \ \right]
\ ,
\end{equation}
where $\alpha_{\parallel\mu}$ is given in 
Eq.~(\ref{def: al para}).
Other building blocks of the Lagrangian are exactly same as that in
the matter field method:
The building blocks are 
$V^{\rm(T)}_{\mu\nu}$,
$\alpha_{\perp\mu}$,
$\widehat{\cal V}_{\mu\nu}$ and $\widehat{\cal A}_{\mu\nu}$
together with the above covariant derivative.~\footnote{%
  The quantities $u_\mu$, $\Gamma_\mu$, $f_{+}^{\mu\nu}$
  and $f_{-}^{\mu\nu}$
  used in Ref.~\cite{Eck:89a}
  are related to $\alpha_{\perp\mu}$, $\alpha_{\parallel\mu}$,
  $\widehat{\cal V}_{\mu\nu}$ and $\widehat{\cal A}_{\mu\nu}$
  by
  $u_\mu = 2 \alpha_{\perp\mu}$,
  $\Gamma_\mu = - i \alpha_{\parallel\mu}$,
  $f_{+}^{\mu\nu} = 2 \widehat{\cal V}^{\mu\nu}$
  and $f_{-}^{\mu\nu} = - 2 \widehat{\cal A}^{\mu\nu}$.
}

For constructing the Lagrangian we should note that
the field $V^{\rm(T)}_{\mu\nu}$ contains six degrees of freedom.
To reduce them to three
the mass and kinetic terms of $V^{\rm(T)}_{\mu\nu}$ must have the
following form~\cite{Eck:89a}:
\begin{equation}
{\cal L}_{\rm T}^{\rm kin}
= - \frac{1}{2} \mbox{tr}
\left[
  D^{{\rm(T)}\mu} V_{\mu\nu}^{\rm(T)} 
  D_\lambda^{\rm(T)} V^{{\rm(T)}\lambda\nu}
\right]
+ \frac{M_v^2}{4} \mbox{tr}
\left[  V_{\mu\nu}^{\rm(T)} V^{{\rm(T)}\mu\nu} \right]
\ .
\end{equation}
The interaction terms are constructed from the building blocks shown
above.
An example of the Lagrangian is given by~\cite{Eck:89a}
\begin{eqnarray}
{\cal L}_{\rm T} = 
  F_\pi^2 \mbox{tr} 
    \left[ \alpha_{\perp\mu} \alpha_\perp^\mu \right]
  + {\cal L}_{\rm T}^{\rm kin} 
  + \frac{F_V}{\sqrt{2}} \mbox{tr}
    \left[ V^{\rm(T)}_{\mu\nu} \widehat{\cal V}_{\mu\nu} \right]
  + i \sqrt{2} G_V \mbox{tr}
    \biggl[ 
      V^{\rm(T)}_{\mu\nu} 
      \left[ \alpha_\perp^\mu \,,\, \alpha_\perp^\nu\right]
    \biggr]
\ .
\label{Lag:T}
\end{eqnarray}

Now let us rewrite the above Lagrangian in terms of the fields of the
HLS following Ref.~\cite{Tan:96}.
This is done by introducing the HLS gauge field $V_\mu$ 
in the unitary gauge
as an auxiliary field.
The dynamics is not modified by adding the auxiliary field to the
Lagrangian:
\begin{equation}
{\cal L}_{\rm T}^\prime = 
{\cal L}_{\rm T} + 
\frac{1}{2} \kappa^2 \mbox{tr} \left[
  \left(
    V_\mu - \alpha_{\parallel\mu} - \frac{1}{\kappa} 
    D^{{\rm(T)}\nu} V^{\rm(T)}_{\nu\mu}
  \right)
  \left(
    V^\mu - \alpha_{\parallel}^{\mu} - \frac{1}{\kappa} 
    D^{{\rm(T)}}_{\lambda} V^{{\rm(T)}\lambda\mu}
  \right)
\right]
\ ,
\label{def:L prime T}
\end{equation}
where $\kappa$ is an arbitrary parameter.
The terms including the derivative of $V^{\rm(T)}_{\mu\nu}$
in ${\cal L}_{\rm T}^\prime$  can be removed by 
performing the partial integral:
\begin{eqnarray}
&&
\mbox{tr} \left[
  \hat{\alpha}_{\parallel}^{\nu} D^{{\rm(T)}\mu}
  V^{\rm(T)}_{\mu\nu}
\right]
\Rightarrow
- \frac{1}{2}
\mbox{tr} \left[
  \left(
    D^\mu \hat{\alpha}_{\parallel}^\nu
    - D^\nu \hat{\alpha}_{\parallel}^\mu
  \right)
  V^{\rm(T)}_{\mu\nu}
\right]
+ i \mbox{tr}
\biggl[
  \left[ 
    \hat{\alpha}_{\parallel}^\mu \,,\, \hat{\alpha}_{\parallel}^\nu
  \right]
  V^{\rm(T)}_{\mu\nu}
\biggr]
\nonumber\\
&& \qquad\quad
=
\frac{i}{2} \mbox{tr}
\biggl[
  \left[ 
    \hat{\alpha}_{\parallel}^\mu \,,\, \hat{\alpha}_{\parallel}^\nu
  \right]
  V^{\rm(T)}_{\mu\nu}
\biggr]
- \frac{i}{2} \mbox{tr}
\biggl[
  \left[ 
    \hat{\alpha}_{\perp}^\mu \,,\, \hat{\alpha}_{\perp}^\nu
  \right]
  V^{\rm(T)}_{\mu\nu}
\biggr]
- \frac{1}{2} \mbox{tr} 
  \left[ \widehat{\cal V}^{\mu\nu} V^{\rm(T)}_{\mu\nu} \right]
+ \frac{1}{2} \mbox{tr} 
  \left[ V^{\mu\nu} V^{\rm(T)}_{\mu\nu} \right]
\ ,
\nonumber\\
\label{VT:part int}
\end{eqnarray}
where we used the identity in Eq.~(\ref{rel:parallel 0})
to obtain the second expression.
Substituting Eq.~(\ref{VT:part int}) into Eq.~(\ref{def:L prime T}),
we obtain
\begin{eqnarray}
&&
{\cal L}_{\rm T}^\prime = 
  F_\pi^2 \mbox{tr} 
    \left[ \alpha_{\perp\mu} \alpha_\perp^\mu \right]
  + \frac{M_v^2}{4} \mbox{tr}
    \left[  V_{\mu\nu}^{\rm(T)} V^{{\rm(T)}\mu\nu} \right]
\nonumber\\
&&\qquad
{} + \frac{i}{2} \kappa \mbox{tr}
\biggl[
  \left[ 
    \hat{\alpha}_{\parallel}^\mu \,,\, \hat{\alpha}_{\parallel}^\nu
  \right]
  V^{\rm(T)}_{\mu\nu}
\biggr]
- i \left( \frac{1}{2}\kappa - \sqrt{2} G_V \right) \mbox{tr}
\biggl[
  \left[ 
    \hat{\alpha}_{\perp}^\mu \,,\, \hat{\alpha}_{\perp}^\nu
  \right]
  V^{\rm(T)}_{\mu\nu}
\biggr]
\nonumber\\
&&\qquad
{} + \frac{1}{2} \kappa \mbox{tr} 
  \left[ V^{\mu\nu} V^{\rm(T)}_{\mu\nu} \right]
- \left( \frac{1}{2} \kappa - \frac{F_V}{\sqrt{2}} \right) \mbox{tr}
  \left[ \widehat{\cal V}^{\mu\nu} V^{\rm(T)}_{\mu\nu} \right]
+ \frac{1}{2} \kappa^2 \mbox{tr}
  \left[ 
    \hat{\alpha}_{\parallel}^{\mu} \hat{\alpha}_{\parallel\mu} 
  \right]
\ .
\end{eqnarray}
In the above Lagrangian, we can integrate out 
$V^{\rm(T)}_{\mu\nu}$ field.
Then, the Lagrangian becomes
\begin{eqnarray}
&&
{\cal L}_{\rm T}^\prime = 
F_\pi^2 \mbox{tr} 
  \left[ \alpha_{\perp\mu} \alpha_\perp^\mu \right]
+ \frac{1}{2} \kappa^2 \mbox{tr}
  \left[ 
    \hat{\alpha}_{\parallel}^{\mu} \hat{\alpha}_{\parallel\mu} 
  \right]
- \frac{\kappa^2}{2 M_v^2} \mbox{tr}
  \left[ V_{\mu\nu} V^{\mu\nu} \right]
\nonumber\\
&&\qquad
{} + \frac{\kappa(\kappa - \sqrt{2} F_V)}{2M_v^2}\, \mbox{tr}
  \left[ V_{\mu\nu} \widehat{\cal V}^{\mu\nu} \right]
+ i \frac{\kappa(\kappa - 2 \sqrt{2} G_V)}{M_v^2} \, \mbox{tr}
  \left[ 
    V_{\mu\nu} \hat{\alpha}_{\perp}^\mu \hat{\alpha}_{\perp}^\nu
  \right]
+ \cdots
\ ,
\end{eqnarray}
where dots stand for the terms irrelevant to the present analysis.
Comparing the above Lagrangian with the leading order
HLS Lagrangian in Eq.~(\ref{leading Lagrangian 0}) or
Eq.~(\ref{leading Lagrangian}) and
the $z_i$ terms of the HLS in Eq.~(\ref{Lag: z terms}),
we obtain the following relations:
{\arraycolsep = 10pt
\renewcommand{\arraystretch}{1.8}
\begin{eqnarray}
\begin{array}{ll}
\displaystyle
F_\sigma^2 = \frac{\kappa^2}{2} \ , &
\displaystyle
\frac{1}{g^2} = \frac{\kappa^2}{2M_v^2}
\\
\displaystyle
z_3 = \frac{\kappa(\kappa - \sqrt{2} F_V)}{2M_v^2} \ , &
\displaystyle
z_4 = \frac{\kappa(\kappa - 2 \sqrt{2} G_V)}{M_v^2} .
\end{array}
\label{rel:T}
\end{eqnarray}
}
As was discussed for $\zeta$ in Sec.~\ref{sssec:MFM},
the artificial coefficient $\kappa$ is related to the redefinition of
the vector meson field in the HLS~\cite{Tan:96}.
As far as we omit the counting scheme in the HLS and
regard ${\cal L}_{\rm HLS(2+4)}$ as the model Lagrangian,
we eliminate $z_4$ term by the redefinition.
Correspondingly, we fix $\kappa$ to eliminate $z_4$ in
Eq.~(\ref{rel:T}):
\begin{equation}
\kappa = 2 \sqrt{2} G_V \ .
\end{equation}
Then we have the following correspondences between the parameters in
the HLS and those in the anti-symmetric tensor field method:
\begin{eqnarray}
F_\sigma^2 = 4 G_V^2 \ ,
\quad
\frac{1}{g^2} = \frac{4 G_V^2}{M_v^2} \ ,
\quad
z_3 = \frac{2 G_V (2G_V - F_V)}{M_v^2} \ .
\end{eqnarray}
With these relations the Lagrangian in the anti-symmetric tensor field
method in Eq.~(\ref{Lag:T}) is equivalent to the leading order terms
and $z_3$ and $z_4$ terms in the HLS Lagrangian.

We should again note that the above equivalence holds only for the
on-shell amplitudes.
For the off-shell amplitudes the equivalence is lost as we discussed
for the matter filed method in Sec.~\ref{sssec:MFM}.

\subsection{Anomalous processes}
\label{ssec:AP}

In QCD with $N_f=3$ there exists a non-Abelian anomaly
which breaks the chiral symmetry explicitly.
In the effective chiral Lagrangian this anomaly is appropriately
reproduced by introducing the Wess-Zumino 
action~\cite{WZ,Witten}.
This can be generalized so as to incorporate vector mesons as
dynamical gauge bosons of the HLS~\cite{FKTUY}.
In this subsection, following 
Refs.~\cite{FKTUY} and \cite{BKY}, we briefly review
the way of incorporating vector mesons,
and then perform analyses on several physical processes focusing 
whether the vector dominance is satisfied in the electromagnetic form
factors. 
Here we restrict ourselves to the 
$G_{\rm global} \times H_{\rm local}
= \left[\mbox{U($3$)}_{\rm L} \times
\mbox{U($3$)}_{\rm R}\right]_{\rm global}
\times
\left[\mbox{U($3$)}_{\rm V}\right]_{\rm local}$ 
model, with $G_{\rm global}$ being fully gauged by the external
gauge field ${\cal L}_\mu$ and ${\cal R}_\mu$.

Since it is convenient to use the language of differential forms in
the proceeding discussions,
we define the following 1-forms:
\begin{eqnarray}
&& V \equiv V_\mu dx^\mu \ , \quad
   {\cal L} \equiv {\cal L}_\mu dx^\mu \ , \quad
   {\cal R} \equiv {\cal R}_\mu dx^\mu \ , 
\nonumber\\
&& \alpha \equiv \frac{1}{i} 
    \left( \partial_\mu U \right) U^{-1} dx^\mu
   = \frac{1}{i} \left( d U \right) U^{-1} \ ,
\quad
  \beta \equiv U^{-1} d U = U^{-1} \alpha U \ .
\end{eqnarray}
Let $\delta$ denote the transformation of 
$G_{\rm global} \times H_{\rm local}$:
\begin{equation}
\delta \equiv \delta_{\rm L} \left(\varepsilon_{\rm L}\right)
+ \delta_{\rm V} ( v )
+ \delta_{\rm R} \left(\varepsilon_{\rm R}\right),
\end{equation}
such that
\begin{eqnarray}
&&
\xi_{\rm L,R} \rightarrow  e^{iv} \xi_{\rm L,R} 
e^{- i \varepsilon_{\rm L,R} }
\ ,
\nonumber\\
&& \delta V = d v + i \left[ v\,,\, V\right] \ ,
\quad
\delta {\cal L} = d \varepsilon_{\rm L} 
  + i \left[ \varepsilon_{\rm L} \,,\, {\cal L} \right] \ ,
\quad
\delta {\cal L} = d \varepsilon_{\rm L} 
  + i \left[ \varepsilon_{\rm L} \,,\, {\cal L} \right] \ .
\end{eqnarray}

The essential point of the Wess-Zumino idea~\cite{WZ} is to notice
that the anomaly at composite level should coincide with that at
quark level.
Therefore the effective action $\Gamma$ which describes low energy
phenomena must satisfy the same anomalous Ward identity as that in
QCD,
\begin{equation}
\delta \Gamma 
\left[
  U , {\cal L}, {\cal R}
\right]
=
- \frac{N_c}{24\pi^2}
\int_{M^4} \mbox{tr}
\left[
  \varepsilon
  \left\{
    (d{\cal L})^2 - \frac{1}{2} i d {\cal L}^3
  \right\}
\right]
- (\mbox{L} \leftrightarrow \mbox{R}) ,
\label{eq:AnomEq}
\end{equation}
where $N_c$ ($=3$) is the number of colors.
Hereafter, we refer Eq.~(\ref{eq:AnomEq}) as the Wess-Zumino anomaly
equation.
The so-called Wess-Zumino action, which is a solution to the
Wess-Zumino anomaly equation in Eq.~(\ref{eq:AnomEq}), is given 
by~\cite{WZ,Witten}
\begin{eqnarray}
  \Gamma_{\rm WZ} 
  \left[
    U, {\cal L}, {\cal R}
  \right]
  =
  \frac{N_{\rm c}}{240\pi^2}
  \int_{M^5} \mbox{tr}\, (\alpha^5) + (\mbox{covariantization}) ,
\end{eqnarray}
where the integral is over a five-dimensional manifold $M^5$ whose
boundary is ordinary Minkowski space $M^4$, and (covariantization)
denotes the terms containing the external gauge fields ${\cal L}$ and
${\cal R}$~\cite{KRS}.
The explicit form of the above action is given by~\cite{KRS,FKTUY}
\begin{eqnarray}
  \Gamma_{\rm WZ} 
  \left[
    U, {\cal L}, {\cal R}
  \right]
&=&
  C \int_{M^5} \mbox{tr}\, (\alpha^5) 
  - 5 C i \int_{M^4} \mbox{tr} 
  \left[ {\cal L} \alpha^3 + {\cal R} \beta^3 \right]
\nonumber\\
&&
  {}- 5 C \int_{M^4} \mbox{tr} 
  \left[
    \left( d {\cal L} {\cal L} + {\cal L} d {\cal L} \right) \alpha
    +
    \left( d {\cal R} {\cal R} + {\cal R} d {\cal R} \right) \beta
  \right]
\nonumber\\
&&
  {}- 5 C i \int_{M^4} \mbox{tr} 
  \left[
    d {\cal L} d U {\cal R} U^{-1} - d {\cal R} d U^{-1} {\cal L} U
  \right]
\nonumber\\
&&
  {}+ 5 C i \int_{M^4} \mbox{tr} 
  \left[
    {\cal R} U^{-1} {\cal L} U \beta^2 
    - {\cal L} U {\cal R} U^{-1} \alpha^2
  \right]
\nonumber\\
&&
  {}+ \frac{5}{2} C i \int_{M^4} \mbox{tr} 
  \left[
    \left( {\cal L} \alpha \right)^2 - 
    \left( {\cal R} \beta \right)^2
  \right]
  {}+ 5 C i \int_{M^4} \mbox{tr} 
  \left[
    {\cal L}^3 \alpha + {\cal R}^3 \beta
  \right]
\nonumber\\
&&
  {}+ 5 C \int_{M^4} \mbox{tr} 
  \left[
    \left( d {\cal R} {\cal R} + {\cal R} d {\cal R} \right)
    U^{-1} {\cal L} U
    -
    \left( d {\cal L} {\cal L} + {\cal L} d {\cal L} \right)
    U {\cal R} U^{-1}
  \right]
\nonumber\\
&&
  {}+ 5 C i \int_{M^4} \mbox{tr} 
  \left[
    {\cal L} U {\cal R} U^{-1} {\cal L} \alpha +
    {\cal R} U^{-1} {\cal L} U {\cal R} \beta
  \right]
\nonumber\\
&&
  {}- 5 C i \int_{M^4} \mbox{tr} 
  \left[
    {\cal R}^3 U^{-1} {\cal L} U - {\cal L}^3 U {\cal R} U^{-1}
    + \frac{1}{2} \left( U {\cal R} U^{-1} {\cal L} \right)^2
  \right]
\ ,
\end{eqnarray}
where 
\begin{equation}
C = \frac{N_c}{240\pi^2} \ .
\end{equation}

In the model possessing the HLS 
$H_{\rm local} = \left[\mbox{U($3$)}_{\rm V}\right]_{\rm local}$
the general solution to 
the
Wess-Zumino anomaly equation in Eq.~(\ref{eq:AnomEq})
is given by~\cite{FKTUY}
\begin{equation}
\Gamma
\left[
  \xi_{\rm L}^{\dag} \xi_{\rm R} , V, {\cal L}, {\cal R}
\right]
=
\Gamma_{\rm WZ} 
\left[
  \xi_{\rm L}^{\dag} \xi_{\rm R} , {\cal L}, {\cal R}
\right]
+
\frac{N_c}{16\pi^2}
\int_{M^4} \sum_{i=1}^4
c_i {\cal L}_i ,
\label{Lag:Anom}
\end{equation}
where $c_i$ are arbitrary constants~\footnote{%
  The normalization of $c_i$ here is different by the factor
  $N_c/(16\pi^2)$ from that in Eq.~(7.49) of Ref.~\cite{BKY}.
}
and ${\cal L}_i$ are
gauge invariant 4-forms 
which conserve parity and charge conjugation but 
violate the intrinsic parity\footnote{%
  The intrinsic parity of a particle is
  defined to be even if its parity equals $(-1)^{\rm spin}$,
  and odd otherwise.
}:
\begin{eqnarray}
  {\cal L}_1
&=&
  i \, \mbox{tr}
  \left[
    \hat{\alpha}_{\rm L}^3 \hat{\alpha}_{\rm R}
    - \hat{\alpha}_{\rm R}^3 \hat{\alpha}_{\rm L}
  \right] 
\ ,
\label{L1 anom}
\\
  {\cal L}_2
&=&
  i \, \mbox{tr}
  \left[
    \hat{\alpha}_{\rm L} \hat{\alpha}_{\rm R}
    \hat{\alpha}_{\rm L} \hat{\alpha}_{\rm R}
  \right] 
\ ,
\label{L2 anom}
\\
  {\cal L}_3
&=&
  \mbox{tr}
  \left[
    F_{\rm V}
    \left(
      \hat{\alpha}_{\rm L} \hat{\alpha}_{\rm R}
      - \hat{\alpha}_{\rm R} \hat{\alpha}_{\rm L}
    \right)
  \right] 
\ ,
\label{L3 anom}
\\
  {\cal L}_4
&=&
  \frac{1}{2}\,
  \mbox{tr}
  \left[
    \hat{F}_{\rm L} 
    \left( 
      \hat{\alpha}_{\rm L} \hat{\alpha}_{\rm R}
      - \hat{\alpha}_{\rm R} \hat{\alpha}_{\rm L}
    \right)
    - \hat{F}_{\rm R} 
    \left(
      \hat{\alpha}_{\rm R} \hat{\alpha}_{\rm L}
      - \hat{\alpha}_{\rm L} \hat{\alpha}_{\rm R}
    \right)
  \right] 
\ ,
\label{L4 anom}
\end{eqnarray}
where the gauge covariant building blocks are given by
\begin{eqnarray}
&&
  \hat{\alpha}_{\rm L}
  \equiv
  \frac{1}{i} D \xi_{\rm L} \cdot \xi_{\rm L}^{\dag}
  = \alpha_{\rm L} - V + \hat{\cal L} \ ,
\quad
  \hat{\alpha}_{\rm R}
  \equiv
  \frac{1}{i} D \xi_{\rm R} \cdot \xi_{\rm R}^{\dag}
  = \alpha_{\rm L} - V + \hat{\cal R} \ ,
\nonumber\\
&&
  F_{\rm V}
  \equiv
  d V - i V^2 \ ,
  \qquad\qquad
  \hat{F}_{\rm L,R}
  =
  \xi_{\rm L,R} F_{\rm L,R} \xi_{\rm L,R}^{\dag} \ ,
\end{eqnarray}
with
\begin{eqnarray}
&&
  \alpha_{\rm L,R}
  =
  \frac{1}{i} d \xi_{\rm L,R} \cdot \xi_{\rm L,R}^{\dag} \ ,
  \quad
  \hat{\cal L}
  =
  \xi_{\rm L} {\cal L} \xi_{\rm L}^{\dag} \ ,
  \quad
  \hat{\cal R}
  =
  \xi_{\rm R} {\cal R} \xi_{\rm R}^{\dag} \ ,
\nonumber\\
&&
  F_{\rm L}
  =
  d {\cal L} - i {\cal L}^2 \ ,
  \quad
  F_{\rm R}
  =
  d {\cal R} - i {\cal R}^2 \ .
\end{eqnarray}
Other possible terms are written in terms of a linear combination of
${\cal L}_1$ to ${\cal L}_4$.~\footnote{%
  In the original version in Ref.~\cite{FKTUY}, six terms were
  included.
  However, two of them turned out to be charge-conjugation 
  odd and should be omitted~\cite{FKU,JJMPS}.  
  This point was corrected in Ref.~\cite{BKY} with resultant four
  terms in Eqs.~(\ref{L1 anom})--(\ref{L4 anom}).
}
We should note that the low energy theorem for anomalous process is
automatically satisfied, since the additional terms other than
$\Gamma_{\rm WZ}$ in Eq.~(\ref{Lag:Anom}) are gauge invariant and thus
do not contribute to the low energy amplitude governed by the anomaly.

We now read from the Lagrangian in Eq.~(\ref{Lag:Anom})
the $VV\pi$, $V\gamma\pi$ and $\gamma\gamma\pi$ vertices.
These are given by
\begin{eqnarray}
  {\cal L}_{VV\pi}
&=&
  - \frac{N_c}{4\pi^2 F_\pi} c_3 \,
  \varepsilon^{\mu\nu\lambda\sigma}
  \mbox{tr} \Bigl[ 
    \partial_\mu V_\nu \partial_\lambda V_\sigma \pi 
  \Bigr]
\nonumber\\
&=&
  g_{\omega\rho\pi} \varepsilon^{\mu\nu\lambda\sigma}
  \partial_\mu \omega_\nu \partial_\lambda \vec{\rho}_\sigma
  \cdot \vec{\pi}
  + \cdots
\ ,
\\
  {\cal L}_{V{\cal V}\pi}
&=&
  - \frac{N_c}{8\pi^2 F_\pi} (c_4 - c_3) \,
  \varepsilon^{\mu\nu\lambda\sigma}
  \mbox{tr} \Bigl[
    \left\{
      \partial_\mu V_\nu , \partial_\lambda {\cal V}_\sigma   
    \right\}
    \pi
  \Bigr]
\nonumber\\
&=&
  e g_{\omega\gamma\pi} \varepsilon^{\mu\nu\lambda\sigma}
  \partial_\lambda A_\sigma 
  \left( 
    \partial_\mu \omega_\nu \pi^0 
    + \frac{1}{3} \partial_\mu \vec{\rho}_\nu \cdot \vec{\pi}
  \right)
  + \cdots
\ ,
\\
  {\cal L}_{{\cal V}{\cal V}\pi}
&=&
  - \frac{N_c}{4\pi^2 F_\pi} (1-c_4)\,
  \varepsilon^{\mu\nu\lambda\sigma}
  \mbox{tr} \Bigl[
    \partial_\mu {\cal V}_\nu \partial_\lambda {\cal V}_\sigma \pi
  \Bigr]
\nonumber\\
&=&
  e^2 g_{\gamma\gamma\pi}
  \varepsilon^{\mu\nu\lambda\sigma}
  \partial_\mu A_\nu \partial_\lambda A_\sigma \pi^0
  + \cdots
\ ,
\end{eqnarray}
where
\begin{eqnarray}
&&
  g_{\omega\rho\pi} = - \frac{N_c g^2}{8 \pi^2 F_\pi} \, c_3 
\ ,
\\
&&
  g_{\omega\gamma\pi} = - \frac{N_c g}{16 \pi^2 F_\pi} (c_4 - c_3) 
\ ,
\\
&&
  g_{\gamma\gamma\pi} 
  = - \frac{N_c}{24\pi^2 F_\pi} (1-c_4) 
\ .
\end{eqnarray}
The $\gamma\pi^3$ and $V\pi^3$ vertices are given by
\begin{eqnarray}
  {\cal L}_{{\cal V}\pi^3}
&=&
  - i\, \frac{N_c}{3\pi^2 F_\pi^3}
  \left[
    1 - \frac{3}{4} \left( c_1 - c_2 + c_4 \right)
  \right]
  \varepsilon^{\mu\nu\lambda\sigma}
  \mbox{tr} \Bigl[
    {\cal V}_\mu \partial_\nu \pi \partial_\lambda \pi 
    \partial_\sigma \pi
  \Bigr]
\nonumber\\
&=&
  i e g_{\gamma\pi^3} \varepsilon^{\mu\nu\lambda\sigma}
  A_\mu \partial_\nu \pi^0 \partial_\lambda \pi^+ 
  \partial_\sigma \pi^-
  + \cdots
\ ,
\\
  {\cal L}_{V\pi^3}
&=&
  - i \, \frac{N_c}{4\pi^2 F_\pi^3} (c_1 - c_2 - c_3) \,
  \varepsilon^{\mu\nu\lambda\sigma}
  \mbox{tr}
  \left[
    V_\mu \partial_\nu \pi \partial_\lambda \pi
    \partial_\sigma \pi
  \right]
\nonumber\\
&=&
  i g_{\omega\pi^3} \varepsilon^{\mu\nu\lambda\sigma}
  \omega_\mu \partial_\nu \pi^0 \partial_\lambda \pi^+
  \partial_\sigma \pi^-
\ ,
\end{eqnarray}
where
\begin{eqnarray}
&&
  g_{\gamma\pi^3} = - \frac{N_c}{12\pi^2 F_\pi^3}
  \left[
    1 - \frac{3}{4} \left( c_1 - c_2 + c_4 \right)
  \right]
\\
&&
  g_{\omega\pi^3} = - \frac{3 N_c\, g}{16\pi^2 F_\pi^3}
  \left( c_1 - c_2 - c_3 \right)
\ .
\end{eqnarray}

\begin{figure}[htbp]
\begin{center}
\epsfxsize = 14cm
\ \epsfbox{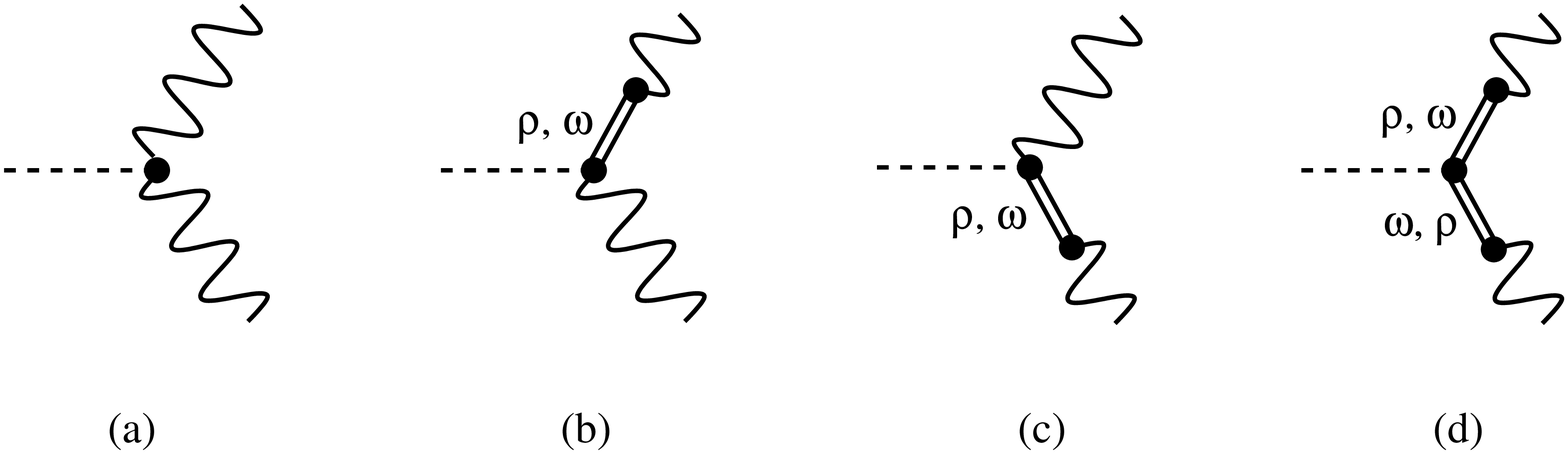}
\end{center}
\caption[effective $\pi^0\gamma^\ast\gamma^\ast$ vertex]{%
Effective $\pi^0\gamma^\ast\gamma^\ast$ vertex:
(a) direct $\pi^0\gamma\gamma$ interaction $\propto (1-c_4)$;
(b) and (c) through $\pi^0 \omega\gamma$ and $\pi^0\rho^0\gamma$
interactions $\propto (c_4-c_3)$;
(d) through $\omega\rho^0\pi^0$ interaction $\propto c_3$.
}
\label{fig:p0gg}
\end{figure}

\begin{figure}[htbp]
\begin{center}
\epsfxsize = 7cm
\ \epsfbox{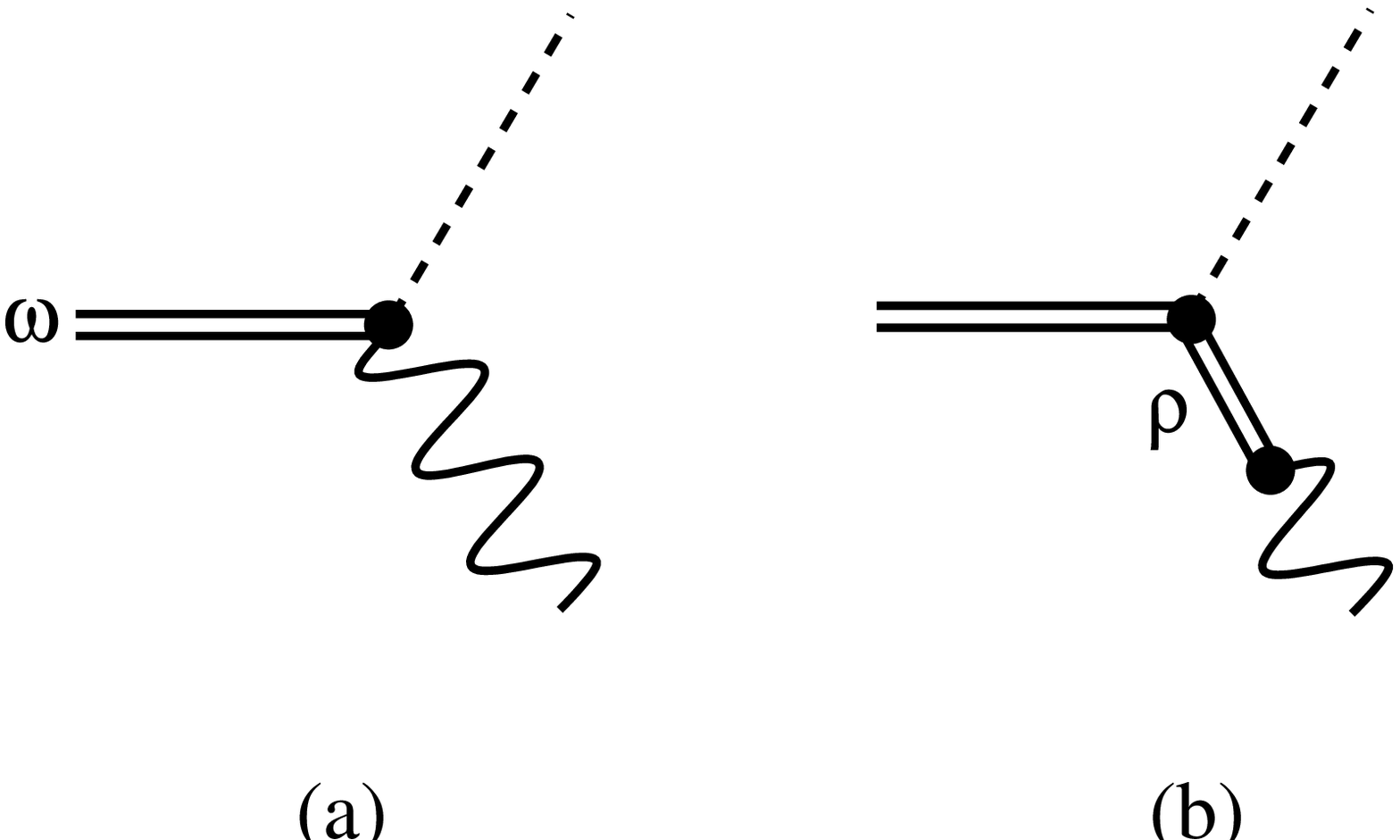}
\end{center}
\caption[effective $\omega\pi^0\gamma^\ast$ vertex]{%
Effective $\omega\pi^0\gamma^\ast$ vertex:
(a) direct $\omega\pi^0\gamma$ interaction $\propto (c_3-c_4)$;
(b) through $\omega\rho^0\pi^0$ interaction $\propto c_3$.
}
\label{fig:wp0g}
\end{figure}

\begin{figure}[htbp]
\begin{center}
\epsfxsize = 13cm
\ \epsfbox{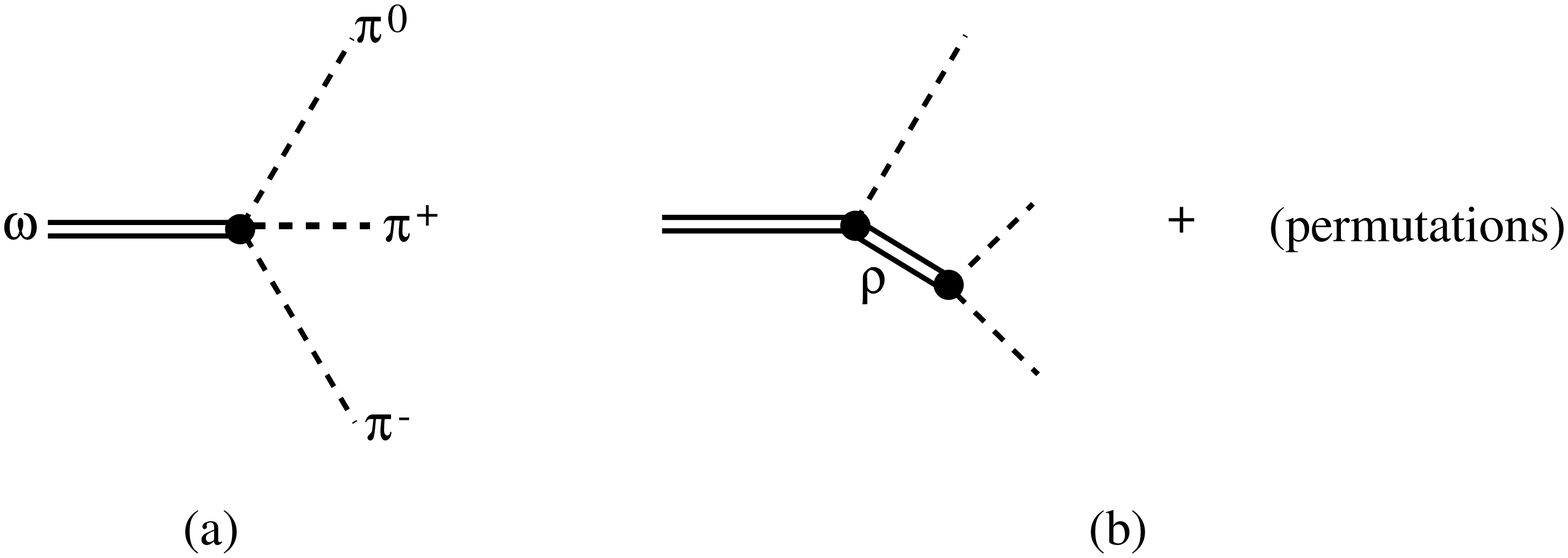}
\end{center}
\caption[effective $\omega\pi^0\pi^+\pi^-$ vertex]{%
Effective $\omega\pi^0\pi^+\pi^-$ vertex:
(a) direct $\omega\pi^0\pi^+\pi^-$ interaction $\propto
(c_1-c_2-c_3)$;
(b) through $\omega\rho\pi$ interaction $\propto c_3$.
}
\label{fig:w3p}
\end{figure}

\begin{figure}[htbp]
\begin{center}
\epsfxsize = 13cm
\ \epsfbox{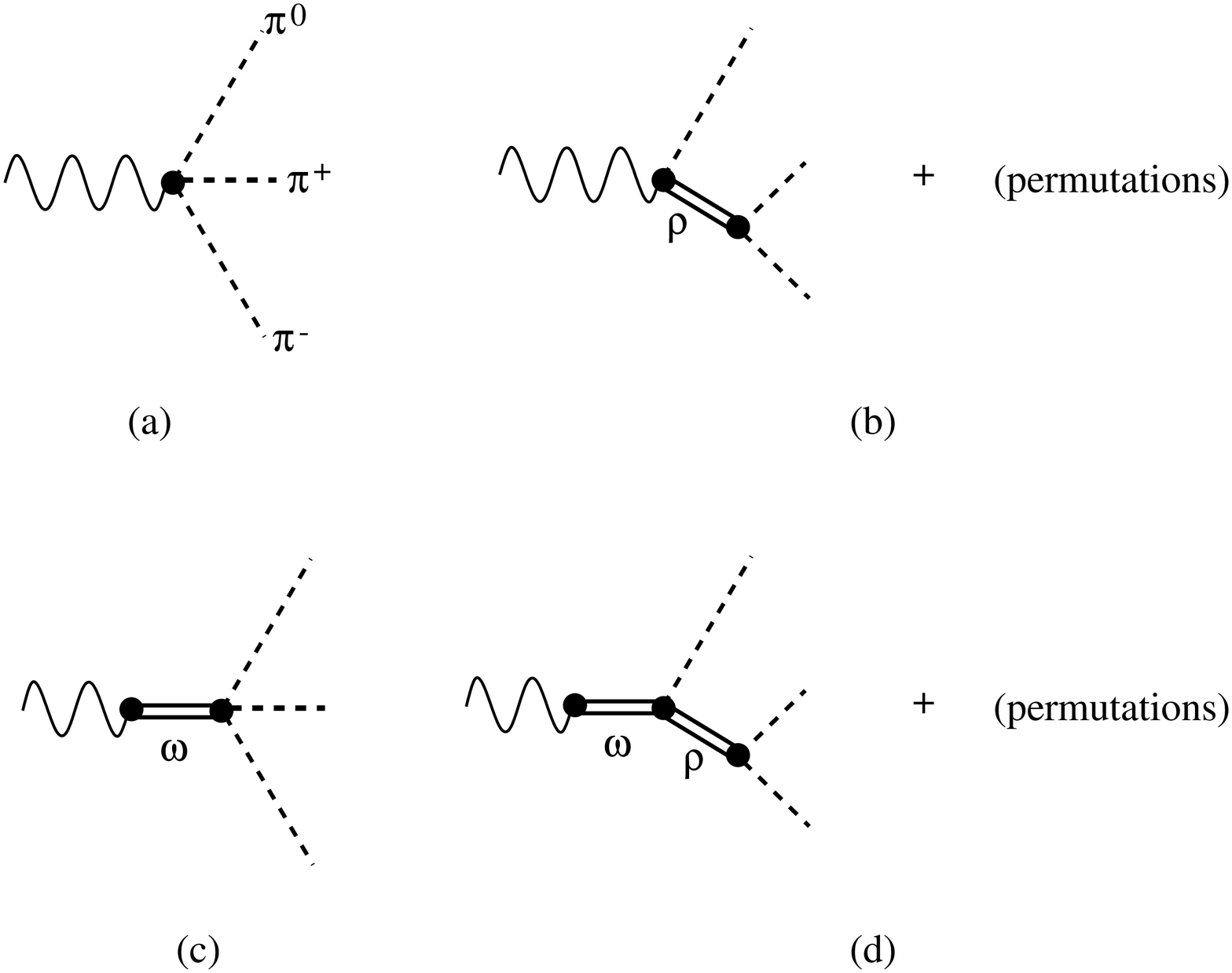}
\end{center}
\caption[effective $\gamma^\ast\pi^0\pi^+\pi^-$ vertex]{%
Effective $\gamma^\ast\pi^0\pi^+\pi^-$ vertex:
(a) direct $\gamma^\ast\pi^0\pi^+\pi^-$ interaction $\propto
g_{\gamma\pi^3}$;
(b) through $\gamma^\ast\rho\pi$ interaction $\propto (c_4-c_3)$;
(c) through $\omega\pi^0\pi^+\pi^-$ interaction $\propto
(c_1-c_2-c_3)$;
(d) through $\omega\rho\pi$ interaction $\propto c_3$.
}
\label{fig:g3p}
\end{figure}

{}From the above vertices we construct the effective vertices
for $\pi^0 \gamma^\ast  \gamma^\ast$,
$\omega \pi^0 \gamma^\ast$,
$\omega \pi^0 \pi^+ \pi^-$ and
$\gamma^\ast \pi^0 \pi^+ \pi^-$.
The relevant diagrams are shown in Figs.~\ref{fig:p0gg},
\ref{fig:wp0g}, \ref{fig:w3p} and \ref{fig:g3p}.
The effective vertices are given by
\begin{eqnarray}
&&
\Gamma^{\mu\nu}
\left[\pi^0, \gamma^\ast (q_1,\mu), \gamma^\ast(q_1,\nu) \right]
=
e^2 \frac{N_c}{12\pi^2 F_\pi} \, \varepsilon^{\mu\nu\alpha\beta}
q_{1\alpha} q_{2\beta}
\Biggl[
  \left\{ 1 - c_4 \right\} 
\nonumber\\
&& \qquad\qquad\qquad\qquad
  {}+ \frac{c_4-c_3}{4}
  \left\{ 
    D_\rho(q_1^2) + D_\rho(q_2^2) + D_\omega(q_1^2) + D_\omega(q_2^2)
  \right\}
\nonumber\\
&& \qquad\qquad\qquad\qquad
  {}+ \frac{c_3}{2}
  \left\{
    D_\rho(q_1^2) D_\omega(q_2^2) + D_\omega(q_1^2) D_\rho(q_2^2)
  \right\}
\Biggr]
\ ,
\label{p0gg vertex}
\\
&&
\Gamma^{\mu\nu}\left[\omega(p,\mu),\pi^0,\gamma^\ast(k,\nu)\right]
=
e g \, \frac{N_c}{8\pi^2 F_\pi}
\, \varepsilon^{\mu\nu\alpha\beta} p_{\alpha} k_{\beta}
\left[ \frac{c_4-c_3}{2} + c_3 D_\rho(q^2) \right]
\ ,
\label{wp0g vertex}
\\
&&
\Gamma_{\mu}
\left[\omega(p,\mu),\pi^0(q_0),\pi^+(q_+),\pi^-(q_-)\right]
=
- g \, \frac{N_c}{16\pi^2 F_\pi^3}
\, \varepsilon_{\mu\nu\alpha\beta} q_0^\nu q_+^\alpha q_-^\beta
\,
\Biggl[
  3 \left( c_1 - c_2 - c_3 \right)
\nonumber\\
&& \qquad\qquad
  {} + 2 c_3
  \left\{
    D_\rho\left( (q_+ + q_-)^2 \right) 
    + D_\rho\left( (q_- + q_0)^2 \right)
    + D_\rho\left( (q_0 + q_+)^2 \right)
  \right\}
\Biggr]
\ ,
\label{w3p vertex}
\\
&&
\Gamma_{\mu}
\left[\gamma^\ast(p,\mu),\pi^0(q_0),\pi^+(q_+),\pi^-(q_-)\right]
=
- e \, \frac{N_c}{12\pi^2 F_\pi^3}
\, \varepsilon_{\mu\nu\alpha\beta} q_0^\nu q_+^\alpha q_-^\beta
\nonumber\\
&& \qquad
\times
\Biggl[
  1 - \frac{3}{4} \left( c_1 - c_2 + c_4 \right)
  {}+ \frac{9}{4} \left( c_1 - c_2 - c_3 \right) D_\omega (p^2)
\nonumber\\
&& \qquad\qquad
  {} + 
  \left\{
    \frac{c_4-c_3}{4} + \frac{3}{2} c_3 D_\omega(p^2)
  \right\}
\nonumber\\
&& \qquad\qquad\qquad
\times
  \left\{
    D_\rho\left( (q_+ + q_-)^2 \right) 
    + D_\rho\left( (q_- + q_0)^2 \right)
    + D_\rho\left( (q_0 + q_+)^2 \right)
  \right\}
\Biggr]
\ ,
\nonumber\\
&&
\label{g3p vertex}
\end{eqnarray}
where $D_\rho(q^2)$ and $D_\omega(q^2)$ are $\rho$ meson
and $\omega$ meson
propagators normalized to one in the low-energy limit:
\begin{equation}
D_\rho(0) = D_\omega(0) = 1 \ .
\end{equation}
In this subsection we use the vector meson propagators at leading
order:
\begin{eqnarray}
D_\rho(q^2) = \frac{m_\rho^2}{m_\rho^2-q^2} \ ,
\quad
D_\omega(q^2) = \frac{m_\omega^2}{m_\omega^2-q^2}
\ .
\label{leading prop}
\end{eqnarray}

Now let us perform several phenomenological analyses.
Below we shall especially focusing whether the vector dominance (VD)
is satisfied in each form factor.
Here we summarize the values of the parameters for VD:
\begin{eqnarray}
&& \mbox{(a)}\ \mbox{VD in $\pi^0\gamma\gamma^\ast$ :}
   \quad \frac{c_3 + c_4}{2} = 1 \ ,
\nonumber\\
&& \mbox{(b)}\ \mbox{VD in $\omega\pi^0\gamma^\ast$ :}
   \quad c_3 = c_4 \ ,
\nonumber\\
&& \mbox{(c)}\ \mbox{VD in $\pi^0\gamma^\ast\gamma^\ast$ :}
   \quad c_3 = c_4 = 1 \ ,
\nonumber\\
&& \mbox{(d)}\ \mbox{VD in $\gamma^\ast\pi^0\pi^+\pi^-$ :}
   \quad c_1 - c_2 + c_4 = \frac{4}{3} \ \mbox{and} \ c_3 = c_4 \ .
\label{VDs}
\end{eqnarray}
When all the above VD's are satisfied
(complete VD), the values of $c_1 - c_2$, $c_3$ and $c_4$ are fixed:
\begin{equation}
\mbox{(e)}\ \mbox{complete VD :}
   \quad c_3 = c_4 = 1 \ \mbox{and} \ c_1 - c_2 = \frac{1}{3} \ .
\label{cVD}
\end{equation}

We first study the decay width of $\pi^0 \rightarrow \gamma \gamma$.
When we take $q_1^2=q_2^2=0$ in the effective vertex in 
Eq.~(\ref{p0gg vertex}), 
terms including $c_3$ and $c_4$ vanish irrespectively of the
detailed forms of the $\rho$ and $\omega$ propagators.
The resultant vertex is identical with the one by the current 
algebra~\cite{BJ,Adler,WZ,ALTZ,AZ}.~\footnote{%
We should note that the low energy theorem for
$\gamma \rightarrow 3\pi$ is also intact.
}
The predicted~\cite{FKTUY} decay width is now given by
\begin{equation}
\Gamma (\pi^0 \rightarrow \gamma\gamma)
=
\frac{\alpha^2}{64\pi^3} \frac{m_{\pi^0}^3}{F_\pi^2} \ .
\end{equation}
Using the values~\cite{PDG:02}
\begin{eqnarray}
&& m_{\pi^0} = (134.9766 \pm 0.0006) \, \mbox{MeV} \ ,
\label{exp mp0}
\\
&& \alpha = 1/137.03599976 \ ,
\label{exp al}
\end{eqnarray}
and
$F_\pi$ in Eq.~(\ref{exp Fp})
we obtain
\begin{equation}
\left. \Gamma (\pi^0 \rightarrow \gamma\gamma) \right\vert_{\rm theo}
=
(7.73 \pm 0.04) \, \mbox{eV} \ .
\end{equation}
This excellently agrees with the experimental value 
estimated from the $\pi^0$ life time and the branching fraction of
$\pi^0\rightarrow \gamma \gamma$:
\begin{equation}
\left. \Gamma (\pi^0 \rightarrow \gamma\gamma) \right\vert_{\rm exp}
=
(7.7 \pm 0.6) \, \mbox{eV} \ .
\end{equation}

Second, we study the $\pi^0$ electromagnetic form factor
($\pi^0\gamma\gamma^\ast$ form factor).
{}From Eq.~(\ref{p0gg vertex}) this form factor is given by
\begin{equation}
F_{\pi^0\gamma} (q^2) = 
 \left( 1 - \frac{c_3+c_4}{2} \right)
 + \frac{c_3+c_4}{4}
   \left[ D_\rho(q^2) + D_\omega (q^2) \right]
\ .
\end{equation}
In the low energy region, by using the explicit forms of $\rho$
and $\omega$ propagators, this is approximated as
\begin{equation}
F_{\pi^0\gamma} (q^2) = 
 1 + \lambda\, \frac{q^2}{m_{\pi^0}^2} + \cdots \ ,
\end{equation}
where the linear coefficient $\lambda$ is given by
\begin{equation}
\lambda =
\frac{c_3+c_4}{4} \left[
  \frac{m_{\pi^0}^2}{m_\rho^2} + \frac{m_{\pi^0}^2}{m_\omega^2}
\right]
\ .
\end{equation}
Using the experimental value of this $\lambda$~\cite{PDG:02}
\begin{equation}
\left. \lambda \right\vert_{\rm exp} = 0.032 \pm 0.004 \ ,
\end{equation}
and the values of masses~\cite{PDG:02}
\begin{eqnarray}
&& m_{\pi^0} = 134.9766 \pm 0.0006 \, \mbox{MeV} \ , \nonumber\\
&& m_\rho = 771.1 \pm 0.9 \, \mbox{MeV} \ , \nonumber\\
&& m_\omega = 782.57 \pm 0.12 \, \mbox{MeV} \ ,
\label{exp masses 1}
\end{eqnarray}
we estimated the value of $(c_3+c_4)/2$:
\begin{equation}
\frac{c_3+c_4}{2} = 1.06 \pm 0.13 \ .
\label{c34 val}
\end{equation}
This implies that the VD (a) in Eq.~(\ref{VDs}) is well satisfied.

Next we calculate the $\omega\rightarrow \pi^0 \gamma$ decay width.
{}From the effective vertex in Eq.~(\ref{wp0g vertex}),
the decay width is expressed as
\begin{equation}
\Gamma (\omega\rightarrow \pi^0 \gamma) =
\left( g\, \frac{c_3+c_4}{2} \right)^2
\frac{3\alpha}{64\pi^4 F_\pi^2}
\left( \frac{ m_\omega^2 - m_{\pi^0}^2}{2m_\omega} \right)^3
\ .
\end{equation}
Using the values of masses in Eq.(\ref{exp masses 1})
and the parameters $F_\pi$, $g$ and $(c_3+c_4)/2$ in 
Eqs.~(\ref{exp Fp}), (\ref{gval:tree}) and (\ref{c34 val}), 
we obtain
\begin{equation}
\Gamma (\omega\rightarrow \pi^0 \gamma)
= 0.85 \pm 0.34 \, \mbox{MeV} \ .
\end{equation}
This agrees with the experimental 
value~\footnote{%
  This value is
  estimated from the $\omega$ total decay
  width and $\omega\rightarrow \pi^0 \gamma$ shown in
  Ref.~\cite{PDG:02}.
}
\begin{equation}
\left. \Gamma (\omega\rightarrow \pi^0 \gamma) \right\vert_{\rm exp}
=
0.73 \pm 0.03 \, \mbox{MeV} \ .
\label{exp wp0g}
\end{equation}
On the other hand, when we use the above experimental value
and the value of $g$ in Eq.~(\ref{gval:tree}),
we obtain
\begin{equation}
\left\vert \frac{c_3+c_4}{2} \right\vert = 
0.99 \pm 0.16 \ ,
\end{equation}
which is consistent with the value in Eq.~(\ref{c34 val}).

We further study the $\omega \rightarrow \pi^0 \mu^+ \mu^-$ decay,
which is suitable for testing the VD (b) in Eq.~(\ref{VDs}).
{}From the effective vertex in Eq.~(\ref{wp0g vertex}),
this decay width is expressed as
\begin{eqnarray}
&&
  \Gamma ( \omega \rightarrow \pi^0 \mu^+ \mu^- )
  = \int^{(m_\omega - m_\pi)^2}_{4m_\mu^2} d q^2
  \frac{\alpha}{3\pi} 
  \frac{\Gamma(\omega \rightarrow \pi^0 \gamma)}{q^2}
  \left( 1 + \frac{2m_\mu^2}{q^2} \right)
  \sqrt{ \frac{ q^2 - 4 m_\mu^2 }{q^2} }
\nonumber\\
&& \qquad\qquad
  \times
  \left[
    \left( 1 + \frac{q^2}{m_\omega^2-m_\pi^2} \right)^2
    - \frac{4 m_\omega^2 q^2}{(m_\omega^2 - m_\pi^2)^2}
  \right]^{3/2}
  \left\vert F_{\omega\pi^0}(q^2) \right\vert
\ ,
\end{eqnarray}
where $q^2$ is the intermediate photon momentum and 
$F_{\omega\pi^0}(q^2)$ is the $\omega\pi^0$ transition form factor.
In the HLS this $F_{\omega\pi^0}(q^2)$ is given 
by~\cite{BH:PTP,BH:PRD}
\begin{equation}
  F_{\omega\pi^0}(q^2) = - \tilde{c} 
  + ( 1 + \tilde{c} ) D_\rho(q^2) 
\ ,
\label{wp0 form}
\end{equation}
where
\begin{equation}
\tilde{c} = \frac{c_4-c_3}{c_3 + c_4} \ .
\end{equation}
Using the $\rho$ propagator
in Eq.~(\ref{leading prop}) 
with the experimental values of masses 
and the ratio of two decay widths
\begin{equation}
\left.
\frac{\Gamma ( \omega \rightarrow \pi^0 \mu^+ \mu^- )}%
{\Gamma(\omega \rightarrow \pi^0 \gamma)}
\right\vert_{\rm exp}
= \left( 1.10 \pm 0.27 \right) \times 10^{-3} \ ,
\end{equation}
we estimated the value of $\tilde{c}$ as
\begin{equation}
\tilde{c} = 0.42 \pm 0.56 \quad \mbox{or}
\quad -7.04 \pm 0.56 \ .
\end{equation}
The second solution is clearly excluded by comparing the 
$\omega\pi^0$ transition form factor in Eq.~(\ref{wp0 form}) with
experiment (see, e.g., Refs.~\cite{BH:PTP,BH:PRD}).
Since the error is huge in the first solution,
the first solution is consistent with the VD (b) in Eq.~(\ref{VDs}).
However, the comparison of the form factor itself with experiment
prefers non-zero value of $\tilde{c}$~\cite{BGP,BH:PTP,BH:PRD},
and thus the VD (b) is violated.

Finally, we study the $\omega \rightarrow \pi^0 \pi^+ \pi^-$ decay
width to check the validity of the complete VD (e) in 
Eq.~(\ref{cVD}).
By using the effective $\omega\pi^0 \pi^+ \pi^-$ vertex in 
Eq.~(\ref{w3p vertex}),
the decay width is expressed as~\cite{FKTUY}
\begin{equation}
\Gamma( \omega \rightarrow \pi^0 \pi^+ \pi^- )
= \int \int E_+ E_-
\left[
  \vert \vec{q}_- \vert^2 \vert \vec{q}_+ \vert^2
  - ( \vec{q}_+ \cdot \vec{q}_- )^2
\right]
\left\vert F_{\omega\rightarrow 3\pi} \right\vert^2
\ ,
\end{equation}
where $E_+$ and $E_-$ are the energies of $\pi^+$ and $\pi^-$ in the
rest frame of $\omega$, $\vec{q}_+$ and $\vec{q}_-$ are the momenta of
them, and
\begin{eqnarray}
&&
  F_{\omega\rightarrow 3\pi}
  = - g\, \frac{N_c}{16\pi^2 F_\pi^3}
  \Biggl[
    3 (c_1 - c_2 - c_3 )
\nonumber\\
&& \qquad\qquad
    {} + 2 c_3
    \left\{
      D_\rho\left( (q_+ + q_-)^2 \right) 
      + D_\rho\left( (q_- + q_0)^2 \right)
      + D_\rho\left( (q_0 + q_+)^2 \right)
    \right\}
  \Biggr]
\ .
\end{eqnarray}
When we use $c_1 - c_2 =1$ and $c_3=1$~\footnote{%
  This is obtained by requiring the VD (c) in Eq.~(\ref{VDs}) 
  [for $c_3=1$]
  and no direct $\omega \pi^0 \pi^+ \pi^-$ vertex
  [for $c_1 - c_2 =1$].
},
we obtain
\begin{equation}
\Gamma( \omega \rightarrow \pi^0 \pi^+ \pi^- )
=
6.9 \pm 2.2 \, \mbox{MeV} 
\quad (c_1 - c_2 = 1 \ \mbox{and} \ c_3 = 1)
\ ,
\end{equation}
where the error mainly comes from the error of $g$ in 
Eq.~(\ref{gval:tree}).
This is consistent with the experimental value~\cite{PDG:02}
\begin{equation}
\left.
\Gamma( \omega \rightarrow \pi^0 \pi^+ \pi^- )
\right\vert_{\rm exp}
=
7.52 \pm 0.10 \, \mbox{MeV} 
\ .
\label{exp w3p}
\end{equation}
On the other hand,
if we assume that the complete VD (e) in Eq.~(\ref{cVD}) were
satisfied, we would have $c_1 - c_2 - c_3 = - 2/3$ and 
$c_3=1$~\cite{FKTUY}.
Then we would obtain
\begin{equation}
\Gamma( \omega \rightarrow \pi^0 \pi^+ \pi^- )
=
4.4 \pm 1.4 \, \mbox{MeV} 
\quad \mbox{(complete vector dominance)}
\ .
\end{equation}
Comparing this value with the experimental value
in Eq.~(\ref{exp w3p}),
{\it we conclude that the complete VD (e) in Eq.~(\ref{cVD}) is
excluded by the experiment}~\cite{FKTUY}.~\footnote{%
  After Ref.~\cite{FKTUY} the experimental value of the $\omega$ width
  was substantially changed (see page 16 ``History plots'' 
  of Ref.~\cite{PDG:02}).
  Then the experimental value of the partial width
  $\Gamma( \omega \rightarrow \pi^0 \pi^+ \pi^- )$ becomes smaller
  than that referred in Ref.~\cite{FKTUY}.
  Nevertheless the prediction of the complete vector dominance is
  still excluded by the new data.
}

\newpage

\section{Chiral Perturbation Theory with HLS}
\label{sec:CPHLS}

In this section we
review the chiral perturbation in the hidden local symmetry (HLS) at
one loop.

First we show that, {\it thanks to the gauge invariance} of the HLS,
we can perform the 
{\it systematic derivative expansion with including
vector mesons} in addition to the pseudoscalar Nambu-Goldstone bosons
(Sec.~\ref{ssec:DEHLS}).
Then, we give the ${\cal O}(p^2)$ Lagrangian with including the
external fields in Sec.~\ref{ssec:OP2L}, 
and then present a complete list of 
the ${\cal O}(p^4)$ terms following
Ref.~\cite{Tanabashi} (Sec.~\ref{ssec:OP4L}).

Explicit calculation is done by using the background field
gauge~\cite{Tanabashi,HY:matching} (The background field gauge
is explained in Sec.~\ref{ssec:BGFM}, and the
calculation is done in Sec.~\ref{ssec:TPFOL}).
Since the effect of quadratic divergences is important in the
analyses in the next sections (see Secs.~\ref{sec:WM} and
\ref{sec:VM}), we explain meaning of the quadratic
divergence in our approach in Sec.~\ref{ssec:QD}.
We briefly summarize a role of the quadratic divergence in 
the phase transition in Sec.~\ref{sssec:RQDPT}.
Then, we show that the chiral symmetry restoration by the mechanism 
shown in Ref.~\cite{HY:letter}
also takes place even in the ordinary nonlinear sigma model
when we include the effect of quadratic divergences
(Sec.~\ref{sssec:CRQDPL}).
We present a way to include the quadratic divergences consistently
with the chiral symmetry in Sec.~\ref{sssec:QDSPR}.

The low-energy theorem of the HLS, 
$g_\rho = 2 g_{\rho\pi\pi}F_\pi^2$ [KSRF (I)]~\cite{KSRF:KS,KSRF:RF},
was shown to be satisfied at one-loop level in Ref.~\cite{HY} by
using the ordinary quantization procedure in the Landau gauge.
Section~\ref{ssec:LETOL} is devoted to show that
the low-energy theorem remains intact in the present 
background field gauge more transparently.

{}From the one-loop corrections calculated in
Sec.~\ref{ssec:TPFOL} we will obtain the RGEs
in the Wilsonian sense, i.e., including quadratic
divergences, in Sec.~\ref{ssec:RGEWS}.

As was shown in Ref.~\cite{Tanabashi}, the relations (matching)
between the
parameters of the HLS and the ${\cal O}(p^4)$ ChPT parameters should
be obtained by including one-loop corrections in both theories, since
one-loop corrections from ${\cal O}(p^2)$ Lagrangian generate 
${\cal O}(p^4)$ contributions.  In Sec.~\ref{ssec:MHC} we show
some examples of the relations.

Finally in Sec.~\ref{ssec:PSH} we study phase structure of
the HLS, following Ref.~\cite{HY:VD}.

We note that convenient formulas and Feynman rules used in this
section are summarized in Appendices~\ref{sec:CF} and \ref{app:FR}.
A complete list of the divergent corrections to the ${\cal O}(p^4)$
terms is shown in Appendix~\ref{app:RHKE}.

\subsection{Derivative expansion in the HLS model}
\label{ssec:DEHLS}

In the chiral perturbation theory (ChPT)~\cite{Wei:79,Gas:84,Gas:85a}
(see Sec.~\ref{sec:BRCPT} for a brief review)
the derivative expansion is systematically done by 
using the fact that the pseudoscalar meson masses are small compared
with the chiral symmetry breaking scale $\Lambda_\chi$.
The chiral symmetry breaking scale is considered as the scale where
the derivative expansion breaks down.
{}According to the naive dimensional analysis
(NDA)~\cite{Man:84}
the loop correction (without quadratic divergece) generally
appears with the factor
\begin{equation}
\frac{p^2}{\left(4\pi F_\pi\right)^2}
\ .
\label{NDA loop}
\end{equation}
For the consistency with the derivative expansion, the above
factor must be smaller than one, which implies that the systematic
expansion breaks down around the energy scale of $4\pi F_\pi$.
Then, the chiral symmetry breaking scale 
$\Lambda_\chi$ is
estimated as [see Eq.~(\ref{lam chi})]
\begin{equation}
\Lambda_\chi \simeq 4 \pi F_\pi \sim 1.1 \, \mbox{GeV} \ ,
\label{NDA}
\end{equation}
where we used $F_\pi = 86.4\,\mbox{MeV}$ estimated in the chiral
limit~\cite{Gas:84,Gas:85b}.
Since the $\rho$ meson and its flavor partners are lighter than this
scale, one can expect that the derivative expansion with including
vector mesons are possible in such a way that
the physics in the energy region
slightly higher than the vector meson mass scale can well
be studied.
On the other hand, axialvector mesons ($a_1$ and its flavor partners)
should not be included since their masses are larger than
$\Lambda_{\chi}$.

It was first pointed by Georgi~\cite{Georgi:1,Georgi:2}
that, {\it thanks to the gauge invariance},
the HLS makes possible the systematic expansion including the
vector meson loops, particularly when the vector meson mass is light.
It turns out that 
such a limit can actually be realized in QCD
when the number of massless flavors $N_f$
becomes large as was demonstrated in Refs.~\cite{HY:letter,HY:VM}.
Then one can perform the derivative expansion with including the
vector mesons under such an extreme condition
where the vector meson masses are
small, and extrapolate the results to the real world $N_f=3$
where the vector meson
masses take the experimental values.

The first one-loop calculation based on this notion was done in
Ref.~\cite{HY}.  
There it was shown that the low-energy theorem of the 
HLS~\cite{BKY:NPB,BKY:PTP} holds at one loop.
This low-energy theorem was proved to hold at any loop order
in Refs.~\cite{HKY:PRL,HKY:PTP} (see Sec.~\ref{sec:RALOLET}).
Moreover, a systematic counting scheme in the framework of the HLS
was proposed in Ref.~\cite{Tanabashi}.
These analyses show that,
although the expansion parameter in the real-life QCD is not very
small: 
\begin{equation}
\frac{m_\rho^2}{\Lambda_\chi^2} \sim 0.5 \ ,
\label{expar}
\end{equation}
the procedure seems to work in the real world. (See, e.g., a
discussion in Refs.~\cite{HKY:PRL,HKY:PTP}.)

Now, let us summarize the counting rule of the present analysis.
As in the ChPT in Ref.~\cite{Gas:84,Gas:85a},
the derivative and the external gauge fields ${\cal L}_\mu$ and 
${\cal R}_\mu$ are counted as ${\cal O}(p)$, while the 
external source fields $\chi$ is
counted as ${\cal O}(p^2)$ since the VEV of $\chi$
in Eq.~(\ref{trans prop ChPT})
is the
square of the pseudoscalar meson mass, $\langle\chi\rangle \sim
m_\pi^2$
[see Eqs.~(\ref{QMM:Nf}), (\ref{GMOR rel}) and 
(\ref{trans prop ChPT})].
Then we obtain the following order assignment:
\begin{eqnarray}
&& \partial_\mu \sim {\cal L}_\mu \sim 
  {\cal R}_\mu \sim {\cal O}(p) \ , \nonumber\\
&& \chi \sim {\cal O}(p^2) \ .
\end{eqnarray}
The above counting rules are the same as those in the ChPT.

Differences appear in the counting rules for the vector mesons between
the HLS and a version of the ChPT~\cite{Eck:89a} where the
vector mesons are introduced by anti-symmetric tensor fields
(``tensor field method''). 
[A brief review of ``tensor field method''
and its relation to the HLS are given in Sec.~\ref{sssec:ATFM}.]
In the ``tensor field method'' the vector meson fields are counted as 
${\cal O}(1)$.
On the other hand, 
for the consistency of the covariant derivative shown in
Eq.~(\ref{covder}) HLS forces us to assign ${\cal O}(p)$ to 
$V_\mu \equiv g \rho_\mu$:
\begin{equation}
V_\mu = g \rho_\mu \sim {\cal O}(p) \ .
\label{V:order}
\end{equation}
Another essential difference between the counting rule in the HLS and 
that in the ``tensor field method''
is in the counting rule for the vector meson
mass.
In the 
latter 
the vector meson mass is counted as
${\cal O}(1)$.
However, as discussed around Eq.~(\ref{expar}), we are performing the
derivative expansion in the HLS by regarding the vector meson
as light.
Thus, similarly to the square of the pseudoscalar meson mass,
we assign ${\cal O}(p^2)$ to the square of the vector meson mass:
\begin{equation}
m_\rho^2 = g^2 F_\sigma^2 \sim {\cal O}(p^2) \ ,
\end{equation}
which is contrasted to $m_\rho^2 \sim {\cal O}(1)$ in the
``tensor field method''.
Since the vector meson mass becomes small in the limit of small
HLS gauge coupling,
we should assign ${\cal O}(p)$ to the
HLS gauge coupling $g$, not to $F_\sigma$~\cite{Tanabashi}:
\begin{equation}
g \sim {\cal O}(p) \ .
\label{g:order}
\end{equation}
This is the most important part in the counting rules in the HLS.
By comparing the order for $g$ in Eq.~(\ref{g:order}) with that for
$g \rho_\mu$ in Eq.~(\ref{V:order}), the $\rho_\mu$ field should be
counted as ${\cal O}(1)$.
Then the kinetic term of the HLS gauge boson is counted as 
${\cal O}(p^2)$ which is of the same order as the kinetic term of the
pseudoscalar meson:
\begin{equation}
- \frac{1}{2} \, \mbox{tr} 
  \left[ \rho_{\mu\nu} \rho^{\mu\nu} \right]
\ \sim \  {\cal O}(p^2) \ .
\end{equation}

We stress that it is the existence of the gauge invariance that
makes the above systematic expansion
possible~\cite{Georgi:1,Georgi:2}.
To clarify this point, let us consider a Lagrangian
including a massive spin-1 field as Lorentz vector field,
which is invariant under the chiral symmetry.
An example is the Lagrangian including the vector meson field as a
matter field in the sense of CCWZ~\cite{CWZ,CCWZ}
(``matter filed method'').
[A brief review of this ``matter filed method'' and its relation to
the HLS are given in Sec.~\ref{sssec:MFM}.]
The kinetic and the mass terms of the vector meson field 
$\rho^{\rm(C)}_\mu$ is given by
[see Eq.~(\ref{Lag:E0})]
\begin{equation}
{\cal L}_{C} =
  - \frac{1}{2} \mbox{tr} 
  \left[
    \rho^{\rm(C)}_{\mu\nu} \rho^{{\rm(C)}\mu\nu} 
  \right]
  + M_\rho^2 \mbox{tr} 
    \left[ \rho^{\rm(C)}_\mu \rho^{{\rm(C)}\mu} \right]
\ ,
\end{equation}
where $\rho^{\rm(C)}_{\mu\nu}$ is defined in 
Eq.~(\ref{def:rhomn:E}).
The vector meson field
$\rho^{\rm(C)}_\mu$ transforms as [see Eq.~(\ref{trans rho C})]
\begin{equation}
\rho^{\rm(C)}_\mu \rightarrow
  h(\pi,g_{\rm R},g_{\rm L}) \cdot \rho^{\rm(C)}_\mu 
  \cdot h^\dag (\pi,g_{\rm R},g_{\rm L}) 
\ ,
\end{equation}
where $h \left( \pi, g_{\rm R}, g_{\rm L}\right)$ is an element 
of $\mbox{SU($N_f$)}_{\rm V}$ as given in Eq.~(\ref{com trans}).
The form of the propagator of the vector meson is given by
\begin{equation}
\frac{1}{p^2-m_\rho^2} 
\left[ g_{\mu\nu} - \frac{p_\mu p_\nu}{m_\rho^2} \right] \ ,
\label{rho:prop0}
\end{equation}
which coincides with the vector meson propagator in the unitary gauge
of the HLS (Weinberg's $\rho$ meson~\cite{Wei:68}).
The longitudinal part ($p_\mu p_\nu$-part) carries the factor of
$1/m_\rho^2$ which may generate quantum corrections
proportional to some powers of $1/m_\rho^2$.
Appearance of a factor
$1/m_\rho^2$ is a disaster in the loop calculations, particularly when
the vector meson mass is light.
Namely, the derivative expansion discussed above breaks down.
We note that the situation is similar in the ``Massive Yang-Mills''
approach and the ``tensor field method'' reviewed in
Sec.~\ref{sssec:ATFM}.

In the HLS, however, the gauge invariance prevent such a $1/m_\rho^2$
factor from appearing.
This can be easily seen by the following vector meson propagator in an
$R_\xi$-like gauge fixing~\cite{HY}:
\begin{equation}
\frac{1}{p^2-m_\rho^2} 
\left[ 
  g_{\mu\nu} - (1-\alpha) \frac{p_\mu p_\nu}{p^2- \alpha m_\rho^2}
\right] \ ,
\label{rho:prop1}
\end{equation}
where $\alpha$ is the gauge fixing parameter.
The propagator in Eq.~(\ref{rho:prop1}) is well defined in the limit
of $m_\rho\rightarrow0$ except for the unitary gauge 
($\alpha =\infty$), 
while the propagator in
Eq.~(\ref{rho:prop0}) is ill-defined in such a limit.
In addition, the gauge invariance guarantees that all the
interactions never include a factor of $1/g^2 \propto 1/m_\rho^2$,
while it may exist for the lack of the gauge invariance.
Then all the loop corrections are well defined
even in the limit of $m_\rho\rightarrow0$.
Thus the HLS gauge invariance is essential to performing the above
derivative expansion.
This makes the HLS most powerful among various methods 
(see Sec.~\ref{ssec:ROMVM})
for including the
vector mesons based on the chiral symmetry.

In the above discussion
we explained the systematic expansion in the HLS 
based on the naive dimensional analysis (NDA).
Here we refine the argument in order to study the 
large $N_f$ QCD.
First, we note that 
the loop corrections generally have an additional
factor $N_f$ in front of the contribution.
Then, the general expression for the
loop correction in Eq.~(\ref{NDA loop}) is rewritten as
\begin{equation}
N_f \frac{ p^2}{\left(4\pi F_\pi(0) \right)^2}
\ ,
\label{FNDA loop}
\end{equation}
where we used $F_\pi(0)$ for expressing the $\pi$ decay constant
at the low-energy limit (i.e., on mass-shell of $\pi$).
Hence when $N_f$ is crucial, we cannot ignore the factor $N_f$,
and
the chiral symmetry breaking scale in Eq.~(\ref{NDA}) should be
changed to
\begin{equation}
\Lambda_\chi \simeq \frac{ 4\pi F_\pi(0) }{ \sqrt{N_f} } \ ,
\label{FNDA lam chi 0}
\end{equation}
which yields 
$\Lambda_\chi \simeq 4 \pi F_\pi(0)/ \sqrt{3} \sim m_\rho$
for $N_f=3$ case.
This implies that the systematic
expansion for $N_f=3$ QCD is valid 
in the energy region 
around and less than
the $\rho$ meson mass.
For large $N_f$,
existence of $\sqrt{N_f}$ in the denominator in
Eq.~(\ref{FNDA lam chi 0}) indicates that
$\Lambda_\chi$ decrease with $N_f$ increased.
Furthermore,
in the large $N_f$ QCD,
as we will study in detail in Sec.~\ref{sec:VM},
the chiral symmetry is expected to be restored
at a certain number of flavor
$N_f^{\rm crit}$,
and $F_\pi(0)$ will vanish.
One might think that there would be no applicable 
energy region near the critical point.
However, this is not the case:
In the present analysis,
we include the effect of quadratic divergences which is necessary
for realizing the chiral restoration (see Sec.~\ref{ssec:QD})
as well as for matching the HLS with underlying QCD in $N_f=3$ QCD
(see Sec.~\ref{sec:WM}).
The inclusion of the quadratic divergence
implies that the loop corrections
are given in terms of the bare parameter $F_\pi(\Lambda)$ instead
of the on-shell decay constant $F_\pi(0)$.
Then,
the scale at which the theory breaks down 
in Eq.~(\ref{FNDA lam chi 0}) is further changed to
\begin{equation}
\Lambda_\chi \simeq \frac{ 4 \pi F_\pi(\Lambda) }{ \sqrt{N_f} }
\ .
\label{FNDA lam chi 1}
\end{equation}
This is somewhat higher than the chiral symmetry breaking scale
in Eq.~(\ref{FNDA lam chi 0}),
$F_\pi(\Lambda) > F_\pi(0) = 86.4 \, {\rm MeV}$ for $N_f=3$
(see Sec.~\ref{ssec:DBPHL}), and even dramatically higher
$F_\pi(\Lambda) \gg F_\pi(0) \rightarrow 0$
near the phase transition point.

One might still think that the above systematic
expansion would break down in such a case,
since the quadratic divergences from higher loops can in principle
contribute to the $O(p^2)$ terms.
However, 
even when
the quadratic divergences are explicitly included,
we think that the systematic expansion is still valid 
in the following sense:
The 
quadratically divergent correction to the $O(p^2)$ term 
at $n$th loop order takes the form of
$[\Lambda^2/\Lambda_\chi^2]^{n}$, where
$\Lambda_\chi$ is defined in Eq.~(\ref{FNDA lam chi 1}).
Then, by requiring
the cutoff $\Lambda$ be smaller than $\Lambda_\chi$,
$\Lambda^2/\Lambda_\chi^2 < 1$,
we can perform the 
systematic expansion 
even when the effect of quadratic divergences are included.
It should be noticed that the condition
$\Lambda^2/\Lambda_\chi^2 < 1$ is essentially the same as the one
needed for the derivative expansion being valid up until the
energy scale $\Lambda$:
$p^2 / \Lambda_\chi^2 < 1$ for $p^2 < \Lambda^2$.

Now, the question is whether the requirement
$\Lambda\ll\Lambda_\chi$ can be satisfied in some limit of QCD.
One possible limit is the large $N_c$ limit of QCD.
As is well known, the mesonic loop corrections are suppressed 
in this limit and tree diagrams give donimant
contributions.
Actually, in the large $N_c$ limit, $F_\pi^2(0)$ scales
as $N_c$ and thus it is natural to assume that
the bare parameter $F_\pi^2(\Lambda)$ has the same scaling 
property, $F_\pi^2(\Lambda)\sim N_c$~\footnote{%
  In Sec.~\ref{sec:WM} we will derive this scaling property
  using the Wilsonian matching condition.
},
which implies that $\Lambda_\chi$ becomes
large in the large $N_c$ limit.
On the other hand, the meson masses such as the vector meson mass 
$m_\rho$ do not scale,
so that we can introduce the $\Lambda$ which has no large $N_c$
scaling property.
Then in the large $N_c$ limit (with fixed $N_f$),
the quadratically divergent correction at $n$th loop order
is suppressed by 
$[\Lambda^2/\Lambda_\chi^2]^n \sim [1/N_c]^n$.
As a result,
we can perform the loop expansion with
quadratic divergences included in the large $N_c$ limit,
and extrapolate the results to the real-life QCD as well as to
the large $N_f$ QCD.
We will give a quantitative argument on this point in 
Sec.~\ref{sec:WM} by determining the value fot the
bare $\pi$ decay constant $F_\pi(\Lambda)$ from QCD
through the Wilsonian matching condition,
and show that
the phenomenological analysis based on the ChPT with HLS
can be done in remarkable agreement with the experiments
in much the same sense as the phenomenological analysis
in the ordinary ChPT
is successfully extended to the energy 
region higher than the pion mass scale, which is 
logically beyond the
validity region of the ChPT.

\subsection{${\cal O}(p^2)$ Lagrangian}
\label{ssec:OP2L}

For complete analysis at one-loop, we need to include terms
including the external scalar and pseudoscalar source fields 
${\cal S}$ and ${\cal P}$, as shown in Ref.~\cite{Tanabashi}.
In this subsection we present a complete ${\cal O}(p^2)$ Lagrangian of
the HLS 
with including the external source fields ${\cal S}$ and ${\cal P}$
in addition to the lowest
derivative Lagrangian~(\ref{leading Lagrangian 0}).

The external source field $\chi$, which is introduced in the ChPT,
transforms linearly under the chiral symmetry as in 
Eq.~(\ref{trans prop ChPT}), and does not transform at all
under the HLS.
Since $\widehat{\alpha}_{\perp}^{\mu}$ as well as
$\widehat{\alpha}_{\parallel}^{\mu}$ transforms as the
adjoint representation of the HLS,
it is convenient 
to convert $\chi$ into a field $\hat{\chi}$ in the adjoint
representation of the HLS
for constructing the HLS Lagrangian.
This is done by using the ``converters'' $\xi_L$ and $\xi_R$
as
\begin{equation}
\hat{\chi} 
= \xi_{\rm L} \, \chi \, \xi_{\rm R}^{\dag} 
= 2 B \xi_{\rm L} \left( {\cal S} + i {\cal P} \right) 
\xi_{\rm R}^{\dag} 
\ ,
\end{equation}
which transforms homogeneously under the HLS
[see Eq.~(\ref{xi:trans}) for the transformation properties of 
$\xi_L$ and $\xi_R$]:
\begin{equation}
\hat{\chi} \rightarrow h(x)\, \cdot \hat{\chi} \cdot h^\dag(x) \ .
\end{equation}
Then the lowest order term is given by
\begin{equation}
{\cal L}_{\chi} = \frac{1}{4} F_\chi^2 \mbox{tr}
\left[ \hat{\chi} + \hat{\chi}^\dag \right ]
\ .
\end{equation}
This source is needed to absorb the point-like transformations of
the $\pi$ and $\sigma$ fields~\cite{Tanabashi}, as was the case for
the $\chi$ field
introduced in the ChPT~\cite{Gas:84,Gas:85a}.
When we include an explicit chiral symmetry breaking due to the current
quark mass, we may introduce it as the vacuum expectation value (VEV)
of the external scalar source field:
\begin{equation}
\langle {\cal S} \rangle = {\cal M} =
\left( \begin{array}{ccc}
m_1 & & \\
 & \ddots & \\
 & & m_{N_f} \\
\end{array} \right) \ .
\label{quark mass matrix}
\end{equation}
However, in the present paper, we work in the
chiral limit, so that we take the VEV to zero 
$\langle {\cal S} \rangle = 0$.

Now, the complete leading order Lagrangian is given
by~\cite{BKUYY,BKY,Tanabashi}
\begin{eqnarray}
{\cal L}_{(2)} &=& 
{\cal L}_{\rm A} + a {\cal L}_{\rm V} + 
{\cal L}_{\rm kin}(V_\mu) + {\cal L}_\chi
\nonumber\\
&=&
F_\pi^2 \, \mbox{tr} 
\left[ \hat{\alpha}_{\perp\mu} \hat{\alpha}_{\perp}^\mu \right]
+ F_\sigma^2 \, \mbox{tr}
\left[ 
  \hat{\alpha}_{\parallel\mu} \hat{\alpha}_{\parallel}^\mu
\right]
- \frac{1}{2g^2} \, \mbox{tr} 
\left[ V_{\mu\nu} V^{\mu\nu} \right] 
+ \frac{1}{4} F_\chi^2 \mbox{tr}
\left[ \hat{\chi} + \hat{\chi}^\dag \right ]
\ ,
\label{leading Lagrangian}
\end{eqnarray}
where
$F_\chi$ in the fourth term 
is introduced to renormalize the quadratically divergent
correction to the fourth term~\cite{HY:matching}.
In the present analysis we introduced this parameter in such a way
that the field $\hat{\chi}$ does not get any renormalization effect.
We note that this $F_\chi$ agrees with $F_\pi$ at tree level.

\subsection{${\cal O}(p^4)$ Lagrangian}
\label{ssec:OP4L}

In this subsection we present a complete list of 
the ${\cal O}(p^4)$ Lagrangian, following Ref.~\cite{Tanabashi}.
We should note that, as in the ChPT (see Sec.~\ref{sec:BRCPT}),
the one-loop contributions calculated from the 
${\cal O}(p^2)$ Lagrangian are counted as ${\cal O}(p^4)$,
and thus the divergences appearing at one loop
are renormalized by the
coefficients of the ${\cal O}(p^4)$ terms listed below.

To construct ${\cal O}(p^4)$ Lagrangian
we need to include field strengths of the external gauge fields
${\cal L}_\mu$ and ${\cal R}_\mu$ in addition to the building blocks
appearing in the leading order Lagrangian in 
Eq.~(\ref{leading Lagrangian}):
\begin{eqnarray}
&& {\cal L}_{\mu\nu} = \partial_\mu {\cal L}_\nu -
\partial_\nu {\cal L}_\mu - 
i \left[ {\cal L}_\mu \,,\, {\cal L}_\nu \right] \ ,
\nonumber\\
&&
{\cal R}_{\mu\nu} = \partial_\mu {\cal R}_\nu -
\partial_\nu {\cal R}_\mu - 
i \left[ {\cal R}_\mu \,,\, {\cal R}_\nu \right] \ .
\end{eqnarray}
We again convert these into the fields which transform as adjoint
representations under the HLS:
\begin{equation}
\widehat{\cal L}_{\mu\nu} \equiv 
  \xi_{\rm L} {\cal L}_{\mu\nu} \xi_{\rm L}^\dag \ ,
\quad
\widehat{\cal R}_{\mu\nu} \equiv 
  \xi_{\rm R} {\cal R}_{\mu\nu} \xi_{\rm R}^\dag \ ,
\end{equation}
which transform as
\begin{eqnarray}
&&
  \widehat{\cal L}_{\mu\nu} \rightarrow 
  h(x) \cdot \widehat{\cal L}_{\mu\nu} \cdot h^{\dag}(x) \ ,
\nonumber\\
&&
  \widehat{\cal R}_{\mu\nu} \rightarrow 
  h(x) \cdot \widehat{\cal R}_{\mu\nu} \cdot h^{\dag}(x) \ .
\end{eqnarray}
Moreover, it is convenient to introduce the following combinations of
the above quantities:
\begin{eqnarray}
&&
\widehat{\cal V}_{\mu\nu} \equiv \frac{1}{2}
  \left[ \widehat{\cal R}_{\mu\nu} + \widehat{\cal L}_{\mu\nu} \right]
\ ,
\nonumber\\
&&
\widehat{\cal A}_{\mu\nu} \equiv \frac{1}{2}
  \left[ \widehat{\cal R}_{\mu\nu} - \widehat{\cal L}_{\mu\nu} \right]
\ .
\label{A V def}
\end{eqnarray}
A complete list of the ${\cal O}(p^4)$ Lagrangian for 
general $N_f$
was given in Ref.~\cite{Tanabashi}.~\footnote{
  We note that there are errors in the divergent corrections to
  $w_i$ in Table~1 of Ref.~\cite{Tanabashi}.
  In this report we list corrected ones in Table~\ref{tab:zwy div}
  in appendix~\ref{app:RHKE}.
}
For general $N_f$
there are 35 ${\cal O}(p^4)$ terms compared with 13 terms 
in the ChPT
[$L_0$, $L_1$, $\ldots$, $L_{10}$, $H_1$ and $H_2$ terms;
see Eq.~(\ref{ChPT:Lag:4})]:
\begin{eqnarray}
{\cal L}_{(4)y} 
&=&
y_1 \, \mbox{tr} \left[ 
  \hat{\alpha}_{\perp\mu} \hat{\alpha}_\perp^\mu
  \hat{\alpha}_{\perp\nu} \hat{\alpha}_\perp^\nu
\right]
+ y_2 \, \mbox{tr} \left[
  \hat{\alpha}_{\perp\mu} \hat{\alpha}_{\perp\nu}
  \hat{\alpha}^\mu_\perp \hat{\alpha}^\nu_\perp
\right]
\nonumber\\
&& 
{}+ y_3 \, \mbox{tr} \left[
  \hat{\alpha}_{\parallel\mu} \hat{\alpha}_\parallel^\mu
  \hat{\alpha}_{\parallel\nu} \hat{\alpha}_\parallel^\nu
\right]
+ y_4 \, \mbox{tr} \left[
  \hat{\alpha}_{\parallel\mu} \hat{\alpha}_{\parallel\nu}
  \hat{\alpha}^\mu_\parallel \hat{\alpha}^\nu_\parallel
\right]
\nonumber\\
&&
{}+ y_5 \, \mbox{tr} \left[
  \hat{\alpha}_{\perp\mu} \hat{\alpha}_\perp^\mu
  \hat{\alpha}_{\parallel\nu} \hat{\alpha}_\parallel^\nu
\right]
+ y_6 \, \mbox{tr} \left[
  \hat{\alpha}_{\perp\mu} \hat{\alpha}_{\perp\nu}
  \hat{\alpha}^\mu_\parallel \hat{\alpha}^\nu_\parallel
\right]
{}+ y_7 \, \mbox{tr} \left[
  \hat{\alpha}_{\perp\mu} \hat{\alpha}_{\perp\nu}
  \hat{\alpha}^\nu_\parallel \hat{\alpha}^\mu_\parallel
\right]
\nonumber\\
&&
{}+ y_8 \, \left\{
\mbox{tr} \left[ 
  \hat{\alpha}_{\perp\mu} \hat{\alpha}_\parallel^\mu
  \hat{\alpha}_{\perp\nu} \hat{\alpha}_\parallel^\nu
\right]
+ \mbox{tr} \left[
  \hat{\alpha}_{\perp\mu} \hat{\alpha}_{\parallel\nu} 
  \hat{\alpha}_\perp^\nu \hat{\alpha}_\parallel^\mu
\right] \right\}
{}+ y_9 \, \mbox{tr} \left[
  \hat{\alpha}_{\perp\mu} \hat{\alpha}_{\parallel\nu}
  \hat{\alpha}^\mu_\perp \hat{\alpha}^\nu_\parallel
\right]
\nonumber\\
&&
{}+ y_{10} \left(
\mbox{tr} \left[
  \hat{\alpha}_{\perp\mu} \hat{\alpha}_\perp^\mu
\right] \right)^2
+ y_{11} \, \mbox{tr} \left[
  \hat{\alpha}_{\perp\mu} \hat{\alpha}_{\perp\nu}
\right] 
\mbox{tr} \left[
  \hat{\alpha}^\mu_\perp \hat{\alpha}^\nu_\perp
\right]
\nonumber\\
&&
{}+ y_{12} \left( \mbox{tr} \left[ 
  \hat{\alpha}_{\parallel\mu} \hat{\alpha}_\parallel^\mu
\right] \right)^2
+ y_{13} \, \mbox{tr} \left[
  \hat{\alpha}_{\parallel\mu} \hat{\alpha}_{\parallel\nu}
\right]
\mbox{tr} \left[
  \hat{\alpha}^\mu_\parallel \hat{\alpha}^\nu_\parallel
\right]
\nonumber\\
&&
{}+ y_{14} \, \mbox{tr} \left[
  \hat{\alpha}_{\perp\mu} \hat{\alpha}_\perp^\mu
\right]
\mbox{tr} \left[
  \hat{\alpha}_{\parallel\nu} \hat{\alpha}_\parallel^\nu
\right]
{}+ y_{15} \, \mbox{tr} \left[
  \hat{\alpha}_{\perp\mu} \hat{\alpha}_{\perp\nu}
\right]
\mbox{tr} \left[
  \hat{\alpha}^\mu_\parallel \hat{\alpha}^\nu_\parallel
\right]
\nonumber\\
&&
{}+ y_{16} \left( \mbox{tr} \left[
  \hat{\alpha}_{\perp\mu} \hat{\alpha}_\parallel^\mu
\right] \right)^2
+ y_{17} \, \mbox{tr} \left[
  \hat{\alpha}_{\perp\mu} \hat{\alpha}_{\parallel\nu}
\right]
\mbox{tr} \left[
  \hat{\alpha}^\mu_\perp \hat{\alpha}^\nu_\parallel
\right]
\nonumber\\
&&
{}+ y_{18} \, \mbox{tr} \left[
  \hat{\alpha}_{\perp\mu} \hat{\alpha}_{\parallel\nu}
\right]
\mbox{tr} \left[
  \hat{\alpha}^\mu_\parallel \hat{\alpha}^\nu_\perp 
\right]
\ ,
\label{Lag: y terms}
\\
{\cal L}_{(4)w} 
&=&
w_1 \, \frac{F_\chi^2}{F_\pi^2} \, \mbox{tr} \left[
  \hat{\alpha}_{\perp\mu} \hat{\alpha}_\perp^\mu
  \left( \hat{\chi} + \hat{\chi}^\dag \right)
\right]
{}+ w_2 \, \frac{F_\chi^2}{F_\pi^2} \, \mbox{tr} \left[
  \hat{\alpha}_{\perp\mu} \hat{\alpha}_\perp^\mu
\right]
\mbox{tr} \left[
  \hat{\chi} + \hat{\chi}^\dag
\right]
\nonumber\\
&&
{}+ w_3 \, \frac{F_\chi^2}{F_\pi^2} \, \mbox{tr} \left[
  \hat{\alpha}_{\parallel\mu} \hat{\alpha}_\parallel^\mu
  \left( \hat{\chi} + \hat{\chi}^\dag \right)
\right]
{}+ w_4 \, \frac{F_\chi^2}{F_\pi^2} \, \mbox{tr} \left[
  \hat{\alpha}_{\parallel\mu} \hat{\alpha}_\parallel^\mu
\right]
\mbox{tr} \left[
  \hat{\chi} + \hat{\chi}^\dag
\right]
\nonumber\\
&&
{}+ w_5 \, \frac{F_\chi^2}{F_\pi^2} \, \mbox{tr} \left[ 
  \left( 
    \hat{\alpha}_\parallel^\mu \hat{\alpha}_{\perp\mu}
    - \hat{\alpha}_{\perp\mu} \hat{\alpha}_\parallel^\mu
  \right)
  \left( \hat{\chi} - \hat{\chi}^\dag \right)
\right]
\nonumber\\
&&
{}+ w_6 \, \frac{F_\chi^4}{F_\pi^4} \, \mbox{tr} \left[
  \left( \hat{\chi} + \hat{\chi}^\dag \right)^2
\right]
+ w_7 \, \frac{F_\chi^4}{F_\pi^4} \left( \mbox{tr} \left[
  \hat{\chi} + \hat{\chi}^\dag
\right] \right)^2
\nonumber\\
&&
{}+ w_8 \, \frac{F_\chi^4}{F_\pi^4} \, \mbox{tr} \left[
  \left( \hat{\chi} - \hat{\chi}^\dag \right)^2
\right]
+ w_9 \, \frac{F_\chi^4}{F_\pi^4} \left( \mbox{tr} \left[
  \hat{\chi} - \hat{\chi}^\dag
\right] \right)^2
\ ,
\label{Lag: w terms}
\\
{\cal L}_{(4)z} 
&=&
z_1 \,\mbox{tr}
 \left[ \hat{\cal V}_{\mu\nu} \hat{\cal V}^{\mu\nu} \right]
+ z_2 \,\mbox{tr}
 \left[ \hat{\cal A}_{\mu\nu} \hat{\cal A}^{\mu\nu} \right]
{}+ z_3 \,\mbox{tr}\left[ \hat{\cal V}_{\mu\nu} V^{\mu\nu} \right]
\nonumber\\
&& 
{} + i z_4 \,\mbox{tr}\left[ 
  V_{\mu\nu} \hat{\alpha}_\perp^\mu \hat{\alpha}_\perp^\nu 
\right]
+ i z_5 \,\mbox{tr}\left[ 
  V_{\mu\nu} \hat{\alpha}_\parallel^\mu \hat{\alpha}_\parallel^\nu 
\right]
\nonumber\\
&&
{} + i z_6 \,\mbox{tr}\left[ 
  \hat{\cal V}_{\mu\nu} \hat{\alpha}_\perp^\mu \hat{\alpha}_\perp^\nu 
\right]
+ i z_7 \,\mbox{tr}\left[ 
  \hat{\cal V}_{\mu\nu} \hat{\alpha}_\parallel^\mu \hat{\alpha}_\parallel^\nu 
\right]
\nonumber\\
&&
{} - i z_8 \,\mbox{tr}\left[ 
  \hat{\cal A}_{\mu\nu} 
  \left( \hat{\alpha}_\perp^\mu \hat{\alpha}_\parallel^\nu 
         + \hat{\alpha}_\parallel^\mu \hat{\alpha}_\perp^\nu \right)
\right]
\ ,
\label{Lag: z terms}
\end{eqnarray}
where use was made of
the equations of motion:
\begin{eqnarray}
D_\mu \hat{\alpha}_{\perp}^\mu 
&=&
- i \left( a- 1 \right)
\left[
  \hat{\alpha}_{\parallel\mu} \,,\,
  \hat{\alpha}_\perp^\mu
\right]
{}- \frac{i}{4} \frac{F_\chi^2}{F_\pi^2} 
\left(
  \hat{\chi} - \hat{\chi}^\dag
  - \frac{1}{N_f} \, \mbox{tr}
  \left[ \hat{\chi} - \hat{\chi}^\dag \right]
\right)
+ {\cal O}(p^4)
\ ,
\label{EOM Npi2}
\\
D_\mu \hat{\alpha}_{\parallel}^\mu 
&=& {\cal O}(p^4)
\ ,
\label{EOM Nsig2}
\\
D_\nu {V}^{\nu\mu}
&=&
g^2 f_\sigma^2 
\hat{\alpha}_{\parallel}^\mu 
+ {\cal O}(p^4)
\ ,
\label{EOM Nvec2}
\end{eqnarray}
and the identities:
\begin{eqnarray}
D_\mu \hat{\alpha}_{\perp\nu}
- D_\nu \hat{\alpha}_{\perp\mu}
&=&
i \left[ 
  \hat{\alpha}_{\parallel\mu} \,,\, \hat{\alpha}_{\perp\nu}
\right]
+
i \left[ 
  \hat{\alpha}_{\perp\mu} \,,\, \hat{\alpha}_{\parallel\nu}
\right]
+ \widehat{\cal A}_{\mu\nu}
\ ,
\label{rel:perp}
\\
D_\mu \hat{\alpha}_{\parallel\nu}
- D_\nu \hat{\alpha}_{\parallel\mu}
&=&
i \left[ 
  \hat{\alpha}_{\parallel\mu} \,,\, \hat{\alpha}_{\parallel\nu}
\right]
+
i \left[ 
  \hat{\alpha}_{\perp\mu} \,,\, \hat{\alpha}_{\perp\nu}
\right]
+ \widehat{\cal V}_{\mu\nu}
- V_{\mu\nu}
\ ,
\label{rel:parallel}
\end{eqnarray}
with $\widehat{\cal A}_{\mu\nu}$
and $\widehat{\cal V}_{\mu\nu}$ being defined in 
Eq.~(\ref{A V def}).

We note that for $N_f = 3$, similarly to the relation (\ref{rel:3:p0})
for the ChPT, using the identity:
\begin{equation}
\mbox{tr} \left[ A B A B \right]
= - 2 \, \mbox{tr} \left[ A^2 B^2 \right] 
+ \frac{1}{2} \, \mbox{tr} \left[ A^2 \right] 
  \mbox{tr} \left[ B^2 \right]
+ \left( \mbox{tr} \left[ A B \right] \right)^2
\end{equation}
valid for any pair of traceless, hermitian $3\times3$ matrices $A$
and $B$,
we have the following relations:
\begin{eqnarray}
\mbox{tr} \left[
  \hat{\alpha}_{\perp\mu} \hat{\alpha}_{\perp\nu}
  \hat{\alpha}^\mu_\perp \hat{\alpha}^\nu_\perp
\right]
&=&
- 2 \, \mbox{tr} \left[ 
  \hat{\alpha}_{\perp\mu} \hat{\alpha}_\perp^\mu
  \hat{\alpha}_{\perp\nu} \hat{\alpha}_\perp^\nu
\right]
\nonumber\\
&&
{} + \frac{1}{2} \left(
\mbox{tr} \left[
  \hat{\alpha}_{\perp\mu} \hat{\alpha}_\perp^\mu
\right] \right)^2
+ \mbox{tr} \left[
  \hat{\alpha}_{\perp\mu} \hat{\alpha}_{\perp\nu}
\right] 
\mbox{tr} \left[
  \hat{\alpha}^\mu_\perp \hat{\alpha}^\nu_\perp
\right]
\ ,
\nonumber\\
\mbox{tr} \left[
  \hat{\alpha}_{\parallel\mu} \hat{\alpha}_{\parallel\nu}
  \hat{\alpha}^\mu_\parallel \hat{\alpha}^\nu_\parallel
\right]
&=&
- 2 \mbox{tr} \left[
  \hat{\alpha}_{\parallel\mu} \hat{\alpha}_\parallel^\mu
  \hat{\alpha}_{\parallel\nu} \hat{\alpha}_\parallel^\nu
\right]
\nonumber\\
&&
{} + \frac{1}{2} \left( \mbox{tr} \left[ 
  \hat{\alpha}_{\parallel\mu} \hat{\alpha}_\parallel^\mu
\right] \right)^2
+ \mbox{tr} \left[
  \hat{\alpha}_{\parallel\mu} \hat{\alpha}_{\parallel\nu}
\right]
\mbox{tr} \left[
  \hat{\alpha}^\mu_\parallel \hat{\alpha}^\nu_\parallel
\right]
\ ,
\nonumber\\
\mbox{tr} \left[
  \hat{\alpha}_{\perp\mu} \hat{\alpha}_{\parallel\nu}
  \hat{\alpha}^\mu_\perp \hat{\alpha}^\nu_\parallel
\right]
&=&
- 2 \, \mbox{tr} \left[
  \hat{\alpha}_{\perp\mu} \hat{\alpha}_\perp^\mu
  \hat{\alpha}_{\parallel\nu} \hat{\alpha}_\parallel^\nu
\right]
\nonumber\\
&&
{} + \frac{1}{2} \mbox{tr} \left[
  \hat{\alpha}_{\perp\mu} \hat{\alpha}_\perp^\mu
\right]
\mbox{tr} \left[
  \hat{\alpha}_{\parallel\nu} \hat{\alpha}_\parallel^\nu
\right]
+ \mbox{tr} \left[
  \hat{\alpha}_{\perp\mu} \hat{\alpha}_{\parallel\nu}
\right]
\mbox{tr} \left[
  \hat{\alpha}^\mu_\perp \hat{\alpha}^\nu_\parallel
\right]
\ .
\end{eqnarray}
Then there are 32 independent terms in the ${\cal O}(p^4)$ Lagrangian
of the HLS 
in contrast to 12 terms 
in
the ChPT Lagrangian
[$L_1$, $\ldots$, $L_{10}$, $H_1$ and $H_2$ terms;
see Eq.~(\ref{p4:ChPT})].

For $N_f =2$, on the other hand, we have the following identity valid
for traceless, hermitian $2\times2$ matrices $A$, $B$, $C$ and $D$:
\begin{equation}
\mbox{tr} \left[ A B C D \right]
=
\frac{1}{2}\, \mbox{tr} \left[ A B \right] 
   \mbox{tr} \left[ C D \right]
- \frac{1}{2}\, \mbox{tr} \left[ A C \right] 
   \mbox{tr} \left[ B D \right]
+ \frac{1}{2}\, \mbox{tr} \left[ A D \right] 
   \mbox{tr} \left[ B C \right]
\ .
\end{equation}
Then, each of $y_1$- through $y_9$-terms is rewritten into a
combination of $y_{10}$- through $y_{18}$-terms:
\begin{eqnarray}
&&
  \mbox{tr} \left[
    \hat{\alpha}_{\perp\mu} \hat{\alpha}_{\perp\nu}
    \hat{\alpha}^\mu_\perp \hat{\alpha}^\nu_\perp
  \right]
  =
  \mbox{tr} \left[
    \hat{\alpha}_{\perp\mu} \hat{\alpha}_{\perp\nu}
  \right] 
  \mbox{tr} \left[
    \hat{\alpha}^\mu_\perp \hat{\alpha}^\nu_\perp
  \right]
  {} - \frac{1}{2} \left(
  \mbox{tr} \left[
    \hat{\alpha}_{\perp\mu} \hat{\alpha}_\perp^\mu
  \right] \right)^2
\ ,
\nonumber\\
&&
  \mbox{tr} \left[ 
    \hat{\alpha}_{\perp\mu} \hat{\alpha}_\perp^\mu
    \hat{\alpha}_{\perp\nu} \hat{\alpha}_\perp^\nu
  \right]
  =
  \frac{1}{2} \left(
  \mbox{tr} \left[
    \hat{\alpha}_{\perp\mu} \hat{\alpha}_\perp^\mu
  \right] \right)^2
\ ,
\nonumber\\
&&
  \mbox{tr} \left[
    \hat{\alpha}_{\parallel\mu} \hat{\alpha}_{\parallel\nu}
    \hat{\alpha}^\mu_\parallel \hat{\alpha}^\nu_\parallel
  \right]
  =
  \mbox{tr} \left[
    \hat{\alpha}_{\parallel\mu} \hat{\alpha}_{\parallel\nu}
  \right]
  \mbox{tr} \left[
    \hat{\alpha}^\mu_\parallel \hat{\alpha}^\nu_\parallel
  \right]
  - \frac{1}{2} \left( \mbox{tr} \left[ 
    \hat{\alpha}_{\parallel\mu} \hat{\alpha}_\parallel^\mu
  \right] \right)^2
\ ,
\nonumber\\
&&
  \mbox{tr} \left[
    \hat{\alpha}_{\parallel\mu} \hat{\alpha}_\parallel^\mu
    \hat{\alpha}_{\parallel\nu} \hat{\alpha}_\parallel^\nu
  \right]
  =
  \frac{1}{2} \left( \mbox{tr} \left[ 
    \hat{\alpha}_{\parallel\mu} \hat{\alpha}_\parallel^\mu
  \right] \right)^2
\ ,
\nonumber\\
&&
  \mbox{tr} \left[
    \hat{\alpha}_{\perp\mu} \hat{\alpha}_\perp^\mu
    \hat{\alpha}_{\parallel\nu} \hat{\alpha}_\parallel^\nu
  \right]
  =
  \frac{1}{2} \mbox{tr} \left[
    \hat{\alpha}_{\perp\mu} \hat{\alpha}_\perp^\mu
  \right]
  \mbox{tr} \left[
    \hat{\alpha}_{\parallel\nu} \hat{\alpha}_\parallel^\nu
  \right]
\ ,
\nonumber\\
&&
  \mbox{tr} \left[
    \hat{\alpha}_{\perp\mu} \hat{\alpha}_{\perp\nu}
    \hat{\alpha}^\mu_\parallel \hat{\alpha}^\nu_\parallel
  \right]
  =
  \frac{1}{2}\, \mbox{tr} \left[
    \hat{\alpha}_{\perp\mu} \hat{\alpha}_{\perp\nu}
  \right]
  \mbox{tr} \left[
    \hat{\alpha}^\mu_\parallel \hat{\alpha}^\nu_\parallel
  \right]
  - \frac{1}{2}
  \left( \mbox{tr} \left[
    \hat{\alpha}_{\perp\mu} \hat{\alpha}_\parallel^\mu
  \right] \right)^2
\nonumber\\
&& \qquad\qquad\qquad\qquad
  {} + \frac{1}{2}\, 
  \mbox{tr} \left[
    \hat{\alpha}_{\perp\mu} \hat{\alpha}_{\parallel\nu}
  \right]
  \mbox{tr} \left[
    \hat{\alpha}^\mu_\parallel \hat{\alpha}^\nu_\perp 
  \right]
\ ,
\nonumber\\
&&
  \mbox{tr} \left[
    \hat{\alpha}_{\perp\mu} \hat{\alpha}_{\perp\nu}
    \hat{\alpha}^\nu_\parallel \hat{\alpha}^\mu_\parallel
  \right]
  =
  \frac{1}{2}\, \mbox{tr} \left[
    \hat{\alpha}_{\perp\mu} \hat{\alpha}_{\perp\nu}
  \right]
  \mbox{tr} \left[
    \hat{\alpha}^\mu_\parallel \hat{\alpha}^\nu_\parallel
  \right]
  - \frac{1}{2}\,
  \mbox{tr} \left[
    \hat{\alpha}_{\perp\mu} \hat{\alpha}_{\parallel\nu}
  \right]
  \mbox{tr} \left[
    \hat{\alpha}^\mu_\parallel \hat{\alpha}^\nu_\perp 
  \right]
\nonumber\\
&& \qquad\qquad\qquad\qquad
  {} + \frac{1}{2}
  \left( \mbox{tr} \left[
    \hat{\alpha}_{\perp\mu} \hat{\alpha}_\parallel^\mu
  \right] \right)^2
\ ,
\nonumber\\
&&
  \mbox{tr} \left[ 
    \hat{\alpha}_{\perp\mu} \hat{\alpha}_\parallel^\mu
    \hat{\alpha}_{\perp\nu} \hat{\alpha}_\parallel^\nu
  \right]
  + \mbox{tr} \left[
    \hat{\alpha}_{\perp\mu} \hat{\alpha}_{\parallel\nu} 
    \hat{\alpha}_\perp^\nu \hat{\alpha}_\parallel^\mu
  \right]
  =
  \frac{1}{2}
  \left( \mbox{tr} \left[
    \hat{\alpha}_{\perp\mu} \hat{\alpha}_\parallel^\mu
  \right] \right)^2
\nonumber\\
&& \qquad\qquad\qquad\qquad
  {}- \frac{1}{2}\, \mbox{tr} \left[
    \hat{\alpha}_{\perp\mu} \hat{\alpha}_{\perp\nu}
  \right]
  \mbox{tr} \left[
    \hat{\alpha}^\mu_\parallel \hat{\alpha}^\nu_\parallel
  \right]
  {} + \frac{1}{2}\,
  \mbox{tr} \left[
    \hat{\alpha}_{\perp\mu} \hat{\alpha}_{\parallel\nu}
  \right]
  \mbox{tr} \left[
    \hat{\alpha}^\mu_\parallel \hat{\alpha}^\nu_\perp 
  \right]
\ ,
\nonumber\\
&&
  \mbox{tr} \left[
    \hat{\alpha}_{\perp\mu} \hat{\alpha}_{\parallel\nu}
    \hat{\alpha}^\mu_\perp \hat{\alpha}^\nu_\parallel
  \right]
  =
  \mbox{tr} \left[
    \hat{\alpha}_{\perp\mu} \hat{\alpha}_{\parallel\nu}
  \right]
  \mbox{tr} \left[
    \hat{\alpha}^\mu_\perp \hat{\alpha}^\nu_\parallel
  \right]
  - \frac{1}{2} \mbox{tr} \left[
    \hat{\alpha}_{\perp\mu} \hat{\alpha}_\perp^\mu
  \right]
  \mbox{tr} \left[
    \hat{\alpha}_{\parallel\nu} \hat{\alpha}_\parallel^\nu
  \right]
\ .
\end{eqnarray}
Furthermore, similarly to the relation
(\ref{rel:2:p5}) for the ChPT,
we have the following relations:
\begin{eqnarray}
&&
\mbox{tr} \left[
  \hat{\alpha}_{\perp\mu} \hat{\alpha}_\perp^\mu
  \left( \hat{\chi} + \hat{\chi}^\dag \right)
\right]
=
\frac{1}{2} \,
\mbox{tr} \left[
  \hat{\alpha}_{\perp\mu} \hat{\alpha}_\perp^\mu
\right]
\mbox{tr} \left[
  \hat{\chi} + \hat{\chi}^\dag
\right]
\ ,
\nonumber\\
&&
\mbox{tr} \left[
  \hat{\alpha}_{\parallel\mu} \hat{\alpha}_\parallel^\mu
  \left( \hat{\chi} + \hat{\chi}^\dag \right)
\right]
=
\frac{1}{2} \,
\mbox{tr} \left[
  \hat{\alpha}_{\parallel\mu} \hat{\alpha}_\parallel^\mu
\right]
\mbox{tr} \left[
  \hat{\chi} + \hat{\chi}^\dag
\right]
\ .
\end{eqnarray}
Thus, there are 24 independent terms in the ${\cal O}(p^4)$ Lagrangian
of the HLS in contrast to 10 terms in the ChPT Lagrangian
[$L_1$, $L_2$, $L_4$, $L_6$, $L_7$, $L_8$, $L_9$,
$L_{10}$, $H_1$ and $H_2$ terms;
see Eq.~(\ref{ChPT:Lag:2})].

At first sight, so many proliferated
terms look untractable and one might think
that the ChPT with HLS would be useless.
However, it is not the case:
In the above ${\cal O}(p^4)$ Lagrangian
all the terms in ${\cal L}_{(4)y}$
generate vertices with at least four legs.
In other words,
all the $y_i$ terms do not contribute to two or three point
functions.
In the chiral limit $\langle \hat{\chi} \rangle =
\langle \hat{\chi}^\dag \rangle = 0$
(no explicit chiral symmetry breaking due to the
current quark masses),
the terms in ${\cal L}_{(4)w}$ do not contribute to the Green
functions of the vector and axialvector currents.
In the terms in ${\cal L}_{(4)z}$ the 
$z_1$ $\sim$ $z_3$ terms contribute to two-point function, while 
the contributions from $z_4$ $\sim$ $z_8$ terms 
are operative only for $N$($\geq3$)-point function.
Thus, as far as
we consider the two-point functions of the vector and
axialvector current, only $z_1$, $z_2$ and $z_3$ terms 
in the entire ${\cal O}(p^4)$ Lagrangian contribute.

Let us study the correspondence between the parameters in the
HLS and the ${\cal O}(p^4)$ ChPT parameters 
at tree level.
By using the method used in Sec.~\ref{ssec:VMSLEC},
the correspondence for $N_f=3$ is obtained 
as~\cite{HY:matching}~\footnote{%
  We note that in Ref.~\cite{HY:matching} the contributions from $z_4$
  to $L_1$, $L_2$ and $L_3$ are missing, and the sign in front of
  $(z_4+z_6)/8$ in $L_9$ was wrong.
  They are corrected in Eq.~(\ref{rel HLS ChPT tree}).
}
\begin{eqnarray}
&& 
L_1 \mathop{\Longleftrightarrow}_{tree}
\frac{1}{32 g^2} - \frac{1}{32} z_4
+ \frac{1}{32} y_2 + \frac{1}{16} y_{10}
\ ,
\nonumber\\
&& 
L_2 \mathop{\Longleftrightarrow}_{tree}
\frac{1}{16 g^2} - \frac{1}{16} z_4
+ \frac{1}{16} y_2 + \frac{1}{16} y_{11}
\ ,
\nonumber\\
&& 
L_3 \mathop{\Longleftrightarrow}_{tree}
- \frac{3}{16 g^2} + \frac{3}{16} z_4
+ \frac{1}{16} y_1 - \frac{1}{8} y_2
\ ,
\nonumber\\
&& 
L_4 \mathop{\Longleftrightarrow}_{tree} \frac{1}{4} w_2
\ ,
\nonumber\\
&& 
L_5 \mathop{\Longleftrightarrow}_{tree} \frac{1}{4} w_1
\ ,
\nonumber\\
&& 
L_6 \mathop{\Longleftrightarrow}_{tree} w_7
\ ,
\nonumber\\
&& 
L_7 \mathop{\Longleftrightarrow}_{tree} w_9
\ ,
\nonumber\\
&& 
L_8 \mathop{\Longleftrightarrow}_{tree} \left( w_6 + w_8 \right)
\ ,
\nonumber\\
&& 
L_9 \mathop{\Longleftrightarrow}_{tree} 
\frac{1}{4} \left( \frac{1}{g^2} - z_3 \right)
- \frac{1}{8} \left( z_4 + z_6 \right)
\ ,
\nonumber\\
&& 
L_{10} \mathop{\Longleftrightarrow}_{tree} 
- \frac{1}{4g^2} + \frac{1}{2} \left( z_3 - z_2 + z_1 \right)
\ ,
\nonumber\\
&& 
H_1 \mathop{\Longleftrightarrow}_{tree} 
- \frac{1}{8g^2} + \frac{1}{4} \left( z_3 + z_2 + z_1 \right)
\ ,
\nonumber\\
&& 
H_2 \mathop{\Longleftrightarrow}_{tree} 2 \left( w_6 - w_8 \right)
\ ,
\label{rel HLS ChPT tree}
\end{eqnarray}
where we took $F_\chi = F_\pi$.
It should be noticed that the above relations are valid only at tree
level.
As discussed in Ref.~\cite{Tanabashi}, since one-loop corrections from
${\cal O}(p^2)$ Lagrangian ${\cal L}_{(2)}$ generate ${\cal O}(p^4)$
contribution, 
we have to relate these
at one-loop level where
finite order corrections appear in several relations.
We will show, as an example, the inclusion of such finite corrections
to the relation for $L_{10}$ in Sec.~\ref{ssec:MHC}
[see Eq.~(\ref{l10})].

\subsection{Background field gauge}
\label{ssec:BGFM}

We adopt the background field gauge
to obtain quantum corrections to the parameters.
[For calculation in other gauges, see Ref.~\cite{HY} for the
$R_\xi$-like gauge and Refs.~\cite{HKY:PRL,HKY:PTP} for the covariant
gauge.]
This subsection is for a preparation to calculate the
quantum corrections at one loop in proceeding subsections.
The background field gauge was used in the ChPT in 
Refs.~\cite{Gas:84,Gas:85a},
and was applied to the HLS in Ref.~\cite{Tanabashi}.
In this gauge we can easily obtain the vector meson propagator,
which is gauge covariant even at off-shell, from the two-point
function.~\footnote{%
  Note that, in the $R_\xi$-like gauge fixing~\cite{HY},
  the propagator obtained from the two-point function
  by naive resummation is not gauge covariant at off-shell,
  since the two-point function at one loop is not gauge covariant 
  at off-shell
  due to the existence of non-Abelian vertex.
  This is well-known in defining the electroweak gauge boson
  propagators in the standard model, which is solved by including
  a part of the vertex correction into the propagator through
  so-called pinch technique (see, e.g., 
  Refs.~\cite{Degrassi-Sirlin:1,Degrassi-Sirlin:2,%
  Degrassi-Kniehl-Sirlin}).
  In the background field gauge, on the other hand, 
  the gauge invariance (or covariance) is manifestly kept, so that
  the resultant two-point function and then the propagator obtained 
  by resumming it are gauge covariant even at off-shell
  (see, e.g., Ref.~\cite{Denner-Weiglein-Dittmaier}).
}
Thus, in the background field gauge, we can easily perform the 
off-shell extrapolation of the gauge field.
Furthermore, while in the covariant or $R_\xi$-like gauge we need to
consider the point transformation of the pion field in addition to the
counter terms included in the Lagrangian
to renormalize the
divergence appearing only in the off-shell 
amplitude of more than two pions
(see, e.g., Ref.~\cite{Appelquist-Bernard})~\footnote{%
  In the analysis done in Sec.~\ref{sec:RALOLET} in the 
  covariant gauge, the point transformation needed in the field
  renormalization in Eq.~(\ref{field renorm LET})
  is expressed by a certain function $F^i(\phi)$
  in Eq.~(\ref{SolY}).
},
we do not need to consider such a transformation separately in the
background field gauge:  The occurrence of the external source field
$\hat{\chi}$ (especially the terms quadratic in $\hat{\chi}$)
in the counter terms is related to the 
point transformation in the covariant or $R_\xi$-like gauge
(see e.g., Ref.~\cite{Gas:84}).

Now, following Ref.~\cite{Tanabashi}
we introduce the background fields $\overline{\xi}_{\rm L}$
and $\overline{\xi}_{\rm R}$ as
\begin{equation}
\xi_{\rm L,R} = \check{\xi}_{\rm L,R} \overline{\xi}_{\rm L,R} \ ,
\end{equation}
where $\check{\xi}_{\rm L,R}$ denote the quantum fields.
It is convenient to write
\begin{eqnarray}
&&
\check{\xi}_{\rm L} = 
\check{\xi}_{\rm S} \cdot
\check{\xi}_{\rm P}^\dag \ ,
\qquad
\check{\xi}_{\rm R} = 
\check{\xi}_{\rm S} \cdot
\check{\xi}_{\rm P} \ ,
\nonumber\\
&& \quad
\check{\xi}_{\rm P} =
\exp \left[ i\, \check{\pi}^a T_a /F_\pi\right] \ ,
\quad
\check{\xi}_{\rm S} =
\exp \left[ i \, \check{\sigma}^a T_a/F_\sigma \right] \ ,
\end{eqnarray}
with $\check{\pi}$ and $\check{\sigma}$ being the
quantum fields corresponding to the NG boson $\pi$ and the would-be NG
boson $\sigma$. 
The background field $\overline{V}_\mu$ and the quantum field
$\check{\rho}_\mu$ of the HLS gauge boson are introduced as
\begin{eqnarray}
&&
V_\mu = \overline{V}_\mu + g \check{\rho}_\mu \ .
\end{eqnarray}
We use the following notations for the background fields
including $\overline{\xi}_{\rm L,R}$:
\begin{eqnarray}
\overline{\cal A}_\mu &\equiv&
\frac{1}{2i} \left[
  \partial_\mu \overline{\xi}_{\rm R} \cdot 
    \overline{\xi}_{\rm R}^\dag
  - \partial_\mu \overline{\xi}_{\rm L} \cdot 
    \overline{\xi}_{\rm L}^\dag
\right]
+ \frac{1}{2} \left[
  \overline{\xi}_{\rm R} {\cal R}_\mu 
    \overline{\xi}_{\rm R}^\dag
  - \overline{\xi}_{\rm L} {\cal L}_\mu 
    \overline{\xi}_{\rm L}^\dag
\right]
\ ,
\nonumber\\
\overline{\cal V}_\mu &\equiv&
\frac{1}{2i} \left[
  \partial_\mu \overline{\xi}_{\rm R} \cdot 
    \overline{\xi}_{\rm R}^\dag
  + \partial_\mu \overline{\xi}_{\rm L} \cdot 
    \overline{\xi}_{\rm L}^\dag
\right]
+ \frac{1}{2} \left[
  \overline{\xi}_{\rm R} {\cal R}_\mu 
    \overline{\xi}_{\rm R}^\dag
  + \overline{\xi}_{\rm L} {\cal L}_\mu 
    \overline{\xi}_{\rm L}^\dag
\right]
\ ,
\end{eqnarray}
which correspond to $\hat{\alpha}_{\perp\mu}$ and 
$\hat{\alpha}_{\parallel\mu} + V_\mu$, respectively.
The field strengths of $\overline{\cal A}_\mu$ and 
$\overline{\cal V}_\mu$ are
defined as
\begin{eqnarray}
\overline{\cal V}_{\mu\nu} &=&
\partial_\mu \overline{\cal V}_\nu 
- \partial_\nu \overline{\cal V}_\mu
- i \left[ \overline{\cal V}_\mu \,,\, \overline{\cal V}_\nu \right]
- i \left[ \overline{\cal A}_\mu \,,\, \overline{\cal A}_\nu \right]
\ ,
\nonumber\\
\overline{\cal A}_{\mu\nu} &=&
\partial_\mu \overline{\cal A}_\nu 
- \partial_\nu \overline{\cal A}_\mu
- i \left[ \overline{\cal V}_\mu \,,\, \overline{\cal A}_\nu \right]
- i \left[ \overline{\cal A}_\mu \,,\, \overline{\cal V}_\nu \right]
\ .
\end{eqnarray}
Note that both $\overline{\cal V}_{\mu\nu}$ and
$\overline{\cal A}_{\mu\nu}$ do not include any derivatives of the
background field $\overline{\xi}_{\rm R}$ and 
$\overline{\xi}_{\rm L}$:
\begin{eqnarray}
\overline{\cal V}_{\mu\nu} &=&
  \frac{1}{2}
  \left[
    \overline{\xi}_{\rm R} {\cal R}_{\mu\nu}
      \overline{\xi}_{\rm R}^\dag
    + \overline{\xi}_{\rm L} {\cal L}_{\mu\nu}
      \overline{\xi}_{\rm L}^\dag
  \right]
\ ,
\nonumber\\
\overline{\cal A}_{\mu\nu} &=&
  \frac{1}{2}
  \left[
    \overline{\xi}_{\rm R} {\cal R}_{\mu\nu}
      \overline{\xi}_{\rm R}^\dag
    - \overline{\xi}_{\rm L} {\cal L}_{\mu\nu}
      \overline{\xi}_{\rm L}^\dag
  \right]
\ .
\end{eqnarray}
Then, $\overline{\cal V}_{\mu\nu}$ and
$\overline{\cal A}_{\mu\nu}$
correspond to $\hat{V}_{\mu\nu}$ and $\hat{A}_{\mu\nu}$
in Eq.~(\ref{A V def}),
respectively.
In addition, we use $\overline{\chi}$ for the background field
corresponding to $\hat{\chi}$:
\begin{equation}
\overline{\chi} \equiv 2 B \overline{\xi}_{\rm L}
\left( {\cal S} + i {\cal P} \right) \overline{\xi}_{\rm R}^\dag
\ .
\end{equation}

It should be noticed that the quantum fields as well as the background
fields $\overline{\xi}_{\rm R,L}$ transform homogeneously
under the background gauge transformation, while the background gauge
field $\overline{V}_\mu$ transforms inhomogeneously:
\begin{eqnarray}
&& \overline{\xi}_{\rm R,L} \rightarrow 
  h(x) \cdot \overline{\xi}_{\rm R,L} \cdot g_{\rm R,L}^\dag \ ,
\nonumber\\
&& \overline{V}_\mu \rightarrow 
  h(x) \cdot \overline{V}_\mu \cdot h^\dag(x)
  - i \partial_\mu h(x) \cdot h^\dag(x) \ ,
\nonumber\\
&& \check{\pi} \rightarrow 
  h(x) \cdot \check{\pi} \cdot h^\dag(x) \ ,
\nonumber\\
&& \check{\sigma} \rightarrow 
  h(x) \cdot \check{\sigma} \cdot h^\dag(x) \ ,
\nonumber\\
&& \check{\rho}_\mu \rightarrow 
  h(x) \cdot \check{\rho}_\mu \cdot h^\dag(x) \ .
\end{eqnarray}
Thus, the expansion of the Lagrangian in terms of the quantum field
manifestly keeps the HLS of the background field
$\overline{V}_\mu$~\cite{Tanabashi}.

We adopt 
the background gauge fixing
in 't\,Hooft-Feynman gauge:
\begin{eqnarray}
{\cal L}_{\rm GF} =
- \mbox{tr}\,
\biggl[
\left(
  \overline{D}^\mu \check{\rho}_\mu + M_\rho \check{\sigma} 
\right)^2
\biggr]
\ ,
\label{gauge fixing}
\end{eqnarray}
where $\overline{D}_\mu$ is the covariant derivative on the background
field:
\begin{equation}
\overline{D}^\mu \check{\rho}_\nu = \partial^\mu \check{\rho}_\nu
- i \left[ \overline{V}^\mu , \check{\rho}_\nu \right] \ ,
\end{equation}
and
\begin{equation}
 M_\rho = g F_\sigma
\end{equation}
is the vector meson mass parameter
which at the loop order should be distinguished from the on-shell mass
$m_\rho$ defined in Eq.~(\ref{on-shell condition}).
The Faddeev-Popov ghost term associated with the gauge 
fixing~(\ref{gauge fixing}) is
\begin{equation}
{\cal L}_{\rm FP} = 
2 i \, \mbox{tr} \, 
\left[
  \overline{C}
  \left( 
    \overline{D}^\mu \overline{D}_\mu + M_\rho^2
  \right)
  C
\right]
+ \cdots
\ ,
\end{equation}
where dots stand for the interaction terms of the quantum fields
$\check{\pi}$, $\check{\sigma}$, $\check{\rho}_\mu$
and the FP ghosts $C$ and $\overline{C}$.

Now, the complete ${\cal O}(p^2)$ Lagrangian,
${\cal L}_{(2)} + {\cal L}_{\rm GF} + {\cal L}_{\rm FP}$, is
expanded in terms of the quantum fields, $\check{\pi}$, 
$\check{\sigma}$, $\check{\rho}$ and $C$, $\overline{C}$.
The terms which do not include the quantum fields are nothing but the
original ${\cal O}(p^2)$ Lagrangian with the fields replaced by the
corresponding background fields.
The terms which are of first order in the quantum fields 
lead to the equations of motions for the background fields:
\begin{eqnarray}
&&
\overline{D}_\mu \overline{\cal A}^\mu =
- i \left( a- 1 \right)
\left[
  \overline{\cal V}_\mu - \overline{V}_\mu \,,\,
  \overline{\cal A}^\mu
\right]
- \frac{i}{4} \frac{F_\chi^2}{F_\pi^2} 
\left(
  \overline{\chi} - \overline{\chi}^\dag
  - \frac{1}{N_f} \, \mbox{tr}
  \left[ \overline{\chi} - \overline{\chi}^\dag \right]
\right)
+ {\cal O}(p^4)
\ ,
\label{EOM Npi B}
\\
&&
\overline{D}_\mu 
\left( \overline{\cal V}^\mu - \overline{V}^\mu \right)
= {\cal O}(p^4)
\ ,
\label{EOM Nsig B}
\\
&&
\overline{D}_\nu \overline{V}^{\nu\mu}
=
g^2 F_\sigma^2 
\left( \overline{\cal V}^\mu - \overline{V}^\mu \right)
+ {\cal O}(p^4)
\ ,
\label{EOM Nvec B}
\end{eqnarray}
which correspond to Eqs.~(\ref{EOM Npi2}), (\ref{EOM Nsig2}) and
(\ref{EOM Nvec2}), respectively.

To write down the terms which are of quadratic order in the quantum
fields in a compact and unified way,
let us define the following ``connections'':
\begin{eqnarray}
\Gamma^{(\pi\pi)}_{\mu,ab} &\equiv&
i \, \mbox{tr} \Biggl[
  \biggl(
    \left( 2 - a \right) \overline{\cal V}_\mu
    + a \overline{V}_\mu 
  \biggr)
  \left[ T_a \,,\, T_b \right]
\Biggr]
\ ,
\label{Xpp}
\\
\Gamma^{(\sigma\sigma)}_{\mu,ab} &\equiv&
i \, \mbox{tr} \Biggl[
  \biggl(
    \overline{\cal V}_\mu + \overline{V}_\mu 
  \biggr)
  \left[ T_a \,,\,T_b \right]
\Biggr]
\ ,
\label{Xss}
\\
\Gamma^{(\pi\sigma)}_{\mu,ab} &\equiv&
i \sqrt{a} \, \mbox{tr} \biggl[
  \overline{\cal A}_\mu
  \left[ T_a \,,\, T_b \right]
\biggr]
\ ,
\label{Xps}
\\
\Gamma^{(\sigma\pi)}_{\mu,ab} &\equiv&
i \sqrt{a} \, \mbox{tr} \biggl[
  \overline{\cal A}_\mu
  \left[ T_a \,,\, T_b \right]
\biggr]
\ ,
\label{Xsp}
\\
\Gamma^{(V_\alpha V_\beta)}_{\mu,ab} &\equiv&
- 2 i \, \mbox{tr} \biggl[
  \overline{V}_\mu \left[ T_a \,,\,T_b \right]
\biggr] \, g^{\alpha\beta}
\ .
\label{Xvv}
\end{eqnarray}
Here one might doubt the minus sign in
front of $\Gamma_\mu^{(V_\alpha V_\beta)}$ compared with
$\Gamma_\mu^{(SS)}$
($S = \pi,\sigma$).  However, since $g^{\alpha\beta} = -
\delta_{\alpha\beta}$ for $\alpha = 1$, 2, 3, the minus sign is a
correct one.
Correspondingly, we should use an unconventional metric 
$- g_{\alpha\beta}$ to change the upper indices to the lower ones:
\begin{eqnarray}
{\Gamma_\mu}^{V_\beta)}_{(V_\alpha,ab} &\equiv&
\sum_{\alpha'} \left( - g_{\alpha \alpha'} \right)
{\Gamma_\mu}^{(V_{\alpha'} V_{\beta})}_{ab}
\label{Xv lu}
\end{eqnarray}
Further we define the following quantities corresponding to the
``mass'' part:
\begin{eqnarray}
\Sigma^{(\pi\pi)}_{ab} &\equiv&
-\frac{4-3a}{2}
\, \mbox{tr} \biggl[
  \left[ \overline{\cal A}^\mu \,,\, T_a \right]
  \left[ \overline{\cal A}_\mu \,,\, T_b \right]
\biggr]
{}- \frac{a^2}{2} \mbox{tr} \biggl[
  \left[ \overline{\cal V}^\mu - \overline{V}^\mu \,,\, T_a \right]
  \left[ \overline{\cal V}_\mu - \overline{V}_\mu \,,\, T_b
  \right]
\biggr]
\nonumber\\
&&
{} + \frac{F_\chi^2}{2F_\pi^2} \, \mbox{tr}
\biggl[
  \left( 
    \overline{\chi} + \overline{\chi}^\dag - 2 {\cal M}_\pi
  \right)
  \left\{ T_a \,,\, T_b \right\}
\biggr]
\ ,
\label{Ypp}
\\
\Sigma^{(\sigma\sigma)}_{ab} &\equiv&
- \frac{1}{2} \, \mbox{tr} \biggl[
  \left[ 
    \overline{\cal V}^\mu - \overline{V}^\mu \,,\, T_a
  \right]
  \left[ 
    \overline{\cal V}_\mu - \overline{V}_\mu \,,\, T_b
  \right]
\biggr]
{}-
\frac{a}{2} 
\, \mbox{tr} \biggl[
  \left[ 
    \overline{\cal A}^\mu \,,\, T_a
  \right]
  \left[ 
    \overline{\cal A}_\mu \,,\, T_b
  \right]
\biggr]
\ ,
\label{Yss}
\\
\Sigma^{(\pi\sigma)}_{ab} &\equiv&
- i \sqrt{a}\, \mbox{tr} \biggl[
  \overline{D}^\mu \overline{\cal A}_\mu
  \left[ T_a \,,\, T_b \right]
\biggr]
{}- \frac{1}{2} \sqrt{a}\, \mbox{tr} \biggl[
  \left[ 
    \overline{\cal A}_\mu \,,\, T_a
  \right]
  \left[ 
    \overline{\cal V}^\mu - \overline{V}^\mu \,,\, T_b
  \right]
\biggr]
\nonumber\\
&&
{}- \left( 1 - \frac{a}{2} \right) 
\sqrt{a}\, \mbox{tr}
\biggl[
  \left[ 
    \overline{\cal V}_\mu - \overline{V}_\mu \,,\, T_a
  \right]
  \left[ 
    \overline{\cal A}^\mu \,,\, T_b
  \right]
\biggr]
\ ,
\\
\Sigma^{(\sigma\pi)}_{ab} &\equiv& \Sigma^{(\pi\sigma)}_{ba} 
\ ,
\\
\Sigma^{(V_\alpha V_\beta)}_{ab} &\equiv&
- 4 i \mbox{tr} 
\biggl[
  \overline{V}^{\alpha\beta} \left[ T_a \,,\, T_b \right]
\biggr]
\ ,
\label{Yvv}
\\
\Sigma^{(\pi V_\beta)}_{ab} &\equiv&
- 2 i a g F_\pi \, \mbox{tr} \biggl[
  \overline{\cal A}^\beta
  \left[ 
     T_a \,,\, T_b
  \right]
\biggr]
\ ,
\label{Ypv}
\\
\Sigma^{(V_\alpha \pi)}_{ab} &\equiv&
2 i a g F_\pi \, \mbox{tr} \biggl[
  \overline{\cal A}^\alpha
  \left[ 
     T_a \,,\, T_b
  \right]
\biggr]
\ ,
\label{Yvp}
\\
\Sigma^{(\sigma V_\beta)}_{ab} &\equiv&
2 i g F_\sigma \mbox{tr} \biggl[
  \left( \overline{\cal V}^\beta - \overline{V}^\beta \right)
  \left[ 
     T_a \,,\, T_b
  \right]
\biggr]
\ ,
\label{Ysv}
\\
\Sigma^{(V_\alpha \sigma)}_{ab} &\equiv&
- 2 i g F_\sigma \mbox{tr} \biggl[
  \left( \overline{\cal V}^\alpha - \overline{V}^\alpha \right)
  \left[ 
     T_a \,,\, T_b
  \right]
\biggr]
\ ,
\label{Yvs}
\end{eqnarray}
where
\begin{equation}
{\cal M}_\pi \equiv 2 B {\cal M}
\ ,
\end{equation}
with the quark mass matrix ${\cal M}$ being defined in 
Eq.~(\ref{quark mass matrix}).
Here by using the equation of motion in Eq.~(\ref{EOM Npi B}),
$\Sigma^{(\pi\sigma)}_{ab}$ is rewritten into
\begin{eqnarray}
\Sigma^{(\pi\sigma)}_{ab}
&=&
- \sqrt{a} (1-a) \, \mbox{tr}
\biggl[
  \left[ 
    \overline{\cal A}_\mu \,,\,
    \overline{\cal V}^\mu - \overline{V}^\mu
  \right]
  \left[ T_a \,,\, T_b \right]
\biggr]
{}- \frac{1}{2} \sqrt{a}\, \mbox{tr} \biggl[
  \left[ 
    \overline{\cal A}_\mu \,,\, T_a
  \right]
  \left[ 
    \overline{\cal V}^\mu - \overline{V}^\mu \,,\, T_b
  \right]
\biggr]
\nonumber\\
&&
{}- \left( 1 - \frac{a}{2} \right) 
\sqrt{a}\, \mbox{tr}
\biggl[
  \left[ 
    \overline{\cal V}_\mu - \overline{V}_\mu \,,\, T_a
  \right]
  \left[ 
    \overline{\cal A}^\mu \,,\, T_b
  \right]
\biggr]
\nonumber\\
&&
{}- \frac{\sqrt{a}}{4} \, \frac{F_\chi^2}{F_\pi^2} 
\, \mbox{tr}
\biggl[
  \left( \overline{\chi} - \overline{\chi}^\dag \right)
  \left[ T_a \,,\, T_b \right]
\biggr]
\ .
\label{Yps:eom}
\end{eqnarray}

To achieve more unified treatment let us introduce the following
quantum fields:
\begin{equation}
\check{\Phi}_A \equiv
\left( \check{\pi}^a \,,\,
\check{\sigma}^a \,,\,
\check{\rho}_{\alpha}^a 
\right)
\ ,
\end{equation}
where the lower and upper indices of
$\check{\Phi}$ should be distinguished as in Eq.~(\ref{Xv lu}).
Thus the metric acting on the indices of $\check{\Phi}$ is defined
by
\begin{eqnarray}
\eta^{AB} &\equiv&
\left( \begin{array}{ccc}
\delta_{ab} & & \\
& \delta_{ab} & \\
& & - g^{\alpha\beta} \delta_{ab}
\end{array} \right)
\ ,
\nonumber\\
\eta^{A}_{B} &\equiv&
\left( \begin{array}{ccc}
\delta_{ab} & & \\
& \delta_{ab} & \\
& & g^{\alpha}_{\beta} \delta_{ab}
\end{array} \right)
\ ,
\nonumber\\
\eta_{AB} &\equiv&
\left( \begin{array}{ccc}
\delta_{ab} & & \\
& \delta_{ab} & \\
& & - g_{\alpha\beta} \delta_{ab}
\end{array} \right)
\ .
\end{eqnarray}
The tree-level mass matrix is defined by
\begin{equation}
\widetilde{\cal M}^{AB} \equiv
\left( \begin{array}{ccc}
M_{\pi,a} \delta_{ab} & & \\
& M_\rho^2 \delta_{ab} & \\
& & - g^{\alpha\beta} M_\rho^2 \delta_{ab}
\end{array} \right)
\ ,
\end{equation}
where the pseudoscalar meson mass $M_{\pi,a}$ is defined by
\begin{equation}
M_{\pi,a}^2 \delta_{ab} \equiv
\frac{F_\chi^2}{F_\pi^2} \, \mbox{tr}
\left[
  {\cal M}_\pi 
  \left\{ T_a \,,\, T_b \right\}
\right]
\ .
\end{equation}
Here the generator $T_a$ is defined in such a way
that the above masses are diagonalized when we introduce the explicit
chiral symmetry breaking due to the current quark masses.
It should be noticed that we work in the chiral limit in this paper,
so that we take 
\begin{equation}
{\cal M}_\pi = 0 \ , \quad \mbox{or} \quad
M_{\pi,a}  = 0 \ .
\end{equation}

Let us further define
\begin{eqnarray}
\left( \widetilde{\Gamma}_\mu \right)^{AB} 
&\equiv&
\left( \begin{array}{ccc}
\Gamma_{\mu,ab}^{(\pi\pi)} & \Gamma_{\mu,ab}^{(\pi\sigma)} & 0 \\
\Gamma_{\mu,ab}^{(\sigma\pi)} & \Gamma_{\mu,ab}^{(\sigma\sigma)} & 0 \\
0 & 0 & \Gamma^{(V_\alpha V_\beta)}_{\mu,ab}
\end{array} \right)
\ ,
\\
\widetilde{\Sigma}^{AB} 
&\equiv&
\left( \begin{array}{ccc}
\Sigma_{ab}^{(\pi\pi)} & \Sigma_{ab}^{(\pi\sigma)} & 
   \Sigma_{ab}^{(\pi V_\beta)} \\
\Sigma_{ab}^{(\sigma\pi)} & \Sigma_{ab}^{(\sigma\sigma)} & 
   \Sigma_{ab}^{(\sigma V_\beta)} \\
\Sigma_{ab}^{(V_\alpha\pi)} & \Sigma_{ab}^{(V_\alpha\sigma)} & 
   \Sigma_{ab}^{(V_\alpha V_\beta)}
\end{array} \right)
\ ,
\end{eqnarray}
and
\begin{equation}
\left( \widetilde{D}_\mu \right)^{AB} \equiv
\eta^{AB} \partial_\mu + 
\left( \widetilde{\Gamma}_\mu \right)^{AB} \ .
\end{equation}
It is convenient to consider the FP ghost contribution separately.
For the FP ghost part we define similar quantities:
\begin{eqnarray}
\Gamma^{(CC)}_{\mu,ab} &\equiv&
2 i \, \mbox{tr} \biggl[
  \overline{V}_\mu \left[ T_a \,,\,T_b \right]
\biggr]
\ ,
\label{XCC}
\\
\left( \widetilde{D}_\mu\right)_{ab}^{(CC)} &\equiv&
\delta_{ab} \partial_\mu + \Gamma^{(CC)}_{\mu,ab}
\ ,
\label{def:DCC}
\\
\widetilde{\cal M}^{(CC)}_{ab} &\equiv&
\delta_{ab} M_\rho^2
\ .
\end{eqnarray}
By using the above quantities the terms quadratic in terms of the
quantum fields in the total Lagrangian
are rewritten into
\begin{eqnarray}
&&
\int \! d^4x\, 
\left[ 
  {\cal L}_{(2)} + {\cal L}_{\rm GF} + {\cal L}_{\rm FP}
\right]
=
\nonumber\\
&&\ 
- \frac{1}{2} \sum_{A,B} \int \! d^4x \,
\check{\Phi}_A
\biggl[
  \left( \widetilde{D}_\mu \cdot \widetilde{D}^\mu \right)^{AB}
  + \widetilde{\cal M}^{AB}
  + \widetilde{\Sigma}^{AB} 
\biggr]
\check{\Phi}_B
\nonumber\\
&& \ 
{}+ i \sum_{a,b} \int \! d^4x \, \overline{C}^a
\biggl[
  \left( 
     \widetilde{D}_{\mu} \cdot \widetilde{D}^{\mu} 
  \right)_{ab}^{(CC)}
  + \widetilde{\cal M}^{(CC)}_{ab}
\biggr]
C^b
\ ,
\end{eqnarray}
where
\begin{eqnarray}
&&
\left( \widetilde{D}_\mu \cdot \widetilde{D}^\mu \right)^{AB}
\equiv
\sum_{A'} \left( \widetilde{D}_\mu \right)^{AA'}
\left( \widetilde{D}^\mu \right)^{B}_{A'}
\ ,
\\
&&
\left( 
   \widetilde{D}_{\mu} \cdot \widetilde{D}^{\mu} 
\right)_{ab}^{(CC)}
\equiv
\sum_{c}
\left( \widetilde{D}_\mu \right)^{(CC)}_{ac}
\left( \widetilde{D}_\mu \right)^{(CC)}_{cb}
\ .
\label{Lag:BGF}
\end{eqnarray}
The Feynman rules obtained from the above Lagrangian relevant to the
present analysis are shown in Appendix~\ref{app:FR}.

\subsection{Quadratic divergences}
\label{ssec:QD}

In the usual phenomenological study in the ChPT of the pseudoscalar
mesons~\cite{Wei:79,Gas:84,Gas:85b} as well as 
the calculations in the early stage of the ChPT with
HLS~\cite{HY,HKY:PRL,HKY:PTP,Tanabashi},
only the logarithmic divergence was included.
As far as the bare theory is not referred to, the quadratic divergence
is simply absorbed into redefinitions of the parameters.
In other words,
when we take only the logarithmic divergence into account
and make the phenomenological analysis in the energy region around
the vector meson mass without referring to the underlying theory,
the systematic expansion explained in Sec.~\ref{ssec:DEHLS}
perfectly works in the idealized world where
the vector meson mass is small.
Furthermore, according to the phenomenological analysis
done so far (e.g, in Refs.~\cite{HY,Tanabashi}), the results
can be extended to the real world in which the vector meson mass
takes the experimental value.
However, as was shown in Refs.~\cite{HY:letter,HY:VM,HY:VD},
the inclusion of the quadratic divergence is essential to studying
the phase transition with referring to the bare theory.
Moreover, it was shown~\cite{HY:matching} that inclusion of the
quadratic divergence is needed to match the HLS with the underlying
QCD even for phenomenological reason.
One might think that the systematic expansion breaks down
when the effect of quadratic divergences is included.
However, as we discussed in Sec.~\ref{ssec:DEHLS},
the systematic expansion still works as far as we regard the cutoff
is smaller than the scale at which the effective field theory
breaks down,
$\Lambda < \Lambda_\chi \simeq 4 \pi F_\pi(\Lambda)/\sqrt{N_f}$.

In this subsection,
before starting one-loop calculations in the ChPT with HLS,
we explain meaning of the quadratic
divergence in our approach.
First, we explain ``physical meaning'' of the quadratic
divergence in our approach in Sec.~\ref{sssec:RQDPT}:
In Sec.~\ref{ssssec:NJL} we show the role of the quadratic
divergence in the phase transition using the Nambu-Jona-Lasinio (NJL)
model;
in Sec.~\ref{ssssec:SM}
we show that the inclusion of quadratic divergence is essential
even in the standard model when we match it with models
beyond standard model;
and in Sec.~\ref{ssssec:CPN}
we review the phase transition in the $CP^{N-1}$ model in $D (\leq 4)$
dimensions, in
which the power divergence $\Lambda^{D-2}$ is responsible for the
restoration of the symmetry.
Then, in Sec.~\ref{sssec:CRQDPL}
we show that the chiral symmetry restoration by the mechanism 
shown in Ref.~\cite{HY:letter}
also takes place even in the ordinary nonlinear sigma model
when we include the effect of quadratic divergences.

As is well known the naive momentum cutoff violates the chiral
symmetry.
Then, it is important to use a way to
include quadratic divergences 
consistently with the chiral symmetry.
We adopt the dimensional regularization and identify the quadratic
divergences with the presence of poles of ultraviolet origin at
$n=2$~\cite{Veltman}:
\begin{equation}
\int \frac{d^n k}{i (2\pi)^n} \frac{1}{-k^2} \rightarrow 
\frac{\Lambda^2} {(4\pi)^2} \ ,
\qquad
\int \frac{d^n k}{i (2\pi)^n} 
\frac{k_\mu k_\nu}{\left[-k^2\right]^2} \rightarrow 
- \frac{\Lambda^2} {2(4\pi)^2} g_{\mu\nu} \ .
\label{regularization 0}
\end{equation}
In Sec.~\ref{sssec:QDSPR}, we discuss a problem in the naive
cutoff regularization, and show that the above regularization in 
Eq.~(\ref{regularization 0}) solves the problem.

\subsubsection{Role of quadratic divergences in the phase transition}
\label{sssec:RQDPT}

\paragraph{NJL model}
\label{ssssec:NJL}
\ \par

\noindent
For explaining the ``physical meaning'' of the quadratic divergence in
our approach, we first discuss the quadratic divergence in the 
Nambu-Jona-Lasinio (NJL) model in four dimensions,
which actually plays precisely the {\it same role} as our quadratic
divergence in HLS model in the chiral phase transition. 

Let us start with the NJL model with the fermion field
carrying the color index:
\begin{equation}
{\cal L}_{\rm NJL} = \bar{\psi} i \gamma^\mu \partial_\mu \psi
+ \frac{G}{2N_c} \biggl[
  \left( \bar{\psi} \psi \right)^2
  + \left( \bar{\psi} i \gamma_5 \psi \right)^2
\biggr]
\ ,
\label{Lag}
\end{equation}
which is invariant under U(1)$_{\rm L}\times$U(1)$_{\rm R}$ rotation.
We should note that we consider only the case of the attractive
interaction $G > 0$.
It is convenient to introduce auxiliary fields
$\varphi \sim - 2 (G/N_c) \bar{\psi} \psi$ and
$\pi \sim - 2(G/N_c) \bar{\psi} i\gamma_5 \psi$, 
and rewrite Eq.~(\ref{Lag})
into
\begin{equation}
{\cal L}_{\rm Aux} = \bar{\psi} i \gamma^\mu \partial_\mu \psi
- \frac{N_c}{2G} \left( \varphi^2 + \pi^2 \right)
- \bar{\psi} \left( \varphi + i \gamma_5 \pi \right) \psi
\ .
\label{Lag A}
\end{equation}
Then the effective potential in the $1/N_c$-leading 
approximation
is given by
\begin{equation}
V\left(\varphi,\pi\right) = 
\frac{N_c}{2G} \left(\varphi^2+\pi^2\right) - 2 N_c
\int \frac{d^4k}{i(2\pi)^4} \ln
\left( \frac{\varphi^2 + \pi^2 - k^2}{-k^2} \right)
+ V(\varphi=\pi=0)
\ .
\label{eff pot}
\end{equation}

The gap equation is derived from the stationary condition of the
effective potential in Eq.~(\ref{eff pot}).
By setting $\pi=0$ and writing $m\equiv \langle \varphi \rangle$ for
the solution for $\varphi$,
it is expressed as
\begin{equation}
m = 4 m G \int \frac{d^4k}{i(2\pi)^4}
\frac{1}{m^2-k^2} \ , 
\label{Gap eq}
\end{equation}
where $m$ is the dynamical mass of the fermion.
The right-hand-side of Eq.~(\ref{Gap eq}) is divergent, so we need to
use some regularizations.

The first one is the naive cutoff regularization, which seems easy to
understand the physical meaning.
When we use the naive cutoff regularization, Eq.~(\ref{Gap eq})
becomes
\begin{equation}
m = m \frac{G}{4\pi^2} \left[ \Lambda^2 - m^2 
\ln \left( \frac{\Lambda^2 + m^2}{m^2} \right) \right]
\ .
\label{Gap cut}
\end{equation}

The second one is the proper time regularization (heat kernel
expansion),  in which the integral is regularized via
\begin{equation}
\frac{1}{m^2-k^2} \rightarrow \int^\infty_{1/\Lambda^2} d\tau
\exp \left[ - \tau (m^2-k^2) \right] \ .
\end{equation}
By using this regularization the gap equation (\ref{Gap eq}) becomes
\begin{equation}
m = m \frac{G}{4\pi^2} \, m^2 
\, \Gamma\left( -1 \,,\, m^2/\Lambda^2 \right)
\ ,
\label{Gap proper}
\end{equation}
where $\Gamma( n \,,\, \epsilon )$ is the incomplete gamma function
defined in Eq.~(\ref{def:IGF}):
\begin{equation}
\Gamma( n \,,\, \epsilon ) \equiv
\int^\infty_\epsilon \frac{dz}{z} z^n e^{-z} \ .
\end{equation}
Noting that $\Gamma\left( -1 \,,\, m^2/\Lambda^2 \right)$ is
approximated as [see Eq.~(\ref{approx:IGF 1})]
\begin{equation}
\Gamma\left( -1 \,,\, m^2/\Lambda^2 \right) \simeq
\frac{\Lambda^2}{m^2} - \ln \frac{\Lambda^2}{m^2} \ ,
\end{equation}
we can show that Eq.~(\ref{Gap cut}) is essentially equivalent to 
Eq.~(\ref{Gap proper}) for large $\Lambda\gg m$.

The third one is the dimensional regularization, in which 
Eq.~(\ref{Gap eq}) becomes
\begin{equation}
m = 4 m G
\frac{ \Gamma\left( 1 - n/2 \right)}{(4\pi)^{n/2} (m^2)^{1-n/2}}
\ ,
\label{Gap dim}
\end{equation}
where $\Gamma(x)$ is the gamma function.
We note here that $\Gamma\left( 1 - n/2 \right)$ generates pole for
$n=2$ as well as that for $n=4$, which correspond to the quadratic
divergence and the logarithmic divergence, respectively  in four
space-time dimensions. 
These correspondences are seen as follows:
In the dimensional regularization we can separate the 
pole for $n=2$ with that for $n=4$ using the identity:
\begin{eqnarray}
\int \frac{d^nk}{i(2\pi)^n} \frac{1}{m^2-k^2}
&=&
\frac{ \Gamma\left( 1 - n/2 \right)}{(4\pi)^{n/2} (m^2)^{1-n/2}}
\nonumber\\
&=&
\frac{1}{(4\pi)^{n/2} (m^2)^{1-n/2}}
\frac{\Gamma\left( 2 - n/2 \right)}{1-n/2}
\nonumber\\
&=&
\frac{1}{(4\pi)^{n/2} (m^2)^{1-n/2}}
\frac{\Gamma\left( 2 - n/2 \right)}{1-n/2}
\left[ \left( 2 - n/2 \right) - \left( 1 - n/2 \right) \right]
\nonumber\\
&=&
\frac{1}{(4\pi)^{n/2} (m^2)^{1-n/2}}
\left[
  \frac{\Gamma\left( 3 - n/2 \right)}{1-n/2}
  - \Gamma\left( 2 - n/2 \right)
\right]
\nonumber\\
&=&
\frac{1}{1-n/2}
\frac{\Gamma\left( 3 - n/2 \right)}{(4\pi)^{n/2} (m^2)^{1-n/2}}
- \frac{1}{ 2 - n/2}
\frac{\Gamma\left( 3 - n/2 \right)}{(4\pi)^{n/2} (m^2)^{1-n/2}}
\nonumber\\
&=&
\frac{1}{4\pi}
\frac{1}{1-n/2}
- 
\frac{m^2}{(4\pi)^2} \frac{1}{ 2 - n/2}
+ \cdots
\ ,
\label{eval:dim}
\end{eqnarray}
where dots stands for the finite terms.
In the naive cutoff regularization,
on the other hand, the same integral is evaluated as
\begin{eqnarray}
\int^\Lambda \frac{d^4k}{i(2\pi)^4} \frac{1}{m^2-k^2}
&=&
\int^\Lambda \frac{d^4k}{i(2\pi)^4}
\left[ \frac{1}{-k^2}  - \frac{m^2}{[m^2-k^2][-k^2]} \right]
\nonumber\\
&=& 
\frac{\Lambda^2}{(4\pi)^2}
- 
\frac{m^2}{(4\pi)^2}
\, \ln \Lambda^2
+ \cdots
\ ,
\label{eval:cut}
\end{eqnarray}
where dots stands for the finite terms.
Comparing Eq.~(\ref{eval:dim}) with Eq.~(\ref{eval:cut}),
we see that
the first term in Eq.~(\ref{eval:dim})
corresponds to the quadratic divergence in Eq.~(\ref{eval:cut}),
while
the second term in Eq.~(\ref{eval:dim}),
as is well-known,
does to the logarithmic divergence:
\begin{eqnarray}
&& \frac{1}{1 - n/2 } \rightarrow
\frac{\Lambda^2} {4\pi} \ ,
\label{quadrepl:1}
\\
&&\frac{1}{2 - n/2 } \rightarrow \ln \Lambda^2 \ .
\label{logrepl}
\end{eqnarray}
By using this, Eq.~(\ref{Gap dim}) gives
the same gap equation as that in Eq.~(\ref{Gap cut})
up to the terms of order $m^2/\Lambda^2$:
\begin{eqnarray}
m =
m\frac{G}{4\pi^2} \left[
  \Lambda^2 - m^2 \, \ln \frac{\Lambda^2}{m^2}
\right]
\ .
\label{Gap dim2}
\end{eqnarray}

{}From the above argument we can conclude that the three
regularization 
methods are equivalent as far as the gap equation is concerned.
Namely, the $1/(n-2)$ pole has exactly the same meaning as the
quadratic 
divergence in the naive cutoff regularization.
So, after the above replacement, the cutoff $\Lambda$ in three
regularizations  
can be understood as the physical cutoff 
above which the theory is not applicable. 

Now, let us study the phase structure of the NJL model.
The gap equation 
of the NJL model in the form given in Eq.~(\ref{Gap dim2})
is rewritten into 
\begin{equation}
m^3 \cdot \frac{1}{4\pi^2} \, \ln \frac{\Lambda^2}{m^2}
=
- m \left( \frac{1}{G} - \frac{1}{G_{\rm cr}} \right)
\ ,
\label{NJL:gap eq}
\end{equation}
where
\begin{eqnarray}
\frac{1}{G_{\rm cr}}  = 
\frac{\Lambda^2}{4\pi^2}
\ .
\label{G crit}
\end{eqnarray}
{}From this we easily see that
$m$ can be non-zero (symmetry breaking solution)
only if $1/G - 1/G_{\rm cr} < 0$.
It should be noticed that without quadratic divergence 
the spontaneous symmetry breaking cannot occur, since 
the bare theory ($1/G >0$) is in the symmetric phase.

This phase structure can be also seen 
by studying the sign of the coefficient of the $\varphi^2$ term
in the effective potential Eq.~(\ref{eff pot}).
By expanding the effective potential 
in Eq.~(\ref{eff pot})
around $\varphi=0$ (we set
$\pi=0$), we have
\begin{eqnarray}
V(\varphi,\pi=0) - V(\varphi=\pi=0) 
&=&
\frac{N_c}{2G} \varphi^2 - 2 N_c
\int \frac{d^4k}{i(2\pi)^4} \ln
\left( \frac{\varphi^2 - k^2}{-k^2} \right)
\nonumber\\
&\simeq&
\frac{1}{2} M_\varphi^2 \varphi^2 + \cdots \ ,
\end{eqnarray}
where $M_\varphi^2$ is evaluated as (in all three regularizations)
\begin{eqnarray}
M_\varphi^2 
=
N_c \left( \frac{1}{G} - \frac{1}{G_{\rm cr}} \right) \ .
\label{sig}
\end{eqnarray}
It should be noticed that 
the first term $1/G$
in the right-hand-side of Eq.~(\ref{sig}) is a bare
mass of $\varphi$ and positive, while the second term 
$1/G_{\rm cr} = \Lambda^2/(4\pi^2)$ [Eq.~(\ref{G crit})]
arises from the quadratic divergence and can change the sign of
$M_\varphi^2$.
By using this we can determine the phase as
\begin{eqnarray}
&&
M_\varphi^2 < 0 \rightarrow \mbox{broken phase} \ ,
\nonumber\\
&&
M_\varphi^2 > 0 \rightarrow \mbox{symmetric phase} \ .
\label{NJL cond}
\end{eqnarray}
Namely, although the bare theory looks as if it were
in the symmetric phase,
the quantum theory can be in the broken phase due to the quadratic
divergence:
The phase change is triggered by the quadratic divergence.

\paragraph{Standard model}
\ \par
\label{ssssec:SM}

In this subsection we consider the effect of quadratic divergence
in the standard model (SM), in which
there exists a quadratically divergent correction to the Higgs
mass parameter.
When we
make phenomenological analysis within the framework of
the SM without referring to the model beyond the SM, 
we can absorb 
the effect of quadratic divergence into the mass 
parameter, and the effect does not enter the phenomenological
analysis.
However, as many people are thinking,
the SM may not be an ultimate theory describing
the real world, and it is just a low-energy 
effective field theory of some underlying theory.
In such a case, the bare Higgs mass parameter should be 
determined from the underlying theory and must be
tuned to be
canceled with the quadratic divergence of order $\Lambda^2$ to yield
an observed value $(250\mbox{GeV})^2$, which is however an enormous
fine-tuning if the cutoff is very big, say the Planck scale 
$10^{19}\,\mbox{GeV}$, 
$(250\mbox{GeV})^2/(10^{19}\mbox{GeV})^2 \sim 10^{-33} \ll 1$.
This is a famous naturalness problem.
Here we study how 
the effect of quadratic divergence enters into the relation
between the bare Higgs mass parameter and the order parameter
(on the order of $250\,\mbox{GeV}$) in the SM,
and show that
the bare Higgs mass
parameter is actually relevant to the phase structure of the SM.

To explain the essential point
we switch off all the gauge interactions since they are
small at the weak scale $250\,\mbox{GeV}$.
Furthermore,
we switch off all the Yukawa couplings except the one related to
the top quark mass.
Then, 
the relevant part of the Lagrangian is given by
\begin{equation}
{\cal L} = {\cal L}_{\rm kinetic}
- y \left( \bar{\psi}_L t_R \phi + \mbox{h.c.} \right)
+ \partial_\mu \phi^\dag \partial^\mu \phi - M^2 \phi^\dag \phi
- \lambda \left( \phi^\dag \phi \right)^2
\ ,
\label{Lag:SM}
\end{equation}
where 
$\bar{\psi}_L = (\bar{t}_L,\bar{b}_L)$ is $\mbox{SU}(2)_{\rm L}$ 
doublet field for left-handed top and bottom quarks, 
$t_R$ is the singlet field for the right-handed top quark,
$\phi$ is the Higgs field, $y$ is the top Yukawa coupling
and $\lambda$ the Higgs self coupling.
The ${\cal L}_{\rm kinetic}$ is the kinetic terms for
$\psi_L$ and $t_R$:
\begin{equation}
{\cal L}_{\rm kinetic} = 
\bar{\psi}_L \gamma^\mu i \partial_\mu \psi_L
+ \bar{t}_R \gamma^\mu i \partial_\mu t_R \ .
\label{L kin SM}
\end{equation}
Note that both $\psi_L$ and $t_R$ are in the fundamental
representation
of $\mbox{SU}(3)_{\rm c}$.
Here we adopt the large $N_c$ approximation to calculate the
effective potential for the Higgs field $\phi$
with regarding the SM as a cutoff theory.
In this approximation we need to take account of 
only the top quark loop, and the resultant effective potential
for the Higgs field is given by
\begin{eqnarray}
V(\phi) - V(0) 
&=&
M_\phi^2 \, \phi^\dag \phi
+ \left(
  \lambda_{\rm bare} + \frac{N_c}{(4\pi)^2} y_{\rm bare}^4
  \ln \frac{\Lambda^2}{y_{\rm bare}^2 \phi^\dag \phi}
\right)
\left( \phi^\dag \phi \right)^2
\ ,
\label{eff pot SM}
\end{eqnarray}
where 
\begin{equation}
M_\phi^2 =
  M_{\rm bare}^2 - \frac{2N_c}{(4\pi)^2} y_{\rm bare}^2 \Lambda^2
\label{rel M0 MLam SM}
\end{equation}
received a correction of quadratic divergence.
Note that 
we put the subscript ``bare'' to clarify that the parameters
are those of the bare Lagrangian
and set $M_{\rm bare}^2 >0$, the sign opposite to the usual Higgs
potential (see the footnote below).
{}From the effective potential in Eq.~(\ref{eff pot SM})
we can determine the phase as
\begin{eqnarray}
\frac{M_{\rm bare}^2}{y_{\rm bare}^2} <
\left(\frac{M^2}{y^2}\right)_{\rm cr}
\ \ &\Rightarrow&\ \   M_\phi^2 < 0 \ \,\mbox{(broken phase)} \ ,
\nonumber\\
\frac{M_{\rm bare}^2}{y_{\rm bare}^2} >
\left(\frac{M^2}{y^2}\right)_{\rm cr}
\ \ &\Rightarrow&\ \   M_\phi^2 > 0 \ \,\mbox{(symmetric phase)} \ ,
\label{phase struc SM}
\end{eqnarray}
where
\begin{equation}
\left(\frac{M^2}{y^2}\right)_{\rm cr} = 
\frac{N_c}{8\pi^2} \, \Lambda^2 \ .
\label{Mofy crit SM}
\end{equation}
This shows that
there exists the critical value for the bare Higgs mass
parameter which distinguishes the broken phase
($\mbox{SU}(2)_{\rm L}\times U(1)_{\rm Y}$ is spontaneously 
broken into $U(1)_{\rm em}$) from the symmetric one.
When the SM is applicable
all the way up to the Planck scale
$\Lambda \sim 10^{19}\,\mbox{GeV}$,
Eq.~(\ref{rel M0 MLam SM}) implies that
the bare Higgs mass parameter must be tuned to be canceled with the
quadratic divergence of order $\Lambda^2$
to yield an observed value of order $(250\mbox{GeV})^2$,
which is an enormous fine-tuning:
$(250\mbox{GeV})^2/(10^{19}\mbox{GeV})^2 \sim 10^{-33} \ll 1$.
This is a 
different version of the famous
naturalness problem.~\footnote{%
  In the usual explanation of the naturalness problem,
  the top Yukawa coupling is neglected and the quadratically divergent
  correction to 
  the Higgs mass parameter is proportional to the Higgs self-coupling
  in the one-loop approximation.
  Then, the relation between the bare Higgs mass parameter
  and the order parameter in Eq.~(\ref{rel M0 MLam SM}) is 
  modified appropriately.
  Note that the sign in front of the quadratic divergence coming 
  from the Higgs self-interaction is plus instead
  of minus in Eq.~(\ref{rel M0 MLam SM}) and we set 
  $M_{\rm bare}^2 <0$ as usual.
  Note also that, when we switch on the gauge interaction of 
  $\mbox{SU}(2)_{\rm L}\times \mbox{U}(1)_{\rm Y}$,
  the gauge boson loop generates 
  the quadratically divergent correction 
  to the Higgs mass parameter which has also positive
  sign (and again $M_{\rm bare}^2 <0$ in contrast to the top 
  Yukawa case).  
}

We should stress that the above phase structure in
Eq.~(\ref{phase struc SM}) implies that
the quadratic divergence in the SM model has the same physical 
meaning as the quadratic divergence of the NJL model explained
in the previous subsection has.
To clarify the physical meaning of the quadratic divergence,
let us regard the SM as an effective field theory of
some more fundamental theory,
and consider the matching condition between 
the SM 
and the underlying theory.
Below we shall adopt
the top quark condensate model
(Top-mode standard model)~\cite{MTY:1,MTY:2} 
as an example of underlying theory
(see, for a review, e.g., 
Ref.~\cite{Yamawaki:96}),
and show that 
it is essential to
include the quadratic divergence in the effective
field theory (i.e., the SM) when we match it with 
the underlying theory (i.e., the top quark condensate model).

The top quark condensate model, which was
proposed by Miransky, Tanabashi
and Yamawaki~\cite{MTY:1,MTY:2} and by Nambu~\cite{Nambu}
independently, provides a natural understanding of the heavy
top quark mass: The mass of top quark is roughly on the order of
weak scale $250\,\mbox{GeV}$.
In this model, the standard Higgs doublet is entirely replaced by a
composite one formed by a strongly coupled short range dynamics
(four-fermion interaction) which triggers the top quark condensate.
The Higgs boson emerges as $\bar{t} t$ bound state and hence is 
deeply connected with the top quark itself.
The model was further developed by the renormalization group 
method~\cite{Marciano:1,Marciano:2,BHL}.
For illustration of the essential point,
we switch off all the gauge interactions, and
furthermore, we keep only the four-fermion coupling for
the top quark.
Then, the relevant Lagrangian is expressed as~\cite{MTY:1,MTY:2,BHL}
\begin{equation}
{\cal L}_{\rm TMSM} = {\cal L}_{\rm kinetic}
+ G_t \bigl(\bar{\psi}_L t_R \bigr) \bigl(\bar{t}_R \psi_L \bigr)
\ ,
\label{Lag:TMSM}
\end{equation}
where
the kinetic terms for $\psi_L$ and $t_R$ are
given in Eq.~(\ref{L kin SM}).
To obtain the gap equation, we adopt the large $N_c$ approximation.
Then, as we obtained in the previous subsection, the
gap equation is given by~\footnote{%
  Extra factor $1/2$ in Eq.~(\ref{gap eq TMSM}) compared
  with Eq.~(\ref{Gap dim2}) comes from the
  projection operators $(1\pm\gamma_5)/2$ of right- and left-handed
  fermions.
}
\begin{equation}
m_t = m_t G_t \frac{N_c}{8\pi^2} 
\left[
  \Lambda^2 - m_t^2 \ln \frac{\Lambda^2}{m_t^2}
\right]
\ ,
\label{gap eq TMSM}
\end{equation}
where $m_t$ is the top quark mass.
This gap equation shows that the model has two phases distinguished by
the value of the four-fermion coupling constant $G_t$:
\begin{eqnarray}
\frac{1}{G_t} < \frac{1}{G_t^{\rm cr}}
\ &\Rightarrow&\  \mbox{broken phase}\ ,
\nonumber\\
\frac{1}{G_t} > \frac{1}{G_t^{\rm cr}}
\ &\Rightarrow&\  \mbox{symmetric phase}\ ,
\label{phase struc TMSM}
\end{eqnarray}
where
\begin{equation}
\frac{1}{ G_t^{\rm cr} } = \frac{N_c}{8\pi^2} \Lambda^2 
\end{equation}
is given by the quadratic divergence as before.

Let us now obtain the effective field theory of the above 
top quark condensate model following Ref.~\cite{BHL}.
For this purpose
it is convenient to introduce 
auxiliary fields $\phi_0 = G_t^{-1} \bar{t}_R \psi$,
and rewrite
the Lagrangian in Eq.~(\ref{Lag:TMSM}) as
\begin{equation}
{\cal L}_{\rm eff} = {\cal L}_{\rm kinetic}
- \left( \bar{\psi}_L t_R \phi_0 + \mbox{h.c.} \right)
- \frac{1}{G_t} \phi_0^\dag \phi_0
\ .
\label{Lag:TMSM2}
\end{equation}
To obtain the effective Lagrangian in the low-energy scale $\mu$
{\it in the Wilsonian sense},
we integrate out the high energy mode $\mu < E < \Lambda$.
In the large $N_c$ approximation the effective Lagrangian at
scale $\mu$ is obtained as~\cite{BHL}
\begin{eqnarray}
{\cal L}_{\rm eff} 
&=& {\cal L}_{\rm kinetic}
- \left( \bar{\psi}_L t_R \phi_0 + \mbox{h.c.} \right)
\nonumber\\
&&{}
+ Z_\phi(\mu) \,\partial_\mu \phi_0^\dag \partial^\mu \phi_0
- M_0^2(\mu) \,\phi_0^\dag \phi_0
- \lambda_0(\mu) \left( \phi_0^\dag \phi_0 \right)^2
\ ,
\label{Lag:TMSM3}
\end{eqnarray}
where
\begin{eqnarray}
Z_\phi(\mu) 
&=& \frac{N_c}{(4\pi)^2} \ln \frac{\Lambda^2}{\mu^2} \ ,
\\
M_0^2(\mu) 
&=& \frac{1}{G_t} - \frac{2N_c}{(4\pi)^2}
  \left( \Lambda^2 - \mu^2 \right) \ ,
\\
\lambda_0(\mu) 
&=&  \frac{2N_c}{(4\pi)^2} \ln \frac{\Lambda^2}{\mu^2} \ .
\end{eqnarray}
Note that the bare mass term $M_0^2(\Lambda)=\frac{1}{G_t} (>0)$ 
has received a quantum correction of the quadratic divergence $(<0)$
in accord with the gap equation~(\ref{gap eq TMSM}), 
and the kinetic term of $\phi_0$ and the quartic coupling $\lambda$
have been generated as quantum corrections.
 
By rescaling the Higgs field as
\begin{equation}
\phi_0 = \frac{1}{ \sqrt{ Z_\phi } } \,\phi \ ,
\end{equation}
the Lagrangian (\ref{Lag:TMSM3}) is rewritten as
\begin{eqnarray}
{\cal L}_{\rm eff} = {\cal L}_{\rm kinetic}
- y(\mu) \left( \bar{\psi}_L t_R \phi + \mbox{h.c.} \right)
+ \partial_\mu \phi^\dag \partial^\mu \phi 
- M^2(\mu)\, \phi^\dag \phi
- \lambda(\mu) \left( \phi^\dag \phi \right)^2
\ ,
\label{Lag:TMSM4}
\end{eqnarray}
where
\begin{eqnarray}
y(\mu) &=& \frac{1}{ \sqrt{ Z_\phi(\mu) } } \ ,
\nonumber\\
M^2(\mu)
&=&
\frac{ M_0^2(\mu) }{ Z_\phi(\mu) } \ ,
\nonumber\\
\lambda(\mu) 
&=&  \frac{ \lambda_0(\mu) }{ Z_\phi^2(\mu) } \ .
\end{eqnarray}
The Lagrangian (\ref{Lag:TMSM4}) has the same form as the SM
Lagrangian (\ref{Lag:SM})
with the parameters renormalized at scale $\mu$
in the Wilsonian sense ({\it including the quadratic divergence}), 
except that we are not free to renormalize the parameters:
By taking $\mu \rightarrow \Lambda$, we have $M_0^2(\Lambda)=\frac{1}{G_t}$,
$Z_\phi(\Lambda)=0$ and $\lambda(\Lambda)=0$ and we get back to the original
top-mode Lagrangian in Eq.~(\ref{Lag:TMSM}) or Eq.(\ref{Lag:TMSM2}).
Then the parameters must satisfy the following matching conditions
(``compositeness condition''~\cite{BHL};
see also 
Refs.~\cite{Bando-Kugo-Maekawa-Nakano,Harada-Kikukawa-Kugo-Nakano}):
\begin{eqnarray}
&&
\frac{1}{y^2(\mu)}
\ \ \mathop{\longrightarrow}_{\mu \rightarrow \Lambda}\ \ 
\frac{1}{y_{\rm bare}^2} = 0 \ ,
\label{match y SM}
\\
&&
\frac{\lambda(\mu)}{y^4(\mu)}
\ \ \mathop{\longrightarrow}_{\mu \rightarrow \Lambda}\ \ 
\frac{ \lambda_{\rm bare} }{y_{\rm bare}^4} = 0 \ ,
\label{match lam SM}
\\
&&
\frac{M^2(\mu)}{y^2(\mu)}
\ \ \mathop{\longrightarrow}_{\mu \rightarrow \Lambda}\ \ 
\frac{M_{\rm bare}^2}{y_{\rm bare}^2}
= \frac{1}{G_t} \ ,
\label{comp cond M2}
\end{eqnarray}
where,
as usual, we identified the parameters renormalized at scale $\Lambda$
in the Wilsonian sense with the bare parameters.~\footnote{%
  Another way to obtain the matching conditions 
  in Eqs.~(\ref{match y SM}), (\ref{match lam SM}) and
  (\ref{comp cond M2})
  is as follows:
  By rescaling the Higgs field as $\phi = \frac{1}{y} \phi_0$,
  the SM Lagrangian in Eq.~(\ref{Lag:SM}) is expressed as
  \begin{eqnarray}
  {\cal L}_{\rm eff} = {\cal L}_{\rm kinetic}
  - \left( \bar{\psi}_L t_R \phi_0 + \mbox{h.c.} \right)
  + \frac{1}{y_{\rm bare}^2} 
  \partial_\mu \phi_0^\dag \partial^\mu \phi_0 
  - \frac{M_{\rm bare}^2}{y_{\rm bare}^2} \phi_0^\dag \phi_0
  - \frac{\lambda_{\rm bare}}{y_{\rm bare}^4} 
    \left( \phi_0^\dag \phi_0 \right)^2
  \ ,
  \nonumber
  \end{eqnarray}
  where we put the subscript ``bare'' to clarify that the matching
  must be done for the bare effective field theory.
  Comparing this Lagrangian
  with the auxiliary field Lagrangian in 
  Eq.~(\ref{Lag:TMSM2}),
  we obtain the following matching conditions
  in Eqs.~(\ref{match y SM}), (\ref{match lam SM}) and
  (\ref{comp cond M2}).
}
Provided the matching condition for the Higgs mass parameter
in Eq.~(\ref{comp cond M2}),
we can easily see that
the phase structure of the effective field theory (SM)
shown in Eq.~(\ref{phase struc SM}) completely agrees with
that of the underlying theory (top quark condensate model)
shown in Eq.~(\ref{phase struc TMSM}).
This shows that the quadratic divergence in the effective field
theory (SM) has the same physical meaning as the underlying
theory (top quark condensate model) has:
The effect of quadratic divergence can trigger the phase change in the
quantum theory.

\paragraph{$CP^{N-1}$ model}
\label{ssssec:CPN}
\ \par

\noindent
Next we review the phase transition in the
$CP^{N-1}$ model in
which the power divergence $\Lambda^{D-2}$ is responsible for the
restoration of the symmetry (see Chapter 5 of Ref.~\cite{BKY}).
The $CP^{N-1}$ model in $D (\leq 4)$ dimensions is a nonlinear sigma
model based on the coset space
$\mbox{SU}(N)/\mbox{SU}(N-1)\times\mbox{U}(1)$.
In its popular form the basic field variable is expressed by an
$N$-component scalar field $\mbox{\boldmath$\phi$}$:
\begin{equation}
^t\hbox{\boldmath$\phi$} \equiv
\left(\phi^1, \phi^2, \ldots , \phi^N\right) \ ,
\quad \phi^a \in \mbox{\bf C} \ ,
\end{equation}
with the constraint
\begin{equation}
\hbox{\boldmath$\phi$}^\dag \hbox{\boldmath$\phi$} = N/g 
\quad \left( g \, : \, \mbox{coupling constant} \right) \ .
\label{CPN:cond}
\end{equation}
The Lagrangian is given by
\begin{equation}
{\cal L}_\phi = D_\mu \hbox{\boldmath$\phi$}^\dag
D^\mu \hbox{\boldmath$\phi$} - \lambda
\left( \hbox{\boldmath$\phi$}^\dag \hbox{\boldmath$\phi$} - N/g 
\right) \ ,
\label{CPN:Lag}
\end{equation}
where the field $\lambda$ is a Lagrange multiplier and the
$\mbox{U}(1)$ covariant derivative 
$D_\mu\hbox{\boldmath$\phi$}$ is given by
\begin{equation}
D_\mu\hbox{\boldmath$\phi$} = \left( \partial_\mu - i g A_\mu \right)
\hbox{\boldmath$\phi$} \ .
\end{equation}
The Lagrangian in Eq.~(\ref{CPN:Lag}) is clearly invariant under
$\mbox{SU}(N)_{\rm global}\times\mbox{U}(1)_{\rm local}$.
The $\mbox{U}(1)_{\rm local}$ gauge field $A_\mu$ has no kinetic
term in Eq.~(\ref{CPN:Lag}) and is an auxiliary field, which can be
eliminated by using the equation of motion for $A_\mu$,
\begin{equation}
A_\mu = - \frac{i}{2N} \hbox{\boldmath$\phi$}^\dag
\mathop{\partial_\mu}^{\leftrightarrow}
\hbox{\boldmath$\phi$}
\quad
\left( f \mathop{\partial_\mu}^{\leftrightarrow} g =
f \partial_\mu g - f \mathop{\partial_\mu}^{\leftarrow} g \right)
\ .
\end{equation}
Then the Lagrangian (\ref{CPN:Lag}) is equivalent to
\begin{equation}
{\cal L}_\phi = \partial_\mu \hbox{\boldmath$\phi$}^\dag
\partial^\mu \hbox{\boldmath$\phi$} 
+ \frac{g}{4N}
\left( 
  \hbox{\boldmath$\phi$}^\dag
  \mathop{\partial_\mu}^{\leftrightarrow} \hbox{\boldmath$\phi$}
\right)^2
- \lambda
\left( \hbox{\boldmath$\phi$}^\dag \hbox{\boldmath$\phi$} - N/g 
\right) \ .
\end{equation}
In this form it still retains the $\mbox{U}(1)_{\rm local}$ invariance
under the transformation 
$\hbox{\boldmath$\phi$}^{\prime}(x) = e^{i\varphi(x)}
\hbox{\boldmath$\phi$}(x)$.
Since $\hbox{\boldmath$\phi$}$ has $2N$ real components and is
constrained by one real condition in Eq.~(\ref{CPN:cond}),
one might think that the field variable $\hbox{\boldmath$\phi$}$
includes $2N-1$ degrees of freedom.
But the system actually possesses the $\mbox{U}(1)_{\rm local}$ gauge
invariance and so we can gauge away one further component of 
$\hbox{\boldmath$\phi$}$, leaving $2N-2$ degrees of freedom which are
exactly the dimension of the manifold
$CP^{N-1} = \mbox{SU}(N)/\mbox{SU}(N-1)\times\mbox{U}(1)$.

Let us consider the effective action for the Lagrangian
(\ref{CPN:Lag}).
In the leading order of the $1/N$ expansion it is evaluated as
\begin{equation}
\Gamma\left[ \hbox{\boldmath$\phi$} , \lambda \right]
= \int d^Dx \left[
  D_\mu \hbox{\boldmath$\phi$}^\dag D^\mu \hbox{\boldmath$\phi$} 
  - \lambda
  \left( 
    \hbox{\boldmath$\phi$}^\dag \hbox{\boldmath$\phi$} - N/g 
  \right) 
\right]
+ i N \,\mbox{Tr} \mbox{Ln} \left(
  - D_\mu D^\mu - \lambda
\right)
\ .
\label{CPN:efac}
\end{equation}
Because of the $\mbox{SU}(N)$ symmetry, the VEV of 
$\hbox{\boldmath$\phi$}$ can be written in the form
\begin{equation}
\left\langle
  ^t \hbox{\boldmath$\phi$}(x)
\right\rangle
= \left( 0, 0, \ldots , \sqrt{N} v \right) \ .
\end{equation}
Then the effective action (\ref{CPN:efac}) gives
the effective potential for $v$ and $\lambda$ as
\begin{equation}
\frac{1}{N} V \left( v, \lambda \right) = 
\lambda \left( v^2 - 1/g \right) +
\int \frac{d^Dk}{i(2\pi)^D} \, \ln \left( k^2 - \lambda \right) \ .
\end{equation}
The stationary conditions of this effective potential are given by
\begin{eqnarray}
&& \frac{1}{N} \frac{\partial V}{\partial v} 
= 2 \lambda v = 0 \ ,
\label{CPN:sta1}
\\
&& \frac{1}{N} \frac{\partial V}{\partial \lambda} 
=
v^2 - \frac{1}{g} + 
\int \frac{d^Dk}{i(2\pi)^D} \, \frac{1}{\lambda- k^2}
= 0 \ .
\label{CPN:sta2}
\end{eqnarray}
The first condition (\ref{CPN:sta1}) is realized in either of the
cases
\begin{equation}
\left\{ \begin{array}{ll}
  \lambda=0 \, (v\neq0) \ , & \mbox{case (i)} \ , \\
  v=0 \, (\lambda\neq0) \ , & \mbox{case (ii)} \ .
\end{array}\right.
\label{CPN:two phases}
\end{equation}
The case (i) corresponds to the broken phase
of the $\mbox{U}(1)$ and $\mbox{SU}(N)$ symmetries,
and case (ii) does to the unbroken phase.
The second stationary condition (\ref{CPN:sta2}) gives relation
between $\lambda$ and $v$.
By putting $\lambda=v=0$ in Eq.~(\ref{CPN:sta2}), the critical point
$g=g_{\rm cr}$ separating the two phases in Eq.~(\ref{CPN:two phases})
is determined as
\begin{equation}
\frac{1}{g_{\rm cr}} = 
\int \frac{d^Dk}{i(2\pi)^D} \, \frac{1}{- k^2}
= \frac{1}{\left(D/2 - 1 \right) \Gamma(D/2) }
  \frac{\Lambda^{D-2}}{(4\pi)^{D/2}}
\ .
\label{CPN:gcr}
\end{equation}
Substituting Eq.~(\ref{CPN:gcr}) into the second stationary condition
(\ref{CPN:sta2}), we obtain
\begin{equation}
v^2 - 
\int \frac{d^Dk}{i(2\pi)^D} \, 
\left( \frac{1}{- k^2} - \frac{1}{\lambda- k^2} \right)
= 
\frac{1}{g} - \frac{1}{g_{\rm cr}} \ .
\label{CPN:sta3}
\end{equation}
We should note that the power divergence in $1/g_{\rm cr}$ in 
Eq.~(\ref{CPN:gcr}) becomes quadratic
divergence in four-dimension ($D=4$):
\begin{equation}
\frac{1}{g_{\rm cr}} = 
\frac{\Lambda^2}{(4\pi)^2} 
\quad
\mbox{for} \  D = 4 \ ,
\label{CPN:gcr2}
\end{equation}
and that the stationary condition in Eq.~(\ref{CPN:sta3}) 
is rewritten into
\begin{equation}
v^2 - \frac{\lambda}{(4\pi)^2} \, \ln
 \left(\frac{\Lambda^2+\lambda}{\lambda}\right) 
= 
\frac{1}{g} - \frac{1}{g_{\rm cr}}
\quad
\mbox{for} \  D = 4
\ ,
\label{CPN:sta4}
\end{equation}
or, combined with Eq.~(\ref{CPN:sta1}),
\begin{equation}
v^3 = 
v \left( \frac{1}{g} - \frac{1}{g_{\rm cr}} \right) \ ,
\label{CPN:gap ep}
\end{equation}
which is compared with Eq.~(\ref{NJL:gap eq}) in the NJL model up to
sign.

{}From Eqs.~(\ref{CPN:sta1}) and (\ref{CPN:sta3}) it turns out that
cases (i) and (ii) in Eq.~(\ref{CPN:two phases}) correspond,
respectively, to
\begin{eqnarray}
\mbox{(i)} \ g < g_{\rm cr} &\Rightarrow&
  v\neq0 \ , \ \lambda = 0 \ , \quad 
(\mbox{broken phase of SU$(N)$}) \ ,
\nonumber\\
\mbox{(ii)} \ g > g_{\rm cr} &\Rightarrow&
  v=0 \ , \ \lambda \neq 0 \ , \quad
(\mbox{symmetric phase of SU$(N)$}) \ .
\label{CPN:res}
\end{eqnarray}
The case (ii) in Eq.~(\ref{CPN:res}) implies that
due to the power divergence in $1/g_{\rm cr}$
from the dynamics of the $CP^{N-1}$ model,
the quantum theory can be in the symmetric phase of $\mbox{SU}(N)$,
even if the bare theory with $1/g >0$
is written as if it were in the broken phase.

\subsubsection{Chiral restoration in the nonlinear chiral Lagrangian}
\label{sssec:CRQDPL}

\paragraph{Quadratic divergence and phase transition}
\ \par
\label{ssssec:QDPT}

Here
we show that the chiral symmetry restoration actually takes place
even in the usual nonlinear chiral Lagrangian when
we include quadratic divergences from the $\pi$ loop
effect~\cite{HY:letter,HY:VD}.

The Lagrangian of 
the nonlinear sigma model associated with
$\mbox{SU}(N_f)_{\rm L} \times \mbox{SU}(N_f)_{\rm R}
\rightarrow \mbox{SU}(N_f)_{\rm V}$ symmetry breaking
is given by 
[the first term of Eq.~(\ref{leading ChPT})]
\begin{equation}
{\cal L} = \frac{1}{4} [F_\pi^{(\pi)}]^2 \,
\mbox{tr} \left[ \nabla_\mu U \nabla^\mu U^\dag \right] \ ,
\label{Lag ChPT quad}
\end{equation}
where we used $F_\pi^{(\pi)}$ for the NG boson decay constant
in the nonlinear chiral Lagrangian to distinguish it from the one
in the HLS.
The covariant derivative $\nabla_\mu U$ is defined by
[see Eq.~(\ref{covdel ChPT})]
\begin{equation}
\nabla_\mu U  = \partial_\mu U - i {\cal L}_\mu U
+ i U {\cal R}_\mu \ .
\end{equation}
In the chiral perturbation theory (ChPT)~\cite{Wei:79,Gas:84,Gas:85a}
explained in Sec.~\ref{sec:BRCPT}
the effect of quadratic divergences is dropped by using the
dimensional regularization.
In other words,
the effect of quadratic divergences is assumed to be subtracted,
and thus $F_\pi^{(\pi)}$ does not get any renormalization effects.
As far as the bare theory is not referred to, the quadratic divergence
is simply absorbed into a redefinition of $F_\pi^{(\pi)}$.
As is done in other cases,
this treatment is enough and convenient 
for the usual phenomenological analysis assuming no phase change.
However, as we discussed in Sec.~\ref{sssec:RQDPT},
when we study the phase structure
with referring to the {\it bare theory},
we have to include the effect of quadratic divergences.

The effect of quadratic divergences is included through the
renormalization group equation (RGE)
in the Wilsonian sense.
Let us calculate the quadratically divergent correction to the
pion decay constant and obtain the RGE for
$[F_\pi^{(\pi)}]^2$.
The field $U$ in the Lagrangian~(\ref{Lag ChPT quad}) includes the
pion field $\pi$ as $U = \exp\left( 2 i \pi/F_\pi^{(\pi)} \right)$.
Then, the Lagrangian is expanded in terms of $\pi$ field as
\begin{equation}
{\cal L} = \mbox{tr} \left[ \partial_\mu \pi \partial^\mu \pi \right]
+ [F_{\pi,{\rm bare}}^{(\pi)}]^2 \, 
  \mbox{tr} \left[ {\cal A}_\mu {\cal A}^\mu \right]
+ \mbox{tr} \Bigl[
  \left[ {\cal A}_\mu \,,\, \pi \right]
  \left[ {\cal A}^\mu \,,\, \pi \right]
\Bigr]
+ \cdots \ ,
\end{equation}
where $F_{\pi,{\rm bare}}^{(\pi)}$ denotes the bare parameter
and 
the axialvector external field ${\cal A}_\mu$ is defined by
\begin{equation}
{\cal A}_\mu = \frac{1}{2} \left( {\cal R}_\mu - {\cal L}_\mu \right)
\ .
\end{equation}
Then the contributions at tree level and one-loop level 
to the ${\cal A}_\mu$-${\cal A}_\nu$ two-point
function are calculated as
\begin{eqnarray}
&&
\Pi_{{\cal A}{\cal A}}^{{\mbox{\scriptsize(tree)}}\mu\nu}
=
g^{\mu\nu} [F_{\pi,{\rm bare}}^{(\pi)}]^2 \ ,
\nonumber\\
&&
\Pi_{{\cal A}{\cal A}}^{{\mbox{\scriptsize(1-loop)}}\mu\nu}
=
- g^{\mu\nu} N_f A_0(0)
=
- g^{\mu\nu} N_f \frac{\Lambda^2}{(4\pi)^2}
\ ,
\end{eqnarray}
where the function $A_0$ is defined in Eq.~(\ref{def:A0}).
The renormalization is done by requiring the following is finite:
\begin{equation}
[F_{\pi,{\rm bare}}^{(\pi)}]^2 -
N_f \frac{\Lambda^2}{(4\pi)^2} = \mbox{(finite)}
\ .
\label{quad div to Fpi pi 1}
\end{equation}
{}From this the RGE for $[F_\pi^{(\pi)}]^2$ is calculated as
\begin{equation}
\mu \frac{ d }{d\mu} \left[F_\pi^{(\pi)}(\mu)\right]^2
= \frac{2N_f}{(4\pi)^2} \mu^2 \ .
\label{RGE for Fpi pi 0}
\end{equation}
This is readily solved as
\begin{equation}
\left[ F_\pi^{(\pi)}(\mu) \right]^2
= 
\left[ F_\pi^{(\pi)}(\Lambda) \right]^2- 
\frac{N_f}{(4\pi)^2} \left( \Lambda^2 - \mu^2 \right)\ ,
\label{sol fpi2 for chpt 0}
\end{equation}
where the cutoff $\Lambda$ is the scale 
at which the bare theory is defined.
By taking $\mu = 0$, this is rewritten as~\footnote{%
  As we will show in Secs.~\ref{ssec:PSH} and
  \ref{sec:VM},
  the chiral symmetry restoration in the HLS takes place by
  essentially
  the same mechanism.
  There is an extra factor $1/2$ in the second term in 
  Eq.~(\ref{RGE for fpi2 at vector limit}) compared with that 
  in Eq.~(\ref{sol fpi2 for chpt 1}).
  This factor comes from the $\rho$ loop contribution.
}
\begin{equation}
\left[ F_\pi^{(\pi)}(0) \right]^2
=
\left[ F_\pi^{(\pi)}(\Lambda) \right]^2- 
\left[ F_\pi^{(\pi),{\rm cr}}(\Lambda) \right]^2
\ ,
\label{sol fpi2 for chpt 1}
\end{equation}
where
\begin{equation}
\left[ F_\pi^{(\pi),{\rm cr}}(\Lambda) \right]^2
= \frac{N_f}{(4\pi)^2} \Lambda^2
\ .
\label{Fp crit ChPT}
\end{equation}

We here stress that 
the quadratic divergence in 
Eq.~(\ref{sol fpi2 for chpt 1})
is
nothing but the same kind of 
the quadratic divergences in Eqs.~(\ref{G crit}), 
(\ref{sig}), (\ref{Mofy crit SM})
and (\ref{CPN:gcr2}).
Equation~(\ref{sol fpi2 for chpt 1}) resembles
Eqs.~(\ref{NJL:gap eq}), (\ref{sig}), (\ref{rel M0 MLam SM})
and (\ref{CPN:gap ep}).
The phase is determined by the order parameter, which is given by
$\left[ F_\pi^{(\pi)}(0) \right]^2$, the pole residue of $\pi$.
Then $[F_\pi^{(\pi)}(0)]^2$ corresponds to $M_\varphi^2$ in
Eq.~(\ref{sig})
[although the broken phase corresponds to opposite
sign],
$M^2(0)$ in Eq.~(\ref{rel M0 MLam SM}),
or the left-hand-side in Eq.~(\ref{CPN:sta4}).
The first term in the right-hand-side (RHS) of 
Eq.~(\ref{sol fpi2 for chpt 1})
($[ F_\pi^{(\pi)}(\Lambda) ]^2$) corresponds to
the first term ($1/G$) of the RHS in Eq.~(\ref{sig}),
the first term ($M_D^2(\Lambda)/y^2(\Lambda)$) of the RHS in 
Eq.~(\ref{rel M0 MLam SM}),
or
the first term ($1/g$) of the RHS in Eq.~(\ref{CPN:sta4}).
The second term in Eq.~(\ref{sol fpi2 for chpt 1}) does
to the second term of the RHS in Eq.~(\ref{sig}),
the second term of the RHS in Eq.~(\ref{rel M0 MLam SM})
or
the second term of the RHS in Eq.~(\ref{CPN:sta4}).
Thus,
{\it the quadratic divergence}
[second term in Eq.~(\ref{sol fpi2 for chpt 1})]
{\it of the $\pi$ loop can give rise to chiral symmetry restoration 
$F_\pi^{(\pi)}(0) = 0$}~\cite{HY:letter,HY:VD}.
Furthermore, we immediately see that
there is a critical value for $F_\pi^{(\pi)}(\Lambda)$ which
distinguishes the broken phase from the symmetric one:
\begin{eqnarray}
&& \mbox{(i)} \ 
  \left[ F_\pi^{(\pi)}(\Lambda) \right]^2 >
  \left[ F_\pi^{(\pi),{\rm cr}}(\Lambda) \right]^2 
  \quad \Rightarrow \quad
  \left[ F_\pi^{(\pi)}(0) \right]^2 > 0 \ 
  (\mbox{broken phase}) \ ,
\nonumber\\
&& \mbox{(ii)} \ 
  \left[ F_\pi^{(\pi)}(\Lambda) \right]^2 =
  \left[ F_\pi^{(\pi),{\rm cr}}(\Lambda) \right]^2 
  \quad \Rightarrow \quad
  \left[ F_\pi^{(\pi)}(0) \right]^2 = 0 \ 
  (\mbox{symmetric phase}) \ .
\label{ChPT crit}
\end{eqnarray}
Although the bare theory looks as if it were in the broken phase
(opposite to the NJL model),
the quantum theory can actually be in the
symmetric phase for certain value of the bare parameter 
$F_\pi^{(\pi)}(\Lambda)$.

We also note that Eq.~(\ref{sol fpi2 for chpt 1}) takes a form similar
to that in the chiral restoration by the pion loop for the finite
temperature ChPT~\cite{Gasser-Leutwyler:87}:
\begin{equation}
\left[ F_\pi^{(\pi)}(T) \right]^2 =
\left[ F_\pi^{(\pi)}(0) \right]^2 - \frac{N_f}{12} T^2 \ ,
\end{equation}
with the replacement $\Lambda \rightarrow T$.
Actually, the term is from precisely 
the same diagrammatic
origin as that of our quadratic divergence
Eq.(\ref{Fp crit ChPT}).
This point will also be discussed in Sec.~\ref{sec:THDMC}.

\paragraph{Quadratic divergence in the systematic expansion}
\ \par
\label{ssssec:QDSE}

Here we discuss the validity of the derivative expansion of
ChPT when we
include quadratic divergences.
As we discussed in Secs.~\ref{ssec:DE} and \ref{ssec:DEHLS},
the derivative expansion in the ChPT is the expansion in terms of
\begin{equation}
\frac{N_f p^2}{(4\pi F_\pi)^2} \ .
\end{equation}
When we include quadratic divergences,
the correction at one-loop is given by
$N_f \Lambda^2/(4\pi F_\pi)^2$, that at two-loop by
$\left[N_f \Lambda^2/(4\pi F_\pi)^2\right]^2$, etc.
Then the derivative expansion becomes obscure when we include
quadratic divergences:
There is no longer exact correspondence between the 
derivative expansion and the loop expansion.
Nevertheless, we expect that
we can perform the systematic expansion when
\begin{equation}
\frac{N_f \Lambda^2}{(4\pi F_\pi(\Lambda))^2} < 1 \ .
\label{valid cond}
\end{equation}
The above result in Eq.~(\ref{ChPT crit})
is based on the one-loop RGE.
Though the condition in Eq.~(\ref{valid cond}) is
satisfied in the broken phase (away from the critical point) where
we expect that the expansion works well,
the expansion becomes less reliable near the
critical point since at critical value 
$N_f \Lambda^2/ [ 4 \pi F_\pi^{(\pi),{\rm cr}}(\Lambda) ]^2 = 1$
holds.

Nevertheless, for $N_f=2$ the model is nothing but 
the $O(4)$ nonlinear sigma model, and it is well known from the
lattice analyses
(See, for example, Ref.~\cite{Mon:book}, and references cited
therein.)
that there exists a phase transition (symmetry restoration) for 
a certain critical value of the hopping parameter
which corresponds to 
$[ F_\pi^{(\pi)}(\Lambda) ]^2$.
This is precisely what we obtained in the above.
Thus, we expect that the above result based on the one-loop RGE is
reliable at least qualitatively
even though a precise value of the 
$F_\pi^{(\pi),{\rm cr}}(\Lambda)$ 
might be changed by the higher loop effects.

It should be emphasized again that the role of quadratic divergence in
the chiral  
Lagrangian is just to decide which phase the theory is in. Once we
know the phase, 
we can simply forget about the quadratic divergence, and then the
whole analysis 
is simply reduced to the ordinary ChPT with only logarithmic divergence
so that 
the systematic expansion is perfect.

The same comments also apply to the ChPT with HLS to be discussed
later: Once we decide 
the phase by choosing the bare parameters relevant to the quadratic
divergence, we can  
forget about the quadratic divergence as far as we do not make a
matching with the QCD 
(as in Sec.~\ref{sec:WM}), the situation being reduced precisely
back to the ChPT with HLS 
without quadratic divergence fully discussed in 
Sec.~\ref{ssec:DEHLS}.
Then the systematic expansion becomes perfect.

\subsubsection{Quadratic divergence in symmetry preserving
regularization}
\label{sssec:QDSPR}

Let us here discuss a problem 
which arises for the naive cutoff regularization
when we consider, for example, the Feynman integral
for the vector current correlator. 
(See, for example, section 6 of Ref.~\cite{Bijnens}.)

A main point can be explained by the following Feynman integral:
\begin{equation}
I^{\mu\nu} \equiv \int \frac{d^4k}{i(2\pi)^4}
\frac{k^\mu k^\nu}{[m^2-k^2]^2} \ .
\end{equation}
When we use the naive cutoff, this integral is evaluated as
\begin{eqnarray}
I^{\mu\nu} &=& \frac{g^{\mu\nu}}{4} \int \frac{d^4k}{i(2\pi)^4}
\frac{k^2}{[m^2-k^2]^2} 
\nonumber\\
&=& 
- \frac{g^{\mu\nu}}{4} 
 \int \frac{d^4k}{i(2\pi)^4} \frac{1}{m^2-k^2} 
+ \frac{g^{\mu\nu}}{4} 
 \int \frac{d^4k}{i(2\pi)^4} \frac{m^2}{[m^2-k^2]^2} 
\nonumber\\
&=& 
\frac{g^{\mu\nu}}{(4\pi)^2} 
\left[ - \frac{1}{4}\Lambda^2 + 
   \frac{1}{2} m^2 \ln\left( \Lambda^2 / m^2 \right) \right]
+ (\mbox{finite terms})
\ ,
\label{I cut}
\end{eqnarray}
where the first term of the second line generates the $\Lambda^2$-term
of the third line and the second term of the second line does the 
$\ln(\Lambda^2/m^2)$-term.
However, in the dimensional regularization $I^{\mu\nu}$ is rewritten
into
\begin{eqnarray}
I^{\mu\nu} &=& \frac{g^{\mu\nu}}{n} \int \frac{d^n k}{i(2\pi)^n}
\frac{k^2}{[m^2-k^2]^2} 
\nonumber\\
&=& 
- \frac{g^{\mu\nu}}{n} 
 \int \frac{d^n k}{i(2\pi)^n} \frac{1}{m^2-k^2} 
+ \frac{g^{\mu\nu}}{n} 
 \int \frac{d^nk}{i(2\pi)^n} \frac{m^2}{[m^2-k^2]^2} 
\ .
\end{eqnarray}
The coefficient of $n=2$ pole in the first term is $1/n=1/2$ instead
of $1/4$.
Then the result after replacement (\ref{regularization 0}) is
\begin{eqnarray}
I^{\mu\nu} 
&=& 
\frac{g^{\mu\nu}}{(4\pi)^2} 
\left[ - \frac{1}{2}\Lambda^2 
+ \frac{1}{2} m^2 \ln\left( \Lambda^2 / m^2 \right) \right]
+ (\mbox{finite terms})
\ .
\label{I dim}
\end{eqnarray}
The coefficients of the quadratic divergences 
in Eqs.~(\ref{I cut}) and (\ref{I dim}) 
are different from each other by a factor 2.
The proper time regularization, which has an explicit cutoff $\Lambda$,
agrees with the dimensional one but not with the naive cutoff
regularization.

Now, when we apply the above results to the calculation of the vector
current correlator (see Eq.~(70) of Ref.~\cite{Bijnens}),
the result from the cutoff regularization in Eq.~(\ref{I cut})
violates the Ward-Takahashi identity, while the one from the dimensional
regularization in Eq.~(\ref{I dim}) as well as the proper time one 
is consistent with it.

Thus the following replacement in the dimensional
regularization is suitable to identify the quadratic divergence:
\begin{equation}
\int \frac{d^n k}{i (2\pi)^n} 
\frac{k_\mu k_\nu}{\left[-k^2\right]^2} \rightarrow 
- \frac{\Lambda^2} {2(4\pi)^2} g_{\mu\nu} \ .
\label{quadrepl:2}
\end{equation}
[This can be seen in 
Eq.~(6.5) of Ref.~\cite{Veltman} and discussions below Eq.~(6.6).]
The $1/(n-2)$ pole is essentially the same as the naive cutoff
up to a numerical factor. 
When we use the proper-time regularization (heat kernel expansion),
we have explicit cutoff to be interpreted physically in the naive 
sense of cutoff and of course consistent with the invariance,
the result being the same as the dimensional regularization with the
above replacement.

We also note that
the same phenomenon is observed (although not for the quadratic
divergence) when we calculate the NG boson
propagator in the NJL model.
In the naive cutoff regularization we must carefully
choose a ``correct'' routing of the loop momentum in order to get
the chiral-invariant result, namely a massless pole for the NG boson.
In both the dimensional and the
proper-time regularizations the invariant
result is automatic.

\subsection{Two-point functions at one loop}
\label{ssec:TPFOL}

In this subsection, we calculate the contributions to
the two-point functions of the background fields, 
$\overline{\cal A}_\mu$,
$\overline{\cal V}_\mu$ and $\overline{V}_\mu$
up until ${\cal O}(p^4)$.
The Lagrangian relevant to two-point functions
contains three parameters $F_\pi^2$, $a$ and $g$ at ${\cal O}(p^2)$ and
three parameters $z_1$, $z_2$ and $z_3$ at ${\cal O}(p^4)$
[see Eqs.~(\ref{leading Lagrangian}) and (\ref{Lag: z terms})]:
\begin{eqnarray}
\left. {\cal L}_{(2)} \right\vert_{\hat{\chi}=0}
&=& 
F_\pi^2 \, \mbox{tr} 
\left[ \hat{\alpha}_{\perp\mu} \hat{\alpha}_{\perp}^\mu \right]
+ F_\sigma^2 \, \mbox{tr}
\left[ 
  \hat{\alpha}_{\parallel\mu} \hat{\alpha}_{\parallel}^\mu
\right]
- \frac{1}{2g^2} \, \mbox{tr} 
\left[ V_{\mu\nu} V^{\mu\nu} \right] 
\ ,
\label{leading Lagrangian 2}
\\
{\cal L}_{(4)z_1,z_2,z_3} 
&=&
z_1 \,\mbox{tr}
 \left[ \hat{\cal V}_{\mu\nu} \hat{\cal V}^{\mu\nu} \right]
+ z_2 \,\mbox{tr}
 \left[ \hat{\cal A}_{\mu\nu} \hat{\cal A}^{\mu\nu} \right]
{}+ z_3 \,\mbox{tr}\left[ \hat{\cal V}_{\mu\nu} V^{\mu\nu} \right]
\ .
\label{Lag: z123}
\end{eqnarray}
The tree-level contribution from 
$\left. {\cal L}_{(2)} \right\vert_{\hat{\chi}=0}$
is counted as ${\cal O}(p^2)$,
while the one-loop
effect calculated from the ${\cal O}(p^2)$ Lagrangian
as well as the tree-level one from $z_1$, $z_2$ and $z_3$ terms 
are counted as ${\cal O}(p^4)$.
The relevant Feynman rules to calculate the one-loop corrections
are listed in Appendix~\ref{app:FR}.

In the present analysis it is important to include the quadratic
divergences to obtain the RGEs in the Wilsonian sense.
Since a naive momentum cutoff violates the chiral symmetry,
we need a careful treatment of the quadratic divergences.
Thus we adopt the dimensional regularization
and identify the
quadratic divergences with the presence of poles
of ultraviolet origin at $n=2$~\cite{Veltman}.
As discussed in the previous subsection,
this can be done by the following replacement in the Feynman
integrals [see Eq.~(\ref{regularization 0})]:
\begin{equation}
\int \frac{d^n k}{i (2\pi)^n} \frac{1}{-k^2} \rightarrow 
\frac{\Lambda^2} {(4\pi)^2} \ ,
\qquad
\int \frac{d^n k}{i (2\pi)^n} 
\frac{k_\mu k_\nu}{\left[-k^2\right]^2} \rightarrow 
- \frac{\Lambda^2} {2(4\pi)^2} g_{\mu\nu} \ .
\end{equation}
On the other hand, 
the logarithmic divergence is identified with the pole at 
$n=4$ [see Eqs.~(\ref{logrepl}) and (\ref{logrepl2})]:
\begin{equation}
\frac{1}{\bar{\epsilon}} + 1 \rightarrow
\ln \Lambda^2
\ ,
\label{logrepl:2}
\end{equation}
where
\begin{equation}
\frac{1}{\bar{\epsilon}} \equiv
\frac{2}{4 - n } - \gamma_E + \ln (4\pi)
\ ,
\end{equation}
with $\gamma_E$ being the Euler constant.~\footnote{
  In Eq.~(\ref{logrepl}) we did not include
  the finite part associated with logarithmic divergence.
  In Eq.~(\ref{logrepl:2}) we determine the finite part
  by evaluating a logarithmically divergent integral in the
  dimensional regularization and the cutoff regularization.
  In the dimensional regularization we have
  \begin{eqnarray}
  \int \frac{d^n k}{i(2\pi)^n} \frac{1}{[ M^2 - k^2 ]^2}
  = \frac{1}{(4\pi)^2}
  \left[ \frac{1}{\bar{\epsilon}} - \ln M^2 \right]
  \ .
  \nonumber
  \end{eqnarray}
  In the cutoff regularization, on the other hand, the same integral is
  evaluated as
  \begin{eqnarray}
  \int^\Lambda \frac{d^4 k}{i(2\pi)^4} \frac{1}{[ M^2 - k^2 ]^2}
  = \frac{1}{(4\pi)^2}
  \left[ \ln \Lambda^2 - \ln M^2 - 1 \right]
  \ ,
  \nonumber
  \end{eqnarray}
  where we drop ${\cal O}\left(M^2/\Lambda^2\right)$ contributions.
  Comparing the above two equations, we obtain the replacement
  in Eq.~(\ref{logrepl:2}).
\label{foot:finite:repl}
}

It is convenient to define the following Feynman integrals to
calculate the one-loop corrections to the two-point function:
\begin{eqnarray}
A_0(M^2) &\equiv&
\int \frac{d^nk}{i(2\pi)^n}
\frac{1}{M^2-k^2}
\ ,
\label{def:A0 2}
\\
B_0(p^2;M_1,M_2) &\equiv&
\int \frac{d^nk}{i(2\pi)^n}
\frac{1}{ [M_1^2-k^2] [M_2^2-(k-p)^2] }
\ ,
\label{def:B0 2}
\\
B^{\mu\nu}(p;M_1,M_2) &\equiv&
\int \frac{d^nk}{i(2\pi)^n}
\frac{\left(2k-p\right)^\mu \left(2k-p\right)^\nu}{%
 [M_1^2-k^2] [M_2^2-(k-p)^2] }
\ .
\label{def:Bmunu 2}
\end{eqnarray}
These are evaluated in Appendix~\ref{ssec:FFI}.
Here we just show the divergent parts of the above integrals
[see Eqs.~(\ref{div:A0}), (\ref{div:B0}) and (\ref{div:Bmunu})]:
\begin{eqnarray}
  \left. A_0(M^2) \right\vert_{\rm div} 
&=&
  \frac{\Lambda^2}{(4\pi)^2} - 
  \frac{M^2}{(4\pi)^2} \ln \Lambda^2 \ ,
\label{div:A0 2}
\\
  \left. B_0(p^2;M_1,M_2) \right\vert_{\rm div} 
&=&
  \frac{1}{(4\pi)^2} \ln \Lambda^2 \ ,
\label{div:B0 2}
\\
  \left. B^{\mu\nu}(p;M_1,M_2) \right\vert_{\rm div} 
&=&
  - g^{\mu\nu} \frac{1}{(4\pi)^2}
    \left[ 2 \Lambda^2 - ( M_1^2 + M_2^2 ) \ln \Lambda^2 \right]
\nonumber\\
&& \ 
  - \left( g^{\mu\nu}p^2 - p^\mu p^\nu \right) 
    \frac{1}{3(4\pi)^2} \ln \Lambda^2 \ .
\label{div:Bmunu 2}
\end{eqnarray}

\begin{figure}[htbp]
\begin{center}
\epsfxsize = 12cm
\ \epsfbox{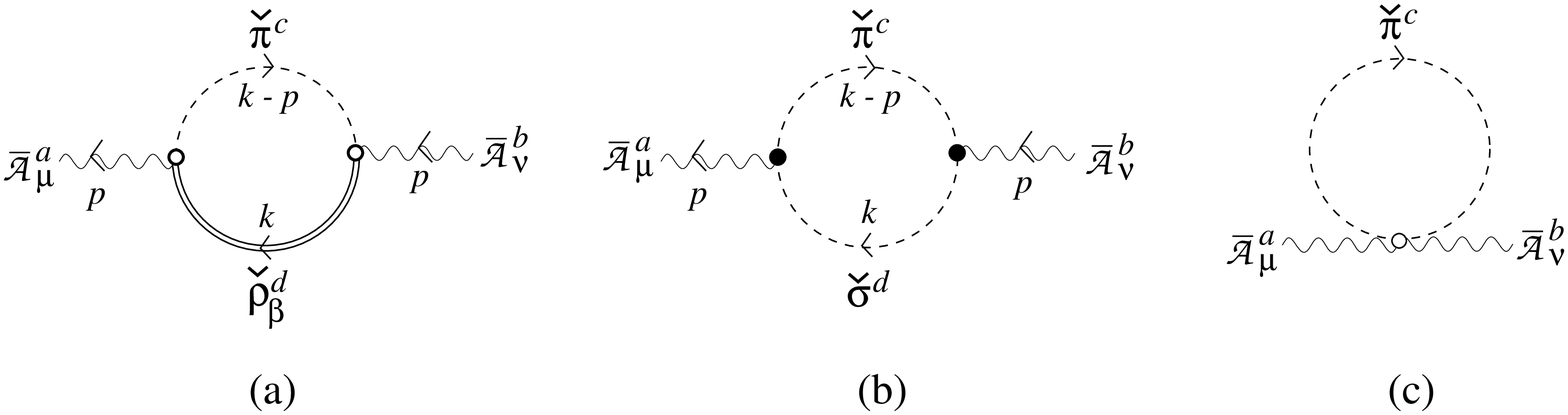}
\end{center}
\caption[$\overline{\cal A}_\mu$-$\overline{\cal A}_\nu$ Two-Point 
function]{%
One-loop corrections to the two-point function
$\overline{\cal A}_\mu$-$\overline{\cal A}_\nu$.
Vertex with a dot ($\bullet$) implies that the derivatives
acting on the quantum fields, while that with a circle ($\circ$)
implies that no derivatives are included.
Feynman rule for each vertex is shown in appendix~\ref{app:FR}.
}\label{fig:aa}
\end{figure}

Let us start with the one-loop correction to the
two-point function
$\overline{\cal A}_\mu$-$\overline{\cal A}_\nu$.
The relevant diagrams are shown in Fig.~\ref{fig:aa}.
By using the Feynman rules given in Appendix~\ref{app:FR},
it immediately follows that the contributions from the
diagrams in Fig.~\ref{fig:aa}(a)--(c) are evaluated as
\begin{eqnarray}
&&
\Pi_{\overline{\cal A}^a\overline{\cal A}^b}^{{\rm(a)}\mu\nu}(p)
\nonumber\\
&& \quad
=
\sum_{c,d}
\int \frac{d^nk}{i(2\pi)^n} 
\left( - \sqrt{a} M_\rho f_{cda} g^{\mu\beta} \right)
\frac{1}{k^2-M_\rho^2}
\left( - \sqrt{a} M_\rho f_{cdb} g^\nu_\beta \right)
\frac{1}{-(k-p)^2}
\ ,
\nonumber\\
&&
\Pi_{\overline{\cal A}^a\overline{\cal A}^b}^{{\rm(b)}\mu\nu}(p)
\nonumber\\
&& \quad
=
\sum_{c,d}
\int \frac{d^nk}{i(2\pi)^n} 
\left[ - i \frac{1}{2} \sqrt{a} (2k-p)^\mu f_{cda} \right]
\frac{1}{M_\rho^2-k^2}
\left[ - i \frac{1}{2} \sqrt{a} (-2k+p)^\nu f_{cdb} \right]
\frac{1}{-(k-p)^2}
\ ,
\nonumber\\
&&
\Pi_{\overline{\cal A}^a\overline{\cal A}^b}^{{\rm(c)}\mu\nu}(p)
\nonumber\\
&& \quad
=
\frac{1}{2}
\sum_{c}
\int \frac{d^nk}{i(2\pi)^n} 
\left[ - (1-a) \sum_{d} 
  \left( f_{acd}f_{bcd} + f_{bcd} f_{acd} \right) g^{\mu\nu}
\right]
\frac{1}{-k^2}
\ .
\end{eqnarray}
Then from the definitions in 
Eqs.~(\ref{def:A0 2})--(\ref{def:Bmunu 2})
and
\begin{equation}
\sum_{c,d} f_{acd} f_{bcd} = N_f \delta_{ab} \ ,
\end{equation}
these are written as 
[Note that
$
\Pi_{\overline{\cal A}^a\overline{\cal A}^b}^{{\rm(a)}\mu\nu}(p)
=
\Pi_{\overline{\cal A}\overline{\cal A}}^{{\rm(a)}\mu\nu}(p)
\delta_{ab}
$.]
\begin{eqnarray}
\Pi_{\overline{\cal A}\overline{\cal A}}^{{\rm(a)}\mu\nu}(p)
&=& 
- N_f\, a M_\rho^2\, g^{\mu\nu}\,
B_0\left(p^2;M_\rho,0\right) 
\ ,
\nonumber\\
\Pi_{\overline{\cal A}\overline{\cal A}}^{{\rm(b)}\mu\nu}(p)
&=&
N_f\, \frac{a}{4} \, B^{\mu\nu}\left(p;M_\rho,0\right)
\ ,
\nonumber\\
\Pi_{\overline{\cal A}\overline{\cal A}}^{{\rm(c)}\mu\nu}(p)
&=&
N_f\, (a-1) g^{\mu\nu}\,A_0(0)
\ .
\end{eqnarray}
Then by using Eqs.~(\ref{div:A0 2}), (\ref{div:B0 2})
and (\ref{div:Bmunu 2}), the divergent contributions are given by
\begin{eqnarray}
\left.
\Pi_{\overline{\cal A}\overline{\cal A}}^{{\rm(a)}\mu\nu}(p)
\right\vert_{\rm div}
&=& 
- g^{\mu\nu}\, N_f\, \frac{a M_\rho^2}{(4\pi)^2} \ln \Lambda^2
\ ,
\nonumber\\
\left.
\Pi_{\overline{\cal A}\overline{\cal A}}^{{\rm(b)}\mu\nu}(p)
\right\vert_{\rm div}
&=&
  - g^{\mu\nu} N_f\, \frac{a}{4(4\pi)^2}
    \left[ 2 \Lambda^2 - M_\rho^2 \ln \Lambda^2 \right]
  - \left( g^{\mu\nu}p^2 - p^\mu p^\nu \right) N_f\, 
    \frac{a}{12(4\pi)^2} \ln \Lambda^2
\ ,
\nonumber\\
\left.
\Pi_{\overline{\cal A}\overline{\cal A}}^{{\rm(c)}\mu\nu}(p)
\right\vert_{\rm div}
&=&
g^{\mu\nu}\, N_f\, \frac{(a-1)}{(4\pi)^2} \Lambda^2
\ .
\end{eqnarray}
By summing up these parts,
the divergent contribution to 
$\overline{\cal A}_\mu$-$\overline{\cal A}_\nu$
two-point function
is given by
\begin{eqnarray}
&&
\left.
\Pi_{\overline{\cal A}\overline{\cal A}}^{\mu\nu}(p)
\right\vert_{\rm div}
=
- \frac{N_f}{4(4\pi)^2} \left[
  2 (2-a) \Lambda^2 + 3 a^2 g^2 F_\pi^2 \ln \Lambda^2
\right]
g^{\mu\nu}
\nonumber\\
&& \qquad
{}- \frac{N_f}{(4\pi)^2} \, \frac{a}{12} \ln \Lambda^2
\, \left( g^{\mu\nu} p^2 - p^\mu p^\nu \right)
\ .
\label{div:aa}
\end{eqnarray}
These divergences are renormalized by the bare parameters in the
Lagrangian.
The tree level contribution with the bare parameters
is given by
\begin{equation}
\Pi_{\overline{\cal A}\overline{\cal A}}^{{\rm(tree)}\mu\nu}(p^2)
= F_{\pi,{\rm bare}}^2 \, g^{\mu\nu} 
+ 2 z_{2,{\rm bare}} \left( p^2 g^{\mu\nu} - p^\mu p^\nu \right) \ .
\end{equation}
Thus the renormalization is done by requiring the followings are
finite:
\begin{eqnarray}
&&
F_{\pi,{\rm bare}}^2 
- \frac{N_f}{4(4\pi)^2} \left[
  2 (2-a) \Lambda^2 + 3 a^2 g^2 F_\pi^2 \ln \Lambda^2
\right]
= \mbox{(finite)} \ ,
\label{ren:Fp}
\\
&&
z_{2,{\rm bare}} 
- 
\frac{N_f}{2(4\pi)^2} \, \frac{a}{12} \ln \Lambda^2
= \mbox{(finite)}
\ .
\label{ren:z2}
\end{eqnarray}

\begin{figure}[htbp]
\begin{center}
\epsfxsize = 14cm
\  \epsfbox{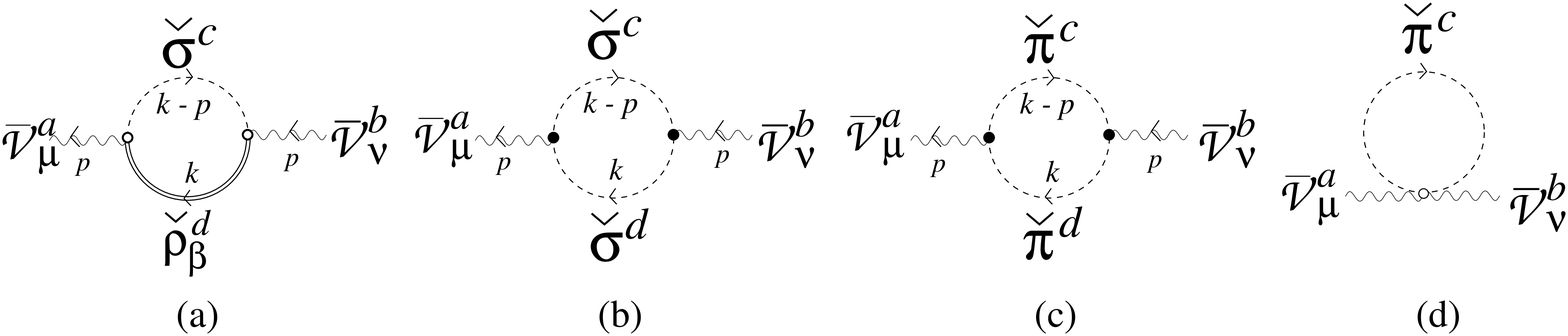}
\end{center}
\caption[$\overline{\cal V}_\mu$-$\overline{\cal V}_\nu$ Two-Point 
function]{%
One-loop corrections to the two-point function
$\overline{\cal V}_\mu$-$\overline{\cal V}_\nu$.
}\label{fig:vv}
\end{figure}
Next we calculate one-loop correction to the
two-point function
$\overline{\cal V}_\mu$-$\overline{\cal V}_\nu$.
The relevant diagrams are shown in Fig.~\ref{fig:vv}.
By using the Feynman rules given in Appendix~\ref{app:FR}
and Feynman integrals in Eqs.~(\ref{def:A0 2}),
(\ref{def:B0 2}) and (\ref{def:Bmunu 2}),
these are evaluated as
\begin{eqnarray}
\Pi_{\overline{\cal V}\overline{\cal V}}^{{\rm(a)}\mu\nu}(p)
&=& 
- N_f\, M_\rho^2\, g^{\mu\nu}\,
B_0\left(p^2;M_\rho,M_\rho\right) 
\ ,
\nonumber\\
\Pi_{\overline{\cal V}\overline{\cal V}}^{{\rm(b)}\mu\nu}(p)
&=&
\frac{1}{8} \, N_f\, B^{\mu\nu}\left(p;M_\rho,M_\rho\right)
\ ,
\nonumber\\
\Pi_{\overline{\cal V}\overline{\cal V}}^{{\rm(c)}\mu\nu}(p)
&=&
\frac{(2-a)^2}{8} \, N_f\, B^{\mu\nu}\left(p;0,0\right)
\ ,
\nonumber\\
\Pi_{\overline{\cal V}\overline{\cal V}}^{{\rm(d)}\mu\nu}(p)
&=&
- (a-1) \,N_f\, g^{\mu\nu}\,A_0(0)
\ .
\label{Pi vv : form 1}
\end{eqnarray}
{}From Eqs.~(\ref{div:A0 2}), (\ref{div:B0 2}) and 
(\ref{div:Bmunu 2})
the divergent parts of the above integrals are evaluated as
\begin{eqnarray}
&&
\left.
\Pi_{\overline{\cal V}\overline{\cal V}}^{{\rm(a)}\mu\nu}(p)
\right\vert_{\rm div}
=
g^{\mu\nu}\,\frac{N_f}{2(4\pi)^2}
\left[
  - 2 a M_\rho^2\, \ln \Lambda^2
\right]
\ ,
\nonumber\\
&&
\left.
\Pi_{\overline{\cal V}\overline{\cal V}}^{{\rm(b)}\mu\nu}(p)
\right\vert_{\rm div}
=
g^{\mu\nu}\,\frac{N_f}{2(4\pi)^2}
\left[
  - \frac{1}{2} \Lambda^2
  + \frac{1}{2} M_\rho^2\, \ln \Lambda^2
\right]
{}- \left( g^{\mu\nu} p^2 - p^\mu p^\nu \right)
\, \frac{N_f}{2(4\pi)^2} \frac{1}{12} \ln \Lambda^2
\ ,
\nonumber\\
&&
\left.
\Pi_{\overline{\cal V}\overline{\cal V}}^{{\rm(c)}\mu\nu}(p)
\right\vert_{\rm div}
=
g^{\mu\nu}\,\frac{N_f}{2(4\pi)^2}
\left[
  - \frac{(2-a)^2}{2} \Lambda^2
\right]
{}- \left( g^{\mu\nu} p^2 - p^\mu p^\nu \right)
\, \frac{N_f}{2(4\pi)^2} \frac{(2-a)^2}{12} \ln \Lambda^2
\ ,
\nonumber\\
&&
\left.
\Pi_{\overline{\cal V}\overline{\cal V}}^{{\rm(d)}\mu\nu}(p)
\right\vert_{\rm div}
=
g^{\mu\nu}\,\frac{N_f}{2(4\pi)^2}
\left[
  - 2 (a-1) \Lambda^2
\right]
\ .
\end{eqnarray}
Then 
the divergent contribution to 
$\overline{\cal V}_\mu$-$\overline{\cal V}_\nu$
two-point function
is given by
\begin{eqnarray}
&&
\left.
\Pi_{\overline{\cal V}\overline{\cal V}}^{\mu\nu}(p)
\right\vert_{\rm div}
=
- \frac{N_f}{4(4\pi)^2} \left[
  (1+a^2) \Lambda^2 + 3 a g^2 F_\pi^2 \ln \Lambda^2
\right]
g^{\mu\nu}
\nonumber\\
&& \qquad
{}- \frac{N_f}{(4\pi)^2} \, \frac{5-4a+a^2}{24} \ln \Lambda^2
\, \left( g^{\mu\nu} p^2 - p^\mu p^\nu \right)
\ .
\label{div:vv}
\end{eqnarray}
The above divergences are renormalized by the bare parameters in the
tree contribution:
\begin{equation}
\Pi_{\overline{\cal V}\overline{\cal V}}^{{\rm(tree)}\mu\nu}(p^2)
= F_{\sigma,{\rm bare}}^2 \, g^{\mu\nu} 
+ 2 z_{1,{\rm bare}} \left( p^2 g^{\mu\nu} - p^\mu p^\nu \right) \ .
\end{equation}
Thus we 
require the followings quantities are finite:
\begin{eqnarray}
&&
F_{\sigma,{\rm bare}}^2 
- \frac{N_f}{4(4\pi)^2} \left[
  (1+a^2) \Lambda^2 + 3 a g^2 F_\pi^2 \ln \Lambda^2
\right]
= \mbox{(finite)} \ ,
\label{ren:Fs}
\\
&&
z_{1,{\rm bare}} 
- 
\frac{N_f}{2(4\pi)^2} \, \frac{5-4a+a^2}{12} \ln \Lambda^2
= \mbox{(finite)}
\ .
\label{ren:z1}
\end{eqnarray}

\begin{figure}[htbp]
\begin{center}
\epsfxsize = 14cm
\  \epsfbox{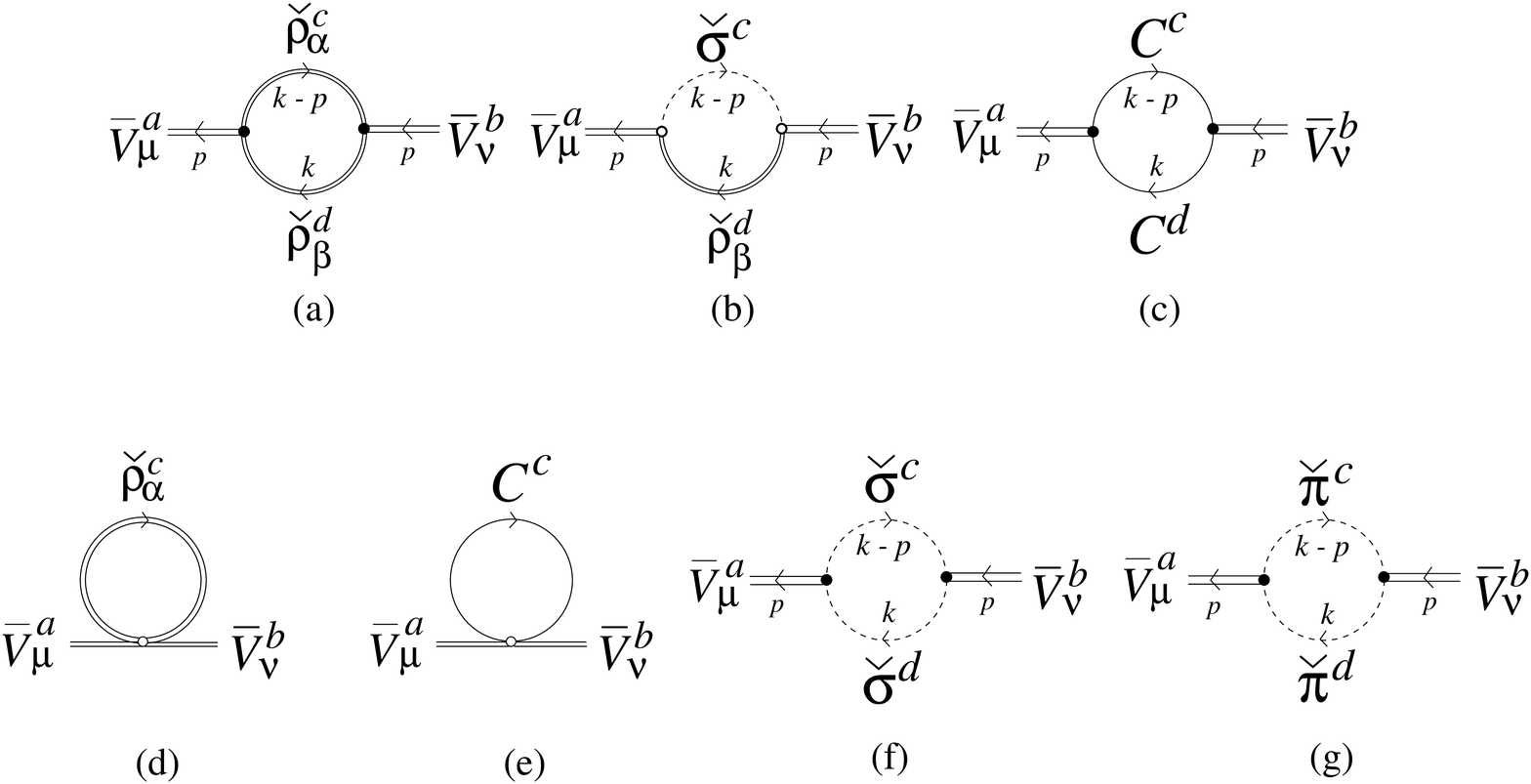}
\end{center}
\caption[$\overline{V}_\mu$-$\overline{V}_\nu$ Two-Point 
function]{%
One-loop corrections to the two-point function
$\overline{V}_\mu$-$\overline{V}_\nu$.
}\label{fig:rr}
\end{figure}
Now, we calculate the one-loop correction to the
two-point function
$\overline{V}_\mu$-$\overline{V}_\nu$.
The relevant diagrams are shown in Fig.~\ref{fig:rr}.
These are evaluated as
\begin{eqnarray}
\Pi_{\overline{V}\overline{V}}^{{\rm(a)}\mu\nu}(p)
&=&
\frac{n}{2} N_f\, B^{\mu\nu}(p;M_\rho,M_\rho)
+ 4 N_f \,( g^{\mu\nu} p^2 - p^\mu p^\nu ) 
B_0(p^2;M_\rho,M_\rho) 
\ ,
\nonumber\\
\Pi_{\overline{V}\overline{V}}^{{\rm(b)}\mu\nu}(p)
&=& 
- N_f\, M_\rho^2\, g^{\mu\nu}\,
B_0\left(p^2;M_\rho,M_\rho\right) 
\ ,
\nonumber\\
\Pi_{\overline{V}\overline{V}}^{{\rm(c)}\mu\nu}(p)
&=& 
- N_f\, B^{\mu\nu}(p;M_\rho,M_\rho)
\ ,
\nonumber\\
\Pi_{\overline{V}\overline{V}}^{{\rm(d)}\mu\nu}(p)
&=& 
n\, N_f\, g^{\mu\nu} \, A_0(M_\rho^2)
\ ,
\nonumber\\
\Pi_{\overline{V}\overline{V}}^{{\rm(e)}\mu\nu}(p)
&=& 
- 2 N_f\, g^{\mu\nu} \, A_0(M_\rho^2)
\ ,
\nonumber\\
\Pi_{\overline{V}\overline{V}}^{{\rm(f)}\mu\nu}(p)
&=& 
\frac{1}{8} \, N_f\, B^{\mu\nu}\left(p;M_\rho,M_\rho\right)
\nonumber\\
\Pi_{\overline{V}\overline{V}}^{{\rm(g)}\mu\nu}(p)
&=& 
\frac{a^2}{8} \, N_f\, B^{\mu\nu}\left(p;0,0\right)
\ ,
\label{V V cont}
\end{eqnarray}
where $n$ is the dimension of the space-time.
Here we need a careful treatment of $n$, since we identify the
quadratic divergence with a pole at $n=2$.
Then, $n$ in front of the quadratic divergence is regarded as $2$,
while $n$ in front of the logarithmic divergence as $4$:
In addition to Eqs.~(\ref{div:A0 2}) and
(\ref{div:Bmunu 2}) we have
\begin{eqnarray}
  \left. n \, A_0(M^2) \right\vert_{\rm div} 
&=&
  2\, \frac{\Lambda^2}{(4\pi)^2} - 
  4\, \frac{M^2}{(4\pi)^2} \ln \Lambda^2 \ ,
\label{div:n A0 2}
\\
  \left. n \, B^{\mu\nu}(p;M_1,M_2) \right\vert_{\rm div} 
&=&
  - g^{\mu\nu} \frac{1}{(4\pi)^2}
    \left[ 4 \Lambda^2 - 4 ( M_1^2 + M_2^2 ) \ln \Lambda^2 \right]
\nonumber\\
&& \ 
  - \left( g^{\mu\nu}p^2 - p^\mu p^\nu \right) 
    \frac{4}{3(4\pi)^2} \ln \Lambda^2 \ .
\label{div:n Bmunu 2}
\end{eqnarray}
{}From Eqs.~(\ref{div:A0 2}), (\ref{div:B0 2}),
(\ref{div:Bmunu 2}), (\ref{div:n A0 2}) and (\ref{div:n Bmunu 2})
the divergent parts of the above contributions in 
Eq.~(\ref{V V cont}) are evaluated
as~\footnote{
  We should note that when the contributions from (a) and (d) are
  added before evaluating the integrals, the sum does not include 
  the quadratic divergence.
  In such a case, we can regard $n$ in front of the sum as $4$.
  Note also
  that the sum of (c) and (e) does not include
  the quadratic divergence.
}
\begin{eqnarray}
&&
\left.
\Pi_{\overline{V}\overline{V}}^{{\rm(a)}\mu\nu}(p)
\right\vert_{\rm div}
=
g^{\mu\nu}\,\frac{N_f}{2(4\pi)^2}
\left[
  - 4 \Lambda^2
  + 8 M_\rho^2\, \ln \Lambda^2
\right]
{}+ \left( g^{\mu\nu} p^2 - p^\mu p^\nu \right)
\, \frac{N_f}{2(4\pi)^2} \frac{20}{3} \ln \Lambda^2
\ ,
\nonumber\\
&&
\left.
\Pi_{\overline{V}\overline{V}}^{{\rm(b)}\mu\nu}(p)
\right\vert_{\rm div}
=
g^{\mu\nu}\,\frac{N_f}{2(4\pi)^2}
\left[
  - 2 M_\rho^2\, \ln \Lambda^2
\right]
\ ,
\nonumber\\
&&
\left.
\Pi_{\overline{V}\overline{V}}^{{\rm(c)}\mu\nu}(p)
\right\vert_{\rm div}
=
g^{\mu\nu}\,\frac{N_f}{2(4\pi)^2}
\left[
  4 \Lambda^2
  - 4 M_\rho^2\, \ln \Lambda^2
\right]
{}+ \left( g^{\mu\nu} p^2 - p^\mu p^\nu \right)
\, \frac{N_f}{2(4\pi)^2} \frac{2}{3} \ln \Lambda^2
\ ,
\nonumber\\
&&
\left.
\Pi_{\overline{V}\overline{V}}^{{\rm(d)}\mu\nu}(p)
\right\vert_{\rm div}
=
g^{\mu\nu}\,\frac{N_f}{2(4\pi)^2}
\left[
  4 \Lambda^2
  - 8 M_\rho^2\, \ln \Lambda^2
\right]
\ ,
\nonumber\\
&&
\left.
\Pi_{\overline{V}\overline{V}}^{{\rm(e)}\mu\nu}(p)
\right\vert_{\rm div}
=
g^{\mu\nu}\,\frac{N_f}{2(4\pi)^2}
\left[
  - 4 \Lambda^2
  + 4 M_\rho^2\, \ln \Lambda^2
\right]
\ ,
\nonumber\\
&&
\left.
\Pi_{\overline{V}\overline{V}}^{{\rm(f)}\mu\nu}(p)
\right\vert_{\rm div}
=
g^{\mu\nu}\,\frac{N_f}{2(4\pi)^2}
\left[
  - \frac{1}{2} \Lambda^2
  + \frac{1}{2} M_\rho^2\, \ln \Lambda^2
\right]
{}- \left( g^{\mu\nu} p^2 - p^\mu p^\nu \right)
\, \frac{N_f}{2(4\pi)^2} \frac{1}{12} \ln \Lambda^2
\nonumber\\
&&
\left.
\Pi_{\overline{V}\overline{V}}^{{\rm(g)}\mu\nu}(p)
\right\vert_{\rm div}
=
g^{\mu\nu}\,\frac{N_f}{2(4\pi)^2}
\left[
  - \frac{a^2}{2} \Lambda^2
\right]
{}- \left( g^{\mu\nu} p^2 - p^\mu p^\nu \right)
\, \frac{N_f}{2(4\pi)^2} \frac{a^2}{12} \ln \Lambda^2
\ .
\label{VV2p}
\end{eqnarray}
Summing up the above contributions, we obtain 
\begin{eqnarray}
&&
\left.
\Pi_{\overline{V}\overline{V}}^{%
{\mbox{\scriptsize(1-loop)}}\mu\nu}(p)
\right\vert_{\rm div}
=
- \frac{N_f}{4(4\pi)^2} \left[
  (1+a^2) \Lambda^2 + 3 a g^2 F_\pi^2 \ln \Lambda^2
\right]
g^{\mu\nu}
\nonumber\\
&& \qquad
{} + 
\frac{N_f}{2(4\pi)^2} \, \frac{87-a^2}{12} \ln \Lambda^2
\left( p^2 g^{\mu\nu} - p^\mu p^\nu \right)
\ .
\label{VV:div}
\end{eqnarray}
On the other hand, 
the tree contribution is given by
\begin{equation}
\Pi_{\overline{V}\overline{V}}^{{\rm(tree)}\mu\nu}(p^2)
= F_{\sigma,{\rm bare}}^2 \, g^{\mu\nu} 
- \frac{1}{g^2_{\rm bare}}
\left( p^2 g^{\mu\nu} - p^\mu p^\nu \right) \ .
\end{equation}
The first term in Eq.~(\ref{VV:div}) which is proportional to
$g^{\mu\nu}$ is renormalized by $F_{\sigma,{\rm bare}}^2$ by using the
requirement in Eq.~(\ref{ren:Fs}).
The second term in Eq.~(\ref{VV:div}) is renormalized by 
$g_{\rm bare}$ by requiring
\begin{equation}
\frac{1}{g^2_{\rm bare}} - 
\frac{N_f}{2(4\pi)^2} \, \frac{87-a^2}{12} \ln \Lambda^2
= \mbox{(finite)}
\ .
\label{ren:g2}
\end{equation}

\begin{figure}[htbp]
\begin{center}
\epsfxsize = 13cm
\  \epsfbox{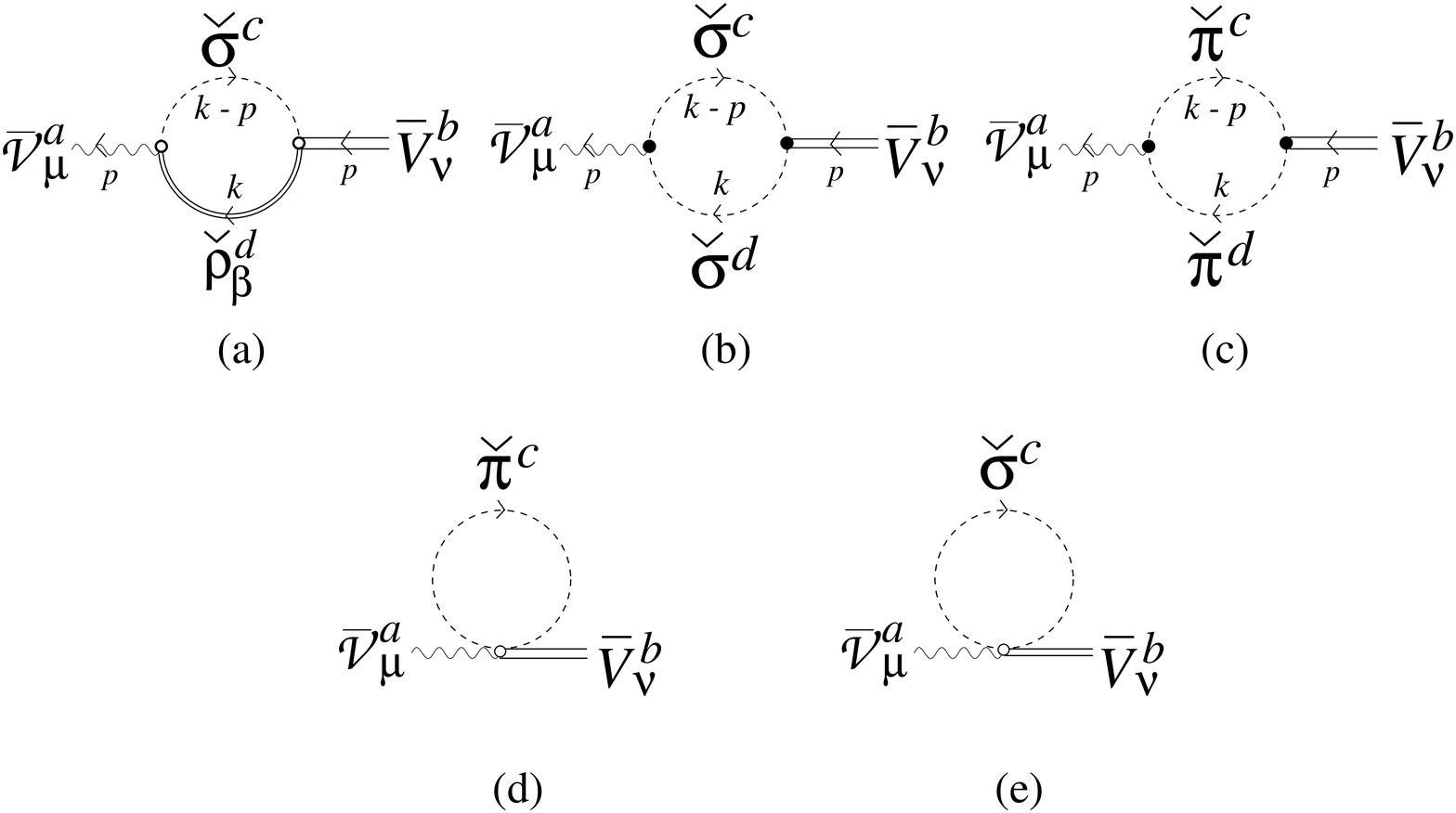}
\end{center}
\caption[$\overline{\cal V}_\mu$-$\overline{V}_\nu$ Two-Point 
function]{%
One-loop corrections to the two-point function
$\overline{\cal V}_\mu$-$\overline{V}_\nu$.
}\label{fig:vr}
\end{figure}
We also calculate the one-loop correction to the two-point
function $\overline{\cal V}_\mu$-$\overline{V}_\nu$ to determine the
renormalization of $z_3$.
The relevant diagrams are shown in Fig.~\ref{fig:vr}.
These are evaluated as
\begin{eqnarray}
\Pi_{\overline{\cal V}\overline{V}}^{{\rm(a)}\mu\nu}(p)
&=&
N_f\, M_\rho^2\, g^{\mu\nu}\,
B_0\left(p^2;M_\rho,M_\rho\right) 
\ ,
\nonumber\\
\Pi_{\overline{\cal V}\overline{V}}^{{\rm(b)}\mu\nu}(p)
&=& 
\frac{1}{8} \, N_f\, B^{\mu\nu}\left(p;M_\rho,M_\rho\right)
\ ,
\nonumber\\
\Pi_{\overline{\cal V}\overline{V}}^{{\rm(c)}\mu\nu}(p)
&=& 
\frac{a(2-a)}{8} \, N_f\, B^{\mu\nu}\left(p;0,0\right)
\ ,
\nonumber\\
\Pi_{\overline{\cal V}\overline{V}}^{{\rm(d)}\mu\nu}(p)
&=& 
\frac{a}{2} N_f\, g^{\mu\nu} \, A_0(0)
\ ,
\nonumber\\
\Pi_{\overline{\cal V}\overline{V}}^{{\rm(e)}\mu\nu}(p)
&=& 
\frac{1}{2} N_f\, g^{\mu\nu} \, A_0(M_\rho^2)
\ .
\label{Pi V v cont}
\end{eqnarray}
{}From Eqs.~(\ref{div:A0 2}), (\ref{div:B0 2}) and 
(\ref{div:Bmunu 2})
the divergent parts of the above contributions are evaluated as
\begin{eqnarray}
&&
\left.
\Pi_{\overline{\cal V}\overline{V}}^{{\rm(a)}\mu\nu}(p)
\right\vert_{\rm div}
=
g^{\mu\nu}\,\frac{N_f}{2(4\pi)^2}
\left[
  2 M_\rho^2\, \ln \Lambda^2
\right]
\ ,
\nonumber\\
&&
\left.
\Pi_{\overline{\cal V}\overline{V}}^{{\rm(b)}\mu\nu}(p)
\right\vert_{\rm div}
=
g^{\mu\nu}\,\frac{N_f}{2(4\pi)^2}
\left[
  - \frac{1}{2} \Lambda^2
  + \frac{1}{2} M_\rho^2\, \ln \Lambda^2
\right]
{}- \left( g^{\mu\nu} p^2 - p^\mu p^\nu \right)
\, \frac{N_f}{2(4\pi)^2} \frac{1}{12} \ln \Lambda^2
\ ,
\nonumber\\
&&
\left.
\Pi_{\overline{\cal V}\overline{V}}^{{\rm(c)}\mu\nu}(p)
\right\vert_{\rm div}
=
- g^{\mu\nu}\,\frac{N_f}{2(4\pi)^2} \, \frac{a(2-a)}{2} \Lambda^2
{}- \left( g^{\mu\nu} p^2 - p^\mu p^\nu \right)
\, \frac{N_f}{2(4\pi)^2} \frac{a(2-a)}{12} \ln \Lambda^2
\ ,
\nonumber\\
&&
\left.
\Pi_{\overline{\cal V}\overline{V}}^{{\rm(d)}\mu\nu}(p)
\right\vert_{\rm div}
=
g^{\mu\nu}\,\frac{N_f}{2(4\pi)^2} a \Lambda^2
\ ,
\nonumber\\
&&
\left.
\Pi_{\overline{\cal V}\overline{V}}^{{\rm(e)}\mu\nu}(p)
\right\vert_{\rm div}
=
g^{\mu\nu}\,\frac{N_f}{2(4\pi)^2}
\left[
  \Lambda^2
  - 2 M_\rho^2\, \ln \Lambda^2
\right]
\ .
\end{eqnarray}
Thus
\begin{eqnarray}
&&
\left.
\Pi_{\overline{\cal V}\overline{V}}^{%
{\mbox{\scriptsize(1-loop)}}\mu\nu}(p)
\right\vert_{\rm div}
=
\frac{N_f}{4(4\pi)^2} \left[
  (1+a^2) \Lambda^2 + 3 a g^2 F_\pi^2 \ln \Lambda^2
\right]
g^{\mu\nu}
\nonumber\\
&& \qquad
{} -
\frac{N_f}{2(4\pi)^2} \, \frac{1+2a-a^2}{12} \ln \Lambda^2
\left( p^2 g^{\mu\nu} - p^\mu p^\nu \right)
\ .
\label{Vr:div}
\end{eqnarray}
The tree contribution is given by
\begin{equation}
\Pi_{\overline{\cal V}\overline{V}}^{{\rm(tree)}\mu\nu}(p^2)
= - F_{\sigma,{\rm bare}}^2 \, g^{\mu\nu} 
+ z_{3,{\rm bare}}
\left( p^2 g^{\mu\nu} - p^\mu p^\nu \right) \ .
\label{div: 2 V r}
\end{equation}
The first term in Eq.~(\ref{Vr:div}) which is proportional to
$g^{\mu\nu}$ is already
renormalized by $F_{\sigma,{\rm bare}}^2$ by using the
requirement in Eq.~(\ref{ren:Fs}).
The second term in Eq.~(\ref{Vr:div}) is renormalized by 
$z_{3,{\rm bare}}$ by requiring
\begin{equation}
z_{3,{\rm bare}} 
- 
\frac{N_f}{2(4\pi)^2} \, \frac{1+2a-a^2}{12} \ln \Lambda^2
= \mbox{(finite)}
\ .
\label{ren:z3}
\end{equation}

To summarize, Eqs.~(\ref{ren:Fp}), (\ref{ren:Fs}), (\ref{ren:g2}),
(\ref{ren:z1}), (\ref{ren:z2}) and (\ref{ren:z3}) are all what we need
to renormalize the Lagrangians in Eqs.~(\ref{leading Lagrangian 2})
and (\ref{Lag: z123}).

To check the above calculations we calculate the divergent
contributions at one loop also by
using the heat kernel expansion with the
proper time regularization in Appendix~\ref{app:RHKE}.

\subsection{Low-energy theorem at one loop}
\label{ssec:LETOL}

In this subsection, we show that the low-energy theorem of the HLS
in Eq.~(\ref{KSRF I}) is intact at one-loop level in the 
low-energy limit.
It was first shown in Landau gauge without including the quadratic
divergences~\cite{HY}.
Here we demonstrate it in the background field gauge 
including the quadratic divergences.
The proof of the low energy theorem at any loop 
order~\cite{HKY:PRL,HKY:PTP}
will be shown in Sec.~\ref{sec:RALOLET}.

In the HLS the off-shell extrapolation of the vector meson fields are
well defined, since they are introduced as a gauge field.
Then we can naturally define the $\rho$-$\gamma$ mixing strength and
the $\rho\pi\pi$ coupling for the off-shell $\rho$.
Although 
these $g_\rho$ and $g_{\rho\pi\pi}$ do not have any momentum
dependences
at the leading order ${\cal O}(p^2)$,
they generally depend on the momenta of $\rho$ and $\pi$
when we include the loop corrections.
We write these dependences on the momenta explicitly
by $g_\rho(p_\rho^2 )$ and
$g_{\rho\pi\pi}( p_\rho^2 ; q_{1}^2 , q_{2}^2 )$,
where $p_\rho$ is the $\rho$ momentum and
$q_{1}$ and $q_{2}$ are the pion momenta.
By using these,
the low-energy theorem in Eq.~(\ref{LET tree})
is now expressed as
\begin{equation}
g_\rho(p_\rho^2 = 0) = 2 
g_{\rho\pi\pi}( p_\rho^2 =0; q_{1}^2 = 0, q_{2}^2 = 0 )
\, F_\pi^2(0) \ ,
\label{LET loop}
\end{equation}
where $F_\pi(0)$ implies that it is also defined at low-energy limit
(on-shell of the {\it massless} pion).

In Ref.~\cite{HY} the explicit calculation of the
one-loop corrections to the $\rho$-$\gamma$ mixing strength and
the $\rho\pi\pi$ coupling was performed in the Landau gauge with an
ordinary quantization procedure.
It was shown that by a suitable renormalization of the field and
parameters the low-energy theorem was satisfied at one-loop level,
and that there were no one-loop corrections in the
low-energy limit.
In the calculation in Ref.~\cite{HY}
the effect from quadratic divergences was disregarded.
Here we include them in the background field gauge.

In the background field gauge
adopted in the present analysis
the background fields
$\overline{\cal A}_\nu$ and $\overline{\cal V}_\nu$
include the photon field
$A_\mu$ and the background pion field $\overline{\pi}$ as
\begin{eqnarray}
&&
\overline{\cal A}_\nu = 
\frac{1}{F_\pi} \partial_\nu \overline{\pi} 
+ \frac{i e}{F_\pi} A_\nu 
\left[ Q \,,\, \overline{\pi} \right]
+ \cdots \ ,
\nonumber\\
&&
\overline{\cal V}_\nu = e Q A_\nu - 
\frac{i}{2F_\pi^2} 
\left[ \partial_\nu \overline{\pi} \,,\, \overline{\pi} \right]
+ \cdots \ ,
\end{eqnarray}
where $F_\pi$ in these expressions should be regarded as $F_\pi(0)$
(residue of the pion pole) to identify the field
$\overline{\pi}$ with the on-shell pion field.
On the other hand, the background field $\overline{V}_\mu$ include the
background $\rho$ field as
\begin{equation}
\overline{V}_\mu = g \bar{\rho}_\mu \ ,
\end{equation}
where $g$ is renormalized in such a way that
the kinetic term of the field $\bar{\rho}_\mu$ is normalized 
to be one.~\footnote{
  When we use other renormalization scheme for $g$,
  the finite wave function renormalization constant $Z_\rho$ 
  appears in this relation as
  $\overline{V}_\mu = g Z_\rho^{1/2} \bar{\rho}_\mu $.
  Accordingly, $g$ in Eqs.~(\ref{grho one loop}) and
  (\ref{grpp one loop}) is replaced with $g Z_\rho^{1/2}$.
  Note that the explicit form of $Z_\rho$ depends on the
  renormalization scheme for $g$
  as well as the renormalization scale, but it is irrelevant to the
  proceeding analysis, since the same factor $Z_\rho^{1/2}$
  appears in both Eqs.~(\ref{grho one loop}) and 
  (\ref{grpp one loop}).
}
The contribution to the $\rho$-$\gamma$ mixing strength
is calculated as that to the 
$\overline{V}_\mu$-$\overline{\cal V}_\nu$ two-point function.
Then the contribution in the low-energy limit is expressed as
\begin{equation}
g_\rho(p_\rho^2=0 ) =
g \,
\Pi_{\overline{\cal V}\overline{V}}^{S}(p_\rho^2=0)
\ ,
\label{grho one loop}
\end{equation}
where $p_\rho$ is the $\rho$ momentum and
the scalar component 
$\Pi_{\overline{\cal V}\overline{V}}^{S}(p^2)$
is defined by
\begin{equation}
\Pi_{\overline{\cal V}\overline{V}}^{S}(p^2) \equiv
\frac{p_\mu p_\nu}{p^2}
\Pi_{\overline{\cal V}\overline{V}}^{\mu\nu} (p)
\ .
\end{equation}
On the other hand,
the correction to the $\rho\pi\pi$ coupling
is calculated from the
$\overline{V}_\mu$-$\overline{\cal V}_\nu$ two-point function
and
$\overline{V}_\mu$-$\overline{\cal A}_\alpha$-%
$\overline{\cal A}_\beta$ three point function.
We can easily show that the correction from the three point function
vanishes at low-energy limit as follows:
Let $\Gamma_{\mu\alpha\beta}$ denotes the
$\overline{V}_\mu$-$\overline{\cal A}_\alpha$-%
$\overline{\cal A}_\beta$ three point function.
Then the $\rho\pi\pi$ coupling is proportional to
$q_1^\alpha q_2^\beta \Gamma_{\mu\alpha\beta}$, where $q_1$ and $q_2$
denote the momenta of two pions.
Since the legs $\alpha$ and $\beta$ of 
$\Gamma_{\mu\alpha\beta}$ are carried by $q_1$ or $q_2$,
$q_1^\alpha q_2^\beta\Gamma_{\mu\alpha\beta}$ generally proportional
to two of $q_1^2$, $q_2^2$ and $q_1\cdot q_2$.
Since the loop integral does not generate any massless poles,
this implies that
$q_1^\alpha q_2^\beta \Gamma_{\mu\alpha\beta}$ vanishes in the
low-energy limit $q_1^2=q_2^2=q_1\cdot q_2=0$, and 
\begin{equation}
g_{\rho\pi\pi}(p_\rho^2=0;q_1^2=0,q_2^2=0) = 
g \,
\frac{\Pi_{\overline{\cal V}\overline{V}}^{S}(p_\rho^2=0)}
{2F_\pi^2(0)}
\ .
\label{grpp one loop}
\end{equation}
Combined with Eq.~(\ref{grho one loop}),
Eq.~(\ref{grpp one loop}) leads to
\begin{equation}
g_\rho(p_\rho^2 = 0) = 2 
g_{\rho\pi\pi}( p_\rho^2 =0; q_{1}^2 = 0, q_{2}^2 = 0 )
\, F_\pi^2(0) \ ,
\label{LET loop 2}
\end{equation}
which is nothing but the low energy theorem in Eq.~(\ref{LET loop}).
Note that the quadratic divergences are included in the above
discussions:
The scalar component
$\Pi_{\overline{\cal V}\overline{V}}^{S}$
includes the effect of quadratic divergences
[see Eq.~(\ref{Vr:div})].
Therefore,
both the corrections to the $V$-$\gamma$ mixing strength and the
$V\pi\pi$ coupling in the low-energy limit
come from only the scalar component of the two-point function
$\overline{V}_\mu$-$\overline{\cal V}_\nu$, and thus
{\it the low-energy theorem remains intact at
one-loop level even including quadratic divergences}.

\subsection{Renormalization group equations in the Wilsonian sense}
\label{ssec:RGEWS}

The RGEs for $g$ and $a$ above the $\rho$ mass scale
with including only the logarithmic divergences were given in
Ref.~\cite{HY}.
We need RGEs in the Wilsonian sense
to study the phase structure.
In Ref.~\cite{HY:letter}
the quadratic divergences are further included for this purpose.
In this subsection we calculate the RGEs
for the parameters $F_\pi$, $F_\sigma$ 
(and $a \equiv F_\sigma^2/F_\pi^2$), $g$, $z_1$, $z_2$ and $z_3$
of the Lagrangians in Eqs.~(\ref{leading Lagrangian 2})
and (\ref{Lag: z123})
from the renormalization conditions derived in 
Sec.~\ref{ssec:TPFOL}.

The renormalization conditions for $F_\pi$ and $F_\sigma$ in 
Eqs~(\ref{ren:Fp}) and (\ref{ren:Fs})
lead to the RGEs for $F_\pi$ and $F_\sigma$ as
\begin{eqnarray}
\mu \frac{d F_\pi^2}{d\mu} &=&
\frac{N_f}{2(4\pi)^2}
 \left[ 3 a^2 g^2 F_\pi^2 + 2 (2-a) \mu^2 \right] \ ,
  \label{RGE for Fpi2}
\\
\mu \frac{d F_\sigma^2}{d\mu} &=&
\frac{N_f}{2(4\pi)^2}
 \left[ 3 a g^2 F_\pi^2 + (a^2+1) \mu^2 \right] \ ,
  \label{RGE for Fsig2}
\end{eqnarray}
where $\mu$ is the renormalization scale.
Combining these two RGEs we obtain the RGE for 
$a = F_\sigma^2/F_\pi^2$ as
\begin{eqnarray}
\mu \frac{d a}{d\mu} &=& - 
\frac{N_f}{2(4\pi)^2}
(a-1)
\left[ 3 a (a+1) g^2 - (3a-1) \frac{\mu^2}{F_\pi^2} \right] \ .
  \label{RGE for a}
\end{eqnarray}
We note here that the above RGEs agree with 
those obtained in Ref.~\cite{HY}
when we neglect the quadratic divergences.
From the renormalization condition for $g$ in Eq.~(\ref{ren:g2})
the RGE for $g$ is calculated as
\begin{eqnarray}
\mu \frac{d g^2}{d\mu} &=& - 
\frac{N_f}{2(4\pi)^2}
\frac{87 - a^2}{6} g^4 \ ,
  \label{RGE for g2}
\end{eqnarray}
which exactly agrees with the RGE obtained in Ref.~\cite{HY}.
It should be noticed
that the values
\begin{equation}
g = 0 \ , \quad a = 1
\label{fp:g0a1}
\end{equation}
are the fixed points of 
the RGEs for $g$ and $a$ in Eqs.~(\ref{RGE for g2}) and 
(\ref{RGE for a}).
These fixed points were first found through the RGEs without
quadratic divergences~\cite{HY}, which actually survive inclusion of
the quadratic divergences~\cite{HY:letter}.

The RGEs for $z_1$,
$z_2$ and $z_3$ are calculated from the renormalization conditions in
Eqs.~(\ref{ren:z1}), (\ref{ren:z2}) and 
(\ref{ren:z3}):~\cite{Tanabashi,HY:matching}
\begin{eqnarray}
\mu \frac{d z_1}{d \mu} &=&
\frac{N_f}{(4\pi)^2} \frac{5-4a+a^2}{24} \ ,
\label{RGE:z1}
\\
\mu \frac{d z_2}{d \mu} &=&
\frac{N_f}{(4\pi)^2} \frac{a}{12} \ ,
\label{RGE:z2}
\\
\mu \frac{d z_3}{d \mu} &=&
\frac{N_f}{(4\pi)^2} \frac{1+2a-a^2}{12} \ .
\label{RGE:z3}
\end{eqnarray}
We note here that the RGE for $z_1$ exactly agrees with that for $z_2$
when $a=1$
[$a=1$ is also the fixed point of RGE (\ref{RGE for a})].
Then
\begin{equation}
z_1-z_2 = (\mbox{constant})
\label{fp:z1z2}
\end{equation}
is the fixed point of the above RGEs when $a=1$.

The mass of $\rho$
is determined by the on-shell condition:
\begin{equation}
m_\rho^2 = a(m_\rho) g^2(m_\rho) F_\pi^2(m_\rho) \ .
\label{on-shell condition}
\end{equation}
Below the $m_\rho$ scale, 
$\rho$ decouples 
and hence $F_\pi^2$ runs by the $\pi$-loop effect 
alone.
The quadratically divergent correction to $F_\pi^2$ 
with including only the
$\pi$-loop effect is obtained as in 
Eq.~(\ref{quad div to Fpi pi 1}).
{}From this the resultant RGE for $F_\pi$ below the $m_\rho$ scale
is obtained as in Eq.~(\ref{RGE for Fpi pi 0}):
\begin{equation}
\mu \frac{ d }{d\mu} \left[F_\pi^{(\pi)}(\mu)\right]^2
= \frac{2N_f}{(4\pi)^2} \mu^2 \ ,
\qquad \ (\mu < m_\rho) \ ,
\label{RGE for Fpi pi}
\end{equation}
where $F_\pi^{(\pi)}(\mu)$ runs by the loop effect of $\pi$ alone for
$\mu<m_\rho$.
This is readily solved analytically, and
the solution is given by
[see Eq.~(\ref{sol fpi2 for chpt 0})]
\begin{equation}
\left[ F_\pi^{(\pi)}(\mu) \right]^2
= 
\left[ F_\pi^{(\pi)}(m_\rho) \right]^2- 
\frac{N_f}{(4\pi)^2} \left( m_\rho^2 - \mu^2 \right)\ .
\label{sol fpi2 for chpt}
\end{equation}
Unlike the parameters renormalized in a mass independent scheme, the
parameter $F_\pi^{(\pi)}(\mu)$ ($\mu<m_\rho$) does not smoothly
connect to $F_\pi(\mu)$ ($\mu > m_\rho$) at $m_\rho$ scale.
We need to include an effect of finite renormalization.
The relation between 
$\left[ F_\pi^{(\pi)}(m_\rho) \right]^2$ and
$F_\pi^2(m_\rho)$ based on the matching of the HLS with the ChPT at
$m_\rho$ scale will be obtained in the next subsection 
[see Eq.~(\ref{finite remalization 0})].
Here we use another convenient way to evaluate 
the dominant contribution:
Taking quadratic divergence
proportional to $a$ ($\rho$ contributions specific to the HLS)
in Eq.~(\ref{div:aa}) and replacing $\Lambda$ by $m_\rho$,
we obtain~\cite{HY:matching}
\begin{equation}
\left[ F_\pi^{(\pi)}(m_\rho) \right]^2 =
F_\pi^2(m_\rho) + 
\frac{N_f}{(4\pi)^2} \frac{a(m_\rho)}{2} m_\rho^2 \ ,
\label{finite renormalization}
\end{equation}
which is actually the same relation as that in 
Eq.~(\ref{finite remalization 0}).
Combining Eq.~(\ref{sol fpi2 for chpt}) with 
Eq.~(\ref{finite renormalization}), we obtain the following relation
between $\left[ F_\pi^{(\pi)}(\mu) \right]^2$ for $\mu < m_\rho$ and 
$F_\pi^2(m_\rho)$
\begin{equation}
\left[ F_\pi^{(\pi)}(\mu) \right]^2
= 
F_\pi^2(m_\rho) 
-
\frac{N_f}{(4\pi)^2} 
\left[ \left( 1 - \frac{a(m_\rho)}{2} \right) m_\rho^2 
- \mu^2 \right]
\quad \mbox{for}\ \mu < m_\rho
\ .
\label{rel: Fp mu Fp mr}
\end{equation}
Then the on-shell decay constant is expressed as
\begin{equation}
F_\pi^2(0) 
=
\left[ F_\pi^{(\pi)}(0) \right]^2
= 
F_\pi^2(m_\rho) 
-
\frac{N_f}{(4\pi)^2} 
\left( 1 - \frac{a(m_\rho)}{2} \right) m_\rho^2 
\ .
\label{rel: Fp 0 Fp mr}
\end{equation}

\subsection{Matching HLS with ChPT}
\label{ssec:MHC}

In Sec.~\ref{ssec:OP4L} we obtained correspondence between
the parameters of the HLS and the ${\cal O}(p^4)$ ChPT parameters at
tree level.  However, one-loop corrections from the ${\cal O}(p^2)$
Lagrangian ${\cal L}_{(2)}$ generate ${\cal O}(p^4)$ contributions,
and then the correct relations should be determined by including the
one-loop effect as was done in Ref.~\cite{Tanabashi}.
[Note that in Ref.~\cite{Tanabashi} effects of quadratic
divergences are not included.]
In this subsection
we match the axialvector and vector current
correlators obtained in the HLS with those in the ChPT
at one loop, 
and obtain the relations among several parameters,
by {\it including quadratic divergences}.

Let us start with the two-point functions of the 
non-singlet axialvector and vector currents:
\begin{eqnarray}
&&
i \int d^4 x e^{i p x}
\left\langle 0 \left\vert T\, J_{5\mu}^a (x) J_{5\nu}^b (0)
\right\vert 0 \right\rangle 
= \delta^{ab} \left( p_\mu p_\nu - g_{\mu\nu} p^2 \right)
\Pi_A (p^2) \ ,
\nonumber\\
&&
i \int d^4 x e^{iqx}
\left\langle 0 \left\vert T\, J_{\mu}^a (x) J_{\nu}^b (0)
\right\vert 0 \right\rangle 
= \delta^{ab} \left( p_\mu p_\nu - g_{\mu\nu} p^2 \right)
\Pi_V (p^2) \ .
\label{A V correlators 4}
\end{eqnarray}

In the HLS the axialvector current
correlator is expressed as
\begin{equation}
\Pi_A^{\rm (HLS)} (p^2) = 
\frac{\Pi^S_{\perp}(p^2)}{-p^2} - \Pi^T_{\perp}(p^2)
\ ,
\label{Pi A HLS full}
\end{equation}
where $\Pi^S_{\perp}(p^2)$ and $\Pi^T_{\perp}(p^2)$ are defined from
the $\overline{\cal A}_\mu$-$\overline{\cal A}_\nu$ two-point function
shown in Sec.~\ref{ssec:TPFOL}
by
\begin{equation}
\Pi^{\mu\nu}_{\overline{\cal A}\overline{\cal A}}(p^2) =
g^{\mu\nu} \Pi^S_{\perp}(p^2) 
+ ( g^{\mu\nu}p^2 - p^\mu p^\nu ) \Pi^T_{\perp}(p^2)
\ .
\label{AA two point def}
\end{equation}
In the ChPT, on the other hand, the same correlator is expressed 
as~\cite{Gas:84,Gas:85a,Tanabashi}
\begin{equation}
\Pi_A^{\rm (ChPT)} (p^2) = 
\frac{\Pi^{{\rm(ChPT)}S}_\perp(p^2)}{-p^2} 
+ 2 L_{10}^r(m_\rho) - 4 H_1^r(m_\rho)
\ ,
\label{Pi A ChPT}
\end{equation}
where we set $\mu = m_\rho$ in the parameters $L_{10}^r$
and $H_1^r$, and
$\Pi^{{\rm(ChPT)}S}_\perp$ is defined in a way similar to 
$\Pi^S_{\perp}$ in Eq.~(\ref{AA two point def}).
Note that $\Pi^{{\rm(ChPT)}S}_\perp(p^2)$ does not depend on the
momentum $p$ at one-loop level, then
\begin{equation}
\Pi^{{\rm(ChPT)}S}_\perp(p^2) =
\Pi^{{\rm(ChPT)}S}_\perp(0)
\ .
\end{equation}

As we stated in Sec.~\ref{ssec:NVM} for general case,
it is not suitable to extrapolate the form of 
$\Pi_A^{\rm (ChPT)} (p^2)$ in Eq.~(\ref{Pi A ChPT}), 
which is derived at one-loop level,
to the energy region around the $\rho$ mass.
Instead, we take the low-energy limit of 
$\Pi_A^{\rm (HLS)} (p^2)$,
and match $\Pi_A^{\rm (HLS)}(p^2)$ in the low-energy limit with
$\Pi_A^{\rm (ChPT)} (p^2)$ in Eq.~(\ref{Pi A ChPT}).

In the HLS for $p^2 \ll m_\rho^2$ the axialvector current
correlator is expressed as~\cite{Tanabashi}
\begin{equation}
\Pi_A^{\rm (HLS)} (p^2) = 
\frac{\Pi^S_{\perp}(0)}{-p^2} 
- \Pi^{S \prime}_{\perp}(0) - \Pi^T_{\perp}(0)
+ {\cal O}\left( \frac{p^2}{m_\rho^2} \right)
\ ,
\label{Pi A HLS small p2}
\end{equation}
where
$\Pi^{S \prime}_{\perp}(0)$ is defined by
\begin{equation}
\Pi^{S \prime}_{\perp}(0) = 
\left.
\frac{d}{dp^2}
\Pi^S_{\perp}(p^2) 
\right\vert_{p^2=0}
\ .
\end{equation}
We match the $\Pi_A^{\rm (HLS)}(p^2)$ in 
Eq.~(\ref{Pi A HLS small p2}) with 
$\Pi_A^{\rm (ChPT)} (p^2)$ in Eq.~(\ref{Pi A ChPT})
for $p^2 \ll m_\rho^2$.
We should note that we can match the pion pole residue
$\Pi^S_{\perp}(0)$ with $\Pi^{{\rm(ChPT)}S}_\perp(0)$ separately 
from the remaining terms.
It should be noticed that $\Pi_A^{\rm (HLS)}(p^2)$ in 
Eq.~(\ref{Pi A HLS small p2}) includes terms higher than 
${\cal O}(p^4)$ in the counting scheme of the ChPT.

Let us first match the $\Pi^S_{\perp}(0)$ 
in Eq.~(\ref{Pi A HLS small p2})
with $\Pi^{{\rm(ChPT)}S}_\perp(0) = \Pi^{{\rm(ChPT)}S}_\perp(p^2)$ in
Eq.~(\ref{Pi A ChPT}).
In the HLS, $\Pi^S_{\perp}(p^2)$ is calculated as
\begin{eqnarray}
  \Pi^{S}_{\perp}(p^2) = 
  F_\pi^2(\mu) - \frac{N_f}{4(4\pi)^2}
  \left\{
    2 (2-a) \mu^2 - 3 a M_\rho^2 \ln \frac{M_\rho^2}{\mu^2}
  \right\}
  +N_f\, a \, \Omega_\pi(p^2;M_\rho,0)
\ ,
\label{Pi S HLS}
\end{eqnarray}
where $\Omega_\pi(p^2;M_\rho,0)$ is defined by
\begin{eqnarray}
  \Omega_\pi(p^2;M_\rho,0) &\equiv&
  \frac{M_\rho^2}{(4\pi)^2} 
  \Biggl[
    \left\{
      F_0( p^2; M_\rho, 0 ) - F_0( 0; M_\rho, 0 )
    \right\}
\nonumber\\
&& 
    {}+ \frac{1}{4}
    \left\{
      F_A( p^2; M_\rho, 0 ) -  F_A( 0; M_\rho, 0 )
    \right\}
  \Biggr]
\ ,
\end{eqnarray}
with the functions
$F_0$ and $F_A$ given in Appendix~\ref{ssec:FPI}.
The renormalized $F_\pi(\mu)$ is determined by the following
renormalization condition:
\begin{eqnarray}
&&
  F_{\pi,{\rm bare}}^2 - \frac{N_f}{4(4\pi)^2}
  \left[ 
    2(2-a) \Lambda^2 + 3 a M_\rho^2
    \left( \ln \Lambda^2 - \frac{1}{6} \right)
  \right]
\nonumber\\
&& \quad
  =
  F_\pi^2(\mu) - \frac{N_f}{4(4\pi)^2}
  \left[ 
    2(2-a) \mu^2 + 3 a M_\rho^2 \ln \mu^2
  \right]
  \ ,
\label{cond Fp}
\end{eqnarray}
where the finite part associated with the logarithmic divergence
is determined in such a way that
the renormalized $F_\pi$
without quadratic divergence at $\mu=M_\rho$
becomes pole residue~\cite{HY,Tanabashi}.
On the other hand, the one-loop corrections to the 
$\overline{\cal A}_\mu$-$\overline{\cal A}_\nu$ two-point function
in the ChPT are calculated 
from the diagram 
in Fig.~\ref{fig:aa}(c) with $a=0$ taken in the vertex
(see also Sec.~\ref{sssec:CRQDPL}).
Then, the 
$\Pi^{{\rm(ChPT)}S}_\perp(p^2)$ is given by
\begin{equation}
\Pi^{{\rm(ChPT)}S}_\perp(p^2) = 
  \left[F_\pi^{(\pi)}(\mu)\right]^2 - \frac{N_f}{(4\pi)^2} \mu^2
\ ,
\end{equation}
where we adopted the following renormalization condition:
\begin{equation}
  \left[ F_{\pi,{\rm bare}}^{(\pi)}\right]^2 
  - \frac{N_f}{(4\pi)^2} \Lambda^2 
=
  \left[F_\pi^{(\pi)}(\mu)\right]^2 - \frac{N_f}{(4\pi)^2} \mu^2
\ .
\label{ren cond Fpi pi 4}
\end{equation}
It is suitable to match $\Pi^S_{\perp}(0)$ with
$\Pi^{{\rm(ChPT)}S}_\perp(0)$ with taking $\mu = m_\rho$:
\begin{eqnarray}
&&
  \Pi^S_{\perp}(0) = 
  F_\pi^2(m_\rho) - \frac{N_f}{2(4\pi)^2} (2-a) m_\rho^2
\nonumber\\
&=&
  \Pi^{{\rm(ChPT)}S}_\perp(0) =
  \left[F_\pi^{(\pi)}(m_\rho)\right]^2 
  - \frac{N_f}{(4\pi)^2} m_\rho^2
\ .
\end{eqnarray}
{}From this we obtain the following parameter relation:
\begin{equation}
\left[F_\pi^{(\pi)}(m_\rho)\right]^2  = 
  F_\pi^2(m_\rho) + \frac{N_f}{(4\pi)^2} \frac{a(m_\rho)}{2} m_\rho^2
\ ,
\label{finite remalization 0}
\end{equation}
where we also took the renormalization point $\mu = m_\rho$
for $a$.
It should be noticed that this is understood as an effect of the
finite renormalization when we include the effect
of quadratic divergences.

Next we match the non-pole terms in $\Pi_A^{\rm (HLS)}(p^2)$ 
in Eq.~(\ref{Pi A HLS small p2})
with those in $\Pi_A^{\rm (ChPT)} (p^2)$
in Eq.~(\ref{Pi A ChPT}).
Since $z_2(\mu)$ does not run for $\mu < m_\rho$,
the transverse part $\Pi^T_{\perp}(p^2)$ 
for $p^2 \le m_\rho^2$ 
is well approximated by $2 z_2(m_\rho)$, then we have
\begin{equation}
- \Pi^T_{\perp}(0) \simeq - 2 z_2(m_\rho) \quad
\ .
\label{Pi T HLS approx}
\end{equation}
By using the explicit form in Eq.~(\ref{Pi S HLS}),
$\Pi^{S \prime}_{\perp}(0)$ is given by
\begin{equation}
- \Pi^{S \prime}_{\perp}(0) = 
\frac{N_f}{(4\pi)^2} \frac{11a}{24} \ .
\label{Pi S prime HLS}
\end{equation}
The sum of Eq.~(\ref{Pi T HLS approx}) and
Eq.~(\ref{Pi S prime HLS}) should be matched with 
$2 L_{10}^r(m_\rho) - 4 H_1^r(m_\rho)$ in
Eq.~(\ref{Pi A ChPT}).
Thus, we obtain
\begin{equation}
2 L_{10}^r(m_\rho) - 4 H_1^r(m_\rho)
= - 2 z_2(m_\rho) + \frac{N_f}{(4\pi)^2} \frac{11a(m_\rho)}{24}
\ .
\label{match L10 A}
\end{equation}
We should note that the second term from 
$\Pi^{S \prime}_{\perp}(0)$
is the finite correction coming from the $\rho$-$\pi$
loop contribution~\cite{Tanabashi,HY:matching}.

We further perform the matching for the vector current correlators.
The vector current
correlator 
in the HLS 
is expressed as
\begin{equation}
\Pi_V^{\rm (HLS)} (p^2) = 
\frac{\Pi_{V}^S(p^2)}{\Pi_{V}^S(p^2) + p^2 \Pi_{V}^T(p^2)}
\left[
- \Pi_{V}^T(p^2) - 2 \Pi_{V \parallel}^T(p^2)
\right]
- \Pi_{\parallel}^T(p^2)
\ ,
\label{Pi V HLS p2}
\end{equation}
where
\begin{eqnarray}
&&
  \Pi^{\mu\nu}_{\overline{\cal V}\overline{\cal V}}(p^2) =
  g^{\mu\nu} \Pi^S_{V}(p^2) 
  + ( g^{\mu\nu}p^2 - p^\mu p^\nu ) \Pi^T_{\parallel}(p^2)
\ ,
\nonumber\\
&&
  \Pi^{\mu\nu}_{\overline{V}\overline{V}}(p^2) =
  g^{\mu\nu} \Pi^S_{V}(p^2) 
  + ( g^{\mu\nu}p^2 - p^\mu p^\nu ) \Pi^T_{V}(p^2)
\ ,
\nonumber\\
&&
  \Pi^{\mu\nu}_{\overline{\cal V}\overline{V}}(p^2) =
  g^{\mu\nu} \Pi^S_{V}(p^2) 
  + ( g^{\mu\nu}p^2 - p^\mu p^\nu ) \Pi^T_{V \parallel}(p^2)
\ .
\end{eqnarray}
Around the $\rho$ mass scale $p^2 \simeq m_\rho^2$,
$\Pi^T_{\parallel}(p^2)$, $\Pi^T_{V}(p^2)$ and
$\Pi^T_{V \parallel}(p^2)$ are dominated by
$2 z_1(m_\rho)$, $-1/g^2(m_\rho)$ and
$z_3(m_\rho)$, respectively.
In the low-energy limit, we need to include the
chiral logarithms from the pion loop in addition.
These chiral logarithms in $\Pi^T_{\parallel}(p^2)$, 
$\Pi^T_{V}(p^2)$ and $\Pi^T_{V \parallel}(p^2)$ are evaluated from the
diagrams in Fig.~\ref{fig:vv}(c),
Fig.~\ref{fig:rr}(g) and Fig.~\ref{fig:vr}(c), respectively:
We should note that the chiral logarithm is included in 
$B^{\mu\nu}(p^2;0,0)$ as
[see Eq.~(\ref{rel: Bmn})]
\begin{eqnarray}
B^{\mu\nu}(p;0,0) &=&
- 2 g^{\mu\nu} A_0(0) 
  - \left( g^{\mu\nu} p^2 - p^\mu p^\nu \right)
  \left[ B_0(p^2;0,0) - 4 B_3(p^2;0,0) \right]
\nonumber\\
&=&
  - 2 g^{\mu\nu} \frac{\Lambda^2}{(4\pi)^2}
  - \left( g^{\mu\nu} p^2 - p^\mu p^\nu \right)
  \frac{1}{3(4\pi)^2}
  \left[
    \frac{1}{\bar{\epsilon}} + \frac{8}{3}
    - \ln ( -p^2)
  \right]
\ .
\end{eqnarray}
Then, the chiral logarithms are obtained by multiplying 
$\frac{1}{3(4\pi)^2} \ln \left( -p^2 / m_\rho^2 \right)$
by the coefficients of $B^{\mu\nu}(p;0,0)$ in 
$\Pi_{\overline{\cal V}\overline{\cal V}}^{{\rm(c)}\mu\nu}(p)$
in Eq.~(\ref{Pi vv : form 1}),
$\Pi_{\overline{V}\overline{V}}^{{\rm(g)}\mu\nu}(p)$
in Eq.~(\ref{V V cont})
and $\Pi_{\overline{\cal V}\overline{V}}^{{\rm(c)}\mu\nu}(p)$
in Eq.~(\ref{Pi V v cont}), respectively.
Noting that $2 z_1(\mu)$, $g(\mu)$ and
$z_3(\mu)$ do not run for $\mu < m_\rho$,
we obtain the following approximate forms
for $p^2 \ll m_\rho^2$:
\begin{eqnarray}
&&
  \Pi^T_{\parallel}(p^2) \simeq 2 z_2(m_\rho) 
  + \frac{(2-a)^2}{24} \frac{N_f}{(4\pi)^2} 
  \ln \frac{-p^2}{m_\rho^2}
\ ,
\\
&&
  \Pi^T_{V}(p^2) \simeq - \frac{1}{g^2(m_\rho)}
  + \frac{a^2}{24} \frac{N_f}{(4\pi)^2} 
  \ln \frac{-p^2}{m_\rho^2}
\ ,
\\
&&
  \Pi^T_{V \parallel}(p^2) \simeq z_3(m_\rho)
  + \frac{a(2-a)}{24} \frac{N_f}{(4\pi)^2} 
  \ln \frac{-p^2}{m_\rho^2}
\ .
\end{eqnarray}
Thus, for $p^2 \ll m_\rho^2$, 
$\Pi_V^{\rm (HLS)} (p^2)$ in Eq.~(\ref{Pi V HLS p2}) is approximated
as~\cite{Tanabashi}
\begin{eqnarray}
\Pi_V^{\rm (HLS)} (p^2) &\simeq&
- \Pi_{V}^T(p^2) - 2 \Pi_{V \parallel}^T(p^2)
- \Pi_{\parallel}^T(p^2)
\nonumber\\
&\simeq&
 \frac{1}{g^2(m_\rho)} - 2 z_3(m_\rho) - 2 z_1(m_\rho)
  - \frac{1}{6} \frac{N_f}{(4\pi)^2} 
  \ln \frac{-p^2}{m_\rho^2}
\ .
\label{Pi V HLS small p2}
\end{eqnarray}
In the ChPT at one loop
the same correlator is 
expressed as~\cite{Gas:84,Gas:85a,Tanabashi}
\begin{equation}
\Pi_V^{\rm (ChPT)} (p^2) = 
- 2 L_{10}^r(m_\rho) - 4 H_1^r(m_\rho) 
- \frac{1}{6} \frac{N_f}{(4\pi)^2} 
\left[  \ln \frac{-p^2}{m_\rho^2} - \frac{5}{3} \right]
\ ,
\label{Pi V ChPT}
\end{equation}
where the last term is the finite correction.\footnote{%
This finite correction was not included in Ref.~\cite{HY:matching}.
}
It should be noticed that the coefficient of the chiral logarithm 
$\ln \left( -p^2/m_\rho^2 \right)$ in Eq.~(\ref{Pi V HLS small p2})
exactly agrees with that in Eq.~(\ref{Pi V ChPT}).
Then, 
matching Eq.~(\ref{Pi V ChPT}) with Eq.~(\ref{Pi V HLS small p2}),
we obtain
\begin{equation}
- 2 L_{10}^r(m_\rho) - 4 H_1^r(m_\rho)
+ \frac{5N_f}{18(4\pi)^2}
= 
\frac{1}{g^2(m_\rho)} - 2 z_3(m_\rho) - 2 z_1(m_\rho)
\ .
\label{match L10 V}
\end{equation}

Finally, 
combining Eq.~(\ref{match L10 V}) with Eq.~(\ref{match L10 A}),
we obtain the following relation:
\begin{eqnarray}
&&
L_{10}^r(m_\rho) =
- \frac{1}{4g^2(m_\rho)} 
+ \frac{z_3(m_\rho) - z_2(m_\rho) + z_1(m_\rho)}{2}
+ \frac{N_f}{(4\pi)^2} \frac{11a(m_\rho)}{96} 
+ \frac{N_f}{(4\pi)^2} \frac{5}{72} 
\ .
\label{l10}
\end{eqnarray}

\subsection{Phase structure of the HLS}
\label{ssec:PSH}

In this subsection,
following Ref.~\cite{HY:VD},
we study the phase structure of the HLS using the 
RGEs for $F_\pi$, $a$ and $g$ derived in 
Sec.~\ref{ssec:RGEWS}
[see Eqs.~(\ref{RGE for Fpi2}), (\ref{RGE for a}) and 
(\ref{RGE for g2})].

As we demonstrated for the Lagrangian of the nonlinear sigma model in
Sec.~\ref{sssec:CRQDPL},
even if the bare Lagrangian is written as if in the broken
phase, the quantum theory can be in the symmetric phase.
As shown in Eq.~(\ref{ChPT crit}),
the phase of the quantum theory is determined from the 
on-shell $\pi$ decay constant $[F_\pi^{(\pi)}(0)]^2$
(order parameter):
\begin{eqnarray}
\mbox{(i)} &\ &
  \left[ F_\pi^{(\pi)}(0) \right]^2 > 0 \ 
  \quad \Leftrightarrow \quad
  \mbox{broken phase} \ ,
\nonumber\\
\mbox{(ii)} &\ &
  \left[ F_\pi^{(\pi)}(0) \right]^2 = 0 \ 
  \quad \Leftrightarrow \quad
  \mbox{symmetric phase} \ .
\label{ChPT phase}
\end{eqnarray}
In the HLS, we can determine the phase from 
the order parameter $F_\pi^2(0)$
in a similar manner:
\begin{eqnarray}
\mbox{(i)} &\ &
  F_\pi^2(0) > 0\ 
  \quad \Leftrightarrow \quad
  \mbox{broken phase} \ ,
\nonumber\\
\mbox{(ii)} &\ &
  F_\pi^2(0) = 0 \ 
  \quad \Leftrightarrow \quad
  \mbox{symmetric phase} \ .
\label{HLS phase}
\end{eqnarray}

Before going into the detailed study of the phase structure of the HLS,
we here demonstrate, 
by taking $g=0$ and $a=1$~\footnote{
  As we shall discuss later (see Sec.~\ref{sssec:VM as a limit}), 
  the point $(g,a)\equiv (0,1)$ should be regarded 
  {\it only as a limit} $g\rightarrow 0,
  a\rightarrow 1$, where
  the essential feature of the arguments below  
  still remains intact.
},
that the phase change similar to that in the nonlinear sigma model 
actually takes place in the HLS~\cite{HY:letter}.
Since the value $g=0$ is the fixed point of the RGE for $g$ in 
Eq.~(\ref{RGE for g2}) and $a=1$ is the one for $a$ in
Eq.~(\ref{RGE for a})~\cite{HY:letter},
the RGE for $F_\pi^2$ in Eq.~(\ref{RGE for Fpi2}) becomes
\begin{equation}
\mu \frac{ d }{d\mu} F_\pi^2(\mu)
= \frac{N_f}{(4\pi)^2} \mu^2 \ .
\label{RGE for Fpi HLS g0 a1}
\end{equation}
Since $m_\rho =0$ for $g=0$,
the RGE~(\ref{RGE for Fpi HLS g0 a1})
is valid all the way down to the low energy limit, 
$\mu \ge m_\rho = 0$.
We should note that there is an extra factor $1/2$ in the
right-hand-side of the RGE~(\ref{RGE for Fpi HLS g0 a1})
compared with the RGE~(\ref{RGE for Fpi pi 0}) in the nonlinear sigma
model.
This is because the $\sigma$ (longitudinal $\rho$)
is the real NG boson in the limit of
$(g,a)=(0,1)$
and it does contribute even for $(g,a)=(0,1)$.
Solution of the RGE~(\ref{RGE for Fpi HLS g0 a1}) is given by
\begin{eqnarray}
&&F_\pi^2(0)
= 
F_\pi^2(\Lambda) - [F_\pi^{\rm cr}]^2\ , 
\nonumber\\
&&   [F_\pi^{\rm cr}]^2 \equiv 
\frac{N_f}{2(4\pi)^2} \Lambda^2 \ .
\label{sol fpi2 for HLS g0 a1}
\end{eqnarray}
We should note again the extra factor $1/2$ compared with 
Eq.~(\ref{Fp crit ChPT}). 
As in the case for Eq.~(\ref{sol fpi2 for chpt 1}),
Eq.~(\ref{sol fpi2 for HLS g0 a1}) implies that 
{\it even if the bare theory of the HLS is written as if 
it were in the broken phase} ($F_\pi^2(\Lambda)>0$),
{\it the quantum theory is actually in the symmetric phase}, 
when we tune the bare parameter as:
$F_\pi^2(\Lambda) = [F_\pi^{\rm cr}]^2$.
We should stress that this can occur only if we use the Wilsonian
RGEs, i.e., the RGEs including the quadratic divergences.\footnote{
  In the case of large $N_f$ QCD to be discussed in 
  Sec.~\ref{ssec:VMLNQ}, we shall determine the bare parameter 
  $F_\pi^2(\Lambda)$ by the underlying theory through the 
  Wilsonian matching (see Sec.~\ref{sec:WM}) and hence
  $F_\pi^2(\Lambda)$  is no longer an adjustable parameter, 
  whereas the  value of $[F_\pi^{\rm cr}]^2$ 
 instead of $F_\pi^2(\Lambda)$ can be tuned by adjusting $N_f$.
}

For studying the phase structure of the HLS through
the RGEs
it is convenient to use the following quantities:
\begin{eqnarray}
&&
X(\mu) \equiv \frac{N_f}{2(4\pi)^2} \frac{\mu^2}{F_\pi^2(\mu)}
\ ,
\label{def X}
\\
&& G(\mu) \equiv \frac{N_f}{2(4\pi)^2} g^2(\mu) 
\quad
(\mu \ge m_\rho)
\ .
\label{def G}
\end{eqnarray}
By using these $X(\mu)$ and $G(\mu)$,
the RGEs in Eqs.~(\ref{RGE for Fpi2}),
(\ref{RGE for a}) and (\ref{RGE for g2})
are rewritten as
\begin{eqnarray}
&& \mu \frac{d X}{d\mu} = ( 2 - 3 a^2 G) X - 2 (2-a) X^2 
\ , 
\label{RGE for X}
\\
&& \mu\frac{d a}{d\mu} = - (a-1) \left[ 3a (a+1) G - (3a-1) X \right]
\ ,
\label{RGE for a 2}
\\
&& \mu\frac{d G}{d \mu} = - \frac{ 87 - a^2}{6} G^2 
\ .
\label{RGE for G}
\end{eqnarray}
As we stated in Sec.~\ref{ssec:RGEWS},
the above RGEs 
are valid above the $\rho$
mass scale $m_\rho$, where $m_\rho$ is defined by 
the on-shell condition in Eq.~(\ref{on-shell condition}).
In terms of $X$, $a$ and $G$, the on-shell condition becomes
\begin{equation}
a(m_\rho) G(m_\rho) = X(m_\rho) \ .
\label{on-shell condition 2}
\end{equation}
Then the region where the RGEs in 
Eqs.~(\ref{RGE for X})--(\ref{RGE for G})
are valid is specified by the condition
$a(\mu) G(\mu) \le X(\mu)$.

We first obtain the the fixed points of the RGEs in 
Eqs.~(\ref{RGE for X})--(\ref{RGE for G}).
This is done by 
seeking the parameters for which all right-hand-sides of three RGEs
vanish {\it simultaneously}.
As a result, there are
{\it three fixed points} and {\it one fixed line
in the physical region} and one fixed point in the
unphysical region (i.e., $a<0$ and $X<0$).
Those in the physical region
are given by~\cite{HY:VD}
\begin{eqnarray}
&& 
  \left(X^\ast_1,\, a^\ast_1,\, G^\ast_1\right)
  = \left( 0,\, \mbox{any},\, 0 \right) \ , 
\nonumber\\
&&
  \left(X^\ast_2,\, a^\ast_2,\, G^\ast_2\right) =
  \left( 1,\, 1,\, 0 \right) \ ,
\nonumber\\
&&
  \left(X^\ast_3,\, a^\ast_3,\, G^\ast_3\right) =
  \left( \frac{3}{5},\, \frac{1}{3},\, 0 \right) \ ,
\nonumber\\
&&
  \left(X^\ast_4,\, a^\ast_4,\, G^\ast_4\right) =
  \left( 
    \frac{2(2+45\sqrt{87})}{4097},\, 
    \sqrt{87},\, 
    \frac{2(11919-176\sqrt{87})}{1069317}
  \right)
\ ,
\label{fixed points}
\end{eqnarray}
and it in the unphysical region is given by
\begin{equation}
  \left(X^\ast_5,\, a^\ast_5,\, G^\ast_5\right) =
  \left( 
    \frac{2(2-45\sqrt{87})}{4097},\, 
    - \sqrt{87},\, 
    \frac{2(11919+176\sqrt{87})}{1069317}
  \right)
\ .
\label{fixed point in unphys}
\end{equation}
We should note that $G=0$ is a fixed point of the RGE for $G$, and
$a=1$ is the one for $a$.
Hence RG flows on $G=0$ plane and $a=1$ plane are confined in the
respective planes.

Now, let us study the phase structure of the HLS.
Below we shall first study the phase structure on $G=0$ plane, 
second on $a=1$ plane, and then on whole $(X,a,G)$ space.
Note also that RG flows in the region of $X<0$ 
(unphysical region) is confined in that
region since
$X=0$ is the fixed point of the RGE for $X$
in Eq.~(\ref{RGE for X}).

\begin{figure}[htbp]
\begin{center}
\epsfxsize = 10cm
\ \epsfbox{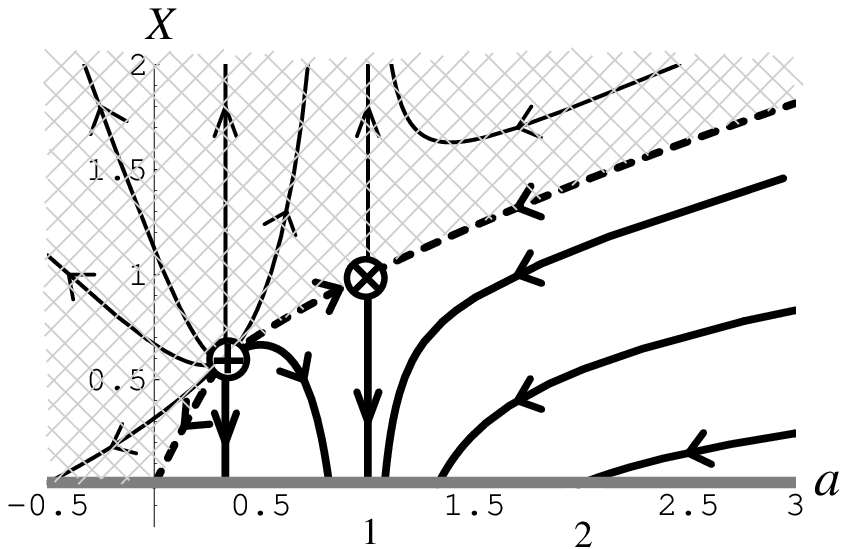}
\end{center}
\caption[Phase diagram of the HLS on $G=0$ plane]{%
Phase diagram on $G=0$ plane.
Arrows on the flows
are written from the ultraviolet to the infrared.
Gray line denotes the
fixed line $( X^\ast_1, a^\ast_1, G^\ast_1 ) = 
( 0, \mbox{any}, 0 )$.
Points indicated by $\oplus$ and $\otimes$ 
(VM point; see Sec.~\ref{sec:VM})
denote
the fixed points
$( 3/5, 1/3, 0 )$ and 
$( 1, 1, 0 )$, respectively.
Dashed lines divide the broken phase (lower side) and the
symmetric phase (upper side; cross-hatched area):
Flows drawn by thick lines are in the broken phase,
while those by thin lines are in the symmetric phase.
}\label{fig:flows G0}
\end{figure}

We first study the phase structure of the HLS for $G(m_\rho)=0$ 
($g^2(m_\rho) = 0$). 
In this case $m_\rho$ vanishes and the RGEs 
(\ref{RGE for X}), (\ref{RGE for a 2}) and (\ref{RGE for G}) are
valid all the way down to the low energy limit, $\mu \ge m_\rho = 0$.
Then the conditions in Eq.~(\ref{HLS phase}) are rewritten into the
following conditions for $X(0)$:
\begin{eqnarray}
\mbox{(A-i)} &\ &
  X(0) = 0\ \ (m_\rho = 0)
  \quad \Leftrightarrow \quad
  \mbox{broken phase} \ ,
\nonumber\\
\mbox{(A-ii)} &\ &
  X(0) \neq 0 \ \ (m_\rho = 0)
  \quad \Leftrightarrow \quad
  \mbox{symmetric phase} \ .
\label{HLS phase X}
\end{eqnarray}
We show the phase diagram on $G=0$ plane in
Fig.~\ref{fig:flows G0}.
There are one fixed line and two fixed points:
$(X^\ast_1,a^\ast_1,G^\ast_1)$,
$(X^\ast_2,a^\ast_2,G^\ast_2)$ and
$(X^\ast_3,a^\ast_3,G^\ast_3)$.
As we showed in Eq.~(\ref{HLS phase X}),
the phase is determined by the value of $X(\mu)$ at the infrared limit
$\mu= 0$.
In particular,
the phase boundary is specified by $F_\pi^2(0)=0$,
namely, governed by the infrared fixed point such that 
$X(0)\neq0$ [see Eq.~(\ref{HLS phase X})].
Such a fixed point is the point
$( X^\ast_2,a^\ast_2,G^\ast_2 ) = (1,1,0)$,
which is nothing but the point corresponding to the
vector manifestation (VM)~\cite{HY:VM} 
(see Sec.~\ref{sec:VM}).
Then the phase boundary is given by 
the RG flows entering 
$( X^\ast_2,a^\ast_2,G^\ast_2 )$.
Since $a=1/3$ is a fixed point of the RGE for $a$ in 
Eq.~(\ref{RGE for a 2}) for $G=0$,
the RG flows for $a<1/3$ cannot enter 
$( X^\ast_2,a^\ast_2,G^\ast_2 )$.
Hence there is no phase boundary specified by 
$F_\pi^2(0)=0$ in $a<1/3$ region.
Instead,
$F_\sigma^2(0)$ vanishes even though $F_\pi^2(0)\neq0$,
namely $a(0)=X(0)=0$.
Then the phase boundary for $a<1/3$
is given by the RG flow entering the point
$(X,a,G) = (0,0,0)$.
In Fig.~\ref{fig:flows G0}
the phase boundary is drawn by the dashed line,
which divides the phases into 
the symmetric phase~\footnote{
  Here ``symmetric phase'' means that
  $F_\pi^2(\mu) = 0$ or $F_\sigma^2(\mu) =0$,
  namely $1/X(\mu)= F_\pi^2(\mu)/C\mu^2 = 0$ or $a(\mu)=0$ 
  for non-zero (finite) $\mu$.
}
(upper side; cross-hatched area)
and the broken one (lower side).
Here we should stress that the exact $G\equiv 0$ plane does not actually
correspond to the underlying QCD as we shall demonstrate in 
Sec.~\ref{sssec:VR} and
Sec.~\ref{sssec:VM as a limit} and
hence Fig.~\ref{fig:flows G0} is only for illustration of the section
at $G=0$ of the phase diagram in entire parameter space $(X,a,G)$.

In the case of $G(m_\rho)>0$ ($g^2(m_\rho) >  0$), 
on the other hand,
the $\rho$ generally becomes massive ($m_\rho\neq 0$),
and thus decouples at $m_\rho$ scale.
As we said in subsection~\ref{ssec:RGEWS},
below the $m_\rho$ scale
$a$ and $G = g^2 \cdot N_f /[2 (4\pi)^2]$ 
no longer run, while $F_\pi$ still runs 
by the $\pi$ loop effect.
The running of $F_\pi$ for $\mu < m_\rho$ 
(denoted by $F_\pi^{(\pi)}$) is given in 
Eq.~(\ref{sol fpi2 for chpt}).
{}From this we should note that
{\it the quadratic divergence} 
(second term in Eq.~(\ref{sol fpi2 for chpt}))
{\it of the $\pi$ loop can give rise to chiral symmetry restoration}
$F_\pi^{(\pi)}(0) = 0$~\cite{HY:letter,HY:VD}.
The resultant relation between the order parameter $F_\pi^2(0)$ and
the $F_\pi^2(m_\rho)$ is given by Eq.~(\ref{rel: Fp 0 Fp mr}),
which in terms of $X(m_\rho)$ is rewritten as
\begin{equation}
F_\pi^2(0) 
=
\frac{N_f}{2(4\pi)^2} m_\rho^2
\left[  X^{-1}(m_\rho) - 2 + a(m_\rho) \right]
\ .
\end{equation}
Thus, the phase is determined by the following conditions:
\begin{eqnarray}
\mbox{(B-i)} &\ &
  X^{-1}(m_\rho) > 2 - a(m_\rho) \ \ (m_\rho > 0)
  \quad \Leftrightarrow \quad
  \mbox{broken phase} \ ,
\nonumber\\
\mbox{(B-ii)} &\ &
  X^{-1}(m_\rho) = 2 - a(m_\rho) \ \ (m_\rho > 0)
  \quad \Leftrightarrow \quad
  \mbox{symmetric phase} \ .
\label{HLS phase X a}
\end{eqnarray}
Then, the phase boundary is specified by the condition
\begin{equation}
2-a(m_\rho) = \frac{1}{X(m_\rho)} \ .
\label{phase boundary}
\end{equation}
Combination of this with the on-shell condition in 
Eq.~(\ref{on-shell condition 2}) determines a line,
which is nothing but an edge of the phase boundary surface:
The phase boundary surface is given by the collection of the RG flows
entering points on the line specified by
Eqs.~(\ref{on-shell condition 2}) and 
(\ref{phase boundary})~\cite{HY:VD}.

\begin{figure}[htbp]
\begin{center}
\epsfxsize = 10cm
\ \epsfbox{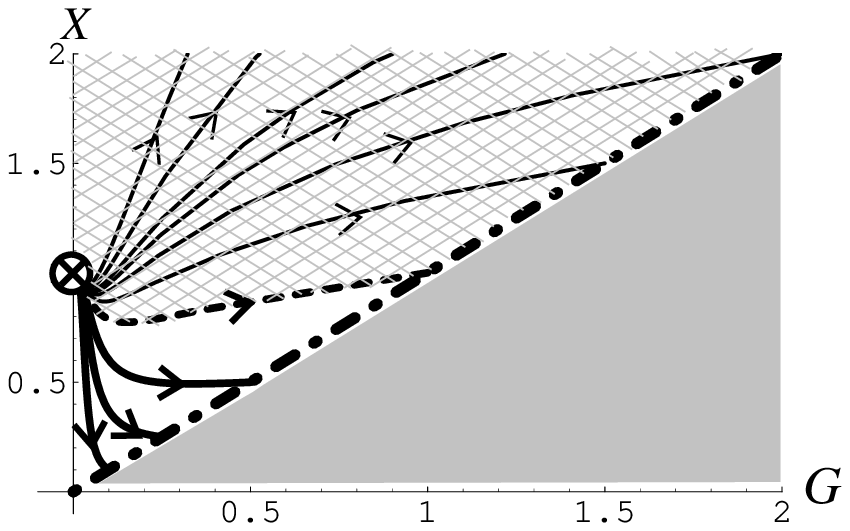}\\
\end{center}
\caption[Phase diagram of the HLS on $a=1$ plane]{%
Phase diagram on $a=1$ plane.
Arrows on the flows
are written from the ultraviolet to the infrared.
Point indicated by $\otimes$ denotes
the VM fixed point 
$(X^\ast_2,a^\ast_2,G^\ast_2) = \left( 1, 1, 0 \right)$.
(See Sec.~\ref{sec:VM}.)
Flows drawn by thick lines are in the broken phase,
while those by thin lines are in the symmetric phase
(cross-hatched area).
Dot-dashed line corresponds to the on-shell condition $G=X$.
In the shaded area the RGEs (\ref{RGE for X}), (\ref{RGE for a 2})
and (\ref{RGE for G}) are not valid
since $\rho$ has already decoupled.
}\label{fig:flows a1}
\end{figure}
We now study the $a=1$ plane
(see Fig.~\ref{fig:flows a1}).
The flows stop at the on-shell
of $\rho$ ($G=X$; dot-dashed line in Fig.~\ref{fig:flows a1})
and should be switched over to RGE of $F_\pi^{(\pi)}(\mu)$ as
mentioned above.
{}From Eqs.~(\ref{on-shell condition 2}) and (\ref{phase boundary})
with $a=1$
the flow entering $(X,G) = (1,1)$
(dashed line) is the phase boundary
which distinguishes the broken phase (lower side) from the
symmetric one (upper side; cross-hatched area).

For $a<1$, RG flows approach to the fixed point
$\left(X^\ast_3,a^\ast_3,G^\ast_3\right)=(3/5, 1/3, 0)$
in the idealized high energy limit ($\mu\rightarrow\infty$).

\begin{figure}[thbp]
\begin{center}
\epsfxsize=7cm
\ \epsfbox{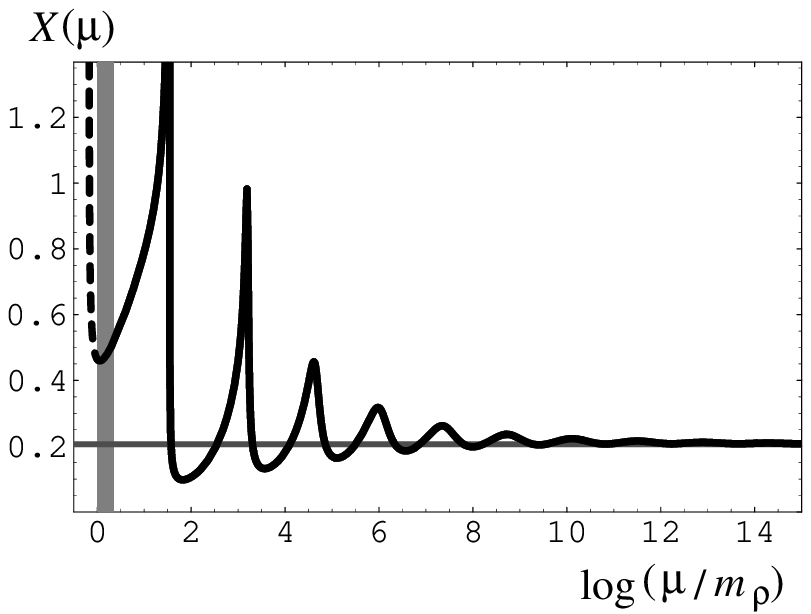}
\epsfxsize=7cm
\ \ \ \epsfbox{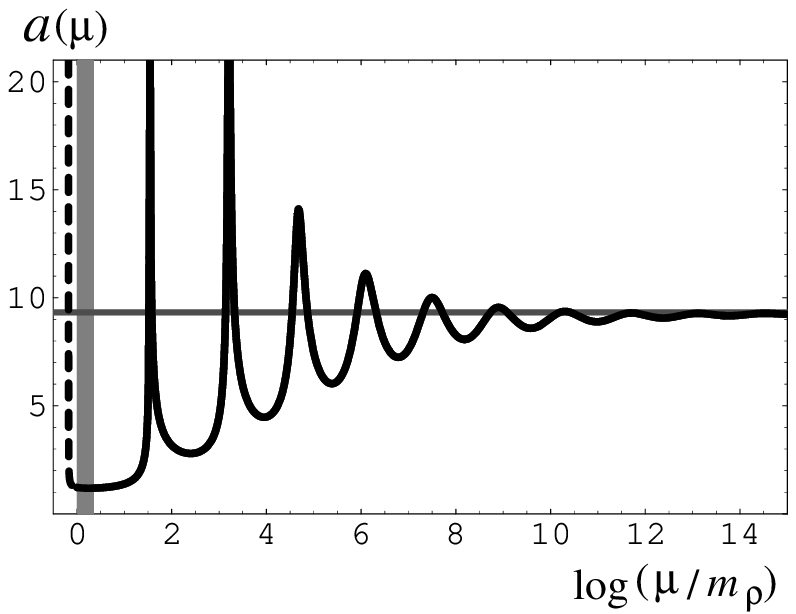}\\
\hspace*{2cm} \  (a) \ \hspace{6cm} \ (b) \hspace*{2cm}
\\
\ \\
\epsfxsize=7cm
\ \epsfbox{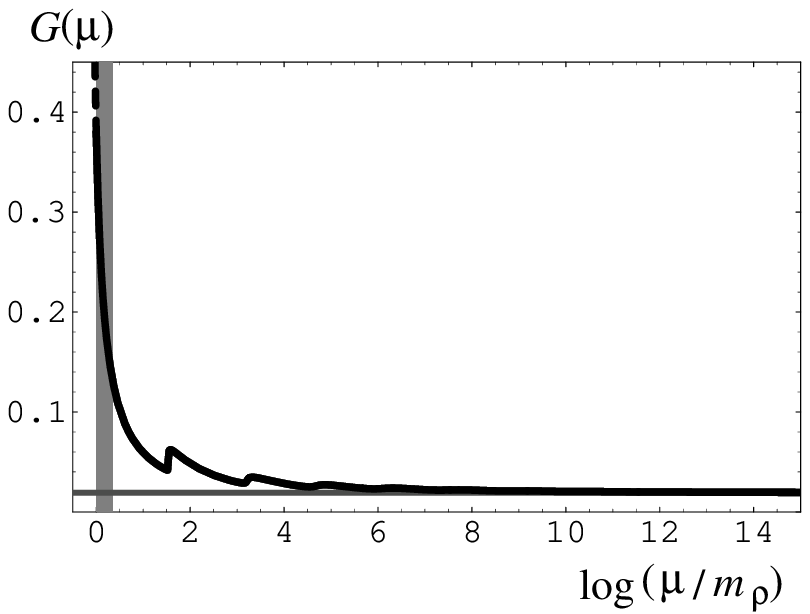}\\
(c)\\
\end{center}
\caption[Scale dependences of the parameters of the HLS]{%
Scale dependences of 
(a) $X(\mu)$, (b) $a(\mu)$ and (c) $G(\mu)$ in QCD with $N_f=3$.
Shaded area denotes the physical region, 
$m_\rho \le \mu \le \Lambda$.
Flow shown by the dashed line are obtained by
extending it to the (unphysical) infrared region by taking literally
the RGEs in Eq.~(\ref{RGE for X}), (\ref{RGE for a 2})
and (\ref{RGE for G}).
In an idealized high energy limit
the flow approaches to the fixed point
$(X^\ast_4,a^\ast_4,G^\ast_4) \simeq (0.2,9.3,0.02)$.
}\label{fig:running}
\end{figure}

For $a>1$, RG flows in the broken phase approach to 
$\left(X^\ast_4,a^\ast_4,G^\ast_4\right)\simeq(0.2, 9.3, 0.02)$,
which is precisely the fixed point 
that {\it the RG flow of the $N_f=3$ QCD belongs to}.
To see how the RG flow of $N_f=3$ QCD
approaches to this fixed point,
we show the $\mu$-dependence of $X(\mu)$, $a(\mu)$ and $G(\mu)$
in Fig.~\ref{fig:running}
where values of the parameters at $\mu = m_\rho$ are set to be 
$\left( X(m_\rho), a(m_\rho), G(m_\rho)\right)
\simeq \left( 0.46,  1.22, 0.38 \right)$ through Wilsonian
matching with the underlying QCD~\cite{HY:matching}
[see Sec.~\ref{sec:WM}].
The values of $X$ close to $1/2$
in the physical region ($m_\rho \le \mu \le \Lambda$) are
{\it very unstable against RGE flow, and hence
$X \sim 1/2$ is realized in a very accidental way}.
We shall return to this point in Sec.~\ref{sssec:VDLNQ}.

\begin{figure}[bhtp]
\begin{center}
\epsfysize = 6.6cm
\ \epsfbox{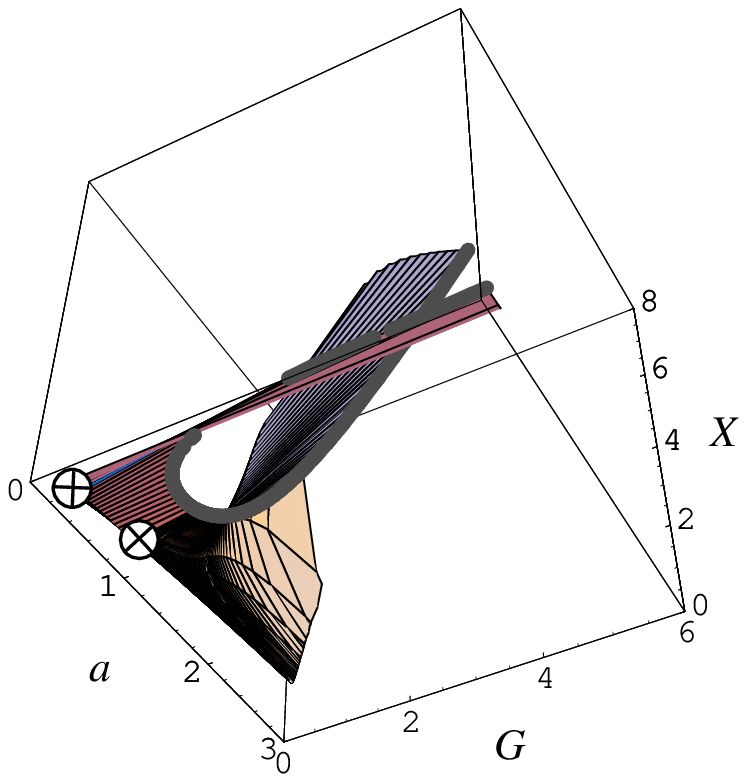}
\hspace{0.5cm}
\epsfxsize = 6.5cm
\ \epsfbox{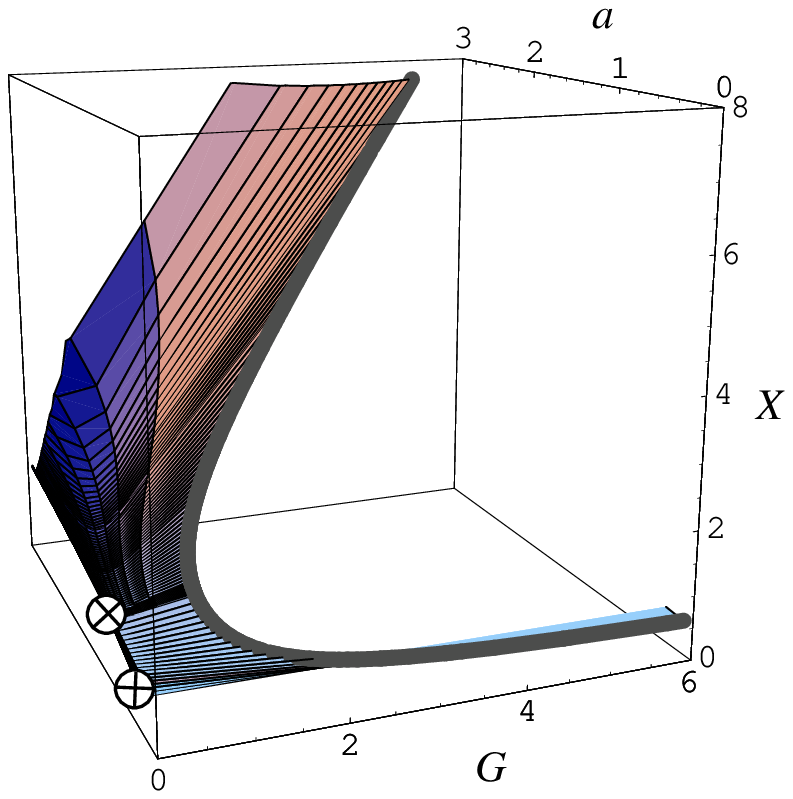}\\
\hspace*{2cm} \  (a) \ \hspace{6cm} \ (b) \hspace*{2cm}
\\
\ \\
\epsfxsize = 6.5cm
\ \epsfbox{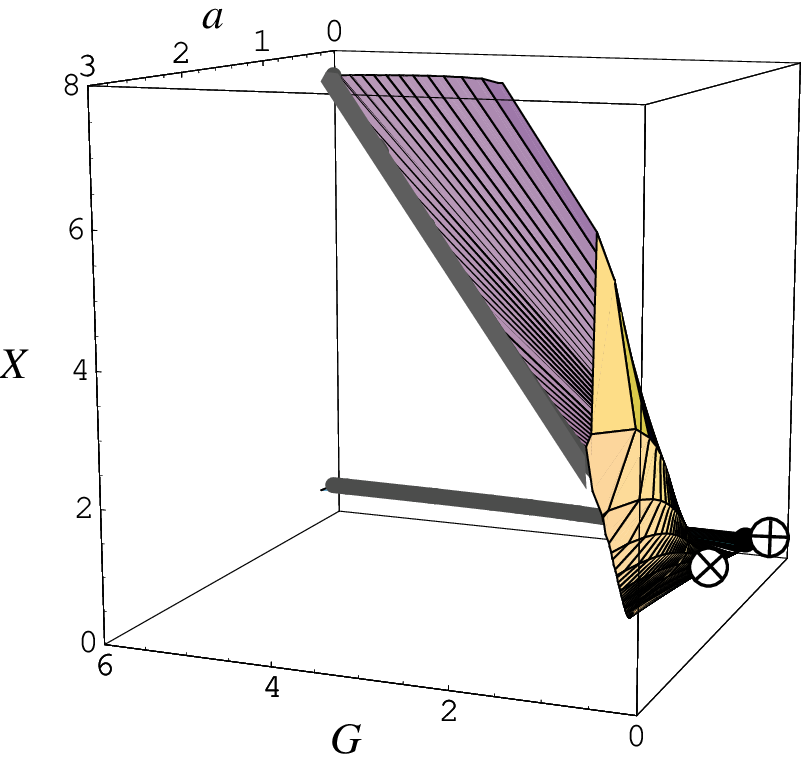}\\
(c)
\end{center}
\caption[Phase boundary surface of the HLS]{%
Phase boundary surface from three different view points.
Points indicated by $\oplus$  and $\otimes$ (VM point)
denote
the fixed points
$( 3/5, 1/3, 0 )$ and $(1,1,0)$, respectively.
Gray line denotes the line specified by 
Eqs.~(\ref{on-shell condition 2}) and (\ref{phase boundary}).
}\label{fig:phase boundary}
\end{figure}

Finally, we show the phase boundary surface in
the whole $(X,a,G)$ space
in Fig.~\ref{fig:phase boundary}
from three different view points.
This shows that the phase boundary spreads in a wide region of the
parameter space.
When we take the HLS model literally,
the chiral symmetry restoration can occur at any point on this phase
boundary.
However, 
{\it 
when we match the HLS with the underlying QCD,
only the point 
$( X^\ast_2,a^\ast_2,G^\ast_2 ) = (1,1,0)$, VM point},
on the phase boundary surface is selected,
since the axialvector and vector current correlators in HLS can be
matched with those in QCD only at that point~\cite{HY:VM}
(see Sec.~\ref{sec:VM}).

Here again we mention that
as we will discuss in Sec.~\ref{sssec:VM as a limit},
we should consider the VM only as a limit (``VM limit'') with
the bare parameters approaching the VM fixed point 
{\it from the broken phase}:
$(X(\Lambda),a(\Lambda),G(\Lambda)) \rightarrow
( X^\ast_2,a^\ast_2,G^\ast_2 ) = (1,1,0)$,
particularly $G(\Lambda)\rightarrow 0$. Setting $G(\Lambda)\equiv 0$ 
would contradict the symmetry of the underlying QCD (see Sec.~\ref{sssec:VR}.)
Also note that,
since the VM fixed point is not an infrared stable fixed point 
as can be seen in Figs.~\ref{fig:flows G0} and \ref{fig:flows a1},
the parameters in the infrared region do not generally approach this
fixed point:
In the case of $G=0$ with $(X(\Lambda),a(\Lambda))\rightarrow(1,1)$,
we can easily see from Fig.~\ref{fig:flows G0} that the 
infrared parameters behave as
$(X(0),a(0)) \rightarrow (0,1)$.
In the case of $a=1$ with
$(X(\Lambda),G(\Lambda))\rightarrow(1,0)$, on the other hand,
without extra fine tuning,
we expect that $G(\Lambda)\rightarrow0$ leads to 
$G(m_\rho)\rightarrow0$.  This together with the on-shell condition
in Eq.~(\ref{on-shell condition 2}) implies that
\begin{equation}
X(m_\rho)=\frac{m_\rho^2}{F_\pi^2(m_\rho^2)}\rightarrow 0,
\end{equation} 
which will be explicitly shown by solving
the RGEs later in Sec.~\ref{sssec:CB}.

\newpage

\section{Wilsonian Matching}
\label{sec:WM}

In the previous section we derived the renormalization group equations
(RGEs) in the Wilsonian sense
for several parameters of the HLS.
In the RGEs we included the quadratic divergence in addition to the
logarithmic divergences.
In Ref.~\cite{HY:matching} it was shown that
quadratic divergences have the physical meaning of phenomenological
relevance besides phase transition,
when we match the bare theory of the HLS with the
underlying QCD (``Wilsonian matching'').
In this section we review the Wilsonian matching proposed in
Ref.~\cite{HY:matching}.

\begin{figure}[htbp]
\begin{center}
\epsfxsize = 14cm
\ \epsfbox{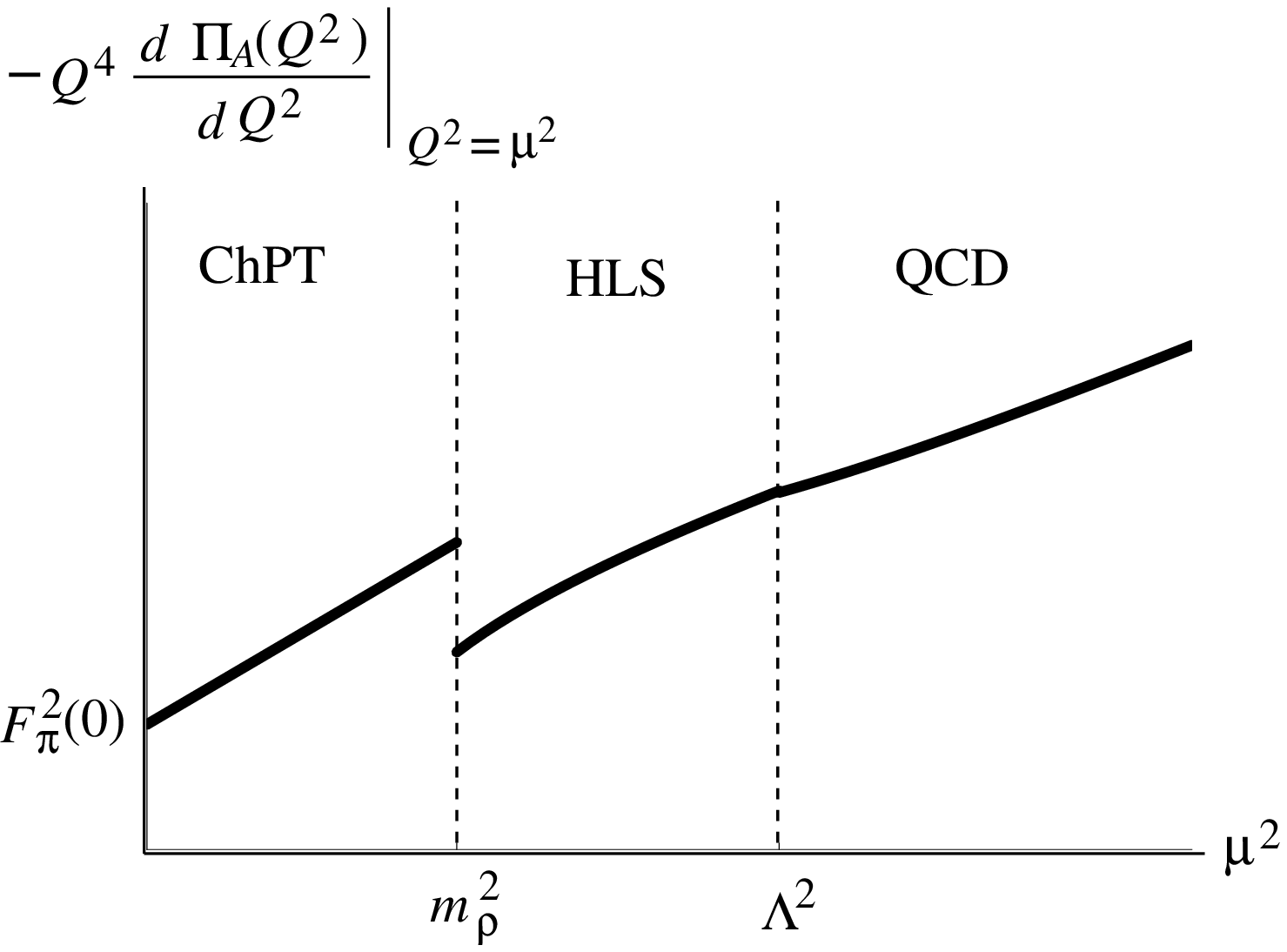}
\end{center}
\caption[Schematic view of matching]{%
Schematic view of our matching procedure:
In the region  ``QCD'' the solid line shows 
$\left. -Q^4 \frac{d \Pi_A(Q^2)}{d Q^2} \right\vert_{Q^2 = \mu^2}$
calculated from the OPE in QCD, while 
in the region ``HLS'' it shows 
$F_\pi^2(\mu)$ determined by the RGE (\ref{RGE for Fpi2}),
and in the region ``ChPT'' it shows 
$\left[ F_\pi^{(\pi)}(\mu) \right]^2$
in Eq.~(\ref{sol fpi2 for chpt}).
[See text for details.]
}
\label{fig:runA}
\end{figure}
Let us explain our basic strategy of the Wilsonian matching
using the axialvector current correlator $\Pi_A(Q^2)$ defined
in Eq.~(\ref{A V correlators 4}) or (\ref{A V correlators}).
We plot the $\mu$-dependence of
$\left. -Q^4 \frac{d \Pi_A(Q^2)}{d Q^2} \right\vert_{Q^2 = \mu^2}$
in Fig.~\ref{fig:runA}.
In the high energy region this current correlator can be calculated
from the operator product expansion (OPE) in QCD.
As we shall show in Sec.~\ref{ssec:MHUQ}
the $\mu$-dependence (for $N_c=3$) is determined by the main term $1 +
\alpha_s/\pi$ as $( \mu^2/8\pi^2) ( 1 + \alpha_s(\mu)/\pi )$,
which is plotted in the region indicated by ``QCD'' in 
Fig.~\ref{fig:runA}.

At the scale $\Lambda$ around 1\,GeV, which we call the matching
scale, we integrate out the quarks and gluons since they are not
well-defined degrees of freedom in the low energy region.
We assume that by
integrating out the quarks and gluons
we obtain the {\it bare Lagrangian} of the effective field theory, 
i.e., the HLS.
This Lagrangian
includes the hadrons lighter than the matching scale $\Lambda$
which are well-defined degrees of freedom in the low energy region.
Note that, as we discussed in 
Secs.~\ref{ssec:DEHLS} and \ref{ssec:QD},
for the consistency of the systematic derivative expansion
in the HLS the matching scale $\Lambda$ must be smaller than
the chiral scale $\Lambda_\chi = 4\pi F_\pi(\Lambda)/\sqrt{N_f}$ 
determined
from the {\it bare} parameter $F_\pi(\Lambda)$.
When the momentum is around the matching scale, $Q^2 \sim \Lambda^2$,
the current correlator is
well described by the tree contributions with including
${\cal O}(p^4)$ terms.
Then
we have
$\left. -Q^4 \frac{d \Pi_A^{\rm(HLS)}(Q^2)}{d Q^2} 
\right\vert_{Q^2 = \Lambda^2} = F_\pi^2(\Lambda)$,
where $F_\pi^2(\Lambda)$ is the {\it bare parameter} of the Lagrangian
corresponding to the $\pi$ decay constant.

The current correlator below $\Lambda$ is calculated from the
{\it bare} HLS Lagrangian defined at $\Lambda$ by including
loop corrections with the effect of quadratic divergences.
Then, we expect that 
$\left. -Q^4 \frac{d \Pi_A^{\rm(HLS)}(Q^2)}{d Q^2} 
\right\vert_{Q^2 = \mu^2}$ is dominated by
$F_\pi^2(\mu)$.
The running of $F_\pi^2(\mu)$ is determined by the RGE 
(\ref{RGE for Fpi2}), and it is shown by the line in the region
indicated by ``HLS'' in Fig.~\ref{fig:runA}.
The important point here is that the bare parameter $F_\pi^2(\Lambda)$
is determined by matching it with the current correlator in OPE,
as we shall show in Sec.~\ref{ssec:MHUQ}.
In the present procedure we equate
$\left. -Q^4 \frac{d \Pi_A^{\rm(QCD)}(Q^2)}{d Q^2} 
\right\vert_{Q^2 = \Lambda^2}$ with $F_\pi^2(\Lambda)$, 
so that the line
in the region ``QCD'' connects with the line in the region
``HLS''.

At the scale of $\rho$ mass $m_\rho$, $\rho$ decouples.
Then, we expect that 
$\left. -Q^4 \frac{d \Pi_A(Q^2)}{d Q^2} 
\right\vert_{Q^2 = \mu^2}$ is dominated by
$\left[ F_\pi^{(\pi)}(\mu) \right]^2$
which runs by the effect of 
quadratic divergence from the $\pi$-loop
effect alone as shown in Eq.~(\ref{sol fpi2 for chpt}).
The solid line in the region indicated by ``ChPT''
shows the $\mu$-dependence of $\left[ F_\pi^{(\pi)}(\mu) \right]^2$.
Since
the ordinary ChPT without HLS
is not applicable around the $\rho$ mass scale $m_\rho$,
the solid line in the ``ChPT'' does not connect with the one
in the ``HLS''.
The difference is understood as the effect of the finite 
renormalization at the scale $\mu = m_\rho$
as shown in Eq.~(\ref{rel: Fp mu Fp mr}) or
Eq.~(\ref{finite remalization 0}).
Through the procedure, which we called the ``Wilsonian matching'' in
Ref.~\cite{HY:matching}, 
the physical quantity $F_\pi^2(0)$
is related to the current correlator calculated in the OPE in QCD.

In Sec.~\ref{ssec:MHUQ},
we introduce the Wilsonian matching conditions 
which are derived by matching the vector and axialvector correlators
in the HLS with those obtained by the OPE in QCD.
Then, we determine the bare parameters of the HLS using the Wilsonian
matching conditions in Sec.~\ref{ssec:DBPHL}.
Physical predictions are made in Sec.~\ref{ssec:RWM} for
$N_f=3$ QCD which is close to the real world.
In Sec.~\ref{ssec:PQNf2}, 
we consider QCD with $N_f=2$
to show how the $N_f$-dependences of the physical quantities
appear.
Finally, in Sec.~\ref{ssec:SR},
we study the sum rules related to the vector and axialvector current
correlators.

\subsection{Matching HLS with the underlying QCD}
\label{ssec:MHUQ}

As is well known, the parameters in the bare theory can be
identified with that at the 
cutoff scale in the Wilsonian renormalization scheme.
In this subsection following Ref.~\cite{HY:matching}
we will present a way to determine the bare parameters of the HLS
by matching the axialvector and vector 
current correlators in the HLS with those obtained by
the operator product expansion (OPE) in QCD.
This is contrasted with the usual renormalization where the bare
theory is never referred to.

Let us start with the two-point functions of the 
non-singlet axialvector and vector currents:
\begin{eqnarray}
&&
i \int d^4 x e^{ipx}
\left\langle 0 \left\vert T\, J_{5\mu}^a (x) J_{5\nu}^b (0)
\right\vert 0 \right\rangle 
= \delta^{ab} \left( p_\mu p_\nu - g_{\mu\nu} p^2 \right)
\Pi_A (Q^2) \ ,
\nonumber\\
&&
i \int d^4 x e^{ipx}
\left\langle 0 \left\vert T\, J_{\mu}^a (x) J_{\nu}^b (0)
\right\vert 0 \right\rangle 
= \delta^{ab} \left( p_\mu p_\nu - g_{\mu\nu} p^2 \right)
\Pi_V (Q^2) \ ,
\label{A V correlators}
\end{eqnarray}
where $Q^2 = -p^2$.
In the HLS these two-point functions are 
well described by the tree contributions with including
${\cal O}(p^4)$ terms 
when the momentum is around the matching scale, $Q^2 \sim \Lambda^2$.
By combining ${\cal O}(p^4)$ terms in Eq.~(\ref{Lag: z terms}) with
the leading terms in Eq.~(\ref{leading Lagrangian})
the correlators in the HLS are given by~\cite{HY:matching}
\begin{eqnarray}
\Pi_A^{\rm(HLS)}(Q^2) &=&
\frac{F_\pi^2(\Lambda)}{Q^2} - 2 z_2(\Lambda)
\ ,
\label{Pi A HLS}
\\
\Pi_V^{\rm(HLS)}(Q^2) &=&
\frac{
  F_\sigma^2(\Lambda)
}{
  M_\rho^2(\Lambda) + Q^2
} 
\left[ 1 - 2 g^2(\Lambda) z_3(\Lambda) \right]
- 2 z_1(\Lambda)
\ ,
\label{Pi V HLS}
\end{eqnarray}
where we defined
\begin{equation}
M_\rho^2(\Lambda) \equiv g^2(\Lambda) F_\sigma^2(\Lambda)
\ .
\label{on-shell cond 5}
\end{equation}
The same correlators are evaluated by the OPE 
up until ${\cal O}(1/Q^6)$~\cite{SVZ:1,SVZ:2}:
\begin{eqnarray}
\Pi_A^{\rm(QCD)}(Q^2) &=& \frac{1}{8\pi^2}
\left( \frac{N_c}{3} \right)
\Biggl[
  - \left( 
      1 +  \frac{3(N_c^2-1)}{8N_c}\, \frac{\alpha_s}{\pi} 
  \right) \ln \frac{Q^2}{\mu^2}
\nonumber\\
&& \qquad
  {}+ \frac{\pi^2}{N_c} 
    \frac{
      \left\langle 
        \frac{\alpha_s}{\pi} G_{\mu\nu} G^{\mu\nu}
      \right\rangle
    }{ Q^4 }
  {}+ \frac{\pi^3}{N_c} \frac{96(N_c^2-1)}{N_c^2}
    \left( \frac{1}{2} + \frac{1}{3N_c} \right)
    \frac{\alpha_s \left\langle \bar{q} q \right\rangle^2}{Q^6}
\Biggr]
\ ,
\label{Pi A OPE}
\\
\Pi_V^{\rm(QCD)}(Q^2) &=& \frac{1}{8\pi^2}
\left( \frac{N_c}{3} \right)
\Biggl[
  - \left( 
      1 +  \frac{3(N_c^2-1)}{8N_c}\, \frac{\alpha_s}{\pi} 
  \right) \ln \frac{Q^2}{\mu^2}
\nonumber\\
&& \qquad
  {}+ \frac{\pi^2}{N_c} 
    \frac{
      \left\langle 
        \frac{\alpha_s}{\pi} G_{\mu\nu} G^{\mu\nu}
      \right\rangle
    }{ Q^4 }
  {}- \frac{\pi^3}{N_c} \frac{96(N_c^2-1)}{N_c^2}
    \left( \frac{1}{2} - \frac{1}{3N_c} \right)
    \frac{\alpha_s \left\langle \bar{q} q \right\rangle^2}{Q^6}
\Biggr]
\ ,
\label{Pi V OPE}
\end{eqnarray}
where $\mu$ is the renormalization scale of QCD
and we
wrote the $N_c$-dependences explicitly
(see, e.g., Ref.~\cite{Bardeen-Zakharov}).

We require that current correlators in the HLS
in Eqs.~(\ref{Pi A HLS}) and (\ref{Pi V HLS})
can be matched with those in QCD in 
Eqs.~(\ref{Pi A OPE}) and (\ref{Pi V OPE}).
Of course,
this matching cannot be made for any value of $Q^2$, 
since the $Q^2$-dependences of the current correlators 
in the HLS are completely
different from those in the OPE:
In the HLS the derivative expansion (in {\it positive} power of $Q$)
is used, and the expressions for
the current correlators are valid in the low energy region.
The OPE, on the other hand, is an asymptotic expansion
(in {\it negative} power of $Q$), and it is
valid in the high energy region.
Since we calculate the current correlators in the HLS including the
first non-leading order [${\cal O}(p^4)$], we expect that we can match
the correlators with those in the OPE up until the first derivative.
Note that both $\Pi_A^{\rm(QCD)}$ and $\Pi_V^{\rm(QCD)}$ explicitly
depend on $\mu$.~\footnote{
  It should be noticed that the
  $\alpha_s/\pi$ term and $\alpha_s \left\langle \bar{q} q
  \right\rangle^2$ term in the right-hand-sides of the
  matching conditions [Eqs.~(\ref{match z}), (\ref{match A}) and
  (\ref{match V})]
  depend on the renormalization point $\mu$ of
  QCD, 
  and that those generate a small dependence of the bare parameters of
  the HLS on $\mu$.
  This $\mu$ is taken to be the matching scale in the QCD sum rule
  shown in Refs.~\cite{SVZ:1,SVZ:2}.  
  Here we take $\mu$ to be equal to the
  matching scale $\Lambda$.
}
Such dependences are assigned to the parameters $z_2(\Lambda)$ and
$z_1(\Lambda)$.
This situation is similar to that for the parameters 
$H_i$ in the ChPT~\cite{Gas:84,Gas:85a} [see, e.g.,
Eq.~(\ref{p4:ChPT})].
However, the difference between two correlators has no
explicit dependence on $\mu$.
Thus our first Wilsonian matching condition is given
by~\cite{HY:matching}
\begin{eqnarray}
&&
\frac{F_\pi^2(\Lambda)}{\Lambda^2} -
\frac{F_\sigma^2(\Lambda)}{\Lambda^2 + M_\rho^2(\Lambda)}
\left[ 1 - 2 g^2(\Lambda) z_3(\Lambda) \right]
- 2 \left[ z_2(\Lambda) - z_1(\Lambda) \right]
\nonumber\\
&& \qquad
=
\frac{4\pi(N_c^2-1)}{N_c^2}
\frac{\alpha_s \left\langle \bar{q} q \right\rangle^2}{\Lambda^6}
\ .
\label{match z}
\end{eqnarray}

We also require that the first derivative of $\Pi_A^{\rm(HLS)}$ in 
Eq.~(\ref{Pi A HLS}) matches that of $\Pi_A^{\rm(QCD)}$ in 
Eq.~(\ref{Pi A OPE}), and similarly for $\Pi_V$'s in 
Eqs.~(\ref{Pi V HLS}) and (\ref{Pi V OPE}).
This requirement gives the following two Wilsonian matching
conditions~\cite{HY:matching}:~\footnote{%
  One might think that there appear corrections from $\rho$ 
  and/or $\pi$ loops in the left-hand-sides of 
  Eqs.~(\ref{match A}) and (\ref{match V}).
  However, such corrections are of higher order in the present counting
  scheme, and thus we neglect them here.
}
\begin{eqnarray}
&&
\frac{F_\pi^2(\Lambda)}{\Lambda^2} 
= \left. - Q^2 \frac{d}{dQ^2} \Pi_A^{\rm(QCD)}(Q^2) 
  \right\vert_{Q^2 = \Lambda^2}
\nonumber\\
&& \qquad\quad
= \frac{1}{8\pi^2} \left( \frac{N_c}{3} \right) (1+\delta_A) \ ,
\nonumber\\
&&\delta_A \equiv
  \frac{3(N_c^2-1)}{8N_c}\, \frac{\alpha_s}{\pi} 
  + \frac{2\pi^2}{N_c} 
    \frac{
      \left\langle 
        \frac{\alpha_s}{\pi} G_{\mu\nu} G^{\mu\nu}
      \right\rangle
    }{ \Lambda^4 }
\nonumber\\
&& \qquad\qquad\qquad\qquad\qquad
  {}+ \frac{288\pi(N_c^2-1)}{N_c^3}
    \left( \frac{1}{2} + \frac{1}{3N_c} \right)
    \frac{\alpha_s \left\langle \bar{q} q \right\rangle^2}{\Lambda^6}
\ ,
\label{match A}
\\
&&
\frac{F_\sigma^2(\Lambda)}{\Lambda^2} 
\frac{\Lambda^4 \left[ 1 - 2 g^2(\Lambda) z_3(\Lambda) \right]
}{\left[ \Lambda^2 + M_\rho^2(\Lambda) \right]^2}
= \left. - Q^2 \frac{d}{dQ^2} \Pi_V^{\rm(QCD)}(Q^2) 
  \right\vert_{Q^2 = \Lambda^2}
\nonumber\\
&& \qquad\quad
= \frac{1}{8\pi^2} \left( \frac{N_c}{3} \right)(1+\delta_V) \ ,
\nonumber
\\
&&\delta_V \equiv
  \frac{3(N_c^2-1)}{8N_c}\, \frac{\alpha_s}{\pi} 
  + \frac{2\pi^2}{N_c} 
    \frac{
      \left\langle 
        \frac{\alpha_s}{\pi} G_{\mu\nu} G^{\mu\nu}
      \right\rangle
    }{ \Lambda^4 }
\nonumber\\
&& \qquad\qquad\qquad\qquad\qquad
  {}- \frac{288\pi(N_c^2-1)}{N_c^3}
    \left( \frac{1}{2} - \frac{1}{3N_c} \right)
    \frac{\alpha_s \left\langle \bar{q} q \right\rangle^2}{\Lambda^6}
\ .
\label{match V}
\end{eqnarray}

The above three equations (\ref{match z}), (\ref{match A}) and 
(\ref{match V}) are the Wilsonian matching conditions
proposed in Ref.~\cite{HY:matching}.
These determine several bare parameters of the HLS without much
ambiguity.  Especially, the second condition (\ref{match A})
determines the ratio $F_\pi(\Lambda)/\Lambda$ directly from QCD.
It should be
noticed that the above Wilsonian matching conditions
determine the absolute value and the 
explicit dependence of bare parameters of HLS
on the parameters of underlying QCD such as $N_c$ (not just
scaling properties in the large $N_c$ limit) and $\Lambda_{\rm QCD}$,
which would never have been obtained without matching and in fact
has never been achieved for the EFT before.

Now we discuss the large $N_c$ behavior of the bare parameters:
As we will show explicitly in Sec.~\ref{ssec:VMLNQ},
it is natural to assume that the matching scale $\Lambda$ has
no large $N_c$-dependence.
Then, the condition (\ref{match A}), together with the fact
that each term in 
$\delta_A$ in Eq.~(\ref{match A}) has only small 
$N_c$-dependence~\footnote{%
  Note that $\alpha_s$ scales as $1/N_c$ in the large $N_c$
  counting, and that both 
  $\langle \frac{\alpha}{\pi} G_{\mu\nu} G^{\mu\nu} \rangle$
  and $\langle \bar{q} q \rangle$ scale as $N_c$.
},
shows that
the bare parameter $F_\pi^2(\Lambda)$ scales as $N_c$.
This is consistent with the ordinary large $N_c$ counting of the
on-shell $\pi$ decay constant, $F_\pi^2(0) \sim N_c$.

In the Wilsonian matching condition (\ref{match V})
it is plausible to assume that the bare $\rho$ mass
parameter $M_\rho(\Lambda)$ does not scale in the large $N_c$
since the on-shell $\rho$ mass $m_\rho$ does not.
The second term inside the square bracket in
the numerator of the left-hand-side,
$g^2(\Lambda) z_3(\Lambda)$,
cannot increase with increasing
$N_c$ for the consistency with the chiral counting, and then
we require that this does not have the large $N_c$ scaling.
These scaling properties
together with the fact that the right-hand-side of 
Eq.~(\ref{match V}) scales as $N_c$ imply that the bare 
parameter $F_\sigma^2(\Lambda)$ scales as $N_c$, and then
the bare parameter $a(\Lambda)=F_\sigma^2(\Lambda)/F_\pi^2(\Lambda)$
does not have large $N_c$ dependence.

Noting that $M_\rho^2(\Lambda) = a(\Lambda) g^2(\Lambda)
F_\pi^2(\Lambda)$, we see, from the scaling properties of
$F_\pi^2(\Lambda)$, $a(\Lambda)$ and $M_\rho^2(\Lambda)$ determined
above,
that the HLS gauge coupling $g(\Lambda)$ scales as $1/\sqrt{N_c}$
which is consistent with the fact that $g$ is the coupling of
the interaction among three $\rho$ mesons.
This scaling property of $g(\Lambda)$ with the requirement that
$g^2(\Lambda) z_3(\Lambda)$ does not have large $N_c$ dependence
leads to $z_3(\Lambda) \sim {\cal O}(N_c)$.
Finally, in 
the Wilsonian matching condition (\ref{match z})
the first and second terms in the left-hand-side as well as
the right-hand-side scale as $N_c$, so that
$z_2(\Lambda)-z_1(\Lambda)$ also scales as $N_c$.

To summarize the Wilsonian matching conditions lead to the
following large $N_c$ scaling properties of the bare parameters
of the HLS Lagrangian:
\begin{eqnarray}
&& F_\pi(\Lambda) \ \sim \ {\cal O}\left(\sqrt{N_c}\right) 
  \ , \nonumber\\
&& a(\Lambda) \ \sim \ {\cal O}(1) \ , \nonumber\\
&& g(\Lambda) \ \sim \ {\cal O}\left(1/\sqrt{N_c}\right) 
  \ , \nonumber\\
&& z_3(\Lambda) \ \sim \ {\cal O}(N_c) \ , \nonumber\\
&& z_2(\Lambda) - z_1(\Lambda) \ \sim \ {\cal O}(N_c) \ .
\end{eqnarray}
Note that the above scaling properties under the large $N_c$
can be also obtained by counting the number of traces in the
Lagrangian as was done in Ref.~\cite{Gas:85a} to determine
the scaling properties of the low-energy constants of the
ordinary chiral perturbation theory.

\subsection{Determination of the bare parameters of the HLS Lagrangian}
\label{ssec:DBPHL}

In this subsection we determine the bare parameters related to the
two-point functions of the axialvector and vector current correlators
from QCD through the Wilsonian matching conditions
shown in the previous subsection.

The right-hand-sides in Eqs.~(\ref{match z}), (\ref{match A}) and
(\ref{match V}) are directly determined from QCD.
First note that the matching scale $\Lambda$ must be smaller than the
mass of 
$a_1$ meson which is not included in our effective theory,
whereas $\Lambda$ has to be big enough for the OPE to be valid.
Here we use a typical value:
\begin{equation}
\Lambda = 1.1 \, \mbox{GeV} \ .
\end{equation}
In order to check the sensitivity of our result to the input value
we also study the cases for the following wide range of the values:
\begin{equation}
\Lambda = 1.0 \ \sim \  1.2\,\mbox{GeV} \ .
\label{matching scale}
\end{equation}

For definiteness of the proceeding analysis let us first 
determine the current correlators from the OPE.
For the value of the gluonic condensate
we use
\begin{eqnarray}
&&
\left\langle \frac{\alpha_s}{\pi} G_{\mu\nu} G^{\mu\nu}
\right\rangle = 0.012 \,\mbox{GeV}^4
\label{val GG}
\end{eqnarray}
shown in Ref.~\cite{SVZ:1,SVZ:2} as a typical value.
In Ref.~\cite{GL:PRep}
the value of quark condensate is estimated as~\footnote{
In the previous paper~\cite{HY:matching} 
we used the SVZ value~\cite{SVZ:1,SVZ:2}
$\langle \bar{q} q \rangle_{\rm 1\,GeV} = - 
(250\,\mbox{MeV})^3
$ 
and hence the numerical analysis here is slightly different from
the previous one, although consistent with it within the error.
}
\begin{eqnarray}
&&\left\langle \bar{q} q \right\rangle_{\rm 1\,GeV} = - 
\left(225\pm25\,\mbox{MeV}\right)^3
\ .
\label{val qq}
\end{eqnarray}
We use the center value and study the dependence of the result
on the quark condensate by including the error shown above.
There are some ambiguities for the value of 
$\Lambda_{\rm QCD}$ (see, e.g., Ref.~\cite{Buras}).
Here we use 
\begin{equation}
\Lambda_{\rm QCD} = 400\, \mbox{MeV} \ ,
\end{equation}
but again we also study the cases
\begin{equation}
\Lambda_{\rm QCD} = 300\ , 350\ , 400\ , 450 \mbox{MeV} \ ,
\end{equation}
to check the sensitivity of our result to the input value.
Furthermore,
we use the one-loop running to
estimate $\alpha_s(\Lambda)$ and 
$\left\langle \bar{q} q \right\rangle_\Lambda$:
\begin{eqnarray}
&&
  \alpha_s(\Lambda) = 
  \frac{4\pi}{ \beta_0 \ln( \Lambda^2/\Lambda^2_{\rm QCD} )}
  \ ,
\nonumber\\
&&
  \left\langle \bar{q} q \right\rangle_\Lambda =
  \left\langle \bar{q} q \right\rangle_{\rm 1\,GeV}
  \left(
    \frac{\alpha_s(\mbox{1\,GeV})}{\alpha_s(\Lambda)}
  \right)^{A/2}
\ ,
\end{eqnarray}
where
\begin{eqnarray}
&&
  \beta_0 = \frac{11N_c - 2 N_f}{3} \ ,
\nonumber\\
&&
  A = \frac{3C_2}{\beta_0} = \frac{9 (N_c^2-1)}{N_c (11N_c-2N_f)}
\ .
\end{eqnarray}
Note that our typical choice $\Lambda = 1.1\, \mbox{GeV}$ and $\Lambda_{\rm QCD}=400 \, \mbox{MeV}$ corresponds to
\begin{equation}
\alpha_s(\Lambda=1.1 \, \mbox{GeV}; \Lambda_{\rm QCD}=400\, \mbox{MeV})
\simeq 0.69 \ .
\label{Lambdatoalpha}
\end{equation} 
{}From the above inputs we evaluate the current correlators in the
OPE, and have for $N_c=N_f=3$:
\begin{equation}
\delta_{A/V} = 0.220+0.054 + (0.089)/(- 0.057)
\sim 0.363/0.217
\label{delta}
\end{equation}
for the respective terms  
$\alpha_s/\pi$,
$\frac{2\pi^2}{3} 
       \frac{\left\langle \frac{\alpha_s}{\pi} 
                G_{\mu\nu} G^{\mu\nu} \right\rangle }
             {\Lambda^4}$ and
$(\pi^3 \frac{1408}{27} 
       \frac{ \alpha_s \langle \bar{q} q \rangle^2}
             { \Lambda^6 })/ 
(-\pi^3 \frac{896}{27} 
       \frac{ \alpha_s \langle \bar{q} q \rangle^2}
             { \Lambda^6 })$, 
appearing in the right-hand-sides of the Wilsonian matching
conditions (\ref{match A}) and (\ref{match V}).
It implies 
that the terms $1$ and $\frac{\alpha_s}{\pi}$ (first
term of $\delta_{A/V}$)
give dominant contributions over the gluonic and the 
quark condensate terms in the 
right-hand-sides of Eqs.~(\ref{match A}) and (\ref{match V}).

We also list 
in Table~\ref{tab:cutOPE} the results for other parameter
choices 
of $\Lambda_{\rm QCD}$ and
$\Lambda$ together with the ambiguities
coming from that of the quark condensate shown in 
Eq.~(\ref{val qq}).
While the gluonic condensate gives very small correction
for any choice of the matching scale,
the quark condensate gives a non-negligible
correction 
for small matching scale ($\Lambda \simeq 1\,\mbox{GeV}$).
\begin{table}[htbp]
\begin{center}
\begin{tabular}{|c|c||c|c|c|c|}
\hline
 $\Lambda_{\rm QCD}$ & $\Lambda$ & $\alpha_s/\pi$ 
   & ($GG$) & 
    ($\bar{q}q$-A) & ($\bar{q}q$-V)
 \\
\hline
 $0.30$
 & $1.00$ & $0.185$ & $0.079$ 
  & $0.122\pm0.081$ & $-0.077\pm0.052$ \\
 & $1.10$ & $0.171$ & $0.054$ 
  & $0.068\pm0.045$ & $-0.043\pm0.029$ \\
 & $1.20$ & $0.160$ & $0.038$ 
  & $0.040\pm0.027$ & $-0.026\pm0.017$ \\
\hline
 $0.35$
 & $1.00$ & $0.212$ & $0.079$ 
  & $0.140\pm0.093$ & $-0.089\pm0.059$ \\
 & $1.10$ & $0.194$ & $0.054$ 
  & $0.078\pm0.052$ & $-0.050\pm0.033$ \\
 & $1.20$ & $0.180$ & $0.038$ 
  & $0.046\pm0.031$ & $-0.029\pm0.020$ \\
\hline
 $0.40$
 & $1.00$ & $0.243$ & $0.079$ 
  & $0.160\pm0.107$ & $-0.102\pm0.068$ \\
 & $1.10$ & $0.220$ & $0.054$ 
  & $0.089\pm0.060$ & $-0.057\pm0.038$ \\
 & $1.20$ & $0.202$ & $0.038$ 
  & $0.053\pm0.035$ & $-0.033\pm0.022$ \\
\hline
 $0.45$
 & $1.00$ & $0.278$ & $0.079$ 
  & $0.183\pm0.122$ & $-0.117\pm0.078$ \\
 & $1.10$ & $0.249$ & $0.054$ 
  & $0.102\pm0.068$ & $-0.065\pm0.043$ \\
 & $1.20$ & $0.227$ & $0.038$ 
  & $0.060\pm0.040$ & $-0.038\pm0.026$ \\
\hline
\end{tabular}\\
\end{center}
\caption[Terms of the current correlators from the OPE.]{%
Values of the terms of the axialvector and vector current
correlators derived from the OPE.
Values in the fourth column indicated by ($GG$) are the values
of 
$\frac{2\pi^2}{3} 
       \frac{\left\langle \frac{\alpha_s}{\pi} 
                G_{\mu\nu} G^{\mu\nu} \right\rangle }
             {\Lambda^4}$,
and those in the fifth [indicated by ($\bar{q}q$-A)] and
the sixth [indicated by ($\bar{q}q$-V)] columns
are of 
$\pi^3 \frac{1408}{27} 
       \frac{ \alpha_s \langle \bar{q} q \rangle^2}
             { \Lambda^6 }$
and
  $-\pi^3 \frac{896}{27} 
       \frac{ \alpha_s \langle \bar{q} q \rangle^2}
             { \Lambda^6 }$,            
respectively.
Units of $\Lambda_{\rm QCD}$ and $\Lambda$ are GeV.
Errors in fifth and sixth columns are from the error 
in the quark condensate 
$\langle\bar{q}q\rangle=-(225\pm25\,\mbox{MeV})^3$ shown in
Eq.~(\ref{val qq}).
}\label{tab:cutOPE}
\end{table}

Now that we have determined the current correlators in the OPE,
we can determine the bare parameters of the HLS through the 
Wilsonian matching conditions.
Especially, the Wilsonian matching condition (\ref{match A})
determines directly the value of
the bare $\pi$ decay constant $F_\pi(\Lambda)$.
Before discussing details, we here give a rough estimation to get
an essential point of our analysis: 
\begin{eqnarray}
F_\pi^2(\Lambda) 
&=&
\frac{\Lambda^2}{(4\pi)^2} \, \frac{N_c}{3} \,
2 \left( 1 + \delta_A \right)
\nonumber\\
&\sim&
3 \left(\frac{\Lambda}{4\pi}\right)^2 \, \frac{N_c}{3} \ ,
\label{valueFpi}
\end{eqnarray}
where $\delta_A$ was estimated in Eq. (\ref{delta}) for $N_c=N_f=3$
and very roughly
\begin{eqnarray}
\delta_A 
\sim 0.5 \ .
\end{eqnarray}
Note again that each term in Eq.~(\ref{delta}) for $\delta_A$ is 
rather independent of $N_c$.

First of all 
Eq.~(\ref{valueFpi}) implies the derivative expansion parameter
can be very small in the large $N_c$ limit (with fixed $N_f$): 
\begin{equation}
N_f \left(\frac{\Lambda}{4\pi F_\pi(\Lambda)}\right)^2
\sim
\frac{N_f}{3} \left(\frac{3}{N_c} \right)
=\frac{N_f}{N_c} \ll 1 \quad (N_c\gg 1)\ . 
\end{equation}
As we discussed in the previous section, we make the systematic
expansion in the large $N_c$ limit, and extrapolate the results to the
real world.
In QCD with $N_c=N_f=3$ the above expansion parameter becomes of order
one, so that one might think that the systematic expansion breaks
down.
However, 
as can be seen in, e.g., Eq.~(\ref{ren:Fp}) with $a\sim1$,
the quadratically divergent loop contributions to $F_\pi^2$ get an
extra factor $1/2$ due to the additional $\rho$ loop and hence the
loop 
expansion would be valid up till
\begin{equation}
\Lambda \sim
\frac{ 4\pi F_\pi(\Lambda) }{ \sqrt{ N_f/2} }
\ ,
\end{equation}
or
\begin{equation}
\frac{N_f}{2 N_c} \sim 1 \ .
\end{equation}
Furthermore,
as we will show below in this section, the analysis based on
the systematic expansion reproduces the experiment in good agreement.
This shows that the extrapolation of the systematic expansion from the
large $N_c$ limit to the real world works very well.

Now, by choosing the matching scale as
$\Lambda = 1.1 \,\mbox{GeV}$, 
or $\frac{\Lambda}{4\pi} \simeq 86.4 \, \mbox{MeV}$,
the value of 
$F_\pi(\Lambda)$ is estimated as
\begin{equation}
F_\pi^2(\Lambda) \sim 3\, 
(86.4 \, \mbox{MeV})^2 \sim (150 \,\mbox{MeV})^2\ .
\label{Fpi Lambda rough estimate}
\end{equation}
Then the Wilsonian matching predicts 
$F_\pi^2(\Lambda)$ in terms of the QCD
parameters and the value definitely disagrees with the on-shell 
value $86.4 \pm9.7\, \mbox{MeV}$
in the chiral limit~\cite{Gas:84,Gas:85b}.
Were it not for the quadratic divergence, we would
have met with a serious discrepancy between the QCD prediction and the 
physical value!  How does the quadratic divergence save the situation?
The key is the Wilsonian 
RGE derived in Sec.~\ref{ssec:RGEWS} which incorporated
 quadratic divergence
(as well as logarithmic one) for the running of $F_\pi^2$.
To perform a crude estimate let us neglect the effect
of logarithmic divergence (by taking 
$g(\Lambda) \rightarrow 0$)
and include the effect of quadratic
divergence only in the RGE for $F_\pi^2$ in Eq.~(\ref{RGE for Fpi2}).
Furthermore, as it turns out that the Wilsonian matching 
implies the bare value $a(\Lambda) \simeq 1$, we take $a=1$, 
which is the fixed point of RGE,
so that the analytical solution of RGE becomes very simple:
$
F_\pi^2(m_\rho) = F_\pi^2(\Lambda) -
\frac{N_f}{2(4\pi)^2} \left( \Lambda^2 - m_\rho^2 \right)
$. 
This together with the relation (\ref{rel: Fp 0 Fp mr})
yields the approximate relation between the bare parameter
$F_\pi^2(\Lambda)$ and the on-shell $\pi$ decay constant
$F_\pi^2(0)$ as
\begin{eqnarray}
F_\pi^2(0) 
&\sim& F_\pi^2(\Lambda) - \frac{N_f}{2(4\pi)^2} \Lambda^2 
\nonumber\\
&\sim&
\frac{\Lambda^2}{8\pi^2} 
\left[ \frac{N_c}{3} \left( 1 + \delta_A \right) - \frac{N_f}{4} \right]
\nonumber\\
&\sim&
1.5 \,
\left(\frac{\Lambda}{4\pi}\right)^2 \sim 
\frac{1}{2} F_\pi^2 (\Lambda) 
\sim (100\,\mbox{MeV})^2 \
\ ,
\label{Fpi 0 rough estimate}
\end{eqnarray}
where we adopted $N_c=N_f=3$ and $\delta_A \sim 0.5$ to obtain the
last line.
Then, 
the on-shell $\pi$ decay constant $F_\pi(0)$ is 
now
close to the value 
$F_\pi(0) = 86.4\pm9.7\,\mbox{MeV}$.
The small deviation from $86.4 \, \mbox{MeV}$ will be resolved by 
taking account of  the logarithmic correction with $g(\Lambda)\ne 0$ 
and the correction by $a(\Lambda) \ne 1$ 
(and more precise value $\delta_A \sim 0.363$) for the  
realistic case $N_c=N_f=3$. At any rate
this already shows that the Wilsonian matching works 
well and quadratic 
divergence palys a vital role.

Let us now determine the precise value of $F_\pi(\Lambda)$
for given 
values of $\Lambda_{\rm QCD}$ and the matching scale $\Lambda$
in the case of $N_c =N_f =3$.
We list the resultant values of $F_\pi(\Lambda)$ 
obtained from the Wilsonian matching condition (\ref{match A})
together with the ambiguity from that of the quark condensate
$\langle\bar{q}q\rangle=-(225\pm25\,\mbox{MeV})^3$
in Table~\ref{tab:cutval0}.
\begin{table}[htbp]
\begin{center}
\begin{tabular}{|c|c||c|}
\hline
 $\Lambda_{\rm QCD}$ & $\Lambda$  & $F_\pi(\Lambda)$ \\
\hline
 $0.30$
 & $1.00$ & $0.132\pm0.004$ \\
 & $1.10$ & $0.141\pm0.002$ \\
 & $1.20$ & $0.150\pm0.002$ \\
\hline
 $0.35$
 & $1.00$ & $0.135\pm0.004$ \\
 & $1.10$ & $0.143\pm0.003$ \\
 & $1.20$ & $0.152\pm0.002$ \\
\hline
\end{tabular}
\begin{tabular}{|c|c||c|}
\hline
 $\Lambda_{\rm QCD}$ & $\Lambda$ & $F_\pi(\Lambda)$ \\
\hline
 $0.40$
 & $1.00$ & $0.137\pm0.005$ \\
 & $1.10$ & $0.145\pm0.003$ \\
 & $1.20$ & $0.154\pm0.002$ \\
\hline
 $0.45$
 & $1.00$ & $0.140\pm0.006$ \\
 & $1.10$ & $0.147\pm0.004$ \\
 & $1.20$ & $0.155\pm0.002$ \\
\hline
\end{tabular}\\
\end{center}
\caption[Values of the bare pion decay constant for $N_c=N_f=3$.]{%
Values of the bare $\pi$ decay constant $F_\pi(\Lambda)$ determined
through the Wilsonian matching condition (\ref{match A})
for given $\Lambda_{\rm QCD}$ and the matching scale $\Lambda$.
Units of $\Lambda_{\rm QCD}$, $\Lambda$ and $F_\pi(\Lambda)$ are
GeV.
Note that
error of $F_\pi(\Lambda)$ is from the error 
in the quark condensate 
$\langle\bar{q}q\rangle=-(225\pm25\,\mbox{MeV})^3$ shown in
Eq.~(\ref{val qq}).
}\label{tab:cutval0}
\end{table}
This shows that the bare $\pi$ decay constant is determined
from the matching condition without much ambiguity:
It is almost determined by the main term $1+\alpha_s/\pi$ in 
the right-hand-side of Eq.~(\ref{match A}), and the
ambiguity of the quark condensate 
$\langle\bar{q}q\rangle=-(225\pm25\,\mbox{MeV})^3$ shown in
Eq.~(\ref{val qq}) does not affect to the bare $\pi$ decay constant
very much.

There are four parameters $a(\Lambda)$,
$g(\Lambda)$, $z_3(\Lambda)$ and $z_2(\Lambda)-z_1(\Lambda)$
other than $F_\pi(\Lambda)$, 
which are
relevant to the low energy phenomena related to two correlators
analyzed in the previous subsection.~\footnote{%
  As we noted in the previous subsection,
  although each of $z_1(\Lambda)$ and $z_2(\Lambda)$ depends on the
  renormalization point $\mu$ of QCD,
  the difference $z_2(\Lambda)-z_1(\Lambda)$ does not.
  Actually,
  $z_2(\Lambda)+z_1(\Lambda)$ corresponds to the parameter
  $H_i$ in the ChPT~\cite{Gas:84,Gas:85a} 
  [see $H_1$ of Eq.~(\ref{rel HLS ChPT tree})].
  Thus, the difference $z_2(\Lambda)-z_1(\Lambda)$ is relevant to the
  low energy phenomena, while
  $z_2(\Lambda)+z_1(\Lambda)$ is irrelevant.
}
We have already used 
one Wilsonian matching condition~(\ref{match A})
to determine one of the bare parameters
$F_\pi(\Lambda)$ for a given
matching scale $\Lambda$.
The remaining two Wilsonian matching
conditions in Eqs.~(\ref{match z}) and (\ref{match V})
are not enough to determine other four relevant bare
parameters.  
We therefore use the on-shell pion decay constant
$F_\pi(0)=86.4\pm9.7$\,MeV estimated
in the chiral limit~\cite{Gas:84,Gas:85a,Gas:85b} and the
$\rho$ mass $m_\rho = 771.1$\,MeV as inputs:
We chose $a(\Lambda)$ and $g(\Lambda)$ which,
combined with $F_\pi(\Lambda)$ determined from the Wilsonian matching
condition~(\ref{match A}),
reproduce $F_\pi(0)$ and $m_\rho$ through the Wilsonian RGEs 
in Eqs.~(\ref{RGE for Fpi2}), (\ref{RGE for a}) and 
(\ref{RGE for g2}).
Then, we use the matching condition~(\ref{match V}) to determine
$z_3(\Lambda)$.
Finally  $z_2(\Lambda) - z_1(\Lambda)$ is fixed by the matching
condition~(\ref{match z}).

The resultant values of five bare parameters of the HLS are shown
in Tables~\ref{tab:WM bare} and \ref{tab:WM bare 2}
for $\Lambda = 1.0$, $1.1$ and $1.2$\,GeV.
\begin{table}[htbp]
\begin{center}
\begin{tabular}{|c|c||c|c|c|}
\hline
 $\Lambda_{\rm QCD}$ & $\Lambda$ & $F_\pi(\Lambda)$ 
  & $a(\Lambda)$ & $g(\Lambda)$ \\
\hline
 $0.30$
 & $1.00$ & $0.132\pm0.004$ & $1.41\pm0.29\pm0.16$
 & $4.05\pm0.16\pm0.01$ \\
 & $1.10$ & $0.141\pm0.002$ & $1.49\pm0.30\pm0.11$
 & $3.68\pm0.11\pm0.00$ \\
 & $1.20$ & $0.150\pm0.002$ & $1.49\pm0.30\pm0.08$
 & $3.42\pm0.09\pm0.00$ \\
\hline
 $0.35$
 & $1.00$ & $0.135\pm0.004$ & $1.32\pm0.28\pm0.18$
 & $4.06\pm0.18\pm0.03$ \\
 & $1.10$ & $0.143\pm0.003$ & $1.41\pm0.29\pm0.12$
 & $3.68\pm0.12\pm0.01$ \\
 & $1.20$ & $0.152\pm0.002$ & $1.42\pm0.29\pm0.09$
 & $3.42\pm0.10\pm0.00$ \\
\hline
 $0.40$
 & $1.00$ & $0.137\pm0.005$ & $1.22\pm0.28\pm0.21$
 & $4.09\pm0.20\pm0.06$ \\
 & $1.10$ & $0.145\pm0.003$ & $1.33\pm0.28\pm0.14$
 & $3.69\pm0.13\pm0.02$ \\
 & $1.20$ & $0.154\pm0.002$ & $1.34\pm0.28\pm0.09$
 & $3.43\pm0.10\pm0.01$ \\
\hline
 $0.45$
 & $1.00$ & $0.140\pm0.006$ & $1.10\pm0.28\pm0.24$
 & $4.13\pm0.22\pm0.10$ \\
 & $1.10$ & $0.147\pm0.004$ & $1.23\pm0.27\pm0.16$
 & $3.71\pm0.14\pm0.03$ \\
 & $1.20$ & $0.155\pm0.002$ & $1.26\pm0.26\pm0.10$
 & $3.44\pm0.11\pm0.01$ \\
\hline
\end{tabular}
\end{center}
\caption[Bare parameters of the HLS (1)]{%
Leading order parameters of the HLS 
at $\mu=\Lambda$ for
several values of $\Lambda_{\rm QCD}$ and $\Lambda$.
Units of $\Lambda_{\rm QCD}$, $\Lambda$ and $F_\pi(\Lambda)$
are GeV.
The error of $F_\pi(\Lambda)$ comes only from 
$\langle\bar{q}q\rangle=-(225\pm25\,\mbox{MeV})^3$.
The firsr error for $a(\Lambda)$ and $g(\Lambda)$ comes from 
$F_\pi(0) = 86.4\pm9.7\,\mbox{MeV}$ 
and the second error from 
$\langle\bar{q}q\rangle=-(225\pm25\,\mbox{MeV})^3$.
Note that $0.00$ in the error of $g(\Lambda)$ implies that the error
is smaller than $0.01$.
}\label{tab:WM bare}
\end{table}
\begin{table}[htbp]
\begin{center}
\begin{tabular}{|c|c||c|c|}
\hline
 $\Lambda_{\rm QCD}$ & $\Lambda$ 
  & $z_3(\Lambda)$ 
  & $z_2(\Lambda)-z_1(\Lambda)$ \\
\hline
 $0.30$
 & $1.00$ & $-6.10\pm4.36\pm0.63$ & $-2.21\pm0.37\pm0.84$ \\
 & $1.10$ & $-3.14\pm5.04\pm0.19$ & $-2.01\pm0.34\pm0.47$ \\
 & $1.20$ & $-1.27\pm5.92\pm0.12$ & $-1.76\pm0.30\pm0.27$ \\
\hline
 $0.35$
 & $1.00$ & $-7.20\pm4.73\pm0.41$ & $-2.05\pm0.38\pm0.97$ \\
 & $1.10$ & $-4.35\pm5.38\pm0.04$ & $-1.90\pm0.34\pm0.54$ \\
 & $1.20$ & $-2.66\pm6.22\pm0.23$ & $-1.69\pm0.29\pm0.31$ \\
\hline
 $0.40$
 & $1.00$ & $-8.65\pm5.19\pm0.05$ & $-1.85\pm0.39\pm1.12$ \\
 & $1.10$ & $-5.84\pm5.78\pm0.18$ & $-1.79\pm0.34\pm0.61$ \\
 & $1.20$ & $-4.31\pm6.56\pm0.39$ & $-1.61\pm0.29\pm0.35$ \\
\hline
 $0.45$
 & $1.00$ & $-10.6\pm5.79\pm0.56$ & $-1.61\pm0.41\pm1.29$ \\
 & $1.10$ & $-7.73\pm6.27\pm0.52$ & $-1.65\pm0.35\pm0.70$ \\
 & $1.20$ & $-6.29\pm6.96\pm0.61$ & $-1.52\pm0.29\pm0.40$ \\
\hline
\end{tabular}\\
\end{center}
\caption[Bare parameters of the HLS (2)]{%
Two of next-leading order parameters of the HLS 
at $\mu=\Lambda$ for
several values of $\Lambda_{\rm QCD}$ and $\Lambda$.
Units of $\Lambda_{\rm QCD}$ and $\Lambda$
are GeV.
Values of $z_3(\Lambda)$ and $z_2(\Lambda)-z_1(\Lambda)$
are scaled by a factor of $10^3$.
The firsr error comes from 
$F_\pi(0) = 86.4\pm9.7\,\mbox{MeV}$ 
and the second error from 
$\langle\bar{q}q\rangle=-(225\pm25\,\mbox{MeV})^3$.
}\label{tab:WM bare 2}
\end{table}
Typical values of the bare parameters for 
$(\Lambda_{\rm QCD},\,\Lambda) = (0.40,\,1.10)\,\mbox{GeV}$
are
\begin{eqnarray}
&&
F_\pi(\Lambda) = 145 \pm 3 \,\mbox{MeV} \ ,
\nonumber\\
&&
a(\Lambda) = 1.33 \pm 0.28 \pm 0.14 \ ,
\nonumber\\
&&
g(\Lambda) = 3.69 \pm 0.13 \pm 0.02 \ ,
\nonumber\\
&&
z_3(\Lambda) = ( -5.84\pm 5.78\pm 0.18)\times 10^{-3} \ ,
\nonumber\\
&&
z_2(\Lambda) - z_1(\Lambda) = 
  ( -1.79 \pm 0.34 \pm 0.61) \times 10^{-3} \ ,
\label{bare typical}
\end{eqnarray}
where
the error of $F_\pi(\Lambda)$ comes only from 
$\langle\bar{q}q\rangle=-(225\pm25\,\mbox{MeV})^3$,
while 
the firsr error for $a(\Lambda)$, $g(\Lambda)$,
$z_3(\Lambda)$ and $z_2(\Lambda) - z_1(\Lambda)$
comes from 
$F_\pi(0) = 86.4\pm9.7\,\mbox{MeV}$ 
and the second error from 
$\langle\bar{q}q\rangle=-(225\pm25\,\mbox{MeV})^3$.
By using the above values, the bare $\rho$ mass defined by
$M_\rho^2(\Lambda) = a(\Lambda) g^2(\Lambda) F_\pi^2(\Lambda)$
is estimated as
\begin{equation}
M_\rho(\Lambda) = 614 \pm 44 \pm 16 \,\mbox{MeV}
\ .
\end{equation}
These values show that the ambiguities of the bare parameters
coming from that of the quark condensate are small for
the leading order parameters as well as the parameter $z_3$,
while it is rather large for $z_2-z_1$.
This is because the leading order parameters are almost
determined by the $1+\alpha_s/\pi$ term of the current correlators
derived from the OPE through the Wilsonian matching, while 
$z_2-z_1$ is directly related to the quark condensate as
in Eq.~(\ref{match z}).

Now, one might 
suspect that the inclusion of the $A_1$ ($a_1$ meson and
its flavor partners) 
would affect the above matching 
result,
since the mass of $a_1$ is 
$m_{a_1}=1.23\pm0.04\,\mbox{GeV}$~\cite{PDG:02} 
close to our
matching scale $\Lambda = 1.1\,\mbox{GeV}$.
Especially, it might give a large contribution in determining 
the value of $F_\pi^2(\Lambda)$ 
so as to pull it down close to the 
$F_\pi^2(0) \simeq (86.4 \mbox{MeV})^2$, 
and hence the large amount of
the quadratic divergence might be an artifact of simply neglecting the
$A_1$ 
contribution. 
However, this is not the case. 
Inclusion of $A_1$ does not affect the large value of $F_\pi(\Lambda)$.
 
This is seen as follows:
We can include the effect of $A_1$ by using the effective field
theory such as the Generalized 
HLS~\cite{BKY:NPB,Bando-Fujiwara-Yamawaki}.
Although a 
complete list of the ${\cal O}(p^4)$ terms has
not yet been given, on the analogy of the $\rho$ contribution
to the vector current correlator given in Eq.~(\ref{Pi V HLS})
it is reasonable to write the axialvector current correlator
around the matching scale
with the $A_1$ contribution included as
\begin{equation}
\Pi_A^{\rm(GHLS)}(Q^2) =
\frac{F_\pi^2(\Lambda)}{Q^2} 
+ \frac{ F_{A_1}^2(\Lambda) }{ M_{A_1}^2(\Lambda) + Q^2 } 
- 2 z_2^\prime(\Lambda)
\ ,
\label{Pi A GHLS}
\end{equation}
where $M_{A_1}(\Lambda)$ is the bare $A_1$ mass,
$F_{A_1}(\Lambda)$ the bare $A_1$ decay constant analog to
$F_\rho(\Lambda) \equiv \sqrt{
F_\sigma^2(\Lambda)\left[ 1 - 2 g^2(\Lambda) z_3(\Lambda) \right]
}$ and $z_2^\prime(\Lambda)$ exhibits the contribution from
higher modes analog to $z_2(\Lambda)$.
By using the above correlator, the Wilsonian matching 
condition~(\ref{match A}) 
would be changed to
\begin{eqnarray}
\frac{F_\pi^2(\Lambda)}{\Lambda^2} 
+ \frac{ \Lambda^2 F_{A_1}^2(\Lambda) 
  }{ 
   [ M_{A_1}^2(\Lambda) + \Lambda^2 ]^2 } 
= \frac{1}{8\pi^2} \left( \frac{N_c}{3} \right) (1+\delta_A) \ .
\label{match A GHLS}
\end{eqnarray}
For determining the value of $F_\pi(\Lambda)$ from the above
Wilsonian matching condition we need to know the values
of $M_{A_1}(\Lambda)$ and $F_{A_1}(\Lambda)$.
Since the matching scale $\Lambda$ is close to the $A_1$ mass,
the on-shell values give a good approximation, $M_{A_1}(\Lambda)\simeq m_{A_1}$,\, $F_{A_1}(\Lambda)\simeq F_{A_1}(m_{A_1})=F_{A_1}$.
Although the experimental value of the $a_1$ mass is known as
$m_{a_1}=1.23\pm0.04\,\mbox{GeV}$~\cite{PDG:02},
the on-shell value of its decay constant $F_{a_1}$ is not known.
However, we could  
use the
pole saturated version of the first Weinberg's sum rule~\cite{Wei:SR}
\begin{equation}
F_\rho^2 - F_{A_1}^2 = F_\pi^2(0)
\ ,
\label{WSR 2 pole}
\end{equation}
together with $F_\rho = g_\rho/m_\rho =0.154\pm0.001\,\mbox{GeV}$
and $F_\pi(0)=86.4\pm9.7\,\mbox{MeV}$,
which yields $F_{A_1}=0.127\pm0.007\,\mbox{GeV}$, and hence roughly 
$F_{A_1}^2(\Lambda) \simeq (130 \, \mbox{MeV})^2$.
Here, instead of this value, we use 
$F_{A_1}^2(\Lambda) = F_\rho^2 \sim (150\,\mbox{MeV})^2
\sim 3 \left( \frac{\Lambda}{4\pi} \right)^2$
and set $M_{A_1}(\Lambda) \sim \Lambda$
to include a possible {\it maximal} $A_1$ contribution to the
Wilsonian matching condition (\ref{match A GHLS}).
The resultant value 
(a possible {\it minimum} value) of $F_\pi(\Lambda)$
with $N_c=3$ 
is estimated as
\begin{eqnarray}
F_\pi^2(\Lambda) &=&
\frac{\Lambda^2}{8\pi^2} ( 1 + \delta_A )
- 
\frac{ \Lambda^4 F_{A_1}^2(\Lambda) 
  }{ 
   [ M_{A_1}^2(\Lambda) + \Lambda^2 ]^2 } 
\nonumber\\
&\sim&
\frac{\Lambda^2}{(4\pi)^2}
\left[ 2 ( 1 + \delta_A ) - \frac{3}{4} \right]
\nonumber\\
&\sim&
\frac{9}{4} \frac{\Lambda^2}{(4\pi)^2}
\sim \left( \frac{3}{2} \times 86.4\,\mbox{MeV}\right)^2 
\sim ( 130 \,\mbox{MeV} )^2
\ ,
\end{eqnarray}
where we again adopted a very rough estimate 
$\delta_A \sim 0.5$ to obtain the last line.
This value $F_\pi(\Lambda) \sim 130\,\mbox{MeV}$ (possible {\it minimum} value)
is still much larger than
the on-shell value $86.4\pm9.7\,\mbox{MeV}$ and close to the
value $150\,\mbox{MeV}$ 
obtained in Eq.~(\ref{Fpi Lambda rough estimate})
by the Wilsonian matching without including the
effect of $A_1$.

\subsection{Results of the Wilsonian matching}
\label{ssec:RWM}

\subsubsection{Full analysis}
\label{fullanalysis}

In the previous subsection we have completely specified the bare
Lagrangian through the Wilsonian matching conditions~(\ref{match z}),
(\ref{match A}) and (\ref{match V}) together with the physical inputs
of the pion decay constant $F_\pi(0)$ and the rho mass $m_\rho$.
Using the Wilsonian RGEs for the parameters obtained in 
Sec.~\ref{ssec:RGEWS} 
[Eqs.~(\ref{RGE for Fpi2}),
(\ref{RGE for a}), (\ref{RGE for g2}), (\ref{RGE:z1}), 
(\ref{RGE:z2}) and (\ref{RGE:z3})],
we obtain the values of five parameters
at $\mu = m_\rho$.
In Table~\ref{tab:WM mrho}
we list several typical values of five parameters at 
$\mu = m_\rho$ 
for several values of 
$\Lambda_{\rm QCD}$ with $\Lambda=1.1\,\mbox{GeV}$.
\begin{table}[htbp]
\begin{center}
\begin{tabular}{|c|c||c|c|c|}
\hline
 $\Lambda_{\rm QCD}$ & $\Lambda$ & $F_\pi(m_\rho)$ 
  & $a(m_\rho)$ & $g(m_\rho)$ \\
\hline
 $0.30$
 & $1.10$ & $0.0995\pm0.0012\pm0.0036$ & $1.57\pm0.34\pm0.13$
  & $6.19\pm0.59\pm0.03$ \\
\hline
 $0.35$
 & $1.10$ & $0.102\pm0.001\pm0.004$ & $1.48\pm0.33\pm0.14$
  & $6.22\pm0.64\pm0.06$ \\
\hline
 $0.40$
 & $1.10$ & $0.105\pm0.001\pm0.004$ & $1.38\pm0.32\pm0.16$
  & $6.27\pm0.69\pm0.11$ \\
\hline
 $0.45$
 & $1.10$ & $0.108\pm0.001\pm0.005$ & $1.27\pm0.32\pm0.18$
  & $6.36\pm0.76\pm0.17$ \\
\hline
\end{tabular}
\\
\begin{tabular}{|c|c||c|c|}
\hline
 $\Lambda_{\rm QCD}$ & $\Lambda$ 
  & $z_3(m_\rho)$ 
  & $z_2(m_\rho)-z_1(m_\rho)$ \\
\hline
 $0.30$
 & $1.10$ & $-4.13\pm5.20\pm0.13$ & $-2.49\pm0.59\pm0.56$ \\
\hline
 $0.35$
 & $1.10$ & $-5.38\pm5.52\pm0.01$ & $-2.32\pm0.60\pm0.64$ \\
\hline
 $0.40$
 & $1.10$ & $-6.90\pm5.89\pm0.23$ & $-2.13\pm0.61\pm0.74$ \\
\hline
 $0.45$
 & $1.10$ & $-8.83\pm6.34\pm0.56$ & $-1.89\pm0.62\pm0.86$ \\
\hline
\end{tabular}
\end{center}
\caption[Parameters of the HLS at $\mu=m_\rho$]{%
Five parameters of the HLS at $\mu=m_\rho$ for 
several values of 
$\Lambda_{\rm QCD}$ with $\Lambda=1.1\,\mbox{GeV}$.
Units of $\Lambda_{\rm QCD}$, $\Lambda$ and $F_\pi$ are GeV.
Values of $z_3(m_\rho)$ and $z_2(m_\rho)-z_1(m_\rho)$
are scaled by a factor of $10^3$.
Note that the first error comes from 
$F_\pi(0) = 86.4\pm9.7\,\mbox{MeV}$ and
the second error from 
$\langle\bar{q}q\rangle=-(225\pm25\,\mbox{MeV})^3$.
}\label{tab:WM mrho}
\end{table}
Typical values for 
$(\Lambda_{\rm QCD},\,\Lambda) = (0.40,\,1.10)\,\mbox{GeV}$
are
\begin{eqnarray}
&&
F_\pi(m_\rho) = 105 \pm 1 \pm 4 \,\mbox{MeV} \ ,
\nonumber\\
&&
a(m_\rho) = 1.38 \pm 0.32 \pm 0.16 \ ,
\nonumber\\
&&
g(m_\rho) = 6.27 \pm 0.69 \pm 0.11 \ ,
\nonumber\\
&&
z_3(m_\rho) = ( -6.90\pm 5.89\pm 0.23)\times 10^{-3} \ ,
\nonumber\\
&&
z_2(m_\rho) - z_1(m_\rho) = 
  ( -2.13 \pm 0.61 \pm 0.74) \times 10^{-3} \ ,
\label{para mr typical}
\end{eqnarray}
where
the firsr error comes from 
$F_\pi(0) = 86.4\pm9.7\,\mbox{MeV}$ 
and the second error from 
$\langle\bar{q}q\rangle=-(225\pm25\,\mbox{MeV})^3$.
It should be noticed that, comparing the above value of $a(m_\rho)$
with that of $a(\Lambda)$ in Eq.~(\ref{bare typical}),
we see that the parameter $a$ does not change its value by the 
running from the matching scale to the scale of $\rho$ on-shell.
Furthermore, the value itself is close to one.
Nevertheless, the parameter $a$ at the low-energy limit becomes
closer to $2$ which leads to the vector dominance of the
electromagnetic form factor of the pion [see the analysis
around Eq.~(\ref{a0 = 2})].

Now that we have determined the five parameters at $\mu = m_\rho$,
we make several physical predictions.
The typical ``physical'' quantities derived from the five parameters
are~\cite{HY:matching}
$\rho$-$\gamma$ mixing strength,
Gasser-Leutwyler's parameter $L_{10}$~\cite{Gas:85a},
$\rho$-$\pi$-$\pi$ coupling constant $g_{\rho\pi\pi}$,
Gasser-Leutwyler's parameter $L_9$~\cite{Gas:85a}
and
the parameter $a(0)$
which parameterizes the validity of the vector 
dominance.
Below we shall list the relations of the five parameters of the HLS to
these ``physical'' quantities following Ref.~\cite{HY:matching}.
The resultant predictions are listed in Tables~\ref{tab:res}
and \ref{tab:res 2}
for several values of $\Lambda_{\rm QCD}$ and
$\Lambda$.

\subsubsection*{$\rho$-$\gamma$ mixing strength}

The second term in Eq.~(\ref{leading Lagrangian}) gives the mass
mixing between $\rho$ and the external field of $\gamma$ (photon
field). 
The $z_3$-term in Eq.~(\ref{Lag: z terms}) gives the kinetic mixing.
Combining these two at the on-shell of $\rho$ leads to the
$\rho$-$\gamma$ mixing strength~\cite{HY:matching}:
\begin{equation}
g_\rho = g(m_\rho) F_\sigma^2(m_\rho) 
\left[ 1 - g^2(m_\rho) z_3(m_\rho) \right] \ ,
\label{g rho}
\end{equation}
which should be compared with the quantity derived from the
experimental data of the $\rho \rightarrow e^+ e^-$ decay width.
As we have shown in Eq.~(\ref{grho exp val}),
$\Gamma(\rho \rightarrow e^+ e^-) = (6.85\pm0.11)\times
10^{-3}$\,MeV~\cite{PDG:02} leads to 
$\left. g_\rho \right\vert_{\rm exp} = 0.119\pm0.001\,\mbox{GeV}^2$.
The typical predicted value of $g_\rho$ for 
$(\Lambda_{\rm QCD}\,,\,\Lambda) = (0.40\,,\,1.10)\,\mbox{GeV}$
is 
\begin{equation}
\left. g_\rho \right\vert_{\rm theo} = 
0.121\pm0.014\pm0.0003\,\mbox{GeV}^2
\ ,
\end{equation}
where the first error comes from the ambiguity of the input
value of $F_\pi(0)$ and the second one from that of the
quark condensate $\langle\bar{q} q\rangle$.
The central value of 
this as well as that for
$(\Lambda_{\rm QCD}\,,\,\Lambda) = 
(0.30\,,\,1.00)$\,GeV 
shown in Table~\ref{tab:res}
are very close to the experimental value.
These values are improved from the tree prediction 
$\left. g_{\rho} \right\vert_{\rm tree} = 0.103\,\mbox{GeV}^2$
in Eq.~(\ref{grhoval:tree}) where
$g_{\rho\pi\pi}$ in addition to $F_\pi(0)$ and $m_\rho$
was used as an input.
It should be noticed that
most predicted values are consistent with the experiment
within the error 
of input values of $F_\pi(0)$ and
$\langle\bar{q} q\rangle$.

\subsubsection*{Gasser-Leutwyler's parameter $L_{10}$~\cite{Gas:85a}}

As we have done in Sec.~\ref{ssec:MHC}
the relation between the 
Gasser-Leutwyler's parameter $L_{10}$ and the parameters of HLS is
obtained by matching the axialvector current correlator 
in the low energy limit.
The resultant relation is given by 
[see Eq.~(\ref{l10})]~\footnote{%
  Note that the finite correction appearing as the last term
  in the right-hand-side of 
  Eq.~(\ref{l10 5}) was not included in Ref.~\cite{HY:matching}.
}
\begin{eqnarray}
&&
L_{10}^r(m_\rho) =
- \frac{1}{4g^2(m_\rho)} 
+ \frac{z_3(m_\rho) - z_2(m_\rho) + z_1(m_\rho)}{2}
+ \frac{N_f}{(4\pi)^2} \frac{11a(m_\rho)}{96} 
+ \frac{N_f}{(4\pi)^2} \frac{5}{72} 
\ .
\label{l10 5}
\end{eqnarray}
The `experimental' value of $L_{10}$ is estimated
as
[see Eq.~(\ref{L10:val}) in Sec.~\ref{ssec:l10}]
$\left. L_{10}^r(m_\rho) \right\vert_{\rm exp} = 
(-5.1\pm0.7)\times 10^{-3}$.
A typical value of the prediction is
\begin{equation}
\left. L_{10}^r(m_\rho) \right\vert_{\rm theo}
=(-4.43\pm2.54\pm0.62)\times 10^{-3}
\ ,
\end{equation}
for $(\Lambda_{\rm QCD}\,,\,\Lambda) = 
(0.40\,,\,1.10)$\,GeV (see Table~\ref{tab:res 2})
where the first error comes from the ambiguity of the input
value of $F_\pi(0)$ and the second one from that of the
quark condensate.
There are large ambiguities mainly from that of $F_\pi(0)$,
and all the predicted values shown in Table~\ref{tab:res 2}
are consistent with the experimental value.
We should note that the central value of the prediction
is somewhat improved from the tree value
$L_{10}^V = (-7.4\pm2.3)\times 10^{-3}$ in Table~\ref{tab:L from V}
in Sec.~\ref{ssec:VMSLEC}.

\subsubsection*{$\rho$-$\pi$-$\pi$ coupling constant $g_{\rho\pi\pi}$}

Strictly speaking, we have to include a higher derivative type
$z_4$-term listed in Eq.~(\ref{Lag: z terms}).
However, a detailed analysis~\cite{HSc}
using a similar model~\cite{Kay:85}
does not require its existence.\footnote{%
  Note that the existence of the
  kinetic type $\rho$-$\gamma$ mixing from $z_3$-term was needed to
  explain the experimental data of 
  $\Gamma(\rho\rightarrow e^+e^-)$.~\cite{HSc}%
}
Hence we neglect the $z_4$-term.
If we simply read the $\rho$-$\pi$-$\pi$ interaction
from Eq.~(\ref{leading Lagrangian}), 
we would obtain 
$g_{\rho\pi\pi} =
g(m_\rho) F_\sigma^2(m_\rho) / 2 F_\pi^2(m_\rho)$.
However, $g_{\rho\pi\pi}$
should be defined for on-shell $\rho$ and $\pi$'s.
While $F_\sigma^2$ and $g^2$ do not 
run for $\mu<m_\rho$,  $F_\pi^2$ does run.
The on-shell pion decay constant is given by $F_\pi(0)$.  Thus we have
to use $F_\pi(0)$ to define the on-shell $\rho$-$\pi$-$\pi$ coupling
constant.  
The resulting expression is given by~\cite{HY:matching}
\begin{equation}
g_{\rho\pi\pi} = \frac{g(m_\rho)}{2}
\frac{F_\sigma^2(m_\rho)}{F_\pi^2(0)} \ .
\label{g rho pi pi}
\end{equation}
As we have shown in Eq.~(\ref{grpp exp val}),
the experimental value of $g_{\rho\pi\pi}$ is estimated as
$\left. g_{\rho\pi\pi} \right\vert_{\rm exp} = 6.00\pm0.01$.
A typical value of the prediction is 
\begin{equation}
\left. g_{\rho\pi\pi} \right\vert_{\rm theo}
= 6.35\pm0.72\pm0.11
\ ,
\end{equation}
for $(\Lambda_{\rm QCD}\,,\,\Lambda) = 
(0.40\,,\,1.10)$\,GeV (see Table~\ref{tab:res}).
The error of the prediction mainly comes from the ambiguity of 
the input of $F_\pi(0)$, and all the predictions shown in 
Table~\ref{tab:res} are consistent with the experiment.
Note that, in the tree-level analysis done in
subsection~\ref{ssec:PPLO},
$g_{\rho\pi\pi}$ was used as an input.

\subsubsection*{Gasser-Leutwyler's parameter $L_9$~\cite{Gas:85a}}

Similarly to the $z_4$-term contribution
to $g_{\rho\pi\pi}$ we neglect the
contribution from the higher derivative type
$z_6$-term.
The resultant relation between $L_9$ and the parameters of the HLS is
given by~\cite{Tanabashi,HY:matching}
\begin{equation}
L_9^r(m_\rho) 
+ \frac{N_f}{(4\pi)^2} \frac{5}{72} 
= 
\frac{1}{4}
\left( \frac{1}{g^2(m_\rho)} - z_3(m_\rho) \right) \ ,
\label{l9}
\end{equation}
where the second term in the left-hand-side is the finite correction
derived in the ChPT~\cite{Gas:84,Gas:85a,Tanabashi}.\footnote{%
  This finite correction in the ChPT 
  was not included in Ref.~\cite{HY:matching}.
}
The `experimental value' of $L_9$ is estimated
as
[see Eq.~(\ref{L9:val}) in subsection~\ref{ssec:l9}]
$\left. L_9^r(m_\rho) \right\vert_{\rm exp} 
= (6.5\pm0.6)\times10^{-3}$,
and the typical prediction is
\begin{equation}
\left. L_9^r(m_\rho) \right\vert_{\rm theo}
= (6.77\pm0.07\pm0.16) \times 10^{-3} 
\ ,
\end{equation}
for 
$(\Lambda_{\rm QCD}\,,\,\Lambda)=(0.4\,,\,1.1)\,\mbox{GeV}$.
The ambiguity in the theoretical prediction from the input value of
$F_\pi(0)$ is not so large as that for $L_{10}$. But
the experimental error is about 10\%, so that
most predictions are consistent with the experiment
(see Table~\ref{tab:res 2}).

\subsubsection*{Parameter $a(0)$}

\begin{figure}[htbp]
\begin{center}
\ \epsfbox{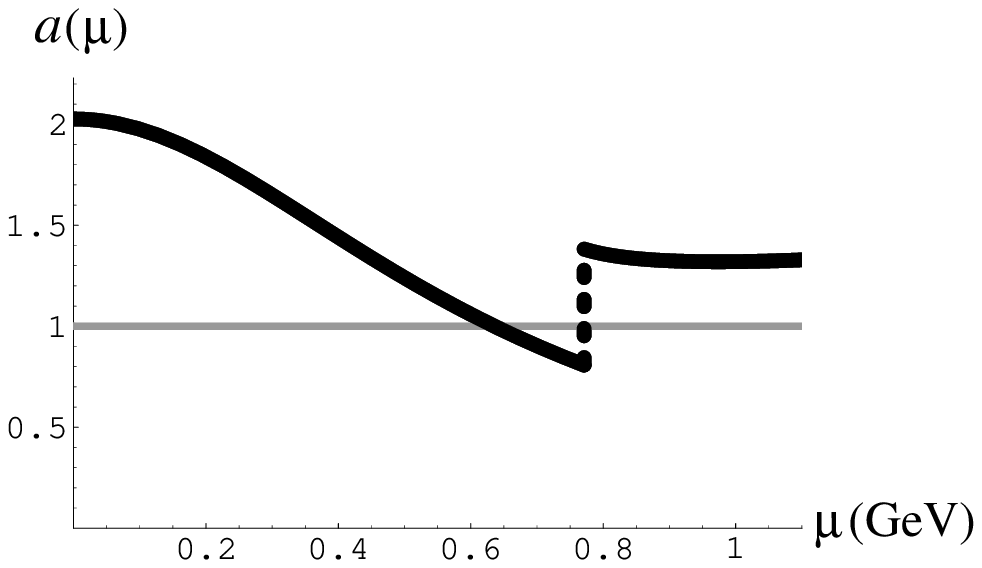}
\end{center}
\caption[Running of $a(\mu)$]{%
Running of $a(\mu)$ for $(\Lambda_{\rm QCD}\,,\,\Lambda) = 
(0.40\,,\,1.10)$\,GeV.
Gap at $m_\rho$ is due to the effect of finite renomalization
between $F_\pi^2(m_\rho)$ and 
$\left[ F_\pi^{(\pi)}(m_\rho) \right]^2$ given 
in Eq.~(\ref{finite renormalization}).
}
\label{fig:arun}
\end{figure}

We further define the parameter $a(0)$ by the direct
$\gamma$-$\pi$-$\pi$ interaction in the second term in 
Eq.~(\ref{leading Lagrangian}).  
As we stated above,
$F_\sigma^2$ does not run for $\mu< m_\rho$
while $F_\pi^2$ does. Thus we have~\cite{HY:matching,HY:VD}
\begin{equation}
a(\mu)
\equiv
\left\{\begin{array}{ll}
 F_\sigma^2(\mu)/F_\pi^2(\mu)
& (\mu > m_\rho ) \ ,
\\
 F_\sigma^2(m_\rho)/\left[ F_\pi^{(\pi)}(\mu) \right]^2 
& (\mu < m_\rho ) \ .
\end{array}\right.
\label{def:a mu}
\end{equation}
This parameter for on-shell pions becomes
\begin{equation}
a(0) = \frac{F_\sigma^2(m_\rho)}{F_\pi^2(0)} \ ,
\label{a0}
\end{equation}
which should be compared with the parameter $a$ used in the
tree-level analysis, $a=2$ corresponding to the 
vector dominance (VD)~\cite{BKUYY,BKY}.
Most values of the prediction are close to $2$:
We obtained
\begin{equation}
a(0) \simeq 2 \ ,
\label{a0 = 2}
\end{equation}
although $a(\Lambda)\simeq a(m_\rho) \simeq1$.
We show the running of $a(\mu)$ for $(\Lambda_{\rm QCD}\,,\,\Lambda) = 
(0.40\,,\,1.10)$\,GeV in Fig.~\ref{fig:arun}.
This shows that although
$a(\mu) \simeq 1$ for $m_\rho < \mu < \Lambda$,
$a(0) \simeq 2$ is realized by the running of 
$\left[ F_\pi^{(\pi)}(\mu) \right]^2$.

\begin{table}[bhtp]
\begin{center}
\begin{tabular}{|c|c||c|c|c|}
\hline
 $\Lambda_{\rm QCD}$ & $\Lambda$ & $g_\rho$ & $g_{\rho\pi\pi}$ 
  & $a(0)$ \\
\hline
 $0.30$
 & $1.00$ & $0.121\pm0.009\pm0.003$ & $6.41\pm0.78\pm0.08$
 & $2.07\pm0.04\pm0.05$ \\
 & $1.10$ & $0.111\pm0.011\pm0.001$ & $6.44\pm0.83\pm0.03$
 & $2.08\pm0.07\pm0.02$ \\
 & $1.20$ & $0.105\pm0.014\pm0.000$ & $6.43\pm0.82\pm0.02$
 & $2.08\pm0.06\pm0.02$ \\
\hline
 $0.35$
 & $1.00$ & $0.125\pm0.011\pm0.003$ & $6.36\pm0.72\pm0.15$
 & $2.03\pm0.00\pm0.09$ \\
 & $1.10$ & $0.116\pm0.013\pm0.001$ & $6.41\pm0.78\pm0.06$
 & $2.06\pm0.04\pm0.04$ \\
 & $1.20$ & $0.110\pm0.015\pm0.000$ & $6.40\pm0.78\pm0.04$
 & $2.06\pm0.04\pm0.03$ \\
\hline
 $0.40$
 & $1.00$ & $0.129\pm0.013\pm0.002$ & $6.26\pm0.63\pm0.24$
 & $1.97\pm0.04\pm0.15$ \\
 & $1.10$ & $0.121\pm0.014\pm0.000$ & $6.35\pm0.72\pm0.11$
 & $2.03\pm0.01\pm0.07$ \\
 & $1.20$ & $0.116\pm0.017\pm0.001$ & $6.35\pm0.74\pm0.07$
 & $2.03\pm0.02\pm0.04$ \\
\hline
 $0.45$
 & $1.00$ & $0.136\pm0.016\pm0.001$ & $6.10\pm0.52\pm0.38$
 & $1.87\pm0.10\pm0.23$ \\
 & $1.10$ & $0.127\pm0.017\pm0.000$ & $6.26\pm0.65\pm0.17$
 & $1.97\pm0.03\pm0.11$ \\
 & $1.20$ & $0.123\pm0.019\pm0.001$ & $6.28\pm0.69\pm0.10$
 & $1.98\pm0.01\pm0.06$ \\
\hline
\multicolumn{2}{|c||}{Exp.} & $0.119\pm0.001$ & $6.00\pm0.01$ & \\
\hline
\end{tabular}
\end{center}
\caption[Physical predictions of the Wilsonian matching]{%
Physical quantities predicted by the Wilsonian matching
conditions and the Wilsonian RGEs.
Units of $\Lambda_{\rm QCD}$ and $\Lambda$ are GeV, and
that of $g_\rho$ is GeV$^2$.
Experimental values of $g_\rho$ and $g_{\rho\pi\pi}$ are derived in
Sec.~\ref{ssec:PPLO}.
Note that the first error comes from 
$F_\pi(0) = 86.4\pm9.7\,\mbox{MeV}$ and
the second error from 
$\langle\bar{q}q\rangle=-(225\pm25\,\mbox{MeV})^3$.
$0.000$ in the error of $g_\rho$ and $0.00$ in
$a(0)$
imply that the errors
are smaller than $0.001$ and $0.01$, respectively.
}
\label{tab:res}
\end{table}
\begin{table}[bhtp]
\begin{center}
\begin{tabular}{|c|c||c|c|}
\hline
 $\Lambda_{\rm QCD}$ & $\Lambda$ 
  & $L_9(m_\rho)$ & $L_{10}(m_\rho)$ \\
\hline
 $0.30$
 & $1.00$ & $6.88\pm0.21\pm0.30$ & $-4.07\pm1.94\pm0.46$ \\
 & $1.10$ & $6.24\pm0.05\pm0.09$ & $-2.62\pm2.38\pm0.43$ \\
 & $1.20$ & $5.82\pm0.26\pm0.01$ & $-1.94\pm2.75\pm0.38$ \\
\hline
 $0.35$
 & $1.00$ & $7.04\pm0.22\pm0.38$ & $-4.87\pm2.02\pm0.56$ \\
 & $1.10$ & $6.50\pm0.05\pm0.12$ & $-3.46\pm2.45\pm0.52$ \\
 & $1.20$ & $6.12\pm0.27\pm0.02$ & $-2.84\pm2.82\pm0.44$ \\
\hline
 $0.40$
 & $1.00$ & $7.22\pm0.21\pm0.48$ & $-5.82\pm2.13\pm0.71$ \\
 & $1.10$ & $6.77\pm0.07\pm0.16$ & $-4.43\pm2.54\pm0.62$ \\
 & $1.20$ & $6.45\pm0.30\pm0.03$ & $-3.84\pm2.91\pm0.52$ \\
\hline
 $0.45$
 & $1.00$ & $7.41\pm0.17\pm0.59$ & $-6.98\pm2.31\pm0.93$ \\
 & $1.10$ & $7.07\pm0.10\pm0.20$ & $-5.57\pm2.68\pm0.76$ \\
 & $1.20$ & $6.81\pm0.35\pm0.04$ & $-4.99\pm3.03\pm0.63$ \\
\hline
\multicolumn{2}{|c||}{Exp.} &  $6.5\pm0.6$ & $-5.1\pm0.7$ \\
\hline
\end{tabular}
\end{center}
\caption[Predicted values of $L_9$ and $L_{10}$ from 
  the Wilsonian matching]{%
Values of Gasser-Leutwyler's parameters
$L_9$ and $L_{10}$ 
predicted by the Wilsonian matching
conditions and the Wilsonian RGEs.
Units of $\Lambda_{\rm QCD}$ and $\Lambda$ are GeV.
Values of $L_9^r(m_\rho)$ and $L_{10}^r(m_\rho)$ are 
scaled by a factor of $10^3$.
Experimental values of 
$L_9^r(m_\rho)$ and $L_{10}^r(m_\rho)$ are derived 
in Secs.~\ref{ssec:l9} and \ref{ssec:l10}.
Note that the first error comes from 
$F_\pi(0) = 86.4\pm9.7\,\mbox{MeV}$ and
the second error from 
$\langle\bar{q}q\rangle=-(225\pm25\,\mbox{MeV})^3$.
}
\label{tab:res 2}
\end{table}

\subsubsection*{KSRF relations}

The KSRF (I) relation 
$g_\rho = 2 g_{\rho\pi\pi} F_\pi^2$ \cite{KSRF:KS,KSRF:RF}
holds as a low energy theorem
of the HLS~\cite{BKY:NPB,BKY:PTP,HY,HKY:PRL,HKY:PTP}.
Here this is satisfied as follows~\cite{HY:matching}:
As we have shown in Sec.~\ref{ssec:LETOL},
higher derivative terms like $z_3$ do not contribute
in the low energy limit, and
the $\rho$-$\gamma$ mixing strength becomes $g_\rho(0) = g(m_\rho)
F_\sigma^2(m_\rho)$. 
Comparing this with $g_{\rho\pi\pi}$ in 
Eq.~(\ref{g rho pi pi})~\footnote{%
  The contribution from the higher derivative term is neglected
  in the expression of $g_{\rho\pi\pi}$ given in 
  Eq.~(\ref{g rho pi pi}), i.e.,
  $g_{\rho\pi\pi} = g_{\rho\pi\pi}(m_\rho^2;0,0) =
  g_{\rho\pi\pi}(0;0,0)$.
},
we can easily read that the 
{\it low energy theorem is satisfied}.
As to the on-shell $\rho$, on the other hand, 
using the $\pi$ decay constant at the chiral limit,
$F_\pi(0) = 86.4\pm9.7\,\mbox{MeV}$, together 
with the experimental values of the $\rho$-$\gamma$ mixing
strength, $g_\rho = 0.119\pm0.001\,\mbox{GeV}^2$, and
the $\rho$-$\pi$-$\pi$ coupling, $g_{\rho\pi\pi}=6.00\pm0.01$,
we have
\begin{equation}
\left. \frac{g_\rho}{2 g_{\rho\pi\pi} F_\pi^2(0) } 
\right\vert_{\rm exp}
=
1.32 \pm 0.30
\ .
\label{KSRF I exp}
\end{equation}
This implies that there is about 30\% deviation of the experimental
value
from the KSRF (I) relation.~\footnote{%
  Note that, in Eq.~(\ref{LET val}), we used the experimental 
  value of the
  $\pi$ decay constant, $F_{\pi,{\rm phys}} = 92.42 \pm 0.23$,
  and obtained 
  $\left. \frac{g_\rho}{ 2 g_{\rho\pi\pi} F_\pi^2 } 
  \right\vert_{\rm exp}= 1.15 \pm 0.01$.
  Strictly speaking, we may 
  have to include the effect of explicit chiral
  symmetry breaking due to the current quark masses into
  $g_\rho$ as well as $g_{\rho\pi\pi}$.  
  However, according to the
  analysis done by the similar model at tree level in Ref.~\cite{HSc},
  the corrections from the explicit chiral symmetry breaking
  to them are small.
  So we neglect the effect in the present analysis.
}
As we have studied in Sec.~\ref{ssec:PPLO},
at the leading order this ratio is predicted as $1$:
$\left. \frac{g_\rho}{ 2 g_{\rho\pi\pi} F_\pi^2(0) } 
  \right\vert_{\rm tree}= 1$.
When the next order correction generated by
the loop effect and the ${\cal O}(p^4)$ term is included,
the
combination of $g_\rho$ in Eq.~(\ref{g rho}) with
$g_{\rho\pi\pi}$ in Eq.~(\ref{g rho pi pi}) provides
\begin{equation}
\left. \frac{g_\rho}{2 g_{\rho\pi\pi} F_\pi^2(0) } 
\right\vert_{\rm theo}
=
1 - g^2(m_\rho) z_3(m_\rho) 
= 
1.27 \pm 0.29 \pm 0.02
\ ,
\end{equation}
where the value is obtained for 
$(\Lambda_{\rm QCD}\,,\,\Lambda) = (0.40\,,\,1.1)$.
This shows that the 30\% diviation of experimental value
from the KSRF (I) relation as in Eq.~(\ref{KSRF I exp})
is actually explained by the existence of the $z_3$ term together
with the loop effect included through the Wilsonian RGEs.

The KSRF (II) relation
$m_\rho^2 = 2 g_{\rho\pi\pi}^2 F_\pi^2$~\cite{KSRF:KS,KSRF:RF}
is approximately satisfied by
the on-shell quantities even though $a(m_\rho) \simeq 1$.
This is seen as follows~\cite{HY:matching}:
Equation~(\ref{g rho pi pi}) with Eq.~(\ref{a0}) and $m_\rho^2 =
g^2(m_\rho) F_\sigma^2(m_\rho)$ leads to $2 g_{\rho\pi\pi}^2
F_\pi^2(0) = m_\rho^2 \left( a(0)/2 \right)$.  Thus $a(0) \simeq 2$
leads to the approximate KSRF (II) relation.
Furthermore, $a(0)\simeq2$
implies that the direct $\gamma$-$\pi$-$\pi$ coupling is suppressed
(vector dominance).
We shall return to this point later
(see Sec.~\ref{sssec:VDLNQ}).\\

To summarize,
the predicted values of $g_\rho$, $g_{\rho\pi\pi}$,
$L_9^r(m_\rho)$ and $L_{10}^r(m_\rho)$ remarkably
agree with the experiment,
although $L_{10}^r(m_\rho)$ is somewhat sensitive to the values of
$\Lambda_{\rm QCD}$ and $\Lambda$.~\footnote{%
  One might think of the matching by the Borel
  transformation of the correlators.
  However, agreement of the predicted values, especially $g_\rho$, 
  are not as remarkably good as that for the present case.%
}
There are 
considerable ambiguities from the input value of $F_\pi(0)$,
and most predicted values are consistent with the experiment.
Furthermore, we have $a(0)\simeq2$, although $a(\Lambda)\simeq
a(m_\rho) \simeq1$. The KSRF (I) relation is reproduced better than the
tree-level result and KSRF (II) relation holds even for $a(m_\rho) \simeq 1$.

\subsubsection{``Phenomenology'' with $a(\Lambda)=1$}
\label{a1phenomenology}

As we have seen above,
$a(\Lambda)=1$ is already close to the reality.
Here it is worth emphasizing this fact by demonstrating
more explicitly, since $a(\Lambda)=1$ is a fixed point of the RGE and
of direct relevance to the Vector Manifestation we shall fully discuss
in Sec.~\ref{sec:VM}. We shall  
show the result of the same analysis
as that already done above except a point that one of the input
data, 
$F_\pi(0) = 86.4\pm9.7\,\mbox{MeV}$, is replaced by $a(\Lambda)=1$.

First, the bare parameters in the case $a(\Lambda)=1$ for 
$\left( \Lambda \,,\, \Lambda_{\rm QCD} \right) = 
(1.1\,,\,0.4)\,\mbox{GeV}$ are given by
\begin{eqnarray}
&&
  g(\Lambda) = 3.86 \pm 0.04 \ ,
\nonumber\\
&&
  z_3(\Lambda) = ( -13.8\pm 2.8) \times 10^{-3} \ ,
\nonumber\\
&&
  z_2(\Lambda) - z_1(\Lambda) = (-1.37 \pm 0.43) \times 10^{-3}
\ .
\end{eqnarray}
In Tables~\ref{cutval 1} and \ref{cutval 2}
we show the values of the bare parameters for
several choices of $\Lambda$ and $\Lambda_{\rm QCD}$.
\begin{table}[htbp]
\begin{center}
\begin{tabular}{|c|c||c|c|c|}
\hline
 $\Lambda_{\rm QCD}$ & $\Lambda$ & $F_\pi(\Lambda)$ 
  & $g(\Lambda)$ & $M_\rho(\Lambda)$ \\
\hline
 $0.30$
 & $1.00$ & $0.132 \pm0.004$
 & $4.33 \pm0.07$
 & $0.574 \pm0.008$
 \\
 & $1.10$ & $0.141 \pm0.002$
 & $3.91 \pm0.03$
 & $0.551 \pm0.005$
 \\
 & $1.20$ & $0.150 \pm0.002$
 & $3.61 \pm0.02$
 & $0.542 \pm0.003$
 \\
\hline
 $0.35$
 & $1.00$ & $0.135 \pm0.004$
 & $4.30 \pm0.07$
 & $0.578 \pm0.009$
 \\
 & $1.10$ & $0.143 \pm0.003$
 & $3.89 \pm0.04$
 & $0.554 \pm0.006$
 \\
 & $1.20$ & $0.152 \pm0.002$
 & $3.59 \pm0.02$
 & $0.545 \pm0.004$
 \\
\hline
 $0.40$
 & $1.00$ & $0.137 \pm0.005$
 & $4.26 \pm0.08$
 & $0.583 \pm0.010$
 \\
 & $1.10$ & $0.145 \pm0.003$
 & $3.86 \pm0.04$
 & $0.558 \pm0.006$
 \\
 & $1.20$ & $0.154 \pm0.002$
 & $3.57 \pm0.02$
 & $0.548 \pm0.004$
 \\
\hline
 $0.45$
 & $1.00$ & $0.140 \pm0.006$
 & $4.21 \pm0.09$
 & $0.588 \pm0.011$
 \\
 & $1.10$ & $0.147 \pm0.004$
 & $3.84 \pm0.05$
 & $0.563 \pm0.007$
 \\
 & $1.20$ & $0.155 \pm0.002$
 & $3.55 \pm0.02$
 & $0.552 \pm0.004$
 \\
\hline
\end{tabular}\\
\end{center}
\caption[Bare parameters of HLS for $a(\Lambda)=1$ (1)]{%
The parameters $F_\pi(\Lambda)$, $g(\Lambda)$ 
and $M_\rho(\Lambda)$ in the case of $a(\Lambda)=1$ for
several values of $\Lambda_{\rm QCD}$ and $\Lambda$.
Units of $\Lambda_{\rm QCD}$, $\Lambda$, $F_\pi(\Lambda)$
and $M_\rho(\Lambda)$ are GeV.
The errors come from 
$\langle\bar{q}q\rangle=-(225\pm25\,\mbox{MeV})^3$.}
\label{cutval 1}
\end{table}
\begin{table}[htbp]
\begin{center}
\begin{tabular}{|c|c||c|c|}
\hline
 $\Lambda_{\rm QCD}$ & $\Lambda$ & $z_3(\Lambda)$ 
  & $z_2(\Lambda)-z_1(\Lambda)$ \\
\hline
 $0.30$
 & $1.00$ & $-13.7 \pm3.1$
 & $-1.63 \pm0.60$
 \\
 & $1.10$ & $-14.0 \pm2.2$
 & $-1.41 \pm0.32$
 \\
 & $1.20$ & $-14.3 \pm1.6$
 & $-1.24 \pm0.19$
 \\
\hline
 $0.35$
 & $1.00$ & $-13.4 \pm3.6$
 & $-1.58 \pm0.69$
 \\
 & $1.10$ & $-13.9 \pm2.5$
 & $-1.39 \pm0.37$
 \\
 & $1.20$ & $-14.3 \pm1.8$
 & $-1.24 \pm0.22$
 \\
\hline
 $0.40$
 & $1.00$ & $-13.2 \pm4.0$
 & $-1.52 \pm0.80$
 \\
 & $1.10$ & $-13.8 \pm2.8$
 & $-1.37 \pm0.43$
 \\
 & $1.20$ & $-14.3 \pm2.0$
 & $-1.23 \pm0.25$
 \\
\hline
 $0.45$
 & $1.00$ & $-12.9 \pm4.5$
 & $-1.45 \pm0.92$
 \\
 & $1.10$ & $-13.7 \pm3.2$
 & $-1.34 \pm0.49$
 \\
 & $1.20$ & $-14.2 \pm2.2$
 & $-1.23 \pm0.28$
 \\
\hline
\end{tabular}\\
\end{center}
\caption[Bare parameters of HLS for $a(\Lambda)=1$ (2)]{%
The parameters 
$z_3(\Lambda)$ and  $z_2(\Lambda) - z_1(\Lambda)$ in the case of 
$a(\Lambda)=1$ for 
several values of $\Lambda_{\rm QCD}$ and $\Lambda$.
Units of $\Lambda_{\rm QCD}$ and $\Lambda$ are GeV. 
Values of $z_3(\Lambda)$ and $z_2(\Lambda)-z_1(\Lambda)$
are scaled by a factor of $10^3$.
The errors come from 
$\langle\bar{q}q\rangle=-(225\pm25\,\mbox{MeV})^3$.}
\label{cutval 2}
\end{table}

Now we present prediction of the several physical quantities
for $a(\Lambda)=1$ 
using the above bare parameters with
the Wilsonian RGEs. 
The resultant values for 
$\left( \Lambda \,,\, \Lambda_{\rm QCD} \right) = 
(1.1\,,\,0.4)\,\mbox{GeV}$ are
\begin{eqnarray}
&&
  F_\pi(0) = 73.6 \pm 5.7 \,\mbox{MeV} \ ,
  \quad
  \left( 
    \left. F_\pi(0) \right\vert_{\rm exp} 
    = 86.4 \pm 9.7 \,\mbox{MeV}
  \right)
\ ,
\nonumber\\
&&
  g_\rho = 0.146 \pm 0.012 \,\mbox{GeV}^2 \ ,
  \quad
  \left(
    \left. g_\rho \right\vert_{\rm exp} 
    = 0.119 \pm 0.001 \,\mbox{GeV}^2 
  \right)
\ ,
\nonumber\\
&&
  g_{\rho\pi\pi} = 7.49 \pm 0.88  \ ,
  \quad
  \left(
    \left. g_{\rho\pi\pi} \right\vert_{\rm exp}
    = 6.00 \pm 0.01
  \right)
\ ,
\nonumber\\
&&
  L_9(m_\rho) = (7.07 \pm 0.35)\times 10^{-3} \ ,
  \quad
  \left(
    \left. L_9(m_\rho) \right\vert_{\rm exp}
    = (6.5 \pm 0.6 ) \times 10^{-3}
  \right)
\ ,
\nonumber\\
&&
  L_{10}(m_\rho) = (-7.94 \pm 0.84)\times 10^{-3} \ ,
  \quad
  \left(
    \left. L_{10}(m_\rho) \right\vert_{\rm exp}
    = (-5.1 \pm 0.7 ) \times 10^{-3}
  \right)
\ ,
\nonumber\\
&&
  a(0) = 2.04 \pm 0.16 \ .
\label{phys pre a1}
\end{eqnarray}
We show the dependences of the results on the several choices of
$\Lambda$ and $\Lambda_{\rm QCD}$ in Tables~\ref{tab:phys val 1}
and \ref{tab:phys val 2}.
\begin{table}[htbp]
\begin{center}
\begin{tabular}{|c|c||c|c|c|c|}
\hline
 $\Lambda_{\rm QCD}$ & $\Lambda$ 
 & $F_\pi(0)$ & $g_\rho$  & $g_{\rho\pi\pi}$ & $a(0)$ \\
\hline
 $0.30$
 & $1.00$ & $70.5 \pm6.8$
 & $0.144 \pm0.013$
 & $8.00 \pm1.19$
 & $2.14 \pm0.22$
 \\
 & $1.10$ & $66.6 \pm4.8$
 & $0.147 \pm0.010$
 & $8.73 \pm0.98$
 & $2.27 \pm0.18$
 \\
 & $1.20$ & $65.5 \pm3.3$
 & $0.149 \pm0.007$
 & $8.95 \pm0.71$
 & $2.31 \pm0.13$
 \\
\hline
 $0.35$
 & $1.00$ & $74.2 \pm7.5$
 & $0.143 \pm0.014$
 & $7.40 \pm1.12$
 & $2.03 \pm0.21$
 \\
 & $1.10$ & $70.0 \pm5.2$
 & $0.146 \pm0.011$
 & $8.09 \pm0.93$
 & $2.15 \pm0.17$
 \\
 & $1.20$ & $68.7 \pm3.7$
 & $0.149 \pm0.008$
 & $8.32 \pm0.68$
 & $2.20 \pm0.13$
 \\
\hline
 $0.40$
 & $1.00$ & $78.2 \pm8.2$
 & $0.143 \pm0.016$
 & $6.84 \pm1.06$
 & $1.92 \pm0.19$
 \\
 & $1.10$ & $73.6 \pm5.7$
 & $0.146 \pm0.012$
 & $7.49 \pm0.88$
 & $2.04 \pm0.16$
 \\
 & $1.20$ & $72.0 \pm4.0$
 & $0.149 \pm0.008$
 & $7.74 \pm0.65$
 & $2.09 \pm0.12$
 \\
\hline
 $0.45$
 & $1.00$ & $82.6 \pm8.9$
 & $0.143 \pm0.017$
 & $6.31 \pm0.99$
 & $1.83 \pm0.18$
 \\
 & $1.10$ & $77.5 \pm6.2$
 & $0.146 \pm0.013$
 & $6.93 \pm0.83$
 & $1.94 \pm0.15$
 \\
 & $1.20$ & $75.6 \pm4.4$
 & $0.149 \pm0.009$
 & $7.19 \pm0.62$
 & $1.99 \pm0.12$
 \\
\hline
\multicolumn{2}{|c||}{Exp.} &  
  $86.4\pm9.7$ & $0.119\pm0.001$ & $6.00\pm0.01$ & \\
\hline
\end{tabular}\\
\end{center}
\caption[Physical predictions of the Wilsonian matching 
  for $a(\Lambda)=1$]{
Physical quantities predicted by the Wilsonian matching
conditions and the Wilsonian RGEs for $a(\Lambda)=1$.
Units of $\Lambda_{\rm QCD}$ and $\Lambda$ are GeV. 
Unit of $F_\pi(0)$ is MeV and that of $g_\rho$ is GeV$^2$.
The errors come from
$\langle\bar{q}q\rangle=-(225.\pm25.\,\mbox{MeV})^3$. 
}\label{tab:phys val 1}
\end{table}
\begin{table}[htbp]
\begin{center}
\begin{tabular}{|c|c||c|c|c|}
\hline
 $\Lambda_{\rm QCD}$ & $\Lambda$ 
 & $L_9(m_\rho)$ & $L_{10}(m_\rho)$  
 & $g_\rho/(2g_{\rho\pi\pi}F_\pi^2(0)$ \\
\hline
 $0.30$
 & $1.00$ & $6.77 \pm0.38$
 & $-7.40 \pm0.86$
 & $1.81 \pm0.25$
 \\
 & $1.10$ & $6.71 \pm0.28$
 & $-7.62 \pm0.67$
 & $1.89 \pm0.19$
 \\
 & $1.20$ & $6.79 \pm0.21$
 & $-7.93 \pm0.50$
 & $1.94 \pm0.13$
 \\
\hline
 $0.35$
 & $1.00$ & $6.94 \pm0.42$
 & $-7.54 \pm0.96$
 & $1.76 \pm0.27$
 \\
 & $1.10$ & $6.88 \pm0.32$
 & $-7.77 \pm0.75$
 & $1.85 \pm0.20$
 \\
 & $1.20$ & $6.97 \pm0.23$
 & $-8.10 \pm0.56$
 & $1.90 \pm0.14$
 \\
\hline
 $0.40$
 & $1.00$ & $7.13 \pm0.47$
 & $-7.70 \pm1.07$
 & $1.71 \pm0.28$
 \\
 & $1.10$ & $7.07 \pm0.35$
 & $-7.94 \pm0.84$
 & $1.80 \pm0.21$
 \\
 & $1.20$ & $7.16 \pm0.25$
 & $-8.28 \pm0.63$
 & $1.86 \pm0.16$
 \\
\hline
 $0.45$
 & $1.00$ & $7.37 \pm0.51$
 & $-7.90 \pm1.19$
 & $1.66 \pm0.29$
 \\
 & $1.10$ & $7.28 \pm0.39$
 & $-8.13 \pm0.93$
 & $1.76 \pm0.22$
 \\
 & $1.20$ & $7.37 \pm0.28$
 & $-8.48 \pm0.70$
 & $1.82 \pm0.17$
 \\
\hline
\multicolumn{2}{|c||}{Exp.} &  $6.5\pm0.6$ & $-5.1\pm0.7$ & \\
\hline
\end{tabular}\\
\end{center}
\caption[Predicted values of $L_9$ and $L_{10}$ from 
  the Wilsonian matching for $a(\Lambda)=1$]{%
Values of Gasser-Leutwyler's parameters
$L_9$ and $L_{10}$ 
predicted by the Wilsonian matching
conditions and the Wilsonian RGEs
for $a(\Lambda)=1$.
Units of $\Lambda_{\rm QCD}$ and $\Lambda$ are GeV. 
Values of $L_9(m_\rho)$ and $L_{10}(m_\rho)$ are scaled by
a factor of $10^3$.
The errors come from 
$\langle\bar{q}q\rangle=-(225\pm25\,\mbox{MeV})^3$.
}\label{tab:phys val 2}
\end{table}
These show that
the choice $a(\Lambda)=1$ reproduces the experimental values
in reasonable agreement.\\

To close this section, we should emphasize that the 
inclusion of the quadratic divergences into the RGEs 
was essential in the present analysis.
{\it The RGEs with logarithmic divergence alone
would not be consistent with the matching to QCD.}
The bare parameter $F_\pi(\Lambda)=132 \sim 155$\,MeV listed in
Table~\ref{tab:cutval0}, which is derived by the matching condition
(\ref{match A}), is about double of the physical value
$F_\pi(0)=86.4$\,MeV.
The logarithmic running by the first term of 
Eq.~(\ref{RGE for Fpi2}) is not enough to change the value of $F_\pi$.
Actually, 
in 
the present procedure with logarithmic running
for $(\Lambda_{\rm QCD} \,,\,\Lambda)=(0.4\,,\,1.1)\,\mbox{GeV}$
we cannot find the parameters $a(\Lambda)$ and $g(\Lambda)$ which
reproduce $F_\pi(0)=86.4\,\mbox{MeV}$ and
$m_\rho=771.1\,\mbox{MeV}$.~\footnote{%
  For 
  $(\Lambda_{\rm QCD} \,,\,\Lambda)=(0.35\,,\,1.0)\,\mbox{GeV}$
  we find
  $a(\Lambda)$ and $g(\Lambda)$ which
  reproduce $F_\pi(0)=86.4\,\mbox{MeV}$ and
  $m_\rho=771.1\,\mbox{MeV}$ as
  $a(\Lambda)=0.27\pm0.70\pm0.49$ and
  $g(\Lambda)=4.93\pm1.00\pm0.69$.
  Then, we obtain
  $g_\rho = 0.53$\,GeV$^2$, $g_{\rho\pi\pi}=2.9$,
  $L_9^r(m_\rho)=15\times10^{-3}$ and 
  $L_{10}^r(m_\rho)=-30\times10^{-3}$.
  These badly disagree with experiment.
  Note that 
  the parameter choice $\Lambda=m_\rho$ does not work, either.
}

We should also
stress that the above success of the Wilsonian matching is
due to the existence of $\rho$ in the HLS.
If we did not include $\rho$ and used the current correlators in the
ChPT, we would have failed to match the effective field theory with
the underlying QCD.

\subsection{Predictions for QCD with $N_f=2$}
\label{ssec:PQNf2}

As we have stressed in Sec.~\ref{ssec:MHUQ},
the Wilsonian matching conditions determine the absolute values
and the explicit dependence of bare parameters of the HLS
on the parameters of underlying QCD such as $N_c$ as well as
$N_f$.
Especially, the current correlators derived from the OPE
has only small $N_f$-dependence, which implies that the
bare parameters of the HLS have also small $N_f$-dependence.
Then, the dependence of the physical quantities such as
the on-shell $\pi$ decay constant on $N_f$
mainly appears
through the Wilsonian RGEs which do depend on $N_f$.
In this subsection,
to show how the $N_f$-dependences of the physical quantities
appear in our framework,
we consider QCD with $N_f=2$.
This should be regarded as 
a {\it prediction} for an idealized world in the 
infinite strange quark mass limit ($m_s \rightarrow \infty$)
of the real world.

Before making a concrete analysis, let us make a rough estimation
as we have done around the beginning of Sec.~\ref{ssec:DBPHL}.
As we stressed, the Wilsonian matching condition (\ref{match A})
determines the value of bare $\pi$ decay constant at the
matching scale, $F_\pi(\Lambda)$.
Since the dominant contribution in the right-hand-side (RHS) of
Eq.~(\ref{match A}) is given by $1+\alpha_s/\pi$ term,
the $N_f$-dependence of the RHS is small:
The ratio $F_\pi^2(\Lambda)/\Lambda^2$ has small
dependence on $N_f$.
Then,
by using the matching scale as $\Lambda = 1.1\,\mbox{GeV}$,
the value of $F_\pi(\Lambda)$ for $N_f=2$ roughly
takes the same value
as that for $N_f=3$ as in 
Eq.~(\ref{Fpi Lambda rough estimate}):
\begin{equation}
F_\pi^2(\Lambda;N_f=2) \sim 3\, 
(86.4 \, \mbox{MeV})^2 \sim (150 \,\mbox{MeV})^2\ .
\label{Fpi Lambda rough estimate 2}
\end{equation}
Similarly to what we have done in Eq.~(\ref{Fpi 0 rough estimate}),
we neglect the logarithmic divergence with taking $a=1$ in
the RGE for $F_\pi^2$ in Eq.~(\ref{RGE for Fpi2})
to perform a crude estimate of the on-shell $\pi$ decay
constant $F_\pi(0;N_f=2)$.
The result is given by
\begin{eqnarray}
F_\pi^2(0;N_f=2) 
&\sim& F_\pi^2(\Lambda) - \frac{N_f}{2(4\pi)^2} \Lambda^2 
\nonumber\\
&\sim&
\frac{\Lambda^2}{8\pi^2} 
\left[ 
  \frac{N_c}{3} \left( 1 + \delta_A \right) - \frac{N_f}{4} 
\right]
\nonumber\\
&\sim&
2 \,
\left(\frac{\Lambda}{4\pi}\right)^2 \sim 
\frac{2}{3} F_\pi^2 (\Lambda) 
\sim (120\,\mbox{MeV})^2 \
\ ,
\end{eqnarray}
where we adopted $\delta_A\sim 0.5$ and $N_c=3$ as in 
Eq.~(\ref{Fpi 0 rough estimate}) but $N_f=2$ to obtain the
last line.
This implies that the on-shell $\pi$ decay constant for $N_f=2$
is about 20\% bigger than that for $N_f=3$ even though the
bare ones have the same values.

For determining all the bare parameters 
through the Wilsonian matching and making more precise
predictions
we need to determine the current correlators in the OPE.
In addition to three Wilsonian matching conditions shown in
Eqs.~(\ref{match A}), (\ref{match V}) and (\ref{match z}),
we need two inputs to determine five relevant
bare parameters.
As we discussed above, the current correlators in the OPE have
only small $N_f$-dependence.
So, 
we assume that the bare parameters for $N_f=2$
are the same as those 
obtained in Sec.~\ref{ssec:DBPHL} for $N_f=3$ QCD.
Then, we
obtain the parameters in the low-energy region
through the Wilsonian RGEs
with $N_f=2$ 
and give predictions on several physical quantities.
We expect that the predictions will not be so much different 
from the ``physical quantities'' obtained in the idealized
QCD with $N_f=2$, which can be checked by, e.g., the lattice
simulation.
Note that the $\rho$ mass $m_\rho(N_f=2)$ here is 
not an input, but an output determined from the
on-shell condition 
in Eq.~(\ref{on-shell condition}).
Similarly, the on-shell $\pi$ decay constant $F_\pi(0;N_f=2)$ is
also an output derived by Eq.~(\ref{rel: Fp 0 Fp mr}).

For definiteness of the analysis, let us use the
bare parameters determined in $N_f=3$ QCD for
$(\Lambda_{\rm QCD}\,,\,\Lambda) = (0.4\,,\,1.1)\,\mbox{GeV}$.
We pick up the values from Tables~\ref{tab:WM bare}
and \ref{tab:WM bare 2}, and show them
in Table~\ref{tab:Nf2 cutdat}.
\begin{table}[htbp]
\begin{center}
\begin{tabular}{|c|c|c|}
\hline
 $F_\pi(\Lambda)$ & $a(\Lambda)$ & $g(\Lambda)$ \\
\hline
 $0.145\pm0.003$ & $1.33\pm0.28\pm0.14$ & $3.69\pm0.13\pm0.02$ \\
\hline
\end{tabular}
\begin{tabular}{|c|c|}
 $z_3(\Lambda)$ & $z_2(\Lambda)-z_1(\Lambda)$ \\
\cline{1-2}
 $-5.84\pm5.78\pm0.18$ & $-1.79\pm0.34\pm0.61$ \\
\cline{1-2}
\end{tabular}
\end{center}
\caption[Bare parameters for $N_f=2$ QCD]{%
Bare parameters used in the present analysis for $N_f=2$ QCD.
These are obtained in $N_f=3$ QCD for
$(\Lambda_{\rm QCD}\,,\,\Lambda) = (0.4\,,\,1.1)\,\mbox{GeV}$
(see Tables~\ref{tab:WM bare} and \ref{tab:WM bare 2}).
}\label{tab:Nf2 cutdat}
\end{table}
In Table~\ref{Nf 2 predictions}
we show the
physical predictions obtained from these bare parameters
through the Wilsonian RGEs 
(\ref{RGE for Fpi2}),
(\ref{RGE for a}), (\ref{RGE for g2}), (\ref{RGE:z1}), 
(\ref{RGE:z2}) and (\ref{RGE:z3})
together with the on-shell condition
(\ref{on-shell condition}) and the relation (\ref{rel: Fp 0 Fp mr})
for the on-shell $\pi$ decay constant.
\begin{table}[htbp]
\begin{center}
\begin{tabular}{|c|c|}
\hline
 $F_\pi(0;N_f=2)$ & $m_\rho(N_f=2)$ \\
\hline
 $0.106\pm0.005\pm0.002$ & $0.719\pm0.012\pm0.004$ \\
\hline
\end{tabular}
\begin{tabular}{|c|c|c|}
\hline
 $g_\rho(N_f=2)$ & $g_{\rho\pi\pi}(N_f=2)$ & $a(0;N_f=2)$ \\
\hline
 $0.116\pm0.005\pm0.002$ & $4.30\pm0.18\pm0.27$  
  & $1.61\pm0.23\pm0.13$ \\
\hline
\end{tabular}
\begin{tabular}{|c|c|}
\hline
 $L_9^r(m_\rho;N_f=2)$ & $L_{10}^r(m_\rho;N_f=2)$ \\
\hline
 $9.59\pm0.28\pm0.27$ & $-8.26\pm1.92\pm0.36$ \\
\hline
\end{tabular}
\end{center}
\caption[Predictions for $N_f=2$ QCD]{%
Several predictions for physical quantities in QCD with $N_f=2$
done in the present analysis from the bare parameters listed
in Table~\ref{tab:Nf2 cutdat} through the Wilsonian RGEs.
Units of $F_\pi(0;N_f=2)$ and $m_\rho(N_f=2)$ are GeV, and
that of $g_\rho$ is GeV$^2$.
Values of $L_9^r(m_\rho)$ and $L_{10}^r(m_\rho)$ are 
scaled by a factor of $10^3$.
Note that the first and second errors correspond to
those of the bare parameters in Table~\ref{tab:Nf2 cutdat}.
}\label{Nf 2 predictions}
\end{table}
As we discussed above, the value of the $\pi$ decay constant,
predicted as 
\begin{equation}
F_\pi(0;N_f=2)=106\,\mbox{MeV} \ ,
\end{equation}
is about 20\% larger than that for $N_f=3$ 
QCD, $F_\pi(0;N_f=3)=86.4\,\mbox{MeV}$.
One might think that the value $F = 88\,\mbox{MeV}$ estimated
in Ref.~\cite{Gas:84} is the value of the pion decay constant
for $N_f=2$ QCD at chiral limit.
However, this value is estimated from the experimental value
by taking the limit of $m_\pi=0$ with $m_K\simeq500\,\mbox{MeV}$ 
kept unchanged.
Here we mean by $N_f=2$ QCD the QCD with $m_u=m_d=0$ but
$m_s=\infty$, i.e., $m_\pi=0$ but $m_K=\infty$.
In our best knowledge, there is no estimation done before
for the pion decay constant in $N_f=2$ QCD.
But the fact that
the value $F = 88\,\mbox{MeV}$ for $m_\pi=0$ but 
$m_K\simeq500\,\mbox{MeV}$ 
is slightly larger than 
$F_\pi(0;N_f=3)=86.4\,\mbox{MeV}$ for $m_\pi=m_K=0$
indicates that increase of 20\% may be possible when we
change the value of $m_s$ (thus $m_K$) from zero to infinity.

On the other hand,
the $\rho$ mass is predicted as
\begin{equation}
m_\rho(N_f=2) = 719\,\mbox{MeV}\ ,
\end{equation}
which is about 10\% smaller than
the experimental value $m_\rho(N_f=3)=771.1\,\mbox{MeV}$.
This is mainly due to the smallness
of
the HLS gauge coupling $g(m_\rho)$:
The present analysis provides 
$g(m_\rho;N_f=2) = 5.33\pm0.53\pm0.10$ to be compared with
$g(m_\rho;N_f=3) = 6.27\pm0.69\pm0.11$ in Table~\ref{tab:WM mrho}.
Accordingly, the absolute values of $L_9$ and $L_{10}$ becomes
larger since their main parts are determined by
$1/g^2(m_\rho)$.
Finally, the predicted value of $a(0)$ shows that
there exists the deviation from 2, which implies that
the vector dominance (VD) is violated in $N_f=2$ QCD.
This also implies that
the VD in the real world (QCD with $N_f=3$)
can be realized only accidentally
(see Secs.~\ref{ssec:PSH} and \ref{sssec:VDLNQ}).

\subsection{Spectral function sum rules}
\label{ssec:SR}

In this subsection we study 
the spectral function sum rules
(the Weinberg sum rules and 
Das-Mathur-Okubo sum rule), which are related to the vector and
axialvector current correlators.

The spectral function sum rules are given by
\begin{eqnarray}
&&
  \int_0^\infty \frac{ds}{s} \left[ \rho_V(s) - \rho_A(s) \right]
=
- 4 \bar{L}_{10} 
\ ,
\label{DMO SR}
\\
&&
  \int_0^\infty ds
  \left[ \rho_V(s) - \rho_A(s) \right] 
  = F_\pi^2 \ ,
\label{WSR 1}
\\
&&
  \int_0^\infty ds \, s
  \left[ \rho_V(s) - \rho_A(s) \right] 
  = 0 \ ,
\label{WSR 2}   
\end{eqnarray}
where $\bar{L}_{10}$ is a constant which corresponds to
the so-called $S$ parameter in the electroweak 
theory~\cite{Holdom-Terning,Peskin-Takeuchi:PRL,Peskin-Takeuchi:PRD,%
Marciano-Rosner,Altarelli-Barbieri} 
as $\bar{L}_{10} \rightarrow - S/(16\pi)$.
The relations in Eqs.~(\ref{WSR 1}) and (\ref{WSR 2}) are called
the Weinberg's first and second sum rules~\cite{Wei:SR}, 
respectively,
and we call 
the relation in Eq.~(\ref{DMO SR})
Das-Mathur-Okubo (DMO) sum rule~\cite{DMO:1}.
In the above expressions, 
$\rho_V(s)$ and $\rho_A(s)$ are the spin $1$ parts of the
spectral functions of the
vector and axialvector currents.
These spectral functions are defined by
\begin{eqnarray}
&&
  \frac{1}{2\pi} \int d^4x e^{i p x}
  \langle 0 \vert J_\mu^a(x) J_\nu^b (0) \vert 0 \rangle
  = \delta^{ab} ( p_\mu p_\nu - g_{\mu\nu} p^2 )
  \rho_V(p^2)
  \ ,
\\
&&
  \frac{1}{2\pi} \int d^4x e^{i p x}
  \langle 0 \vert J_{5\mu}^a(x) J_{5\nu}^b (0) \vert 0 \rangle
  = \delta^{ab} p_\mu p_\nu \rho_{A}^0(p^2)
  + \delta^{ab} ( p_\mu p_\nu - g_{\mu\nu} p^2 )
  \rho_A(p^2)
  \ ,
\end{eqnarray}
where $\rho_A^0(p^2)$ is the spin $0$ part.
By using these spectral functions, 
the $g^{\mu\nu}$-term and the $p^\mu p^\nu$-term of
the $VV-AA$ current correlator
are expressed as
\begin{eqnarray}
\Pi_{V-A}^{(1)}(-p^2)
&=&
\frac{1}{- p^2}
\int d s \frac{
  s \left\{ \rho_V(s) - \rho_A (s) \right\}
}{ s - p^2 - i \epsilon }
\ ,
\label{spec Pi 1}
\\
\Pi_{V-A}^{(2)}(-p^2)
&=&
\int d s \frac{
  \rho_V(s) - \rho_A (s) - \rho_A^0(s)
}{ s - p^2 - i \epsilon }
\ ,
\label{spec Pi 2}
\end{eqnarray}
where $\Pi_{V-A}^{(1)}$ and $\Pi_{V-A}^{(2)}$ are related to the
$VV-AA$ current correlator as
\begin{eqnarray}
&&
i \int d^4 x e^{ipx}
\left[
  \left\langle 0 \left\vert T\, J_{\mu}^a (x) J_{\nu}^b (0)
  \right\vert 0 \right\rangle 
  -
  \left\langle 0 \left\vert T\, J_{5\mu}^a (x) J_{5\nu}^b (0)
  \right\vert 0 \right\rangle 
\right]
\nonumber\\
&& \quad
=
\delta^{ab} 
\left[ 
  p_\mu p_\nu \Pi_{V-A}^{(2)}(-p^2) 
  - g_{\mu\nu} p^2 \Pi_{V-A}^{(1)}(-p^2)
\right]
\ .
\end{eqnarray}
Note that 
both
$\Pi_{V-A}^{(1)}$ and $\Pi_{V-A}^{(2)}$ agree with
$\Pi_V - \Pi_A$ defined in Eq.~(\ref{A V correlators})
when the current conservation is satisfied in the chiral limit
(massless current quark):
\begin{equation}
\Pi_{V-A}^{(1)}(-q^2) = \Pi_{V-A}^{(2)}(-p^2)
= \Pi_V(-p^2) - \Pi_A(-p^2) \ ,
\quad (\mbox{for}\ m_q = 0) \ .
\end{equation}

For the convergence of the above sum rules
a crucial role is played by 
the asymptotic behavior of the spectral functions
which is rephrased by the requirement for the
high energy behavior of the $VV - AA$ current correlator:
The convergence of the sum rules in 
Eqs.~(\ref{DMO SR}), (\ref{WSR 1}) and (\ref{WSR 2}), respectively,
requires that
the $VV - AA$ current correlator
must satisfy
\begin{eqnarray}
&&
\Pi_{V-A}^{(1)}(Q^2) =
\Pi_V(Q^2) - \Pi_A(Q^2) 
\ \mathop{\longrightarrow}_{Q^2 \rightarrow \infty}\ 
0 \ ,
\label{conv cond 0}
\\
&&
Q^2 \Pi_{V-A}^{(2)}(Q^2) =
Q^2 \left[ \Pi_V(Q^2) - \Pi_A(Q^2) \right]
\ \mathop{\longrightarrow}_{Q^2 \rightarrow \infty}\ 
0 \ ,
\label{conv cond 1}
\\
&&
Q^4 \Pi_{V-A}^{(1)}(Q^2) =
Q^4 \left[ \Pi_V(Q^2) - \Pi_A(Q^2) \right]
\ \mathop{\longrightarrow}_{Q^2 \rightarrow \infty}\ 
0 
\ ,
\label{conv cond 2}
\end{eqnarray}
where $Q^2 = - p^2$.~\footnote{%
  When we wrote the dispersive form as in Eq.~(\ref{spec Pi 1})
  with no subtraction,
  we implicitly assumed the converge condition in 
  Eq.~(\ref{conv cond 0}).  Then, the form in Eq.~(\ref{spec Pi 1})
  automatically satisfies Eq.~(\ref{conv cond 0}).
}
It should be noticed that 
the $VV-AA$ current correlator obtained by the OPE in QCD
satisfies in the chiral limit all the above convergence conditions as
[see Eqs.~(\ref{Pi A OPE}) and (\ref{Pi V OPE})]
\begin{equation}
\Pi_V^{\rm(QCD)}(Q^2) - \Pi_A^{\rm(QCD)}(Q^2)
\ \mathop{\longrightarrow}_{Q^2 \rightarrow \infty}\ 
- \frac{4\pi(N_c^2-1)}{ N_c^2 }
\frac{\alpha_s \left\langle \bar{q}q \right\rangle^2}{Q^6}
\ .
\end{equation}

Provided the above convergence conditions, 
the DMO sum rule and the first Weinberg sum rule are rewritten as
the following relations in the low-energy region:
\begin{eqnarray}
&&
\left.
  \frac{d}{d Q^2} \left[ Q^2 \Pi_V(Q^2) - Q^2 \Pi_A(Q^2) \right]
\right\vert_{Q^2=0}
= - 4 \bar{L}_{10}
\ ,
\label{DMO b}
\\
&&
\left.
  \left[ - Q^2 \Pi_V(Q^2) + Q^2 \Pi_A(Q^2) \right]
\right\vert_{Q^2=0}
= F_\pi^2
\ .
\label{WSR 1 b}
\end{eqnarray}
It should be noticed that there is an infrared divergence
in Eq.~(\ref{DMO b}) coming from the $\pi$-loop contribution.
To regularize the infrared divergence we introduce the $\pi$ mass
when we consider the DMO sum rule.~\footnote{%
  As can be seen in, e.g., 
  Refs.~\cite{Bernard-Duncan-LoSecco-Weinberg,Yamawaki:82},
  introduction of the $\pi$ mass, or equivalently the current
  quark mass, changes the higher energy behavior of the current
  correlators in such a way that the convergence conditions
  in Eqs.~(\ref{conv cond 1}) and (\ref{conv cond 2}) are not 
  satisfied 
  while that in Eq.~(\ref{conv cond 0}) is still satisfied.
  Thus, we do not include the $\pi$ mass when we consider the
  first and second Weinberg's sum rules, while for the DMO sum rule
  we include it as an infrared regulator.
}
In such a case, the constant $\bar{L}_{10}$ is related to 
the axialvector form factor 
$F_A$ of 
$\pi\rightarrow \ell^+\nu \gamma$ studied in
Sec.~\ref{ssec:l10}
and 
the charge radius of pion 
$\left\langle r^2 \right\rangle_V^{\pi^\pm}$
studied in Sec.~\ref{ssec:l9} as~\cite{DMO:1}
\begin{equation}
- 4 \bar{L}_{10} = - \frac{F_\pi F_A}{\sqrt{2}m_{\pi^\pm}}
  + \frac{F_\pi^2}{3} \left\langle r^2 \right\rangle_V^{\pi^\pm}
\ ,
\label{DMU l10}
\end{equation}
which is related to the ChPT parameter $L_{10}^r(\mu)$ in
Sec.~\ref{ssec:l10} as~\footnote{%
  We can check the validity of Eq.~(\ref{l10 rel}) for $N_f=3$, 
  especially the second term in the square bracket by substituting
  the expression of $F_A$ in Eq.~(\ref{FA l9 l10}) and
  that of $\langle r^2 \rangle_V^{\pi^{\pm}}$ in
  Eq.~(\ref{charge radius pi}) with $m_K = m_\pi$ 
  into Eq.~(\ref{DMU l10}).
}
\begin{equation}
- 4 \bar{L}_{10} = - 4 L_{10}(\mu)
- \frac{N_f}{6(4\pi)^2} 
  \left[ \ln \frac{m_\pi^2}{\mu^2} + 1  \right]
\ .
\label{l10 rel}
\end{equation}

Let us show how the DMO sum rule and the first and second
Weinberg's sum rules are satisfied in the present
approach.
As we have shown above, the spectral function sum rules under
consideration
are
equivalent to the combination of the 
convergence conditions (\ref{conv cond 0})--(\ref{conv cond 2})
and the low-energy relations (\ref{DMO b}) and (\ref{WSR 1 b}).
In the following, therefore,
we consider only the current correlators.

In the present approach
we switch the theory from the HLS to QCD at the matching scale
$\Lambda$.
In other words,
in the energy region below $\Lambda$ we use the HLS, while in the 
energy region above $\Lambda$ we use QCD.
Then, the vector and axialvector current correlators may be  expressed 
as~\footnote{
  More precisely, our matching conditions Eqs.~(\ref{match z}), 
  (\ref{match A}) and (\ref{match V}) read:
  \begin{eqnarray}
    \lim_{Q^2 \rightarrow \Lambda^2 - \epsilon}
    \left[ \Pi_V(Q^2) - \Pi_A(Q^2) \right]
    &=&
    \lim_{Q^2 \rightarrow \Lambda^2 + \epsilon}
    \left[ \Pi_V(Q^2) - \Pi_A(Q^2) \right]
    \ ,
  \nonumber\\
    \lim_{Q^2 \rightarrow \Lambda^2 - \epsilon}
    \frac{d}{d Q^2} \Pi_{V,A}(Q^2)
    &=&
    \lim_{Q^2 \rightarrow \Lambda^2 + \epsilon}
    \frac{d}{d Q^2} \Pi_{V,A}(Q^2)
    \ .
  \nonumber
  \end{eqnarray}
  Note that a low-energy expansion of the $\Pi_{V,A}^{\rm(HLS)}(Q^2)$
  is in  
  {\it positive} powers of $Q^2$, while the high-energy 
  expansion or the OPE of the $\Pi_{V,A}^{\rm(QCD)}(Q^2)$ is 
  in 
  {\it negative} power of $Q^2$. Our matching condition thus 
  is a best compromise between these two with different $Q^2$ 
  behaviors.
}
\begin{equation}
\Pi_{V,A}(Q^2) = 
   \theta( \Lambda^2 - Q^2 ) \Pi_{V,A}^{\rm(HLS)}(Q^2)
 + \theta( Q^2 - \Lambda^2 ) \Pi_{V,A}^{\rm(QCD)}(Q^2)
\ ,
\label{Pi VA HLS QCD}
\end{equation}
where $\Pi_{V,A}^{\rm(HLS)}(Q^2)$ are the correlators obtained
by the HLS in the energy region below $\Lambda$ and
$\Pi_{V,A}^{\rm(QCD)}(Q^2)$ are those obtained by QCD
in the energy region above $\Lambda$.
Thus we have
\begin{equation}
\Pi_V(Q^2) - \Pi_A(Q^2)
\ \mathop{\longrightarrow}_{Q^2 \rightarrow \infty}\ 
\Pi_V^{\rm(QCD)}(Q^2) - \Pi_A^{\rm(QCD)}(Q^2)
\ \mathop{\longrightarrow}_{Q^2 \rightarrow \infty}\ 
- \frac{4\pi(N_c^2-1)}{ N_c^2 }
\frac{\alpha_s \left\langle \bar{q}q \right\rangle^2}{Q^6}
\ .
\end{equation}
This implies that the current correlators in the present
approach satisfy 
all the convergence conditions
(\ref{conv cond 0})--(\ref{conv cond 2}).
The fact that the $VV-AA$ current correlator satisfies the
convergence condition in Eq.~(\ref{conv cond 2}) already
implies that the second Weinberg's sum rule is satisfied
in the present approach.

The next issue for showing the DMO sum rule and the
first Weinberg's sum rule
is to check whether or not
the low-energy relations 
(\ref{DMO b}) and (\ref{WSR 1 b}) are satisfied.
For this purpose
we consider the vector and axialvector current correlators in the
HLS.
In the HLS at one loop the vector and axialvector current correlators
are given by [see also Eqs.~(\ref{Pi V HLS p2}) and
(\ref{Pi A HLS full})]
\begin{eqnarray}
&&
\Pi_V^{\rm (HLS)} (-p^2) = 
\frac{\Pi_{V}^S(p^2)}{\Pi_{V}^S(p^2) + p^2 \Pi_{V}^T(p^2)}
\left[
- \Pi_{V}^T(p^2) - 2 \Pi_{V \parallel}^T(p^2)
\right]
- \Pi_{\parallel}^T(p^2)
\ ,
\label{Pi V HLS p2 2}
\\
&&
\Pi_A^{\rm (HLS)} (-p^2) = 
\frac{\Pi^S_{\perp}(0)}{-p^2} 
- \tilde{\Pi}^{S}_{\perp}(p^2) - \Pi^T_{\perp}(p^2)
\ ,
\label{Pi A HLS p2 2}
\end{eqnarray}
where
\begin{equation}
\tilde{\Pi}^{S}_{\perp}(p^2) = 
\frac{ \Pi_{\perp}^S (p^2) - \Pi_{\perp}^S (0)}{p^2} \ .
\end{equation}
Since
$\Pi^S_{\perp}(0) = F_\pi^2(0)$,
the low-energy relation (\ref{WSR 1 b}) is satisfied,
which together with the convergence condition in 
Eq.~(\ref{conv cond 1}) implies that
the first Weinberg's sum rule is actually satisfied in the 
present approach.
By using Eq.~(\ref{Pi A HLS small p2}) and
Eq.~(\ref{Pi V HLS small p2}) together with
Eq.~(\ref{Pi T HLS approx}),
the DMO sum rule Eq. (\ref{DMO b}) takes the form:
\begin{equation}
- 4 \bar{L}_{10} \simeq
 \frac{1}{g^2(m_\rho)} - 2 z_3(m_\rho) - 2 z_1(m_\rho)
 + \Pi^{S \prime}_{\perp}(0) + 2 z_2(m_\rho)
  - \frac{1}{6} \frac{N_f}{(4\pi)^2} \ln \frac{m_\pi^2}{m_\rho^2}
\ ,
\label{DMO HLS}
\end{equation}
where we put the $\pi$ mass $m_\pi$ to regularize the infrared
divergence.
By putting the HLS parameters determined 
in Sec.~\ref{fullanalysis}
into the right-hand-side of Eq.~(\ref{DMO HLS}),
the value of $\bar{L}_{10}$ for 
$(\Lambda\,,\,\Lambda_{\rm QCD}) = (1.1\,,\,0.4)\,\mbox{GeV}$
is estimated as
\begin{equation}
\bar{L}_{10} = ( -8.5 \pm 2.5 \pm 0.6 ) \times 10^{-3}
\ ,
\end{equation}
where the first error comes from the error of the input value 
of $F_\pi(0)$; $F_\pi(0) = 86.4\pm 9.7\,\mbox{MeV}$, and the second
error from that of $\langle\bar{q}q\rangle$,
$\left\langle\bar{q}q\right\rangle_{\rm 1GeV} = 
 (-225\pm25\,\mbox{MeV})^3$.
This is to be compared with the experimental value
\begin{equation}
\left. \bar{L}_{10} \right\vert_{\rm exp}
= (-7.0 \pm 0.2 ) \times 10^{-3} \ ,
\end{equation}
obtaied by substituting the experimental values of 
$F_A$ given in Eq.~(\ref{exp FA}) and 
$\left\langle r^2 \right\rangle_V^{\pi^\pm} = 0.455 \pm 0.005 
 \, (\mbox{fm})^2$ from the most recent data~\cite{Erk:87}
in Table~\ref{tab:radii}
into Eq.~(\ref{DMU l10}).

Here, 
let us consider the pole-saturated form of the 
sum rules which are usually saturated by $\pi$, $\rho$ and $a_1$.
When we assume that the vector and the axialvector current correlators
are saturated by $\pi$, $\rho$ and $a_1$, they are expressed as
\begin{eqnarray}
\Pi_V^{\rm(pole)}(-p^2) 
&=&
\frac{ \left(g_\rho/m_\rho\right)^2 }
  { m_\rho^2 - p^2 - i\epsilon }
\ ,
\nonumber\\
\Pi_A^{\rm(pole)}(-p^2) 
&=&
\frac{ F_\pi^2 }{ -p^2 - i\epsilon}
+ 
\frac{ \left(g_{a_1}/m_{a_1}\right)^2 }
  { m_{a_1}^2 - p^2 - i\epsilon }
\ ,
\label{Pi VA pole saturation}
\end{eqnarray}
where $F_\pi$, $m_\rho$, $m_{a_1}$,
$g_\rho$ and $g_{a_1}$ are the parameters at the on-shell of
corresponding particles.
Note that
the above forms 
written by the on-shell parameters
are valid only around the on-shell of the relevant particles,
and that
we have no guarantee to use the same forms 
in the off-shell region, especially in the high-energy region.
Nevertheless, as customarily done, we may assume that the above forms
are valid even in the high-energy region.
In such a case, the above correlators must satisfy
the convergence conditions in 
Eqs.~(\ref{conv cond 0})--(\ref{conv cond 2}) as well as the
low-energy relations in Eqs.~(\ref{DMO b}) and (\ref{WSR 1 b}).
As we can see easily,
the above correlators satisfy
the convergence condition (\ref{conv cond 0}) 
corresponding to the DMO sum rule.
On the other hand,
the convergence conditions (\ref{conv cond 1}) and (\ref{conv cond 2})
corresponding to the first and second Weinberg's sum rule
require that the parameters in Eq.~(\ref{Pi VA pole saturation})
must satisfy
\begin{eqnarray}
&&
\frac{g_\rho^2}{m_\rho^2} = F_\pi^2 + \frac{g_{a_1}^2}{m_{a_1}^2}
\ ,
\label{WSR 1 p}
\\
&&
g_\rho^2 = g_{a_1}^2 
\ .
\label{WSR 2 p}
\end{eqnarray}
Equation~(\ref{WSR 1 p}) implies that 
the 
low-energy relation (\ref{WSR 1 b}) corresponding to the first
Weinberg's sum rule is already satisfied.
The low-energy relation (\ref{DMO b}) corresponding to the
DMO sum rule
is satisfied as
\begin{equation}
- 4 \bar{L}_{10}=
\frac{g_\rho^2}{ m_\rho^4} - \frac{ g_{a_1}^2 }{ m_{a_1}^4 }
- \frac{1}{6} \frac{N_f}{(4\pi)^2} \ln \frac{m_\pi^2}{m_\rho^2}
\ ,
\label{DMO pole}
\end{equation}
where we added the last term to include the possible contribution
from the $\pi$ loop with the infrared regularization.
Equations (\ref{DMO pole}), (\ref{WSR 1 p}) and (\ref{WSR 2 p})
are the pole saturated forms of the DMO sum rule and the
first and the second Weinberg's sum rules.

One might think that 
the spectral function
sum rules in Eqs.~(\ref{DMO SR}), (\ref{WSR 1}) and (\ref{WSR 2})
always lead to
the above relations in 
Eqs.~(\ref{DMO pole}), (\ref{WSR 1 p}) and (\ref{WSR 2 p}),
and hence the existence of $a_1$ meson is inevitable.
However, it is not true: It is merely the peculiarity of
the assumption of the pole saturation.
In our approach, on the other hand,
we have demonstrated thet the sum rules are saturated in a different
manner 
without $a_1$ meson which is heavier than the scale $\Lambda$.
Then, 
it does not make sense to consider the above relations in
Eqs.~(\ref{DMO pole}), (\ref{WSR 1 p}) and (\ref{WSR 2 p})
in the framework of the present approach.
Nevertheless, 
it may be worth showing
how the $a_1$ contribution in the pole saturated form
is numerically reproduced in the present approach.
Using the definition of $g_\rho$ given in Eq.~(\ref{g rho}) 
together with the definition of the on-shell $\rho$ mass
$m_\rho^2 = g^2(m_\rho) F_\sigma^2(m_\rho)$,
we obtain
\begin{equation}
\frac{g_\rho^2}{m_\rho^4} \simeq
\frac{1}{g^2(m_\rho)} - 2 z_3(m_\rho) \ ,
\label{rho cont DMO}
\end{equation}
where we neglected the higher order corrections.
Comparing Eq.~(\ref{DMO pole}) with
Eq.~(\ref{DMO HLS}) and using Eq.~(\ref{rho cont DMO}),
we see the following correspondence:
\begin{equation}
\frac{g_{a_1}^2}{m_{a_1}^4} 
\ \Leftrightarrow\ 
2 \left[ z_1(m_\rho) - z_2(m_\rho) \right] - 
\Pi^{S \prime}_{\perp}(0) 
\ .
\end{equation}
This implies that the $a_1$ contribution in 
the pole saturated form 
of the DMO sum rule in Eq.~(\ref{DMO pole})
is numerically 
imitated by especially the $\pi$-$\rho$ loop 
contribution~\cite{Tanabashi}
expressed by $\Pi^{S \prime}_{\perp}(0)$ 
in the present approach.
In a similar way,
the $\rho$-$\pi$ loop contribution
does yield additional
contribution to the axialvector correlator,
as shown by $\tilde{\Pi}^{S}_{\perp}(p^2)$ in 
Eq.~(\ref{Pi A HLS p2 2}).
This actually gives an imaginary part
(i.e., the additional contribution to the spectral function) 
above the $\rho$-$\pi$ threshold and hence mimic 
the $a_1$ pole effects in the first and second Weinberg's sum
rules.

As we have shown above,
while the current correlators obtained
in our approach within the framework
of the HLS do satisfy the spectral function sum rules,
the 
pole saturated form of the first and second Weinberg's
sum rules
are not generally
reproduced as it stands
since $a_1$ is not explicitly included
in our approach.
Nevertheless,
there is a special limit where the pole saturated forms without
$a_1$ contribution, i.e.,
\begin{eqnarray}
&&
\frac{g_\rho^2}{m_\rho^2} = F_\pi^2(0)\ ,
\label{WSR 1 VM}
\\
&&
g_\rho^2 = 0
\ ,
\label{WSR 2 VM}
\end{eqnarray}
are well reproduced.
This in fact occurs at the limit of 
the Vector Manifestation (VM)
which will be 
studied in detail in Sec.~\ref{sec:VM}.
In the VM, 
the chiral symmetry is restored at the critical point by the massless
degenerate $\pi$ and the $\rho$ as the chiral partner, which is
characterized by [see Eq.~(\ref{VM def}) as well as Eq.~(\ref{a0})]
\begin{eqnarray}
&& F_\pi^2(0) \rightarrow 0 \ , \quad
m_\rho^2 \rightarrow m_\pi^2=0\ , \quad
a(0) = 
F_\sigma^2(m_\rho) / F_\pi^2(0) \rightarrow 1 \ ,
\label{VM def sec5}
\end{eqnarray}
where $F_\sigma(m_\rho)$ is the decay constant of $\sigma$
(longitudinal $\rho$) at $\rho$ on-shell.
As we will show in Sec.~\ref{sec:VM},
the VM is realized within the framework of the HLS
due to the fact that,
at the chiral restoration point,
the bare parameters of the HLS determined from the 
Wilsonian matching satisfy the
VM conditions 
given in 
Eqs.~(\ref{vector condition:g})--(\ref{vector condition:fp})
which lead to the following condition
for the
parameter $g(m_\rho)$ at the on-shell of $\rho$
[see Eq.~(\ref{vector condition:on-shell g})]:
\begin{eqnarray}
&& g(m_\rho) \rightarrow 0 \ .
\label{v cond:g:sec5}
\end{eqnarray}
Using the expression of $g_\rho$
in terms of the parameters of the HLS
given in Eq.~(\ref{g rho})
and the $\rho$ on-shell condition
$m_\rho^2 = g^2(m_\rho) F_\sigma^2(m_\rho)$
together with Eqs.~(\ref{v cond:g:sec5}) and
(\ref{VM def sec5})
we obtain
\begin{equation}
\frac{g_\rho^2}{m_\rho^2} \, \frac{1}{F_\pi^2(0)}
=
\frac{F_\sigma^2(m_\rho)}{F_\pi^2(0)}
\left[ 1 - 2 g^2(m_\rho) z_3(m_\rho) \right]
\ \rightarrow\ 
1
\ .
\end{equation}
This implies that the pole saturated form of the first Weinberg's
sum rule without $a_1$ given in Eq.~(\ref{WSR 1 VM}) is 
actually satisfied at the VM limit.
Furthermore,
Eq.~(\ref{v cond:g:sec5}) already implies that the
second Weinberg's sum rule 
is satisfied 
even without $a_1$ contribution
at the VM limit:
\begin{equation}
g_\rho^2 = 
g^2(m_\rho) \left[ 1 - 2 g^2(m_\rho) z_3(m_\rho) \right]
\ \rightarrow \ 0 \ .
\end{equation}

\newpage

\section{Vector Manifestation}
\label{sec:VM}

Chiral symmetry restoration (Wigner realization of chiral symmetry) is
an outstanding phenomenon expected in QCD under extreme conditions
such as the finite temperature and/or density
(for reviews, see, e.g., 
Refs.~\cite{Hatsuda-Kunihiro:94,Pisarski:95,Brown-Rho:96,%
Hatsuda-Shiomi-Kuwabara:96,Wilczek,%
Rapp-Wambach:00,Brown-Rho:01b}),
the 
large $N_f$ ($3< N_f < 33/2$), 
$N_f$ being the number of {\it massless} flavors
(see, e.g., Refs.~\cite{Banks-Zaks,Kogut-Sinclair,BCCDMSV:92,%
IKSY:92,IKSY:92b,IKSY:93,IKSY:94,IKKSY:96,IKKSY:98,%
Damgaard-Heller-Krasnitz-Olesen,%
Appelquist-Terning-Wijewardhana,%
Appelquist-Ratnaweera-Terning-Wijewardhana,%
Miransky-Yamawaki}), etc..
Conventional picture of the chiral symmetry restoration is based on
the 
linear sigma model where the scalar meson (``sigma'' meson) 
denoted by $S$ becomes
massless
degenerate with the pion as the chiral partner: 
\begin{equation}
F_\pi^2(0) \rightarrow 0 \ , \quad
m_S^2 \rightarrow m_\pi^2 =0 \ .
\label{GL manifestation}
\end{equation}
This we shall call ``GL 
manifestation'' after the effective theory of Ginzburg--Landau or
Gell-Mann--Levy. 
However, the GL manifestation is not a unique way where the Wigner
realization 
manifests itself. Recently the present authors~\cite{HY:VM}
proposed ``Vector Manifestation
(VM)'' as a novel manifestation of Wigner realization
of chiral symmetry where the vector meson 
$\rho$ 
becomes massless at the chiral
phase transition point.  
Accordingly, the (longitudinal) $\rho$
becomes the chiral partner of the NG boson 
$\pi$. 
The VM is characterized by
\begin{eqnarray}
&& F_\pi^2(0) \rightarrow 0 \ , \quad
m_\rho^2 \rightarrow m_\pi^2=0\ , \quad
F_\sigma^2(m_\rho) / F_\pi^2(0) \rightarrow 1 \ ,
\label{VM def}
\end{eqnarray}
where $F_\sigma(m_\rho)$ is the decay constant of $\sigma$
(longitudinal $\rho$) at $\rho$ on-shell.

Here we should stress that 
{\it the power counting rule in our derivative 
expansion} developed in Sec.~\ref{ssec:DEHLS},
which presumes $\rho$ mass is conceptutally small in the same sense as
$\pi$ mass, {\it is now literally (not just conceptually) operative 
near the VM phase transition}, although it is not
a priori justified for the case $N_f=3$ where $m_\rho$ is 
actually not very small, except that it happened to work as
demonstrated in 
Sec.~\ref{sec:WM}.

In this section we discuss the VM of chiral symmetry, 
based on the HLS model at one loop developed in the
previous sections:

In Sec.~\ref{ssec:VMCSR}
we first formulate in Sec.~\ref{sssec:FVM}
  what we call {\it ``VM conditions''},
Eqs. (\ref{vector condition:g}) - (\ref{vector condition:fp}),
a part of which coincides with the Georgi's 
``vector limit''~\cite{Georgi:1,Georgi:2}.
The VM conditions are  necessary conditions of the Wigner realization
of  
chiral symmetry of QCD in
terms of the HLS parameters as a direct consequence of 
the Wilsonian matching of the HLS with the underlying QCD at the 
matching scale $\Lambda$.
We then argue that  
we have the chiral restoration $F_\pi^2(0) \rightarrow 0$ 
through the dynamics of
the HLS model istself in a way already discussed in
Sec.~\ref{ssec:PSH},
once the VM conditons are imposed on 
the bare parameters of the HLS model for a 
particular value of $F_\pi(\Lambda)$ and/or $N_f$ such that  
$X(\Lambda)\equiv (N_f\Lambda^2/2(4\pi)^2)/F_\pi^2(\Lambda) 
\rightarrow 1$, where $X(\mu)$ 
was defined by Eq.~(\ref{def X}). Then we show that the VM conditions
in fact lead to VM.
We compare the VM with the conventional manifestation, i.e., GL
manifestation in Sec.~\ref{sssec:VGL}: 
we demonstrate that the GL manifestation
\'a la linear sigma model 
does not satisfies the requirement
on the current correlators from the 
Wilsonian matching  (i.e., VM conditions), and hence 
is excluded by the Wilsonian matching
as a candidate for the chiral restoration of QCD.
In Sec.~\ref{sssec:CPT} we discuss the ``conformal phase
transition''~\cite{Miransky-Yamawaki} as an example of 
non-GL manifestation having the essential-singularity-type
scaling.
In Sec.~\ref{sssec:VR} we distinguish our VM 
as a {\it Wigner realization} 
from a similar but essentially different concept,
the ``Vector Realization''~\cite{Georgi:1,Georgi:2}, 
which was claimed
as a new realization, {\it neither Wigner nor NG realization}. 
In Sec.~\ref{sssec:VM as a limit} we emphasize
that the VM makes sense only as a limit
of the bare parameters approaching the values of VM conditions
(never does the ``Vector Realization'' even as a limit). 
  
In Sec.~\ref{ssec:CPTLNQ}, as an illustration of VM 
we shall discuss the chiral restoration
in the large $N_f$ QCD:
 we first review the arguments on
chiral restoration in the large $N_f$ QCD in terms of the QCD
language, i.e., $\langle \bar q q \rangle \rightarrow 0$.
It is noted that the conformal phase transition was observed
also in the chiral restoration
of the large $N_f$ QCD in the (improved) ladder
approximation~\cite{Appelquist-Terning-Wijewardhana}.

In Sec.~\ref{ssec:VMLNQ} we show that the chiral restoration
in the large $N_f$ QCD in fact takes place also in 
the HLS model, $F_\pi^2(0) \rightarrow 0$, and so does the VM, 
when we tune in a concrete manner  
the bare parameters to satisfy the above condition
$X(\Lambda)\equiv (N_f\Lambda^2/2(4\pi)^2)/F_\pi^2(\Lambda)
\rightarrow 1$.
In Sec.~\ref{sssec:CHRES}
we determine by this 
the critical number of flavors 
$N_f =N_f^{\rm crit} \simeq 5$ above which the chiral symmetry is
restored, 
which is in rough agreement with the recent lattice simulations 
$6 < N_f^{\rm crit} < 7 $~\cite{IKKSY:98}.
The critical behaviors of the parameters in the large $N_f$ QCD
are studied in Sec.~\ref{sssec:CB}.
Full $N_f$-dependences of the parameters are shown in
Sec.~\ref{sssec:NDP} by using a simple ansatz.
In Sec.~\ref{sssec:VDLNQ}
we argue, following Ref.~\cite{HY:VD}, that 
the vector dominance is badly violated near the critical point 
in the large $N_f$ QCD. 

Finally, in Sec.~\ref{ssec:STD} we explain the proposal of  
Ref.~\cite{HY:letter} that the HLS in the broken phase
of chiral symmetry is dual to QCD in the sense of Seiberg 
duality~\cite{Seiberg}.

\subsection{Vector manifestation (VM) of chiral symmetry restoration}
\label{ssec:VMCSR}

\subsubsection{Formulation of the VM}
\label{sssec:FVM}

The essence of VM stems from the new matching 
 of the EFT with QCD 
(Wilsonian matching) proposed by
Ref.~\cite{HY:matching} [see Sec.~\ref{sec:WM}]
in which bare parameters of the
EFT are determined by matching  
 the current correlators in the EFT
with those obtained by the OPE in QCD,
based on the RGE in the Wilsonian 
sense {\it including the quadratic divergence}~\cite{HY:letter}
[see Sec.~\ref{sec:CPHLS}].
Several physical quantities for $\pi$ and $\rho$ were predicted by 
the Wilsonian matching in the framework of the HLS
model~\cite{BKUYY,BKY} as the EFT, in excellent agreement
 with the experiments for $N_f=3$,
where $N_f$ is the number of 
{\it massless flavors}~\cite{HY:matching}.
This encourages
us to perform the analysis for other situations such as 
larger $N_f$ and finite temperature and/or density 
up to near the critical point, based on the Wilsonian matching.

The chiral symmetry restoration in Wigner realization should
be characterized by 
\begin{equation}
F_\pi(0) =0
\label{zerofpi}
\end{equation}
(see also discussions in Sec.~\ref{sssec:VR})
and the equality of the vector and axialvector current
correlators in the underlying QCD:
\begin{equation}
\Pi_V(Q^2) = \Pi_A(Q^2) \ ,
\end{equation}
which is in accord with $\langle \bar{q} q \rangle = 0$ in 
Eqs.~(\ref{Pi A OPE}) and (\ref{Pi V OPE}).
On the other hand, the same current correlators are described 
in terms of the HLS model for energy lower than the cutoff $\Lambda$: 
When we approach to the critical point {\it from the broken phase (NG
phase)},  the axialvector current correlator 
is still dominated by 
the massless $\pi$ as the NG boson, 
while the vector current correlator is
by the massive $\rho$. 
In such a case, there exists a scale $\Lambda$ around which the
current correlators are well described by the forms given in
Eqs.~(\ref{Pi A HLS}) and (\ref{Pi V HLS}):
\begin{eqnarray}
\Pi_A^{\rm(HLS)}(Q^2) &=&
\frac{F_\pi^2(\Lambda)}{Q^2} - 2 z_2(\Lambda)
\ ,
\label{Pi A HLS 2}
\\
\Pi_V^{\rm(HLS)}(Q^2) &=&
\frac{
  F_\sigma^2(\Lambda)
}{
  M_\rho^2(\Lambda) + Q^2
} 
\left[ 1 - 2 g^2(\Lambda) z_3(\Lambda) \right]
- 2 z_1(\Lambda)
\ ,
\label{Pi V HLS 2}
\end{eqnarray}
where 
$M_\rho^2(\Lambda) \equiv g^2(\Lambda) F_\sigma^2(\Lambda)$
is the bare $\rho$ mass parameter
[see Eq.~(\ref{on-shell cond 5})].
Then, through
the Wilsonian matching discussed in Sec.~\ref{sec:WM},
we determine the bare parameters of the HLS.
At the critical point
the quark condensate vanishes,
$\left\langle \bar{q} q \right\rangle \rightarrow 0$, while
the gluonic condensate 
$\left\langle 
\frac{\alpha_s}{\pi} G_{\mu\nu} G^{\mu\nu}
\right\rangle$
is independent of the renormalization point of
QCD and hence  it is expected not to vanish.
Then Eq.~(\ref{match A}) reads
\begin{eqnarray}
&&
F_\pi^2(\Lambda)
\rightarrow (F_\pi^{\rm crit})^2\equiv   
\frac{N_c}{3}\left(\frac{\Lambda}{4\pi}\right)^2
\cdot 2
(  1 + \delta_A^{\rm crit})\ \ne 0 
\nonumber
,\\
&&\delta_A^{\rm crit}
\equiv \delta_A|_{\langle \bar q q \rangle = 0} =
\frac{3(N_c^2-1)}{8N_c} \,\frac{\alpha_s}{\pi}
  + \frac{2\pi^2}{N_c} 
    \frac{
      \left\langle 
        \frac{\alpha_s}{\pi} G_{\mu\nu} G^{\mu\nu}
      \right\rangle
    }{ \Lambda^4 }
>0\quad (\ll 1)
\ ,
\label{match A VM}
\end{eqnarray}
implying that {\it matching with QCD dictates}
\begin{equation}
F_\pi^2(\Lambda) \ne 0
\label{nvf_pi}
\end{equation}
{\it even at the critical point}~\cite{HY:VM} where
\begin{equation}
F_\pi^2(0)=0 \ .
\end{equation}
One might think that this is somewhat strange.
However,
as we have already discuused in Secs.~\ref{sssec:CRQDPL} 
and \ref{ssec:PSH},
we have a possibility~\cite{HY:letter} that 
the order parameter  can become zero $F_\pi(0)\rightarrow 0$,
even when 
$F_\pi(\Lambda) \ne 0$, 
where 
{\it $F_\pi(\Lambda)$ is not an order parameter but just a
parameter of the bare HLS Lagrangian} defined at the cutoff
$\Lambda$ where the
matching with QCD is made.

Let us obtain further 
constraints on other bare parameters of the HLS   
through the Wilsonian matching for the currents correlators.
The constraints on other parameters defined at $\Lambda$
come from the fact that $\Pi_A^{\rm (QCD)}$ 
and $\Pi_V^{\rm (QCD)}$ in Eqs.~(\ref{Pi A OPE}) and (\ref{Pi V OPE})
agree with each other for any value of $Q^2$ when the chiral symmetry
is restored with $\left\langle \bar{q} q \right\rangle \rightarrow 0$.
Thus, we require that
$\Pi_A^{\rm (HLS)}$ and $\Pi_V^{\rm (HLS)}$ in Eqs.~(\ref{Pi A HLS 2})
and (\ref{Pi V HLS 2})
agree with each other for {\it any value of $Q^2$}
(near $\Lambda^2$)\footnote{
  Note that chiral restoration requires equality of
  $\Pi_A^{\rm (HLS)}$ and $\Pi_V^{\rm (HLS)}$ for {\it any} $Q^2$
  (even without referring to QCD), while
  Eqs.~(\ref{Pi A HLS 2}) and (\ref{Pi V HLS 2}) are valid only for 
  $Q^2 \sim \Lambda^2$.
  See the discussions below  Eqs.~(\ref{Pi A OPE}) and 
  (\ref{Pi V OPE}). For instance, 
   the forms in Eqs.~(\ref{Pi A HLS 2}) and  (\ref{Pi V HLS 2})
   might be changed  for $Q^2 < \Lambda^2$ by the corrections to
  $\Pi_V$ and  
  $\Pi_A$ from $\rho$ and/or $\pi$ loop effects 
  which, however, are of higher order 
  in our power counting rule developed in Sec.~\ref{ssec:DEHLS}
  and hence can be neglected. Note that the counting rule actually
  becomes  
  precise near the VM limit satisfying the VM conditions. 
  Also note that the VM limit is the fixed point 
  and hence the ``pole-saturated forms'' of 
  Eqs.~(\ref{Pi A HLS 2}) and  (\ref{Pi V HLS 2}) must be equal 
  for {\it any}
  $Q^2$, once the VM conditions are satisfied at  $Q^2 \sim
  \Lambda^2$:  
  Namely, other possible effects if any should be equal   
  to each other at the VM limit and hence would not affect our
  arguments. 
}.
Under the conditon Eq.~(\ref{nvf_pi}),
this agreement is satisfied only if the following conditions are met:
\begin{eqnarray}
&& 
M_\rho^2(\Lambda) \equiv g^2(\Lambda) F_\sigma^2(\Lambda)
\rightarrow 0 \ , 
\quad
F_\sigma^2(\Lambda) \rightarrow F_\pi^2(\Lambda) 
\neq 0\ , 
\quad
z_1(\Lambda) - z_2(\Lambda) \rightarrow 0 \ ,
\label{vector conditions:mv fs}
\end{eqnarray}
or
\begin{eqnarray}
&& g(\Lambda) \rightarrow 0 \ , 
\label{vector condition:g}
\\
&& a(\Lambda) = \frac{ F_\sigma^2(\Lambda) }{ F_\pi^2(\Lambda) }
\rightarrow 1 \ , 
\label{vector condition:a}
\\
&& z_1(\Lambda) - z_2(\Lambda) \rightarrow 0 \ ,
\label{vector condition:z}
\\
&& 
F_\pi^2(\Lambda) \rightarrow (F_\pi^{\rm crit})^2
=
\frac{N_c}{3}
\left(\frac{\Lambda}{4\pi}\right)^2\cdot 2(1+\delta_A^{\rm crit}) 
\neq 0
\ .
\label{vector condition:fp}
\end{eqnarray}
 These conditions, may be called ``VM conditions'',
follow solely from the requirement of the equality of
the vector and axialvector currents correlators (and the Wilsonian matching)
without explicit requirement of Eq.~(\ref{zerofpi}), 
and are actually a precise expression of the
VM in terms of the {\it bare}
HLS parameters for the Wigner realization in 
QCD~\cite{HY:VM}. 
Note that
the values in Eqs.~(\ref{vector condition:g}) and
(\ref{vector condition:a})
agree with the values in the Georgi's vector 
limit~\cite{Georgi:1,Georgi:2}.

Once the bare HLS parameters satisfy the VM conditions,
Eqs.~(\ref{vector condition:g})--(\ref{vector condition:fp}),
the RGE for $F_\pi^2$ leads to
 Eq.~(\ref{sol fpi2 for HLS g0 a1}), 
\begin{eqnarray}
&&F_\pi^2(0)
= 
F_\pi^2(\Lambda) - 
\frac{N_f\Lambda^2}{2(4\pi)^2} \rightarrow (F_\pi^{\rm crit})^2
- \frac{N_f\Lambda^2}{2(4\pi)^2} 
\ ,
\label{Fpirun}
\end{eqnarray}
 which implies that
we can have 
\begin{equation}
F_\pi^2(0)\rightarrow 0
\end{equation}
by tuning the bare parameters 
$N_f$ and/or $F_\pi^2(\Lambda)$ (which explicitly depends on $N_c$)
in such a way that 
$X(\Lambda)\equiv [N_f\Lambda^2/2(4\pi)^2]
/F_\pi^2(\Lambda) \rightarrow 1$.
Then the chiral restoration $F_\pi^2(0) \rightarrow 0$ is actually
{\it derived within the dynamics 
of the HLS model itself} solely from  the requirement 
of the Wilsonian matching. (We shall discuss a concrete way of tuning 
the bare parameters in the case of large $N_f$ QCD
in Sec.~\ref{ssec:VMLNQ}).

One may wonder what would happen if we tune the HLS parameters such as
$N_f$  
so as to keep $X(\Lambda) \ne 1$ 
even when the bare parameters obey the VM conditions:
in such a case the underlying QCD gives a chiral restoration, 
$\langle \bar q q \rangle =0$, 
while the EFT would have an NG boson pole coupled to the
axialvector current with the strength of a pole residue
$F_\pi^2(0)\ne 0$\,!
This is similar to the Georgi's 
``Vector Realization''~\cite{Georgi:1,Georgi:2}. 
We shall discuss in details 
in Sec.~\ref{sssec:VR} that the ``Vector Realization''
is in contradition with the Ward-Takahashi identity 
for the chiral symmetry and also produces a fake symmetry larger than
the 
underlying QCD and hence is impossible. 
So the parameters of HLS model must choose a choice such that
$X(\Lambda) \rightarrow 1$ or $F_\pi^2(0)\rightarrow 0$.   

Now that we have shown the Wigner realization in the
HLS model, we can show that the VM conditons
actually lead to the VM characterized by
Eq.~(\ref{VM def}):
First note that since the values in 
Eqs.~(\ref{vector condition:g})--(\ref{vector condition:z}) 
coincide with 
those at the fixed points of the RGE's [See Eqs.~(\ref{fp:g0a1}) and 
(\ref{fp:z1z2}).],
the parameters remains the same for any scale,
and hence even at $\rho$ on-shell point:
\begin{eqnarray}
&& g(m_\rho) \rightarrow 0 \ , 
\label{vector condition:on-shell g}
\\
&& a(m_\rho) \rightarrow 1 \ , 
\label{vector condition:on-shell a}
\\
&& z_1(m_\rho) - z_2(m_\rho) \rightarrow 0 \ ,
\label{vector condition :on-shell z}
\end{eqnarray}
where $m_\rho$ is determined from the on-shell condition in 
Eq.~(\ref{on-shell condition}):
\begin{equation}
m_\rho^2 = a(m_\rho) g^2(m_\rho) F_\pi^2(m_\rho) \ .
\label{on-shell condition 6}
\end{equation}
Then, the condition in Eq.~(\ref{vector condition:on-shell g})
together with the above on-shell condition immediately leads to
\begin{eqnarray}
&&
m_\rho^2 \rightarrow 0\ .
\label{VM mrho}
\end{eqnarray}
Equation~(\ref{vector condition:on-shell a}) is rewritten as
$F_\sigma^2(m_\rho) /F_\pi^2(m_\rho) \rightarrow 1$, and 
Eq.~(\ref{VM mrho}) implies $F_\pi^2(m_\rho) \rightarrow F_\pi^2(0)$.
Thus, 
\begin{eqnarray}
&&
F_\sigma^2(m_\rho) /F_\pi^2(0) \rightarrow 1 \ ,
\label{VM fsig}
\end{eqnarray}
namely, the pole residues of $\pi$ and $\rho$ become identical.
Then the VM defined by Eq.~(\ref{VM def}) does follow.
Note that we have used only the requirement of Wigner realization in
QCD 
through the Wilsonian matching 
and {\it arrived uniquely at VM but not GL manifestation \'a la linear 
sigma model}.
The crucial ingredient to exclude the GL manifestation as a chiral
restoration in QCD was the Wilsonian 
matching, particularly Eq.~(\ref{nvf_pi}). 
We shall return to this point later in Sec.~\ref{sssec:VGL}.

Actually, the VM conditions with $X(\Lambda)\rightarrow 1$ are nothing
but 
a limit of bare parameters 
approaching a particular fixed point (what we called ``VM limit'')
 $(X^*_2,a_2^\ast,G_2^\ast)=(1,1,0)$ 
in Eq.~(\ref{fixed points}) which was extensively 
discussed in Sec.~\ref{ssec:PSH}.
Namely, through the VM conditions the 
{\it QCD singles out just one fixed point (as a limit) out of
otherwise  
allowed wide phase boundary surface of HLS model} 
which is given by the collection of the RG flows
entering points on the line specified by
Eqs.~(\ref{on-shell condition 2}) and 
(\ref{phase boundary}).

Now, does it make sense
that Lorentz scalar $\pi$ and Lorentz vector $\rho$ are the chiral
partner? 
It is crucial that only the longitudinal component of $\rho$ becomes
a chiral partner of $\pi$, while the tansverse $\rho$ decouples.

When the VM occurs,
both the axialvector and vector
current correlators in 
Eqs.~(\ref{Pi A HLS 2}) and (\ref{Pi V HLS 2})
take the form~\cite{HY:VM}
\begin{equation}
\Pi_A^{\rm (HLS)}(Q^2) = 
\frac{F_\pi^2(\Lambda)}{Q^2} - 2 z_2(\Lambda) 
= 
\frac{F_\sigma^2(\Lambda)}{Q^2} - 2 z_1(\Lambda) 
=
\Pi_V^{\rm (HLS)}(Q^2) 
\ .
\end{equation}
For the axialvector current
correlator, the first term 
$F_\pi^2(\Lambda)/Q^2 $ ($=F_\sigma^2(\Lambda)/Q^2$) comes from
the $\pi$-exchange contribution, while 
for the vector current correlator it can be easily understood as the
$\sigma$ (would-be NG boson absorbed into $\rho$)-exchange
contribution in the $R_\xi$-like gauge.
Thus only the longitudinal
$\rho$ couples to the vector current, and 
{\it the transverse $\rho$ with the helicity
$\pm 1$
is decoupled from it}~\cite{HY:VM}.
This can be also seen in the unitary gauge as follows:

Let us start with the expression of the vector current correlator in
the chiral broken phase in the unitary gauge of $\rho$:
\begin{eqnarray}
\Pi_{{\rm (V)}\,\mu\nu}^{\rm (HLS)} (p) = 
- \frac{\left(g F_\sigma^2\right)^2}{m_\rho^2-p^2}
\left( g_{\mu\nu} - \frac{p_\mu p_\nu}{m_\rho^2} \right)
+ g_{\mu\nu} F_\sigma^2 \ ,
\label{Pi V unitary}
\end{eqnarray}
where $p_\mu = ( p_0 \,,\, \vec{p} )$ and we have neglected higher
order $z_3$ and $z_1$ terms for simplicity.
The polarization vector for the longitudinal $\rho$ is given by
\begin{equation}
\varepsilon_\mu^{(0)}(P) = \frac{1}{m_\rho} 
\left( \left\vert \vec{p} \right\vert \,,\, E 
       \frac{\vec{p}}{\left\vert \vec{p} \right\vert}
\right)
\ ,
\label{pol l}
\end{equation}
where $P_\mu \equiv ( E\,,\, \vec{p} )$ with
$E = \sqrt{ \left\vert \vec{p} \right\vert^2 + m_\rho^2 }$.
It is given for the transverse $\rho$ by
\begin{equation}
\varepsilon_\mu^{(\pm)}(P) = 
\left( 0 \,,\, \vec{e^{(\pm)}}(\vec{p}) \right)
\ ,
\label{pol t}
\end{equation}
where $\vec{e^{(\pm)}}(\vec{p})$ satisfy 
$\vec{e^{(\pm)}}(\vec{p}) \cdot \vec{p} = 0$,
$\vec{e^{(+)}}(\vec{p}) \cdot \vec{e^{(-)}}(\vec{p}) =0$
and 
$\vec{e^{(\pm)}}(\vec{p}) \cdot \vec{e^{(\pm)}}(\vec{p}) = 1$.
Using a relation
\begin{equation}
\sum_{l = \pm,0} \varepsilon_\mu^{(l)}(P) \varepsilon_\nu^{(l)}(P) 
= - \left( g_{\mu\nu} - \frac{P_\mu P_\nu}{m_\rho^2} \right) \ ,
\end{equation}
we can rewrite Eq.~(\ref{Pi V unitary}) into
\begin{eqnarray}
&&
\Pi_{{\rm (V)}\,\mu\nu}^{\rm (HLS)} (p) = 
g_{\mu\nu} F_\sigma^2 + 
\sum_{l = \pm} \varepsilon_\mu^{(l)}(P) \varepsilon_\nu^{(l)}(P) 
\frac{\left(g F_\sigma^2\right)^2}{m_\rho^2-p^2}
\nonumber\\
&& \quad
{}+
\varepsilon_\mu^{(0)}(P) \varepsilon_\nu^{(0)}(P) 
\frac{\left(g F_\sigma^2\right)^2}{m_\rho^2-p^2}
+
\left( p_\mu p_\nu - P_\mu P_\nu \right)
\frac{F_\sigma^2}{m_\rho^2-p^2}
\ .
\end{eqnarray}
Let us consider VM such that $(g,F_\sigma) \rightarrow (0,F_\pi)$.
We can easily show
\begin{equation}
g \varepsilon_\mu^{(\pm)} \rightarrow 0 
\label{trans rho pol}
\end{equation}
from Eq.~(\ref{pol t}).
This implies that the transverse components of $\rho$ decouple from
the vector current.
On the other hand, Eq.~(\ref{pol l}) leads to
\begin{equation}
g \varepsilon_\mu^{(0)} \rightarrow 
\frac{1}{F_\sigma} \left( \left\vert \vec{p} \right\vert \,,\,
\vec{p} \right) =
\frac{1}{F_\sigma} P_\mu \ ,
\label{pol l limit}
\end{equation}
where we used $E \rightarrow \left\vert \vec{p} \right\vert$ as
$m_\rho \rightarrow 0$.
Equation~(\ref{pol l limit}) implies that the longitudinal component
of $\rho$ does couple to the vector current.
The resultant expression of the vector current is given by
\begin{equation}
\Pi_{{\rm (V)}\,\mu\nu}^{\rm (HLS)} (p) = 
\left( p_\mu p_\nu - p^2 g_{\mu\nu} \right)
\frac{F_\pi^2}{-p^2} \ ,
\end{equation}
which agrees with the axial vector current correlator
as it should.

\subsubsection{VM vs. GL (Ginzburg--Landau/Gell-Mann--Levy)
manifestation}
\label{sssec:VGL}

The crucial ingredient 
of the Wilsonian matching is
the {\it quadratic divergence} of HLS model which yields 
the quadratic running of (square of) the 
decay constant $F_\pi^2(\mu)$~\cite{HY:letter}, 
where $\mu$ is the renormalization point. 
Then {\it the $\pi$ contribution to the axialvector current correlator
at $\mu \ne 0 $ persists, $F_\pi(\mu) \ne 0$, \it even at the critical 
point where $F_\pi(0)=0$}.
Thus the only possibility for the equality $\Pi_A = \Pi_V$
to hold at any $\mu \ne 0$ 
is that {\it the $\rho$ contribution to the vector current correlator 
also persists at the critical point in such a way that 
$\rho$ yields a massless pole with the current coupling equal to
that of $\pi$}, i.e.,
the VM occurs:
the chiral restoration is accompanied by 
degenerate massless $\pi$ and (longitudinal) $\rho$ ( 
the would-be NG boson $\sigma$).~\footnote{%
The transverse $\rho$ is
  decoupled from the current correlator in the limit approaching 
  the critical point,
  as we discussed around Eq.~(\ref{trans rho pol}).
  Note that when the theory is put exactly on the critical point,
  then not only the tranverse $\rho$ but also the whole light spectrum
  including the $\pi$ and the longitudinal $\rho$ would dissapear 
  as we shall discuss in Sec.~\ref{sssec:CPT}, \ref{sssec:VR} and 
  \ref{ssec:CPTLNQ}. The effective field theory based on the light
  composite 
  spectrum would break down at the exact critical point.
}
On the contrary, the scalar meson in the linear sigma model does 
not contribute to 
$\Pi_V$ and hence the {\it GL manifestation \`a la linear sigma model 
(without $\rho$) is simply ruled out by Eq.~(\ref{nvf_pi})}:
The Wilsonian matching with QCD definitely favors VM rather 
than GL manifestation.

Let us discuss the difference between the VM and 
GL manifestation in terms of the chiral representation of the mesons
by extending the analyses done in
Refs.~\cite{Gilman-Harari,Weinberg:69} for two flavor QCD.
Since we are approaching the chiral restoration point only {\it from
the broken phase} where the chiral symmetry is realized only nonlinearly, 
it does not make sense to discuss the chiral representation of such a 
spontaneously broken symmetry. One might suspect that 
in the HLS model having the linearlized
symmetry $G_{\rm global}\times H_{\rm local}$, the $\rho$ is an 
adjoint representation of the gauge symmetry $H_{\rm local}$ and is
a singlet of the chiral symmetry $G_{\rm global}$. However, 
the $G_{\rm global}\times H_{\rm local}$ is actually spontaneously 
broken down to $H$, which is a diagonal subgroup of $H_{\rm global} (\subset G_{\rm global})$  and $H_{\rm local}$, and hence
the $\rho$ is no longer subject to the linear representation.
Then we need a tool to formulate the {\it linear representation}
 of the chiral algebra
even {\it in the broken phase}, 
namely the classification algebra valid even in the broken phase, in such a way that it smoothly moves over to the 
original chiral algebra as we go over to the symmetric phase.

Following Ref.~\cite{Weinberg:69},
we define the
axialvector coupling matrix $X_a(\lambda)$ (an analogue of the $g_A$ for
the nucleon matrix)
by giving the matrix elements at zero invariant momentum transfer 
of
the axialvector current between states with collinear momenta
as~\footnote{%
  Note that we adopted the invariant normalization for the state:
  \begin{displaymath}
    \langle \vec{q} \,\lambda^\prime \, \beta \vert
    \vec{p} \,\lambda \, \alpha \rangle = (2\pi)^3 2 p^+ 
    \delta(\vec{q} - \vec{p}) \ ,
  \end{displaymath}
  which is different from the one used in Ref.~\cite{Weinberg:69}.
  Furthermore, the current in this expression is half of the 
  current used in Ref.~\cite{Weinberg:69}.
}
\begin{equation}
\langle \vec{q} \,\lambda^\prime \, \beta \vert
J_{5a}^+ (0)
\vert \vec{p} \,\lambda \, \alpha \rangle
=
2 
p^+\, \delta_{\lambda\lambda^\prime}
\left[ X_a(\lambda) \right]_{\beta\alpha}
\ ,
\label{axial coupling}
\end{equation}
where $J_{5a}^+ = (J_{5a}^{0} + J_{5a}^{3} )/\sqrt{2}$,
and $\alpha$ and $\beta$ are one-particle states
with collinear momentum $\vec{p}\equiv (p^+,p^1,p^2)$ and $\vec{q}
\equiv (q^+,q^1,q^2)$ such that $p^+=q^+$, 
$\lambda$ and $\lambda^\prime$ are their helicities.
It was stressed~\cite{Weinberg:69} that the definition
of the axialvector couplings in Eq.~(\ref{axial coupling}) can be 
used for particles of arbitrary spin, and in arbitrary collinear
reference frames, including both the frames in which $|\alpha\rangle$
is at 
rest and in which it moves with infinite momentum:
The matrix $X_a(\lambda)$ is independent of the reference frame. 
Note that the $X_a(\lambda)$ matrix does not contain the $\pi$ pole
term 
which would behave as $(p^+ -q^+)/[(p-q)^2 -m_\pi^2]$ and hence be
zero for 
kinematical reason, $p^+ = q^+$, even in the chiral limit of 
$m_\pi^2 \rightarrow 0$.

As was done for $N_f=2$ in Ref.~\cite{Weinberg:69},
considering
the forward scattering process
$\pi_a + \alpha(\lambda) \rightarrow \pi_b + \beta(\lambda^\prime)$
and 
requiring the cancellation of the terms in the $t$-channel,
we obtain
\begin{equation}
\left[ \, X_a(\lambda)\,,\, X_b(\lambda) \,\right]
= i f_{abc} T_c
\ ,
\label{algebra}
\end{equation}
where
$T_c$ is the generator of $\mbox{SU}(N_f)_{\rm V}$ and 
$f_{abc}$ is the structure constant. This is nothing but the algebraization
of the Adler-Weisberger sum rule~\cite{Adler:65, Weisberger} 
and the basis of the good-old-days classification of the hadrons by the chiral 
algebra~\cite{Gilman-Harari,Weinberg:69}
or the ``mended  symmetry''~\cite{Weinberg:90}.
It should be noticed that Eq.~(\ref{algebra}) tells us that the
one-particle states of any given helicity must be assembled into
representations of chiral 
$\mbox{SU}(N_f)_{\rm L}\times\mbox{SU}(N_f)_{\rm R}$.
Furthermore, since Eq.~(\ref{algebra}) does not give any relations 
among the states with different helicities, those states can 
generally belong to the different representations even though
they form a single particle such as the longitudinal $\rho$
($\lambda=0$) and the transverse $\rho$ ($\lambda=\pm1$).
Thus, the notion of the chiral partners can be considered
separately for each helicity.

Here we should note that the above axialvector coupling matrix $X_a(\lambda)$ can be
equivalently defined through the light-front (LF) axial charge
$\hat{Q}_{5a} \equiv \int dx^-dx^1dx^2
 J_{5a}^+(x)$
as 
\begin{equation}
\langle \vec{q} \,\lambda^\prime \, \beta \vert
\hat{Q}_{5a}
\vert \vec{p} \,\lambda \, \alpha \rangle
=
(2\pi)^3 2 p^+\, \delta^3 (\vec{p} -\vec{q})
\, \delta_{\lambda\lambda^\prime}
\left[ X_a(\lambda) \right]_{\beta\alpha}
\ .
\label{axial coupling LF}
\end{equation}
The LF axial charge $\hat{Q}_{5a}$ does not contain the $\pi$ pole
term for the 
same reason as the absence of $\pi$ pole contribution in the
$X_a(\lambda)$  
matrix and
is well defined even in the chiral limit in the
broken phase in such a way that the
 vacuum is singlet under the chiral
transformation with $\hat{Q}_{5a}$,
\begin{equation}
\hat{Q}_{5a}|0\rangle =0 \ ,
\end{equation}
 whereas the ordinary axial charge $Q_{5a}$ is not well defined
due to the presence of the $\pi$ pole, or usually phrased as 
$Q_{5a}|0\rangle \ne 0$. However, due to the very 
absence of the $\pi$ pole term, $\hat{Q}_{5a}$ is
not conserved even in the chiral limit $m_\pi^2 \rightarrow 0$ 
in the broken phase: 
\begin{equation}
i \frac{d}{dx^+} \hat{Q}_{5a} =[\hat{Q}_{5a}, P^-]\ne 0 \ ,
\end{equation}
in sharp contrast to the conservation of $Q_{5a}$,
where $x^+=(x^0+x^3)/\sqrt{2}$ is the LF time and 
$P^-=(P^0-P^3)/\sqrt{2}$ 
is the LF Hamiltonian. Then it does not commute with
the $({\rm mass})^2$ operator $M^2=2P^+ P^- -(P^1)^2-(P^2)^2$:
\begin{equation}
[\hat{Q}_{5a}, M^2]\ne 0 \ .
\end{equation}
This implies that the mass eigenstates are in general
admixtures of the representaions of the chiral algebra (LF chiral algebra)
which is formed by the LF axial charge $\hat{Q}_{5a}$ 
together with the LF vector charge $\hat{Q}_a$. This is nothing but 
the representation mixing in the saturation 
scheme~\cite{Gilman-Harari,Weinberg:69,Weinberg:90} 
of the celebrated
Adler-Weisberger sum rule which is actually a physical manifestation
of the LF chiral algebra. 
When the symmetry is restored with vanishing $\pi$ pole, the LF axial
charge  
agrees with the ordinary axial charge, and then the representations
of the algebra with $\hat{Q}_{5a}$ agree with the ones
under the ordinary axial charge.
(For details of the LF charge algebra, see Ref.~\cite{Yamawaki:98}.)

The same is of course true for the algebra formed by the
$X_a(\lambda)$ matrix 
directly related to $\hat{Q}_{5a}$ through 
Eq.~(\ref{axial coupling LF}). 
In the broken phase of chiral symmetry,
the Hamiltonian (or $({\rm mass})^2$) matrix $M_{\alpha\beta}^2$ 
defined by the matrix elements of the
Hamiltonian ($({\rm mass})^2$) between states $|\alpha\rangle$ 
and $|\beta\rangle$ 
does not generally commute with the
axialvector coupling matrix:
\begin{equation}
[X_{5a}(\lambda), M^2]_{\alpha\beta}\ne 0 \ .
\end{equation}
Then, the algebraic representations of the 
axialvector coupling matrix do not always coincide
with the mass eigenstates: There occur representation mixings.

Let us first consider the zero helicity ($\lambda=0$) states
and saturate the algebraic relation in Eq.~(\ref{algebra}) by low
lying mesons; 
the $\pi$, the (longitudinal) $\rho$, the (longitudinal) axialvector
meson denoted by $A_1$ ($a_1$ meson and its flavor partners)
and the scalar meson denoted by $S$, and so on.
The $\pi$ and the longitudinal $A_1$ 
are admixture of $(8\,,\,1) \oplus(1\,,\,8)$ and 
$(3\,,\,3^*)\oplus(3^*\,,\,3)$, since the symmetry is spontaneously
broken~\cite{Weinberg:69,Gilman-Harari}:
\begin{eqnarray}
\vert \pi\rangle &=&
\vert (3\,,\,3^*)\oplus (3^*\,,\,3) \rangle \sin\psi
+
\vert(8\,,\,1)\oplus (1\,,\,8)\rangle  \cos\psi
\ ,
\nonumber
\\
\vert A_1(\lambda=0)\rangle &=&
\vert (3\,,\,3^*)\oplus (3^*\,,\,3) \rangle \cos\psi 
- \vert(8\,,\,1)\oplus (1\,,\,8)\rangle  \sin\psi
\ ,
\label{mix pi A}
\end{eqnarray}
where the experimental value of the mixing angle $\psi$ is 
given by approximately 
$\psi=\pi/4$~\cite{Weinberg:69,Gilman-Harari}.  
On the other hand, the longitudinal $\rho$
belongs to pure $(8\,,\,1)\oplus (1\,,\,8)$
and the scalar meson to 
pure $(3\,,\,3^*)\oplus (3^*\,,\,3)$:
\begin{eqnarray}
\vert \rho(\lambda=0)\rangle &=&
\vert(8\,,\,1)\oplus (1\,,\,8)\rangle  
\ ,
\nonumber
\\
\vert S\rangle &=&
\vert (3\,,\,3^*)\oplus (3^*\,,\,3) \rangle 
\ .
\label{rhoS}
\end{eqnarray}

When the chiral symmetry is restored at the
phase transition point, the axialvector coupling matrix
commutes with the Hamiltonian matrix, and thus the 
chiral representations coincide with the mass eigenstates:
The representation mixing is dissolved.
{}From Eq.~(\ref{mix pi A}) we can easily see~\cite{HY:VM}
that
there are two ways to express the representations in the
Wigner phase of the chiral symmetry:
The conventional GL manifestation
corresponds to 
the limit $\psi \rightarrow \pi/2$ in which
$\pi$ is in the representation
of pure $(3\,,\,3^*)\oplus(3^*\,,\,3)$ 
[$(N_f\,,\,N_f^\ast)\oplus(N_f^\ast\,,\,N_f)$ of
$\mbox{SU}(N_f)_{\rm L} \times\mbox{SU}(N_f)_{\rm R}$ in 
large $N_f$ QCD]
together with the scalar meson, both being the chiral partners:
\begin{eqnarray}
\mbox{(GL)}
\qquad
\left\{
\begin{array}{rcl}
\vert \pi\rangle\,, \vert S\rangle
 &\rightarrow& 
\vert  (N_f\,,\,N_f^\ast)\oplus(N_f^\ast\,,\,N_f)\rangle\ ,
\\
\vert \rho (\lambda=0) \rangle \,,
\vert A_1(\lambda=0)\rangle  &\rightarrow&
\vert(N_f^2-1\,,\,1) \oplus (1\,,\,N_f^2-1)\rangle\ .
\end{array}\right.
\end{eqnarray}
On the other hand, the VM corresponds 
to the limit $\psi\rightarrow 0$ in which the $A_1$ 
goes to a pure 
$(3\,,\,3^*)\oplus (3^*\,,\,3)$
[$(N_f\,,\,N_f^\ast)\oplus(N_f^\ast\,,\,N_f)$], now degenerate with
the scalar meson in the same representation, 
but not with $\rho$ in 
$(8\,,\,1)\oplus (1\,,\,8)$
[$(N_f^2-1\,,\,1) \oplus (1\,,\,N_f^2-1)$]:
\begin{eqnarray}
\mbox{(VM)}
\qquad
\left\{
\begin{array}{rcl}
\vert \pi\rangle\,, \vert \rho (\lambda=0) \rangle
 &\rightarrow& 
\vert(N_f^2-1\,,\,1) \oplus (1\,,\,N_f^2-1)\rangle\ ,
\\
\vert A_1(\lambda=0)\rangle\,, \vert S\rangle  &\rightarrow&
\vert  (N_f\,,\,N_f^\ast)\oplus(N_f^\ast\,,\,N_f)\rangle\ .
\end{array}\right.
\end{eqnarray}
Namely, the
degenerate massless $\pi$ and (longitudinal) $\rho$ at the 
phase transition point are
the chiral partners in the
representation of $(8\,,\,1)\oplus (1\,,\,8)$
[$(N_f^2-1\,,\,1) \oplus (1\,,\,N_f^2-1)$].~\footnote{
We again stress that the VM is realized only as a limit approaching the 
critical point from the broken phase 
but not exactly on the critical point where the light spectrum
including the $\pi$ and the $\rho$ would dissappear altogether.}

Next, we consider the helicity $\lambda=\pm1$. 
As we stressed above,
the transverse $\rho$
can belong to the representation different from the one
for the longitudinal $\rho$ ($\lambda=0$) and thus can have the
different chiral partners.
According to the analysis in Ref.~\cite{Gilman-Harari},
the transverse components of $\rho$ ($\lambda=\pm1$)
in the broken phase
belong to almost pure
$(3^*\,,\,3)$ ($\lambda=+1$) and $(3\,,\,3^*)$ ($\lambda=-1$)
with tiny mixing with
$(8\,,\,1)\oplus(1\,,\,8)$.
Then, it is natural to consider in VM that
they become pure $(N_f\,,\,N_f^\ast)$ and 
$(N_f^\ast\,,\,N_f)$
in the limit approaching the chiral restoration point:
\begin{eqnarray}
\vert \rho(\lambda=+1)\rangle \rightarrow 
  \vert (N_f^*,N_f)\rangle\ ,\quad
\vert \rho(\lambda=-1)\rangle \rightarrow 
  \vert (N_f,N_f^*)\rangle \ .
\end{eqnarray}
As a result,
the chiral partners of the transverse components of $\rho$ 
in the VM
will be  themselves. Near the critial point the longitudinal $\rho$
becomes 
almost $\sigma$, namely the would-be NG boson $\sigma$ almost becomes
a 
true NG boson and hence a different particle than the transverse
$\rho$.

{\it The $A_1$ in the VM is resolved 
and/or decoupled from the axialvector current near the critical
flavor}~\cite{HY:VM}
since there is no contribution in the vector current
correlator to be matched with the axialvector current
correlator.  As to the scalar 
meson~\cite{Harada-Sannino-Schechter:PRD,
Harada-Sannino-Schechter:PRL,Tornqvist-Roos,IITITT:96,%
Morgan-Pennington,Janssen-Pearce-Holinde-Speth}, although the mass
is smaller than the matching scale adopted in
Ref.~\cite{HY:matching} for $N_f=3$\footnote{%
  The scalar meson
  does not couple to the axialvector and vector currents, anyway.%
},
we expect that 
{\it the scalar meson is also resolved and/or decoupled near
the chiral phase transition point}~\cite{HY:VM}, since it is
in the $(N_f,N_f^*)\oplus (N_f^*,N_f)$ representation together
with the $A_1$ in the VM.

We further
show the difference between the VM and GL manifestation
discussed above in the quark contents.
In the chiral broken phase,
the pion and the axialvector meson
couple to both the pseudoscalar density ($\bar{q} \gamma_5 q$)
and the axialvector current ($\bar{q} \gamma_\mu \gamma_5 q$).
On the other hand, the scalar meson couples to the scalar 
density ($\bar{q} q$), and
the vector meson couples to the vector current 
($\bar{q} \gamma_\mu q$).
This situation is schematically expressed as
\begin{eqnarray}
&&
  \bar{q} \gamma_5 q \sim G_\pi \pi \  \oplus \ G_A \ A_\mu \ ,
\nonumber\\
&&
  \bar{q} q \sim G_S \ S \ ,
\nonumber\\
&&
  \bar{q} \gamma_\mu q \sim F_V V_\mu \ ,
\nonumber\\
&&
  \bar{q} \gamma_\mu \gamma_5 q \sim 
  F_\pi \pi \ \oplus \ F_A \ A_\mu \ .
\end{eqnarray}
In the GL manifestation,
$F_\pi$ becomes small and
$G_S$ becomes identical to $G_\pi$ near the restoration point.
Then the scalar meson is a chiral partner of the pion.
On the other hand, in the VM
$G_\pi$ becomes small and 
$F_V$ becomes identical to $F_\pi$.
Thus the vector meson becomes a chiral partner of the pseudoscalar
meson.

The problem is which manifestation the 
QCD would choose.
As we discussed in Sec.~\ref{ssec:VMCSR},
the Wilsonian matching persists $F_\pi(\Lambda) \neq0$,
even at the critical point where $F_\pi(0)=0$.
Thus we conclude~\cite{HY:VM} that
the VM is preferred by the 
QCD chiral restoration.

\subsubsection{Conformal phase transition}
\label{sssec:CPT}

In this sub-subsection, we shall argue 
that there actually exists an example of non-GL manifestation 
in field theoretical models, which is called ``conformal phase
transition'' ~\cite{Miransky-Yamawaki} 
characteried by an essential-singularity-type 
scaling (see below).
 
Following Ref.~\cite{Miransky-Yamawaki},
we here briefly summarize the ``conformal phase transition'',
and demonstrate how the GL (linear sigma model-like) manifestation
breaks down, 
using the Gross-Neveu model~\cite{Gross-Neveu} as an example.

In the linear sigma model-like phase transition, around the critical
point
$z=z_c$ (where $z$ is a generic notation for parameters of a theory, 
as the coupling constant $\alpha$, number of particle flavors $N_f$,
etc), 
an order parameter $\Phi$ takes the form
\begin{equation}
\Phi = \Lambda f(z) \label{2}
\end{equation}
($\Lambda$ is an ultraviolet cutoff),
where $f(z)$ has a non-essential singularity at $z=z_c$ such
that $\lim f(z)=0$ as $z$ goes to $z_c$ both in the symmetric and
broken phases. The standard form for $f(z)$ is 
$f(z) \sim (z - z_c)^{\nu}$, $\nu > 0$, around $z = z_c$.  

The ``conformal phase transtion'' is a very different continuous phase
transition. We define it 
as a phase transition in which an order parameter $\Phi$ is given by
Eq.~(\ref{2})
where however $f(z)$ has an {\it essential} singularity 
at $z=z_c$ in such a way that while 
\begin{equation}
\lim_{z \to z_c} f(z) =0 \label{3}
\end{equation}
as $z$ goes to $z_c$ from the side of the broken phase, 
$\lim f(z) \ne 0 $ 
as $z \to z_c$ from the side of the symmetric phase 
(where $\Phi \equiv 0$). 
Notice that since the relation (\ref{3}) ensures that 
the order parameter $\Phi \to 0$ as $z \to z_c$,
the phase transition is continuous.

A typical example of the conformal phase transition
is given by the phase transition 
in the $(1+1)$-dimensional Gross-Neveu model.
Here
we first consider the dynamics in the $D$-dimensional 
$(2 \leq D < 4)$ Nambu-Jona-Lasinio (Gross-Neveu) model, and
then,
describe the ``conformal phase transtion'' in the Gross-Neveu (GN)
model at $D=2$. 
This will allow to illustrate main features of the ``conformal phase
transtion'' in a very clear way.

The Lagrangian of the $D$-dimensional GN model, with the 
$U(1)_L \times U(1)_R$ chiral symmetry, 
takes the same form as the Lagrangian (\ref{Lag}) for the 
Nambu--Jona-Lasinio model in 4 dimensions:
\begin{equation}
{\cal L}= \bar{\psi}\, i \gamma^{\mu}\partial_{\mu} \psi  
+ \frac{G}{2} \left[ (\bar{\psi} \psi)^2 + 
(\bar{\psi} i \gamma_5\psi)^2  
\right] , \label{9}
\end{equation}
where $\mu=0,1, \cdot \cdot ,D-1$, and the fermion field carries 
an additional ``color'' index $\alpha = 1,2, \cdot \cdot , N_c$. 
As we have shown in Eq.~(\ref{Lag A}),
the theory is equivalent to the theory with the Lagrangian
\begin{equation}
{\cal L}' = \bar{\psi} \, i\gamma^{\mu}\partial_{\mu}\psi 
-\bar{\psi} (\varphi + i \gamma_5 \pi) \psi 
- \frac{1}{2G} (\varphi^2 + \pi^2) 
\ . \label{10}
\end{equation}

Let us look at the effective potential in this theory, which takes the
same form 
as Eq.~(\ref{eff pot}) except that $\int d^4 k$ is replaced by 
$\int d^D k$. 
It is explicitly calculated as~\cite{Kondo-Tanabashi-Yamawaki}:
\begin{equation}
V(\varphi , \pi) = \frac{4 N_c \Lambda^D}{(4\pi)^{D/2} \Gamma(D/2)}
\left[ 
  \left( \frac{1}{g} - \frac{1}{g_{\rm cr}} \right) 
  \frac{\rho^2}{2\Lambda^2} 
  + \frac{2}{4-D} \frac{\lambda_D}{D} 
    \left( \frac{\rho}{\Lambda} \right)^D 
\right] 
+ O \left(\frac{\rho^4}{\Lambda^4}\right) 
\ , \label{13}
\end{equation}
where $\rho = (\varphi^2 + \pi^2 )^{1/2}$, 
$\lambda_D = B(D/2-1,3-D/2)$, 
the dimensionless coupling constant $g$ is defined by 
\begin{equation}
g = \frac{4 N_c \Lambda^{D-2}}{(4\pi)^{D/2}\Gamma(D/2)} G 
\ , \label{14}
\end{equation}
and the critical coupling $g_{\rm cr} = \frac{D}{2}-1$. 

At $D>2$, one finds that 
\begin{equation}
M_\varphi^2 
\equiv 
\left. \frac{d^2 V}{d \rho^2} \right\vert_{\rho=0} \simeq 
\frac{4 N_c \Lambda^{D-2}}{(4\pi)^{D/2}\Gamma(D/2)}
  \frac{g_{\rm cr}-g}{g_{\rm cr}\,g}
\ .  
\label{15}
\end{equation}
As shown for the 4-dimensional NJL model in 
Eq.~(\ref{NJL cond}),
the sign of $M_\varphi^2$ defines two different phases: 
$M_\varphi^2 > 0$ $(g < g_{\rm cr})$ 
corresponds to the symmetric phase
and  
$M_\varphi^2 < 0$ $(g > g_{\rm cr})$ corresponds to the broken phase with
spontaneous 
chiral 
symmetry breaking, $U(1)_L \times U(1)_R \to U(1)_{L+R}$. 
The value $M_\varphi^2 =0$ defines the critical point $g=g_{\rm cr}$.

Therefore at $D>2$, a linear-sigma-model-like phase transition is
realized. 
However the case $D=2$ is special: 
now $g_{\rm cr} \to 0$ and $\lambda_D \to \infty$ as 
$D \to 2.$ In this
case  
the effective potential is the well-known potential
of the Gross-Neveu model~\cite{Gross-Neveu}: 
\begin{equation}
V(\varphi , \pi) = \frac{N_c}{2\pi g} \rho^2 
- \frac{N_c \rho^2}{2\pi}
 \left[ \ln \frac{\Lambda^2}{\rho^2} + 1 \right]
\ . \label{16}
\end{equation}
The parameter $M_\varphi^2$ is now : 
\begin{equation}
M_\varphi^2 = 
\left. \frac{d^2 V}{d \rho^2} \right|_{\rho=0} \to - \infty
\ . 
\end{equation} 
Therefore, in this model, one cannot use $M_\varphi^2$ as a parameter
governing  
the continuous phase transition at $g = g_{\rm cr} =0$ : 
the phase transition is not a linear sigma model-like phase transition 
in this case. 
Indeed, as follows from Eq.~(\ref{16}), 
the order parameter, which is a solution to the gap equation 
$\frac{d V }{d \rho} =0$, is 
\begin{equation}
\bar{\rho} = \Lambda \exp \left( -\frac{1}{2g} \right) 
\label{18}
\end{equation}
in this model. The function $f(z)$, defined in Eq.~(\ref{2}), is now 
$f(g) = \exp (- \frac{1}{2g})$, i.e., $z=g$, and therefore the
conformal phase transition 
takes place in this model at $g=0$: $f(g)$ goes to zero only if
$g \to 0$ from the side of the broken phase ($g>0$). 

Let us discuss this point in more detail. 

At $D \geq 2$, the spectrum of the $\varphi$ and $\pi$ excitations 
in the symmetric solution, with $\bar{\rho}=0$, is defined by the
following  
equation (in leading order in 
$\frac{1}{N_c}$)~\cite{Kondo-Tanabashi-Yamawaki}: 
\begin{equation}
\left(\frac{1}{g} - \frac{1}{g_{\rm cr}}\right) \Lambda^{D-2} 
+ \frac{\lambda_D}{2-D/2} 
(-M_{\pi}^2)^{D/2-1} = 0
\ . \label{19}
\end{equation}
Therefore at $D>2$, there are tachyons with 
\begin{equation}
M_{\pi}^2=M_{\varphi}^2 = M^2_{tch} = 
-\Lambda^2  
\left(\frac{4 -D}{2\lambda_D}\right)^{\frac{2}{D-2}} 
\left(\frac{g-g_{\rm cr}}{g_{\rm cr}g}\right)^{\frac{2}{D-2}}
 \label{20} 
\end{equation}
at $g > g_{\rm cr}$, and  at $g < g_{\rm cr}$ there are ``resonances''
with
\begin{equation}
|M_{\pi}^2| = |M_{\varphi}^2| = 
\Lambda^2 
\left(\frac{4 -D}{2\lambda_D}\right)^{\frac{2}{D-2}} 
\left(\frac{g_{\rm cr}-g}{g_{\rm cr}g}\right)^{\frac{2}{D-2}} 
\ . \label{21} 
\end{equation} 
Equation~(\ref{21}) implies that the limit $D \to 2$ is special. 
One finds from Eq.~(\ref{19}) that at $D=2$
\begin{equation}
M_{\pi}^2=M_{\varphi}^2 = M^2_{tch} = - \Lambda^2 
\exp \left( - \frac{1}{g}\right) 
\label{22}
\end{equation}
at $g> 0$, and 
\begin{equation}
|M_{\pi}^2| = |M_{\varphi}^2| =  \Lambda^2 
\exp \left(  \frac{1}{|g|} \right) 
\label{23}
\end{equation}
at $g< 0$, i.e., in agreement with the main feature of the
conformal phase transition,  
there are no light resonances in the symmetric phase at $D=2$.

The effective potential (\ref{16}) can be rewritten as 
\begin{equation}
V(\varphi , \pi) = \frac{N_c \rho^2}{2\pi}
 \left[ \ln \frac{\rho^2}{\bar{\rho}^2}- 1 \right]  \label{24}
\end{equation}
(with $\bar{\rho}$ given by Eq.~(\ref{18})) in the broken
phase. 
That is, in this phase $V(\varphi , \pi)$ is finite in the continuum
limit 
$\Lambda \to \infty$ after the renormalization of the coupling
constant,  
\begin{equation}
g=\frac{1}{\ln \frac{\Lambda^2}{\bar{\rho}^2}} \label{25}
\end{equation}
[see Eq.~(\ref{18})]. But what is the form of the effective potential 
in the continuum limit in the symmetric phase, with $g<0$ ? 
As Eq.~(\ref{16}) implies, it is infinite as $\Lambda \to \infty$: 
indeed at $g<0$, there is no way to cancel the logarithmic divergence
in $V$.  

It is unlike the case with $D> 2$ : in that case, using
Eq.~(\ref{15}),  
the potential (\ref{13}) can be put in a linear-sigma-model-like form: 
\begin{equation}
V(\varphi , \pi) = \frac{M_\varphi^2}{2} \rho^2 
+ \frac{8 N_c}{(4\pi)^{D/2} \Gamma(D/2)} 
  \frac{\lambda_D}{(4-D)D} \rho^D
\ . 
\label{26}
\end{equation}
However, since $M_\varphi^2 = -\infty$ at $D=2$, 
the linear-sigma-model-like
form for the potential is not available in the Gross-Neveu model. 

What are physical reasons of such a peculiar behavior of the effective 
potential at $D=2$ ? 
Unlike the case with $D>2$, at $D=2$ the Lagrangian
(\ref{9}) defines a conformal theory in the classical limit. 
By using the conventional approach, one can derive the following 
equation for the conformal anomaly in this model 
(see, for detailed derivation, the Appendix of 
Ref.~\cite{Miransky-Yamawaki}): 
\begin{equation}
\partial^{\mu} D_{\mu} = \theta_{\mu}^{\mu} = 
\frac{\pi}{2 N_c} \beta (g)  
\left[ (\bar{\psi}\psi)^2 + (\bar{\psi} i 
\gamma_5 \psi )^2 \right] \label{27} 
\ ,
\end{equation}
where $D_{\mu}$ is the dilatation current, $\theta_{\nu}^{\mu}$ 
is the energy-momentum tensor, and the $\beta(g)$ is the $\beta$ function 
given by 
\begin{equation}
\beta(g) = \frac{\partial g}{\partial \ln \Lambda}
=-g^2
\end{equation} 
both in the broken and 
symmetric phases. 
While the broken phase ($g>0$) corresponds to asymptotically free 
dynamics, the symmetric phase ($g < 0$)
defines infrared free dynamics: 
as $\Lambda \to \infty$, we are led to a free theory of massless 
fermions, which is of course conformal invariant. 

On the other hand, in the broken phase the conformal symmetry 
is broken, even as $\Lambda \to \infty$. 
In particular, Eq.~(\ref{24}) implies that
\begin{equation}
\langle 0 | \theta_{\mu}^{\mu} | 0 \rangle = 4V(\bar{\rho}) =
- \frac{2N_c}{\pi}{\bar{\rho}}^2 \ne 0
\label{28}
\end{equation}
in leading order in $\frac{1}{N_c}$ in that phase. 

The physics underlying this difference between the two phases 
in this model is clear:
while  $g<0$ corresponds to repulsive interactions between 
fermions, attractive interactions at $g>0$ lead to the
formation of  
bound states, thus breaking the conformal symmetry. 
Thus the conformal phase transition
describes the two essentially 
different realizations of the conformal symmetry in the symmetric and
broken phases.

The confromal phase transition is also observed in other field
theoretic models: 
A most notable example is the ordinary QCD (with small $N_f$) which 
exhibits a well-known 
essential-singularity-type scaling at $\alpha(\Lambda)=0$:
\begin{equation}
m \sim \Lambda  e^{- \frac{1}{b \alpha(\Lambda)}} \ ,
\end{equation}
although it has no symmetric phase (corresponding to $\alpha
<0$). Similar 
essential-singularity-type scaling has been observed in the 
ladder QED~\cite{Miransky}, the gauged NJL model  
in the ladder approximation~\cite{Kondo-Mino-Yamawaki,
Appelquist-Soldate-Takeuchi-Wijewardhana}, etc.  We shall discuss in 
Sec.~\ref{ssec:CPTLNQ} a conformal phase transtion observed in
the
large $N_f$ QCD within the ladder approximation.  
(Details are discussed in
Ref.~\cite{Miransky-Yamawaki}).

\subsubsection{Vector Manifestation vs. ``Vector Realization''}
\label{sssec:VR}

The VM in the HLS
is similar to the ``Vector
realization''~\cite{Georgi:1,Georgi:2} also formulated in the HLS, 
in the sense that the chiral symmery gets unbroken in such a way that 
vector meson $\rho$ 
becomes massless $m_\rho \rightarrow 0$ and a chiral partner of $\pi$.
However VM is different from the ``Vector
realization'' in an essential way:
The ``Vector realization'' was claimed to be 
{\it neither the Wigner realization nor the NG realization} 
in such a way that the NG boson
does exist ($F_\pi(0)\ne 0$), while the chiral symmetry is still 
unbroken ($\langle \bar{q} q \rangle = 0$):
\begin{equation}
F_\pi(0)\ne 0, \quad  \langle \bar{q} q \rangle = 0 
.
\label{VR1}
\end{equation}
On the contrary, our  VM is precisely the limit of the
{\it Wigner realization} having 
\begin{equation}
F_\pi(0) =0, \quad \langle \bar{q} q \rangle = 0 
.
\label{VM1}
\end{equation}

A crucial difference between the two comes from the fact that
in VM  the quadratic divergence  of our Wilsonian RGEs  
leads to the Wigner realization with
$F_\pi(0)\rightarrow 
0$ at the low-energy limit (on-shell of NG bosons) in spite of 
$F_\pi(\Lambda) \ne 0$, while in the ``Vector realization''
the quadratic divergence is not included and hence it was presumed that
$F_\pi(0)=F_\pi(\Lambda)$ and thus $F_\pi(0)\ne 0$.

Technically, in the vector limit (or the VM limit with the VM conditions), 
the {\it bare} HLS Lagrangian in the VM
and that of the ``Vector realization'',
{\it formally} approach  the same fixed point Lagrangian 
${\cal L}_{\rm HLS}^{\ast} $ 
which is defined just on the fixed point 
$g(\Lambda)=0$, $a(\Lambda)=1$ and
$F_\pi(\Lambda)\neq 0$ (plus
$z_1(\Lambda) = z_2(\Lambda)$):
\begin{eqnarray}
{\cal L}_{\rm HLS}^{\ast} &=& 
F_\pi^2(\Lambda) \, 
\left\{
 \mbox{tr} 
  \left[ \hat{\alpha}_{\perp\mu} \hat{\alpha}_{\perp}^\mu 
  \right]
  + \mbox{tr}
  \left[ \hat{\alpha}_{\parallel\mu} \hat{\alpha}_{\parallel}^\mu
  \right]
\right\}
+ z_1(\Lambda) \,
\left\{
 \mbox{tr}
 \left[ \hat{\cal V}_{\mu\nu} \hat{\cal V}^{\mu\nu} 
   \right]
  + \mbox{tr}
   \left[ \hat{\cal A}_{\mu\nu} \hat{\cal A}^{\mu\nu} 
   \right]
\right\} 
\nonumber\\
&=&
-\frac{F_\pi^2(\Lambda)}{4} \mbox{tr}
\left\{
 \left[{\cal D}_\mu \xi_L \cdot \xi_L^\dagger
 \right]^2
+ \left[{\cal D}_\mu \xi_R \cdot \xi_R^\dagger
  \right]^2
\right\}
+ \frac{z_1(\Lambda)}{2} \,
\mbox{tr}
\left\{
  \left[ \hat{\cal L}_{\mu\nu} \hat{\cal L}^{\mu\nu} 
   \right]
  + \left[ \hat{\cal R}_{\mu\nu} \hat{\cal R}^{\mu\nu} 
   \right]
\right\}
\ ,
\nonumber\\
\label{fixed point Lagrangian}
\end{eqnarray}
where ${\cal D}_\mu \xi_L\equiv \partial_\mu \xi + i \xi {\cal L}_\mu$
(and $L \leftrightarrow R$).

However, when the external 
gauge fields are 
switched off, it was pointed out~\cite{Georgi:1,Georgi:2}
that  the fixed point Lagrangian 
${\cal L}_{\rm HLS}^{\ast}$ possesses
 a large (global) symmetry based on the manifold
\begin{equation}
\frac{G_1 \times G_2}{G}=\frac{
\left[\mbox{SU}(N_f)_{\rm L}\times \mbox{SU}(N_f)_{\rm R} \right]_1
\times
\left[\mbox{SU}(N_f)_{\rm L} \times \mbox{SU}(N_f)_{\rm R}\right]_2
}
  {
  \mbox{SU}(N_f)_{{\rm L}_1+{\rm L}_2} \times 
  \mbox{SU}(N_f)_{{\rm R}_1 +{\rm R}_2 }
 }
\ ,
\label{fixedpointlagrangian}
\end{equation}
where the residual symmetry 
$G=\mbox{SU}(N_f)_{{\rm L}_1+{\rm L}_2} \times 
\mbox{SU}(N_f)_{{\rm R}_1 +{\rm R}_2 }$ was identified 
in Ref.~\cite{Georgi:1,Georgi:2} with  
the chiral symmetry of the QCD, while 
$G_1 \times G_2$ is a (global) symmetry larger than that of QCD
such that
\begin{equation}
\xi_{\rm L} \rightarrow g_{{\rm L}_1}\,
                                  \xi_{\rm L}\,
                                  g_{{\rm L}_2}^\dagger
\ ,
\end{equation}
with $g_{{\rm L}_1} \in  \mbox{SU}(N_f)_{{\rm L}_1}$
and  $g_{{\rm L}_2} \in \mbox{SU}(N_f)_{{\rm L}_2}$
 (and $L \leftrightarrow R$).
Then the fixed point Lagrangian ${\cal L}_{\rm HLS}^{\ast}$ 
has no connection with the QCD and must be decoupled from QCD!
Even if we are off the point $(a(\Lambda),g(\Lambda))=(1,0)$ by
$a(\Lambda) \ne 1$, we still have a redundant global symmetry
$H \times G$ which is larger than the QCD symmetry by the additional 
global symmetry $H(\subset G_1)$,
where $G_1$ is reduced to the subgroup $H$ by $a(\Lambda)\ne 1$. 

When the HLS coupling is switched on, $g (\Lambda) \ne 0$, on the other hand, the $G_1$
(or $H\subset G_1$ when $a(\Lambda) \ne 1$) becomes a local symmetry, namely 
the HLS $H_{\rm local}=\mbox{SU}(N_f)_{\rm local}$, and hence the larger 
global symmetry 
$G_1 \times G_2$   
is reduced  to the original symmetry of the HLS model, 
$H_{\rm local} \times  G_{\rm global}$ ($G=G_2$), 
as it should, in accord with the QCD 
symmetry. 
 Such a redundant larger (global) symmetry $G_1 \times G_2$ 
 (or $H \times G$) is
specific to just on the fixed point $g\equiv 0, a\equiv 1$
(or $g\equiv 0$). 
Then the point $(a,g)=(1,0)$ must be regarded only as a limit
\begin{equation}
g \, (\ne 0) \rightarrow 0 \,,
\end{equation}
in which case the effective Lagrangian has no such a
redundant  
global symmetry. 
Actually, as was shown in Sec.~\ref{a1phenomenology}, 
the real-life QCD with $N_f=3$ is very close to 
$a(\Lambda)=1$ but $g^2(\Lambda) \gg 1$, which means that
Nature breaks such a redundant $G_1\times G_2$ symmetry only by a strong
coupling gauge interaction of the composite gauge boson $\rho$.
When we approach the chiral restoration point of the underlying QCD,
this strong gauge coupling becomes  vanishingly small, thus forming a weak 
couling composite gauge theory,  but the gauge coupling should never vanish,
however small.
In the next sub-subsection, we shall discuss in
detail that VM must actually be regarded as such a limit.

On the other hand, situation is completely different for the 
``Vector realization'': In order to have the {\it unbroken}
 chiral symmetry of QCD  under the condition $F_\pi(0)\ne 0$, 
 namely existence of NG bosons,
$\pi$ and $\sigma$, it desperately needs a redundant larger 
global symmetry which is to be spontaneously broken down to the
unbroken 
chiral symmetry of QCD. It then must be formulated precisely on the
point $(a,g)\equiv (1,0)$ whose effective  
Lagrangian ${\cal L}_{\rm HLS}^{\ast}$ in 
Eq.~(\ref{fixed point Lagrangian}) 
actually does have such a redundant symmetry.
Then it implies that ``Vector realization'' is decoupled from the QCD!

We now show, based on the general arguments~\cite{Y}
on the chiral Ward-Takahashi (WT) identity, 
that the ``Vector realization'', Eq.~(\ref{VR1}), implies that the NG
bosons are 
actually all decoupled from the QCD. This is consistent with the fact
that 
the fixed point Lagrangian in the ``Vector realization'' has a
different 
symmetry than QCD and is decoupled from the QCD.

Let us start with the symmetry $G$ of a system
including fields $\phi_i$ under the transformation
$\delta^A \phi_i = -i (T^A)^j_i \phi_j= [i Q^A, \phi_i]$, 
with $A=1,2,\ldots,\dim G$, where $T^A$ are the matrix
representations of the generators of the symmetry group $G$
and $Q^A$ the corresponding charge operators.
Let the symmetry be spontaneously broken into a subgroup $H$,
$Q^a |0\rangle\ne 0$, where $Q^a$ are the charges corresponding to
the broken generators $T^a \in {\cal G}-{\cal H}$, with ${\cal G}$ and
${\cal H}$ being algebras of $G$ and $H$,
in such a way that 
\begin{equation}
\delta^a G_n(x_1,\ldots,x_n) =
\bigl\langle
  0
  \bigl\vert
    \left[ i Q^a, T\, \phi_1(x_1)  \cdots \phi_n(x_n)
    \right]
  \Bigr\vert
  0
\bigr\rangle
\ne 0
\ ,
\end{equation}
where 
$\delta^a G_n$ is an $n$-point order parameter
given by the variation 
of the n-point Green function 
\begin{equation}
G_n(x_1, \ldots, x_n) \equiv
\bigl\langle
  0
  \bigl\vert
    T\, \phi_1(x_1)  \cdots \phi_n(x_n)
  \Bigr\vert
  0
\bigr\rangle
\end{equation}
($T$: time-ordered product)
under the transformation corresponding to the broken
generators $T^a$.
Then, we have a general form of the chiral WT identity:
\begin{eqnarray}
\lim_{q_{\mu}\rightarrow 0}q^{\mu} M^a_{\mu}
=\delta^a G_n(x_1,\ldots,x_n) \ ,
\end{eqnarray}
where the current-inserted Green function for broken current
(axialvector current) $J_{5\mu}$ is defined by
\begin{equation}
M_\mu^a(q, x_1, \ldots, x_n)
\equiv \int d^4 z e^{i q z}  
\bigl\langle
  0
  \bigl\vert
    T\, J_{5\mu}^a(z) \phi_1(x_1)  \cdots \phi_n(x_n)
  \Bigr\vert
  0
\bigr\rangle\ .
\end{equation}
Noticing that $\delta^a G_n(x_1,\ldots,x_n)$ is a residue of the 
NG boson pole at $q^2=0$ in $M_\mu^a(q, x_1, \ldots, x_n)$, we have~\cite{Y}
\begin{eqnarray}
\delta^a G_n(x_1,\ldots,x_n)= F_\pi(0) \cdot
\Bigl\langle
  \pi^a (q_\mu=0)
  \Bigl\vert
    T\, \phi_1(x_1) \cdots \phi_n(x_n)
  \Bigr\vert
  0
\Bigr\rangle
\ ,
\label{eq:WT}
\end{eqnarray}
where
$\bigl\langle
  \pi^a (q_\mu)
  \bigl\vert
    T\, \phi_1(x_1) \cdots \phi_n(x_n)
  \bigr\vert
  0
\bigr\rangle
$
is a Bethe-Salpeter amplitude
which plays a role of ``wave function'' of the NG boson $\pi^a$
and the NG boson decay constant $F_\pi(0)$ is defined by
\begin{equation}
\Bigl\langle
  0
  \Bigl\vert
    J_\mu^a(x)
  \Bigr\vert
  \pi^b (q)
\Bigr\rangle
= - i \delta^{ab} F_\pi(0) q_\mu e^{-i q x} \ .
\end{equation}

 A simple example of the relation Eq.~(\ref{eq:WT}) is given by the
linear sigma model: 
$\delta^a \pi^b=  \delta^{ab} \sigma$ and $\delta^a \sigma=- \pi^a$,
$\delta^a G_1 (x)= \delta^a \langle 0| \pi^b(x)|0\rangle
=\delta^{ab} \langle 0| \sigma(x) | 0 \rangle$, while
$\langle \pi^a| \pi^b(x)|0 \rangle =\delta^{ab} Z_\pi^{1/2}$, and
hence 
Eq.~(\ref{eq:WT}) reads $\langle 0| \sigma| 0 \rangle=F_\pi(0) \,
Z_\pi^{1/2}$,  
or  $\langle \sigma \rangle=F_\pi(0)$ at tree level where the $\pi$
wave  
function renormalization is trivial, $Z_\pi^{1/2}=1$.
Another popular example is the (generalized) Goldberger-Treiman
relation for the 
quark propagator $S(p) ={\cal F}{\cal T} G_2(x) ={\cal F}{\cal T}
\langle 0|T q(x) \bar q(0)|\rangle$ 
where ${\cal F}{\cal T}$ stands for the Fourier transform:
Eq.~(\ref{eq:WT}) reads $\delta^a G_2(x)=F_\pi(0) \cdot \langle \pi^a
(q^\mu=0)|T q(x) \bar q(0)|0\rangle$, which after taking Fourier
transform reads  
\footnote{ 
  In the case of $\pi$-nucleon system, 
  $\Gamma_\pi^a(p,0)$ reads $G_{{\rm N}{\rm N}\pi}$ (NN$\pi$ Yukawa
  coupling) 
  and $\Sigma(p^2)$ does $m_N$ (nucleon mass), and hence the
  Goldberger-Treiman relation 
  follows $2 m_N \, g_A = F_\pi(0)\,  G_{{\rm N}{\rm N}\pi}$ with
  $g_A=1$. $g_A\ne 1$ would  
  follow only when we take account of the fact that the nucleon is not
  the irreducible  
  representation of the chiral algebra due to the representation mixing
  in the  
  Adler-Weisberger sum rule.
}
\begin{equation}
2\, \Sigma (p^2) =F_\pi(0) \, \Gamma_\pi^a(p,0)\ ,
\end{equation}
where $\Sigma(p^2)$ is the dynamical mass of the quark parametrized as
$i S^{-1}(p) = Z_\psi^{-1}(\gamma^\mu p_\mu -\Sigma(p^2))$ and 
$\delta^a S=S\, \{ -i \gamma_5 T^a,\, S^{-1}\}\, S =\gamma_5 T^a
Z_\psi^{-1}\, 
 S\cdot 2\, \Sigma\cdot S$ 
for $\delta^a q(x)= -i \gamma_5 T^a q(x)$, while $\Gamma_\pi^a(p,q)$
is 
an amputated (renormalized) Bethe-Salpeter amplitude of $\pi^a$, or
dynamically  
induced $\pi$-$q$-$q$ vertex; 
\begin{equation}
\gamma_5 T^a  Z_\psi^{-1} S(p+q)\, \Gamma_\pi^a(p,q)\, 
S(p)
\equiv {\cal F}{\cal T}   \langle \pi^a (q^\mu)|T q(x) \bar
q(0)|0\rangle  
\ .
\end{equation}

Now, when the broken symmetry is restored,
$Q^a |0\rangle=0$, we simply have  
\begin{equation}
\delta^a G_n (x_1, \ldots ,x_n) =
\bigl\langle
  0
  \bigl\vert
    \left[ i Q^a, T\, \phi_1(x_1)  \cdots \phi_n(x_n)
    \right]
  \bigr\vert
  0
\bigr\rangle
=0
\end{equation} 
for {\it all} Green functions. If one assumed there still exist
NG bosons $F_\pi(0)\neq 0$ as in ``Vector Realization'', 
then Eq.~(\ref{eq:WT}) would dictate    
\begin{equation}
 F_\pi(0) \cdot
\Bigl\langle
  \pi^a (q_\mu=0)
  \Bigl\vert
    T\, \phi_1(x_1) \cdots \phi_n(x_n)
  \Bigr\vert
  0
\Bigr\rangle
=0
\ ,
\end{equation}
and hence 
\begin{eqnarray}
&& 
\Bigl\langle
  \pi^a (q_\mu=0)
  \Bigl\vert
    T\, \phi_1(x_1) \cdots \phi_n(x_n)
  \Bigr\vert
  0
\Bigr\rangle
= 0 
\ , \quad \mbox{for all $n$}
\ .
\label{wave 0 WT}
\end{eqnarray}
This would imply a  situation that the 
NG bosons $\pi$ with $q_\mu =0$ would be  totally decoupled from 
{\it any}
operator, local or nonlocal,
 of the underlying theory, the QCD in the case at hand.
Then the ``Vector realization'' is totally decoupled from the QCD,
 which is also consistent with the fact that the fixed point
 Lagrangian  
 Eq.~(\ref{fixed point Lagrangian}) has a different
 symmetry than QCD.
 
On the other hand,
the VM is simply a limit to a Wigner phase, 
\begin{eqnarray}
&& F_\pi (0) \rightarrow 0 \ ,
\label{fp 0 WT}
\end{eqnarray}
and hence we can have
$\bigl\langle
  \pi^a (q_\mu=0)
  \bigl\vert
    T\, \phi_1(x_1) \cdots \phi_n(x_n)
  \bigr\vert
  0
\bigr\rangle
\neq 0\,
$
although $\pi^a$ in this case are no longer the NG bosons and 
may be no longer light composite spectrum as in the
conformal phase transition~\cite{Miransky-Yamawaki}.
If the light composite spectrum dissapear as in conformal phase
transition, then 
the effective field theory breaks down anyway just at the phase
transition point.

\subsubsection{Vector manifestation only as a limit}
\label{sssec:VM as a limit}

We actually defined the 
VM {\it as a limit} (``VM limit'') with bare parameters  
approaching the fixed point, VM point
$(X(\Lambda),a(\Lambda),G(\Lambda))=(1,1,0)=(X_2^*,a_2^*,G_2^*)$, 
{\it from the broken phase}
but not exactly on the fixed point.  
Since the fixed point Lagrangian has a different symmetry than QCD, 
we must approach the VM limit along the line {\it other than} $G\equiv 0$ 
(Fig.~\ref{fig:flows G0} in Sec.~\ref{ssec:PSH}).
We shall give an example to approach the VM limit from $G\ne 0$ in 
Sec.~\ref{sssec:CB}.

Here we demonstrate through the chiral WT 
identity that a relation precisely on the point $g=0$
contradicts 
the QCD even when  $F_\pi^2(0)\rightarrow 0$, 
while that as a limit $g\rightarrow 0$ is perfectly consistent.
This also gives another example to show that the ``Vector
realization'' is  
decoupled from the QCD.

The chiral WT identity is the same as that in the previous
sub-subsection   
except that two axialvector currents $J_{5\mu}$ and two vector
currents $J_\mu$ 
are involved:
\begin{eqnarray}
M_{\alpha\beta;\mu\nu}^{ab;cd}(x_1,x_2;q_1,q_2) 
 = {\cal FT} \langle 0| T J_{5\mu}^c(z_1) J_{5\nu}^d(z_2) 
                     J_\alpha^a(x_1) J_\beta^b(x_2) |0\rangle
\ ,
\end{eqnarray}
where $\cal{FT}$ stands for Fourier transform with respect to
$z_1$ and $z_2$. 
Then we have
\begin{eqnarray}
&&
\lim_{q_1\rightarrow 0, q_2\rightarrow 0} q_1^\mu q_2^\nu 
  M_{\alpha\beta;\mu\nu}^{ab;cd}(x_1,x_2;q_1,q_2) 
\nonumber\\
&& \qquad\qquad
  = ( f_{ace} f_{bde} + f_{ade} f_{bce} )
    \Bigl(
       \langle 0|T J_{5\alpha}(x_1) J_{5\beta}(x_2) |0\rangle
   -  \langle 0| T J_\alpha(x_1) J_\beta(x_2) |0\rangle 
  \Bigr)
\ ,
\label{2-currents WT}
\end{eqnarray}
where use has been made of 
\begin{eqnarray}
 \langle 0|T J_{5\alpha}^a(x_1) J_{5\beta}^b(x_2)  |0\rangle
  = \delta^{ab}  \langle 0| T J_{5\alpha}(x_1) J_{5\beta}(x_2)  
  |0\rangle
\ ,
\end{eqnarray}
etc.. Looking at the residues of massless poles of two $\pi$'s in 
$M_{\alpha\beta;\mu\nu}^{ab;cd}$, we have
\begin{eqnarray}
&&
  F_\pi^2(0)  \cdot
  \Gamma_{\alpha\beta}^{ab;cd}(x_1,x_2;0,0)
\nonumber\\
&& \qquad
  =
   ( f_{ace} f_{bde} + f_{ade} f_{bce} )
      \Bigl( \langle 0|T J_{5\alpha}(x_1) J_{5\beta}(x_2) |0\rangle
   -  \langle 0| T J_\alpha(x_1) J_\beta(x_2) |0\rangle\Bigr) \ ,
\end{eqnarray}
where
\begin{eqnarray}
\Gamma_{\alpha\beta}^{ab;cd}(x_1,x_2;q_1,q_2) 
=  \langle \pi^c (q_1) | T J_\alpha^a(x_1) J_\beta^b(x_2) | \pi^d (q_2) 
\rangle
\end{eqnarray}
is the amplitude for $\pi\pi\gamma\gamma$ process.
Taking Fourier transform with respect to $x_1-x_2$ and omitting 
$a$, $b$, $c$ and $d$,
we have
\begin{eqnarray}
F_\pi^2(0) \cdot \tilde{\Gamma}_{\alpha\beta} (k)
= ( g_{\alpha\beta} k^2 - k_\alpha k_\beta ) 
    [  \Pi_A(k^2) - \Pi_V(k^2) ]
\ .
\end{eqnarray}
Writing $ \tilde{\Gamma}_{\alpha\beta} (k)=
( g_{\alpha\beta} 
 - k_\alpha k_\beta/k^2 )  
\tilde{\Gamma} (k^2)$, we have
\begin{eqnarray}
F_\pi^2(0)\cdot \tilde{\Gamma} (k^2)= k^2  [  \Pi_A(k^2) - \Pi_V(k^2) ] \ ,
\label{WTk}
\end{eqnarray}
which clearly shows that when the chiral symmetry gets restored as 
$\Pi_A(k^2) - \Pi_V(k^2) \rightarrow 0$ in the underlying QCD, 
we would have a disaster, $\tilde{\Gamma} (k^2) \rightarrow 0$ for any $k^2$,
iff $F_\pi(0) \ne 0$ as in the ``Vector realization''. Actually,
by taking a limit $k^2\rightarrow 0$, we have
\begin{eqnarray}
F_\pi^2(0)\cdot \tilde{\Gamma} (0)= F_\pi^2(0) \ ,
\label{WT2}
\end{eqnarray}
 where use has been made of the Weinberg first sum rules, 
 $\lim_{k^2 \rightarrow 0} k^2 [  \Pi_A(k^2) - \Pi_V(k^2) ] 
  =F_\pi^2(0)$,
 which are valid in QCD for $N_f<33/2$ even in the restoration limit
 $F_\pi(0) \rightarrow 0$.
Then it follows
\begin{eqnarray}
\tilde{\Gamma} (0)= 1 \ ,
\label{twopitwogamma}
\end{eqnarray}
as far as  $F_\pi(0) \ne 0$ (including the limit
$F_\pi(0) \rightarrow 0$).

Now we compute the $\pi\pi\gamma\gamma$ amplitude
$\tilde{\Gamma} (k^2)$ in terms of the HLS model at $O(p^2)$:
\begin{eqnarray}
\tilde{\Gamma} (k^2) =
  (1 - a) + a \frac{ M_\rho^2 }{ M_\rho^2 - k^2 } \ ,
\end{eqnarray}
the first term of which correpsonds to the
direct  coupling of $\pi\pi\gamma\gamma$, while the
second term  does to the vertex $\pi\pi\gamma\rho$
followed by the transition $\rho \rightarrow \gamma$.
(There is no $\pi\pi\rho\rho$ vertex in the HLS model
at leading order.)
             
If we set 
$g\equiv 0$ (hence $M_\rho^2\equiv 0$), then
we would get 
\begin{equation}
\tilde{\Gamma} (k^2) =1-a 
\label{Gammag0}
\end{equation}
for {\it all} $k^2$, which would vanish at $a\rightarrow 1$
 in contradiction with the QCD result, 
Eq.~(\ref{twopitwogamma}). This again implies that the ``Vector
realization'' 
with $F_\pi^2(0)\ne 0$ is inconsistent with QCD. Eq.~(\ref{Gammag0})
implies that 
the VM having $F_\pi^2(0) \rightarrow 0$
is also inconsistent with Eq.~(\ref{twopitwogamma}), 
although not inconsistent with the Eq.~(\ref{WT2}). 
On the other hand, if we 
take a limit $g \rightarrow 0$ at $k^2=0$ for the VM,
we have
\begin{equation}
\tilde{\Gamma} (0) =1 \ ,
\end{equation}
in perfect agreement with Eq.~(\ref{twopitwogamma}). Therefore the VM
must be formulated as a limit 
\begin{equation}
g(\Lambda) \, (\ne 0) \rightarrow 0 
\end{equation}
in such a way that $F_\pi^2(0) \, (\ne 0) \rightarrow 0$,
while we can safely put $a =1$. 

Similar unphysical
situation can be seen for the parameter $X$ defined in
Sec.~\ref{ssec:PSH}:
When the bare parameter $X(\Lambda)$ approaches the one at the
VM point
$(X_2^\ast,a_2^\ast,g_2^\ast) = (1,1,0)$ from the
broken phase as $X(\Lambda)\rightarrow1$, $g(\Lambda) \rightarrow 0$, 
the parameter $X(0)$
approaches $0$ as $X(0)\rightarrow 0$, which implies that
$m_\rho^2/F_\pi^2(0) \rightarrow 0$ [see Sec.~\ref{sssec:CB}].
When the theory is exactly on the VM point,
on the other hand, we have $X(\Lambda)\equiv 1$ which leads to
$X(0)=1$ since $(X_2^\ast,a_2^\ast,g_2^\ast) = (1,1,0)$ is 
the fixed point.

The discussion in this subsection
also implies that presence of gauge 
coupling, however small, can change drastically the pattern of 
symmetry restoration in the nonlinear sigma model: For instance,
the lattice calculation has shown that 
the $N_f=2$ chiral Lagrangian has a $O(4)$ type restoration, i.e., the
linear sigma model-type restoration, while it has not given a definite
answer  
if it is coupled to gauge bosons like $\rho$, namely the lattice
calculation  
has not been inconsistent with the VM,  other than $O(4)$-type
restoration,  
in the limit $g\rightarrow 0$ (not $g\equiv 0$)  even for
$N_f=2$.~\footnote{ 
  We thank Yoshio Kikukawa for discussions on this point
}

\subsection{Chiral phase transition in large $N_f$ QCD}
\label{ssec:CPTLNQ}

In this subsection we summarize the known results of the
chiral symmetry restoration in the 
large $N_f$ QCD, with $N_f$
($N_f < N_f^{**}\equiv \frac{11}{2}N_c$) being number of massless
quark flavors. 
For a certain large $N_f$ the coupling has an infrared fixed point 
which becomes very small near $\frac{11}{2}N_c$~\cite{Banks-Zaks}.
The  two-loop $\beta$ function is given by 
\begin{equation}
\beta(\alpha) = - b \alpha^2 -c \alpha^3 \ , 
  \label{90}  
\end{equation}
where two coefficients are \cite{Jones,Caswell}:
\begin{eqnarray}
b &=& \frac{1}{6\pi} ( 11 N_c - 2 N_f)\ , 
\nonumber \\
c &=& \frac{1}{24\pi^2} 
\left( 34 N_c^2 - 10 N_c N_f - 3 \frac{N_c^2 -1}{N_c} N_f \right)
\ .
   \label{91}  
\end{eqnarray}
There is at least one 
renormalization scheme in which the two-loop $\beta$ function 
is (perturbatively) exact~\cite{tHooft}. 
We will use such a renormalization scheme. 
Then we have an infrared fixed point for 
if $b>0$ and $c<0$
\begin{equation}
\alpha = \alpha^*  = - \frac{b}{c}\ .  \label{92}  
\end{equation}
When $N_f$ is close to (but smaller than) 
$N_f^{**} = \frac{11}{2}N_c$, the value of $\alpha^*$ is small and
hence 
one should expect that the chiral
symmetry is not spontaneously broken: namely, there is a critical 
value of $N_f$, $N_f = N_f^{\rm crit}$ beyond which the
(spontaneous broken) chiral symmetry is restored~\cite{Appelquist-Terning-Wijewardhana}.
 
Actually when we decrease $N_f$, the value of the fixed point
$\alpha^*$  
increases 
and eventually blows up (this fixed point disappears)
 at the value $N_f= N_f^{*}$
when the coefficient $c$ becomes positive 
($N_f^* \simeq 8.05 $ for $N_c = 3$, although this value is not
reliable since 
the perturbation must breaks down for strong coupling). 
However, before reaching $N_f^*$ 
the perturbative infrared fixed point in the $\beta$ function
will disappear at $N_f=N_f^{\rm crit}(>N^*)$ where 
the coupling $\alpha^*$ exceeds a certain critical value $\alpha_c$
so that the chiral symmetry is spontaneously broken; namely, 
 fermions can acquire a dynamical mass and hence decouple 
from the infrared dynamics, 
and only gluons will contribute to the $\beta$ function.

The value $N_f^{\rm crit}$ may be estimated in the (improved) ladder 
Schwinger-Dyson (SD) equation
combined with the perturbative fixed 
point~\cite{Appelquist-Terning-Wijewardhana}. 
It is well known~\cite{Maskawa-Nakajima,Fukuda-Kugo,Fomin} that 
in the (improved) ladder SD equation the spontaneous 
chiral symmetry breaking
would not occur when the gauge coupling 
is less than a critical value 
$\alpha < \alpha^* <\alpha_c = \frac{2N_c}{N_c^2 -1} 
  \cdot \frac{\pi}{3}$. 
Then, the estimate for the critical value $N_f^{\rm crit}$ is 
given by~\cite{Appelquist-Terning-Wijewardhana}:
\begin{equation}
\alpha^*  \left. \right|_{N_f = N_f^{\rm crit}} = \alpha_c
  \label{93}
\end{equation}
or,
\begin{equation}
N_f^{\rm crit} =N_c \left(\frac{100 N_c^2-66}{25 N_c^2 -15}\right)
\simeq 12 \frac{N_c}{3} 
\ .
  \label{94}
\end{equation}

However, the above estimate of $N_f^{\rm crit}$
through the ladder SD equation combined with the perturbative fixed
point 
may not be 
reliable, since besides various uncertainties of 
the ladder approximation for the estimate of 
the critical coupling $\alpha_c$, the {\it perturbative}
estimate of the fixed point value $\alpha^*$ in Eq.~(\ref{92}) 
is far from reliable, when it is equated to $\alpha_c$
which is of order $O(1)$. 

As we said before,
such a chiral symmetry restoration in the large $N_f$ QCD
is actually observed by various
other methods such as the
lattice simulation~\cite{Kogut-Sinclair,BCCDMSV:92,%
IKSY:92,IKSY:92b,IKSY:93,IKSY:94,IKKSY:96,IKKSY:98,%
Damgaard-Heller-Krasnitz-Olesen},
dispersion
relation~\cite{Oehme-Zimmerman:1,Oehme-Zimmerman:2}, 
instanton calculus~\cite{Velkovsky-Shuryak}, etc..
The most recent result of the lattice simulation 
shows~\cite{IKKSY:98}
\begin{equation}
6 < N_f^{\rm crit} < 7 
\ ,
\label{lattcrit}
\end{equation}
which is substantially smaller than the ladder-perturbative estimate 
Eq.~(\ref{94}).

Although the ladder-perturbative 
estimate of $N_f^{\rm crit}$ may not be reliable, it is
worth metioning that the result of the ladder SD equation
has a scaling of an essential-singularity for the dynamical mass $m$
of the 
fermions~\cite{Appelquist-Terning-Wijewardhana}, called 
``Miransky scaling'' as first observed
in the ladder QED~\cite{Miransky}:
\begin{eqnarray}
m \approx \Lambda
\exp\left({{- \pi}\over{\sqrt{{{\alpha^*}\over{\alpha_c}} -1}}}\right) 
=\Lambda 
\exp\left({{- C} \over{\sqrt{1/N_f - 1/N_f^{\rm crit}      }}}\right) 
\ ,
\label{Mscaling}
\end{eqnarray}
with $\Lambda$ being the ``cutoff'' of the dominant momentum region in
the 
integral of the SD equation and 
$C=\sqrt{ (13 N_c^2 N_f - 34 N_c^3 - 3N_f)/(
  100 N_c^3 N_f - 66 N_c N_f ) }$. 
Relatively independent of the
estimate of $N_f^{\rm crit}$, this feature may describe partly the
reality of  
the chiral phase transition of large $N_f$ QCD.
It was further shown%
~\cite{Appelquist-Terning-Wijewardhana,Miransky-Yamawaki} 
in the ladder approximation that the light spectrum does not exist 
in the symmetric 
phase in contrast to the broken phase
where the scalar bound state becomes massless in addition to 
the massless NG boson $\pi$.
As was discussed in Sec.~\ref{sssec:CPT},
these features are in accord with the 
conformal phase transition where the
Ginzburg-Landau (GL) effective theory (linear sigma model-like
manifestation, or GL manifestation) simply breaks down.
It is also to be noted that the (two-loop) running coupling in this theory 
is expected to become walking, $\alpha(Q^2) \simeq \alpha^*$ for
entire  
low energy 
region $Q^2 <\Lambda^2$, so that the condensate scales
with anomalous dimension
$\gamma_m \simeq 1$ as in the walking 
technicolor~\cite{Holdom85,Yamawaki-Bando-Matumoto,Akiba-Yanagida,
Appelquist-Karabali-Wijewardhana,BMSY} 
(For reviews see Ref.~\cite{Y,Hill-Simmons}):
\begin{eqnarray}
\langle \, \bar q q \, \rangle \sim m^2 \Lambda\, ,
\label{anomalous scaling}
\end{eqnarray}  
with $m$ given by Eq. (\ref{Mscaling}).

\subsection{Chiral restoration and VM in the effective field theory of 
large $N_f$ QCD}
\label{ssec:VMLNQ}

In this subsection we show that
the chiral restoration in the large $N_f$ QCD, 
$F_\pi^2(0) \rightarrow 0$,
is also derived in the EFT,
the HLS model, 
when we impose the Wilsonian matching 
to determine the bare parameters by the VM conditions,   
Eqs.~(\ref{vector condition:g})--(\ref{vector condition:fp}).
Once the chiral restoration takes place under the VM conditions,
the VM actually occurs at the critical point as we demonstrated in
Sec.~\ref{sssec:FVM} and so does the VM 
in the large $N_f$ QCD. It is to be noted that 
although the HLS model as it stands 
carries only the information of $N_f$ of the underlying
QCD but no other information such as $N_c$ and 
$\Lambda_{\rm QCD}$, the latter information actually is mediated into
the 
bare parameters of the HLS model,
 $F_\pi^2(\Lambda)$, $g(\Lambda)$, $a(\Lambda)$,
etc., through the Wilsonian matching. Then we can play with $N_c$ and
$\Lambda_{\rm QCD}$ as well as $N_f$ even at the EFT level. 

\subsubsection{Chiral restoration}
\label{sssec:CHRES}

As we have already shown in Sec.~\ref{sssec:FVM}, when the chiral restoration
takes place in the underlying QCD, we have 
the VM conditions which lead to Eq.~(\ref{Fpirun}):
\begin{eqnarray}
F_\pi^2(0)
\rightarrow 
 F_\pi^2(\Lambda)- 
\frac{N_f\Lambda^2}{2(4\pi)^2} \ ,
\label{Fpirun2}
\label{RGE for fpi2 at vector limit}
\end{eqnarray}
with $F_\pi^2(\Lambda)$ being given by Eq.~(\ref{vector condition:fp}):
\begin{equation}
F_\pi^2(\Lambda) \rightarrow (F_\pi^{\rm crit})^2
=
\frac{N_c}{3}
\left(\frac{\Lambda}{4\pi}\right)^2\cdot 2(1+\delta_A^{\rm crit}) 
\ .
\end{equation}
Then the chiral restoration can take place also in the HLS model 
when
\begin{eqnarray}
F_\pi^2(0)   
&=& 
(F_\pi^{\rm crit})^2  - 
\frac{N_f\Lambda^2}{2(4\pi)^2}
\rightarrow 0 \ ,
\end{eqnarray}
or 
\begin{equation}
X(\Lambda) \equiv \frac{N_f \Lambda^2}{2 (4\pi)^2}
\frac{1}{F_\pi^2(\Lambda)}\rightarrow\frac{N_f \Lambda^2}{2 (4\pi)^2}
\frac{1}{(F_\pi^{\rm crit})^2}
\rightarrow 1 \ .
\label{X Lam limit}
\end{equation}

This is actually realized in a concrete manner in the HLS model for the
large $N_f$ QCD. 
In the large $N_f$ QCD at the chiral restoration point, $F_\pi^2(\Lambda)$ 
determined by the underlying QCD is
almost independent of $N_f$ but crucially depends on (and is
proportional to)  
$N_c$, while the quadratic divergence of the HLS model does on
$N_f$. Then $N_f$ is essentially the only explicit parameter of the HLS
model to be adjustable  
after VM conditions 
Eqs.~(\ref{vector condition:a})-(\ref{vector condition:fp})
are imposed and can be
increased for fixed $N_c$ towards the critical $N_f$:
\begin{eqnarray}
N_f &\rightarrow& N_f^{\rm crit}-0
\ ,
\label{toward Nf crit}
\\
N_f^{\rm crit} &=& 2 (4\pi)^2\frac{(F_\pi^{\rm crit})^2}{\Lambda^2}
=\frac{N_c}{3} \cdot 4 (1+\delta_A^{\rm crit}) \ .
\label{Nfcrit0}
\end{eqnarray}
Note that this correponds to $X(\Lambda) \rightarrow 1-0$ in accord with the
flow in  Fig.~\ref{fig:flows a1}: If we take $X(\Lambda) \rightarrow 1+0$,
on the other hand, 
we would enter, before reaching the VM fixed point $(1,1,0)$, 
the symmetric phase where the HLS model breaks down or  
the light composite spectrum would disappear in the underlying QCD
with $N_f >N_f^{\rm crit}$.
As will be discussed below, 
\begin{eqnarray}
\delta_A^{\rm crit}\equiv \delta_A|_{\langle \bar q q \rangle = 0} 
=\frac{3(N_c^2-1)}{8N_c} \frac{\alpha_s}{\pi}
  + \frac{2\pi^2}{N_c} 
    \frac{
      \left\langle 
        \frac{\alpha_s}{\pi} G_{\mu\nu} G^{\mu\nu}
      \right\rangle
    }{ \Lambda_f^4 }
\label{deltacrit}
\end{eqnarray}
is almost
independent of $N_c$ as well as $N_f$, 
and is roughly given by simply neglecting the
quark condensate term (the third term) in $\delta_A|_{N_c=N_f=3} 
(\simeq 0.36)$ in Eq.~(\ref{delta}): 
\begin{equation}
\delta_A^{\rm crit} \simeq 0.27 \pm 0.04 \pm 0.03 \ ,
\label{deltaA crit1}
\end{equation}
which yields
\begin{equation}
N_f^{\rm crit} \simeq 
\left( 5.1 \pm 0.2 \pm 0.1 \right) \times
  \left(\frac{N_c}{3}\right) \ ,
\label{Nfcrit1}
\end{equation}  
where the center values in Eqs.~(\ref{deltaA crit1}) and
(\ref{Nfcrit1}) are given for 
$(\Lambda_3\,,\,\Lambda_{\rm QCD}) = (1.1\,,\,0.4)\,\mbox{GeV}$,
and the first and second errors are obtained 
by allowing $\Lambda_3$ and $\Lambda_{\rm QCD}$ to vary
$\delta \Lambda_3 = 0.1\,\mbox{GeV}$ and 
$\delta \Lambda_{\rm QCD} = 0.05 \,\mbox{GeV}$, respectively.
Hence Eq.~(\ref{Nfcrit0}) implies that
$N_f^{\rm crit} \sim O(N_c)$. This is natural, since both
$F_\pi^2(0)$ and $F_\pi^2(\Lambda)$ are of $O(N_c)$ 
in Eq.~(\ref{Fpirun2}) (see the discussion 
in Sec.~\ref{ssec:MHUQ}) 
and so is the $N_f$ as far as it is to be non-negligible
(near the critical point). Thus the chiral restoration is a peculiar phenomenon
which takes place only when both $N_f$ and $N_c$ are regarded as 
large, with 
\begin{equation}
N_f \sim N_c \gg 1 \ .
\end{equation}

Historically, the chiral restoration in terms of HLS for
the large $N_f$ QCD was first obtained
in Ref.~\cite{HY:letter}, 
based on 
an assumption that the bare parameters take the fixed point
values 
$(a,g)=(1,0)$ and hence Eq.~(\ref{Fpirun}) follows
and also based on a further assumption that
$F_\pi(\Lambda)^2/\Lambda^2$ 
has a small dependence on $N_f$ (for fixed $N_c$). 
These assumptions were justified
later by the Wilsonian matching~\cite{HY:matching,HY:VM}, although
the second assumption got a small correction between $N_f=N_f^{\rm crit}$ 
and $N_f=3$ essentially arising from a factor
$(1+\delta_A^{\rm crit})/(1+\delta_A|_{N_c=N_f=3})\sim
  (1+0.27)/(1+0.36) \sim 0.93$.

Now, we discuss the result of $N_f^{\rm crit}$ in Eq.~(\ref{Nfcrit1})
based on the estimation of Eq.~(\ref{deltacrit}).
First of all we should mention that the OPE is valid only 
for $\delta_A^{\rm crit} (<\delta_A) <1$
and hence in order for our approach to be 
self-consistent, our estimate of $N_f^{\rm crit}$ must be within the range:
\begin{equation}
4\left(\frac{N_c}{3}\right) 
< N_f^{\rm crit}= \frac{N_c}{3}\cdot 4(1+\delta_A^{\rm crit}) < 
8\left(\frac{N_c}{3}\right) \ ,
\end{equation} 
which is consistent with the recent lattice result $6< N_f^{\rm crit}<7$
(for $N_c=3$)~\cite{IKKSY:98} and in sharp contrast to
the ladder-perturbative estimate
$N_f^{\rm crit} \simeq 12 \frac{N_c}{3}$~\cite{Appelquist-Terning-Wijewardhana}.

Let us next discuss some details:
Here we make explicit the $N_f$-dependence of the parameters like  
the matching scale $\Lambda_f\equiv\Lambda(N_f)$ for fixed $N_c$,
$\alpha_s (\Lambda_f, N_f)$, etc.,
since they generally depend on $N_f$ (also on $N_c$ and 
$\Lambda_{\rm QCD}$ from the QCD side).
For the first term in Eq.~(\ref{deltacrit}),
as we will discuss later, $N_c\, \alpha_s(\Lambda_f;N_f)$ is 
independent of $N_f$ and $N_c$, and hence $\Lambda_f$ increases 
with $N_f$ and decreases with $N_c$.  Then we use
$\frac{N_c}{3}\alpha_s(\Lambda_f;N_f)/\pi = 
\left. \alpha_s(\Lambda_3;N_f=3)/\pi\right\vert_{N_c=3} \simeq 0.22$,
again the value obtained in Eq.~(\ref{delta}).
For  the gluonic condensate term, 
$\left\langle \frac{\alpha_s}{\pi} G_{\mu\nu} G^{\mu\nu}
\right\rangle$ is independent of the renormalization point of QCD, so
that it is natural to say that it is independent of $N_f$.
Furthermore, 
$\frac{1}{N_c} \left\langle \frac{\alpha_s}{\pi} G_{\mu\nu} 
G^{\mu\nu} \right\rangle$ is independent of
$N_c$~\cite{Bardeen-Zakharov}.
Although $\Lambda_f$ is somewhat larger than $\Lambda_3$ as mentioned above,
we here make a crude estimate of  the 
second
term by simply
taking $\Lambda_f = \Lambda_3$, $(\Lambda_3\,,\,\Lambda_{\rm QCD})=
(1.1\,,\,0.40)$\,GeV and using
$\frac{3}{N_c} \left\langle \frac{\alpha_s}{\pi} G_{\mu\nu} G^{\mu\nu}
\right\rangle=0.012\,
\mbox{GeV}^4$~\cite{SVZ:1,SVZ:2,Bardeen-Zakharov},
which yields the value, $0.054$, already given for $N_c=N_f=3$ 
in Eq.~(\ref{delta}).
At any rate, the 
gluon condensate term is numerically negligible 
(less than $5 \%$ for $N_f^{\rm crit}$) 
in any estimate and hence does not
give much uncertainty.
Now, we set $(N_c^2-1)/N_c^2 =8/9$ but this factor will yield $1$ for 
large $N_c$ and thus enhance $0.22$ to $0.25$.
In conclusion we have $\delta_A^{\rm crit} \simeq 0.22 (0.25) + 0.054 
\simeq 0.27 (0.30)$, which yields 
\begin{eqnarray}
N_f^{\rm crit} &\simeq& 5.1 \left(\frac{N_c}{3}\right) 
  \quad (N_c \sim 3),
\nonumber
\\
&\simeq&  
5.2\left(\frac{N_c}{3}\right) \quad (N_c \gg 3)\ .
\label{Nf crit val 2}
\end{eqnarray} 

A more precise estimation of $N_f^{\rm crit}$  will be done by
determining the 
$N_f$-dependences of the QCD coupling $\alpha_s$ and $\Lambda_f$ in
Sec.~\ref{sssec:NDP}.
Here we just quote the result 
$N_f^{\rm crit}\simeq 5.0 \frac{N_c}{3}$ (for $N_c =3$),
which is consistent with the above estimate and somewhat similar to
the recent lattice result $6< N_f^{\rm crit}<7$
(for $N_c=3$)~\cite{IKKSY:98},
while much smaller than the ladder-perturbative estimate
$N_f^{\rm crit} \simeq 12
  \frac{N_c}{3}$~\cite{Appelquist-Terning-Wijewardhana}.
It is amusing that our estimate 
coincides with the instanton argument~\cite{Velkovsky-Shuryak}.

If such a relatively small value of $N_f^{\rm crit}$ is indeed the
case, 
it would imply that for some (nonperturbative ?) reason 
the running coupling 
might level off in the infrared
region at smaller $N_f$ than that expected in the perturbation.

At any rate, what we have shown here implies a rather amazing fact:
Recall that the real-life QCD with $N_f=3$ is very close to
$a(\Lambda)=1$ (see Sec.~\ref{a1phenomenology}), which corresponds
to the ideal situation that the bare HLS Lagrangian is 
the fixed point Lagrangian, Eq.~(\ref{fixedpointlagrangian}), 
having a redundant global
symmetry $G_1\times G_2$ which is explicitly broken only by the strong 
$\rho$ gauge coupling. Now in the large $N_f$ QCD with $N_f$ very close
to the critical point $N_f^{\rm crit}$, the $\rho$ coupling becomes 
vanishingly small and the bare HLS Lagrangian
realizes a {\it weak coupling gauge theory} with the $G_1 \times G_2$
symmetry explicitly 
broken only by the ``weak'' coupling of the {\it composite}
 $\rho$ meson.

\subsubsection{Critical behaviors}
\label{sssec:CB}

In this sub-subsection we study the critical behaviors of the
parameters 
and several physical quantities
when $N_f$ approaches to its critical value $N_f^{\rm crit}$
using the RGEs.
As we discussed at the end of Sec.~\ref{ssec:PSH},
since the VM fixed point 
$( X^\ast_2,a^\ast_2,G^\ast_2 ) = (1,1,0)$
is not an infrared stable fixed point,
the VM limit with 
bare parameters approaching the VM fixed point from the
broken phase
does not generally imply that the parameters in the infrared region
approach the same point:
We expect that, without extra fine tuning,
$g^2(m_\rho)\rightarrow0$ is obtained from one of the VM conditions,
$g^2(\Lambda)\rightarrow0$.
Combining this with the on-shell condition 
(\ref{on-shell condition}) leads to 
the infrared parameter $X(m_\rho)$ behaving as
$X(m_\rho)\rightarrow0$, although $X(\Lambda) \rightarrow 1$.
This 
implies $\frac{m_\rho^2}{F_\pi^2(m_\rho)} \rightarrow0$.
{}From this together with Eq.~(\ref{rel: Fp 0 Fp mr})
we 
infer
\begin{equation}
\frac{m_\rho^2 }{F_\pi^2(0)} \rightarrow 0 \ .
\label{mr2 ov fp2 VM}
\end{equation}

Below we shall discuss that this is indeed the case by
examining 
the critical behaviors of the physical parameters 
near the critical point in a more precise manner through the RGEs. 
For that we need to know
how the bare parameters
$g(\Lambda_f;N_f)$ and $a(\Lambda_f;N_f)$ approach to 
the VM limit in Eqs.~(\ref{vector condition:g}) and
(\ref{vector condition:a}).
Taking the limits $g^2(\Lambda) \ll 1$,
$M_\rho^2(\Lambda)/\Lambda^2 = g^2(\Lambda) a(\Lambda) 
F_\pi^2(\Lambda)/\Lambda^2 \ll 1$ and
$F_\sigma^2(\Lambda)/F_\pi^2(\Lambda) - 1 = a(\Lambda) -1 \ll 1$
in the Wilsonian matching condition (\ref{match z}),
we obtain
\begin{eqnarray}
&&
g^2(\Lambda) \, \left(\frac{ F_\pi^2(\Lambda) }{\Lambda^2 }\right)^2
- \left( a(\Lambda)-1 \right) \frac{ F_\pi^2(\Lambda) }{\Lambda^2 }
+ 2 g^2(\Lambda) z_3(\Lambda) \frac{ F_\pi^2(\Lambda) }{\Lambda^2 }
{}- 2 \left[ z_2(\Lambda) - z_1(\Lambda) \right]
\nonumber\\
&& \qquad
= \frac{4(N_c^2-1)\pi}{N_c^2} 
\frac{\alpha_s \langle \bar{q} q \rangle^2}{ \Lambda^6} \ .
\label{z match near VM}
\end{eqnarray}
It is plausible to require that there are no cancellations among the
terms in the left-hand-side (LHS)
of the above matching condition.
Then, we expect that all the terms in the LHS have the same scaling
behavior near the restoration point.
The critical behavior of the HLS gauge coupling
$g^2(\Lambda_f;N_f)$ is then given by
\begin{equation}
g^2(\Lambda_f;N_f) \sim 
\frac{\alpha_s}{N_c^2} \, 
\left\langle \bar{q} q
\right\rangle^2 \ ,
\label{g2 crit}
\end{equation}
where we put the extra $N_c$-dependence coming from 
$[F_\pi^2(\Lambda)]^2 \sim N_c^2$ into the right-hand-side of
the above relation.
Since the quark condensate scales as $N_c$, 
$\langle \bar{q} q \rangle \sim N_c$, and the QCD gauge coupling
scales as $1/N_c$ in the large $N_c$ counting,
the above relation implies that the HLS gauge coupling
scales as $1/N_c$, $g^2(\Lambda_f;N_f) \sim 1/N_c$,
in the large $N_c$ counting.

Now we consider the $N_f$-dependence. We may parameterize the scaling
behavior  
of $g^2$ as
\begin{equation}
g^2(\Lambda_f;N_f) 
= 
\bar{g}^2 f(\epsilon) \ , \quad
\epsilon \equiv
\frac{1}{N_f} - \frac{1}{N_f^{\rm crit}} \ ,
\label{initial value of g}
\end{equation}
where $\bar{g}$ is independent of $N_f$.
$f(\epsilon)$ is a certain function
characterizing the scaling of 
\begin{equation}
\langle \bar q q \rangle^2 \sim m^{6-2\gamma_m} 
\Lambda^{2\gamma_m}\, ,
\label{a-scaling}
\end{equation}
where $\gamma_m$ is the anomalous dimension and  
$m=m(\epsilon)$ is the dynamical mass of the fermion which vanishes
as $\epsilon \rightarrow 0$.  For example, the improved ladder SD equation 
with the two-loop running gauge coupling~\cite{Appelquist-Terning-Wijewardhana} implies the walking gauge 
theory~\cite{Holdom85,Yamawaki-Bando-Matumoto,Akiba-Yanagida,%
Appelquist-Karabali-Wijewardhana,BMSY} which suggests
$\gamma_m \simeq 1$ and $m=\exp{[-C/\sqrt{\epsilon}]}$ as in  
Eq.~(\ref{Mscaling}), so that we have
$f(\epsilon) =\exp{[-4 C/\sqrt{\epsilon}]}$.  
However, since we do not know the reliable estimate of
the scaling function $f(\epsilon)$, we will leave it unspecified in
the below.

To make an argument based on the analytic solution,
we here fix 
\begin{equation}
a(\Lambda_f;N_f) = 1
\end{equation}
 even off the critical point, since
the Wilsonian matching conditions
with the physical inputs $F_\pi(0)=86.4$\,MeV 
and $m_\rho=771.1$\,MeV
leads to $a(\Lambda)\simeq 1$ already  for
$N_f=3$ [see Sec.~\ref{sec:WM} as well as 
Ref.~\cite{HY:matching}].
Recall that putting $a=1$ does not contradict the symmetry of the underlying QCD
though $g=0$ does (See Sec.~\ref{sssec:VM as a limit} ). 
A deviation from $a(\Lambda)=1$ will be discussed in the next
sub-subsection.

Before studying the critical behaviors of the parameters
in the quantum theory,
let us show the solutions of 
the RGEs (\ref{RGE for Fpi2}) and (\ref{RGE for g2}).
We note that these RGEs are solvable analytically
when we take $a=1$ from the beginning.
{}From Eq.~(\ref{RGE for g2}) with $a=1$ the solution $g^2(\mu;N_f)$
is expressed as
\begin{equation}
g^2(\mu;N_f) = \frac{1}{C_f b \ln(\mu/\Lambda_H(N_f))} \ ,
\label{sol g2}
\end{equation}
where
$C_f = N_f/(2 (4\pi)^2)$ and $b = 43/3$.
$\Lambda_H(N_f)$, which generally depends on $N_f$,
is the intrinsic scale of the HLS,
analog to $\Lambda_{\rm QCD}$ of QCD.
To show the solution for $F_\pi(\mu;N_f)$ 
it is convenient to use a cutoff 
scale $\Lambda_f$ as the reference scale.
The solution is given by
\begin{eqnarray}
\frac{F_\pi^2(\mu;N_f)}{\Lambda_f^2} =
\left[ \frac{g^2(\Lambda_f;N_f)}{g^2(\mu;N_f)} \right]^l
\left[
  \frac{F_\pi^2(\Lambda_f;N_f)}{\Lambda_f^2} -
  \frac{N_f}{(4\pi)^2}
  \int^{s}_0 d z \left( \frac{t_\Lambda}{t_\Lambda - z} \right)^l
  e^{ - 2 z }
\right] \ ,
\label{sol fpi2}
\end{eqnarray}
where $l=9/43$, $s=\ln(\Lambda_f/\mu)$ and 
$t_\Lambda=\ln\left(\Lambda_f/\Lambda_H(N_f)\right)$.

Let us now study the critical behaviors of the parameters
in the quantum theory.
The solution (\ref{sol g2}) for $g^2$ with 
Eq.~(\ref{initial value of g}) determines 
the critical behavior of 
the intrinsic scale of the HLS as $N_f \rightarrow N_f^{\rm crit}$:
$\Lambda_H(N_f) \longrightarrow \Lambda 
\exp \left[ - T /f(\epsilon) \right]$,
where $T = 1/(C_f b \bar{g}^2)$.
The intrinsic scale of the HLS goes to zero with an essential
singularity scaling.
Since $m_\rho(N_f) > \Lambda_H(N_f)$,
it is natural to assume that 
the gauge coupling at the scale $m_\rho(N_f)$
approaches to zero showing the same power behavior:
$g^2(m_\rho(N_f);N_f) \rightarrow 
\bar{g'}^2 f(\epsilon)$ as
$N_f \rightarrow N_f^{\rm crit}$.
Replacing $s$ with 
$s_V \equiv \ln \left( \Lambda_f/m_\rho(N_f)\right)$ in
Eq.~(\ref{sol fpi2}) and substituting it into the on-shell 
condition~(\ref{on-shell condition}), we obtain 
\begin{eqnarray}
\frac{m_\rho^2(N_f)}{\Lambda_f^2}
&=&
g^2\left(m_\rho(N_f);N_f\right)
\left[ \frac{g^2(\Lambda_f;N_f)}{
  g^2\left(m_\rho(N_f);N_f\right)} \right]^l
\nonumber\\
&& \times
\Biggl[
  \frac{N_f^{\rm crit}}{2(4\pi)^2}
  - \frac{N_f}{2(4\pi)^2}
  {}- \frac{N_f}{(4\pi)^2}
  \int^{s_V}_0 d z \left\{
      \left( \frac{t_\Lambda}{t_\Lambda - z} \right)^l - 1
  \right\}
  e^{ - 2 z }
\nonumber\\
&& \qquad \quad
  {}+ \frac{N_f}{2(4\pi)^2}\frac{m_\rho^2(N_f)}{\Lambda_f^2}
  {}+ \frac{F_\pi^2(\Lambda_f;N_f)}{\Lambda_f^2} -
  \frac{F_\pi^2(\Lambda_f^{\rm crit};N_f^{\rm crit})}{
     \left(\Lambda_f^{\rm crit}\right)^2} 
\Biggr] \ ,
\label{on-shell solution}
\end{eqnarray}
where inside the bracket we added
\begin{equation}
0 = 
-  \frac{F_\pi^2(\Lambda_f^{\rm crit};N_f^{\rm crit})}{
    \left(\Lambda_f^{\rm crit}\right)^2} 
  + \frac{N_f^{\rm crit}}{2(4\pi)^2} \ .
\end{equation}

To obtain the critical behavior we note 
\begin{eqnarray}
&&
\int^{s_V}_0 d z \left\{
    \left( \frac{t_\Lambda}{t_\Lambda - z} \right)^l - 1
\right\}
e^{ - 2 z }
\rightarrow
\frac{l T}{4} \, f(\epsilon) \ ,
\nonumber\\
&&
\frac{N_f^{\rm crit}}{2(4\pi)^2} - \frac{N_f}{2(4\pi)^2}
\rightarrow 
\frac{\left(N_f^{\rm crit}\right)^2}{2(4\pi)^2} \, \epsilon \ .
\end{eqnarray}
Then Eq.~(\ref{on-shell solution}) behaves as
\begin{eqnarray}
&&
\frac{m_\rho^2(N_f)}{\Lambda_f^2}
\sim
f(\epsilon)
\left[ 
  N_f^{\rm crit} \epsilon 
  - \frac{l T}{2} \, f(\epsilon)
   +
  \frac{m_\rho^2(N_f)}{\Lambda_f^2}
  + \frac{32\pi^2}{N_f^{\rm crit}}
  \left\{
    \frac{F_\pi^2(\Lambda_f;N_f)}{\Lambda_f^2} -
    \frac{F_\pi^2(\Lambda_f^{\rm crit};N_f^{\rm crit})}{
       \left(\Lambda_f^{\rm crit}\right)^2}
  \right\}
\right]
\ .
\nonumber\\
&&
\label{on-shell behavior}
\end{eqnarray}
Since the second term in the square bracket is negative,
this cannot dominate over the other terms.
Thus we have to require
\begin{equation}
f(\epsilon)/\epsilon \ll 1.
\end{equation}
in Eq.~(\ref{initial value of g}).
The behavior of 
$F_\pi^2(\Lambda_f;N_f)/\Lambda_f^2 -
F_\pi^2(\Lambda_f^{\rm crit};N_f^{\rm crit})%
/(\Lambda_f^{\rm crit})^2$
in the fourth term of Eq.~(\ref{on-shell behavior}) is determined by
that of $\left\langle \bar{q} q \right\rangle$ through the Wilsonian
matching condition (\ref{match A}).
Then it is reasonable to assume that this term goes to zero faster
than the first term does.
In addition the third term cannot
dominate over the other terms, of course.
As a result the critical behavior of $m_\rho^2(N_f)/\Lambda_f^2$ is
governed by the first term in the right-hand-side of 
Eq.~(\ref{on-shell behavior}).
This implies that $m_\rho^2(N_f)$ takes the form:
\begin{equation}
m_\rho^2(N_f)/\Lambda_f^2 \sim
\epsilon f(\epsilon)
\rightarrow 0 \ ,
\label{critical of mrho}
\end{equation}
which leads to 
$F_\pi^2\left(m_\rho(N_f);N_f\right) /\Lambda_f^2
\sim \epsilon$.
The second term of RHS of Eq.~(\ref{rel: Fp 0 Fp mr})
approaches to zero faster than the first term does.
Thus we obtain the critical behavior of the order parameter as
\begin{equation}
F_\pi^2(0;N_f)/\Lambda_f^2 \sim \epsilon \rightarrow 0
\ .
\label{critical of fpi}
\end{equation}
Equations~(\ref{critical of mrho}) and (\ref{critical of fpi})
shows that {\it $m_\rho$ approaches to zero faster than
$F_\pi$}~\cite{HY:VM}:
\begin{equation}
\frac{m_\rho^2}{F_\pi^2(0;N_f)} \sim f(\epsilon)  
\rightarrow 0 \ ,
\label{crit mr2 ov fp2}
\end{equation}
as we naively expected in Eq.~(\ref{mr2 ov fp2 VM}).
This is a salient feature of the VM~\cite{HY:VM}.

  Since $F_\pi^2(0)$ is usually expected to scale as 
  $F_\pi^2(0) \sim m^2$,
  Eq.~(\ref{critical of fpi}) implies that 
  $m \sim \sqrt{\epsilon}$, in contrast to the essential-singularity
  type 
  Eq.~(\ref{Mscaling}). This may be 
  a characteristic feature of 
  the one-loop RGEs we are using. 
  However, the essential-singularity scaling is more sensitive to the 
  ladder artifact than the estimate of the anomalous dimension 
  $\gamma_m \simeq 1$ which implies that 
$\langle \bar q q \rangle \sim m^2$.
  Then Eq.~(\ref{critical of fpi}) implies
\begin{equation}
f(\epsilon) \sim \langle \bar q q \rangle^2 
 \sim m^4 \sim \epsilon^2 \, ,
\label{q2}
\end{equation}
which will be later used as an ansatz for explicit computation of
the global  
$N_f$-dependence for $3<N_f <N_f^{\rm crit}$.\footnote{
  We could also assume a case $f(\epsilon) \sim \epsilon$ which is a 
  simple mean field type corresponding to the NJL type scaling with 
  $\gamma_m=2$. Such a behavior may be related to the following large
  $N_f$ 
  argument~\cite{HY:letter}:
  $g$ is the coupling of the three-point interaction of the vector
  mesons.
  Then, as we have shown below Eq.~(\ref{g2 crit}),
  large $N_c$ argument of QCD tells us that
  $g^2$ behaves as $1/N_c$ in the large $N_c$ limit with fixed $N_f$.
  On the other hand,
  to make a large $N_f$ expansion in the HLS consistent the gauge
  coupling $g^2$ falls as $1/N_f$.
  However, the HLS is actually related to 
  QCD, so that the large $N_f$ limit should be taken with
  $N_c/N_f$ finite.
  This situation can be seen by rewriting the RHS of
  Eq.~(\ref{initial value of g}) into
  $\left(\bar{g}^2/N_c\right) 
  \left( N_c/N_f - N_c/N_f^{\rm crit} \right)$.
}

Let us now consider the behaviors of the physical quantities listed in
Sec.~\ref{ssec:RWM} [see also Ref.~\cite{HY:matching}]:

The $\rho$--$\gamma$ mixing strength $g_\rho$ in Eq.~(\ref{g rho})
and the
$\rho$-$\pi$-$\pi$ coupling constant $g_{\rho\pi\pi}$ 
in Eq.~(\ref{g rho pi pi}) go to zero as~\cite{HY:VM}
\begin{eqnarray}
&& g_\rho(m_\rho) = g(m_\rho) F_\pi^2(m_\rho)
\sim \epsilon f^{1/2}(\epsilon) \rightarrow 0 \ ,
\label{eq:grho}
\\
&& g_{\rho\pi\pi}(m_\rho,0,0) = \frac{g(m_\rho)}{2}
\frac{F_\pi^2(m_\rho)}{F_\pi^2(0)} \sim f^{1/2}(\epsilon) 
\rightarrow 0 \ ,
\label{eq:grhopipi}
\end{eqnarray}
where $a(\Lambda)=a(m_\rho)=1$ was used.
As discussed in Ref.~\cite{HY:matching}, the KSRF (I) relation for the
low-energy quantities 
$g_\rho(0) = 2 g_{\rho\pi\pi}^2(0,0,0) F_\pi^2(0)$
holds as a low energy theorem of the 
HLS~\cite{BKY:NPB,BKY:PTP,HY,HKY:PRL,HKY:PTP}
for any $N_f$.
The relation for on-shell quantities is violated by about 15\% for
$N_f=3$ (see Eq.~(\ref{LET val}) as well as Ref.~\cite{HY:matching}).
As $N_f$ goes to $N_f^{\rm crit}$, $g_\rho(m_\rho)$ and 
$g_{\rho\pi\pi}(m_\rho,0,0)$ approach to
$g_\rho(0)$ and $g_{\rho\pi\pi}(0,0,0)$, respectively,
and hence the on-shell KSRF (I) relation becomes more
accurate for larger $N_f$.
On the other hand, the (on-shell) KSRF (II) relation $m_\rho^2 = 2
g_{\rho\pi\pi}^2(m_\rho,0,0) F_\pi^2(0)$ becomes less accurate.  
Near the critical flavor it reads as $m_\rho^2 = 4
g_{\rho\pi\pi}^2(m_\rho,0,0) F_\pi^2(0) \rightarrow 0$.
By substituting 
the critical behaviors in Eqs.~(\ref{critical of mrho}),
(\ref{eq:grho}) and (\ref{eq:grhopipi}) into the expressions
for 
the $\rho\rightarrow\pi\pi$ decay width and
the $\rho\rightarrow e^+e^-$ decay width given in 
Eqs.~(\ref{rho width}) and (\ref{rho ee width}) with putting
$m_e = m_\pi = 0$,
the critical behaviors of the
ratio of the $\rho$ width to the $\rho$ mass
and the
peak value of $e^+ e^- \rightarrow \pi \pi$ cross section
are expressed as~\cite{HY:VM}
\begin{eqnarray}
&&\Gamma/m_\rho \sim g_{\rho\pi\pi}^2 \sim 
  f(\epsilon) \rightarrow 0
\ ,
\label{crit gam/mr}
\\
&&\Gamma_{ee}\Gamma_{\pi\pi}/\Gamma^2
\sim g_\rho^2/(g_{\rho\pi\pi}^2 m_\rho^4)
\sim 1/f^2(\epsilon) \rightarrow \infty \ .
\label{crit ee pipi}
\end{eqnarray}

The parameters $L_{10}^r(m_\rho)$ and $L_9^r(m_\rho)$ defined in
Eqs.~(\ref{l10 5}) and (\ref{l9})~\cite{HY:matching} diverge as
$N_f$ approaches to $N_f^{\rm crit}$.
However, we should note that, even for $N_f=3$, both 
$L_{10}^r(\mu)$ and
$L_9^r(\mu)$ have the infrared logarithmic divergences
when we take $\mu\rightarrow 0$ in
the running obtained by the chiral perturbation 
theory~\cite{Gas:84,Gas:85a}.
Thus we need more careful treatment of these quantities for large
$N_f$.  This is beyond the scope of this report.

\subsubsection{$N_f$-dependence of the parameters for 
  $3 \le N_f < N_f^{\rm crit}$ }
\label{sssec:NDP}

In this subsection we illustrate
how the HLS parameters would change as we vary the $N_f$ from $3$ to
$N_f^{\rm crit}$. For that purpose we need more specific assumption on the
$N_f$-dependence of the QCD parameters in OPE.
Here we adopt a simple ansatz which is consistent with the scaling property 
near the critical point given in the previous subsubsection.

Let us start from the parameters of the QCD appearing in the OPE.
The HLS is matched with the underlying QCD at the
matching scale $\Lambda_f$.
This matching scale can be regarded as the scale where the QCD running
coupling becomes of order one.
Thus it seems natural to require
$\alpha_s(\Lambda_f;N_f)$ to be a constant against the change of $N_f$.
Furthermore, the large-$N_c$ analysis shows that 
$N_c\,\alpha_s(\Lambda_f;N_f)$ is independent of $N_c$.
Here we show how to determine the $N_f$-dependence of the matching 
scale from this requirement.
We note that theories of QCD with different $N_f$ are compared by
fixing $\Lambda_{\rm QCD}$, and that
it is enough to use the one-loop QCD running coupling above the
matching scale since the running coupling is small at the
scale above the matching scale.
The one-loop running coupling is given by
\begin{equation}
\alpha(\mu;N_f) = 
\frac{4\pi}{%
  \beta_0(N_f) \ln \left( \mu^2/\Lambda_{\rm QCD}^2 \right)
}
\ ,
\end{equation}
where
\begin{equation}
\beta_0(N_f) = \frac{1}{3} \left( 11 N_c - 2 N_f \right)\ .
\end{equation}
The requirement 
$(N_c/3)\,\alpha_s(\Lambda_f;N_f)=\mbox{constant}=\alpha_s 
  (\Lambda_3,3)|_{N_c=3}\simeq 0.7$,
with $\Lambda_3=1.1 \, \mbox{GeV}$,
is rewritten into the following form:
\begin{equation}
\frac{3}{N_c}\beta_0(N_f)
\ln \left( \Lambda_f / \Lambda_{\rm QCD} \right)
= \left( 11 - 2 \frac{N_f}{N_c} \right)
\ln \left( \Lambda_f / \Lambda_{\rm QCD} \right)
= \mbox{constant} \ .
\label{Lambda cond}
\end{equation}
This determines the $N_f$-dependence as well as the $N_c$-dependence
of the matching scale
$\Lambda_f$.
Note that the $N_c$-dependence of the ratio
$\Lambda_f/\Lambda_{\rm QCD}$ is
actually very small: The difference between the ratio for 
$N_c = N_f=3$ and that for $N_c=\infty$ and $N_f=3$ is about 2\%.
One might think that the $N_f$-dependence
of the ratio
$\Lambda_f/\Lambda_{\rm QCD}$ is very 
strong  and 
$\Lambda_f/\Lambda_{\rm QCD}$ 
vanishes in
the large $N_f$ limit.
However, the large $N_f$ limit should be taken with $N_f/N_c$ fixed,
so that the ratio $\Lambda_f/\Lambda_{\rm QCD}$ remains as constant in
the large $N_f$ limit. 
Actually, 
the ratio varies at most by 4\% for 
$0< N_f/N_c < N_f^{\rm crit}/N_c \simeq 5/3$.~\footnote{%
  We could use the two-loop running coupling (and the associated
  $\Lambda_{\rm QCD}$~\cite{Appelquist-Terning-Wijewardhana})
  determined by 
  Eqs.~(\ref{90}) and (\ref{91}) which has an
  infrared fixed point for $N_f >N_f^*$ ($\sim 8$ for $N_c=3)$ 
  and would have  more relevance to the ladder/perturbative argument
  which  
  indicates $N_f^{\rm crit} \sim 12\frac{N_c}{3}$.
  However, our rough result $N_f^{\rm crit} \sim 5 \frac{N_c}{3}$ is
  rather 
  different  
  from that and is closer to the lattice result, and hence the
  two-loop running may not be relevant.  
  Actually, 
  the small dependence of $\Lambda_f/\Lambda_{\rm QCD}$ on 
  $N_c$ as well as $N_f$ in the region $0< N_f/N_c < 5/3$ 
  is valid even when we use the solution of the two-loop beta function
  in Eq.~(\ref{Lambda cond}).
  This can be seen from the following explicit form of the solution
  of the two-loop beta
  function~\cite{Appelquist-Ratnaweera-Terning-Wijewardhana}:
  \begin{eqnarray}
    \ln \frac{\Lambda_{\rm QCD}}{\Lambda_f}
    = \frac{1}{b\, \alpha_\ast} 
      \ln \left( \frac{ \alpha_\ast - \alpha(\Lambda_f;N_f) }
         { \alpha(\Lambda_f;N_f)} \right)
     - \frac{ 1 }{ b\, \alpha(\Lambda_f;N_f)}
    \ ,
    \nonumber
  \end{eqnarray}
  where $b$ and $\alpha_\ast$ are defined in
  Eqs.~(\ref{91}) and (\ref{92}).
}

As we mentioned earlier, the gluonic condensate
$\left\langle \frac{\alpha_s}{\pi} G_{\mu\nu} G^{\mu\nu}
\right\rangle$
is independent of the renormalization point of QCD, so
that it is reasonable to assume that it is independent of $N_f$,
and scales as $N_c$~\cite{Bardeen-Zakharov}.
So we assume 
\begin{equation}
\frac{1}{N_c} 
\left\langle \frac{\alpha_s}{\pi} G_{\mu\nu} G^{\mu\nu}
\right\rangle
= \mbox{constant} \ .
\label{GG cond}
\end{equation}

Let us now discuss the more involved 
estimate of the critical value $N_f^{\rm crit}$.
When we estimated the value of $\delta_A^{\rm crit}$ 
in Eq.~(\ref{deltaA crit1})
[and then $N_f^{\rm crit}$ in Eq.~(\ref{Nfcrit1})
or Eq.~(\ref{Nf crit val 2})],
we used the same values of 
$\alpha_s = \alpha_s(\Lambda_f,N_f)$,
$\left\langle \frac{\alpha_s}{\pi} G_{\mu\nu} G^{\mu\nu}
  \right\rangle$
and $\Lambda_f = \Lambda(N_f)$ for $N_f = N_f^{\rm crit}$
as those for $N_f = 3$.
Here, although we assume that $\alpha_s$ and 
$\left\langle \frac{\alpha_s}{\pi} G_{\mu\nu} G^{\mu\nu}
  \right\rangle$
do not depend on $N_f$ as in Eqs.~(\ref{Lambda cond})
and (\ref{GG cond}), $\Lambda_f$ does depend on $N_f$,
which is determined from Eq.~(\ref{Lambda cond}).
Then, the critical number of flavors $N_f^{\rm crit}$ 
is determined by solving
\begin{equation}
N_f^{\rm crit}
=
\frac{3(N_c^2-1)}{8N_c} \frac{\alpha_s}{\pi}
  + \frac{2\pi^2}{N_c} 
    \frac{
      \left\langle 
        \frac{\alpha_s}{\pi} G_{\mu\nu} G^{\mu\nu}
      \right\rangle
    }{ \Lambda^4(N_f^{\rm crit}) }
\ .
\end{equation}
By using $\alpha_s = \alpha_s( \Lambda_3 = 1.1\,\mbox{GeV}, 
N_f =3 ) \simeq 0.69$ and 
$\left\langle \frac{\alpha_s}{\pi} G_{\mu\nu} G^{\mu\nu}
  \right\rangle = 0.012\,\mbox{GeV}^4$,
the value of $N_f^{\rm crit}$
for $N_c=3$ is estimated as~\footnote{%
  The center values in Eqs.~(\ref{crit num Nf}) and
  (\ref{lam del crit}) are given for 
  $(\Lambda_3\,,\,\Lambda_{\rm QCD}) = (1.1\,,\,0.4)\,\mbox{GeV}$,
  and the first and second errors are obtained 
  by allowing $\Lambda_3$ and $\Lambda_{\rm QCD}$ to vary
  $\delta \Lambda_3 = 0.1\,\mbox{GeV}$ and 
  $\delta \Lambda_{\rm QCD} = 0.05 \,\mbox{GeV}$, respectively.
}
\begin{equation}
N_f^{\rm crit} \simeq 5.0 \pm 0.1 \pm 0.1 \ ,
\label{crit num Nf}
\end{equation}
and the values of $\Lambda_f$ and $\delta_A^{\rm crit}$ are
determined as
\begin{equation}
\Lambda(N_f^{\rm crit}) \simeq 1.3 \pm 0.1 \pm 0.01 \,\mbox{GeV}\ ,
\quad
\delta_A^{\rm crit} \simeq 0.25 \pm 0.03 \pm 0.03 \ ,
\label{lam del crit}
\end{equation}
which are compared with the previous rough estimate in 
Sec.~\ref{sssec:CHRES}:
$N_f^{\rm crit} \simeq 5.1 \,(N_c=3)$, 
$\Lambda_f=\Lambda_3 \simeq 1.1\,\mbox{GeV}$ 
and  $\delta_A^{\rm crit} \simeq 0.27$.

Now we discuss the quark condensate.
As we have shown in Eq.~(\ref{critical of fpi}), $F_\pi^2(0)$ in the
present approach scales as $F_\pi^2(0) 
\sim m^2 \sim \epsilon \equiv 1/N_f - 1/N_f^{\rm crit}$
for any choice of the scaling property of 
$\langle \bar{q}q\rangle^2$. On the other hand, 
we have argued below Eq.~(\ref{a-scaling}) that 
the dynamics of large $N_f$ QCD 
will provide  
$\gamma_m\simeq 1$ which implies $\langle \bar{q}q\rangle \sim m^2$. 
Then we here adopt the following ansatz 
for the global $N_f$-dependence of $\langle \bar{q}q\rangle$:
\begin{equation}
\frac{\left\langle \bar{q} q \right\rangle_{\Lambda_f}}%
{\left\langle \bar{q} q \right\rangle_{\Lambda_3}}
=
  \frac{1/N_f - 1/N_f^{\rm crit}}{1/3 - 1/N_f^{\rm crit}}
\ .
\label{qq cond}
\end{equation}

Combination of Eqs.~(\ref{Lambda cond}), (\ref{qq cond}) and
(\ref{GG cond}) determines the $N_f$-dependences of the axialvector
and vector current correlators derived in the OPE.
Through the Wilsonian macthing the $N_f$-dependences of the parameters
in the OPE are transfered to those of the parameters in the HLS.
However, as we discussed in Sec.~\ref{ssec:DBPHL},
three Wilsonian matching conditions in Eqs.~(\ref{match z}), 
(\ref{match A}) and (\ref{match V}) are not enough to determine five
parameters 
$F_\pi(\Lambda_f;N_f)$, $a(\Lambda_f;N_f)$,
$g(\Lambda_f;N_f)$, $z_3(\Lambda_f;N_f)$ and 
$z_2(\Lambda_f;N_f)-z_1(\Lambda_f;N_f)$.
As for the $N_c$-dependence of the HLS gauge coupling 
$g$, as we discussed below Eq.~(\ref{initial value of g}),
$g^2$ scales as $1/N_c$.
Then, from
Eq.~(\ref{z match near VM}) together with the assumpion that
each term in the left-hand-side have the same scaling property,
we see that 
$z_3$ scale as 
$N_c$.
Then we use the following assumptions for the $N_c$- and
$N_f$-dependences
of $g(\Lambda;N_f)$
and $z_3(\Lambda;N_f)$:
\begin{eqnarray}
&&
\frac{g^2(\Lambda_f;N_f)}{g^2(\Lambda_3;3)}
=
\left(
  \frac{1/N_f - 1/N_f^{\rm crit}}{1/3 - 1/N_f^{\rm crit}}
\right)^2
\ ,
\label{g2 cond}
\\
&&
\frac{1}{N_c} z_3(\Lambda_f;N_f) = \mbox{constant} \ .
\end{eqnarray}
Note that the condition in Eq.~(\ref{g2 cond})
is consistent with Eq.~(\ref{q2}) or Eq.~(\ref{qq cond})
through the condition in Eq.~(\ref{g2 crit}).
{}From the above assumptions we can determine the $N_f$-dependences of
other three bare parameters through the Wilsonian matching.

Now that we have determined the $N_f$-dependences of five parameters
$F_\pi(\Lambda_f;N_f)$, $a(\Lambda_f;N_f)$,
$g(\Lambda_f;N_f)$, $z_3(\Lambda_f;N_f)$ and 
$z_2(\Lambda_f;N_f)-z_1(\Lambda_f;N_f)$ in the HLS.
we study the $N_f$-dependences of the physical quantities by soving
the RGEs with $N_c=3$ fixed.
To determine the current correlators in the OPE for 
$N_f=3$ we use
\begin{eqnarray}
&&
\left\langle \frac{\alpha_s}{\pi} G_{\mu\nu} G^{\mu\nu}
\right\rangle = 0.012 \,\mbox{GeV}^4 \ , 
\nonumber\\
&&\left\langle \bar{q} q \right\rangle_{\rm 1\,GeV} = - 
\left(\mbox{0.225\,GeV}\right)^3
\ ,
\end{eqnarray}
as a typical example.
To determine the parameters in the HLS for $N_f=3$ through the
Wilsonian matching we use
\begin{equation}
\Lambda_3 = 1.1 \, \mbox{GeV} \ ,
\quad
\Lambda_{\rm QCD} = 400 \, \mbox{MeV} \ ,
\end{equation}
for illustration.

\begin{figure}[htbp]
\begin{center}
\epsfxsize = 7cm
\ \epsfbox{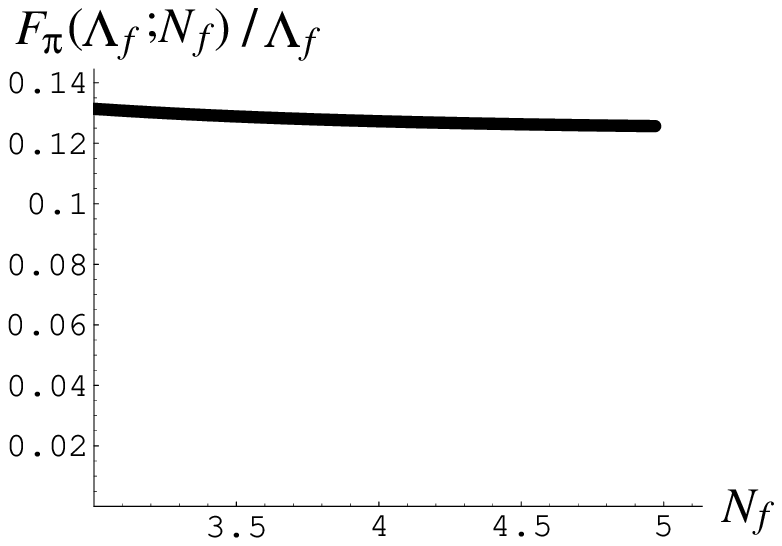}
\epsfxsize = 7cm
\ \epsfbox{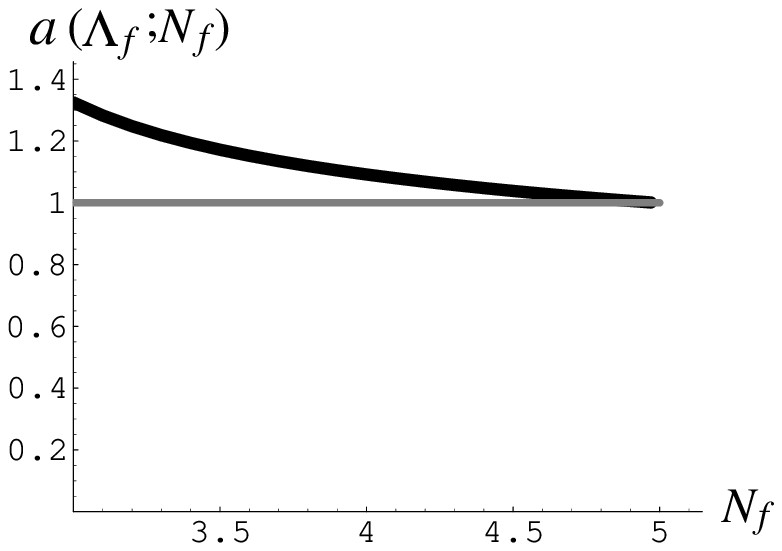}\\
\hspace*{1cm} \  (a) \ \hspace{6.5cm} \ (b) \hspace*{2cm}\\
\ \\
\epsfxsize = 7cm
\ \epsfbox{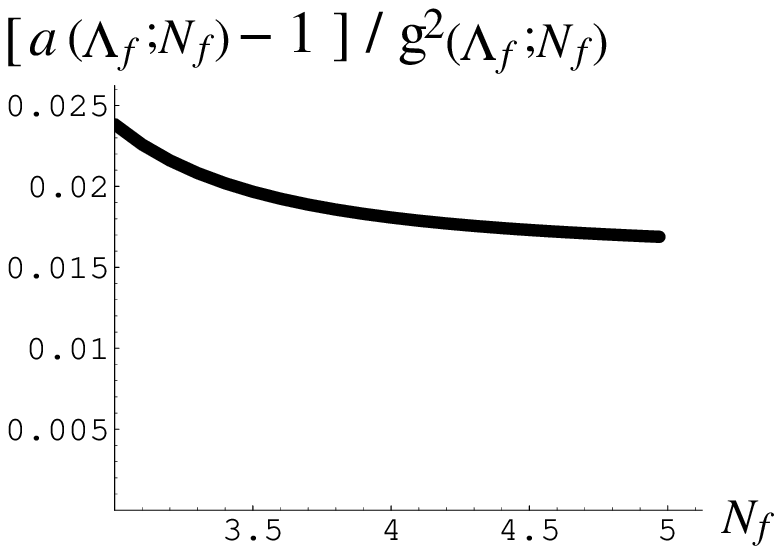}
\epsfxsize = 7cm
\ \epsfbox{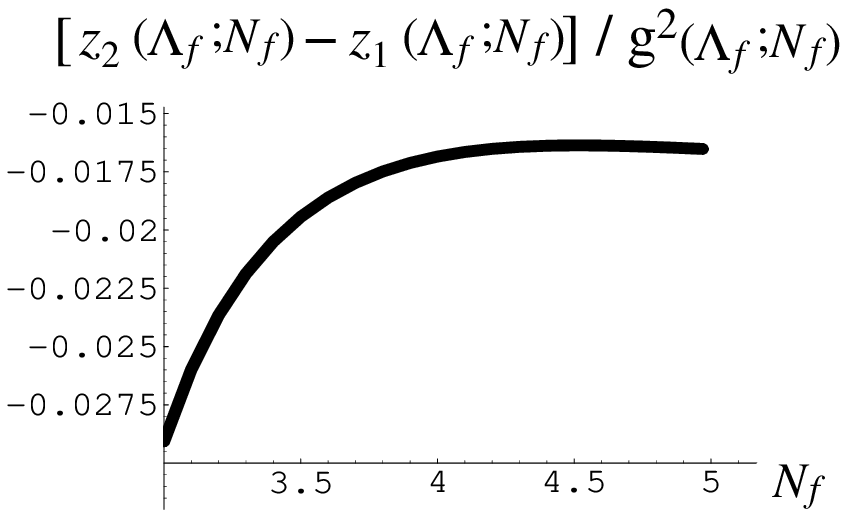}\\
\hspace*{1cm} \  (c) \ \hspace{6.5cm} \ (d) \hspace*{2cm}\\
\end{center}
\caption[$N_f$-dependences of bare parameters]{%
$N_f$-dependences of 
(a) $F_\pi(\Lambda_f;N_f)/\Lambda_f$,
(b) $ a(\Lambda_f;N_f)$,
(c) $ [a(\Lambda_f;N_f)-1]/g^2(\Lambda_f;N_f)$
and 
(d) $[z_2(\Lambda_f;N_f)-z_1(\Lambda_f;N_f)]/g^2(\Lambda_f;N_f)$.
}\label{fig:nfdep bare}
\end{figure}

First, in Fig.~\ref{fig:nfdep bare},
we show the $N_f$-dependences of $F_\pi(\Lambda_f;N_f)/\Lambda_f$
and $a(\Lambda_f;N_f)$ together with those of
$[a(\Lambda_f;N_f)-1]/g^2(\Lambda_f;N_f)$
and $ [z_2(\Lambda_f;N_f)-z_1(\Lambda_f;N_f)]/g^2(\Lambda_f;N_f)$
which are determined through the Wilsonian matching conditions
(\ref{match A}), (\ref{match V}) and (\ref{match z}) together with
the above assumptions of the $N_f$-dependences of other
parameters.
Figure~\ref{fig:nfdep bare}(a) shows that the ratio
$F_\pi(\Lambda_f;N_f)/\Lambda_f$ has only small $N_f$-dependence
as we have discussed before.
{}From Fig.~\ref{fig:nfdep bare}(b) we can see that the value 
of $a(\Lambda_f;N_f)$ is close to one in most region.
Figures~\ref{fig:nfdep bare}(c) and (d) show that
$a(\Lambda_f;N_f)-1$ and
$z_2(\Lambda_f;N_f)-z_1(\Lambda_f;N_f)$ actually
scale as $g^2(\Lambda_f;N_f)$ and 
$\left\langle \bar{q} q \right\rangle^2$ near the critical 
flavor $N_f^{\rm crit} \simeq 5$
as we have discussed below Eq.~(\ref{z match near VM}).

\begin{figure}[htbp]
\begin{center}
\epsfxsize = 7cm
\ \epsfbox{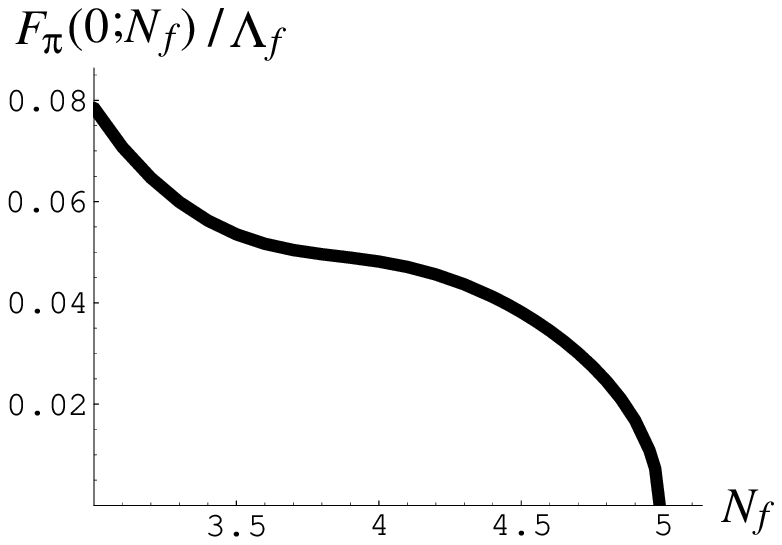}
\epsfxsize = 7cm
\ \epsfbox{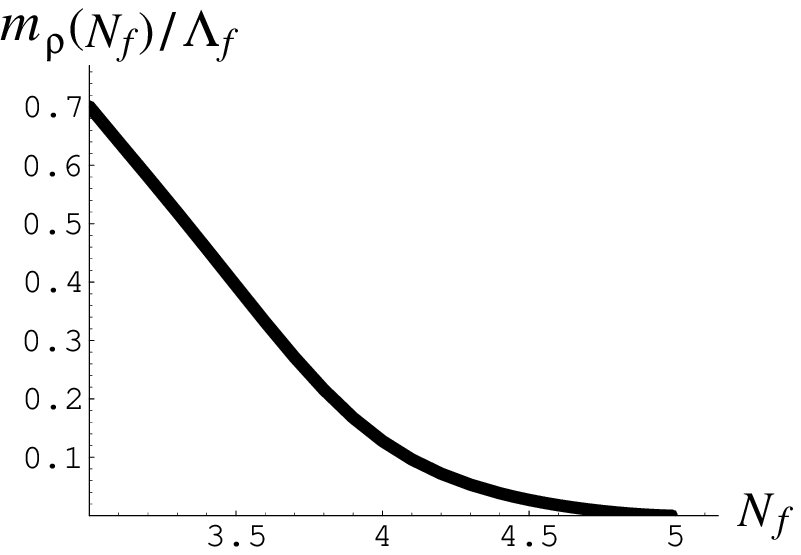}\\
\hspace*{2cm} \  (a) \ \hspace{6cm} \ (b) \hspace*{2cm}
\end{center}
\caption[$N_f$-dependences of $F_\pi$ and $m_\rho$]{%
$N_f$-dependences of (a) $F_\pi(0;N_f)/\Lambda_f$ and
(b) $m_\rho(N_f)/\Lambda_f$.
}\label{fig:nfdep fp mrho}
\end{figure}

\begin{figure}[htbp]
\begin{center}
\epsfxsize = 7cm
\ \epsfbox{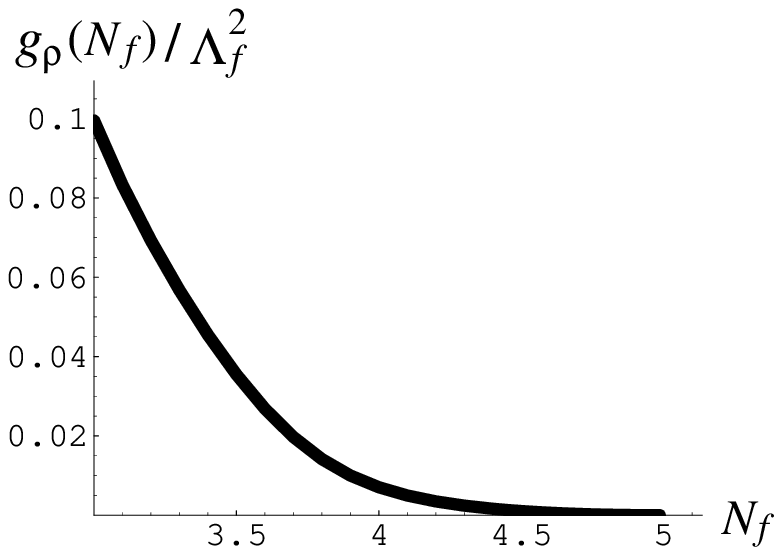}
\epsfxsize = 7cm
\ \epsfbox{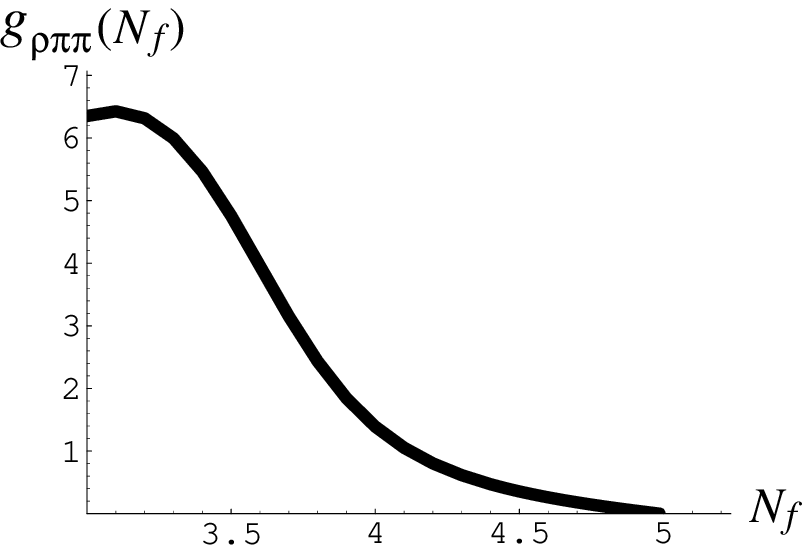}\\
\hspace*{2cm} \  (a) \ \hspace{6cm} \ (b) \hspace*{2cm}\\
\ \\
\epsfxsize = 7cm
\ \epsfbox{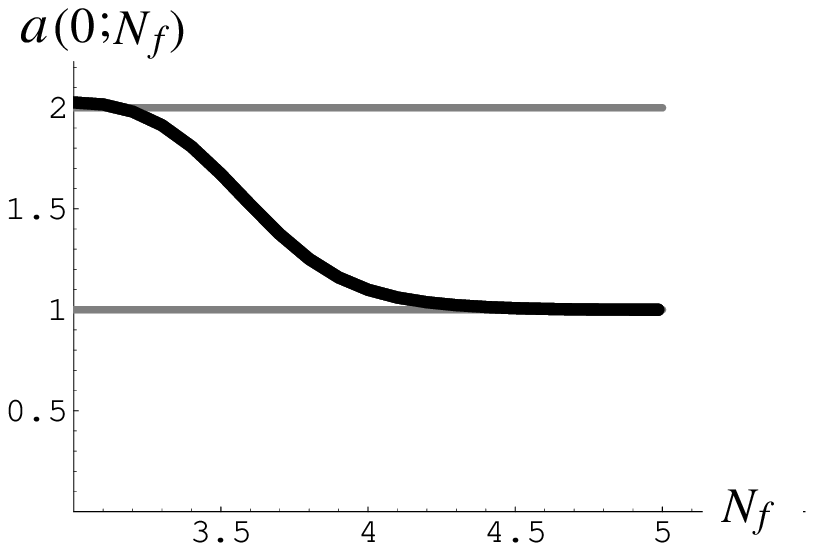}\\
(c)
\end{center}
\caption[$N_f$-dependences of $g_\rho$, $g_{\rho\pi\pi}$ and $a$]{%
$N_f$-dependences of 
(a) $g_\rho$, (b) $g_{\rho\pi\pi}$ and
(c) $a(0;N_f)$.
}\label{fig:nfdep grho grpp a0}
\end{figure}

\begin{figure}[htbp]
\begin{center}
\epsfxsize = 7cm
\ \epsfbox{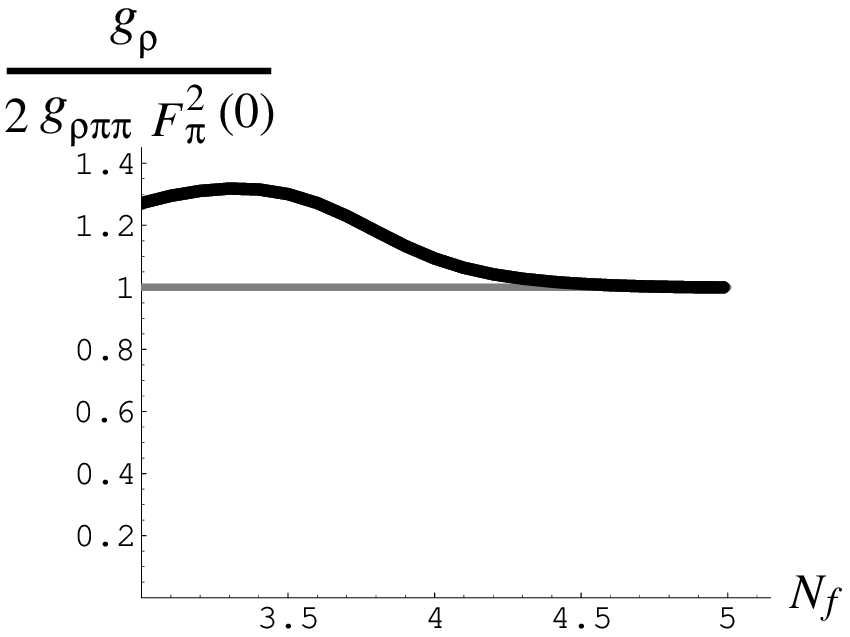}
\epsfxsize = 7cm
\ \epsfbox{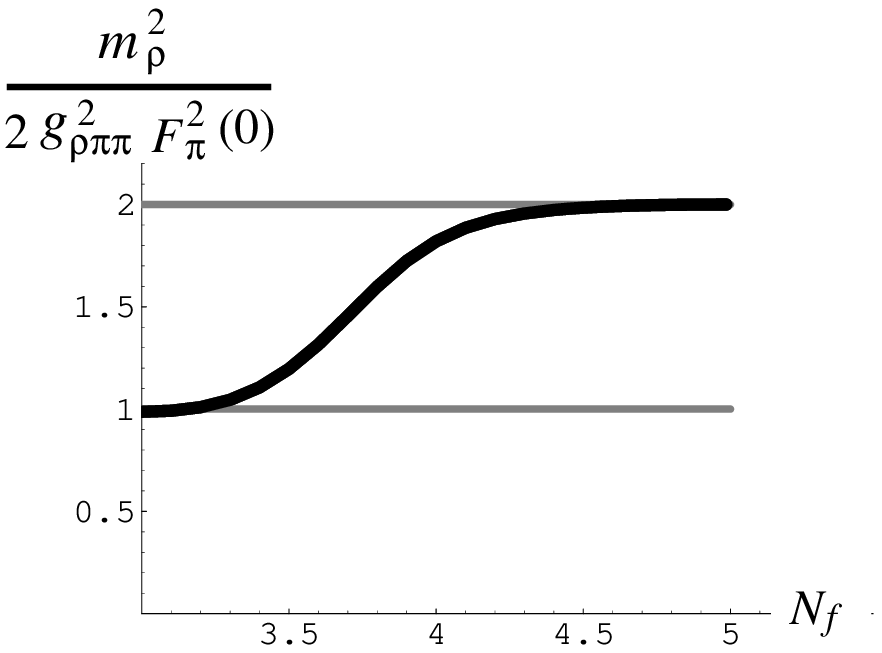}\\
\hspace*{2cm} \  (a) \ \hspace{6cm} \ (b) \hspace*{2cm}\\
\end{center}
\caption[$N_f$-dependences of KSRF relations]{%
$N_f$-dependences of KSRF relations (I) and (II):
(a) $g_\rho/(2g_{\rho\pi\pi}F_\pi^2(0))$
and (b)
$m_\rho^2/(2g_{\rho\pi\pi}F_\pi^2(0))$.
}\label{fig:nfdep ksrf 1 2}
\end{figure}

Next,
we show the $N_f$-dependences of $F_\pi(0;N_f)/\Lambda_f$ and
$m_\rho(N_f)/\Lambda_f$ 
in Fig.~\ref{fig:nfdep fp mrho}.
This shows that $F_\pi(0;N_f)$ and $m_\rho(N_f)$ smoothly go to zero
when $N_f \rightarrow N_f^{\rm crit}$.~\footnote{%
  In Fig.~\ref{fig:nfdep fp mrho}, the value of 
  $m_\rho(N_f)/\Lambda_f$ becomes small already at the off-critical
  point. This is due to the ansatz of $N_f$-dependence of
  $g^2(\Lambda_f;N_f)$ adopted in Eq.~(\ref{g2 cond}).
  If we used the ansatz of essential-singularity-type scaling 
  suggested by the Schwinger-Dyson
  approach~\cite{Appelquist-Terning-Wijewardhana,%
  Appelquist-Ratnaweera-Terning-Wijewardhana,Miransky-Yamawaki},
  on the other hand, 
  the $\rho$ mass $m_\rho$ (and other physical quanties as well) 
  would not change much off the critical point 
  but suddenly approach the critical point value only near the
  critical point.
}
Next we show in Fig.~\ref{fig:nfdep grho grpp a0}
the $N_f$-dependences of $g_\rho$, $g_{\rho\pi\pi}$ and
$a(0;N_f)$ which were defined in Sec.~\ref{ssec:RWM}. 
The $N_f$-dependence of $a(0)$ shows that
the vector dominance is already largely violated even off the
ctritical point.
Finallly, to check the KSRF relations I and II in large $N_f$ QCD
[see Sec.~\ref{ssec:PPLO}],
we show the $N_f$-dependences of 
$g_\rho/(2g_{\rho\pi\pi}F_\pi^2(0))\, [=1-g^2(m_\rho) z_3(m_\rho)]$
and
$m_\rho^2/(2g_{\rho\pi\pi}F_\pi^2(0))\, [=2/a(0)]$
in Fig.~\ref{fig:nfdep ksrf 1 2}, the unity value of which 
correponds to the KSRF relations.
This shows that the KSRF I relation, which is the low energy theorem
of the HLS, approaches to the exact relation near the ctitical point,
while the KSRF II relation is largely violated there as they should
(due to the VM; $a(0)\rightarrow 1$,
$g^2(m_\rho)\rightarrow 0$).

\subsubsection{Vector dominance in large $N_f$ QCD}
\label{sssec:VDLNQ}

Since Sakurai advocated Vector Dominance (VD) as well as vector
meson universality~\cite{Sakurai},
VD has been a widely accepted notion in describing vector meson
phenomena in hadron physics. 
In fact several models such as the gauged sigma 
model (See, e.g., Refs.~\cite{KRS,Mei}.)
are based on VD to introduce 
the photon field into the Lagrangian. Moreover, it is often taken for
granted in analysing the dilepton spectra 
to probe the phase of quark-gluon plasma for the hot
and/or dense QCD (See, e.g., 
Refs.~\cite{Pis:95,Kli:97,Rapp-Wambach:00}.).

As far as the well-established hadron physics for the $N_f=3$ case is
concerned,
it in fact has been extremely successful in many processes such as the
electromagnetic form factor of the pion~\cite{Sakurai} and the 
electromagnetic $\pi\gamma$ transition
form factor (See, e.g., Ref.~\cite{CELLO}.), etc, as studied in
Sec.~\ref{ssec:AP}.
However, there has
been no theoretical justification for VD 
and as it stands might be no more than a mnemonic useful only for the
three-flavored QCD at zero
temperature/density. 
Actually, as studied in Sec.~\ref{ssec:AP},
{\it VD is already violated} for the
three-flavored QCD for the 
anomalous processes such as 
$\gamma \rightarrow 3 \pi/\pi^0 \rightarrow
2\gamma$~\cite{FKTUY,BKY}
and $\omega\pi$ transition form factor (See, e.g.,
Ref.~\cite{BGP,BH:PTP,BH:PRD}.).
This strongly suggests that VD may not be a sacred discipline of
hadron physics but 
may largely be violated in the different parameter space than the
ordinary three-flavored QCD (non-anomalous processes) such as in the
large $N_f$ QCD, $N_f$ being number of massless flavors,
and hot and/or dense QCD
where the chiral symmetry 
restoration is expected to occur. It is rather crucial whether or not
VD is still valid 
when probing such a chiral symmetry restoration  through vector meson
properties~\cite{Pis:82,Brown-Rho:91,Brown-Rho:96,%
Brown-Rho:01a,Brown-Rho:01b}.

Here we emphasize that in the Hidden Local Symmetry (HLS)
model~\cite{BKUYY,BKY}
{\it the vector mesons are formulated 
precisely as gauge bosons}; nevertheless 
{\it VD as well as the universality is merely a dynamical consequence}
characterized by the parameter choice $a=2$
(see Sec.~\ref{ssec:PPLO}).

In this sub-subsection we study the vector dominance (VD) in large
$N_f$ 
QCD following Ref.~\cite{HY:VD}.
Here it is convenient to use the parameters $X(\mu)$ and $G(\mu)$
defined in Eqs.~(\ref{def X}) and (\ref{def G}).

The VD is characterized by $a(0) = 2$, where $a(0)$ is defined in 
Eq.~(\ref{a0}).
Substituting Eqs.~(\ref{sol fpi2 for chpt}) and
(\ref{finite renormalization}) with Eq.~(\ref{def X})
into Eq.~(\ref{a0}),
we obtain
\begin{equation}
a(0)
=
a(m_\rho) / \left[ 1 + a(m_\rho) X(m_\rho) - 2X(m_\rho) \right]
\ .
\label{def:a0}
\end{equation}
This implies that the VD ($a(0) =2$) is only realized
for $( X(m_\rho), a(m_\rho) ) = 
( 1/2, \mbox{any} )$ or
$( \mbox{any}, 2 )$~\cite{HY:VD}.

In $N_f=3$ QCD, the parameters at $m_\rho$ scale,
$\left( X(m_\rho), a(m_\rho), G(m_\rho)\right)
\simeq \left( 0.51,  1.38, 0.37 \right)$,
happen to be near such a VD point.
However, the RG flow actually belongs to the fixed point
$\left(X^\ast_4,a^\ast_4,G^\ast_4\right)$ which is far away from the
VD value.
Thus, {\it the VD in $N_f=3$ QCD is accidentally realized by 
$X(m_\rho)\sim1/2$ which is very unstable against the RG 
flow}~\cite{HY:VD}
(see Fig.~\ref{fig:running}).
For $G=0$ (Fig.~\ref{fig:flows G0})
the VD holds only if the parameters are
(accidentally)
chosen to be on the RG flow entering
$\left( X, a, G \right) = \left( 0, 2, 0 \right)$
which is an end point of the line
$\left( X(m_\rho), a(m_\rho) \right)=(\mbox{any},2)$.
For $a=1$ (Fig.~\ref{fig:flows a1}), 
on the other hand,
the VD point
$\left( X, a, G \right) = \left( 1/2, 1, 1/2 \right)$
lies on the line
$\left( X(m_\rho), a(m_\rho) \right)=(1/2,\mbox{any})$.

Then, phase diagrams in Figs.~\ref{fig:flows G0} and 
\ref{fig:flows a1} 
and their extensions to the entire parameter space
(including Fig.~\ref{fig:running})
show that neither 
$X(m_\rho) = 1/2$ nor $a(m_\rho) = 2$ is a special point in the
parameter space of the HLS.
Thus 
the {\it VD as well as the universality
can be satisfied only accidentally}~\cite{HY:VD}.
Therefore,
when we change the parameter of QCD,
the VD is generally violated.
In particular, neither 
$X(m_\rho) = 1/2$ nor $a(m_\rho) = 2$ is satisfied on the phase
boundary surface characterized by Eq.~(\ref{phase boundary}) where the
chiral restoration takes place in HLS model.
Therefore,
{\it VD is realized nowhere on the chiral restoration
surface}~\cite{HY:VD}.

Moreover, 
{\it when the HLS is matched with QCD},
only the point
$( X^\ast_2, a^\ast_2, G^\ast_2 ) = ( 1, 1, 0 )$, 
the {\it VM point},
on the phase boundary is selected,
since 
the axialvector and vector current correlators in HLS
can be matched with those in QCD only at that point~\cite{HY:VM}.
{\it
Therefore, QCD predicts $a(0)=1$,
i.e., large violation of the VD at chiral restoration}.
Actually, 
as is seen in Fig.~\ref{fig:nfdep grho grpp a0}(c),
for the chiral restoration
{\it in the large $N_f$ 
QCD}~\cite{IKKSY:98,Appelquist-Terning-Wijewardhana}
{\it the VM can in fact
takes place}~\cite{HY:VM}, {\it and thus the VD is badly 
violated}~\cite{HY:VD}.

\subsection{Seiberg-type duality}
\label{ssec:STD}

\begin{table}[htbp]
\begin{center}
\begin{tabular}{|c|l|l|}
\hline
$N_f$ & ``Electric theory''
& ``Magnetic theory'' \\
& $\mbox{SU($N_c$)}$ SQCD
& $\mbox{SU($N_f-N_c$)}$ SQCD \\
\hline
$\uparrow$ &
Free non-Abelian electric theory&Strong no-Abelian magnetic theory\\
$3 N_c$ &IR free &Asymptotic free \\
\hline
$\updownarrow$ & \multicolumn{2}{c|}{%
(Interacting non-Abelian Coulomb phase)} \\
$3N_c/2$ &IR fixed point & IR fixed point \\
\hline
$\updownarrow$ & 
Strong non-Abelian electric theory&Free non-Abelian magnetic theory\\
$N_c+ 2$ &Asymptotic free & IR free \\
\hline
$N_c + 1$ & 
$\begin{array}{l}
\mbox{complete confinement} \\
\mbox{No S$\chi$SB (s-confinement)}
\end{array}$
& completely Higgsed \\
\hline
$N_c$ & 
$\begin{array}{l}
\mbox{complete confinement} \\
\mbox{S$\chi$SB}
\end{array}$
 & completely Higgsed \\
\hline
\end{tabular}
\end{center}
\caption[Duality and conformal window in ${\cal N}=1$ SUSY QCD]{%
Duality and conformal window in ${\cal N}=1$ SUSY QCD.%
}\label{tab:Dual:SUSY}
\end{table}
Increasing attention has been paid to the duality 
in various contexts of modern particle theory. 
Seiberg found the ``electric-magnetic'' 
duality in ${\cal{N}}=1$ Supersymmetric 
(SUSY) QCD with $N_c$ colors and
$N_f$ flavors~\cite{Seiberg}.
The $N_f$-dependence of the theory is summarized in
Table~\ref{tab:Dual:SUSY}.
For the region $\frac{3}{2} N_c < N_f < 3 N_c$
(``conformal window'') in the SUSY QCD,
there exists a ``magnetic theory'' with the SU($N_f-N_c$) 
gauge symmetry which is dual to the original 
SU($N_c$) theory regarded as the ``electric theory''.
Although the origin of the magnetic gauge symmetry 
(``induced at the composite level'') is not obvious from
the original theory,
both theories in fact have the infrared (IR) fixed point 
with exact conformal symmetry and with the same IR physics.
This region is called ``interacting non-Abelian Coulomb phase''.
When $N_f$ decreases, 
the electric theory becomes stronger in IR, while
the magnetic theory gets weaker, with the magnetic
gauge group being reduced through the Higgs mechanism. 
Decreasing $N_f$ further beyond the conformal window, we finally
arrive at $N_f=N_c$ where the magnetic theory is in complete Higgs
phase (reduced to no gauge group), which corresponds to the complete
confinement (and spontaneously broken chiral symmetry) of the electric
theory.

\begin{table}[htbp]
\begin{center}
\begin{tabular}{|c|l|l|}
\hline
$N_f$ & ``Electric Theory'' & ``Magnetic Theory'' \\
      & $\mbox{SU($N_c$)}$ QCD  & $\mbox{SU($N_f$)}$ HLS \\
\hline
$\uparrow$ &Free electric theory &  EFT ?\\
$11N_c/2$ 
    & IR free             &    \\
\hline
$11N_c/2$ & 
Interacting non-Abelian Coulomb phase &   EFT ? \\
$\updownarrow$& IR fixed point (No S$\chi$SB/Confinement) &    \\
\hline
$\sim 5(N_c/3)$ & Conformal phase transition &Vector Manifestation \\
$\updownarrow$ & Confined electric theory (S$\chi$SB) & 
  Higgsed magnetic theory (S$\chi$SB) \\
$N_c$ & ``real world''($\mbox{SU(3)}$ QCD) & $\mbox{SU(3)}$ HLS \\
\hline
\end{tabular}
\end{center}
\caption[Duality and conformal window in QCD]{%
Duality and conformal window 
($33N_c/2 > N_f > N_f^{\rm crit} \sim 5(N_c/3)$)
in QCD.%
}\label{tab:Dual}
\end{table}
Similar conformal window may also exist in the ordinary (non-SUSY)
QCD with massless $N_f$ flavors ($33N_c/2 
> N_f > N_f^{\rm crit} \sim 5(N_c/3)$),
as was discussed in 
Sec.~\ref{ssec:CPTLNQ}. 
Situation including the proposal in Ref.~\cite{HY:letter} is
summarized in Table~\ref{tab:Dual}.

Here we recall that, for small $N_f$, 
the vector mesons such as the $\rho$ meson 
can be regarded as the dynamical gauge bosons of HLS~\cite{BKUYY,BKY}.
The HLS is completely broken 
through the Higgs mechanism as the origin of the 
vector meson mass. 
This gauge symmetry is
induced at the composite level and has nothing to do with the 
fundamental color gauge
symmetry. Instead, the HLS is associated with
the flavor symmetry. 

In Ref.~\cite{HY:letter} we found that the Seiberg duality is realized
also in the ordinary (non-SUSY) QCD through the HLS.
For small $N_c(=3)\le N_f <N_f^{\rm crit} \sim 5(N_c/3)$,
the SU($N_f$) HLS is in complete
Higgs phase and yields the same IR physics as the SU($N_c$) QCD in the
confinement/chiral-symmetry-breaking phase, and plays the role of
the ``Higgsed magnetic gauge theory'' 
dual to the ``Confined electric gauge theory'' (QCD)
in the spirit of Seiberg duality. Then the $\rho$ mesons
can in fact be regarded as the Higgsed ``magnetic gluons''
of the SU($N_f$) HLS.

In order for such a duality between QCD and the HLS
be consistently satisfied,
there should be a way 
that the chiral restoration takes place for large $N_f$
also in the HLS theory 
{\it by its own dynamics}.
We have already seen in Sec.~\ref{ssec:VMLNQ} that 
the HLS can provide the chiral restoration by its own dynamics
for a certain value of $N_f=N_f^{\rm crit} \simeq 5 (N_c/3)$ which
is in rough agreement with $6<N_f^{\rm crit}<7$ found in 
the lattice simulation of the electric theory, the QCD with $N_c=3$.
Thus the Seiberg-type duality does exist also in the ordinary
(non-SUSY) 
QCD at least for 
$N_c(=3)\le N_f <N_f^{\rm crit} \sim 5(N_c/3)$~\cite{HY:letter}. 
We do not know at this moment, however, what the duality 
would be for $11N_c/2 >N_f >N_f^{\rm crit}$ where the EFT like HLS may
not 
exist because of a possible absence of effective fields of
light bound states in the symmetric phase, as was suggested by
the conformal phase transition (see Sec.~\ref{sssec:CPT}).

It should also be emphasized that near the critical point this
Higgsed magnetic gauge theory prodides an example of  
a {\it weakly-coupled composite gauge theory} 
with light gauge boson and NG boson, while the underlying electric gauge 
theory is still in the strongly-coupled 
phase with confinement and chiral symmetry breaking. This unusual
feature may be useful for model building beyond the Standard Model.
 
\newpage

\section{Renormalization at Any Loop Order and the Low Energy
Theorem}
\label{sec:RALOLET}

As was discussed in Sec.~\ref{sec:HLS},
the KSRF relation (version I) [see Eq.~(\ref{KSRF I})],
\begin{equation}
g_\rho = 2 F_\pi^2 g_{\rho\pi\pi} \ ,
\label{HLS LET}
\end{equation}
holds as a ``low energy theorem'' of the HLS~\cite{BKY:NPB},
which was first
proved at the tree level~\cite{BKY:PTP}, then at one-loop
level~\cite{HY} and further at any loop order~\cite{HKY:PRL,HKY:PTP}.

In this section we briefly review the proof of the low energy theorem
of the HLS at any loop order, following Refs.~\cite{HKY:PRL,HKY:PTP}.
Although Refs.~\cite{HKY:PRL,HKY:PTP} presumed only logarithmic divergence,
only the relevant assumption made there was that 
{\it there exists a symmetry preserving regularization}.
As was discussed in Sec.~\ref{ssec:QD}, inclusion of the quadratic
divergence through the replacement in Eq.~(\ref{regularization 0})
is in fact consistent with the gauge invariance.
{\it Then the proposition and the proof below are valid even if we
include the quadratic divergences}.

We restrict ourselves to the chiral symmetric case\footnote{%
See Refs.~\cite{HKY:PRL,HKY:PTP} for the effect of the symmetry
breaking mass terms of NG fields.%
}, so that we take 
$\hat{\chi} = 0$ in the leading order Lagrangian in 
Eq.~(\ref{leading Lagrangian}):
\begin{equation}
{\cal L}_{(2)} = {\cal L}_{\rm A} + a {\cal L}_{\rm V}
+ {\cal L}_{\rm kin}(V_\mu) \ ,
\label{startLag}
\end{equation}
where ${\cal L}_{\rm A}$ and
$a {\cal L}_{\rm V}$ are defined in Eqs.~(\ref{def:LA})
and (\ref{def:LV}), respectively:
\begin{eqnarray}
&& {\cal L}_{\rm A} \equiv F_\pi^2 \, \mbox{tr} 
\left[ \hat{\alpha}_{\perp\mu} \hat{\alpha}_{\perp}^\mu \right]
\ ,
\\
&& a {\cal L}_{\rm V} \equiv F_\sigma^2 \, \mbox{tr} 
\left[ 
  \hat{\alpha}_{\parallel\mu} \hat{\alpha}_{\parallel}^\mu
\right]
\ .
\end{eqnarray}

It should be noticed that in this section we classify 
${\cal L}_{\rm A}$ and ${\cal L}_{\rm V}$ as ``dimension-2 terms''
and ${\cal L}_{\rm kin}$ as  ``dimension-4 term'', based on 
{\it counting the dimension of only the fields and derivatives}.
This is somewhat different from the chiral counting
explained in Sec.~\ref{ssec:DEHLS}
where the HLS gauge coupling carries ${\cal O}(p)$, and thus
${\cal L}_{\rm kin}$ is counted as ${\cal O}(p^2)$.
The counting method adopted in this section is convenient for
classifying the terms with the same chiral order:
The contribution at $n$-th loop order is expected to generate
${\cal O}(p^{2n+2})$ corrections which take the form 
of $(g^2)^n\,{\cal L}_{\rm A}$,
$(g^2)^n\,{\cal L}_{\rm V}$, 
$(g^2)^{n}\,{\cal L}_{\rm kin}$,
and so on.
In 
$(g^2)^n\,{\cal L}_{\rm A}$ and
$(g^2)^n\,{\cal L}_{\rm A}$,
the ${\cal O}(p^2)$ out of ${\cal O}(p^{2n+2})$ is carried by
derivatives and fields, while in 
$(g^2)^{n}\,{\cal L}_{\rm kin}$,
the ${\cal O}(p^4)$ out of ${\cal O}(p^{2n+2})$ is by them.
Then, by counting the dimensions of only the fields and derivatives,
we can extract the terms 
relevant to the low-energy region
out of all the possible $n$-loop corrections.
Note that
we focus on the renormalizability of the terms of dimension two, 
${\cal L}_{\rm A}$
and ${\cal L}_{\rm V}$ terms in Eq.~(\ref{startLag}), 
which is just what we need for proving the low energy theorem.

We introduce the BRS transformation and make the proposition in 
Sec.~\ref{ssec:BRSTP}.
We prove the proposition in Sec.~\ref{ssec:PP}.
Finally, in Sec.~\ref{ssec:LET}, we prove that the 
low-energy theorem in Eq.~(\ref{HLS LET}) holds at any-loop order.

Also note that, in this section, we use the {\it covariant gauge}
instead of the background field gauge, since
the higher order loop calculation 
is well-defined compared with the background field gauge.
Also the off-shell extrapolation is easily done in covariant gauge
compared with the $R_\xi$ gauge (see Sec.~\ref{ssec:LET}).

\subsection{BRS transformation and proposition}
\label{ssec:BRSTP}

Let us take a covariant gauge condition
for the HLS,
and introduce the corresponding gauge-fixing
and Faddeev-Popov (FP) terms:
\begin{equation}
  {\cal L}_{GF} + {\cal L}_{FP}
  =
  B^a \partial^\mu V_\mu^a
  + {1\over2} \alpha B^a B^a
  + i \bar{ C}^a \partial^\mu D_\mu  C^a ,
\label{GFFPlagrangian}
\end{equation}
where $B^a$ is the Nakanishi-Lautrap (NL) field and
$ C^a$ ($\bar{ C}^a$) the FP ghost (anti-ghost) field.
As in the previous sections we do not consider
the radiative corrections due to the external gauge fields
${\cal V}_\mu^i \equiv ({\cal L}_{\mu}^a ,
{\cal R}_{\mu}^a)$,
so that we need not introduce the gauge-fixing terms
for ${\cal V}_\mu$.
Then, the corresponding ghost fields
${\cal C}^i\equiv({\cal C}_{\rm L}^a,{\cal C}_{\rm R}^a)$
are non-propagating.

The infinitesimal form of the
$ G_{\rm global} \times  H_{\rm local}$ transformation
(\ref{xi:trans}) is given by
\begin{eqnarray}
&& \delta \xi(x) = i \theta(x) \xi(x) - i \xi(x) \vartheta(x) \ ,
\nonumber\\
&& \theta(x) \equiv \theta^a(x) T_a \ , \quad
\vartheta(x) \equiv \vartheta^a(x) T_a \ .
\end{eqnarray}
This defines the transformation of the Nambu-Goldstone (NG) field
$\phi^i\equiv(\sigma^a/F_\sigma,\pi^a/F_\pi)$ 
[see Eq.~(\ref{def:xiLR})] in the form
\begin{equation}
  \delta \phi^i
  =
  \theta^a W_a^i(\phi)
  + \vartheta^j {\cal W}_j^i (\phi)
  \left({
    \equiv \theta^A \hbox{\boldmath $W$}_A^i(\phi)
  }\right)
\ ,
\end{equation}
where $A$ denotes a set $(a,i)$ of labels of $ H_{\rm local}$ and
$ G_{\rm global}$.
Accordingly,
the BRS transformation of the NG fields $\phi^i$,
the gauge fields
$\hbox{\boldmath $V$}_\mu^A \equiv (V_\mu^a,{\cal V}_\mu^i)$
and the FP ghost fields
$\hbox{\boldmath $C$}^A\equiv( C^a,{\cal C}^i)$
are respectively given by
\begin{eqnarray}
  \delta_{\rm B} \phi^i
&=&
  \hbox{\boldmath $C$}^A \hat{\hbox{\boldmath $W$}}_{\!\!A} \phi^i
  \quad
  ( \hat{\hbox{\boldmath $W$}}_{\!\!A} \equiv
  \hbox{\boldmath $W$}_{\!A}^i (\phi)
  {\partial \over \partial \phi^i} ) \ ,
\nonumber\\
  \delta_{\rm B} \hbox{\boldmath $V$}_\mu^A
&=&
  \partial_\mu \hbox{\boldmath $C$}^A +
  \hbox{\boldmath $V$}_\mu^B \hbox{\boldmath $C$}^C {f_{BC}}^A \ ,
\nonumber\\
  \delta_{\rm B} \hbox{\boldmath $C$}^A
&=&
  - {1\over2} \hbox{\boldmath $C$}^B
  \hbox{\boldmath $C$}^C {f_{BC}}^A \ .
\end{eqnarray}

For definiteness
we define the dimension of the fields as
\begin{equation}
\dim[\phi^i]=0 \ , \quad
\dim[\hbox{\boldmath $V$}_\mu^A]=1
\ .
\end{equation}
It is also convenient to assign the following dimensions to
the FP-ghosts:
\begin{equation}
\dim[\hbox{\boldmath $C$}^A]=0 \ , \quad
\dim[\bar{ C}^a]=2 \ .
\end{equation}
Then the BRS transformation does not change the dimension.
According to the above dimension counting,
we may divide
the Lagrangian Eq.~(\ref{startLag}) plus
Eq.~(\ref{GFFPlagrangian}) into the following two parts:
\begin{enumerate}
\renewcommand{\labelenumi}{(\theenumi)}
\renewcommand{\theenumi}{\alph{enumi}}
\item
dimension-2 part 
${\cal L}_{\rm A} + a {\cal L}_{\rm V}$,

\item
dimension-4 part
${\cal L}_{\rm kin}(V_\mu) + {\cal L}_{\rm GF} + {\cal L}_{\rm FP}$,
\end{enumerate}
where
we count the dimension of the fields
and derivatives only.

Now, we consider the quantum correction
to this system at any loop order, and
prove the following proposition.

\noindent
{\bf Proposition} :
{\it As far as the dimension-2 operators are concerned,
all the quantum corrections,
including the finite parts as well as the divergent parts,
can be absorbed into the original dimension-2 Lagrangian
${\cal L}_{\rm A} + a {\cal L}_{\rm V}$
by a suitable redefinition (renormalization) of the parameters
$a$, $F_\pi^2$, and the fields $\phi^i$,
$V_\mu^a$.}

\noindent
This implies that the tree-level dimension-2 Lagrangian,
with the parameters and fields
substituted by the ``renormalized" ones,
already
describes the exact action at any loop order,
and therefore that all the ``low energy theorems"
derived from it
receive no quantum corrections at all.

\subsection{Proof of the proposition}
\label{ssec:PP}

We prove our proposition
in the same way as the renormalizability proof
for gauge theories\cite{BRS} and
two dimensional nonlinear sigma 
models~\cite{BlasiCollina:87,BlasiCollina:88}.
We can write down the WT identity
for the effective action $\Gamma$.
The NL fields $B^a$ and
the FP anti-ghost fields $\bar{ C}^a$
can be eliminated from $\Gamma$
by using their equations of motion as usual.
Then the tree level action $S=\Gamma_{\rm tree}$ reads
\begin{eqnarray}
&{}&
  S [ \Phi, {\bf K}; \hbox{\boldmath $a$} ]
  =
  S_2 [\phi,\hbox{\boldmath $V$}] + S_4[\Phi,{\bf K}] ,
\nonumber\\
&{}& \qquad
  S_2 [\phi,\hbox{\boldmath $V$}] = \int d^4x
  \left({
    a_{\perp} {\cal L}_{\rm A} (\phi,\hbox{\boldmath $V$}) +
     a_{\parallel} {\cal L}_{\rm V} (\phi,\hbox{\boldmath $V$})
  }\right) ,
\nonumber\\
&{}& \qquad
  S_4 [\Phi,{\bf K}] = \int d^4x
  \left({
    {\cal L}_{\rm kin}(V_\mu)
     + {\bf K} \cdot \delta_{\rm B} \Phi
  }\right) ,
\label{SfourAction}
\end{eqnarray}
where
$\Phi \equiv (\phi^i,\hbox{\boldmath $V$}_\mu^A,
\hbox{\boldmath $C$}^A)$ are
the field variables and
${\bf K}\equiv(K_i,\hbox{\boldmath $K$}_A^\mu,
\hbox{\boldmath $L$}_A)$
( $ \hbox{\boldmath $K$}_A^\mu \equiv (K_a^\mu,{\cal K}_i^\mu)$,
$\hbox{\boldmath $L$}_A \equiv (L_a,{\cal L}_i)$ )
denote the BRS source fields for the NG field $\phi^i$, the gauge
fields $\hbox{\boldmath $V$}_\mu^A$ and the ghost fields
$\hbox{\boldmath $C$}^A$, respectively; i.e.,
\begin{eqnarray}
&&  \hbox{\boldmath $K$}_A^\mu \delta_{\rm B} 
\hbox{\boldmath $V$}_\mu^A = K_a^\mu \delta_{\rm B} V_\mu^a
+ {\cal K}_i^\mu {\cal V}_\mu^i \ ,
\nonumber\\
&& \hbox{\boldmath $L$}_A \delta_{\rm B} \hbox{\boldmath $C$}^A
= L_a \delta_{\rm B} C^a + {\cal L}_i \delta_{\rm B} 
{\cal C}^i
\ .
\end{eqnarray}
We have rewritten $F_\sigma^2$ and $F_\pi^2$
as
\begin{equation}
a_{\perp} f^2 \equiv F_\pi^2 \ , \quad
a_{\parallel} f^2 \equiv F_\sigma^2 \ ,
\end{equation}
so that the renormalization of $F_\pi^2$ and $F_\sigma^2$ corresponds
to that of $\hbox{\boldmath $a$}\equiv( a_{\parallel},a_{\perp})$.
According to the dimension assignment of the fields,
the dimension of the above BRS source fields ${\bf K}$ is given by
\begin{equation}
\dim[K_i]=\dim[\hbox{\boldmath $L$}_A]=4 \ ,
\quad
\dim[\hbox{\boldmath $K$}_A^\mu]=3
\ .
\end{equation}

The WT identity for the effective action
$\Gamma$ is given by
\begin{equation}
  \Gamma\ast\Gamma=0 ,
\end{equation}
where the $\ast$ operation is defined by
\begin{equation}
  F \ast G
  =
  {(-)}^{\Phi}
  {\overleftarrow{\delta} F \over \delta \Phi }
  {\delta G \over \delta {\bf K}}
  - {(-)}^{\Phi}
  {\overleftarrow{\delta} F \over \delta {\bf K}}
  {\delta G \over \delta \Phi}
\end{equation}
for arbitrary functionals $F[\Phi,{\bf K}]$
and $G[\Phi,{\bf K}]$.
(Here the symbols $\delta$ and $\overleftarrow{\delta}$
denote the derivatives from the left and right, respectively,
and ${(-)}^\Phi$ denotes $+1$ or $-1$
when $\Phi$ is bosonic or fermionic, respectively.)

The effective action is calculated in the loop expansion:
\begin{equation}
\Gamma = S + \hbar \Gamma^{(1)} + \hbar^2 \Gamma^{(2)} + \cdots 
\ .
\end{equation}
The $\hbar^n$ term $\Gamma^{(n)}$
contains contributions not only from the genuine $n$-loop diagrams
but also from the lower loop diagrams
including the counter terms.
We can expand the $n$-th term
$\Gamma^{(n)}$ according to the dimension:
\begin{equation}
  \Gamma^{(n)} =
  \Gamma_0^{(n)} [\phi] + \Gamma_2^{(n)}[\phi,\hbox{\boldmath $V$}]
  + \Gamma_4^{(n)}[\Phi,{\bf K}]
  + \cdots .
\label{dimexp}
\end{equation}
Here again we are counting the dimension only of the fields
and derivatives.
The first dimension-0 term $\Gamma_0^{(n)}$ can contain only
the dimensionless field $\phi^i$ without derivatives.
The two dimensions of the second term $\Gamma_2^{(n)}$ is
supplied by derivative and/or the gauge field
$\hbox{\boldmath $V$}_\mu^A$.
The BRS source field ${\bf K}$ carries dimension 4 or 3,
and hence it can appear only in $\Gamma_4^{(n)}$ and beyond:
the dimension-4 term $\Gamma_4^{(n)}$
is at most linear in ${\bf K}$,
while the dimension-6 term $\Gamma_6^{(n)}$ can
contain a quadratic term in $K_a^\mu$,
the BRS source of the hidden gauge boson $V_\mu^a$.
To calculate $\Gamma^{(n)}$,
we need to use the ``bare" action,
\begin{equation}
  {(S_0)}_n = S
  \left[{
    {(\Phi_0)}_n , {({\bf K}_0)}_n ; {(\hbox{\boldmath $a$}_0)}_n
  }\right]
\ ,
\end{equation}
where the $n$-th loop order ``bare" fields
${(\Phi_0)}_n$, ${({\bf K}_0)}_n$
and parameters ${(\hbox{\boldmath $a$}_0)}_n$ are given by
\begin{eqnarray}
 {(\Phi_0)}_n
&=&
  \Phi + \hbar \delta \Phi^{(1)} +
  \cdots + \hbar^n \delta \Phi^{(n)} ,
\nonumber\\
  {({\bf K}_0)}_n
&=&
  {\bf K} + \hbar \delta {\bf K}^{(1)}
  + \cdots + \hbar^n \delta {\bf K}^{(n)} ,
\nonumber\\
  {(\hbox{\boldmath $a$}_0)}_n
&=&
  \hbox{\boldmath $a$} + \hbar \delta \hbox{\boldmath $a$}^{(1)}
 + \cdots + \hbar^n \delta \hbox{\boldmath $a$}^{(n)} .
\end{eqnarray}

Let us now prove the following by mathematical induction
with respect
to the loop expansion parameter $n$:

\begin{enumerate}
\renewcommand{\labelenumi}{\theenumi}
\renewcommand{\theenumi}{(\Roman{enumi})}

\item $\Gamma_0^{(n)} (\phi) = 0$.
\label{induction:1}

\item
\label{induction:2}
By choosing suitably the $n$-th order counter terms
$\delta\Phi^{(n)}$, $\delta{\bf K}^{(n)}$ and
$\delta\hbox{\boldmath $a$}^{(n)}$,
$\Gamma_2^{(n)}[\phi,A]$ and the ${\bf K}$-linear terms
in $\Gamma_4^{(n)}[\Phi,{\bf K}]$ can be made vanish;\
\begin{displaymath}
 \Gamma^{(n)}_2 [\phi,\hbox{\boldmath $V$}] = 
 \left. \Gamma^{(n)}_4 [\Phi,{\bf K}] %
 \right\vert_{{\bf K}\hbox{-}{\rm linear}} = 0
\ .
\end{displaymath}

\item The field reparameterization (renormalization)
$(\Phi,{\bf K})\rightarrow
\left({ {(\Phi_0)}_n,{({\bf K}_0)}_n }\right)$ is a ``canonical"
transformation which leaves the $\ast$ operation invariant.
\label{induction:3}

\end{enumerate}

Suppose that the above statements are satisfied
for the $(n-1)$-th loop order effective action $\Gamma^{(n-1)}$.
We calculate, for the moment,
the $n$-th loop effective action $\Gamma^{(n)}$
using the $(n-1)$-th loop level ``bare" action
${(S_0)}_{n-1}$,
i.e., without $n$-th loop counter terms.
We expand
the $\hbar^n$ terms
in the WT identity
\begin{equation}
S \ast \Gamma^{(n)} = - {1\over2}
\sum_{l=1}^{n-1} \Gamma^{(l)} \ast \Gamma^{(n-l)}
\ ,
\end{equation}
according to the dimensions
like in Eq.~(\ref{dimexp}).
Then using the above induction assumption,
we find:
\begin{eqnarray}
&&  S_4 \ast \Gamma_0^{(n)} + S_2 \ast \Gamma_2^{(n)} = 0 \ 
\mbox{(dim 0)}\ ,
\label{dim0WT}
\\
&&  S_4 \ast \Gamma_2^{(n)} + S_2 \ast \Gamma_4^{(n)} = 0 \ 
\mbox{(dim 2)} \ ,
\label{dim2WT}
\\
&&  S_4 \ast \Gamma_4^{(n)} + S_2 \ast \Gamma_6^{(n)} = 0 \ 
\mbox{(dim 4)} \ .
\label{dim4WT}
\end{eqnarray}
These three renormalization equations give enough
information for determining possible forms of
$\Gamma_0^{(n)}$, $\Gamma_2^{(n)}$ and
$\Gamma_4^{(n)}\vert_{{\bf K}\hbox{-}{\rm linear}}$
(the ${\bf K}$-linear term in $\Gamma_4^{(n)}$)
which we are interested in.

Noting that the BRS transformation $\delta_{\rm B}$ on the fields
$\Phi \equiv (\phi^i,\hbox{\boldmath $V$}_\mu^A,
\hbox{\boldmath $C$}^A)$ can be written in the form
\begin{equation}
  \delta_{\rm B} =
  \frac{\delta S_4}{\delta {\bf K} }
  \frac{\delta}{\delta \Phi} \ ,
\end{equation}
we see it convenient to define an analogous transformation
$\delta'_\Gamma$ on the fields $\Phi$ by
\begin{equation}
\delta'_\Gamma \equiv
\frac{ \delta \Gamma^{(n)}_4 }{ \delta {\bf K} }
\frac{\delta}{\delta \Phi}
\ .
\end{equation}
Then we can write $\Gamma_4^{(n)}$ in the form
\begin{equation}
  \Gamma_4^{(n)} =
  A_4[\phi,\hbox{\boldmath $V$}] + K_i \delta'_\Gamma \phi^i
  + \hbox{\boldmath $K$}_A^\mu \delta'_\Gamma 
  \hbox{\boldmath $V$}_\mu^A +
  \hbox{\boldmath $L$}_A \delta'_\Gamma \hbox{\boldmath $C$}^A \ .
\end{equation}
In terms of this notation, Eqs.~(\ref{dim0WT})--(\ref{dim4WT})
can be rewritten into
\begin{eqnarray}
&& \delta_{\rm B}\Gamma_0^{(n)}=0 \ ,
\label{DimzeroWTB} 
\\
&&
  \delta_{\rm B} \Gamma_2^{(n)} + \delta'_\Gamma S_2 = 0 \ ,
\label{DimtwoWTB} 
\\
&&
  \delta_{\rm B} \Gamma_4^{(n)} + \delta'_\Gamma S_4
  + {\delta \Gamma_6^{(n)} \over \delta {\bf K}}
    {\delta S_2 \over \delta \Phi} = 0 \ .
\label{DimfourWTB}
\end{eqnarray}

First, let us consider
the dimension-0 part
of the renormalization equation (\ref{DimzeroWTB}).
Since there are no invariants containing no derivatives,
we can immediately conclude $\Gamma_0^{(n)}=0$,
and hence our statement \ref{induction:1} follows.

Next, we consider the 
dimension-2 and dimension-4 parts of the renormalization 
equations~(\ref{DimtwoWTB}) and (\ref{DimfourWTB}).
A tedious but straightforward analysis~\cite{HKY:PRL,HKY:PTP}
of the ${\bf K}$-linear term in Eq.~(\ref{DimfourWTB})
determines
the general form of
the $\Gamma_4^{(n)}\vert_{{\bf K}\hbox{-}{\rm linear}}$
and $\Gamma_6^{(n)}\vert_{{\bf K}\hbox{-}{\rm quadratic}}$ terms:
the solution for $\Gamma_4^{(n)}\vert_{{\bf K}\hbox{-}{\rm linear}}$
or equivalently $\delta'_\Gamma$
is given by
\begin{eqnarray}
&&
  \delta'_\Gamma  C^a = \beta \delta_{\rm B}  C^a ,
\\
&&
  \delta'_\Gamma \phi^i
  =
  \left\{{
     C^a
    \left({
      [ \hat{W}_a,\hat{F} ] + \beta \hat{W}_a
    }\right)
    + {\cal C}^j [\hat{\cal W}_j,\hat{F}]
  }\right\}\phi^i ,
\\
&&
  \delta'_\Gamma V_\mu^a
  =
  \alpha \partial_\mu  C^a
  + \beta \delta_{\rm B} V_\mu^a
  + \gamma \delta_{\rm B}
  \left({ V_\mu^a - \tilde{\cal V}_\mu^a }\right) ,
\end{eqnarray}
where $\alpha$, $\beta$ and $\gamma$ are constants,
\begin{equation}
\tilde{\cal V}_\mu \equiv
  \xi_{\rm L} {\cal L}_{\mu} \xi_{\rm L}^{\dag}
  -i \partial_\mu\xi_{\rm L} \cdot \xi_{\rm L}^{\dag}
  + \xi_{\rm R} {\cal R}_{\mu} \xi_{\rm R}^{\dag}
  -i \partial_\mu\xi_{\rm R} \cdot \xi_{\rm R}^{\dag}
\ ,
\end{equation}
and
$\hat{F} \equiv F^i(\phi) \partial/\partial\phi^i$,
with $F^i(\phi)$ being a certain dimension-0 function.
Note that $\delta'_\Gamma{\cal V}_\mu^i=\delta'_\Gamma{\cal C}^i=0$,
since the external $ G_{\rm global}$-gauge
fields ${\cal V}_\mu^i$ and
their ghosts ${\cal C}^i$ are not quantized
and hence their BRS source fields
${\cal K}_i^\mu$ and ${\cal L}_i$ appear only in the tree action.

Using $\delta'_\Gamma$ thus obtained,
we next solve the above WT identity (\ref{DimtwoWTB})
and easily find
\begin{equation}
  \Gamma_2^{(n)}
  =
  A_{2{\rm GI}} [\phi,\hbox{\boldmath $V$}]-
  \left({
    \hat{F} S_2 + \alpha V_\mu^a
    {\delta \over \delta V_\mu^a} S_2
  }\right),
\end{equation}
where $A_{2{\rm GI}}$ is a dimension-2
gauge-invariant function of $\phi^i$ and
$\hbox{\boldmath $V$}_\mu^A$.

The solutions
are combined into a simple form
\begin{equation}
  \Gamma_2^{(n)} +
  \left. \Gamma_4^{(n)} \right\vert_{{\bf K}\hbox{-}{\rm linear}}
  =
  A_{2{\rm GI}} [\phi,\hbox{\boldmath $V$}] - S \ast Y
\label{Solution}
\end{equation}
up to irrelevant terms
(dimension-6 or ${\bf K}$-independent dimension-4 terms),
where
the functional $Y$ is given by
\begin{eqnarray}
  Y
&=&
  \int d^4x
  \bigl[
    K_i F^{i} (\phi)
    + \alpha K_a^\mu V_\mu^a
    + \beta L_a  C^a
    + \gamma f_{abc} K_a^\mu K_{b\mu}  C^c
  \bigr] \ .
\label{SolY}
\end{eqnarray}

Now, we prove our statements \ref{induction:2} and \ref{induction:3}
in the above.
We have calculated the above effective action $\Gamma^{(n)}$
{\it without} using $n$-th loop level counter terms
$\delta\Phi^{(n)}$, $\delta{\bf K}^{(n)}$ and
$\delta\hbox{\boldmath $a$}^{(n)}$.
If we include those,
we have the additional contribution given by
\begin{equation}
  \Delta \Gamma^{(n)}
  =
  \delta\Phi^{(n)} {\delta S \over \delta \Phi}
  + \delta{\bf K}^{(n)} {\delta S \over \delta {\bf K}}
  + \delta\hbox{\boldmath $a$}^{(n)}
    {\partial S \over \partial \hbox{\boldmath $a$}} ,
\label{counter}
\end{equation}
where $S[\Phi,{\bf K};\hbox{\boldmath $a$}]$
is the tree-level action.
So the true $n$-th loop level effective action is given by
\begin{equation}
\Gamma^{(n)}+\Delta\Gamma^{(n)}
\equiv\Gamma^{(n)}_{{\rm total}} \ .
\end{equation}
The tree-level action $S_2$
is the most general gauge-invariant dimension-2 term,
so that
$A_{2{\rm GI}}[\phi,\hbox{\boldmath $V$}]$ term
in Eq.~(\ref{Solution})
can be canceled
by suitably chosen counter terms,
$\delta\hbox{\boldmath $a$}^{(n)} 
\frac{\partial S}{\partial \hbox{\boldmath $a$}}$.
The
second term
$-S\ast Y$ term in Eq.~(\ref{Solution})
just represents
a ``canonical transformation'' of $S$
generated by $-Y$.
Therefore we choose
the $n$-th
order field counter terms $\delta\Phi^{(n)}$
and $\delta{\bf K}^{(n)}$ to be equal to the canonical
transformations of $\Phi$ and ${\bf K}$ generated by $+Y$;
\begin{equation}
\delta\Phi^{(n)} = \Phi \ast Y \ , \quad
\delta{\bf K}^{(n)} = {\bf K} \ast Y \ .
\label{field renorm LET}
\end{equation}
Then
the first and the second terms in Eq.~(\ref{counter})
just give $S \ast Y$
and precisely cancel
the second term in Eq.~(\ref{Solution}).
Thus we have completed the proof of our statements 
\ref{induction:2} and \ref{induction:3}.

\subsection{Low energy theorem of the HLS}
\label{ssec:LET}

In the previous subsections of this section, 
we have shown in the covariant gauges
that our tree-level dimension-2 action
$\int d^4 x ({\cal L}_{\rm A} + a {\cal L}_{\rm V} ) $,
if written in terms of renormalized parameters and fields,
already gives the exact action $\Gamma_2$
including all the loop effects. This form of the effective action
(in particular the ${\cal L}_{\rm V}$ part) implies that
the previously derived
relation~\cite{BKY:NPB,BKY:PTP}
\begin{equation}
\frac{ g_V (p^2) }{ 
g_{V\pi\pi}(p^2, p_{\pi_1}^2\!\!=\! p_{\pi_2}^2\!\!=\!0)}
\bigg\vert_{p^2=0} = 2F_\pi^2
\label{eqLET}
\end{equation}
{\it is} actually an exact low energy theorem valid at any loop
order. Of course, this theorem concerns off-shell quantities at
$p^2=0$, and hence is not physical as it stands.
However, as discussed in Sec.~\ref{sec:CPHLS}
(see also Refs.~\cite{Georgi:1,Georgi:2}), we can perform the
systematic low energy expansion in the HLS
when the vector meson can be regarded as light.
We expect that the on-shell value of
$g_V/g_{V\pi\pi}$ at $p^2=m_V^2$ can deviate
from the LHS of Eq.~(\ref{eqLET}) only by a quantity of order
$m_\rho^2/\Lambda_\chi^2$, 
since the contributions of the
dimension-4 or higher terms in the effective
action $\Gamma$ (again representing all the loop effects) are
suppressed by a factor of $p^2/\Lambda_\chi^2$ at least.
Therefore as far as the vector mass is light,
our theorem is truly a physical one. 
In the actual world of QCD, the $\rho$ meson mass
is not so light ($m_\rho^2/\Lambda_\chi^2 \sim 0.5$) 
so that the situation
becomes a bit obscure. Nevertheless,
the fact that the KSRF (I) relation
$g_{\rho}/g_{\rho\pi\pi}=2F_\pi^2$
holds on the $\rho$ mass shell with good accuracy
strongly suggests that the $\rho$ meson is
the hidden gauge field
and {\it the KSRF (I) relation is a physical manifestation of our
low energy theorem}.

Our conclusion in this section remains unaltered
even if the action $S$ contains other dimension-4 or higher terms,
as far as they respect the symmetry.
This is because we needed just
$\left({S \ast \Gamma}\right)_2$
and $\left({S \ast \Gamma}\right)_4 %
\vert_{{\bf K}\hbox{-}{\rm linear}}$
parts in the WT identity
to which only
$S_2$ and ${\bf K}\hbox{-}{\rm linear}$ part
of $S_4$ can contribute.

When we regard this HLS model
as a low energy effective field theory of QCD,
we must
take account of
the anomaly
and the corresponding
Wess-Zumino-Witten term $\Gamma_{\rm WZW}$.
The WT identity
now reads
$\Gamma\ast\Gamma=({\rm anomaly})$.
However,
the RHS
is saturated already at the tree level
in this effective Lagrangian
and
so the WT identity
at loop levels,
which we need,
remains the same as before.
The WZW term $\Gamma_{\rm WZW}$ or
any other intrinsic-parity-odd terms~\cite{FKTUY} in $S$
are of dimension-4 or higher and hence
do not change our conclusion as explained above.

Since the low energy theorem concerns off-shell quantities,
we should comment on the gauge choice.
In the covariant gauges which we adopted here,
the $ G_{\rm global}$ and $ H_{\rm local}$ BRS symmetries
are separately preserved.
Accordingly,
the $V_\mu$ field is multiplicatively
renormalized (recall that
$\delta V_\mu^{(n)} = V_\mu \ast Y = \alpha V_\mu$),
and the above (off-shell) low energy theorem (\ref{eqLET}) holds.
However, if we adopt $R_\xi$-gauges
(other than Landau gauge),
these properties are violated;
for instance, $\phi\partial_\mu \phi$ or the external gauge field
${\cal V}_\mu$ gets mixed with our $V_\mu$
through the renormalization,
and our off-shell low energy theorem (\ref{eqLET}) is violated.
This implies that the $V_\mu$ field in the $R_\xi$ gauge
generally does not give a smooth off-shell extrapolation;
indeed, in $R_\xi$ gauge with gauge parameter
$\alpha \equiv 1/\xi$, the correction
to $g_{\rho}/g_{\rho\pi\pi}$ by
the extrapolation from $p^2=m_\rho^2$ to $p^2=0$ is seen
to have a part proportional to $\alpha g^2/16\pi^2$, which diverges
when $\alpha$ becomes very large.
Thus, in particular, the unitary gauge 
[see Sec.~\ref{ssec:LOL}],
which corresponds to
$\alpha \rightarrow \infty$,
gives an ill-defined off-shell field.

Our argument is free from infrared divergences
at least in Landau gauge.
This can be seen as follows.
In this gauge
the propagators of the NG bosons, the hidden gauge bosons
and the FP ghosts
(after rescaling
the FP anti-ghost
$\bar{C}$ into $f_\pi^2\bar{C}$)
are all proportional to $1/f_\pi^2$ in the infrared region.
Therefore, a general $L$-loop diagram,
which includes $V_4$ dimension-4 vertices
and $K$ BRS source vertices,
yields an amplitude proportional to
$\left({1/f_\pi^2}\right)^{(L-1+V_4+K)}$\cite{Wei:79}.
Thus, from dimensional consideration
we see that there is no infrared contribution to
$\Gamma_0^{(n)}[\phi]$,
$\Gamma_2^{(n)}[\phi,\hbox{\boldmath $V$}]$ and
$\Gamma_4^{(n)}[\Phi,{\bf K}]\vert_{{\bf K}\hbox{-}{\rm linear}}$.
In other covariant gauges,
there appears a dipole ghost
in the vector propagator,
which is to be defined by a suitable regularization.

\newpage

\section{Towards Hot and/or Dense Matter Calculation}
\label{sec:THDMC}

In this section we consider an application of the approach introduced
in this report to the hot and/or dense matter calculation.

In hot and/or dense matter,
the chiral symmetry is expected to be restored
(for reviews, see, e.g., 
Refs.~\cite{Hatsuda-Kunihiro:94,Pisarski:95,Brown-Rho:96,%
Hatsuda-Shiomi-Kuwabara:96,Wilczek,%
Rapp-Wambach:00,Brown-Rho:01b}).
The BNL Relativistic Heavy Ion Collider (RHIC) has started to measure
the effects in hot and/or dense matter.
One of the interesting quantities in hot and/or dense matter is the
change of $\rho$-meson mass.
In Refs.~\cite{Brown-Rho:91,Brown-Rho:96} it was proposed that the
$\rho$-meson mass 
scales like the pion decay constant in hot and/or dense matter, and
vanishes at the chiral phase transition point.

The vector manifestation (VM) reviewed in Sec.~\ref{sec:VM}
is a general property in the chiral symmetry restoration
when the HLS can be matched with the underlying QCD at the critical
point.
In Ref.~\cite{HY:VM} the application of the VM to the large $N_f$
chiral restoration was done.
It was then 
suggested~\cite{HY:VM,Brown-Rho:01a,Brown-Rho:01b}
that the VM can be applied to the chiral restoration 
in hot and/or dense matter.
Recently, it was shown the the VM actually occurs in hot matter at
zero density~\cite{Harada-Sasaki} and also in dense matter at zero
temperature~\cite{Harada-Kim-Rho}.
The purpose of this section is to give an outline of the application
of the chiral perturbation, the Wilsonian matching and the VM of the
HLS to the hot and/or dense matter calculation based on these works.

We first consider the hot matter calcuation at zero density.
In the low temperature region,
the temperature dependence of the physical quantities
are expected to be dominated by the hadronic thermal effects.
Inclusion of the hadronic thermal corrections to the $\rho$-meson
mass 
within the framework of the HLS
has been done by several groups
(see, e.g.,
Refs.~\cite{Lee-Song-Yabu:95,Song-Koch:96,%
Harada-Shibata,Rapp-Wambach:00}).
However, most of them included only the thermal effect of pions and
dropped the thermal effects of the $\rho$ meson itself.
In Ref.~\cite{Harada-Shibata},
the first application of the systematic chiral perturbation
reviewed in the previous sections to the hot matter calculation
was made.
There
hadronic thermal effects were included,
based on the systematic chiral perturbation in the HLS,
by calculating the one-loop
corrections in hot matter in the Landau gauge.
We review the chiral perturbation of the HLS in hot matter
following Ref.~\cite{Harada-Shibata}
in Sec.~\ref{ssec:HTE}.
A part of 
the calculation in the background field gauge which we 
introduced in Sec.~\ref{sec:CPHLS}
is shown in Ref.~\cite{Harada-Kim-Rho-Sasaki},
and the complete version 
will be shown in 
Ref.~\cite{Harada-Sasaki:2}.

In Ref.~\cite{Harada-Sasaki},
an application of 
the Wilsonian matching explained in
Sec.~\ref{sec:WM} to hot matter calculation was done.
In Sec.~\ref{ssec:VMNT},
we briefly review the analysis.
The main result in Ref.~\cite{Harada-Sasaki} is that
{\it by imposing the Wilsonian matching
of the HLS with the underlying QCD at the critical temperature,
where the chiral symmetry restoration takes place,
the vector manifestation (VM) necessarily occurs:
The vector meson mass becomes zero}.
Accordingly,
the light vector meson gives a large thermal correction to the pion
decay constant,
and the value of the critical temperature 
becomes larger than the value estimated by 
including only the pion thermal effect.
The result that the vector meson becomes light near the critical
temperature is consistent with the picture shown in 
Refs.~\cite{Brown-Rho:91,Brown-Rho:96,Brown-Rho:01a,Brown-Rho:01b}.

In Sec.~\ref{ssec:Admc}
we briefly review an application of the Wilsonian matching and the VM
to dense matter calculation recently done in
Ref.~\cite{Harada-Kim-Rho}.
It was shown that the VM is
realized in dense matter at the chiral restoration with the $\rho$
mass $m_\rho$ going to zero at the crtical point.

To avoid confusion,
we use $f_\pi(T)$ [$f_\pi(\widetilde{\mu})$]~\footnote{%
  In Ref.~\cite{Harada-Kim-Rho} $\mu$ is used for expressing the
  chemical potential.
  Throughout this report, however, we use $\mu$ for the energy scale,
  and then we use $\widetilde{\mu}$ for the chemical potential.
}
for the physical decay constant of $\pi$
at non-zero temperature [density],
and $F_\pi$ for the parameters of the Lagrangian.
Similarly, 
$M_\rho$ denotes the parameter of the Lagrangian and
$m_\rho$ the $\rho$ pole mass.

\subsection{Hadronic thermal effects}
\label{ssec:HTE}

In this subsection 
we show the hadronic thermal corrections to the pion
decay constant and the vector meson mass
following Ref.~\cite{Harada-Shibata},
where the calculation was performed in the Landau gauge with the
ordinary quantization procedure.

In Ref.~\cite{Harada-Shibata}
the pion decay constant at non-zero temperature was defined through
the axialvector current correlator following the definition given in 
Ref.~\cite{Bochkarev-Kapusta}.
In Eq.~(\ref{A V correlators}),
two-point function of the axialvector current $J^{a}_{5\mu}$
is expressed by one tensor structure.
At non-zero temperature, however,
we can decompose this 
current correlator
into longitudinal and transverse pieces as
\begin{eqnarray}
i \int d^4x e^{i p x} 
\left\langle 0 \left\vert T\, J_{5\mu}^a (x) J_{5\nu}^b (0)
\right\vert \right\rangle_{T}
=
\delta _{ab}
\left[
  P_{T}^{\mu \nu }G_{{\cal A}T}(p_{0},\vec{p};T)
  + P_{L}^{\mu \nu }G_{{\cal A}L}(p_{0},\vec{p};T)
\right]
\ ,
\end{eqnarray}
where polarization tensors $P_{L}$ and $P_{T}$ are defined in 
Eq.~(\ref {polar tensor}) in Appendix~\ref{ssec:polar}.
It is natural to define the pion decay constant at non-zero
temperature through the longitudinal component in the low energy limit%
\footnote{%
  Even when we use the transverse part instead of the longitudinal
  part to 
  define $f_\pi(T)$ in Eq.~(\ref{def: fpi}), 
  we obtain the same result: 
  $G_{{\cal A}T}(p_0,\vec{p}=0) = G_{{\cal A}L}(p_0,\vec{p}=0)$.%
}:~\cite{Bochkarev-Kapusta} 
\begin{equation}
f_{\pi}^{2}(T)
\equiv - \lim_{p_{0}\rightarrow 0}
G_{{\cal A}L}(p_{0},\vec{p}=0;T)
\ .
\label{def: fpi}
\end{equation}
There are two types of contributions to this Green function in the
HLS:
(i) the pion exchange diagrams, and 
(ii) the contact or one-particle irreducible (1PI) diagrams. 
The contribution (i) is proportional to 
$p_{\mu }$ or $p_{\nu }$ at one loop. 
At most only one of
the $J^a_{5\mu}$-$\pi$
coupling can be corrected at one loop, which is not generally
proportional to the four-momentum $p_{\mu }$.
The other coupling is the tree-level one and proportional to $p_\mu$. 
When we act with the projection operator $P_{L\mu \nu}$, 
the term proportional to $p_\mu$ vanishes. 
Because of the current
conservation we have the same kinds of contributions from 1PI
diagrams: 
Those are roughly proportional to $g_{\mu \nu }$ instead of $p_\mu$.
Then we calculate only the 1PI diagrams.

There exist three 1PI diagrams which contribute to 
$G_{{\cal A}L}$ 
at one-loop level: 
(a) $\pi $+$\rho $ loop, 
(b) $\pi $+$\sigma $ loop, (c) $\pi$ tad-pole,
which are the same diagrams as those shown in Fig.~\ref{fig:aa}
with replacing $\overline{\cal A}_\mu$ with $J_{5\mu}$.
[The Feynman rules for the propagators and
the vertices in the Landau gauge are given in 
Appendix~\ref{app:FRLG}.]
These diagrams include ultraviolet divergences, 
which are renormalized by the parameters and fields.
By taking a suitable subtraction scheme
at zero temperature, 
all the divergences including finite corrections in the
low energy limit are absorbed into the redefinitions of the parameters
and fields~\cite{HY,HKY:PRL,HKY:PTP}.
Then the loop diagrams generate only the temperature-dependent part.
By using standard imaginary time formalism~\cite{Matsubara} 
we obtain 
\begin{eqnarray}
  G_{{\cal A}L}^{(a)}(p_{0},\vec{p} =0)
&=&
  \frac{N_f}{2}\frac{a}{\pi ^{2}}
  \left[ 
    \frac{5}{6} I_2(T) - J_1^2(M_\rho;T) 
    + \frac{1}{3M_\rho^2} 
      \left\{ I_4(T) - J_1^4(M_\rho;T) \right\}
  \right] 
\ ,
\nonumber \\
  G_{{\cal A}L}^{(b)}(p_0,\vec{p} =0)
&=&
  \frac{N_f}{2}\frac{a}{6\pi^2}I_2(T)
\ ,
\nonumber\\
  G_{{\cal A}L}^{(c)}(p_0,\vec{p} =0)
&=&
  \frac{N_f}{2}\frac{1-a}{\pi^2}I_2 (T)
\ , 
\label{eq:fpiT}
\end{eqnarray}
where the functions 
$I_n (T)$ and $J_m^n (M_\rho;T)$ are defined in 
Appendix~\ref{ssec:FUNZT}. 
The total contribution is given by
\begin{equation}
f_\pi^2(T)=
F_\pi^2 - \frac{N_f}{2\pi^2} 
\left[
  I_2 (T) - a J_1^2 (M_\rho;T)
  + \frac{a}{3M_\rho^2} \left\{ I_4 (T) - J_1^4(M_\rho;T) \right\}
\right] 
\ .
\label{eq: fpi}
\end{equation}
When we consider the low temperature region $T \ll M_\rho$ 
in the above expression, only the $I_2(T)$ term remains:
\begin{equation}
f_\pi^2(T) \approx
F_\pi^2 - \frac{N_f}{2\pi^2} I_2(T)
= F_\pi^2 - \frac{N_f}{12} T^2
\ .
\label{fpi: ChPT}
\end{equation}
which is consistent with the result given by 
Gasser-Leutwyler~\cite{Gasser-Leutwyler:87}.
Thus, {\it the pion decay constant decreases as $T^2$ dominated by the
effect of the thermal pseudoscalar mesons in the low temperature
region.}
We should note that
when we quantize only the $\pi$ field,
only the diagram (c) in Fig.~\ref{fig:aa} contributes
and the resultant temperature dependence does not agree with the 
result by Gasser-Leutwyler.
The above agreement is obtained from the fact that
each diagram in Fig.~\ref{fig:aa}
does generate the dominant contribution 
$I_2(T) = (\pi^2/6)T^2$, and the terms
proportional to 
$a I_2 (T)$ are completely canceled among three diagrams.
This cancellation is naturally understood as follows:
The term proportional to $a I_2 (T)$ in 
$G_{{\cal A}L}^{(c)}(p_0,\vec{p} =0)$ in Eq.~(\ref{eq:fpiT})
comes from the ${\cal A}$-${\cal A}$-$\pi$-$\pi$ vertex obtained from
$a {\cal L}_{\rm V}$ term in the Lagrangian
in Eq.~(\ref{leading Lagrangian}),
while the other term from the vertex from ${\cal L}_{\rm A}$ term.
The vertices in the diagrams (a) and (b) in Fig.~\ref{fig:aa}
come from $a {\cal L}_{\rm V}$ term.
Then the above cancellation implies that the $a {\cal L}_{\rm V}$ term
does not generate the thermal effect proportional to $T^2$ which is
dominant in the low temperature region.
The cancellation is similar to that occurred in the $\pi\pi$ scattering:
As was shown in Sec.~\ref{ssec:PPLO}, 
the $a {\cal L}_{\rm V}$ term
generates the extra contact 4-$\pi$ interaction of order 
${\cal O}(p^2)$, which appears to violate the low energy theorem of
the $\pi\pi$ scattering amplitude.
However, the $a {\cal L}_{\rm V}$ term also generates the
$\rho$-exchange contribution of order ${\cal O}(p^2)$,
which exactly cancels the contribution from
the extra contact 4-$\pi$ interaction in the low energy region,
$E\ll m_\rho$.
Thus, the $a {\cal L}_{\rm V}$ term does not generate the 
contribution of order ${\cal O}(p^2)$.
The similar cancellation occurs when the temperature is small
enough compared with the $\rho$ meson mass, $T \ll m_\rho$.
As a result,
the hadronic thermal effects is dominated by the contribution
from ${\cal L}_{\rm A}$ term in the low temperature region, thus we
obtained the result consistent with the 
``low temperature theorem''~\cite{Gasser-Leutwyler:87}.

Let us estimate the critical temperature
by naively extrapolating the above results to the higher temperature
region.
{}From Eq.~(\ref{fpi: ChPT}) the critical temperature is well
approximated as
\begin{equation}
T_c^{\rm(had)} \approx \sqrt{ \frac{12}{N_f} } F_\pi(0) \ ,
\label{Tc had 0}
\end{equation}
where $F_\pi(0)$ is the decay constant of $\pi$ at $T=0$.
In Ref.~\cite{Harada-Shibata} the number of light flavors is chosen to
be two, but here for later convenience we fix $N_f=3$.
Then the critical temperature is given by
\begin{equation}
T_c^{\rm(had)} \approx 2 F_\pi(0) \simeq 180\,\mbox{MeV} \ .
\label{Tc had}
\end{equation}
It should be noticed that
the above value of the critical temperature is changed only slightly
even when we include the full effect given in 
Eq.~(\ref{eq: fpi}) as shown in Ref.~\cite{Harada-Shibata},
as far as the vector meson mass is heavy enough:
$T_c^{\rm(had)} \ll M_\rho$.

Next, let us study the corrections from the hadronic thermal effect to
the $\rho$ mass. 
As was shown in Appendix~C of Ref.~\cite{Harada-Shibata},
the $\rho$ and $\sigma$ propagators are separated from each other 
in the Landau gauge, 
and the $\rho$ propagator takes simple form:
\begin{equation}
- i D_{\mu \nu }=
- \frac{P_{T\mu \nu }}{p^{2}-M_{\rho }^{2}+ \Pi_V^T }
-\frac{P_{L\mu \nu }}{p^{2}-M_{\rho }^{2}+\Pi_V^L }
\ .
\label{rho propagator at T}
\end{equation}
It is reasonable to define the 
$\rho$ pole mass by using the longitudinal 
part in the low momentum limit, $\vec{p}=0$:~\footnote{%
  It should be noticed that the transverse polarization agrees with
  the longitudinal one in the low momentum limit: 
  $\Pi_V^T(p_0,\vec{p}=0;T) =
  \Pi_V^L(p_0,\vec{p}=0;T)$.
}
\begin{equation}
m_\rho^2(T) = M_\rho^2 - 
\mbox{Re}\,\Pi_V^L( p_0=M_\rho , \vec{p}=0;T)
\ ,
\end{equation}
where $\mbox{Re}\,\Pi_V^L$ denotes the real
part of $\Pi_V^L$ and
inside the one-loop correction $\Pi_V^L$ we replaced
$m_\rho$ with $M_\rho$, since the difference is of higher order.

\begin{figure}[htbp]
\begin{center}
\ \epsfbox{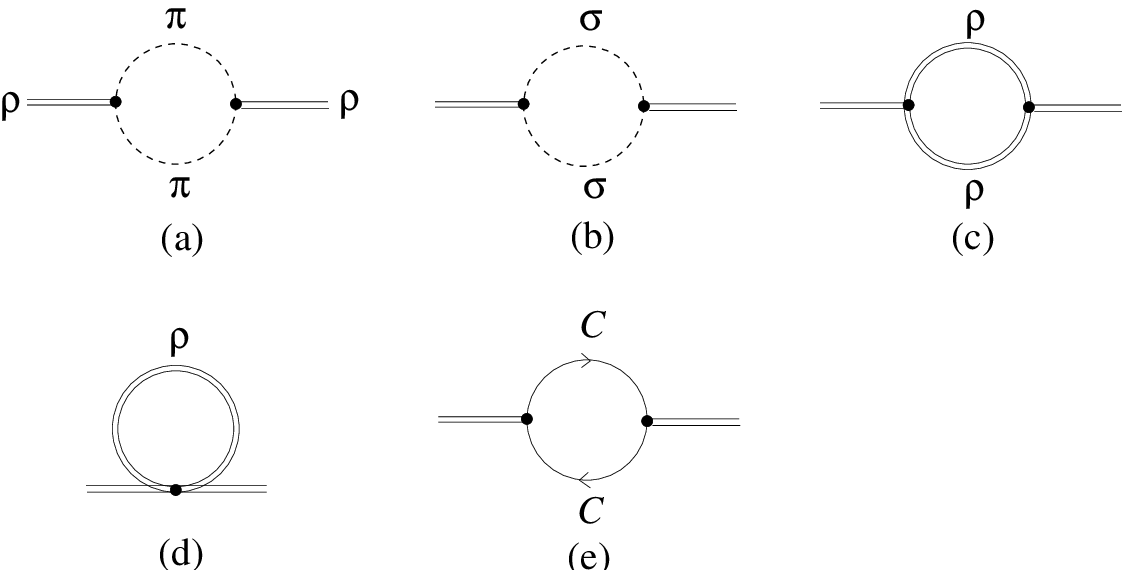}
\end{center}
\caption[Corrections to the vector meson propagator 
in the Landau gauge]{Feynman diagrams contributing to the 
vector meson self-energy in the Landau gauge: a) $\pi$ loop, 
b) $\sigma$ loop, c) $\rho$ loop, 
d) $\rho$ tad-pole and e) ghost loop. }
\label{fig: rho propagator}
\end{figure}
The one-loop diagrams contributing to $\rho$ self-energy 
in the Landau gauge are shown in
Fig.~\ref{fig: rho propagator}. 
Feynman rules for the vertices 
are shown in Appendix~\ref{app:FRLG}.
In Ref.~\cite{Harada-Shibata},
the divergences are renormalized in the on-shell renormalization
scheme and 
the thermal corrections to the vector-meson two
point function from the pseudoscalar and vector mesons 
are calculated.
Thus, the parameter $M_\rho$ in this section is renormalized in a way
that it becomes the pole mass at $T=0$.
Since the calculation was done in the Landau gauge,
the off-shell structure of the propagator is not gauge invariant,
while the result on mass-shell of vector meson is of course gauge
invariant.
Thus, here we show the thermal effects to the
on-shell self-energy 
$\Delta \Pi (p_{0}=M_\rho,\vec{p};T)
\equiv \Pi (p_{0}=M_\rho,\vec{p};T)-\Pi (p_{0}=M_\rho,\vec{p};T=0)$,
in the low-momentum limit ($\vec{p}=0$)
by using the standard imaginary time formalism~\cite{Matsubara}:
\begin{eqnarray}
&& 
  \mbox{\rm Re}\,\Delta 
  \Pi _{V}^{L (a)}(p_{0}=M_\rho,\vec{p}=0;T)
  =\frac{N_f}{2}\frac{g^{2}}{\pi ^{2}}\frac{a^{2}}{12}
  G_2 (M_\rho;T)
\ ,
\nonumber\\
&&
  \mbox{\rm Re}\,\Delta 
  \Pi _{V}^{L (b)}(p_{0}=M_\rho,\vec{p}=0;T)
  =\frac{N_f}{2}\frac{g^{2}}{\pi ^{2}}\frac{1}{12}
  G_2 (M_\rho;T)
\ ,
\nonumber\\
&&
  \mbox{\rm Re}\,\Delta 
  \Pi_V^{L (c)}(p_{0}=M_\rho,\vec{p}=0;T)
\nonumber\\
&&\qquad 
  =
  \frac{N_f}{2}\frac{g^{2}}{\pi ^{2}}
  \Biggl[ 
    - \frac{3}{2} F_3^2(M_\rho;M_\rho;T)
    + \frac{1}{2} F_3^4(M_\rho;M_\rho;T)
    - \frac{1}{3M_\rho^2} F_3^6(M_\rho;M_\rho;T)
\nonumber\\
&&\qquad\qquad\qquad
    {} - \frac{4}{3} K_6(M_\rho;M_\rho;T)
    + \frac{1}{12} G_2(M_\rho;T)
  \Biggr]
\nonumber\\
&&
  \mbox{\rm Re}\,\Delta 
  \Pi_V^{L (d)}(p_{0}=M_\rho,\vec{p}=0;T)
  =\frac{N_f}{2}\frac{g^{2}}{\pi ^{2}}
  \left[ 
    -2J_1^2(M_\rho;T)
    - \frac{1}{3M_\rho^2}
      \left( I_4(T) - J_1^4(M_\rho;T) \right) 
  \right]
\ ,
\nonumber\\
&&
  \mbox{\rm Re}\,\Delta 
  \Pi_V^{L (e)}(p_{0}=M_\rho,\vec{p}=0;T)
  = - \frac{N_f}{2}\frac{g^2}{\pi^2}
  \frac{1}{6} G_2(M_\rho;T) 
\ ,
\label{func: T dep}
\end{eqnarray}
where functions 
$F$, $G$, $H$, $I$, $J$ and $K$ are defined in 
Appendix~\ref{ssec:FUNZT}.
Since the on-shell renormalization scheme implies that
$M_\rho^2 + \mbox{Re} \Pi (p_{0}=M_\rho,\vec{p}=0;T=0) = M_\rho^2 $,
the sum of the above contributions is the thermal correction to the
pole mass of $\rho$ meson.
By noting that
\begin{eqnarray}
&&
  - \frac{1}{3M_\rho^2} F_3^{n+2}(M_\rho;M_\rho;T)
  = \frac{1}{4} F_3^n(M_\rho;M_\rho;T)
    - \frac{1}{3M_\rho^2} J_1^n(M_\rho;T)
\ ,
\nonumber\\
&&
  K_6(M_\rho;M_\rho;T) = - \frac{1}{4M_\rho^2} I_4(T)
\ ,
\end{eqnarray}
the thermal corrections to the vector meson pole mass is summarized
as~\footnote{%
  It should be stressed that this result is intact even when use the
  background field gauge~\cite{Harada-Sasaki:2}.
}
\begin{eqnarray}
&& m_\rho^2(T)
  = M_\rho^2 +
  \frac{N_f g^2}{2\pi^2}
  \Biggl[
    - \frac{a^2}{12} G_2(M_\rho;T)
    + \frac{5}{4} J_1^2(M_\rho;T) 
    {}+ \frac{33}{16} M_\rho^2 \, F_3^2(M_\rho;M_\rho;T)
  \Biggr]
\ .
\label{mrho at T}
\end{eqnarray}

Let us consider the low temperature region $T\ll M_\rho$.
The functions $F$ and $J$ are suppressed by 
$e^{-M_\rho/T}$, and give negligible contributions.
Noting that 
$G_2(M_\rho;T)\approx -\frac{\pi^4}{15}\frac{T^4}{M_\rho^2}$
for $T\ll M_\rho$,
the $\rho$  pole mass becomes
\begin{equation}
m_\rho^2(T) \approx M_\rho^2
- \frac{N_f g^2}{2\pi^2} \frac{a^2}{12} G_2(M_\rho;T)
\approx M_\rho^2 + \frac{N_f \pi^2 a }{360 F_\pi^2} T^4
\ .
\end{equation}
Thus, 
{\it the vector meson pole mass increases as $T^4$ at low
temperature dominated by pion-loop effect.}
The lack of $T^2$-term is consistent with the result by the
current algebra analysis~\cite{Dey-Eletsky-Ioffe}.

\subsection{Vector manifestation at non-zero temperature}
\label{ssec:VMNT}

In the analysis done in Ref.~\cite{Harada-Shibata},
the parameters $F_\pi$, $a$ and $g$ 
were assumed to have no temperature dependences, 
and the values at $T=0$ were used.
When we naively extrapolate the results in the previous subsection
to the critical temperature,
the resultant axialvector and vector current correlators do not agree
with each other.
Disagreement between the axialvector and vector current correlators
is obviously inconsistent with the chiral symmetry restoration in
QCD.
However, the parameters of the HLS Lagrangian
should be determined by the underlying QCD.
As we explained in Sec.~\ref{sec:WM},
{\it the bare parameters of the (bare) HLS Lagrangian are determined
by matching the HLS with the underlying QCD}
at the matching scale $\Lambda$
through the Wilsonian matching conditions.
Since the quark condensate $\left\langle \bar{q} q \right\rangle$
as well as the gluonic condensate
$\left\langle \frac{\alpha_s}{\pi} G_{\mu\nu} G^{\mu\nu}
\right\rangle$
in the right-hand-side of the Wilsonian matching conditions
(\ref{match z}), (\ref{match A}) and (\ref{match V})
generally depends on the temperature, the application of the Wilsonian
matching to the hot matter calculation implies that
the bare parameters of the HLS 
(and hence $M_\rho^2 = a g^2 F_\pi^2$)
do depend on the temperature which are called
the 
{\it intrinsic temperature dependences}~\cite{Harada-Sasaki}
in contrast to the hadronic thermal effects.
As is stressed in Ref.~\cite{Harada-Sasaki},
the above disagreement
is cured by including the intrinsic temperature dependences
of the parameters through the Wilsonian matching conditions.
In this subsection we briefly review the analysis done
in Ref.~\cite{Harada-Sasaki}.

The intrinsic temperature dependences of the bare parameters lead to
those of the on-shell parameters used in the analysis in the previous
subsection through the Wilsonian RGE's.
We write these intrinsic temperature dependences of the on-shell
parameters explicitly as~\footnote{%
  We note that $\mu$ in Eq.~(\ref{on-shell para T}) is the
  renormalization scale, not a chemical potential.
  In the next subsection where we consider the dense matter
  calculation, 
  we use $\widetilde{\mu}$ for expressing the chemical potential.
}
\begin{eqnarray}
&&
  F_\pi = F_\pi(\mu=0;T) \ ,
\nonumber\\
&&
  g = g\mbox{\boldmath$\bigl($} 
  \mu = M_\rho(T);T
  \mbox{\boldmath$\bigr)$}
\ ,
\nonumber\\
&&
  a = a\mbox{\boldmath$\bigl($} 
  \mu = M_\rho(T);T
  \mbox{\boldmath$\bigr)$}
\ ,
\label{on-shell para T}
\end{eqnarray}
where $M_\rho$ is determined from the on-shell condition
\begin{eqnarray}
  M_\rho = M_\rho(T) =
  a\mbox{\boldmath$\bigl($} 
  \mu = M_\rho(T);T
  \mbox{\boldmath$\bigr)$}
  g^2\mbox{\boldmath$\bigl($} 
  \mu = M_\rho(T);T
  \mbox{\boldmath$\bigr)$}
  F_\pi^2\mbox{\boldmath$\bigl($} 
  \mu = M_\rho(T);T
  \mbox{\boldmath$\bigr)$}
\ .
\end{eqnarray}
These intrinsic temperature dependences of the parameters give extra
temperature dependences to the physical quantities
which are not included by the hadronic thermal
effects calculated in the previous subsection.

Let us now apply the Wilsonian matching at the critical temperature
$T_c$ for $N_f = 3$.
Here we assume $\langle \bar{q} q \rangle$ approaches to $0$
continuously for $T \rightarrow T_c$.~\footnote{
  It is known that there is no Ginzburg-Landau type phase transition
  for $N_f=3$ (see, e.g., Refs.~\cite{Wilczek,Brown-Rho:96}).
  There may still be a possibility of non-Ginzburg-Landau type
  continuous phase transition such as the conformal phase
  transition~\cite{Miransky-Yamawaki}.
}
In such a case,
the axialvector and vector current correlators derived from the OPE 
given in Eqs.~(\ref{Pi A OPE}) and (\ref{Pi V OPE}) agree with each
other.
Then the Wilsonian matching requires that the 
axialvector and vector current correlators in the HLS given in 
Eqs.~(\ref{Pi A HLS}) and (\ref{Pi V HLS}) must agree with each
other. 
As we discussed in Sec.~\ref{sec:VM} for large $N_f$ chiral
restoration, 
this agreement is satisfied if the following conditions are 
met~\cite{Harada-Sasaki}:
\begin{eqnarray}
&&
g(\Lambda;T) \mathop{\longrightarrow}_{T \rightarrow T_c} 0 \ ,
\label{g:VMT}
\\
&&
a(\Lambda;T) \mathop{\longrightarrow}_{T \rightarrow T_c} 1 \ ,
\label{a:VMT}
\\
&&
z_1(\Lambda;T) - z_2(\Lambda;T) 
\mathop{\longrightarrow}_{T \rightarrow T_c} 0 \ .
\label{z12:VMT}
\end{eqnarray}

The conditions for the parameters at the matching scale
$g(\Lambda;T_c) =0$ and $a(\Lambda;T_c) = 1$ are converted into the
conditions for the on-shell parameters through the Wilsonian RGEs
in Eqs.~(\ref{RGE for g2}) and (\ref{RGE for a}).
Since $g=0$ and $a=1$ are separately the fixed points of the RGEs for
$g$ and $a$,
the on-shell parameters also satisfy
$(g,a)=(0,1)$, and thus $M_\rho = 0$.
Noting that
\begin{eqnarray}
&&
G_2(M_\rho;T) 
  \mathop{\longrightarrow}_{M_\rho\rightarrow 0} I_2(T) 
  = \frac{\pi^2}{6} T^2 
\ ,
\nonumber
\\
&&
J_1^2(M_\rho;T) 
  \mathop{\longrightarrow}_{M_\rho\rightarrow 0} I_2(T) 
  = \frac{\pi^2}{6} T^2 
\ ,
\nonumber
\\
&&
M_\rho^2 F_3^2(M_\rho;M_\rho;T)
  \mathop{\longrightarrow}_{M_\rho\rightarrow 0} 0 \ ,
\end{eqnarray}
Eq.~(\ref{mrho at T}) 
in the limit $M_\rho \ll T$ 
reduces to
\begin{eqnarray}
&& m_\rho^2(T)
  = M_\rho^2 +
  g^2 \frac{N_f }{2\pi^2} \frac{15 - a^2}{12} I_2(T)
\ .
\label{mrho at T 2}
\end{eqnarray}
Since $a \simeq 1$ near the restoration point,
the second term is positive. 
Then the $\rho$ pole mass $m_\rho$
is bigger than the parameter
$M_\rho$ due to the hadronic thermal corrections.
Nevertheless, 
{\it the intrinsic temperature dependence determined by the
Wilsonian matching requires
that the vector meson becomes massless at the
critical temperature}:
\begin{equation}
m_\rho^2(T)
\mathop{\longrightarrow}_{T \rightarrow T_c} 0 \ ,
\end{equation}
since the first term vanishes as $M_\rho\rightarrow 0$, and the second
term also vanishes since $g\rightarrow 0$ for $T \rightarrow T_c$.
This implies that, as was suggested in 
Refs.~\cite{HY:VM,Brown-Rho:01a,Brown-Rho:01b},
{\it the vector manifestation (VM) actually
occurs at the critical
temperature}~\cite{Harada-Sasaki}.
This is consistent with the picture shown in
Refs.~\cite{Brown-Rho:91,Brown-Rho:96,Brown-Rho:01a,Brown-Rho:01b}.
We should stress here that the above $m_\rho(T)$ is the pole mass of
$\rho$ meson, which is important
for analysing the dilepton spectra in RHIC experiment.
It is noted~\cite{HY:VM} that although conditions for $g(\Lambda;T)$
and $a(\Lambda;T)$ in Eqs.~(\ref{g:VMT}) and (\ref{a:VMT}) coincide
with the Georgi's vector limit~\cite{Georgi:1,Georgi:2}, 
the VM here should be distinguished from Georgi's vector
realization~\cite{Georgi:1,Georgi:2}.

Let us determine the critical temperature.
For $T > 0$ 
the thermal averages of the Lorentz non-scalar operators such as 
$\bar{q} \gamma_\mu D_\nu q$ exist
in the current correlators in the OPE~\cite{Hatsuda-Koike-Lee}.
Since these contributions are small compared with the main term 
$1 + \alpha_s/\pi$,
we expect that they give only
small corrections to the value of the critical temperature,
and neglect them here.
Then, the Wilsonian matching condition to determine the bare parameter 
$F_\pi(\Lambda;T_c)$ is obtained from that in Eq.~(\ref{match A})
by taking $\langle \bar{q} q \rangle=0$ and
including a possible temperature dependence of the gluonic condensate:
\begin{eqnarray}
&&
\frac{F_\pi^2(\Lambda;T_c)}{\Lambda^2} 
= \frac{1}{8\pi^2}
\left[
  1 + \frac{\alpha_s}{\pi}
  + \frac{2\pi^2}{3} 
    \frac{
      \left\langle 
        \frac{\alpha_s}{\pi} G_{\mu\nu} G^{\mu\nu}
      \right\rangle_{T_c}
    }{ \Lambda^4 }
\right]
\ ,
\label{fp Tc WM}
\end{eqnarray}
which determines the on-shell parameter $F_\pi(\mu=0;T_c)$ through the
Wilsonian RGE for $F_\pi$ in Eq.~(\ref{RGE for Fpi2}) with taking
$(g,a)=(0,1)$.
It should be noticed that 
the $F_\pi(\mu;T)$ does run with scale $\mu$ 
by the Wilsonian RGE~\cite{HY:letter,HY:matching} 
even at the critical point.
As we obtained for large $N_f$ in 
Eq.~(\ref{RGE for fpi2 at vector limit}),
the relation between 
$F_\pi(\Lambda;T_c)$ and $F_\pi(\mu=0;T_c)$
is given by
\begin{equation}
\frac{F_\pi^2(0;T_c)}{\Lambda^2} = 
\frac{F_\pi^2\left(\Lambda;T_c\right)}{\Lambda^2}
- \frac{N_f}{2(4\pi)^2} \ .
\label{RGE for fpi2 at vector limit at Tc}
\end{equation}
On the other hand,
the relation between $F_\pi(0;T_c)$ and 
the physical pion decay constant,
which of course vanishes at $T=T_c$,
is given by taking $M_\rho = 0$ and
$a=1$ in Eq.~(\ref{eq: fpi})~\cite{Harada-Sasaki}:
\begin{equation}
0 = f_\pi^2(T_c) =
F_\pi^2(0;T_c) - \frac{N_f}{4\pi^2} I_2(T_c)
= F_\pi^2(0;T_c) - \frac{N_f}{24} T_c^2
\ .
\label{Fp: WM}
\end{equation}
Here we should note that the coefficient of 
$I_2(T_c)$ in the second term
is a half of that in Eq.~(\ref{fpi: ChPT})
which is an approximate form for $T \ll M_\rho$ taken with assuming
that the vector meson does not become light.
The factor $1/2$ appears from 
the contribution of $\sigma$ which becomes the real
NG boson at the critical temperature due to the VM.
This situation is similar to that occurring in the coefficients of the
quadratic divergences in the solution of the RGE for $F_\pi$:
In Eq.~(\ref{sol fpi2 for chpt}) only the quadratic divergence from
the pion loop is included, while in 
Eq.~(\ref{RGE for fpi2 at vector limit}) that from the 
$\rho$ loop ($\sigma$-loop) is also included.
Then the extra factor $1/2$ appears in the second term of 
Eq.~(\ref{RGE for fpi2 at vector limit}) compared with that of
Eq.~(\ref{sol fpi2 for chpt}).
{}From Eq.~(\ref{Fp: WM}) together with Eqs.~(\ref{fp Tc WM})
and (\ref{RGE for fpi2 at vector limit at Tc})
the critical temperature is given by
\begin{eqnarray}
T_c = \sqrt{ \frac{24}{N_f} } F_\pi(0;T_c)
= 
\sqrt{ \frac{3 \Lambda^2}{N_f\pi^2} }
\left[
  1 + \frac{\alpha_s}{\pi}
  + \frac{2\pi^2}{3} 
    \frac{
      \left\langle 
        \frac{\alpha_s}{\pi} G_{\mu\nu} G^{\mu\nu}
      \right\rangle_{T_c}
    }{ \Lambda^4 }
  - \frac{N_f}{4}
\right]^{1/2}
\ .
\label{Tc VM}
\end{eqnarray}

Let us estimate the critical temperature for $N_f=3$.
The value of the gluonic condensate near phase transition point
becomes about half of that at $T=0$~\cite{Miller,Brown-Rho:01b},
so we use 
\begin{eqnarray}
&&
\left\langle \frac{\alpha_s}{\pi} G_{\mu\nu} G^{\mu\nu}
\right\rangle = 0.006 \,\mbox{GeV}^4 \ ,
\label{val GG 2}
\end{eqnarray}
obtained by multiplying
the value at $T=0$ shown in Refs.~\cite{SVZ:1,SVZ:2}
[see Eq.~(\ref{val GG})] by $1/2$.
For the value of the QCD scale
$\Lambda_{\rm QCD}$ we use 
\begin{eqnarray}
\Lambda_{\rm QCD} = 400 \, \mbox{MeV} \ ,
\end{eqnarray}
as a typical example.
For this value of $\Lambda_{\rm QCD}$, 
as we showed in Tables~\ref{tab:res} and \ref{tab:res 2}
in Sec.~\ref{ssec:RWM}, 
the choice of
$\Lambda\simeq1.1\,\mbox{GeV}$
for the values of the matching scale 
provides the predictions in good
agreement with experiment at $T=0$.
However, the matching scale may have the temperature dependence.
In the present analysis
we use 
\begin{eqnarray}
\Lambda = 0.8\,,\, 0.9\,,\, 1.0\ \mbox{and} \ 1.1 \,\mbox{GeV} \ , 
\end{eqnarray}
and determine the value of the critical temperature $T_c$
from Eq.~(\ref{Tc VM}).
We show the resultant values in Table~\ref{tab:Tc}.
\begin{table}[htbp]
\begin{center}
\begin{tabular}{|c||r|r|r|r|}
\hline
 $\Lambda$ &  $0.8$ &  $0.9$ &  $1.0$ &  $1.1$ \\
\hline
 $T_c$     & $0.21$ & $0.22$ & $0.23$ & $0.25$ \\
\hline
\end{tabular}
\end{center}
\caption[Predicted values of the critical temperature]{%
Estimated values of the critical temperature $T_c$ for
several choices of the value of the matching scale $\Lambda$
with $\Lambda_{\rm QCD} = 400\,\mbox{MeV}$.
Units of $\Lambda$ and $T_c$ are GeV.
}\label{tab:Tc}
\end{table}

We note that the estimated values of $T_c$ in Table~\ref{tab:Tc} are 
larger than that in Eq.~(\ref{Tc had})
which is obtained by naively
extrapolating the temperature dependence from the hadronic thermal
effects without including the intrinsic temperature dependences.
This is because the extra factor $1/2$ appears in the second term in
Eq.~(\ref{Fp: WM}) compared with that in 
Eq.~(\ref{fpi: ChPT}).
As we stressed below Eq.~(\ref{Fp: WM}), the factor $1/2$ comes from
the contribution of $\sigma$ (longitudinal $\rho$) which becomes
massless at the chiral restoration point.

The vector dominance in hot matter
and the dependences of the critical temperature 
on other parameter choices 
will be studied in Ref.~\cite{Harada-Sasaki:2}.

\subsection{Application to dense matter calculation}
\label{ssec:Admc}

In this subsection we briefly review the application of the Wilsonian
mathing and the vector manifestion (VM) to the dense matter
calculation done in Ref.~\cite{Harada-Kim-Rho}.

To set up the arguments for the density problem, we consider a
system of hadrons in the background of a filled Fermi sea. For the
moment, we consider the Fermi sea as merely a $background$,
side-stepping the question of how the Fermi sea is formed from a
theory defined in a matter-free vacuum. 
Imagine that mesons -- the
pion and the $\rho$ meson -- are introduced in HLS
with a cutoff set at the scale $\Lambda_\chi$. 
Since we are dealing with dense fermionic matter,
we may need to introduce the degrees of freedom associated with
baryons or alternatively constituent quarks (or quasiquarks)
into the HLS.
At low density, say, $n < \tilde{n}$, with $\tilde{n}$ being some
density greater than $n_0$, the precise value of which cannot be
pinned down at present, we may choose the cutoff $\Lambda_0$ below
the nucleon mass, $m_N\sim 1$ GeV, but above the $\rho$ mass
$m_\rho=770$ MeV and integrate out all the baryons. In this case,
the $bare$ parameters of the HLS Lagrangian will depend upon the
density $n$ (or equivalently Fermi momentum $P_F$) since the
baryons that are integrated out carry information about the baryon
density through their interactions in the full theory with the
baryons within the Fermi sea. 
Once the baryons are integrated out,
we will then be left with the standard HLS Lagrangian theory with
the NG and gauge boson fields only {\it except that the bare
parameters of the effective Lagrangian will be density-dependent}.
It should be noticed that {\it the cutoff can also be density
dependent}. However, in general, the density-dependence of the
cutoff is not related to those of the bare parameters by the RGEs.
For $T> 0$ and $n=0$ this difference appears from the
``intrinsic'' temperature dependence introduced in
Ref.~\cite{Harada-Sasaki} (see previous subsection)
which was essential for the VM to occur at the
chiral restoration point.

As density increases beyond $\tilde{n}$, the fermions may however
start figuring explicitly, that is, the fermion field may be
present below the cutoff $\tilde{\Lambda}$ ($n>\tilde{n}$).
The reason is that as density approaches the chiral restoration point,
the constituent-quark (called quasiquark) picture -- which seems
to be viable even in matter-free space~\cite{Szczepaniak-Swanson} 
-- becomes
more appropriate~\cite{Brown-Rho:01b} and the quasiquark mass drops
rapidly, ultimately vanishing (in the chiral limit) at the
critical point. This picture has been advocated by several authors
in a related context~\cite{Riska-Brown}.

To study the effects of the quasiquark near the critical density
in Ref.~\cite{Harada-Kim-Rho} 
the HLS with the quasiquark
was adopted.
There a systematic counting scheme was introduced into the model
and a systematic derivative expansion similar to the one explained in
Sec.~\ref{sec:CPHLS} was made.
In the HLS with the quasiqurk (constituent quark)
the quasiquark field $\psi$ is introduced in the
Lagrangian in such a way that it transforms homogeneously under
the HLS: 
$\psi \rightarrow h(x)\cdot \psi$ where $h(x) \in H_{\rm local}$. 
Since we consider the model near chiral phase transition
point where the quasiquark mass is expected to become small, we
assign ${\cal O}(p)$ to the quasiquark mass
$m_q$. Furthermore, we assign ${\cal O}(p)$ to the chemical
potential $\widetilde{\mu}$~\footnote{%
  In Ref.~\cite{Harada-Kim-Rho} $\mu$ is used for expressing the
  chemical potential.
  Throughout this report, however, we use $\mu$ for the energy scale,
  and then we use $\widetilde{\mu}$ for the chemical potential in this
  subsection.
}
or the Fermi momentum $P_F$, since we consider
that the cutoff is larger than $\widetilde{\mu}$ even near the phase
transition point. Using this counting scheme we can make the
systematic expansion in the HLS with the quasiquark included. We
should note that this counting scheme is different from the one in
the model for $\pi$ and baryons given in Ref.~\cite{MOR:01} where
the baryon mass is counted as ${\cal O}(1)$. The leading order
Lagrangian including one quasiquark field and one anti-quasiquark
field is counted as ${\cal O}(p)$ and given 
by~\cite{BKY,Harada-Kim-Rho}
\begin{eqnarray}
 \delta {\cal L}_{Q(1)} 
&=& 
 \bar{\psi}(x) \left( 
   iD_\mu \gamma^\mu - \widetilde{\mu} \gamma^0 - m_q 
  \right)\psi(x)
\nonumber\\
&&
  {}+ \bar \psi(x) \left(
    \kappa\gamma^\mu \hat{\alpha}_{\parallel \mu}(x )
    + \lambda\gamma_5\gamma^\mu \hat{\alpha}_{\perp\mu}(x) 
  \right)
  \psi(x) 
\label{lagbaryon}
\end{eqnarray}
where $D_\mu\psi=(\partial_\mu -ig\rho_\mu)\psi$ and $\kappa$ and
$\lambda$ are constants to be specified later. 

At one-loop level
the Lagrangian (\ref{lagbaryon}) generates the ${\cal O}(p^4)$
contributions including hadronic dense-loop effects as well as
divergent effects. The divergent contributions are renormalized by
the parameters, and thus the RGEs for three leading order
parameters $F_\pi$, $a$ and $g$ (and parameters of ${\cal O}(p^4)$
Lagrangian) are modified from those without quasiquark field. In
addition, we need to consider the renormalization group flow for
the quasiquark mass $m_q$~\footnote{%
  The constants $\kappa$ and
  $\lambda$ will also run such that at 
  $\widetilde{\mu}=\widetilde{\mu}_c$,
  $\kappa=\lambda=1$ while at $\widetilde{\mu} < \widetilde{\mu}_c$, 
  $\kappa\neq \lambda$.
  The running will be small near $n_c$, so we will ignore their
  running here.
}. 
Calculating one-loop contributions for RGEs, we
find~\cite{Harada-Kim-Rho}
\begin{eqnarray}
  \mu \frac{dF_\pi^2}{d\mu} 
&=&
  \frac{N_f}{2(4\pi)^2}
  \left[
    3a^2g^2F_\pi^2 +2(2-a)\mu^2
  \right] 
  -\frac{m_q^2}{2\pi^2}\lambda^2 N_c
\ ,
\label{RGE:Fp q}
\\
\mu \frac{da}{d\mu} 
&=&
  -
  \frac{N_f}{2(4\pi)^2}
  (a-1)
  \left[ 3a(1+a)g^2-(3a-1)\frac{\mu^2}{F_\pi^2} \right]
  {} + a\frac{\lambda^2}{2\pi^2}\frac{m_q^2}{F_\pi^2}N_c
\ ,
\label{RGE:a q}
\\
  \mu\frac{d g^2}{d \mu}
&=&
  -
  \frac{N_f}{2(4\pi)^2} \frac{87-a^2}{6}g^4
  +\frac{N_c}{6\pi^2}g^4 (1-\kappa)^2
\label{RGE:g q}
\ ,
\\
  \mu\frac{dm_q}{d\mu}
&=&
  -\frac{m_q}{8\pi^2}
  \Biggl[ 
    (C_\pi-C_\sigma)\mu^2 -m_q^2 (C_\pi-C_\sigma)
    {}+M_\rho^2 C_\sigma -4C_\rho 
  \Biggr]
\ ,
\label{RGE:m q}
\end{eqnarray}
where
\begin{eqnarray}
  C_\pi
&\equiv&
  \left(\frac{\lambda}{F_\pi}\right)^2 
  \frac{N_f^2 - 1}{2 N_f}
\ ,
\nonumber\\
  C_\sigma
&\equiv&
  \left(\frac{\kappa}{F_\sigma}\right)^2
  \frac{N_f^2 - 1}{2 N_f}
\ ,
\nonumber\\
  C_\rho
&\equiv& 
  g^2 (1-\kappa)^2 \frac{N_f^2 - 1}{2 N_f}
\ .
\end{eqnarray}

Hadronic dense corrections from the quasiquark loop 
to the $\pi$ decay constant $f_\pi(\widetilde{\mu})$ and the $\rho$
pole mass $m_\rho(\widetilde{\mu})$ were calculated in
Ref.~\cite{Harada-Kim-Rho}.
Here we will briefly review the analysis.
As for the calculation of the hadronic thermal corrections explained
in the previous subsections,
it is convenient to use 
the following ``on-shell" quantities:
\begin{eqnarray}
&&
  F_\pi = F_\pi(\mu=0;\widetilde{\mu}) \ ,
\nonumber\\
&&
  g = g\mbox{\boldmath$\bigl($}
  \mu = M_\rho(\widetilde{\mu});\widetilde{\mu}
  \mbox{\boldmath$\bigr)$}
\ , \quad
  a = a\mbox{\boldmath$\bigl($}
  \mu = M_\rho(\widetilde{\mu});\widetilde{\mu}
  \mbox{\boldmath$\bigr)$}
\ , \label{on-shell para mu}
\end{eqnarray}
where $M_\rho$ is determined from the ``on-shell condition":
\begin{eqnarray}
&&
  M_\rho^2 = M_\rho^2(\widetilde{\mu}) =
  a\mbox{\boldmath$\bigl($}
  \mu = M_\rho(\widetilde{\mu});\widetilde{\mu}
  \mbox{\boldmath$\bigr)$}
\nonumber\\
&& \quad
  \times
  g^2\mbox{\boldmath$\bigl($}
  \mu = M_\rho(\widetilde{\mu});\widetilde{\mu}
  \mbox{\boldmath$\bigr)$}
  F_\pi^2\mbox{\boldmath$\bigl($}
  \mu = M_\rho(\widetilde{\mu});\widetilde{\mu}
  \mbox{\boldmath$\bigr)$}
\ .
\end{eqnarray}
Then, as in the previous subsection,
the parameter $M_\rho$ in this subsection is renormalized in such
a way that it becomes the pole mass at $\widetilde{\mu}=0$.

For obtaining the dense-loop corrections to the pion decay constant
we should note that
distinction has to be made
between the temporal and spatial components of the pion
decay constants, since the Lorentz invariance is broken in the medium.
We use the following definition~\cite{Pisarski-Tytgat}:
\begin{eqnarray}
&&
\left\langle 0 \left\vert
  J_5^{\mu=0}(0)
\right\vert \pi (\vec{p}) \right\rangle_{\widetilde{\mu}}
=
  - i p_0 f_\pi^t (\widetilde{\mu})
\ ,
\nonumber\\
&&
\left\langle 0 \left\vert
  J_5^{\mu=i}(0)
\right\vert \pi (\vec{p}) \right\rangle_{\widetilde{\mu}}
=
  - i \vec{p}_i f_\pi^s (\widetilde{\mu})
\ .
\end{eqnarray}
In terms of the axialvector-axialvector two-point function
$\Pi_{\overline{\cal A}\overline{\cal A}}^{\mu\nu}$, 
the temporal and spatial components of the
pion decay constant are generally expressed as
\begin{eqnarray}
 f_\pi^t (\widetilde{\mu})
&=&
  \frac{1}{\widetilde{F}}
  \left.
    \frac{ 
      u_\mu 
      \Pi_{\overline{\cal A}\overline{\cal A}}^{\mu\nu} (p_0,\vec{p})
      p_\nu 
     }{ p_0 }
  \right\vert_{p_0 = \tilde{\omega}}
\ ,
\nonumber\\
  f_\pi^s (\widetilde{\mu})
&=&
  \frac{1}{\widetilde{F}}
  \left.
    \frac{
      - p^\alpha ( g_{\alpha\mu} - u_\alpha u_\mu )
      \Pi_{\overline{\cal A}\overline{\cal A}}^{\mu\nu}(p_0,\vec{p})
      p_\nu
    }{ \bar{p}^2 }
  \right\vert_{p_0 = \tilde{\omega}}
\ ,
\label{fpit fpis defs}
\end{eqnarray}
where $\widetilde{F}$ is the $\pi$ wave function renormalization
constant in medium.~\footnote{%
  Note that the backgroud field $\overline{\cal A}$ includes the
  background pion field $\overline{\pi}$ as
  $
  \overline{\cal A}_\mu = {\cal A}_\mu +
  \partial_\mu \overline{\pi}/\widetilde{F}
  + \cdots
  $.
  For $\widetilde{\mu} = 0 $ 
  this $\widetilde{F}$ agrees with $F_\pi$.
}
According to the analysis of Ref.~\cite{MOR:01} in dense
matter, this $\widetilde{F}$ is nothing but $f_\pi^t$:
\begin{equation}
\widetilde{F} = f_\pi^t (\widetilde{\mu}) \ .
\end{equation}
In the HLS with present renormalization scheme, 
this $\Pi_{\overline{\cal A}\overline{\cal A}}^{\mu\nu}$
is expressed as
\begin{equation}
  \Pi_{\overline{\cal A}\overline{\cal A}}^{\mu\nu} (p_0,\vec{p}) 
  =
  g^{\mu\nu} F_\pi^2
  + 2 z_2 \left( g^{\mu\nu} p^2 - p^\mu p^\nu \right)
  + \overline{\Pi}_{\overline{\cal A}\overline{\cal A}}^{\mu\nu}
  (p_0,\vec{p})
\ ,
\end{equation}
where $\overline{\Pi}_{\overline{\cal A}\overline{\cal A}}^{\mu\nu}
(p_0,\vec{p})$ denotes the
hadronic dense corrections of interest.
In Ref.~\cite{Harada-Kim-Rho}
the dense-loop corrections from the interaction Lagrangian 
(\ref{lagbaryon}) were calculated at one loop, and
it was shown that there is no hadronic dense-loop correction to the
$\pi$ decay constants:
\begin{eqnarray}
&&
  \left[ f_\pi^t(\widetilde{\mu}) \right]^2
  = f_\pi^t(\widetilde{\mu}) f_\pi^s(\widetilde{\mu})
  = F_\pi^2(\mu=0;\widetilde{\mu}) 
\ .
\label{fpi: dense}
\end{eqnarray}

Next we calculate the
hadronic dense-loop corrections to the $\rho$ pole mass.
As in the previous subsection there are two pole masses related to the
longitudinal and transverse components of $\rho$ propagator.
In Ref.~\cite{Harada-Kim-Rho} the dense-loop corrections to
them from the Lagrangian (\ref{lagbaryon}) were calculated at one-loop
level.
The results are
\begin{eqnarray}
m_{\rho L}^2 (\widetilde{\mu})
&=&
m_{\rho T}^2 (\widetilde{\mu})
\nonumber\\
&=&
M_\rho^2 +
\frac{2}{3}\, g^2 (1-\kappa)^2
\left[
  \bar{B}_S
  - (M_\rho^2 + 2 m_q^2) \bar{B}_0(p_0=M_\rho,\vec{p}=0)
\right]
\ ,
\label{dense: mrL mrT}
\end{eqnarray}
where
\begin{eqnarray}
  \bar{B}_S
&=& 
  \frac{1}{4\pi^2} 
  \left[
    P_F \omega_F - m_q^2 \, \ln \frac{ P_F + \omega_F }{m_q} 
  \right] 
\ ,
\nonumber\\
  \bar{B}_0(p_0,\vec{p}=0)
&=& 
  \frac{1}{8\pi^2}
  \Biggl[
    - \ln \frac{ P_F + \omega_F }{ m_q }
\nonumber\\
&& 
    {}+ \frac{1}{2} 
    \sqrt{ \frac{ 4m_q^2 -p_0^2 - i \epsilon }{ -p_0^2 - i \epsilon } }
    \ln \frac{
      \omega_F\,\sqrt{ 4m_q^2 -p_0^2 - i \epsilon } 
      + P_F\,\sqrt{ -p_0^2 - i \epsilon }
    }{
      \omega_F\,\sqrt{ 4m_q^2 -p_0^2 - i \epsilon } 
      - P_F\,\sqrt{ -p_0^2 - i \epsilon }
    }
  \Biggr]
\ ,
\label{def: Bs B0}
\end{eqnarray}
with $P_F$ being the Fermi momentum of quasiquark and
$\omega_F = \sqrt{ P_F^2 + m_q^2 }$.

Let us now apply the Wilsonian matching at the critical chemical
potential $\widetilde{\mu}_c$ for $N_f = 3$.
Here we note that the current correlators in the HLS remains unchanged
as the forms given in Eqs.~(\ref{Pi A HLS}) and (\ref{Pi V HLS}) 
except that the bare parameters are density-dependent
even when we include the quasiquark field as explained
above.~\footnote{%
  Since the Lorentz non-invariant terms in the current correlators by
  the OPE are suppressed by some powers of
  $n/\Lambda^3$ (see, e.g. Ref.~\cite{Hatsuda-Lee}),
  we ignore them from both the hadronic and QCD
  sectors.
}
As is done for the chiral restoration in hot matter in
Ref.~\cite{Harada-Sasaki} (see previous subsection)
we assume that $\langle \bar{q} q \rangle$ approaches to $0$ 
continuously for $n \rightarrow n_c$~\footnote{%
  We are assuming that the transition is not strongly first order. If
  it is strongly first order, some of the arguments used here
  may need qualifications.
  However, we should note that, 
  in the presence of the small current quark mass, the quark
  condensate is shown to decrease rapidly but continuously around the
  ``phase transition'' point~\cite{Brown-Rho:96}.
}.
In such a case
the axialvector and vector current correlators 
by OPE in the QCD sector approach each other, and will agree at $n_c$.
Then, 
 through the Wilsonian matching we require that the correlators in
the HLS in Eqs.~(\ref{Pi A HLS}) and (\ref{Pi V HLS}) 
agree with each other.
As in the case of large $N_f$ ~\cite{HY:VM} (see Sec.~\ref{sec:VM})
and in the case of $T\sim T_c$~\cite{Harada-Sasaki} (see 
Sec.~\ref{ssec:VMNT}), 
this agreement can be satisfied also in dense
matter if the following conditions are 
met~\cite{Harada-Kim-Rho}:
\begin{eqnarray}
&& g(\Lambda;n) \mathop{\longrightarrow}_{n \rightarrow n_c} 0 \ ,
\qquad a(\Lambda;n) \mathop{\longrightarrow}_{n \rightarrow n_c} 1
\ ,
\nonumber\\
&& z_1(\Lambda;n) - z_2(\Lambda;n) 
\mathop{\longrightarrow}_{n \rightarrow n_c} 0 
\ .
\label{g a z12:VMn}
\end{eqnarray}

The above conditions for the bare parameters
are converted to the ones
for the on-shell parameters through the Wilsonian RGE's given in 
Eqs.~(\ref{RGE:Fp q})--(\ref{RGE:m q}).
Differently from the cases for large $N_f$ QCD and hot QCD,
Eqs.~(\ref{RGE:a q}) and (\ref{RGE:g q})
show that $(g\,,\,a)=(0\,,\,1)$ is a
fixed point only when $m_q=0$.
Since
the ``on-shell'' quasiquark mass
$m_q$ is expected to vanish at the critical point:
\begin{equation}
  m_q(n) \mathop{\longrightarrow}_{n \rightarrow n_c} 0 
\ ,
\end{equation}
and that $m_q = 0$ is actually a fixed point of the RGE in 
Eq.~(\ref{RGE:m q}),
$(g\,,\,a\,,\,m_q) = (0\,,\,1\,,\,0)$
is a fixed point of the coupled RGEs for $g$, $a$ and $m_q$.
Furthermore and most importantly, {\it $X=1$ becomes the fixed
point of the RGE for} $X$~\cite{HY:VD}. This means that at the
fixed point, $F_\pi (0)=0$ [see Eq.~(\ref{def X})]. What does this
mean in dense matter? To see what this means, we note that for
$T=\widetilde{\mu}=0$, 
this $F_\pi (0)=0$ condition is satisfied for a given
number of flavors $N_f^{\rm cr}\sim 5$ through the Wilsonian
matching~\cite{HY:VM}. For $N_f=3$, $\widetilde{\mu}=0$ 
and $T\neq 0$, this
condition is never satisfied due to thermal hadronic
corrections~\cite{Harada-Sasaki}. 
Remarkably, as was shown in Ref.~\cite{Harada-Kim-Rho} and 
we briefly reviewd above, for
$N_f=3$, $T=0$ and $\mu=\mu_c$, it turns out that dense hadronic
corrections to the pion decay constant
vanish up to ${\cal O}(p^6)$ corrections. 
Therefore
the fixed point $X=1$ (i.e., $F_\pi (0)=0$) does indeed signal
chiral restoration at the critical density.

Let us here
focus on what
happens to hadrons at and very near the critical point 
$\widetilde{\mu}_c$.
As is shown in Eq.~(\ref{fpi: dense}),
there is no hadronic dense-loop corrections to the $\pi$ decay
constants.
Thus
\begin{equation}
f_\pi (\widetilde{\mu}_c)=F_\pi (0;\widetilde{\mu}_c)=0
\ .
\label{fp: muc}
\end{equation}
Since
\begin{equation}
F_\pi^2 (0;\widetilde{\mu}_c)=F_\pi^2
(\Lambda;\widetilde{\mu}_c)-\frac{N_f}{2(4\pi)^2}\Lambda^2
\ ,
\end{equation}
and at the matching scale $\Lambda$, 
$F_\pi^2 (\Lambda;\widetilde{\mu}_c)$ is
given by a QCD correlator at $\widetilde{\mu}=\widetilde{\mu}_c$,
$\widetilde{\mu}_c$ can be computed
from
\begin{equation}
  F_\pi^2 (\Lambda;\widetilde{\mu}_c)
  =\frac{N_f}{2(4\pi)^2}\Lambda^2 
\ .
\end{equation}
Note that in free space, this is the equation that determines
$N_f^{\rm crit}\sim 5$~\cite{HY:VM}. 
In order for this equation to have a
solution at the critical density, it is necessary that $F_\pi^2
(\Lambda;\widetilde{\mu}_c)/F_\pi^2 (\Lambda;0) \sim 3/5$. 
We do not have at
present a reliable estimate of the density dependence of the QCD
correlator to verify this condition but the decrease of $F_\pi$ of
this order in medium looks quite reasonable.

Next we compute the $\rho$ pole mass near $\widetilde{\mu}_c$.
For $M_\rho, m_q \ll P_F$ 
Eq.~(\ref{dense: mrL mrT}) reduces to
\begin{eqnarray}
 m_\rho^2(\widetilde{\mu}) &=& M_\rho^2 (\widetilde{\mu}) + g^2
 \frac{\widetilde{\mu}^2}{6\pi^2} 
 (1-\kappa)^2
\ .
\label{mrho at muc}
\end{eqnarray}
At $\widetilde{\mu}=\widetilde{\mu}_c$, 
we have $g=0$ and $a=1$ so that $M_\rho (\widetilde{\mu})=0$,
and then
$m_\rho(\widetilde{\mu})=0$. 
Thus the fate of
the $\rho$ meson at the critical density is the same as that at
the critical temperature~\cite{Harada-Kim-Rho}:
\begin{equation}
m_\rho^2(\widetilde{\mu})
\mathop{\longrightarrow}_{\widetilde{\mu} 
  \rightarrow \widetilde{\mu}_c} 0 \ .
\label{VM: dense}
\end{equation}
This implies that, as was suggested in 
Ref.~\cite{HY:VM} and then proposed in
Refs.~\cite{Brown-Rho:01a,Brown-Rho:01b},
the vector manifestation (VM) is
realized in dense matter at the chiral restoration with the $\rho$
mass $m_\rho$ going to zero at the crtical point. 
Thus the VM
is {\it universal} in the sense that it occurs at 
$N_f^{\rm crit}$ for
$T=\widetilde{\mu}=0$, at $T_c$ for $N_f < N_f^{\rm crit}$ and
$\widetilde{\mu}=0$ and at $\widetilde{\mu}_c$ 
for $T=0$ and $N_f <N_f^{\rm crit}$.

Detailed calculations of the hadronic dense-loop corrections
are shown in Ref.~\cite{Harada-Kim-Rho},
where
the ${\cal O}(p^2)$ interaction Lagrangian was included in addition to
the Lagrangian in Eq.~(\ref{lagbaryon}) and it was shown that the
results in Eqs.~(\ref{fp: muc}) and (\ref{VM: dense}) are intact.

\newpage

\section{Summary and Discussions}
\label{sec:SD}

In this report we have explained recent developement, 
particularly the loop effects, of
the effective field theory (EFT) of QCD and QCD-like theories
for light pseudoscalar
and vector mesons, based on the hidden local symmetry (HLS) model. 

The HLS model as explained in Sec.~\ref{sec:HLS} is simply reduced 
to the nonlinear chiral Lagrangian in the low
energy region 
where the kinetic term of the vector meson is negligible compared with
the mass term, $p^2 \ll m_\rho^2$, and the gauge symmetry 
(gauge-boson degree of freedom) becomes ``hidden''. 
Although there are many vector meson theories which yield the same
classical 
(tree level) result as that of the HLS model, they may not lead to the
same  
quantum theory. Actually, as was illustrated in the $CP^{N-1}$
model~\cite{BKY}, 
theories being the same at classical (tree) level, the one with
explicit gauge  
symmetry and the other without it, may not be 
the same at quantum 
level. 

In Sec.~\ref{sec:CPHLS} it was emphasized that 
presence of the gauge symmetry of HLS is in fact vital to
the systematic low-energy expansion (chiral perturbation) with loops
of vector as well as pseudoscalar mesons, when their masses can be
regarded as small.
We developed  a systematic expansion of HLS model on the same footing as the
Chiral Perturbation Theory (ChPT) of the ordinary chiral Lagrangian 
without vector mesons (reviewed in Sec.~\ref{sec:BRCPT}), 
based on the order counting of the HLS coupling $g$:
\begin{equation}
g \sim {\cal O}(p) \ .
\end{equation}

Based on this systematic expansion, we developed in 
Sec.~\ref{sec:CPHLS}
analyses of the one-loop 
Renormalization-Group Equations (RGEs) in the sense of Wilson
(``Wilsonian RGE'')
which {\it includes quadratic divergence}. 
The Lagrangian having such running parameters
correponds to the Wilsonian effective action which is obtained from the
bare action (defined at cutoff) by integrating
higher energy modes down to lower energy scale and necessarily contains 
quadratic divergences. 
Here we should emphasize that, {\it as a matter of principle}, 
the bare parameters of EFT are not free 
parameters but are determined by the underlying theory and 
hence the quadratic divergences should not be renormalized out by
cancelling 
with the arbitrary choice of the bare parameters as in the usual
renormalization procedure. Once we determined the bare parameters of
EFT, we necessarily predict the physical quantities through the
Wilsonian RGEs  
including the quadratic divergence.

A novel feature of the approach in this report is the ``Wilsonian
matching'' 
given in Sec.~\ref{sec:WM}, which determines the bare parameters 
(defined at the cutoff scale $\Lambda$)
of the EFT in terms of the underlying theory,
the QCD or QCD-like theories. We wrote down the current correlators at 
$\Lambda$ in terms of the bare parameters of the HLS model, which was
then evaluated in terms of the OPE of the underlying QCD at the same
scale $\Lambda$. 
This 
provides the EFT with otherwise unknown information
of the underlying theory such as the explicit dependence on $N_c$ and
$\Lambda_{\rm QCD}$ as well as the precise value of the bare
parameters. 
Once the bare values were given as the boundary conditions of RGEs, 
the physics below $\Lambda$ was uniquely 
predicted via RGEs through the own dynamics of
the HLS model.

Main issues of this approach were:
\begin{enumerate}

\item Prediction of a very successful phenomenology of $\pi$ and
$\rho$ for the  
realistic case of $N_f=3$ (Sec.~\ref{sec:WM}).

\item  Prediction of chiral symmetry restoration due to quadratic
divergence 
for certain choice of the parameters of the underlying QCD, 
such as the number of colors $N_c$ and of the massless flavors $N_f$ 
such that $N_f/N_c >5$ (Sec.~\ref{sec:VM}). 
The vector meson dominance, though accidentally valid for the $N_f=3$, 
does not hold in general and is largely violated near the chiral
restoration  
point. 

\item  Prediction of ``Vector manifestation (VM)''  as a novel feature
of this  
chiral restoration: The $\rho$ becomes the chiral partner of the $\pi$
in contrast 
to the conventional manifestation of the 
linear sigma model (``GL manifestaion'')
where the scalar meson
becomes the chiral partner of the $\pi$ (Sec.~\ref{sec:VM}).
Similar phenomenon can also take place in the hot/dense QCD 
(Sec.~\ref{sec:THDMC}).

\item The chiral restoration in the HLS model takes place by its 
own dynamics as in the underlying QCD, which suggested that the Seiberg-type 
duality is operative even for the non-SUSY QCD where the HLS plays a role of the ``magnetic gauge theory'' dual to the QCD as the ``electric gauge theory''
(Sec.~\ref{sec:VM}).
\end{enumerate}

It was demonstrated in Sec.~\ref{sec:CPHLS} that 
the quadratic divergences are actually vital to the  
chiral symmetry restoration in the EFT which corresponds
to the underlying QCD 
and QCD-like theories under extreme conditions where such a 
chiral phase transition is expected to take place. 
The point is that the quadratic divergence in the HLS model
gives rise to an essential part of the 
running of the decay constant $F_\pi^2
(\mu)$ 
whose bare value $F_\pi^2(\Lambda)$ is not the order parameter but 
merely a Lagrangian parameter, while the pole residue of the NG boson 
is proportional to the value $F_\pi^2(0)$ at the pole position $p^2=0$ 
which is then the order parameter of the chiral
 symmetry breaking.  
  The chiral restoration is thus idenified with 
 $F_\pi^2(0)=0$, while $F_\pi^2(\Lambda) \ne 0$ in general. 

We gave detailed explanation why the quadratic/power divergence
is so vital to the phase transition of the EFT, based on the
illustration 
of the phase transitions in various well-known models having the
chiral 
phase transition, like the NJL model, the Standard model (SM) and the
$CP^{N-1}$ in $D(\le 4)$ 
dimensions as well as the nonlinear chiral Lagrangian which is of
direct 
relevance to our case. 
The point is that the bare Lagrangian as it stands
does not tell us which phase we are actually living in. The quadratic
divergence is the main driving force to make the quantum theory to
choose a  
different phase than that the bare Lagrangian looks like. 

Now, one might suspect that 
the systematic expansion in our
case might break down when we include the 
quadratic divergence.
Actually the
quadratic divergence carrying no momentum would not be suppressed by
powers of $p$ in the HLS model as well as in the ChPT (with the
quadratic  
divergence included): 
Quadratic divergences from all higher loops would in principle
contribute to the $O(p^2)$ term in powers of 
$[N_f\Lambda^2/(4\pi F_\pi(\Lambda))^2]^{n}$
for $n$-th loop and hence would invalidate the power
counting rule in the systematic expansion 
unless $N_f\Lambda^2/(4\pi F_\pi(\Lambda))^2 <1$.  

However, such a condition is needed even in the usual
ChPT (without quadratic divergence, 
$F_\pi(\Lambda)=F_\pi(0)\equiv F_\pi$) 
where the systematic expansion  
breaks down unless 
$N_f\Lambda^2/(4\pi F_\pi)^2 <1$. 
Inclusion of the quadratic divergence
is actually even better for 
the systematic expansion to work,
\begin{equation}
N_f\frac{\Lambda^2}{\left(4\pi F_\pi(\Lambda)\right)^2} <1\, ,
\end{equation} 
since generally we have 
$F_\pi^2(\Lambda) > F_\pi^2(0)$ due to quadratic divergence,
and in particular near the chiral restoration point where
$F_\pi^2(0) \rightarrow 0$ whereas $F_\pi^2(\Lambda)$ remains finite.

More specifically, 
$F_\pi^2(\Lambda)$ was given by the Wilsonian matching with the QCD
(Eq.(\ref{valueFpi})): 
\begin{equation}
F_\pi^2(\Lambda) =
\frac{N_c}{3}2 (1+\delta_A) \left(\frac{\Lambda}{4\pi}
\right)^2\sim N_c \left(\frac{\Lambda}{4\pi}\right)^2  
\, ,
\end{equation}
where $\delta_A$ stands for the higher order corrections in OPE to the
parton  
(free quark loop) contribution $1$ and hence 
is expected to be $\delta_A \ll 1$. 
Actualy we estimated $\delta_A \sim 0.5$ for $N_f=3$.
Then the systematic expansion would be  valid if
\begin{equation}
N_f \frac{\Lambda^2}{\left(4\pi F_\pi(\Lambda)\right)^2} 
\sim \frac{N_f}{N_c}
< 1 \, .
\end{equation}
Such a situation can be realized, 
if we consider the large $N_c$ limit $N_c \rightarrow \infty$ such
that 
$N_f/N_c \, \ll 1$ and then extrapolate it to the parameter region
$N_f/N_c \sim 1$. 
Moreover, in the HLS model (in contrast to ChPT without vector meson),
the quadratic divergence for $F_\pi^2$ has an additional factor 
$1/2$ (at $a\simeq 1$) and hence the systematic expansion is expected
to 
be valid for
\begin{equation}
\frac{N_f}{2} \frac{\Lambda^2}{(4\pi F_\pi(\Lambda))^2} 
\sim \frac{N_f}{2 N_c}
<1 \, .
\end{equation}
Thus the inclusion of the quadratic divergence does not 
affect the validity of the systematic expansion.
It even improves the scale for the systematic expansion better than
the conventional naive dimensional analysis
(without quadratic divergence).

Note that the edge of the validity region of the systematic 
expansion roughly corresponds  
to the chiral restoration point
where the tree and the loop cancel out each other.
Actually, the phase transition in many cases is 
a phenomenon in which the tree (bare) and the 
loop effects (quadratic divergences)
are becoming comparable and are balanced (cancelled) by each
other. Hence 
this phenomenon is generally 
at the edge of the validity of the systematic expansion,
such as in the usual perturbation (SM),
chiral perturbation (nonlinear sigma model), etc., although in the NJL 
case the loop to be balanced by the tree is treated also as the
leading order  
in  the $1/N$ expansion. 
  
To summarize the roles of the quadratic divergence: It must be
included 
as a matter of principle once the bare parameters are fixed; It is 
crucial to the phase transition; It improves the validity scale of the 
systematic expansion rather than naive dimensional analysis; 
It leads to a very successful 
phenomenology of $\pi$ and $\rho$ system.

Now,
once we matched the EFT, the HLS model, 
with the underlying theory in this way, 
we can play with arbitrary 
$N_c$ and $\Lambda_{\rm QCD}$
as well as $N_f$ in the same sense as dealing with the underlying QCD.
Then we expect that 
the HLS model
by its own dynamics will give rise to the same infrared physics 
as the underlying theory itself for arbitray parameter choice other
than $N_c=N_f=3$ of the real life QCD: When the underlying QCD gets
chiral restoration, 
the HLS model will also get chiral restoration.

In Sec.~\ref{sec:VM} 
we actually formulated conditions of chiral symmetry restoration on
the bare HLS parameters (``VM conditions'') by
matching the current correlators 
with those of the underlying QCD where the chiral symmetry gets
restored,
$\langle \bar q q \rangle \rightarrow 0 $ as $N_f \rightarrow N_f^{\rm crit}$:
\begin{eqnarray}
g(\Lambda) \rightarrow 0, \,  a(\Lambda) \rightarrow 1, \, 
z_1(\Lambda)-z_2(\Lambda)
\rightarrow 0 \,  ,
\nonumber
\\
F_\pi^2(\Lambda)\rightarrow (F_\pi^{\rm crit})^2\equiv 
\frac{N_c}{3}2 ( 1+\delta_A^{\rm crit} )
  \left(\frac{\Lambda}{4 \pi}\right)^2\ ,
\end{eqnarray}
where
\begin{equation}
\delta_A^{\rm crit}
\equiv \delta_A|_{\langle \bar q q \rangle = 0} =
\frac{3(N_c^2-1)}{8N_c} \,\frac{\alpha_s}{\pi}
  + \frac{2\pi^2}{N_c} 
    \frac{
      \left\langle 
        \frac{\alpha_s}{\pi} G_{\mu\nu} G^{\mu\nu}
      \right\rangle
    }{ \Lambda^4 }
\end{equation}
must satisfy $0< \delta_A^{\rm crit} <1$ in order that the OPE makes
sense. 

Although the VM conditions as they stand might not seem to indicate
chiral 
symmetry restoration, they actually lead to the vanishing order
parameter $F_\pi^2(0) \rightarrow 0$  and thus the chiral restoration through
the own dynamics of the 
HLS model as follows:
The RGEs of the HLS model are readily solved for the VM conditions,
since $g=0$ and  
$a=1$ are the fixed points of the RGEs.

By taking
$g \rightarrow 0$ and $a\rightarrow 1$, we had
\begin{equation}
F_\pi^2(0)=F_\pi^2(\Lambda)-\frac{1}{2} N_f \left(\frac{\Lambda}{4 \pi}\right)^2
\rightarrow \left(\frac{N_c}{3} 2(1+\delta_A^{\rm crit}) -\frac{N_f^{\rm crit}}{2}\right)
\left(\frac{\Lambda}{4 \pi}\right)^2 \, ,
\end{equation}
where the first term given by the Wilsonian matching with QCD is
proportional to $N_c$  
and the second term given by the quadratic divergence in the HLS model
is proportional to $N_f$.
Now, the chiral restoration takes place with vanishing 
right-hand-side (RHS),
$F_\pi^2(0)\rightarrow 0$, 
 by precise cancellation between the two terms, namely the 
interplay between $N_c$ and $N_f$ such that $N_f\sim N_c\gg 1$.
Then the chiral restoration takes place
at 
\begin{equation}
N_f=N_f^{\rm crit}= \frac{N_c}{3} 4(1+\delta_A^{\rm crit})\, ,
\end{equation}
where $0< \delta_A^{\rm crit} <1$ in order for the OPE to make sense.
Then we predicted the critical
value $N_f^{\rm crit}$ fairly independently of the detailed input data:
\begin{equation}
4 \left(\frac{N_c}{3}\right) < N_f^{\rm crit}
                              < 8 \left(\frac{N_c}{3}\right) \ ,
\end{equation}
which is consistent with the lattice simulation~\cite{IKKSY:98}
\begin{equation}
6 < N_f^{\rm crit} <7 \quad (N_c=3) \ ,
\end{equation}
but in disagreement with the analysis of ladder 
Schwinger-Dyson (SD) equation combined with
the perturbative infrared fixed 
point~\cite{Appelquist-Ratnaweera-Terning-Wijewardhana,Appelquist-Terning-Wijewardhana};
\begin{equation}
N_f^{\rm crit}
\simeq 12  \left(\frac{N_c}{3}\right) \ .
\end{equation}  
More specifically, we estimated  $\delta_A^{\rm crit} \simeq 0.25$ 
at the QCD chiral restoration point
$\langle \bar q q\rangle=0$ ($\delta_A \sim 0.5$ for 
$N_c=N_f=3$ where $\langle \bar q q\rangle\ne 0$) and hence:
\begin{equation}
N_f^{\rm crit} \simeq 5 \left(\frac{N_c}{3}\right) \, ,
\end{equation}
which coincides with the instanton argument~\cite{Velkovsky-Shuryak}.

It was emphasized that this chiral 
restoration should be regarded as a limit $F_\pi^2(0)
\rightarrow 0$ 
 but not precisely on the critical point $F_\pi^2(0) \equiv 0$ 
where no light composite spectrum would exist and hence the HLS model would 
break down. 
The limit (``VM limit'')
corresponds to the VM conditions for the bare parameters; 
$F_\pi^2(\Lambda) \rightarrow (F_\pi^{\rm crit})^2$, $a(\Lambda) \rightarrow 1$ and 
$g(\Lambda) \rightarrow 0$ as $N_f \rightarrow N_f^{\rm crit}$ in the
underlying QCD, with a special care for the $g(\Lambda) \rightarrow 0$,
in contrast to setting $g(\Lambda) \equiv 0$ which gives the HLS model a
redundant global symmetry, $G_1\times G_2$ with $G=\mbox{SU}(N_f)_{\rm L}\times 
\mbox{SU}(N_f)_{\rm R}$,
larger than that of the underlying QCD and should be avoided. 
On the other hand, there is no peculiarity for setting $a(\Lambda)=1$ as far as we keep $g(\Lambda) \ne 0$, in which case the redundant global 
symmetry $G_1\times G_2$ is explicitly 
broken only by the $\rho$ gauge coupling down to the symmetry of the HLS model, 
$G_{\rm global}\times H_{\rm local}$. In the real-life QCD with $N_f=3$
which we showed is very close to $a(\Lambda) =1$, 
this $\rho$ coupling is rather strong.
It is amazing, however, that by simply setting $N_f \rightarrow N_f^{\rm crit}$
in the underlying QCD, we arrive at the VM limit which does realize 
the weak coupling gauge theory of light composite $\rho$, $g \rightarrow 0$
and $m_\rho \rightarrow 0$,
 in spite of the fact 
that this $\rho$ coupling is dynamically generated 
at composite level from the underlying strong coupling gauge theory.

The salient feature of the above chiral restoration is that the 
$\rho$ becomes the chiral partner of the $\pi$ with its mass
vanishing at the ciritical point:
\begin{eqnarray}
m_\rho^2 \rightarrow m_\pi^2=0\ , \quad
F_\sigma^2(m_\rho) / F_\pi^2(0) \rightarrow 1 \ ,
\end{eqnarray}
as $F_\pi^2(0) \rightarrow 0$,
where $F_\sigma(m_\rho)$ is the decay constant of $\sigma$
(longitudinal $\rho$) at $\rho$ on-shell.
This we called ``Vector Manifestation (VM)'' in contrast to the
conventional 
manifestation \`{a} la linear sigma model 
(``Ginzburg--Landau/Gell-Mann--Levy(GL) 
Manifestation''):
\begin{equation}
m_S^2 \rightarrow m_\pi^2 =0 \, ,
\end{equation}
as $F_\pi^2(0) \rightarrow 0$, 
where $m_S$ stands for the mass of the
scalar meson (``sigma'' meson in the linear sigma model).
The VM implies that $\pi$ belongs to 
$(N_f^2-1\,,\,1) \oplus (1\,,\,N_f^2-1)$  of the chiral 
representation together with the $\rho$, 
while in the GL manifestation $\pi$ does to 
$(N_f\,, N_f^*) \oplus (N_f^*\,,N_f)$ together with the scalar meson.
 
The GL manifestation does not satisfy the Wilsonian matching: since
the vector current correlator has no scalar meson contributions, we would have 
$\Pi_V=0$, were it not for the $\rho$ contribution,
and hence
$-Q^2 \frac{d}{d Q^2} \Pi_V|_{Q^2=\Lambda^2}=0$,  whereas QCD yields non-zero
value
$-Q^2 \frac{d}{d Q^2} 
\Pi_V^{\rm (QCD)}|_{Q^2=\Lambda^2}=
(1+\delta_V^{\rm crit}) \, N_c/(24\pi^2) \ne 0$. The vanishing $\Pi_V$ together with
the restoration requirement $\Pi_A=\Pi_V$ would imply $\Pi_A=0$, which
is in contradiction with the Wilsonian matching for $\Pi_A$:
$Q^2 \frac{d}{d Q^2} 
\Pi_A|_{Q^2=\Lambda^2}=
F_\pi^2(\Lambda)/\Lambda^2 =-Q^2 \frac{d}{d Q^2} 
\Pi_A^{\rm (QCD)}|_{Q^2=\Lambda^2}=(1+\delta_A^{\rm crit})\, N_c/(24\pi^2)
\ne 0$. 

The fact that both the effective theory and the underlying theory give
the same infrared physics  
is an aspect of the duality of Seiberg-type first observed in the SUSY
QCD: 
in the case at hand, non-SUSY QCD, we found that the HLS plays a role
of  
the ``magnetic gauge theory'' dual to the QCD as the ``electric gauge
theory''. 
Here we recall that the phase structure of the SUSY QCD was
revealed by Seiberg only in terms of 
the effective theory in the sense of Wilsonian effective action. 
In this paper we have
demonstrated that the same is true also in the non-SUSY QCD, namely
the  
Wilsonian RGEs (including the quadratic divergence) 
in the effective field theory approach are
 very powerful tool to investigate the phase 
structure of the QCD and the QCD-like gauge theories.

In Sec.~\ref{sec:RALOLET},
we gave a brief review of the proof of the low-energy theorem
of the HLS, $g_\rho = 2 g_{\rho\pi\pi} F_\pi^2$ at any loop order
following Refs.~\cite{HKY:PRL,HKY:PTP}.
We showed that the inclusion of the quadratic divergence
does not change the proof, which
implies that the low-energy theorem of the HLS is valid at any loop
order even under the existence of the quadratic divergence.

Finally in Sec.~\ref{sec:THDMC},
we gave a brief review on the application of the approach
explained in previous sections to the 
hot and/or dense matter calculation based on
Refs.~\cite{Harada-Sasaki,Harada-Kim-Rho}.
We have summarized how
the VM takes place at the chiral restoration point
in hot matter at zero density~\cite{Harada-Sasaki} and also in
dense matter at zero temperature~\cite{Harada-Kim-Rho}.
The picture based on the VM in hot matter would provide
several peculiar 
predictions on, e.g., the vector and 
axialvector susceptibilities~\cite{Harada-Kim-Rho-Sasaki},
the vector dominance of the electromagnetic form factor of 
pion~\cite{Harada-Sasaki:2}, and so on
which can be checked in the
experiments in operation
as well as in future experiments.
These analysis are still developing, so we did not
include the review in this report.
We encourage those who have interest
to read Refs.~\cite{Harada-Kim-Rho-Sasaki,Harada-Sasaki:2}

Several comments are in oder:

One might suspect that the limit of $m_\rho \rightarrow 0$ would be
problematic since the on-shell amplitude would have a factor
$\frac{1}{m_\rho^2}$ in the (longitudinal) polarization tensor
$\epsilon_\mu^{(0)}$ 
and thus divergent in such a limit. In our case such a polarization
factor  
is always accompanied by a gauge coupling $g^2$ of $\rho$, which
yields 
the amplitude a factor 
$\frac{g^2}{m_\rho^2} \sim \frac{1}{F_\sigma^2(m_\rho)}
\sim \frac{1}{F_\pi^2(m_\rho)}$. Then the above problem is not a
peculiarity of 
the massive vector mesons in our HLS model but is simply reduced to
the  
similar problem as in the nonlinear sigma model 
at the chiral restoration
point.  
In the nonlinear chiral Lagrangian without quadratic divergence, 
the on-shell amplitude like $\pi$-$\pi$ scattering behaving like
$A(p^2) = \frac{p^2}{F_\pi^2(0)}$ 
[see Eq.~(\ref{pipi amplitude})]
would be divergent at 
the chiral restoration point $F_\pi(0) \rightarrow 0$, which 
simply implies that 
the EFT is valid only for $p <F_\pi(0)$, namely the
validity 
region is squeezed out at the restoration point 
$F_\pi(0) \rightarrow 0$.
Such a problem does not exist in the linear sigma model (or Higgs
Lagrangian in the SM) thanks to the light scalar meson (Higgs boson)
introduced in addition to the NG boson $\pi$.~\footnote{%
  In the linear sigma model 
  having a scalar meson (Higgs boson) in
  addition to 
  the NG boson $\pi$,
  the $\pi$-$\pi$ scattering amplitude is expressed as
  $A(s) =
  \lambda + \frac{2 (\lambda F_\pi)^2}{ s - M_S^2 }$,
  where $M_S^2 = 2 \lambda F_\pi^2$ with $\lambda$
  being the four-point coupling.
  In the broken phase ($F_\pi\neq 0$) we can easily see that
  $A(s=0)=0$ consistently with the low-energy theorem, which
  holds even we approach the chiral restoration point 
  ($F_\pi\rightarrow 0$).
  In the symmetric phase ($F_\pi \equiv 0$),
  on the other hand, we have $A(s\neq0)=\lambda \neq 0$ which holds
  even at 
  the low-energy limit $s\rightarrow0$: $A(s=0) \ne 0$. 
  In any case the amplitude is non-singular, although the low-energy
  limit amplitude 
  is discontinuous across the phase transition point.
 }
However, in our case where the quadratic divergence is included,
the problem is also solved in a similar way even 
without the additional scalar meson as follows:
The amplitude is expected to behave as $\frac{p^2}{F_\pi^2(p^2)}$ 
$\sim X(\mu^2 = p^2)$, with  $X(\mu)$ defined in Eq.~(\ref{def X}),
which is non-singular in the $m_\rho \rightarrow 0$ limit
as we have
discussed around the end of Sec.~\ref{sssec:VM as a limit}.
Actually,  the amplitude
 $A(s) \sim \frac{s}{F_\pi^2(s)}= X(s)$ has a vanishing low-energy 
limit $X(0)=0$,
as far as we approach the chiral restoration point 
$F_\pi^2(0)\rightarrow0$ 
from the broken phase
$F_\pi^2(0)\neq0$, i.e., 
the VM limit with $m_\rho \rightarrow 0$. 
On the other hand, when the theory is exactly on the VM point,
$X(s)$ is a certain (non-zero) constant, i.e.,
$X(s)\equiv (\mbox{constant})$ which leads to
$X(s) \rightarrow (\mbox{constant}) \neq 0$ even at
the $s\rightarrow0$ limit. 
Although the low-energy limit amplitude $A(0)$
is discontinuous
across the phase transition point, the amplitude is non-singular at
the phase transition point similarly to the linear sigma model, 
in sharp contrast 
to the conventional nonlinear sigma model without quadratic 
divergence. Thus our case is a counter example against the folklore that
the massive vecor meson theory has a problem in the massless limit, unless
the mass is via Higgs mechanism with the additional light 
scalar meson (Higgs boson).

The axialvector mesons $A_1$ including $a_1$ are heavier than the
matching 
scale, $\Lambda= 1.1\sim 1.2$\,GeV, so that we did not include
them in the analysis based on the Wilsonian matching in 
Sec.~\ref{sec:WM}. 
It was checked that even including $A_1$ does not substantially
change the value of $F_\pi^2(\Lambda)$  given by the OPE and hence
does not affect the qualitative 
feature of our analysis for $N_f=3$. We then expect that
{\it the $A_1$ in the VM is resolved 
and/or decoupled from the axialvector current near the critical
point}, since the $\rho$ is already balanced with the $\pi$ and there
is no contribution in the vector current 
correlator to be matched with the additional contribution in the
axialvector current 
correlator.
 
On the other hand,
the recent 
analyses~\cite{Harada-Sannino-Schechter:PRD,
Harada-Sannino-Schechter:PRL,Tornqvist-Roos,IITITT:96,%
Morgan-Pennington,Janssen-Pearce-Holinde-Speth}
show that there exist light 
scalar mesons, some of which has a mass smaller 
than our matching scale $\Lambda \simeq 1.1 {\rm GeV}$.
However, the scalar meson
does not couple to the axialvector and vector currents, anyway.
We expect that 
{\it the scalar meson is also resolved and/or decoupled near
the chiral phase transition point}, since it is in
 the pure $(N_f,N_f^*)\oplus (N_f^*,N_f)$ representation together
with the $A_1$ in the VM limit.

We did not include the loop effects of the nucleon or constituent quarks which
would become massless near the chiral restoration point. Inclusion 
of these would affect the result in this report. Such effects were studied 
by the ladder SD equation
where the meson loop effects were ignored, instead. Since both approaches
yield qualitatively the same result, there might exist some kind of
duality between them.

In this report we applied the VM to the chiral
restoration in the large $N_f$ QCD.
It may be checked by the lattice
simulation:  
As we obtained from a simple expectation
around Eq.~(\ref{mr2 ov fp2 VM})
and explicitly formulated in Sec.~\ref{sssec:CB},
the VM generally implies
\begin{equation}
\frac{m_\rho^2 }{F_\pi^2(0)} \rightarrow 0 \ ,
\label{mr2 ov fp2 VM 2}
\end{equation}
which is a salient feature of the VM~\cite{HY:VM}.
This will be a clear
indication of the VM and may be testable in the
lattice simulations.

The results of Refs.~\cite{Harada-Sasaki,Harada-Kim-Rho} 
shown in
section~\ref{sec:THDMC}
imply that
the position of the $\rho$ peak of the dilepton spectrum will move to
the lower energy region in accord with the picture shown in
Ref.~\cite{Brown-Rho:91,Brown-Rho:96,Brown-Rho:01a,Brown-Rho:01b}.
In the analysis 
we did not study the temperature dependence 
of the $\rho$ width.
However, 
when the scaling properties of the parameters in hot QCD
are equal to those in large flavor QCD,
Eqs.~(\ref{critical of mrho}),
(\ref{eq:grho}) and (\ref{eq:grhopipi})
would further imply smaller $\rho$ width 
 and
larger peak value  near the critical point
[see Eqs.~(\ref{crit gam/mr}) and 
(\ref{crit ee pipi})]:~\cite{HY:VM}
\begin{eqnarray}
&&\Gamma/m_\rho \sim g_{\rho\pi\pi}^2 \sim 
  f(\epsilon)\rightarrow 0 \ ,\\
&&\Gamma_{ee}\Gamma_{\pi\pi}/\Gamma^2
\sim g_\rho^2/(g_{\rho\pi\pi}^2 m_\rho^4)
\sim 1/f^2(\epsilon) \rightarrow \infty \ .
\end{eqnarray} 
If it is really the 
case, these would  be clear signals of VM tested in the future
experiments.

The VM reviewed in section~\ref{sec:VM}
may be applied to the models for the composite
$W$ and $Z$.  Our analysis shows that the mass of the
composite vector boson approaches to zero
faster than the order parameter,
which is fixed to the electroweak symmetry breaking scale,
near the critical point:
\begin{equation}
m_\rho^2 \ll F_\pi^2(0) \simeq (250 \,{\rm GeV})^2\,
\end{equation}
in accord with the mass of $W$ and $Z$ bosons being smaller than $250\, {\rm GeV}$. Moreover, near the VM point 
the composite theory becomes a weakly-coupled gauge theory
of the light gauge and NG bosons, while the underlying gauge theory
is still in the strongly-coupled phase 
with confinement and chiral symmetry breaking.
Such a situation has been hardly realized in the conventional
strongly-coupled dynamics for the composite gauge boson. 
The VM may also be applied to the technicolor 
with light techni-$\rho$.

In the present analysis
we worked in the chiral limit with neglecting the effects
from the current quark masses which 
explicitly break the chiral symmetry.
For comparing the predictions for the system of the mesons other than
the $\rho$ and $\pi$ such as $K^\ast$ and $K$ 
with experiment,
we need to include the effects from the explicit breaking terms.
Such analysis is also important for 
lattice analysis.
In several 
analyses~(see, e.g., Ref.~\cite{CP-PACS:01})
where the chiral limit is usually taken by just the
linear extraporation.
However, the chiral perturbation with systematically
including the vector meson will generate the chiral
logarithms in the chiral corrections to the vector meson masses.
The chiral logarithms 
in the chiral perturbation
theory in the light pseudoscalar meson system
plays an important role, so that
the inclusion of them in the chiral corrections to the vector meson
masses is important to extrapolate the lattice results to the chiral
limit.\\

In conclusion we have developed an effective field theory of QCD and 
QCD-like theories based on the HLS model. In contrast to other vector
meson models which are all equivalent to the HLS model {\it at 
tree level}, we have provided a well-organized quantum field theory
and thus established {\it a theory} as a precise science which goes 
beyond a mere mnemonic of hadron phenomenology. In particular, we have  
presented a novel dynamical possibility for the
chiral phase  
transition which is materialized through the quantum effects of the
HLS model as the effective field theory in such a way that 
the bare parameters of the HLS model are determined through matching with the
underlying QCD-like theories.  
We do hope that it will shed some deeper insights
 into the strong coupling gauge theories
and the concept of the composite gauge boson as well as the various 
possible phases
of the hadronic matter.

\newpage

\mmsectionb{Acknowledgements}

We would like to thank Gerry Brown and Mannque Rho for useful
discussions and continuous encouragements. We appreciate 
discussions with 
Tom Appelquist, Howard Georgi, Kazushi Kanaya, Yoshio Kikukawa, 
Youngman Kim,
Taichiro Kugo,
Ken Lane, Volodya Miransky, Chihiro Sasaki, 
Masaharu Tanabashi,
Scott Thomas and 
Arkady Vainshtein. MH would like to thank Mannque Rho for
his hospitality during the stay at KIAS, Gerry Brown for
his hospitality during the stay at SUNY at Stony Brook
and Dong-Pil Min for his hospitality during the stay at
Seoul National University
where part of this work was done.
Part of this work was done when KY was staying 
at Aspen Center for Physics in 2002 summer. 
The work is supported in part by the
JSPS Grant-in-Aid for Scientific Research (B) (2) 14340072 and
11695030.
The work of MH was supported in part by 
USDOE Grant \#DE-FG02-88ER40388 and
the
Brain Pool program (\#012-1-44) provided by the Korean Federation
of Science and Technology Societies.

\newpage

\appendix

\renewcommand{\theequation}{\Alph{section}.\arabic{equation}}

\section{Convenient Formulae}
\label{sec:CF}

\subsection{Formulae for Feynman integrals}
\label{ssec:FFI}

Let us consider the following Feynman integrals:
\begin{eqnarray}
A_0(M^2) &\equiv&
\int \frac{d^nk}{i(2\pi)^n}
\frac{1}{M^2-k^2}
\ ,
\label{def:A0}
\\
B_0(p^2;M_1,M_2) &\equiv&
\int \frac{d^nk}{i(2\pi)^n}
\frac{1}{ [M_1^2-k^2] [M_2^2-(k-p)^2] }
\ ,
\label{def:B0}
\\
B^\mu(p;M_1,M_2) &\equiv&
\int \frac{d^nk}{i(2\pi)^n}
\frac{k^\mu}{ [M_1^2-k^2] [M_2^2-(k-p)^2] }
\ ,
\label{def:Bmu}
\\
B^{\mu\nu}(p;M_1,M_2) &\equiv&
\int \frac{d^nk}{i(2\pi)^n}
\frac{\left(2k-p\right)^\mu \left(2k-p\right)^\nu}{%
 [M_1^2-k^2] [M_2^2-(k-p)^2] }
\ .
\label{def:Bmunu}
\end{eqnarray}
$A_0(M^2)$ and $B^{\mu\nu}(p;M_1,M_2)$ 
are quadratically divergent.
Since a naive momentum cutoff violates the chiral symmetry,
we need a careful treatment of the quadratic divergences.
As discussed in section~\ref{sec:CPHLS}, we adopt the dimensional
regularization and identify the
quadratic divergences with the presence of poles
of ultraviolet origin at $n=2$~\cite{Veltman}.
This can be done by the following replacement in the Feynman
integrals [see Eq.~(\ref{regularization 0})]:
\begin{equation}
\int \frac{d^n k}{i (2\pi)^n} \frac{1}{-k^2} \rightarrow 
\frac{\Lambda^2} {(4\pi)^2} \ ,
\qquad
\int \frac{d^n k}{i (2\pi)^n} 
\frac{k_\mu k_\nu}{\left[-k^2\right]^2} \rightarrow 
- \frac{\Lambda^2} {2(4\pi)^2} g_{\mu\nu} \ .
\end{equation}
As is usual, the logarithmic divergence is identified with the pole at 
$n=4$ by [see Eq.~(\ref{logrepl})]
\begin{equation}
\frac{1}{\bar{\epsilon}} + 1 \equiv
\frac{2}{4 - n } - \gamma_E + \ln (4\pi) + 1
\rightarrow \ln \Lambda^2 \ ,
\label{logrepl2}
\end{equation}
where $\gamma_E$ is the Euler constant.

Now, $A_0(M^2)$ is evaluated as
\begin{equation}
A_0(M^2) = \frac{\Lambda^2} {(4\pi)^2}
- \frac{M^2}{(4\pi)^2}
\left[
  \frac{1}{\bar{\epsilon}} + 1 - \ln M^2
\right]
\ .
\end{equation}
$B_0(p^2;M_1,M_2)$ and $B^\mu(p;M_1,M_2)$ are evaluated as
\begin{eqnarray}
B_0(p^2;M_1,M_2) &=& \frac{1}{(4\pi)^2}
\left[
  \frac{1}{\bar{\epsilon}} - F_0(p^2;M_1,M_2)
\right]
\ ,
\nonumber\\
B^\mu(p;M_1,M_2) &=& p^\mu B_1(p^2;M_1,M_2) 
\ ,
\end{eqnarray}
where
\begin{equation}
B_1(p^2;M_1,M_2) \equiv
\frac{1}{(4\pi)^2} \left[
  \frac{1}{2} \frac{1}{\bar{\epsilon}} - F_1(p^2;M_1,M_2)
\right] \ .
\end{equation}
$B^{\mu\nu}(p;M_1,M_2)$ is evaluated as
\begin{eqnarray}
B^{\mu\nu}(p;M_1,M_2) &=&
- g^{\mu\nu} \left[ A_0(M_1) + A_0(M_2)  - B_A(p^2;M_1,M_2) \right]
\nonumber\\
&&
{}- \left( g^{\mu\nu} p^2 - p^\mu p^\nu \right)
  \left[ B_0(p^2;M_1,M_2) - 4 B_3(p^2;M_1,M_2) \right]
\ ,
\label{rel: Bmn}
\end{eqnarray}
where
\begin{eqnarray}
B_3(p^2;M_1,M_2) &\equiv&
\frac{1}{(4\pi)^2} \left[
  \frac{1}{6} \frac{1}{\bar{\epsilon}} - F_3(p^2;M_1,M_2)
\right] \ ,
\nonumber\\
B_A(p^2;M_1,M_2) &\equiv&
\frac{1}{(4\pi)^2} \left( M_1^2 - M_2^2 \right)
F_A(p^2;M_1,M_2) 
\ .
\end{eqnarray}
The definitions of $F_0$, $F_A$, $F_1$ and $F_3$ and formulas
are given in Appendix~\ref{ssec:FPI}.

Here we summarize the divergent parts of the Feynman integrals
which are used in Sec.~\ref{ssec:TPFOL} to obtain the divergent
corrections to the parameters $F_\pi$, $F_\sigma$ and $g$:
\begin{eqnarray}
  \left. A_0(M^2) \right\vert_{\rm div} 
&=&
  \frac{\Lambda^2}{(4\pi)^2} - 
  \frac{M^2}{(4\pi)^2} \ln \Lambda^2 \ ,
\label{div:A0}
\\
  \left. B_0(p^2;M_1,M_2) \right\vert_{\rm div} 
&=&
  \frac{1}{(4\pi)^2} \ln \Lambda^2 \ ,
\label{div:B0}
\\
  \left. B^\mu(p;M_1,M_2) \right\vert_{\rm div} 
&=&
  \frac{p^\mu}{2(4\pi)^2} \ln \Lambda^2 \ ,
\label{div:Bmu}
\\
  \left. B^{\mu\nu}(p;M_1,M_2) \right\vert_{\rm div} 
&=&
  - g^{\mu\nu} \frac{1}{(4\pi)^2}
    \left[ 2 \Lambda^2 - ( M_1^2 + M_2^2 ) \ln \Lambda^2 \right]
\nonumber\\
&& \ 
  - \left( g^{\mu\nu}p^2 - p^\mu p^\nu \right) 
    \frac{1}{3(4\pi)^2} \ln \Lambda^2 \ .
\label{div:Bmunu}
\end{eqnarray}

\subsection{Formulae for parameter integrals}
\label{ssec:FPI}

Several parameter integrals are given as follows:
\begin{eqnarray}
F_0 (s;M_1,M_2) &=& \int^1_0 dx \ln 
\left[ (1-x) M_1^2 + x M_2^2 - x (1-x) s \right] \ , \nonumber\\
F_A (s;M_1,M_2) &=& \int^1_0 dx\,(1-2x)\, \ln 
\left[ (1-x) M_1^2 + x M_2^2 - x (1-x) s \right] \ , \nonumber\\
F_1 (s;M_1,M_2) &=& \int^1_0 dx \, x \ln 
\left[ (1-x) M_1^2 + x M_2^2 - x (1-x) s \right] \ , \nonumber\\
F_2 (s;M_1,M_2) &=& \int^1_0 dx \, x^2 \ln 
\left[ (1-x) M_1^2 + x M_2^2 - x (1-x) s \right] \ , \nonumber\\
F_3 (s;M_1,M_2) &=& \int^1_0 dx \, x (1-x) \ln 
\left[ (1-x) M_1^2 + x M_2^2 - x (1-x) s \right] \ , \nonumber\\
F_4 (s;M_1,M_2) &=& \int^1_0 dx \, 
\left( (1-x) M_1^2 + x M_2^2 \right)
\ln \left[ (1-x) M_1^2 + x M_2^2 - x (1-x) s \right] \ , \nonumber\\
F_5 (s;M_1,M_2) &=& \int^1_0 dx \, 
\left( (1-2x) (M_1^2 - M_2^2) + (1-2x)^2 s \right)
\nonumber\\
&& \ \ \ \times
\ln \left[ (1-x) M_1^2 + x M_2^2 - x (1-x) s \right] 
\ ,
\nonumber\\
F_6(s;M_1,M_2) &=& F_4(s;M_1,M_2) - s F_3(s;M_1,M_2)
\ .
\end{eqnarray}
These are given by
\begin{eqnarray}
\lefteqn{
F_0 (s;M_1,M_2) =
\bar{L}(s;M_1,M_2) 
+ \frac{M_1^2-M_2^2}{s}\ln\frac{M_1}{M_2} - 2 + \ln( M_1 M_2) \ ,
}
\nonumber\\
\lefteqn{
  F_A(s;M_1,M_2) =
  - \frac{M_1^2-M_2^2}{s} 
  \biggl[ F_0(s;M_1,M_2) - F_0(0;M_1,M_2) \biggr]
  \ ,
}
\nonumber\\
\lefteqn{
  F_1(s;M_1,M_2) = \frac{1}{2} 
  \biggl[ F_0(s;M_1,M_2) - F_A(s;M_1,M_2) \biggr]
\ ,
}
\nonumber\\
\lefteqn{
  F_2(s;M_1,M_2) =  F_1(s;M_1,M_2) - F_3(s;M_1,M_2)
\ ,
}
\nonumber\\
\lefteqn{
  F_3 (s;M_1,M_2) =
  \frac{1}{4} F_0(s;M_1,M_2) 
}
\nonumber\\
&& \ \ 
{} - \frac{1}{12} \left( 1 - \frac{2(M_1^2+M_2^2)}{s} \right)
    \left[ F_0(s;M_1,M_2)  - F_0(0;M_1,M_2) \right]
\nonumber\\
&& \ \ 
{}- \frac{(M_1^2-M_2^2)^2}{3s^2} 
  \left[ 
    F_0(s;M_1,M_2)  - F_0(0;M_1,M_2) - s F'_0(0;M_1,M_2)
  \right]
\nonumber\\
&& \ \ 
{}- \frac{1}{12} F_0(0;M_1,M_2) + \frac{1}{18}
\ ,
\nonumber\\
\lefteqn{
  F_4(s;M_1,M_2) = 
  \frac{M_1^2+M_2^2}{2} F_0(s;M_1,M_2) 
  + \frac{M_1^2-M_2^2}{2} F_A(s;M_1,M_2) 
  \ ,
}
\nonumber\\
\lefteqn{
  F_5(s;M_1,M_2) = 
  (M_1^2-M_2^2) F_A(s;M_1,M_2) + s F_0(s;M_1,M_2) 
  - 4 s F_3(s;M_1,M_2) 
  \ ,
}
\nonumber\\
\lefteqn{
  F_6(s;M_1,M_2) = 
  F_4(s;M_1,M_2) - s F_3(s;M_1,M_2) 
  \ ,
}
\end{eqnarray}
where
\begin{eqnarray}
\bar{L}(s;M_1,M_2) \equiv
\left\{
\begin{array}{l}
\displaystyle
- \frac{2}{s} \sqrt{ (M_1+M_2)^2-s } \sqrt{ (M_1-M_2)^2-s }
\\
\\
\displaystyle
\ \quad
\times
\ln \frac{ \sqrt{ (M_1+M_2)^2-s } + \sqrt{ (M_1-M_2)^2-s } }
  { 2 \sqrt{ M_1 M_2} }
\ ,
\\
\qquad\qquad\qquad\qquad\qquad\qquad (\mbox{for $s<(M_1-M_2)^2$}) \ ,
\\
\\
\displaystyle
\frac{2}{s} \sqrt{ (M_1+M_2)^2-s } \sqrt{ s - (M_1-M_2)^2 } 
\times
\tan^{-1} \sqrt{ \frac{ s - (M_1-M_2)^2 }{ (M_1+M_2)^2-s } } \ ,
\\
\qquad\qquad\qquad\qquad\qquad 
  (\mbox{for $(M_1-M_2)^2<s<(M_1+M_2)^2$}) \ ,
\\
\\
\displaystyle
\frac{2}{s} \sqrt{ s - (M_1+M_2)^2 } \sqrt{ s - (M_1-M_2)^2 } \\
\\
\displaystyle
\ \ \times
\left[
  \ln \frac{ \sqrt{ s-(M_1+M_2)^2 } + \sqrt{ s-(M_1-M_2)^2 } }
  {2 \sqrt{ M_1 M_2}}
  - i \pi
\right]
\ ,
\\
\qquad\qquad\qquad\qquad\qquad\qquad (\mbox{for $(M_1+M_2)^2<s$}) \ .
\end{array}
\right.
\ ,
\end{eqnarray}
and
\begin{eqnarray}
F_0(0;M_1,M_2) &=& \frac{M_1^2+M_2^2}{M_1^2-M_2^2} 
  \ln \frac{M_1}{M_2}
-1 + \ln ( M_1 M_2 )
\ ,
\\
F_0^\prime (0;M_1,M_2) &=& 
- \frac{M_1^2+M_2^2}{ 2 (M_1^2-M_2^2)^2 }
+ \frac{M_1^2 M_2^2}{(M_1^2-M_2^2)^3} \ln \frac{M_1^2}{M_2^2}
\ .
\end{eqnarray}
The following formulae are convenient:
\begin{eqnarray}
F_0(0;M,M) &=& \ln M^2
\ ,
\\
F_A(0;M_1,M_2) &=&
\frac{M_1^2+M_2^2}{ 2 (M_1^2-M_2^2) }
- \frac{M_1^2 M_2^2}{(M_1^2-M_2^2)^2} \ln \frac{M_1^2}{M_2^2}
\ ,
\\
F_3(0;M,0) &=&
\frac{1}{6} \ln M^2 - \frac{5}{36}
\ .
\end{eqnarray}

\subsection{Formulae for generators}
\label{ssec:FG}

Let me summarize useful formulae for the sum in terms the generators
of SU($N_f$).
In the following the generators are normalized as
\begin{equation}
\mbox{tr} \left[ T_a T_b \right] = \frac{1}{2} \delta_{ab} \ .
\end{equation}

The basic formulae for SU($N_f$) generators are given by
\begin{eqnarray}
&& \sum_{a=1}^{N_f^2-1} \mbox{tr} \left[ T_a A T_a B \right] 
= - \frac{1}{2N_f} \mbox{tr} \left[ A B \right]
+ \frac{1}{2} \mbox{tr} \left[ A \right] \mbox{tr} \left[ B \right] \ ,
\\
&& \sum_{a=1}^{N_f^2-1} \mbox{tr} \left[ T_a A \right]
\mbox{tr} \left[ T_a B \right] 
= \frac{1}{2} \mbox{tr} \left[ A B \right] 
- \frac{1}{2N_f} \mbox{tr} \left[ A \right] \mbox{tr} \left[ B \right] \ ,
\end{eqnarray}
where $A$ and $B$ are arbitrary $N_f\times N_f$ matrices.

Below we list several convenient formulae for generators:
\begin{eqnarray}
&&
\sum_{a = 1}^{N_f^2-1} \mbox{tr} \left[ A T_a T_a \right]
= \frac{N_f^2-1}{2N_f} \mbox{tr} \left[ A \right] \ ,
\\
&&
\sum_{a = 1}^{N_f^2-1} 
\mbox{tr} \biggl[ C \left[ A \,,\, T_a \right] \biggr] 
\mbox{tr} \biggl[ D \left[ B \,,\, T_a \right] \biggr] 
= \frac{1}{2} \mbox{tr} 
\biggl[ \left[ C \,,\, A \right] \left[ D \,,\, B \right] \biggr] 
\ ,
\\
&&
\sum_{a,b = 1}^{N_f^2-1} 
\mbox{tr} \biggl[ T_a \left[ A \,,\, T_b \right] \biggr] 
\mbox{tr} \biggl[ T_a \left[ B \,,\, T_b \right] \biggr] 
\nonumber\\
&& \quad
= - \frac{N_f}{2} \mbox{tr} \left[ A B \right]
+ \frac{1}{2} \mbox{tr} \left[ A \right] \mbox{tr} \left[ B \right]
= - \frac{N_f}{2} \mbox{tr} \left[ 
  \widetilde{A} \widetilde{B} \right]
\ ,
\\
&&
\sum_{a = 1}^{N_f^2-1} 
\mbox{tr} \biggl[ \left\{ A \,,\, B \right\}
\left\{ T_a \,,\, T_a \right\} \biggr] 
= \frac{2(N_f^2-1)}{N_f} \, \mbox{tr} \left[ A B \right]
\ ,
\\
&&
\sum_{a = 1}^{N_f^2-1} 
\mbox{tr} \biggl[ \left\{ A \,,\, T_a \right\}
\left\{ B \,,\, T_a \right\} \biggr] 
= \frac{N_f^2-2}{N_f} \, \mbox{tr} \left[ A B \right]
+ \mbox{tr} \left[ A \right] \mbox{tr} \left[ B \right]
\ ,
\\
&&
\sum_{a = 1}^{N_f^2-1} 
\mbox{tr} \biggl[ \left[ A \,,\, T_a \right]
\left[ B \,,\, T_a \right] \biggr] 
\nonumber\\
&& \quad
= 
- N_f \, \mbox{tr} \left[ A B \right]
+ \mbox{tr} \left[ A \right] \mbox{tr} \left[ B \right]
= 
- N_f \, \mbox{tr} \left[ \widetilde{A} \widetilde{B} \right]
\ ,
\\
&&
\sum_{a,b = 1}^{N_f^2-1} 
\mbox{tr}\, \biggl[ A \left\{ T_a \,,\, T_b \right\} \biggr]
\mbox{tr}\, \biggl[ B \left\{ T_a \,,\, T_b \right\} \biggr]
= 
\frac{N_f^2-4}{2N_f} \mbox{tr}\, \left[ A B \right]
+ \frac{N_f^2+2}{2N_f^2} \mbox{tr}\, \left[ A \right]
\mbox{tr}\, \left[ B \right]
\ ,
\\
&&
\sum_{a,b = 1}^{N_f^2-1} 
\mbox{tr}\, \biggl[ C \left\{ T_a \,,\, T_b \right\} \biggr]
\mbox{tr} \biggl[ \left[ A \,,\, T_a \right]
\left[ B \,,\, T_b \right] \biggr] 
\nonumber\\
&& \quad
=
- \frac{N_f}{4} \mbox{tr}\, \biggl[
  \left\{ A \,,\, B \right\} C
\biggr]
+
\frac{1}{2} \Biggl(
  \mbox{tr} \, \left[ A \right] \mbox{tr} \, \left[ BC \right]
  + \mbox{tr} \, \left[ B \right] \mbox{tr} \, \left[ AC \right]
  - \mbox{tr} \, \left[ C \right] \mbox{tr} \, \left[ AB \right]
\Biggr)
\nonumber\\
&& \quad
=
- \frac{N_f}{4} \mbox{tr}\, \biggl[
  \left\{ \widetilde{A} \,,\, \widetilde{B} \right\} C
\biggr]
- \frac{1}{2}
\mbox{tr} \, \left[ C \right] 
\mbox{tr} \, \left[ \widetilde{A} \widetilde{B} \right]
\end{eqnarray}
where $\widetilde{A}$ and $\widetilde{B}$ are the traceless parts of
$A$ and $B$, respectively:
\begin{eqnarray}
\widetilde{A} &\equiv& A - \frac{1}{N_f} \mbox{tr} \,\left[ A \right]
\ ,
\nonumber\\
\widetilde{B} &\equiv& B - \frac{1}{N_f} \mbox{tr} \,\left[ B \right]
\ .
\end{eqnarray}

\subsection{Incomplete gamma function}
\label{ssec:IGF}

The incomplete gamma function is defined by
\begin{equation}
\Gamma\left( j  , \varepsilon \right)
\equiv
\int^\infty_\varepsilon \frac{dz}{z} e^{-z} z^j \ .
\label{def:IGF}
\end{equation}
For $j = \mbox{integer} \ge 1$ these satisfy
\begin{eqnarray}
\Gamma\left( 1  , \varepsilon \right) &=& e^{-\varepsilon}
\ ,
\\
\Gamma\left( j \ge 2  , \varepsilon \right) &=& 
e^{-\varepsilon} \varepsilon^{j-1} + (j-1) 
\Gamma\left( j -1  , \varepsilon \right)
\ .
\end{eqnarray}
The incomplete gamma functions for $j=0$ are approximately given by
\begin{equation}
\Gamma\left( 0  , \varepsilon \right) \simeq
\ln \left( \frac{1}{\varepsilon} \right) \ .
\end{equation}
For $j = \mbox{integer} < 0$ ($j\ge-2$) the
incomplete gamma functions are given by
\begin{eqnarray}
\Gamma\left( -1  , \varepsilon \right) &=& 
\frac{1}{\varepsilon} e^{-\varepsilon}  - 
\Gamma\left( 0  , \varepsilon \right) \simeq
\frac{1}{\varepsilon} - \ln \left( \frac{1}{\varepsilon} \right)
\ ,
\label{approx:IGF 1}
\\
\Gamma\left( -2  , \varepsilon \right) &=& 
\frac{1}{2}
\left[ 
  \frac{1}{\varepsilon^2} e^{-\varepsilon}  - 
  \Gamma\left( -1  , \varepsilon \right) 
\right]
\simeq
\frac{1}{2}
\left[ 
  \frac{1}{\varepsilon^2} -
  \frac{1}{\varepsilon} + \ln \left( \frac{1}{\varepsilon} \right)
\right]
\ .
\label{approx:IGF 2}
\end{eqnarray}

\subsection{Polarization tensors at non-zero temperature}
\label{ssec:polar}

In this subsection we list the polarization tensor at non-zero
temperature, and give several convenient formulae among them.
These polarization tensors are used in the calculation
at non-zero temperature given in Sec.~\ref{sec:THDMC}.
At non-zero temperature, 
the polarization tensor is no longer restricted to be 
Lorentz covariant, but only $O(3)$ covariant. 
Then the polarization tensors
can be expressed by four independent symmetric $O(3)$ tensors. Here we
list the polarization tensors at non-zero
temperature:~\cite{Toimele,NDoray}
\begin{eqnarray}
P_{T\mu\nu} &=&
  g_{\mu i} \left( 
    \delta_{ij} - 
    \frac{\vec{p}_i\vec{p}_j}{\left\vert\vec{p}\right\vert^2}
  \right)g_{j \nu} \nonumber \\
&=& \left\{
\begin{array}{l}
P_{T00} = P_{T0i} = P_{Ti0} = 0 \ , \\ 
P_{Tij} = \delta_{ij} - \frac{\vec{p}_i\vec{p}_j}{\left\vert\vec{p}%
\right\vert^2} \ ,
\end{array}
\right.
\nonumber \\
  P_{L\mu\nu} 
&\equiv& 
  - \left( g_{\mu\nu} - \frac{p_\mu p_\nu}{p^2} \right) -
  P_{T\mu\nu}   
\nonumber \\
&=&
  \left(g_{\mu 0}-\frac{p_\mu p_0}{p^2} \right) 
  \frac{p^2}{\left\vert\vec{p}\right\vert^2}
  \left(g_{0 \nu}-\frac{p_0 p_\nu}{p^2} \right) 
\ , 
\nonumber \\
  P_{C\mu\nu} 
&\equiv& 
  \frac{1}{\sqrt{2}\left\vert\vec{p}\right\vert} 
  \left[
    \left( g_{\mu0} - \frac{p_\mu p_0}{p^2} \right) p_\nu 
    + p_\mu \left( g_{0\nu} - \frac{p_0 p_\nu}{p^2} \right) 
  \right] 
\ ,
\nonumber \\
  P_{D\mu\nu} 
&\equiv&
  \frac{p_\mu p_\nu}{p^2} 
\ , 
\label{polar tensor}
\end{eqnarray}
where $p^\mu = (p_0,\vec{p})$ is four-momentum.

The following formulas are convenient:~\footnote{%
  There is an error in the third formula 
  in Ref.~\cite{Harada-Shibata}.
}
\begin{eqnarray}
P_{L\mu\alpha} P_{L}^{\alpha\nu} &=& - {P_{L\mu}}^\nu \ ,
\nonumber\\
P_{T\mu\alpha} P_{T}^{\alpha\nu} &=& - {P_{T\mu}}^\nu \ ,
\nonumber\\
P_{C\mu\alpha} P_{C}^{\alpha\nu} &=& \frac{1}{2}
\left( {P_{L\mu}}^\nu - {P_{D\mu}}^\nu \right) \ ,
\nonumber\\
P_{D\mu\alpha} P_{D}^{\alpha\nu} &=& {P_{D\mu}}^\nu \ ,
\nonumber\\
P_{L\mu\alpha} P_{T}^{\alpha\nu} &=& 
P_{C\mu\alpha} P_{T}^{\alpha\nu} =
P_{D\mu\alpha} P_{T}^{\alpha\nu} =
P_{D\mu\alpha} P_{L}^{\alpha\nu} = 0 \ ,
\nonumber\\
P_{C\mu\alpha} P_{L}^{\alpha\nu} &=& 
- P_{D\mu\alpha} P_{C}^{\alpha\nu} =
- \frac{p_\mu}{\sqrt{2}\left\vert\vec{p}\right\vert}
\left( g_{0}^\nu - \frac{p_0 p^\nu}{p^2} \right) \ .
\end{eqnarray}

\subsection{Functions used at non-zero temperature}
\label{ssec:FUNZT}

Here we list the functions used at non-zero temperature
in Sec.~\ref{sec:THDMC}.

Functions used in the expressions of $f_\pi$
in Eq.~(\ref{eq: fpi})
are defined as follows;
\begin{eqnarray}
  I_{n}(T) 
&\equiv &
  \int_{0}^{\infty }d{\rm k} \frac{{\rm k}^{n-1}}{e^{{\rm k}/T}-1}
  =\widetilde{I}_{n}T^{n} 
\ ,
\nonumber\\
&&
  \widetilde{I}_{n}=
  \int_{0}^{\infty }d{y}\frac{y^{n-1}}{e^y-1} =(n-1)! \, \zeta(n)
\ ,
\nonumber \\
&&
  \widetilde{I}_{2}=\frac{\pi ^{2}}{6}\ ,\quad
  \widetilde{I}_{4}=\frac{\pi ^{4}}{15}\ ,\quad
  \widetilde{I}_{6}=\frac{8\pi ^{6}}{63}\ ,   
\nonumber \\
  J_{m}^{n}(M_\rho;T) 
&\equiv &
  \int_{0}^{\infty }d{\rm k}\frac{1}{e^{\omega/T}-1}
  \frac{{\rm k}^{n}}{\omega ^{m}}\qquad ;\quad
  n,m:\,\mbox{\rm integer}\ ,  \nonumber \\
  &&\qquad \omega \equiv \sqrt{{\rm k}^{2}+M_{\rho }^{2}}\ .
\label{function 1}
\end{eqnarray}
We also define 
the functions in the $\rho$-meson propagator as follows:
\begin{eqnarray}
  F_3^n(p_0;M_\rho;T) 
&\equiv&
  \int_0^\infty d{\rm k}{\cal P}
  \frac{1}{e^{\omega/T}-1}
  \frac{4{\rm k}^{n}}{\omega (4\omega ^{2}-p_{0}^{2})} 
\ ,
\nonumber \\
  G_n(p_0;T) 
&\equiv&
  \int_0^\infty d{\rm k}{\cal P}
  \frac{{\rm k}^{n-1}}{e^{{\rm k}/T}-1}
  \frac{4{\rm k}^2}{4{\rm k}^2 - p_0^2}
\nonumber\\
&=&
  I_n(T) +
  \int_0^\infty d{\rm k}{\cal P}
  \frac{{\rm k}^{n-1}}{e^{{\rm k}/T}-1}
  \frac{p_0^2}{4{\rm k}^2 - p_0^2} 
\, 
\nonumber\\
  H_1^n(p_0;M_\rho ;T) 
&\equiv&
  \int_0^\infty d{\rm k}{\cal P}
  \frac{1}{e^{\omega /T}-1}
  \frac{{\rm k}^n}{\omega} 
  \frac{1}{(M_\rho^2 - p_0^2)^2 - 4{\rm k}^2 p_0^2}
\ ,
\nonumber\\
  K_n(p_0;M_\rho ;T) 
&\equiv&
  \int_0^\infty d{\rm k}{\cal P}
  \frac{{\rm k}^{n-1}}{e^{{\rm k}/T}-1}
  \frac{1}{(M_\rho^2 - p_0^2)^2 - 4{\rm k}^2 p_0^2}
\ ,
\label{function 2}
\end{eqnarray}
where ${\cal P}$ denotes the principal part.

\newpage

\section{Feynman Rules in the Background Field Gauge}
\label{app:FR}

In this appendix
we show the Feynman Rules for the propagators of the
quantum fields and the vertices including two quantum fields
in the background field gauge.
The relevant Lagrangian is given in Eq.~(\ref{Lag:BGF}) in 
Sec.~\ref{ssec:BGFM}.
In the following figures 
$f_{abc}$ is the structure constant of the SU($N_f$) group.
Vertices with a dot ($\bullet$) imply that the dirivatives
affect to the quantum fields, while those with a circle ($\circ$)
imply that no derivatives are included.

\subsection{Propagators}

\begin{figure}[htbp]
\begin{center}
\setlength{\unitlength}{1mm}
\begin{picture}(140,110)(0,0)
\put(0,90){\makebox(0,0)[lb]{%
\epsfxsize = 5cm
\epsfbox{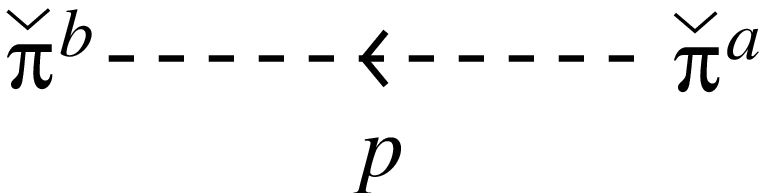}
}}
\put(80,97){\makebox(0,0)[lb]{%
{\Large
$\displaystyle \delta_{ab}\, \frac{1}{-p^2}$
}
}}
\put(0,60){\makebox(0,0)[lb]{%
\epsfxsize = 5cm
\epsfbox{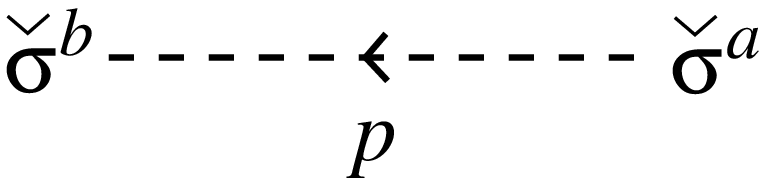}
}}
\put(80,67){\makebox(0,0)[lb]{%
{\Large
$\displaystyle \delta_{ab}\, \frac{1}{M_\rho^2-p^2}$
}
}}
\put(0,30){\makebox(0,0)[lb]{%
\epsfxsize = 5cm
\epsfbox{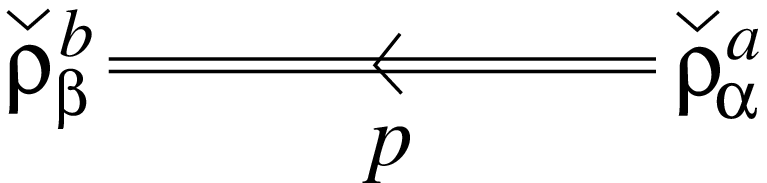}
}}
\put(80,37){\makebox(0,0)[lb]{%
{\Large
$\displaystyle \delta_{ab} \,g_{\alpha\beta}\,
\frac{1}{p^2-M_\rho^2}$
}
}}
\put(0,0){\makebox(0,0)[lb]{%
\epsfxsize = 5cm
\epsfbox{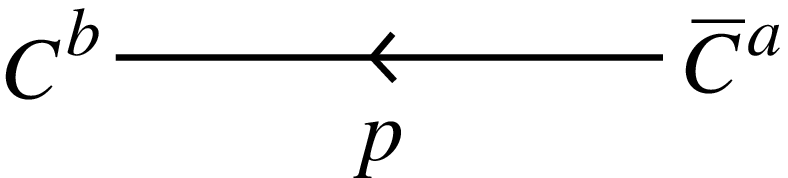}
}}
\put(80,7){\makebox(0,0)[lb]{%
{\Large
$\displaystyle \delta_{ab}\, \frac{i}{M_\rho^2-p^2}$
}
}}
\end{picture}
\end{center}
\caption[Feynman Rule (Propagators)]{Feynman Rules for the propagators}
\label{fig:prop}
\end{figure}

\newpage

\subsection{Three-point vertices}

\begin{figure}[htbp]
\begin{center}
\setlength{\unitlength}{1mm}
\begin{picture}(140,110)(0,-50)
\put(-10,50){\makebox(0,0)[lb]{\large\bf
Vertices with $\overline{\cal A}_\mu$}}
\put(0,20){\makebox(0,0)[lc]{\large (a)}}
\put(10,0){\makebox(0,0)[lb]{
\epsfxsize = 5cm
\epsfbox{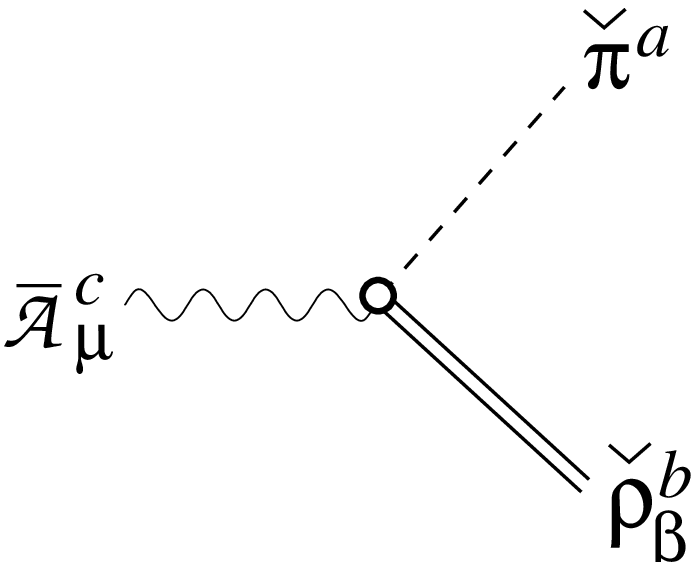}
}}
\put(80,20){\makebox(0,0)[lc]{
{\Large
$\displaystyle
-\sqrt{a} M_\rho f_{abc} \, g^{\mu\beta}$
}
}}
\put(0,-30){\makebox(0,0)[lc]{\large (b)}}
\put(10,-50){\makebox(0,0)[lb]{
\epsfxsize = 5cm
\epsfbox{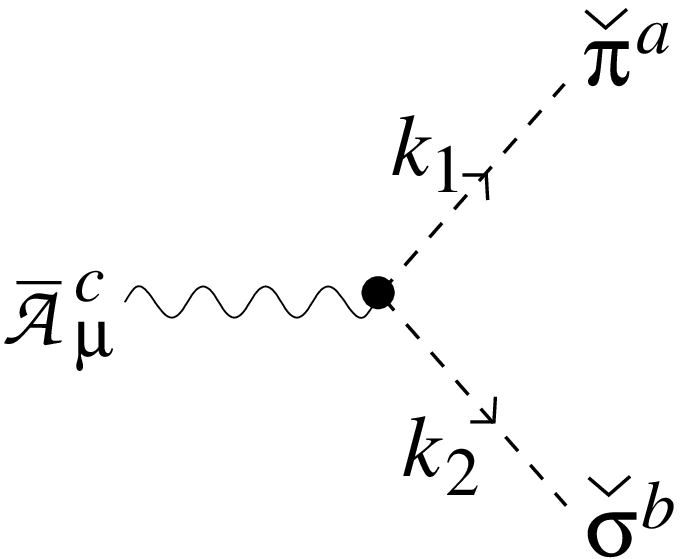}
}}
\put(80,-30){\makebox(0,0)[lc]{
{\Large
$\displaystyle
-i \frac{1}{2} \sqrt{a} f_{abc} \left( k_1 - k_2 \right)^\mu $
}
}}
\end{picture}
\end{center}
\caption[Feynman Rule (vertices with $\overline{\cal A}_\mu$)]{%
Feynman Rules for the vertices which include
one $\overline{\cal A}_\mu$.}
\label{fig:a1}
\end{figure}

\begin{figure}[htbp]
\begin{center}
\setlength{\unitlength}{1mm}
\begin{picture}(140,160)(0,0)
\put(-10,150){\makebox(0,0)[lb]{\large\bf
Vertices with $\overline{\cal V}_\mu$}}
\put(0,120){\makebox(0,0)[lc]{\large (a)}}
\put(10,100){\makebox(0,0)[lb]{
\epsfxsize = 5cm
\epsfbox{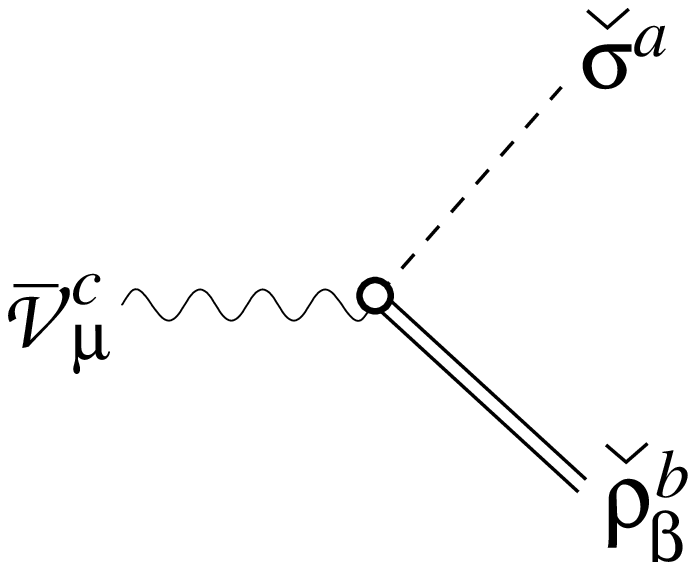}
}}
\put(80,120){\makebox(0,0)[lc]{
{\Large
$\displaystyle
- M_\rho f_{abc} \, g^{\mu\beta} $
}
}}
\put(0,70){\makebox(0,0)[lc]{\large (b)}}
\put(10,50){\makebox(0,0)[lb]{
\epsfxsize = 5cm
\epsfbox{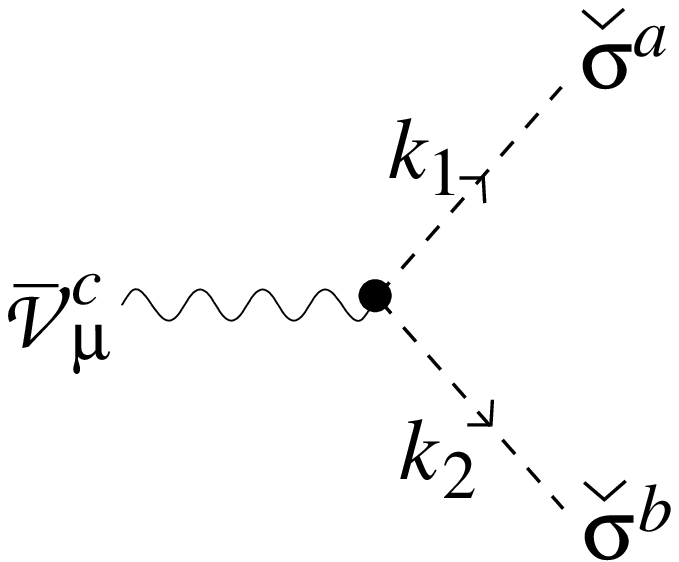}
}}
\put(80,70){\makebox(0,0)[lc]{
{\Large
$\displaystyle
- i \frac{1}{2} f_{abc} \left( k_1 - k_2 \right)^\mu $
}
}}
\put(0,20){\makebox(0,0)[lc]{\large (c)}}
\put(10,0){\makebox(0,0)[lb]{
\epsfxsize = 5cm
\epsfbox{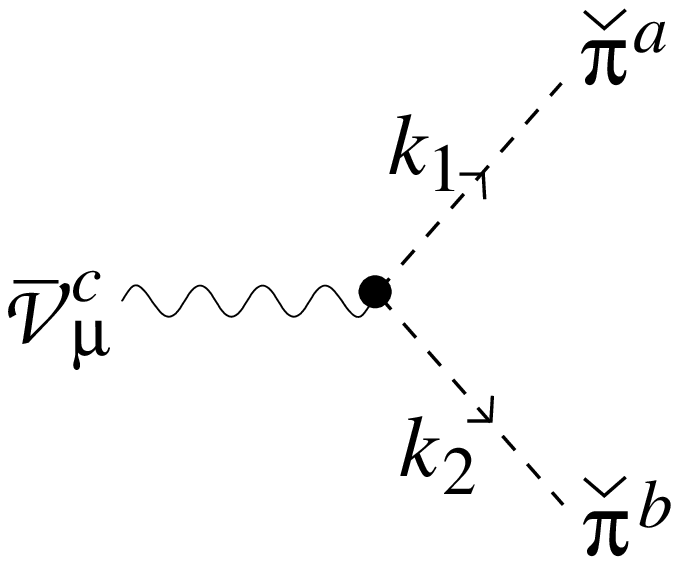}
}}
\put(80,20){\makebox(0,0)[lc]{
{\Large
$\displaystyle
- i \frac{2-a}{2} f_{abc} \left( k_1 - k_2 \right)^\mu $
}
}}
\end{picture}
\end{center}
\caption[Feynman Rule (vertices with $\overline{\cal V}_\mu$)]{%
Feynman Rules for the vertices which include
one $\overline{\cal V}_\mu$.}
\label{fig:v1}
\end{figure}

\begin{figure}[htbp]
\begin{center}
\setlength{\unitlength}{1mm}
\begin{picture}(140,200)(0,0)
\put(-10,190){\makebox(0,0)[lb]{\large\bf
Vertices with $\overline{V}_\mu$}}
\put(0,167){\makebox(0,0)[lc]{\large (a)}}
\put(10,152){\makebox(0,0)[lb]{
\epsfysize = 3.2cm
\epsfbox{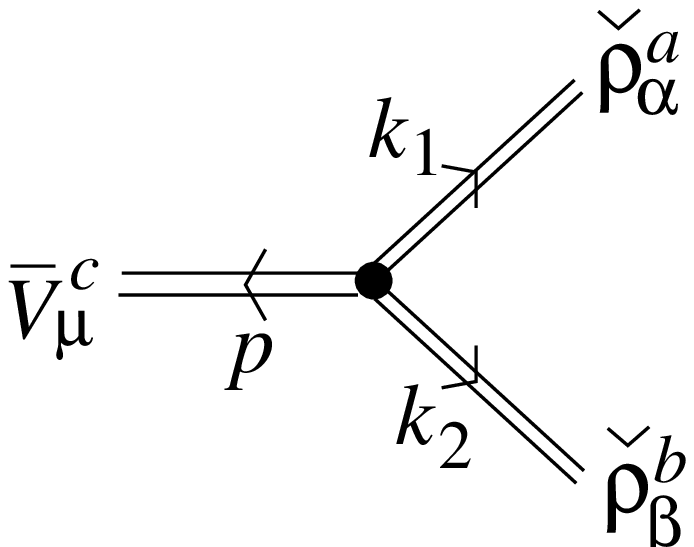}
}}
\put(75,172){\makebox(0,0)[lc]{
{\large
$\displaystyle
i\, f_{abc} 
  \left( k_1 - k_2 \right)^\mu g^{\alpha\beta}
$
}
}}
\put(78,162){\makebox(0,0)[lc]{
{\large
$\displaystyle
- 2 i \, f_{abc} 
\left( p^\alpha g^{\mu\beta} - p^\beta g^{\mu\alpha} \right)
$
}
}}
\put(0,129){\makebox(0,0)[lc]{\large (b)}}
\put(10,114){\makebox(0,0)[lb]{
\epsfysize = 3.2cm
\epsfbox{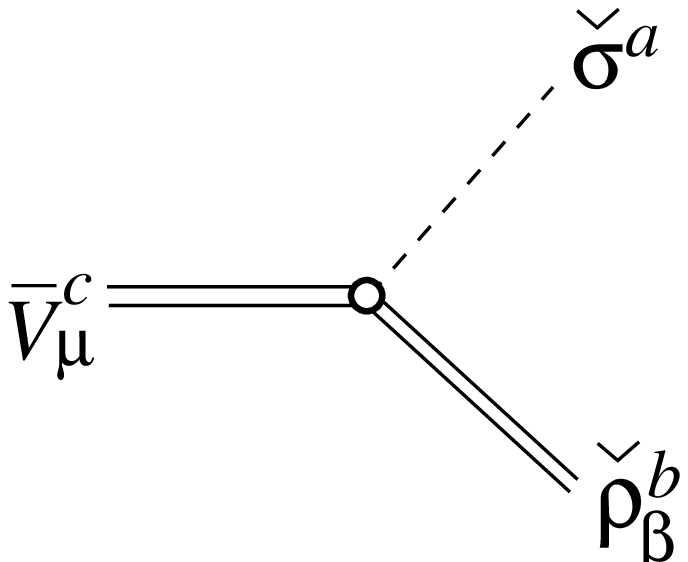}
}}
\put(80,129){\makebox(0,0)[lc]{
{\large
$\displaystyle
M_\rho f_{abc} \, g^{\mu\beta} $
}
}}
\put(0,91){\makebox(0,0)[lc]{\large (c)}}
\put(10,76){\makebox(0,0)[lb]{
\epsfysize = 3.2cm
\epsfbox{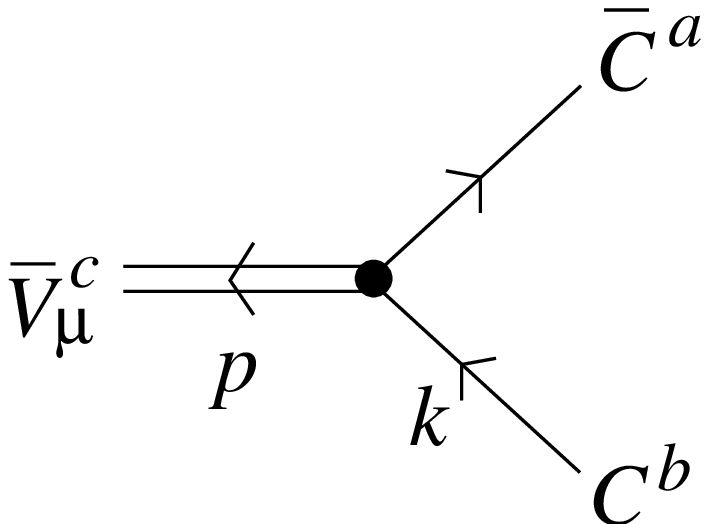}
}}
\put(75,91){\makebox(0,0)[lc]{
{\large
$\displaystyle
- f_{abc} \left( 2 k - p \right)^\mu
$
}
}}
\put(0,53){\makebox(0,0)[lc]{\large (d)}}
\put(10,38){\makebox(0,0)[lb]{
\epsfysize = 3.2cm
\epsfbox{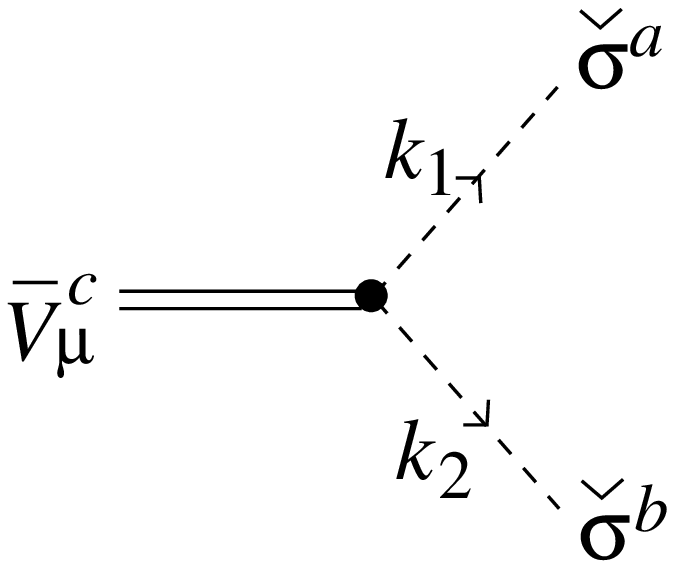}
}}
\put(80,53){\makebox(0,0)[lc]{
{\large
$\displaystyle
- i \frac{1}{2} f_{abc} \left( k_1 - k_2 \right)^\mu $
}
}}
\put(0,16){\makebox(0,0)[lc]{\large (e)}}
\put(10,0){\makebox(0,0)[lb]{
\epsfysize = 3.2cm
\epsfbox{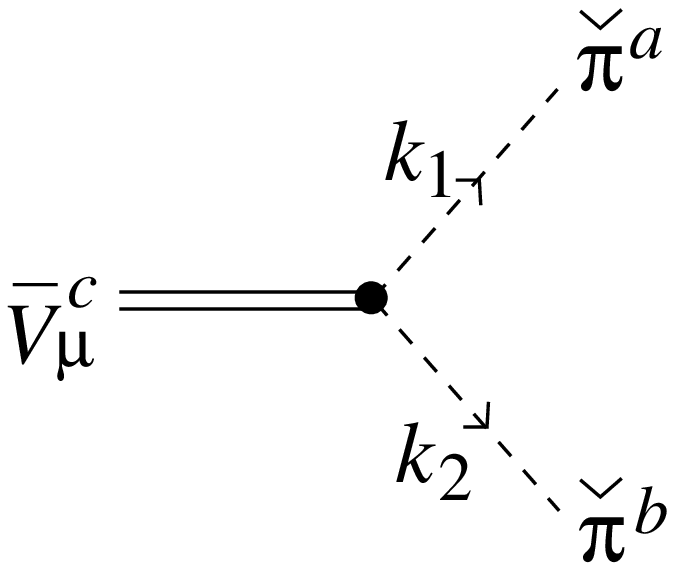}
}}
\put(80,16){\makebox(0,0)[lc]{
{\large
$\displaystyle
- i \frac{a}{2} f_{abc} \left( k_1 - k_2 \right)^\mu $
}
}}
\end{picture}
\end{center}
\caption[Feynman Rule (vertices with $\overline{V}_\mu$)]{%
Feynman Rules for the vertices which include
one $\overline{V}_\mu$.}
\label{fig:r1}
\end{figure}

\newpage

\subsection{Four-point vertices}

\begin{figure}[htbp]
\begin{center}
\setlength{\unitlength}{1mm}
\begin{picture}(140,45)(0,0)
\put(-10,35){\makebox(0,0)[lb]{\large\bf
Vertices with $\overline{\cal A}_\mu\overline{\cal A}_\nu$}}
\put(0,12){\makebox(0,0)[lc]{\large (a)}}
\put(10,0){\makebox(0,0)[lb]{
\epsfxsize = 5cm
\epsfbox{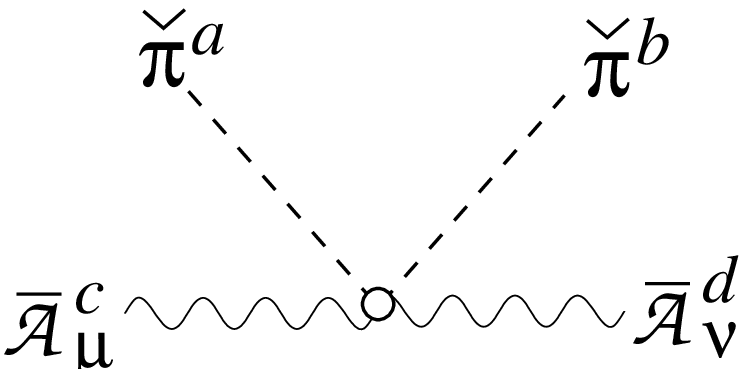}
}}
\put(70,12){\makebox(0,0)[lc]{
{\large
$\displaystyle
- (1-a) \left( f_{cae} f_{dbe} + f_{dae} f_{cbe} \right) g^{\mu\nu} $
}
}}
\end{picture}
\end{center}
\caption[Feynman Rule (vertices with 
$\overline{\cal A}_\mu\overline{\cal A}_\nu$)]{%
Feynman Rules for the vertices which include
$\overline{\cal A}_\mu\overline{\cal A}_\nu$.
Here summation over $e$ is taken.}
\label{fig:a2}
\end{figure}

\begin{figure}[htbp]
\begin{center}
\setlength{\unitlength}{1mm}
\begin{picture}(140,45)(0,0)
\put(-10,35){\makebox(0,0)[lb]{\large\bf
Vertices with $\overline{\cal V}_\mu\overline{\cal V}_\nu$}}
\put(0,12){\makebox(0,0)[lc]{\large (a)}}
\put(10,0){\makebox(0,0)[lb]{
\epsfxsize = 5cm
\epsfbox{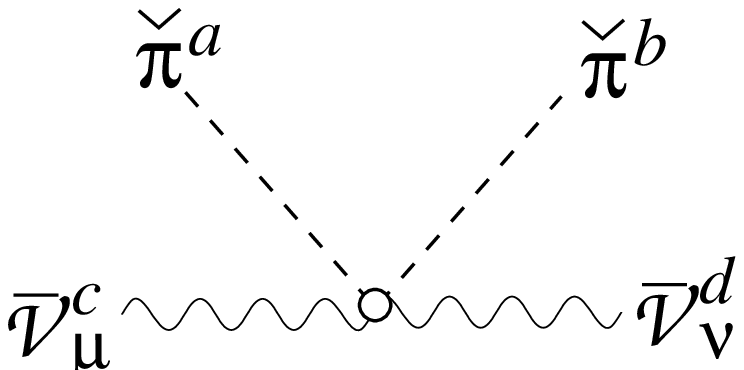}
}}
\put(70,12){\makebox(0,0)[lc]{
{\large
$\displaystyle
(1-a) \left( f_{cae} f_{dbe} + f_{dae} f_{cbe} \right) g^{\mu\nu} $
}
}}
\end{picture}
\end{center}
\caption[Feynman Rule (vertices with 
$\overline{\cal V}_\mu\overline{\cal V}_\nu$)]{%
Feynman Rules for the vertices which include
$\overline{\cal V}_\mu\overline{\cal V}_\nu$.
Here summation over $e$ is taken.}
\label{fig:v2}
\end{figure}

\begin{figure}[htbp]
\begin{center}
\setlength{\unitlength}{1mm}
\begin{picture}(140,80)(0,0)
\put(-10,70){\makebox(0,0)[lb]{\large\bf
Vertices with $\overline{\cal V}_\mu\overline{V}_\nu$}}
\put(0,47){\makebox(0,0)[lc]{\large (a)}}
\put(10,35){\makebox(0,0)[lb]{
\epsfxsize = 5cm
\epsfbox{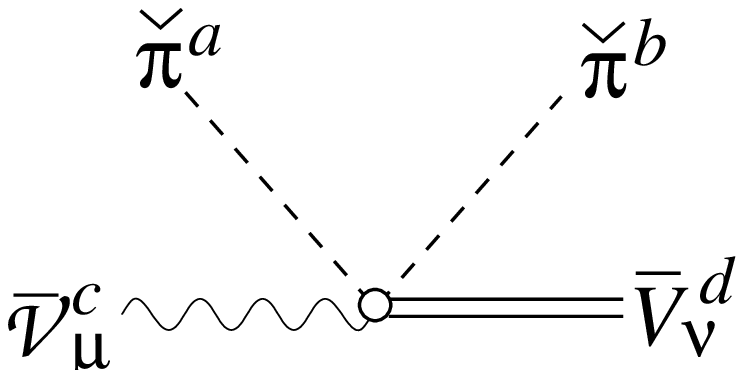}
}}
\put(70,47){\makebox(0,0)[lc]{
{\large
$\displaystyle
\frac{a}{2} 
\left( f_{cae} f_{dbe} + f_{dae} f_{cbe} \right) g^{\mu\nu} $
}
}}
\put(0,12){\makebox(0,0)[lc]{\large (b)}}
\put(10,0){\makebox(0,0)[lb]{
\epsfxsize = 5cm
\epsfbox{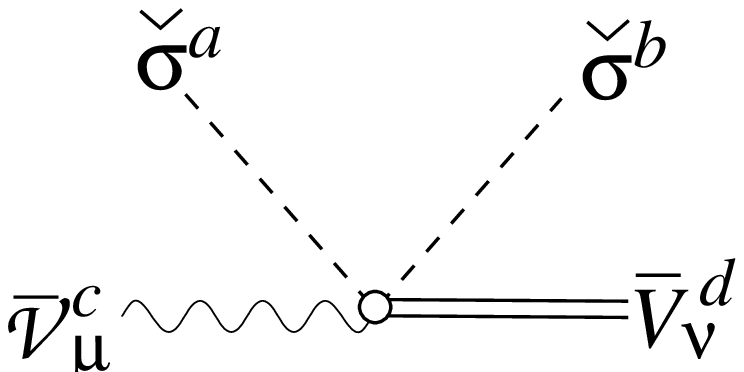}
}}
\put(70,12){\makebox(0,0)[lc]{
{\large
$\displaystyle
\frac{1}{2} 
\left( f_{cae} f_{dbe} + f_{dae} f_{cbe} \right) g^{\mu\nu} $
}
}}
\end{picture}
\end{center}
\caption[Feynman Rule (vertices with 
$\overline{\cal V}_\mu\overline{V}_\nu$)]{%
Feynman Rules for the vertices which include
$\overline{\cal V}_\mu\overline{V}_\nu$.
Here summation over $e$ is taken.}
\label{fig:vr2}
\end{figure}

\begin{figure}[htbp]
\begin{center}
\setlength{\unitlength}{1mm}
\begin{picture}(140,80)(0,0)
\put(-10,70){\makebox(0,0)[lb]{\large\bf
Vertices with $\overline{V}_\mu\overline{V}_\nu$}}
\put(0,47){\makebox(0,0)[lc]{\large (a)}}
\put(10,35){\makebox(0,0)[lb]{
\epsfxsize = 5cm
\epsfbox{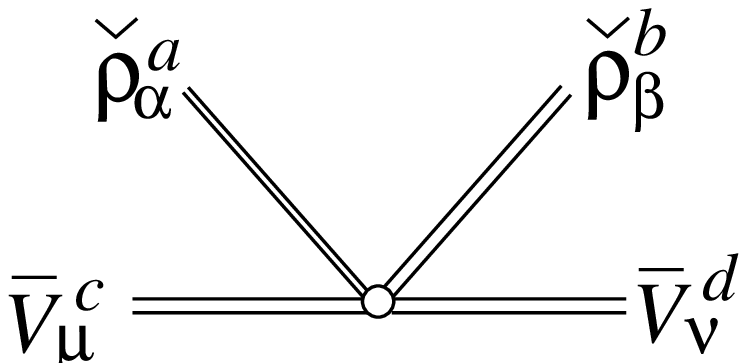}
}}
\put(70,52){\makebox(0,0)[lc]{
{\large
$\displaystyle
- \left( f_{cae} f_{dbe} + f_{dae} f_{cbe} \right) 
g^{\mu\nu} g^{\alpha\beta} $
}
}}
\put(75,42){\makebox(0,0)[lc]{
{\large
$\displaystyle
- 2 f_{abe} f_{cde} 
\left( 
  g^{\mu\alpha} g^{\nu\beta} - g^{\mu\beta} g^{\nu\alpha}
\right)
$
}
}}
\put(0,12){\makebox(0,0)[lc]{\large (b)}}
\put(10,0){\makebox(0,0)[lb]{
\epsfxsize = 5cm
\epsfbox{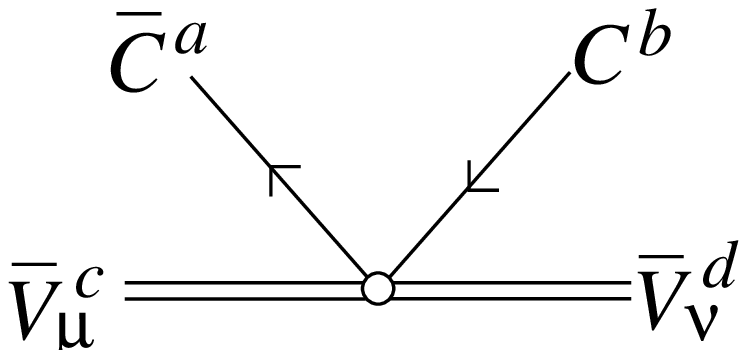}
}}
\put(75,12){\makebox(0,0)[lc]{
{\large
$\displaystyle
- i \, \left( f_{cae} f_{dbe} + f_{dae} f_{cbe} \right)\, g^{\mu\nu}
$
}
}}
\end{picture}
\end{center}
\caption[Feynman Rule (vertices with 
$\overline{V}_\mu\overline{V}_\nu$)]{%
Feynman Rules for the vertices which include
$\overline{V}_\mu\overline{V}_\nu$.
Here summation over $e$ is taken.}
\label{fig:r2}
\end{figure}

\begin{figure}[htbp]
\begin{center}
\setlength{\unitlength}{1mm}
\begin{picture}(140,45)(0,0)
\put(-10,35){\makebox(0,0)[lb]{\large\bf
Vertices with $\overline{\cal A}_\mu\overline{\cal V}_\nu$}}
\put(0,12){\makebox(0,0)[lc]{\large (a)}}
\put(10,0){\makebox(0,0)[lb]{
\epsfxsize = 5cm
\epsfbox{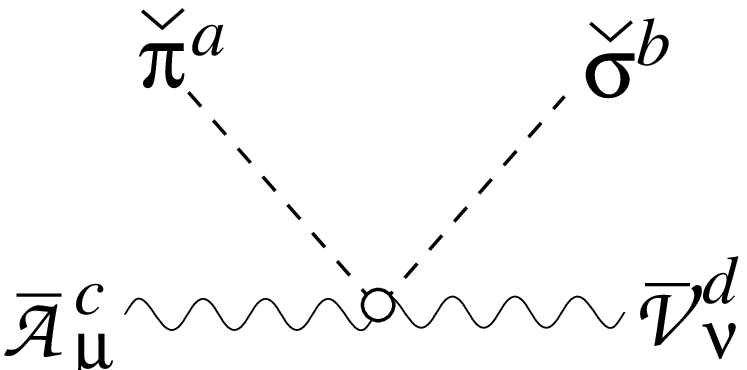}
}}
\put(70,17){\makebox(0,0)[lc]{
{\large
$\displaystyle
-\frac{1}{4} \sqrt{a} f_{cae} f_{dbe} g^{\mu\nu} $
}
}}
\put(75,7){\makebox(0,0)[lc]{
{\large
$\displaystyle
{}- \frac{1}{2} \sqrt{a} (1-a) f_{abe} f_{cde} 
$
}
}}
\end{picture}
\end{center}
\caption[Feynman Rule (vertices with 
$\overline{\cal A}_\mu\overline{\cal V}_\nu$)]{%
Feynman Rules for the vertices which include
$\overline{\cal A}_\mu\overline{\cal V}_\nu$.
Here summation over $e$ is taken.}
\label{fig:av}
\end{figure}

\begin{figure}[htbp]
\begin{center}
\setlength{\unitlength}{1mm}
\begin{picture}(140,45)(0,0)
\put(-10,35){\makebox(0,0)[lb]{\large\bf
Vertices with $\overline{\cal A}_\mu\overline{V}_\nu$}}
\put(0,12){\makebox(0,0)[lc]{\large (a)}}
\put(10,0){\makebox(0,0)[lb]{
\epsfxsize = 5cm
\epsfbox{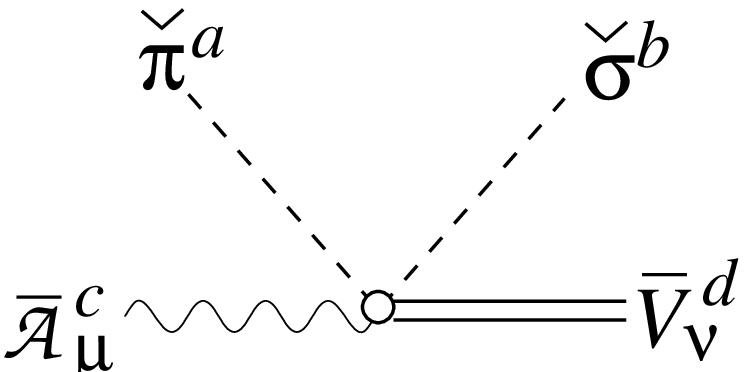}
}}
\put(70,17){\makebox(0,0)[lc]{
{\large
$\displaystyle
\frac{3}{4} \sqrt{a} f_{cae} f_{dbe} g^{\mu\nu} $
}
}}
\put(75,7){\makebox(0,0)[lc]{
{\large
$\displaystyle
{}+ \frac{1}{2} \sqrt{a} (1-a) 
\left( f_{dae} f_{cbe} + f_{abe} f_{cde} \right)
$
}
}}
\end{picture}
\end{center}
\caption[Feynman Rule (vertices with 
$\overline{\cal A}_\mu\overline{V}_\nu$)]{%
Feynman Rules for the vertices which include
$\overline{\cal A}_\mu\overline{V}_\nu$.
Here summation over $e$ is taken.}
\label{fig:ar}
\end{figure}

\newpage

\section{Feynman Rules in the Landau Gauge}
\label{app:FRLG}

In this appendix, for convenience,
we show the Feynman Rules for the propagators 
and the vertices in the Landau gauge with ordinary quantization
procedure.
The relevant Lagrangian is given in Eq.~(\ref{leading Lagrangian}) in 
Sec.~\ref{ssec:OP2L}.
The gauge fixing is done by 
introducing an $R_\xi$-gauge-like gauge-fixing and the
corresponding 
Faddeev-Popov ghost Lagrangian is added~\cite{HY}:
\begin{eqnarray}
  {\cal L}_{{\rm GF}+{\rm FP}} 
&=& 
  - \frac{1}{\alpha} \mbox{tr} 
  \left[ \left( \partial^\mu \rho_\mu \right) \right] 
  + \frac{i}{2} a g F_\pi^2 \, \mbox{tr}
  \left[
    \partial^\mu \rho_\mu 
    \left( 
      \xi_{{\rm L}} - \xi_{{\rm L}}^{\dag} + \xi_{{\rm R}} 
      - \xi_{{\rm R}}^{\dag} 
    \right) 
  \right]  
\nonumber \\
&& 
  {} + \frac{1}{16} \alpha a^2 g^2 F_\pi^4 
  \left\{ 
    \mbox{tr} 
    \left[ 
      \left(
        \xi_{{\rm L}} - \xi_{{\rm L}}^{\dag} + \xi_{{\rm R}} 
        - \xi_{{\rm R}}^{\dag}
      \right)^2 
    \right] 
    - \frac{1}{N_f} 
    \left( 
      \mbox{tr} 
      \left[ 
        \xi_{{\rm L}} - \xi_{{\rm L}}^{\dag} + \xi_{{\rm R}} 
        - \xi_{{\rm R}}^{\dag} 
      \right] 
    \right)^2
  \right\}  
\nonumber \\
&& 
  {} + i \, \mbox{tr} 
  \left[ 
    \bar{C} 
    \left\{ 
       2 \partial^\mu D_\mu C 
       + \frac{1}{2} \alpha a g^2 F_\pi^2 
       \left( 
          C \xi_{{\rm L}} + \xi_{{\rm L}}^{\dag} C 
          + C \xi_{{\rm R}} + \xi_{{\rm R}}^{\dag} C 
       \right) 
    \right\} 
  \right] 
\ ,
\label{Lag: GF FP}
\end{eqnarray}
where $\alpha$ denotes a gauge parameter and $C$ denotes a ghost
field.   
Here  we choose the Landau gauge, $\alpha=0$. 
In this gauge the 
would-be NG boson $\sigma$ is still massless, no other
vector-scalar interactions are created and the ghost field couples
only to the HLS gauge field $\rho_\mu$.
As in the Feynman rules for the background field gauge in 
Appendix~\ref{app:FR},
in the following figures 
$f_{abc}$ is the structure constant of the SU($N_f$) group.
Vertices with a dot ($\bullet$) imply that the dirivatives
are included, while those with a circle ($\circ$)
imply that no derivatives are included.
For calculating the two-point functions at one-loop level,
it is enough to have Feynman rules up until four-point vertices.
In this appendix we do not list the vertices with more than four legs.
It should be noticed that the Feynman rules listed below
except for the $\rho$-$\sigma$-$\sigma$-$\sigma$ and
$\rho$-$\sigma$-$\pi$-$\pi$ vertices
in 
the Landau gauge of the 
$R_\xi$-gauge-like gauge-fixing agrees with those in 
the Landau gauge of the
covariant gauge fixing given in Eq.~(\ref{GFFPlagrangian})
used in Sec.~\ref{sec:RALOLET}.

\newpage

\subsection{Propagtors}

\begin{figure}[htbp]
\begin{center}
\setlength{\unitlength}{1mm}
\begin{picture}(140,85)(0,15)
\put(0,90){\makebox(0,0)[lb]{%
\epsfxsize = 5cm
\epsfbox{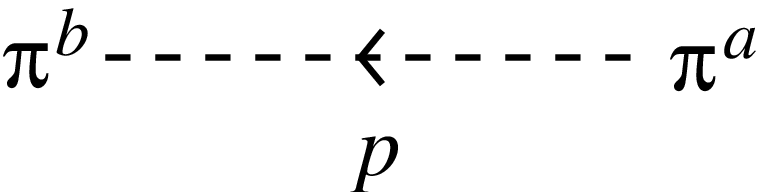}
}}
\put(80,97){\makebox(0,0)[lb]{%
{\Large
$\displaystyle \delta_{ab}\, \frac{1}{-p^2}$
}
}}
\put(0,65){\makebox(0,0)[lb]{%
\epsfxsize = 5cm
\epsfbox{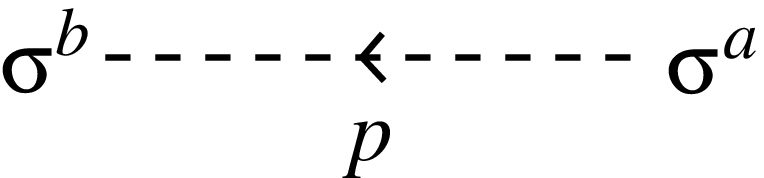}
}}
\put(80,72){\makebox(0,0)[lb]{%
{\Large
$\displaystyle \delta_{ab}\, \frac{1}{-p^2}$
}
}}
\put(0,40){\makebox(0,0)[lb]{%
\epsfxsize = 5cm
\epsfbox{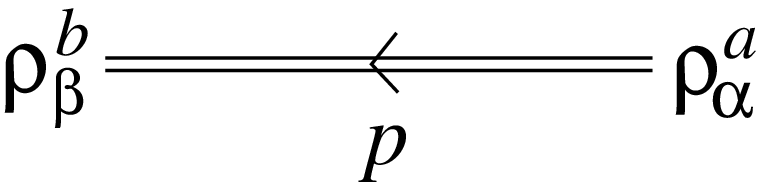}
}}
\put(80,47){\makebox(0,0)[lb]{%
{\Large
$\displaystyle \delta_{ab} 
\frac{1}{p^2-M_\rho^2}
\left[ g_{\alpha\beta} - \frac{p_\alpha p_\beta}{p^2} \right]
$
}
}}
\put(0,15){\makebox(0,0)[lb]{%
\epsfxsize = 5cm
\epsfbox{prop_c.eps}
}}
\put(80,22){\makebox(0,0)[lb]{%
{\Large
$\displaystyle \delta_{ab}\, \frac{i}{-p^2}$
}
}}
\end{picture}
\end{center}
\caption[Feynman rule in the Landau gauge (Propagators)]{%
Feynman rules for the propagators in the Landau gauge.}
\label{fig:propL}
\end{figure}


\subsection{Two-point vertices (mixing terms)}

\begin{figure}[htbp]
\begin{center}
\setlength{\unitlength}{1mm}
\begin{picture}(140,45)(0,20)
\put(0,60){\makebox(0,0)[lc]{\large (a)}}
\put(10,55){\makebox(0,0)[lb]{
\epsfxsize = 5cm
\epsfbox{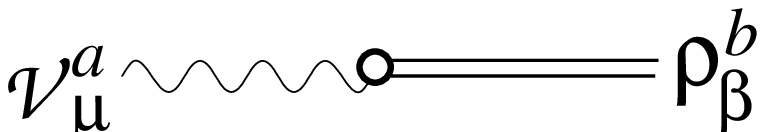}
}}
\put(80,60){\makebox(0,0)[lc]{
{\Large
$\displaystyle
- a g F_\pi^2 \delta_{ab} \, g^{\mu\beta}$
}
}}
\put(0,30){\makebox(0,0)[lc]{\large (b)}}
\put(10,25){\makebox(0,0)[lb]{
\epsfxsize = 5cm
\epsfbox{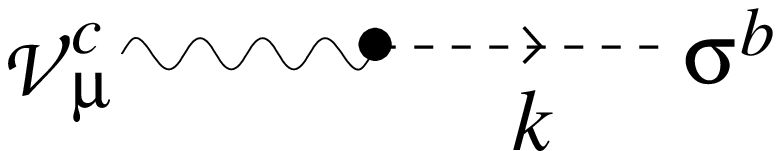}
}}
\put(80,30){\makebox(0,0)[lc]{
{\Large
$\displaystyle
i F_\sigma k_\mu \delta_{ab} $
}
}}
\end{picture}
\end{center}
\caption[Feynman rule in the Landau gauge 
(mixing terms)]{%
Feynman rules in the Landau gauge for two-point 
vertices (mixing terms).}
\label{fig:mixL}
\end{figure}

\newpage

\subsection{Three-point vertices}

\begin{figure}[htbp]
\begin{center}
\setlength{\unitlength}{1mm}
\begin{picture}(140,100)(0,-50)
\put(-10,50){\makebox(0,0)[lb]{\large\bf
Vertices with ${\cal A}_\mu$}}
\put(0,20){\makebox(0,0)[lc]{\large (a)}}
\put(10,0){\makebox(0,0)[lb]{
\epsfxsize = 5cm
\epsfbox{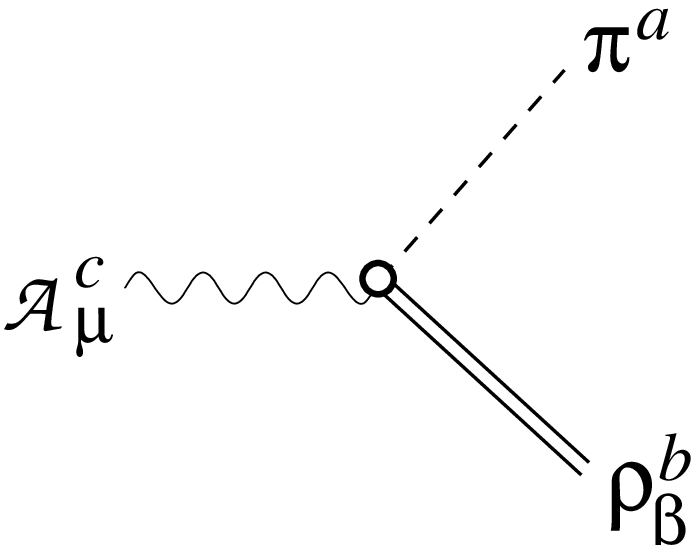}
}}
\put(80,20){\makebox(0,0)[lc]{
{\Large
$\displaystyle
-\sqrt{a} M_\rho f_{abc} \, g^{\mu\beta}$
}
}}
\put(0,-30){\makebox(0,0)[lc]{\large (b)}}
\put(10,-50){\makebox(0,0)[lb]{
\epsfxsize = 5cm
\epsfbox{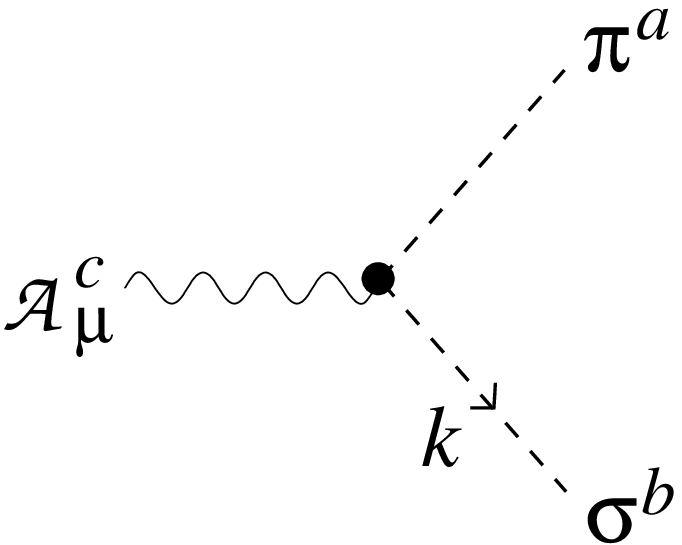}
}}
\put(80,-30){\makebox(0,0)[lc]{
{\Large
$\displaystyle
i \sqrt{a} f_{abc} \, k^\mu $
}
}}
\end{picture}
\end{center}
\caption[Feynman rule in the Landau gauge 
(3-point vertices with $\overline{\cal A}_\mu$)]{%
Feynman rules in the Landau gauge for three-point 
vertices which include one ${\cal A}_\mu$.}
\label{fig:a1L}
\end{figure}

\begin{figure}[htbp]
\begin{center}
\setlength{\unitlength}{1mm}
\begin{picture}(140,160)(0,0)
\put(-10,150){\makebox(0,0)[lb]{\large\bf
Vertices with ${\cal V}_\mu$}}
\put(0,120){\makebox(0,0)[lc]{\large (a)}}
\put(10,100){\makebox(0,0)[lb]{
\epsfxsize = 5cm
\epsfbox{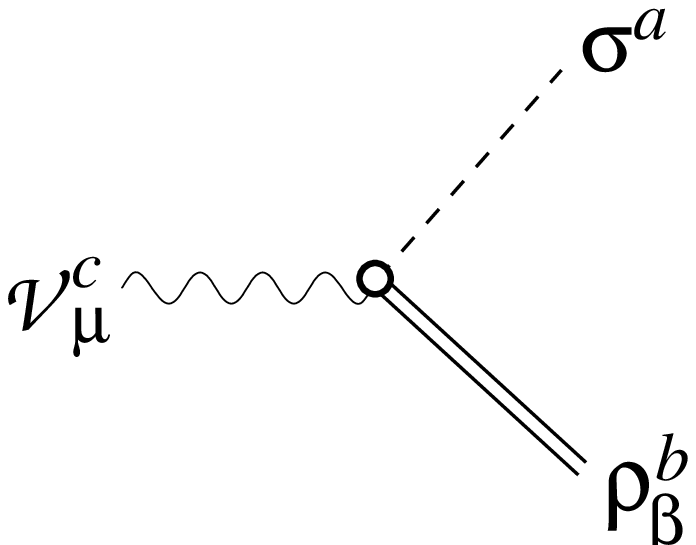}
}}
\put(80,120){\makebox(0,0)[lc]{
{\Large
$\displaystyle
- M_\rho f_{abc} \, g^{\mu\beta} $
}
}}
\put(0,70){\makebox(0,0)[lc]{\large (b)}}
\put(10,50){\makebox(0,0)[lb]{
\epsfxsize = 5cm
\epsfbox{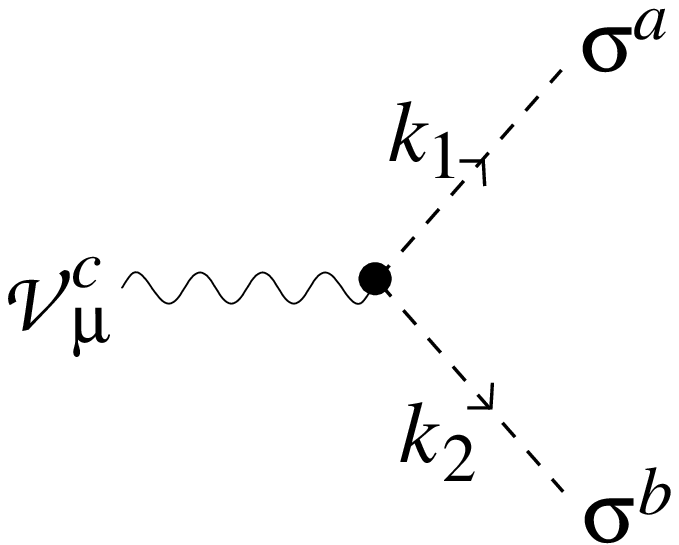}
}}
\put(80,70){\makebox(0,0)[lc]{
{\Large
$\displaystyle
- i \frac{1}{2} f_{abc} \left( k_1 - k_2 \right)^\mu $
}
}}
\put(0,20){\makebox(0,0)[lc]{\large (c)}}
\put(10,0){\makebox(0,0)[lb]{
\epsfxsize = 5cm
\epsfbox{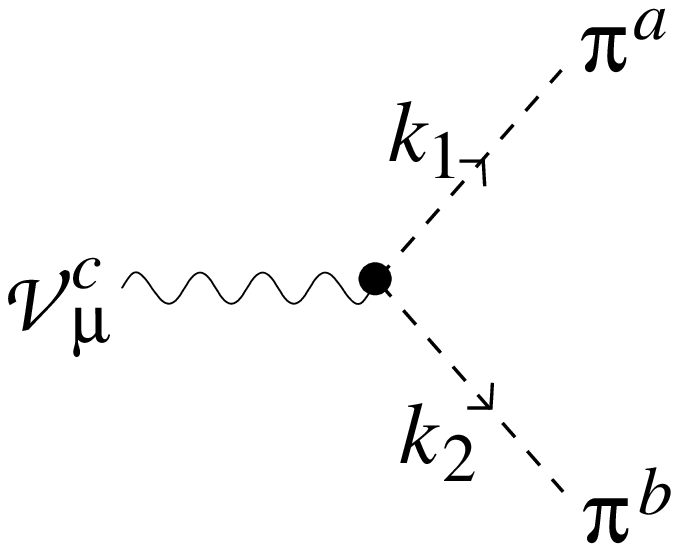}
}}
\put(80,20){\makebox(0,0)[lc]{
{\Large
$\displaystyle
- i \frac{2-a}{2} f_{abc} \left( k_1 - k_2 \right)^\mu $
}
}}
\end{picture}
\end{center}
\caption[Feynman rule in the Landau gauge
(3-point vertices with ${\cal V}_\mu$)]{%
Feynman rules in the Landau gauge
for the vertices which include
one ${\cal V}_\mu$.}
\label{fig:v1L}
\end{figure}

\begin{figure}[htbp]
\begin{center}
\setlength{\unitlength}{1mm}
\begin{picture}(140,200)(0,0)
\put(-10,190){\makebox(0,0)[lb]{\large\bf
Vertices with no ${\cal V}_\mu$ and ${\cal A}_\mu$}}
\put(0,167){\makebox(0,0)[lc]{\large (a)}}
\put(10,152){\makebox(0,0)[lb]{
\epsfysize = 3.2cm
\epsfbox{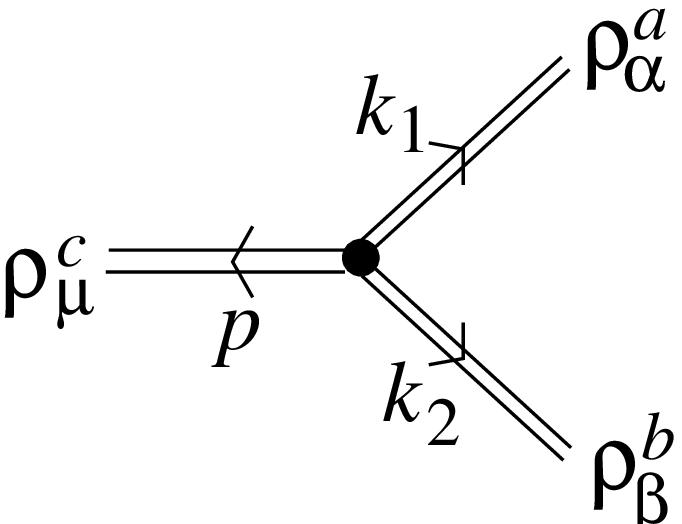}
}}
\put(75,172){\makebox(0,0)[lc]{
{\large
$\displaystyle
  i g f_{abc}
  \biggl[
    ( k_1 - k_2 )^\mu g^{\alpha\beta}
$
}
}}
\put(78,162){\makebox(0,0)[lc]{
{\large
$\displaystyle
    + ( k_2 - p )^\alpha g^{\beta\mu}
    + ( p - k_1 )^\beta g^{\mu\alpha}
  \biggr]
$
}
}}
\put(0,129){\makebox(0,0)[lc]{\large (b)}}
\put(10,114){\makebox(0,0)[lb]{
\epsfysize = 3.2cm
\epsfbox{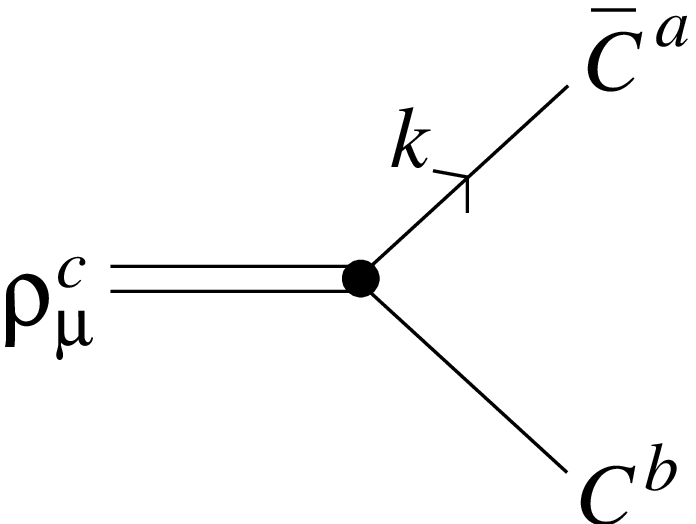}
}}
\put(80,129){\makebox(0,0)[lc]{
{\large
$\displaystyle
g f_{abc} \, k^\mu $
}
}}
\put(0,91){\makebox(0,0)[lc]{\large (c)}}
\put(10,76){\makebox(0,0)[lb]{
\epsfysize = 3.2cm
\epsfbox{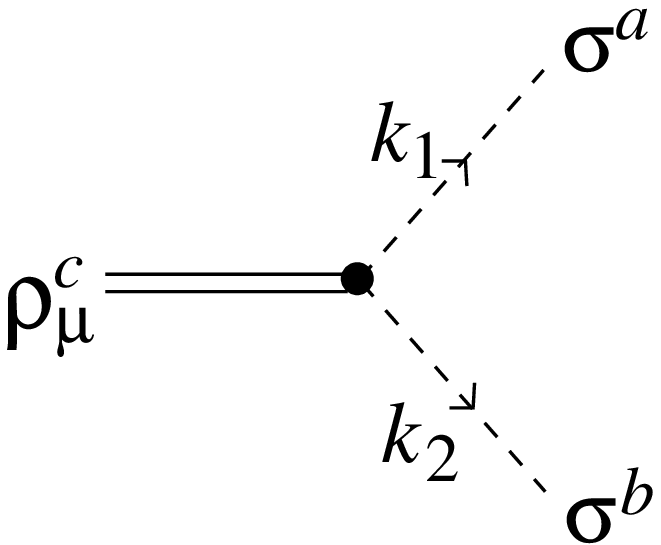}
}}
\put(75,91){\makebox(0,0)[lc]{
{\large
$\displaystyle
- i \frac{1}{2}\, g\, f_{abc} \left( k_1 - k_2 \right)^\mu $
}
}}
\put(0,53){\makebox(0,0)[lc]{\large (d)}}
\put(10,38){\makebox(0,0)[lb]{
\epsfysize = 3.2cm
\epsfbox{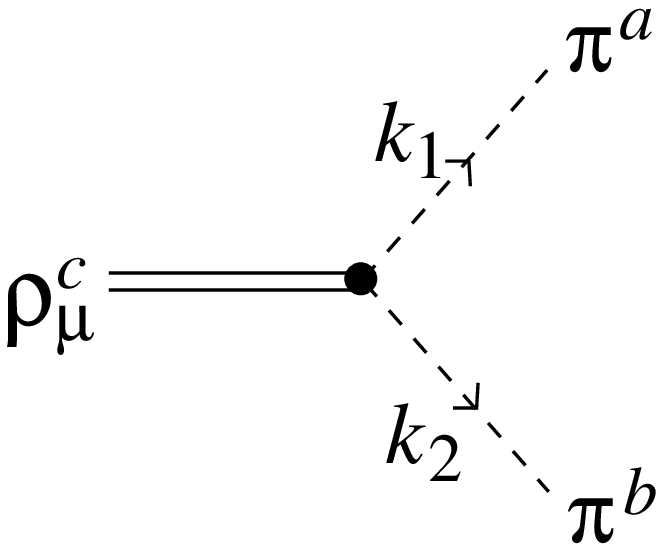}
}}
\put(80,53){\makebox(0,0)[lc]{
{\large
$\displaystyle
- i \frac{a}{2}\, g \, f_{abc} \left( k_1 - k_2 \right)^\mu $
}
}}
\put(0,16){\makebox(0,0)[lc]{\large (e)}}
\put(10,0){\makebox(0,0)[lb]{
\epsfysize = 3.2cm
\epsfbox{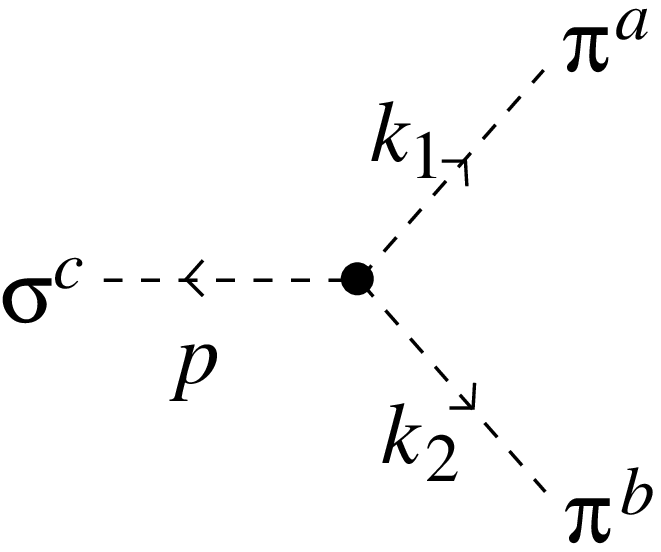}
}}
\put(80,16){\makebox(0,0)[lc]{
{\large
$\displaystyle
- i \frac{\sqrt{a}}{2F_\pi} \, f_{abc} \,
p \cdot \left( k_1 - k_2 \right) $
}
}}
\end{picture}
\end{center}
\caption[Feynman rule in the Landau gauge 
(3-point vertices with no ${\cal V}_\mu$ and ${\cal A}_\mu$)]{%
Feynman rules in the Landau gauge 
for three-point vertices which include
no ${\cal V}_\mu$ and ${\cal A}_\mu$.
Note that there are no $\rho$-$\rho$-$\sigma$ and
$\rho$-$\sigma$-$\sigma$ vertices.}
\label{fig:r1L}
\end{figure}

\newpage

\subsection{Four-point vertices}

\begin{figure}[htbp]
\begin{center}
\setlength{\unitlength}{1mm}
\begin{picture}(140,153)(0,-2)
\put(-10,155){\makebox(0,0)[lb]{\large\bf
Vertices with ${\cal A}_\mu$}}
\put(0,122){\makebox(0,0)[lc]{\large (a)}}
\put(10,100){\makebox(0,0)[lb]{
\epsfxsize = 5cm
\epsfbox{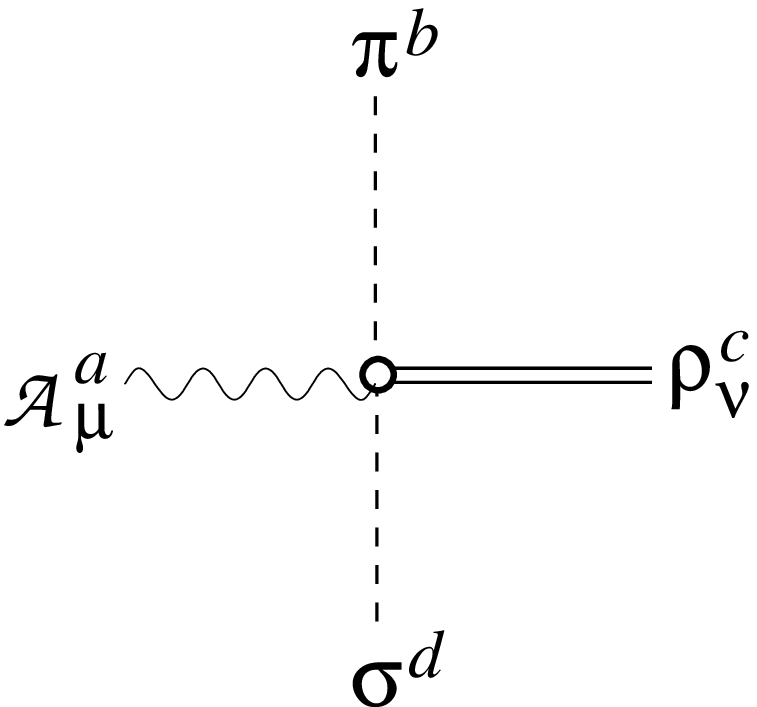}
}}
\put(80,120){\makebox(0,0)[lc]{
{\Large
$\displaystyle
\sqrt{a} \, g \, f_{abe} f_{cde}\, g^{\mu\nu}
$
}
}}
\put(0,69){\makebox(0,0)[lc]{\large (b)}}
\put(10,48){\makebox(0,0)[lb]{
\epsfxsize = 5cm
\epsfbox{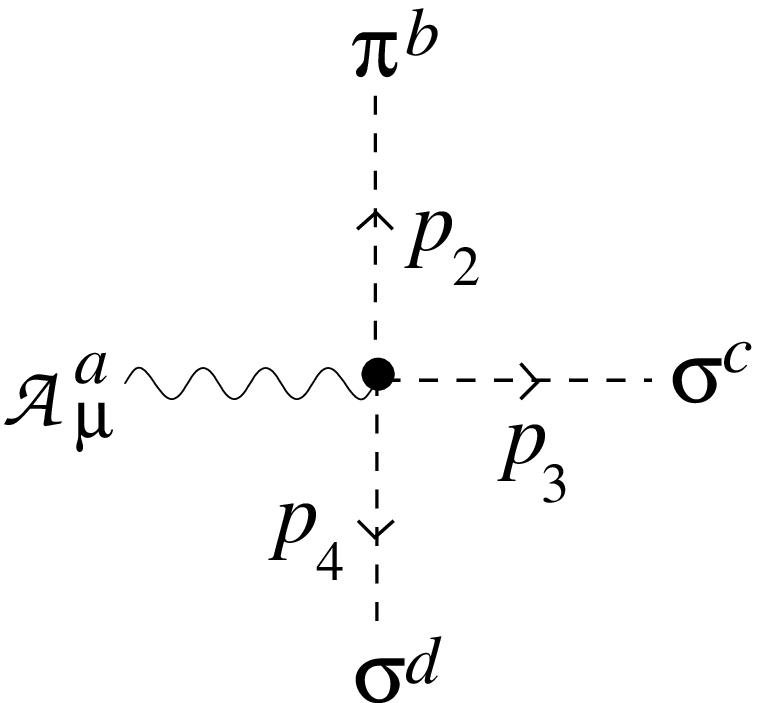}
}}
\put(80,69){\makebox(0,0)[lc]{
{\Large
$\displaystyle
- i \frac{1}{2F_\pi}
\, f_{abe} f_{cde} \, (p_3 - p_4)^\mu
$
}
}}
\put(0,20){\makebox(0,0)[lc]{\large (c)}}
\put(10,-2){\makebox(0,0)[lb]{
\epsfxsize = 5cm
\epsfbox{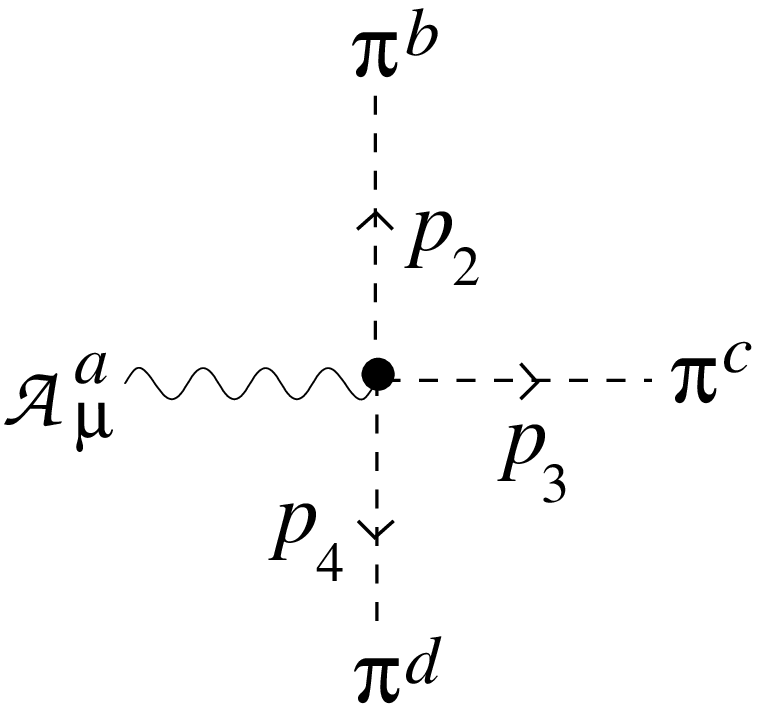}
}}
\put(80,28){\makebox(0,0)[lc]{
{\Large
$\displaystyle
i \frac{3a-4}{6F_\pi}
\biggl[ f_{abe} f_{cde} (p_3 - p_4)^\mu
$
}
}}
\put(85,15){\makebox(0,0)[lc]{
{\Large
$\displaystyle
{} + f_{ace} f_{bde} (p_2 - p_4)^\mu
$
}
}}
\put(85,2){\makebox(0,0)[lc]{
{\Large
$\displaystyle
{} + f_{ade} f_{bce} (p_2 - p_3)^\mu
\biggr]
$
}
}}
\end{picture}
\end{center}
\caption[Feynman rule in the Landau gauge 
(4-point vertices with ${\cal A}_\mu$)]{%
Feynman rules in the Landau gauge for four-point 
vertices which include one ${\cal A}_\mu$.
Here summations over $e$ are taken.
There is no ${\cal A}$-$\rho$-$\rho$-$\pi$ vertex.}
\label{fig:a4L}
\end{figure}

\begin{figure}[htbp]
\begin{center}
\setlength{\unitlength}{1mm}
\begin{picture}(140,155)(0,-25)
\put(-10,155){\makebox(0,0)[lb]{\large\bf
Vertices with ${\cal V}_\mu$}}
\put(0,130){\makebox(0,0)[lc]{\large (a)}}
\put(10,110){\makebox(0,0)[lb]{
\epsfxsize = 4.5cm
\epsfbox{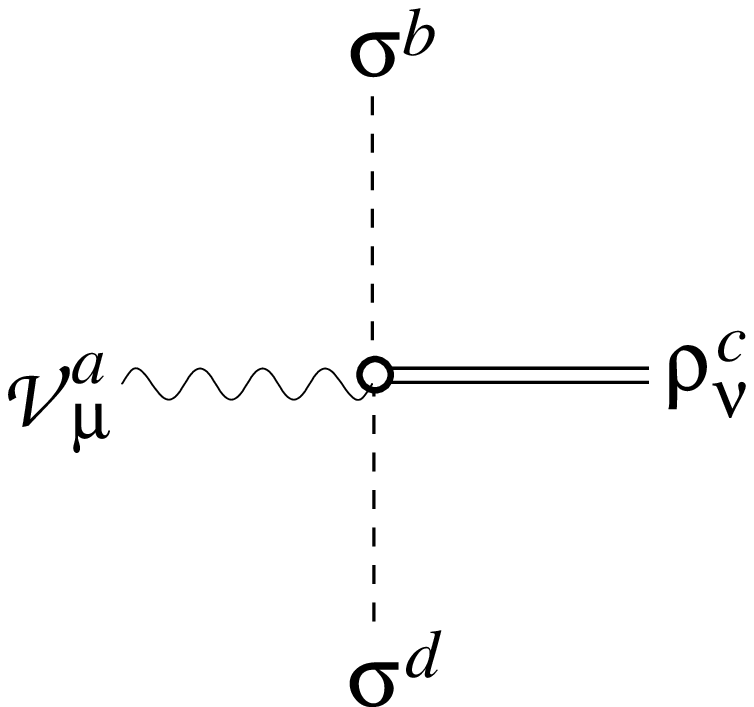}
}}
\put(80,130){\makebox(0,0)[lc]{
{\Large
$\displaystyle
\frac{1}{2} \, g 
\left( f_{abe} f_{cde} + f_{ade} f_{cbe} \right) g^{\mu\nu}
$
}
}}
\put(0,85){\makebox(0,0)[lc]{\large (b)}}
\put(10,65){\makebox(0,0)[lb]{
\epsfxsize = 4.5cm
\epsfbox{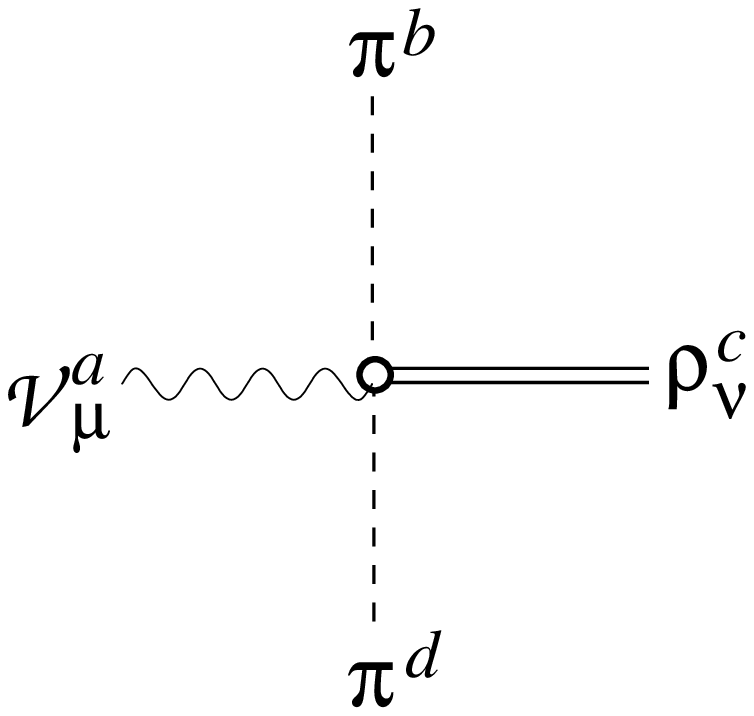}
}}
\put(80,85){\makebox(0,0)[lc]{
{\Large
$\displaystyle
\frac{a}{2} \, g 
\left( f_{abe} f_{cde} + f_{ade} f_{cbe} \right) g^{\mu\nu}
$
}
}}
\put(0,40){\makebox(0,0)[lc]{\large (c)}}
\put(10,20){\makebox(0,0)[lb]{
\epsfxsize = 4.5cm
\epsfbox{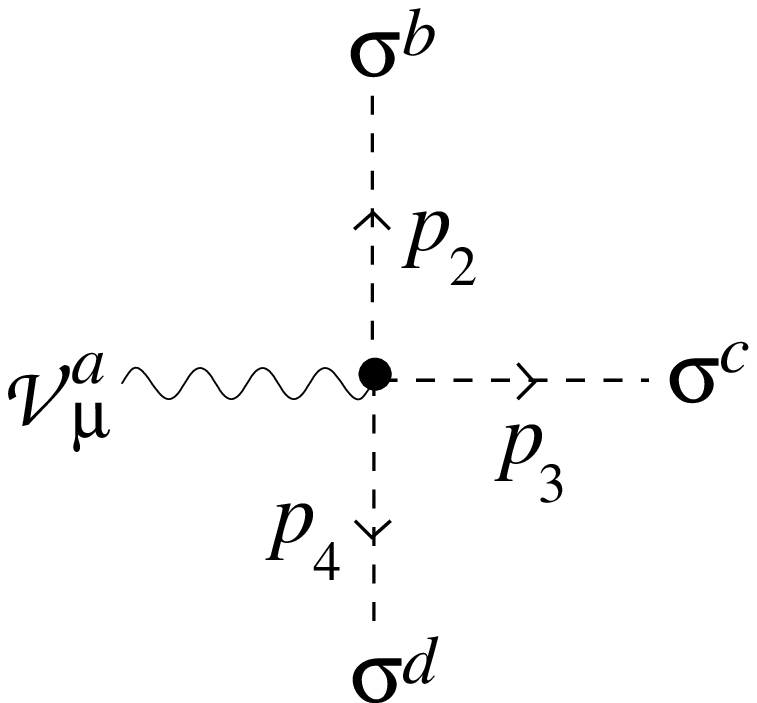}
}}
\put(80,53){\makebox(0,0)[lc]{
{\Large
$\displaystyle
- \frac{i}{6F_\sigma}
\biggl[ f_{abe} f_{cde} (p_3 - p_4)^\mu
$
}
}}
\put(85,40){\makebox(0,0)[lc]{
{\Large
$\displaystyle
{} + f_{ace} f_{bde} (p_2 - p_4)^\mu
$
}
}}
\put(85,27){\makebox(0,0)[lc]{
{\Large
$\displaystyle
{} + f_{ade} f_{bce} (p_2 - p_3)^\mu
\biggr]
$
}
}}
\put(0,-5){\makebox(0,0)[lc]{\large (d)}}
\put(10,-25){\makebox(0,0)[lb]{
\epsfxsize = 4.5cm
\epsfbox{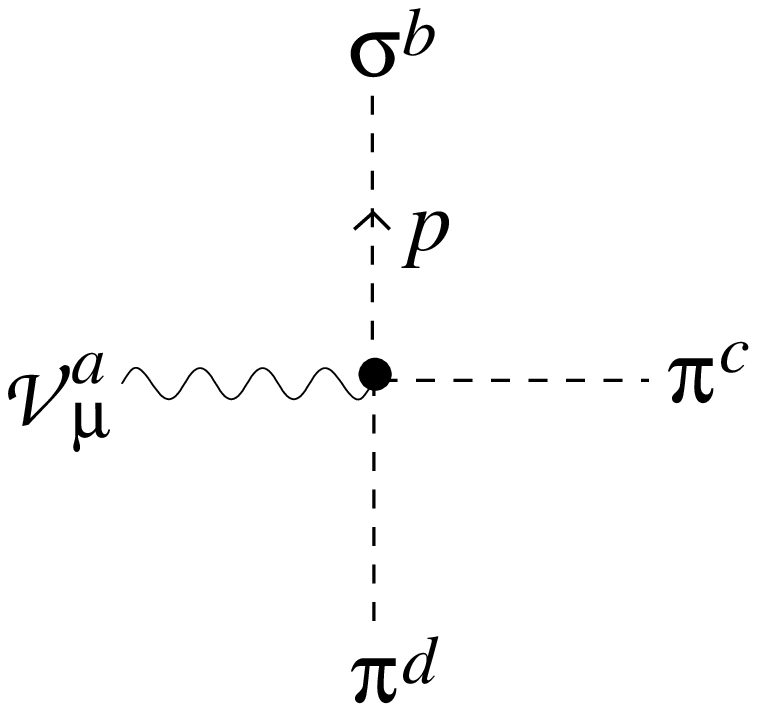}
}}
\put(75,-5){\makebox(0,0)[lc]{
{\Large
$\displaystyle
- i \frac{\sqrt{a}}{2F_\pi}
\left( f_{ace} f_{bde} + f_{ade} f_{bce} \right)
p^\mu
$
}
}}
\end{picture}
\end{center}
\caption[Feynman rule in the Landau gauge 
(4-point vertices with ${\cal V}_\mu$)]{%
Feynman rules in the Landau gauge for four-point 
vertices which include one ${\cal V}_\mu$.
Here summations over $e$ are taken.
There are no ${\cal V}$-$\rho$-$\rho$-$\rho$ 
and ${\cal V}$-$\rho$-$\rho$-$\sigma$ vertices.}
\label{fig:v4L}
\end{figure}

\begin{figure}[htbp]
\begin{center}
\setlength{\unitlength}{1mm}
\begin{picture}(140,45)(0,0)
\put(-10,35){\makebox(0,0)[lb]{\large\bf
Vertices with ${\cal A}_\mu {\cal A}_\nu$}}
\put(0,12){\makebox(0,0)[lc]{\large (a)}}
\put(10,0){\makebox(0,0)[lb]{
\epsfxsize = 5cm
\epsfbox{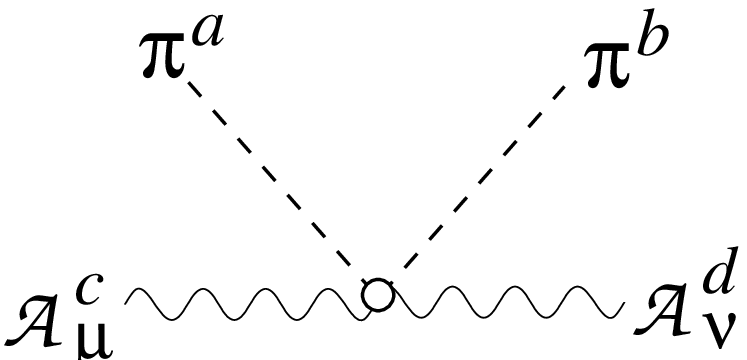}
}}
\put(70,12){\makebox(0,0)[lc]{
{\large
$\displaystyle
- (1-a) \left( f_{cae} f_{dbe} + f_{dae} f_{cbe} \right) g^{\mu\nu} $
}
}}
\end{picture}
\end{center}
\caption[Feynman rule in the Landau gauge (4-point vertex with 
${\cal A}_\mu {\cal A}_\nu$)]{%
Feynman rule for the four-point vertex which includes
${\cal A}_\mu {\cal A}_\nu$.
Here summation over $e$ is taken.
Note that there are no ${\cal A}$-${\cal A}$-$\rho$-$\rho$
and ${\cal A}$-${\cal A}$-$\sigma$-$\sigma$ vertices.
}
\label{fig:a2L}
\end{figure}

\begin{figure}[htbp]
\begin{center}
\setlength{\unitlength}{1mm}
\begin{picture}(140,45)(0,0)
\put(-10,35){\makebox(0,0)[lb]{\large\bf
Vertices with ${\cal V}_\mu {\cal V}_\nu$}}
\put(0,12){\makebox(0,0)[lc]{\large (a)}}
\put(10,0){\makebox(0,0)[lb]{
\epsfxsize = 5cm
\epsfbox{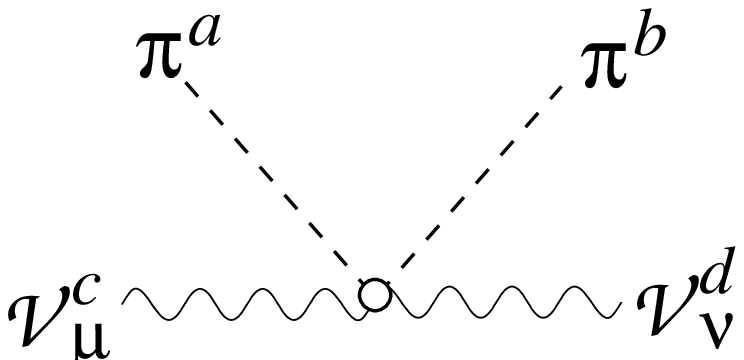}
}}
\put(70,12){\makebox(0,0)[lc]{
{\large
$\displaystyle
(1-a) \left( f_{cae} f_{dbe} + f_{dae} f_{cbe} \right) g^{\mu\nu} $
}
}}
\end{picture}
\end{center}
\caption[Feynman rule in the Landau gauge (4-point vertex with 
${\cal V}_\mu {\cal V}_\nu$)]{%
Feynman rule for the four-point vertex which includes
$\overline{\cal V}_\mu\overline{\cal V}_\nu$.
Here summation over $e$ is taken.
Note that there are no ${\cal V}$-${\cal V}$-$\rho$-$\rho$
and ${\cal V}$-${\cal V}$-$\sigma$-$\sigma$ vertices.
}
\label{fig:v2L}
\end{figure}

\begin{figure}[htbp]
\begin{center}
\setlength{\unitlength}{1mm}
\begin{picture}(140,150)(0,10)
\put(-10,160){\makebox(0,0)[lb]{\large\bf
Vertices with $\rho$ (no ${\cal V}_\mu$ and ${\cal A}_\mu$)
}}
\put(0,130){\makebox(0,0)[lc]{\large (a)}}
\put(10,108){\makebox(0,0)[lb]{
\epsfxsize = 4.5cm
\epsfbox{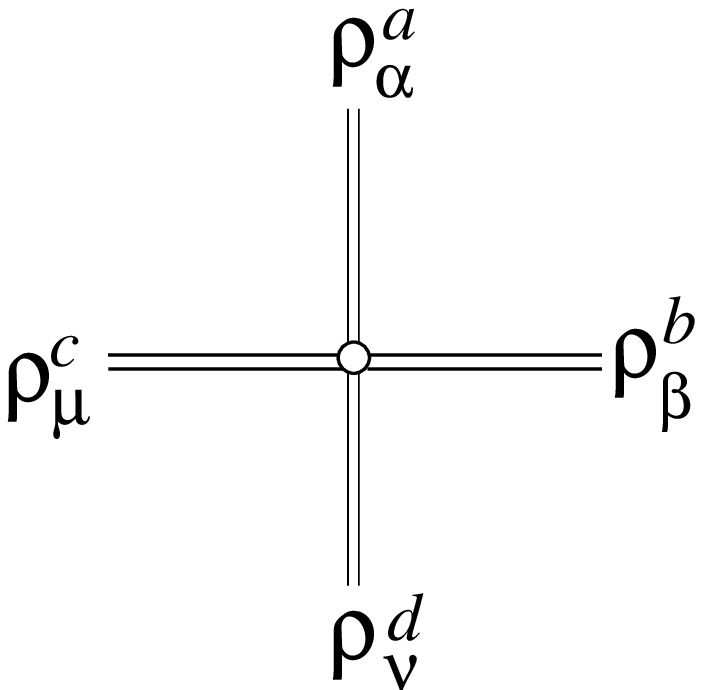}
}}
\put(80,140){\makebox(0,0)[lc]{
{\large
$\displaystyle
  - g^2
  \Biggl[
    f_{e a b}f_{e c d} 
    ( g^{\alpha\mu} g^{\beta\nu} - g^{\alpha\nu} g^{\beta\mu} )
$
}
}}
\put(85,129){\makebox(0,0)[lc]{
{\large
$\displaystyle
    {}+ f_{e a c}f_{e b d} 
    ( g^{\alpha\beta} g^{\mu\nu} - g^{\alpha\nu} g^{\mu\beta} )
$
}
}}
\put(85,118){\makebox(0,0)[lc]{
{\large
$\displaystyle
    {}+ f_{e a d}f_{e b c} 
    ( g^{\alpha\beta} g^{\nu\mu} - g^{\alpha\mu} g^{\nu\beta} )
  \Biggr]
$
}
}}
\put(0,80){\makebox(0,0)[lc]{\large (b)}}
\put(10,60){\makebox(0,0)[lb]{
\epsfxsize = 4.5cm
\epsfbox{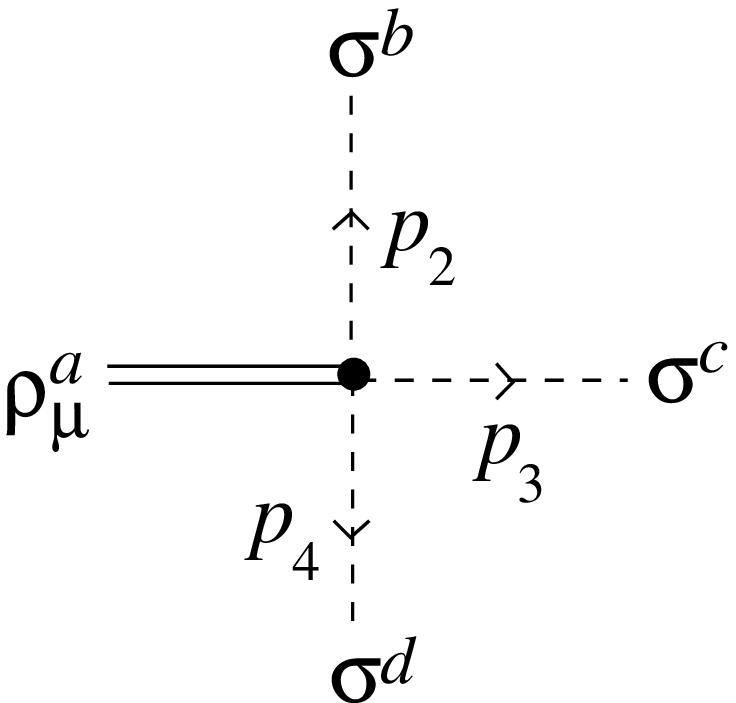}
}}
\put(65,97){\makebox(0,0)[lc]{
{\normalsize
$\displaystyle
i \frac{g}{6F_\sigma}
\biggl( f_{abe} f_{cde} (p_3 - p_4)^\mu
{} + f_{ace} f_{bde} (p_2 - p_4)^\mu
$
}
}}
\put(70,86){\makebox(0,0)[lc]{
{\normalsize
$\displaystyle
{} + f_{ade} f_{bce} (p_2 - p_3)^\mu
\biggr)
$
}
}}
\put(65,75){\makebox(0,0)[lc]{
{\normalsize
$\displaystyle
{} - i \frac{g}{3F_\sigma} 
\left( p_2 + p_3 + p_4 \right)^\mu
\biggl(
\mbox{tr} \bigl[ \{ T_a \,,\, T_b \} \{ T_c \,,\, T_d\} \bigr]
$
}
}}
\put(80,64){\makebox(0,0)[lc]{
{\normalsize
$\displaystyle
{} + \mbox{tr} \bigl[ T_a T_c  T_b T_d \bigr]
{} + \mbox{tr} \bigl[ T_a T_d  T_b T_c \bigr]
\biggr)
$
}
}}
\put(0,30){\makebox(0,0)[lc]{\large (c)}}
\put(10,10){\makebox(0,0)[lb]{
\epsfxsize = 4.5cm
\epsfbox{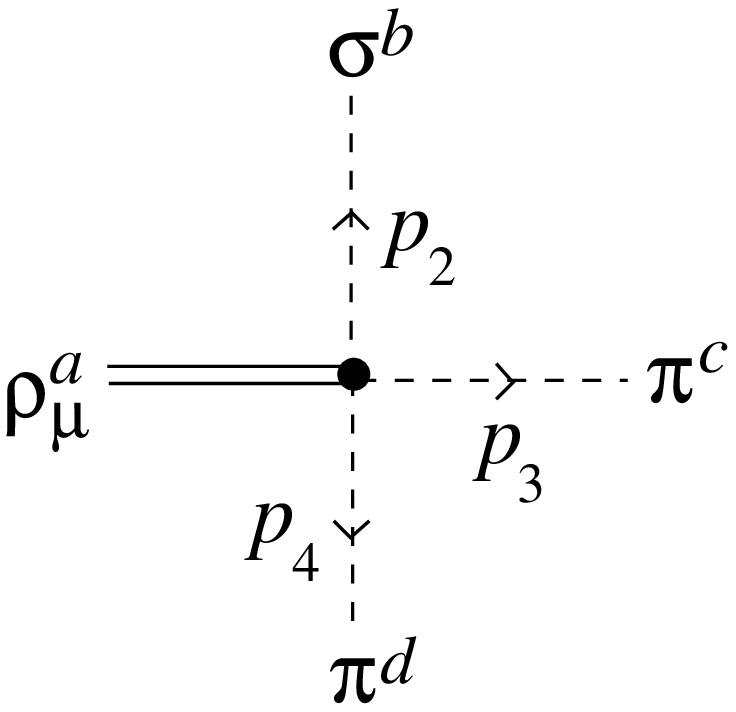}
}}
\put(65,36){\makebox(0,0)[lc]{
{\normalsize
$\displaystyle
i \frac{a g}{2F_\sigma} f_{abe} f_{cde} \,
(p_3 - p_4)^\mu 
$
}
}}
\put(65,23){\makebox(0,0)[lc]{
{\normalsize
$\displaystyle
- i \frac{a g}{2F_\sigma} 
\mbox{tr} \biggl[ \{ T_a \,,\, T_b \} \{ T_c \,,\, T_d\} \biggr]
(p_2 + p_3 + p_4)^\mu
$
}
}}
\end{picture}
\end{center}
\caption[Feynman rule in the Landau gauge 
(4-point vertices with $\rho_\mu$)]{%
Feynman rules in the Landau gauge for four-point 
vertices which include one $\rho$ but no 
${\cal V}_\mu$ and ${\cal A}_\mu$.
Here summations over $e$ are taken.
There are no $\rho$-$\rho$-$\sigma$-$\sigma$
and $\rho$-$\rho$-$\pi$-$\pi$ vertices.
Note that the second term proportional to 
$\left( p_2 + p_3 + p_4 \right)^\mu$ in (b) as well as in (c)
comes from the 
gauge fixing term in the $R_\xi$-like gauge fixing.
}
\label{fig:r4L}
\end{figure}

\begin{figure}[htbp]
\begin{center}
\setlength{\unitlength}{1mm}
\begin{picture}(140,150)(0,10)
\put(-10,160){\makebox(0,0)[lb]{\large\bf
Vertices with no $\rho$, ${\cal V}$ and ${\cal A}$
}}
\put(0,130){\makebox(0,0)[lc]{\large (a)}}
\put(10,108){\makebox(0,0)[lb]{
\epsfxsize = 4.5cm
\epsfbox{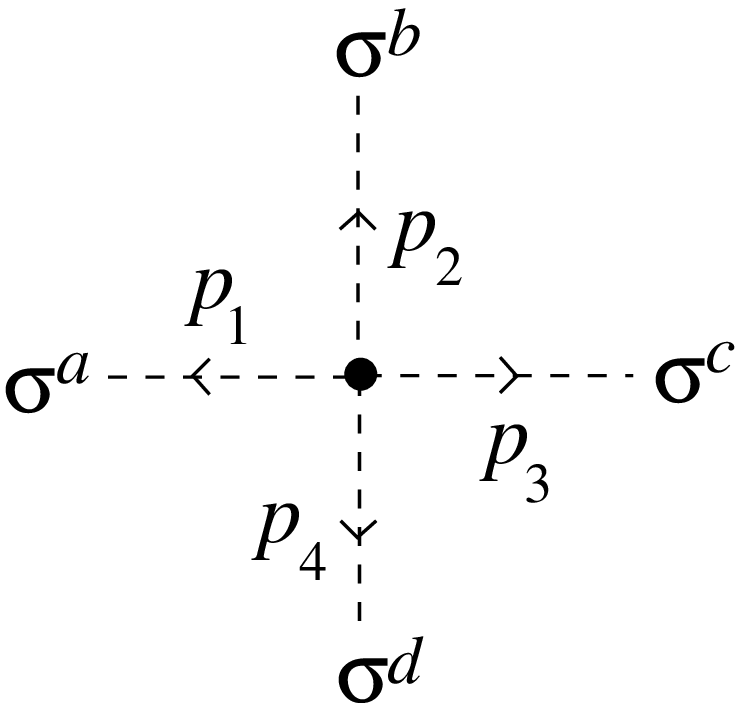}
}}
\put(70,140){\makebox(0,0)[lc]{
{\large
$\displaystyle
  \frac{1}{12 F_\sigma^2}
  \Biggl[
    f_{a c e}f_{b d e} 
    \left( p_1 - p_3 \right) \cdot \left( p_2 - p_4 \right)
$
}
}}
\put(75,129){\makebox(0,0)[lc]{
{\large
$\displaystyle
    {}+ f_{a d e}f_{c b e} 
    \left( p_1 - p_4 \right) \cdot \left( p_3 - p_2 \right)
$
}
}}
\put(75,118){\makebox(0,0)[lc]{
{\large
$\displaystyle
    {}+ f_{a b e}f_{d c e} 
    \left( p_1 - p_2 \right) \cdot \left( p_4 - p_3 \right)
  \Biggr]
$
}
}}
\put(0,80){\makebox(0,0)[lc]{\large (b)}}
\put(10,60){\makebox(0,0)[lb]{
\epsfxsize = 4.5cm
\epsfbox{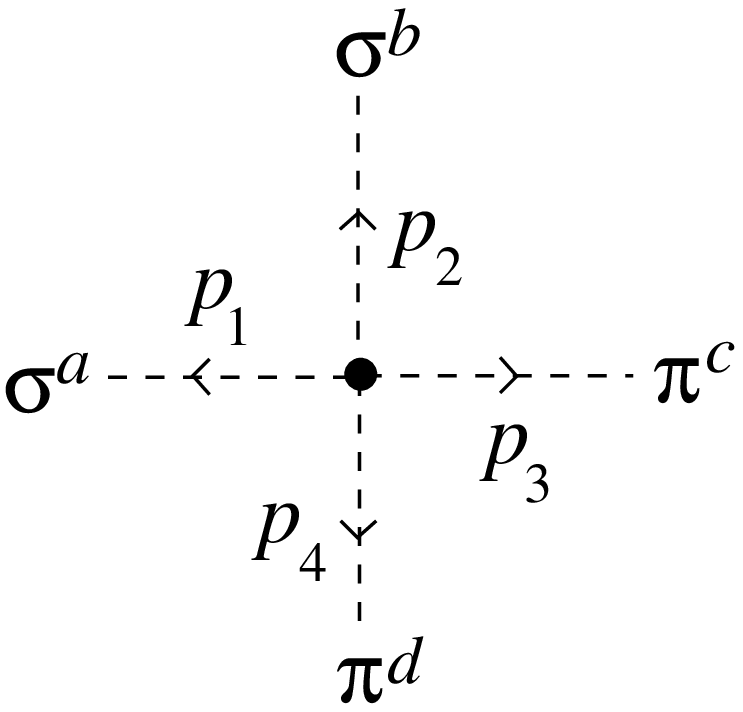}
}}
\put(65,80){\makebox(0,0)[lc]{
{\large
$\displaystyle
\frac{1}{4F_\sigma^2} \,
f_{abe} f_{cde} \, (p_1 - p_2) \cdot (p_3 - p_4)
$
}
}}
\put(0,30){\makebox(0,0)[lc]{\large (c)}}
\put(10,10){\makebox(0,0)[lb]{
\epsfxsize = 4.5cm
\epsfbox{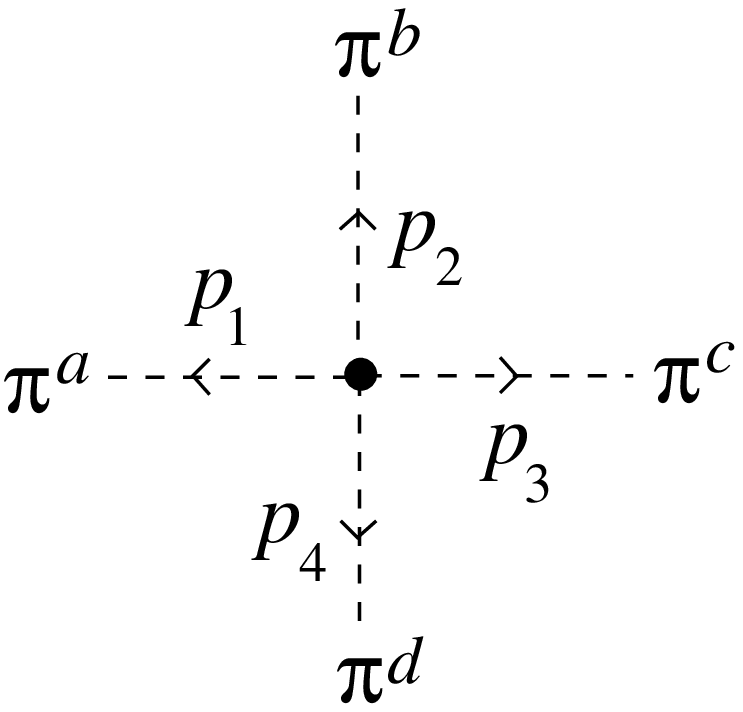}
}}
\put(70,42){\makebox(0,0)[lc]{
{\large
$\displaystyle
  \frac{4-3a}{12 F_\pi^2}
  \Biggl[
    f_{a c e}f_{b d e} 
    \left( p_1 - p_3 \right) \cdot \left( p_2 - p_4 \right)
$
}
}}
\put(75,31){\makebox(0,0)[lc]{
{\large
$\displaystyle
    {}+ f_{a d e}f_{c b e} 
    \left( p_1 - p_4 \right) \cdot \left( p_3 - p_2 \right)
$
}
}}
\put(75,20){\makebox(0,0)[lc]{
{\large
$\displaystyle
    {}+ f_{a b e}f_{d c e} 
    \left( p_1 - p_2 \right) \cdot \left( p_4 - p_3 \right)
  \Biggr]
$
}
}}
\end{picture}
\end{center}
\caption[Feynman rule in the Landau gauge 
(4-point vertices with no $\rho_\mu$, ${\cal V}$ and ${\cal A}$)]{%
Feynman rules in the Landau gauge for four-point 
vertices which include no $\rho$, ${\cal V}_\mu$ and ${\cal A}_\mu$.
Here summations over $e$ are taken.
}
\label{fig:4L}
\end{figure}

\newpage

\section{Renormalization in the Heat Kernel Expansion}
\label{app:RHKE}

In this appendix we 
use the heat kernel expansion and the proper time
regularization to determine the divergent contributions at one loop.

\subsection{Ghost contributions}

Let us first consider the ghost contribution.
The contribution is given by
\begin{equation}
\Gamma_{\rm FP} = - i \, \mbox{Ln} \det \widetilde{\nabla}^{(CC)}
\ ,
\end{equation}
where
\begin{equation}
\widetilde{\nabla}^{(CC)}_{ab}
\equiv
  \left( 
     \widetilde{D}_{\mu} \cdot \widetilde{D}^{\mu} 
  \right)_{ab}^{(CC)}
  + \widetilde{\cal M}^{(CC)}_{ab}
\ .
\end{equation}
By using the proper time integral with a ultraviolet 
cutoff $\Lambda$ this
$\Gamma_{\rm FP}$ is evalated as
\begin{equation}
\Gamma_{\rm FP} = i \, 
\int^{\infty}_{1/\Lambda^2} \frac{dt}{t} \mbox{Tr} \sum_{a}
\, \left( \bar{F}(x,x;t) \right)_{aa}
\ ,
\label{Gam FP}
\end{equation}
where $\left( \bar{F}(x,y;t) \right)_{ab}$ is obtained by solving the
heat equation:
\begin{equation}
\frac{\partial}{\partial t}
\left( \bar{F}(x,y;t) \right)_{ab}
+ \sum_{c}
\widetilde{\nabla}^{(CC)}_{ac}
\left( \bar{F}(x,y;t) \right)_{cb} = 0
\ .
\label{full heat C}
\end{equation}
To solve the above heat equation let us start from the
heat equation for the free field:
\begin{equation}
\frac{\partial}{\partial t}
\bar{F}_0(x,y;t;M_\rho^2) +
\left( \Box + M_\rho^2 \right) \bar{F}_0(x,y;t;M_\rho^2) = 0 
\ .
\end{equation}
The solution is given by
\begin{equation}
\bar{F}_0(x,y;t;M_\rho^2) = 
\frac{i}{\left( 4 \pi t \right)^{n/2}}\,
\exp\left[ \frac{(x-y)^2}{4t} -t M_\rho^2 \right]
\ ,
\end{equation}
where $n$ is the space-time dimension.
The solution of Eq.~(\ref{full heat C}) is expressed as
\begin{equation}
\left( \bar{F}(x,y;t) \right)_{ab} = 
\bar{F}_0(x,y;t;M_\rho^2) 
\left( \bar{H}(x,y:t) \right)_{ab}
\ ,
\end{equation}
where $\left( \bar{H}(x,y:t) \right)_{ab}$ satisfies
\begin{eqnarray}
&&
\frac{\partial}{\partial t}
\left( \bar{H}(x,y;t) \right)_{ab}
+ \frac{(x-y)^\mu}{t} 
\sum_{c}
\left( \widetilde{D}_{\mu} \right)_{ac}^{(CC)}
\left( \bar{H}(x,y;t) \right)_{cb}
\nonumber\\
&& \qquad
+ \sum_{c}
\left(
   \widetilde{D}_{\mu} \cdot \widetilde{D}^{\mu} 
\right)_{ac}^{(CC)}
\left( \bar{H}(x,y;t) \right)_{cb} = 0
\ .
\label{heat H C}
\end{eqnarray}
Equation~(\ref{heat H C}) can be solved by expanding
$\left( \bar{H}(x,y:t) \right)_{ab}$ in terms of $t$:
\begin{equation}
\left( \bar{H}(x,y:t) \right)_{ab} =
\sum_{j=0} t^j \left( \bar{H}_j(x,y) \right)_{ab}
\ .
\label{expand H}
\end{equation}
Substituting this into Eq.~(\ref{Gam FP}) and taking $n=4$ we obtain
\begin{equation}
\Gamma_{\rm FP} = - \frac{1}{(4\pi)^2}
\sum_{j=0} \,
\left( \bar{M}^2_v \right)^{2-j}
\Gamma\left( j - 2 , \varepsilon \right)
\, \mbox{Tr} H_j(x,x)
\ ,
\end{equation}
where
\begin{equation}
\varepsilon = \frac{M_\rho^2}{\Lambda^2} \ ,
\end{equation}
and
$\Gamma\left( j  , \varepsilon \right)$ is the incomplete gamma
function defined by
\begin{equation}
\Gamma\left( j  , \varepsilon \right)
\equiv
\int^\infty_\varepsilon \frac{dz}{z} e^{-z} z^j \ .
\end{equation}
Several convenient formulae for the incomplete gamma function are
summarized in Appendix~\ref{ssec:IGF}.

Equation~(\ref{heat H C}) is solved by 
substituting the expansion in Eq.~(\ref{expand H}).
The results are given by
\begin{eqnarray}
\left( \bar{H}_0(x,y) \right)_{ab} &=& \delta_{ab}
\ , \qquad \mbox{(normalization)} \ ,
\\
\left( \bar{H}_1(x,y) \right)_{ab} &=& 0 \ ,
\\
\left( \bar{H}_2(x,y) \right)_{ab} &=& 
\frac{1}{12} \left( \bar{\Gamma}_{\mu\nu} \cdot
\bar{\Gamma}^{\mu\nu} \right)_{ab}
\ ,
\end{eqnarray}
where 
\begin{equation}
\left( \bar{\Gamma}_{\mu\nu} \right)_{ab}
\equiv
\biggl[ 
  \widetilde{D}_{\mu,(CC)} \ ,\  \widetilde{D}_{\mu}^{(CC)}
\biggr]_{ab}
\ .
\end{equation}
By using the definitions in Eqs.~(\ref{XCC}) and (\ref{def:DCC})
this $\left( \bar{\Gamma}_{\mu\nu} \right)_{ab}$ is expressed as
\begin{equation}
\left( \bar{\Gamma}_{\mu\nu} \right)_{ab}
=
2 i \, \mbox{tr} \biggl[
  \overline{V}_{\mu\nu} \left[ T_a \,,\,T_b \right]
\biggr]
\ .
\end{equation}

Now $\Gamma_{\rm FP}$ is evaluated as
\begin{eqnarray}
\Gamma_{\rm FP}
= \frac{1}{(4\pi)^2} \int d^4x
\Biggl[
  - M_\rho^4 \Gamma\left(-2,\varepsilon\right)
  + \frac{N_f}{6} \Gamma\left(0,\varepsilon\right)
  \, \mbox{tr} \, \left[
    \widetilde{V}_{\mu\nu} \widetilde{V}^{\mu\nu} 
  \right]
\Biggr]
+ \cdots \ ,
\end{eqnarray}
where dots stands for the non-divergent contributions.
Using the formulae for the incomplete gamma function 
given in Appendix~\ref{ssec:IGF}, the divergent contribution to 
$\Gamma_{\rm FP}$ is evaluated as
\begin{eqnarray}
\Gamma_{\rm FP}
= \frac{1}{(4\pi)^2} 
\int d^4x
\Biggl[
  \ln \frac{\Lambda^2}{M_\rho^2}
  \times
  \frac{N_f}{6}
  \, \mbox{tr} \, \left[
    \overline{V}_{\mu\nu} \overline{V}^{\mu\nu} 
  \right]
\Biggr]
\ ,
\label{gam fp}
\end{eqnarray}
where we dropped the constant term.

\subsection{$\pi$, $V$ and $\sigma$ contributions}

Let us calculate the one-loop contributions 
from $\pi$, $V$ and $\sigma$.
These are given by
\begin{eqnarray}
\Gamma_{PV} &=&
\frac{i}{2} \mbox{Ln} \mbox{Det} \widetilde{\nabla} \ ,
\end{eqnarray}
where $\widetilde{\nabla}$ is defined by
\begin{equation}
\widetilde{\nabla}^{AB} \equiv 
  \left( \widetilde{D}_\mu \cdot \widetilde{D}^\mu \right)^{AB}
  + \widetilde{\cal M}^{AB}
  + \widetilde{\Sigma}^{AB} 
\ .
\end{equation}
Similarly to the ghost contributions 
this $\Gamma_{\rm PV}$ is evaluated as
\begin{equation}
\Gamma_{\rm PV} = - \frac{i}{2} \, 
\int^{\infty}_{1/\Lambda^2} \frac{dt}{t} \mbox{Tr} \sum_{A}
\, F_{A}^A(x,x;t)
\ ,
\label{Gam PV}
\end{equation}
where $\left( F(x,y;t) \right)_{A}^B$ is obtained by solving the
heat equation:
\begin{equation}
\frac{\partial}{\partial t} F_A^B(x,y;t)
+ \sum_{C} \widetilde{\nabla}_A^C F_C^B(x,y;t) = 0
\ .
\label{full heat PV}
\end{equation}
This looks similar to the ghost case.
However, it is much more difficult to solve it since a difference
appears in the heat equation in the free fields:
\begin{equation}
\frac{\partial}{\partial t} {F_0}_A^B(x,y;t) +
\sum_{C} \left( \eta_A^C \Box + \widetilde{\cal M}_A^C \right) 
  {F_0}_C^B(x,y;t) = 0 
\ .
\end{equation}
The solution is given by
\begin{equation}
{F_0}_A^B(x,y;t) = \widetilde{F}_0 (x,y;t) P_A^B \ ,
\end{equation}
where
\begin{equation}
P_A^B = 
\left( \begin{array}{ccc}
1 & & \\
 & e^{-tM_\rho^2} & \\
 & & e^{-tM_\rho^2}
\end{array} \right)
\ ,
\end{equation}
and
\begin{equation}
\widetilde{F}_0(x,y;t) \equiv
\frac{i}{(4\pi t)^{n/2}} \exp 
\left[ \frac{(x-y)^2}{4t} \right]
\ .
\end{equation}
Hereafter we suppress the sufixes of the matrices.
It is useful to express the full solution as
\begin{equation}
\bar{F}(x,y;t) = \widetilde{F}_0(x,y;t)
\left[ P \cdot H(x,y;t) \right]
\ ,
\end{equation}
where $H(x,y;t)$ satisfies
\begin{eqnarray}
&&
\frac{\partial}{\partial t} H(x,y;t) 
+ \frac{(x-y)^\mu}{t} 
  P^{-1} \cdot \widetilde{D}_{\mu} \cdot P \cdot H(x,y;t)
\nonumber\\
&& \qquad
+ P^{-1} \cdot
\left(
  \widetilde{D}_{\mu} \cdot \widetilde{D}^{\mu} 
  + \widetilde{\Sigma} 
\right)
\cdot P \cdot
H(x,y;t) = 0
\ .
\label{heat H PV}
\end{eqnarray}
Similarly to the ghost contributions
Eq.~(\ref{heat H PV}) can be solved by expanding
$H(x,y;t)$ in terms of $t$:
\begin{equation}
H(x,y;t) = \sum_{j=0} t^j H_j(x,y) 
\ .
\end{equation}
First three, i.e., $H_0$, $H_1$ and $H_2$ are given by
\begin{eqnarray}
H_0(x,y) &=& 1
\ , \qquad \mbox{(normalization)} \ ,
\\
H_1(x,y) &=& - 
P^{-1} \cdot \widetilde{\Sigma} \cdot P \ ,
\\
H_2(x,y) &=& P^{-1} \left(
  \frac{1}{12} \Gamma_{\mu\nu} \cdot \Gamma^{\mu\nu}
  + \frac{1}{2} \widetilde{\Sigma} \cdot \widetilde{\Sigma}
  + \frac{1}{6} \biggl[
    \widetilde{D}^\mu \,,\, 
    \left[ \widetilde{D}_\mu \,,\, \widetilde{\Sigma} \right]
  \biggr]
\right)
\cdot P
\ ,
\end{eqnarray}
where 
\begin{equation}
\Gamma_{\mu\nu} 
\equiv
\biggl[ \widetilde{D}_\mu \, ,\,  \widetilde{D}_\nu \biggr]
\label{def gam}
\ .
\end{equation}
Then $\Gamma_{\rm PV}$ is formally evaluated by
\begin{equation}
\Gamma_{\rm PV} = \frac{1}{2(4\pi)^2}
\int^{\infty}_{1/\Lambda^2} dt \sum_{j=0}
t^{j-3} \mbox{Tr} \, \biggl( P \cdot H_j(x,x) \biggr)
\ .
\end{equation}
Since this expression includes an infrared divergence coming from the
pion loops, we regularize this by introducing a small mass to pions.
This is done by performing the following replacement:
\begin{equation}
P \rightarrow \widetilde{P} \equiv
\left( \begin{array}{ccc}
 e^{-t\mu^2} & & \\
 & e^{-tM_\rho^2} & \\
 & & e^{-tM_\rho^2}
\end{array} \right)
\ .
\end{equation}

Let us evaluate $\Gamma_{\rm PV}$ step by step.
The contribution for $j=0$ is just a constant:
\begin{eqnarray}
\Gamma_{\rm PV}^{(0)} = \frac{1}{2(4\pi)^2}
\int d^4x
\left[
  \Gamma\left(-2,\widetilde{\varepsilon}\right) 
  + 5 M_\rho^4 \Gamma\left(-2,\varepsilon\right)
\right]
\ ,
\end{eqnarray}
where
\begin{equation}
\varepsilon \equiv \frac{M_\rho^2}{\Lambda^2} \ ,\quad
\widetilde{\varepsilon} \equiv \frac{\mu^2}{\Lambda^2} \ .
\end{equation}
The contribution for $j=1$ is given by
\begin{eqnarray}
\Gamma_{\rm PV}^{(1)} = \frac{1}{2(4\pi)^2}
\sum_{a}
\int d^4x
\left[
  - \mu^2 \Gamma\left(-1,\widetilde{\varepsilon}\right) 
    \Sigma^{(\pi\pi)}_{aa} 
  - M_\rho^2 \Gamma\left(-1,\varepsilon\right)
    \Sigma^{(\sigma\sigma)}_{aa} 
\right]
\ .
\end{eqnarray}
Using the formulas given in Appendix~\ref{ssec:FG}, we obtain
\begin{eqnarray}
\sum_{a=1}^{N_f^2-1} \Sigma^{(\pi\pi)}_{aa}
&=&
\frac{4-3a}{2} \, N_f \, \mbox{tr}
\left[ 
  \overline{\cal A}_{\mu} \overline{\cal A}^{\mu} 
\right]
+
\frac{a^2}{2} \, N_f \, \mbox{tr}
\left[ 
  \left( \overline{\cal V}_{\mu} - \overline{V}_{\mu} \right)
  \left( \overline{\cal V}^{\mu} - \overline{V}^{\mu} \right)
\right]
\nonumber\\
&&
{} +
\frac{N_f^2-1}{N_f} \frac{F_\chi^2}{F_\pi^2} \, \mbox{tr}
\biggl[
  \overline{\chi} + \overline{\chi}^\dag 
\biggr]
\ ,
\\
\sum_{a=1}^{N_f^2-1} \Sigma^{(\sigma\sigma)}_{aa}
&=&
\frac{a}{2} \, N_f \, \mbox{tr}
\left[ 
  \overline{\cal A}_{\mu} \overline{\cal A}^{\mu} 
\right]
+
\frac{1}{2} \, N_f \, \mbox{tr}
\left[ 
  \left( \overline{\cal V}_{\mu} - \overline{V}_{\mu} \right)
  \left( \overline{\cal V}^{\mu} - \overline{V}^{\mu} \right)
\right]
\ .
\end{eqnarray}
By using the formulas for $\Gamma\left(j,\varepsilon\right)$
in
Appendix~\ref{ssec:IGF}, $\Gamma_{\rm PV}^{(1)}$ is evaluated as
\begin{eqnarray}
\Gamma_{\rm PV}^{(1)} &=& \frac{1}{(4\pi)^2}
\int d^4x
\Biggl[
  \biggl(
    - \frac{2-a}{2} \, N_f \, \Lambda^2
    + \frac{a}{4} \,N_f\, 
       M_\rho^2 \ln \frac{\Lambda^2}{M_\rho^2}
  \biggr)
  \mbox{tr} 
  \left[
    \overline{\cal A}_{\mu} \overline{\cal A}^{\mu} 
  \right]
\nonumber\\
&& \quad
  {} + 
  \biggl(
    - \frac{1+a^2}{4} \, N_f \, \Lambda^2 
    + \frac{1}{4} \,N_f\, M_\rho^2 
      \ln \frac{\Lambda^2}{M_\rho^2}
  \biggr)
  \mbox{tr} 
  \left[
    \left( \overline{\cal V}_{\mu} - \overline{V}_{\mu} \right)
    \left( \overline{\cal V}^{\mu} - \overline{V}^{\mu} \right)
  \right]
\nonumber\\
&& \quad
  {} - \frac{N_f^2-1}{2N_f} \Lambda^2
  \frac{F_\chi^2}{F_\pi^2} 
  \, \mbox{tr}
  \left[
    \overline{\chi} + \overline{\chi}^\dag
  \right]
\Biggr]
\ ,
\label{gam pv1}
\end{eqnarray}
where we have taken $\mu=0$.

For identifying the logarithmic divergence in $\Gamma_{\rm PV}^{(2)}$
it is easy and enough to take $\mu=M_\rho$,
so that we can simply take 
$\mbox{Tr}\, \left[ H_2(x,x) \right]$ instead of
$\mbox{Tr}\, \left[ P \cdot H_2(x,x) \right]$.
Thus,
\begin{eqnarray}
\Gamma_{\rm PV}^{(2)} &=& 
\frac{1}{2(4\pi)^2}
\int^{\infty}_{1/\Lambda^2} \frac{dt}{t} \, e^{-tM_\rho^2}\,
\mbox{Tr} \, \biggl( H_2(x,x) \biggr)
\nonumber\\
&=&
\frac{1}{2(4\pi)^2} \Gamma(0,\varepsilon) \,
\int d^4x 
\left(
  \frac{1}{2} \, \mbox{tr} 
    \left[ \widetilde{\Sigma} \cdot \widetilde{\Sigma} \right]
  + \frac{1}{12} \, \mbox{tr} 
    \left[ \Gamma_{\mu\nu} \cdot \Gamma^{\mu\nu} \right]
\right)
\ ,
\end{eqnarray}
where
\begin{equation}
\Gamma(0,\varepsilon) \simeq \ln \frac{\Lambda^2}{M_\rho^2}
\ .
\end{equation}

Let us calculate $\widetilde{\Sigma} \cdot \widetilde{\Sigma}$ 
parts by parts.
$\sum_a \left( \Sigma \cdot \Sigma \right)_{aa}^{(\pi\pi)}$ is given
by
\begin{eqnarray}
&&
\sum_a \left( \Sigma \cdot \Sigma \right)_{aa}^{(\pi\pi)}
=
\sum_{a,b} \left[ 
  \Sigma_{ab}^{(\pi\pi)} \Sigma_{ba}^{(\pi\pi)}
  + \Sigma_{ab}^{(\pi\sigma)} \Sigma_{ba}^{(\sigma\pi)}
  + \sum_{\alpha} \Sigma_{ab}^{(\pi V_\alpha)} 
                  \Sigma_{(V_\alpha ba}^{\pi)}
\right]
\ ,
\end{eqnarray}
where
\begin{eqnarray}
&&
\sum_{a,b} \Sigma_{ab}^{(\pi\pi)} \Sigma_{ba}^{(\pi\pi)}
\nonumber\\
&& \ 
=
\frac{(4-3a)^2}{8} \, N_f
\, \mbox{tr} \left[ 
  \left(
    \overline{\cal A}_\mu \overline{\cal A}^\mu
  \right)^2
\right]
{} + \frac{a^4}{8} \, N_f
\, \mbox{tr} \left[
  \left(
    \left( \overline{\cal V}_\mu - \overline{V}_\mu \right)
    \left( \overline{\cal V}^\mu - \overline{V}^\mu \right)
  \right)^2
\right]
\nonumber\\
&& \quad
{} + \frac{a^2(4-3a)}{4} \, N_f
\, \mbox{tr} \left[
  \overline{\cal A}_\mu \overline{\cal A}^\mu
  \left( \overline{\cal V}_\nu - \overline{V}_\nu \right)
  \left( \overline{\cal V}^\nu - \overline{V}^\nu \right)
\right]
\nonumber\\
&& \quad
{}+ \frac{(4-3a)^2}{8} 
\biggl(
  \mbox{tr} \left[ 
    \overline{\cal A}_\mu \overline{\cal A}^\mu
  \right]
\biggr)^2
{} + \frac{(4-3a)^2}{4} 
\, \mbox{tr} \left[ 
  \overline{\cal A}_\mu \overline{\cal A}_\nu
\right]
\, \mbox{tr} \left[ 
  \overline{\cal A}^\mu \overline{\cal A}^\nu
\right]
\nonumber\\
&& \quad
{} + \frac{a^4}{8}
\biggl(
  \mbox{tr} \left[
    \left( \overline{\cal V}_\mu - \overline{V}_\mu \right)
    \left( \overline{\cal V}^\mu - \overline{V}^\mu \right)
  \right]
\biggr)^2
\nonumber\\
&& \quad
{} + \frac{a^4}{4}
\, \mbox{tr} \left[
  \left( \overline{\cal V}_\mu - \overline{V}_\mu \right)
  \left( \overline{\cal V}_\nu - \overline{V}_\nu \right)
\right]
\, \mbox{tr} \left[
  \left( \overline{\cal V}^\mu - \overline{V}^\mu \right)
  \left( \overline{\cal V}^\nu - \overline{V}^\nu \right)
\right]
\nonumber\\
&& \quad
{} + \frac{a^2(4-3a)}{4}
\, \mbox{tr} \left[
  \overline{\cal A}_\mu \overline{\cal A}^\mu
\right]
\, \mbox{tr} \left[
  \left( \overline{\cal V}_\nu - \overline{V}_\nu \right)
  \left( \overline{\cal V}^\nu - \overline{V}^\nu \right)
\right]
\nonumber\\
&& \quad
{} + \frac{a^2(4-3a)}{2}
\, \mbox{tr} \left[
  \overline{\cal A}_\mu 
  \left( \overline{\cal V}_\nu - \overline{V}_\nu \right)
\right]
\, \mbox{tr} \left[
  \overline{\cal A}^\mu
  \left( \overline{\cal V}^\nu - \overline{V}^\nu \right)
\right]
\nonumber\\
&& \quad
{} + \frac{4-3a}{4} \, N_f \, \frac{F_\chi^2}{F_\pi^2}
\, \mbox{tr} \left[
  \overline{\cal A}_\mu \overline{\cal A}^\mu
  \left( \overline{\chi} + \overline{\chi}^\dag \right)
\right]
{} + \frac{4-3a}{4} \, \frac{F_\chi^2}{F_\pi^2}
\, \mbox{tr} \left[
  \overline{\cal A}_\mu \overline{\cal A}^\mu
\right]
\, \mbox{tr} \left[ \overline{\chi} + \overline{\chi}^\dag \right]
\nonumber\\
&& \quad
{} + \frac{a^2}{4} \, N_f \, \frac{F_\chi^2}{F_\pi^2}
\, \mbox{tr} \left[
  \left( \overline{\cal V}_\mu - \overline{V}_\mu \right)
  \left( \overline{\cal V}^\mu - \overline{V}^\mu \right)
  \left( \overline{\chi} + \overline{\chi}^\dag \right)
\right]
\nonumber\\
&& \quad
{} + \frac{a^2}{4} \, \frac{F_\chi^2}{F_\pi^2}
\, \mbox{tr} \left[
  \left( \overline{\cal V}_\mu - \overline{V}_\mu \right)
  \left( \overline{\cal V}^\mu - \overline{V}^\mu \right)
\right]
\, \mbox{tr} \left[ \overline{\chi} + \overline{\chi}^\dag \right]
\nonumber\\
&& \quad
{} + \frac{ N_f^2-4}{8N_f} \left(\frac{F_\chi^2}{F_\pi^2}\right)^2
\, \mbox{tr} \left[
  \left( \overline{\chi} + \overline{\chi}^\dag \right)^2
\right]
{} + \frac{ N_f^2+2}{8N_f^2} \left(\frac{F_\chi^2}{F_\pi^2}\right)^2
\biggl(
  \mbox{tr} \left[ \overline{\chi} + \overline{\chi}^\dag \right]
\biggr)^2
\ ,
\\
&&
\sum_{a,b} \Sigma_{ab}^{(\pi\sigma)} \Sigma_{ba}^{(\sigma\pi)}
\nonumber\\
&& \ 
=
\frac{a\left(5a^2-12a+9\right)}{8} \, N_f
\, \mbox{tr}
\left[
  \overline{\cal A}_\mu
  \overline{\cal A}_\nu
  \left( \overline{\cal V}^\nu - \overline{V}^\nu \right)
  \left( \overline{\cal V}^\mu - \overline{V}^\mu \right)
\right]
\nonumber\\
&& \quad
{} + \frac{a^2\left(3-2a\right)}{8} \, N_f
\Biggl(
  \mbox{tr}
  \left[
    \overline{\cal A}_\mu
    \left( \overline{\cal V}_\nu - \overline{V}_\nu \right)
    \overline{\cal A}^\nu
    \left( \overline{\cal V}^\mu - \overline{V}^\mu \right)
  \right]
\nonumber\\
&& \qquad\qquad\qquad\qquad\qquad
  {}+
  \, \mbox{tr}
  \left[
    \overline{\cal A}_\mu
    \left( \overline{\cal V}^\mu - \overline{V}^\mu \right)
    \overline{\cal A}_\nu
    \left( \overline{\cal V}^\nu - \overline{V}^\nu \right)
  \right]
\Biggr)
\nonumber\\
&& \quad
{} + \frac{a(9-6a+a^2)}{8}
\, \mbox{tr}
\left[
  \overline{\cal A}_\mu
  \left( \overline{\cal V}^\mu - \overline{V}^\mu \right)
\right]
\, \mbox{tr}
\left[
  \overline{\cal A}_\nu
  \left( \overline{\cal V}^\nu - \overline{V}^\nu \right)
\right]
\nonumber\\
&& \quad
{} + \frac{a(9-6a+a^2)}{8}
\, \mbox{tr}
\left[
  \overline{\cal A}_\mu
  \overline{\cal A}_\nu
\right]
\, \mbox{tr}
\left[
  \left( \overline{\cal V}^\mu - \overline{V}^\mu \right)
  \left( \overline{\cal V}^\nu - \overline{V}^\nu \right)
\right]
\nonumber\\
&& \quad
{} + \frac{a(9-6a+a^2)}{8}
\, \mbox{tr}
\left[
  \overline{\cal A}_\mu
  \left( \overline{\cal V}_\nu - \overline{V}_\nu \right)
\right]
\, \mbox{tr}
\left[
  \left( \overline{\cal V}^\mu - \overline{V}^\mu \right)
  \overline{\cal A}^\nu
\right]
\nonumber\\
&& \quad
{} + \frac{3a(a-1)}{16} \, N_f \, \frac{F_\chi^2}{F_\pi^2}
\, \mbox{tr}
\biggl[
  \left( \overline{\chi} - \overline{\chi}^\dag \right)
  \left[ 
    \overline{\cal A}_\mu \,,\,
    \overline{\cal V}^\mu - \overline{V}^\mu
  \right]
\biggr]
\nonumber\\
&& \quad
{} - \frac{a}{32} \, N_f \, \left(\frac{F_\chi^2}{F_\pi^2}\right)^2
\left(
  \mbox{tr}
  \biggl[
    \left( \overline{\chi} - \overline{\chi}^\dag \right)^2
  \biggr]
  - \frac{1}{N_f}
  \biggl(
    \mbox{tr}
    \left[ \overline{\chi} - \overline{\chi}^\dag \right]
  \biggr)^2
\right)
\ ,
\\
&&
\sum_{a,b} \sum_{\alpha} 
\Sigma_{ab}^{(\pi V_\alpha)} \Sigma_{(V_\alpha ba}^{\pi)}
=
- 2 a M_\rho^2 \, N_f 
\, \mbox{tr} \left[ 
  \overline{\cal A}_\mu \overline{\cal A}^\mu
\right]
\ .
\end{eqnarray}
It should be noticed that 
in the above expression for
$\sum_{a,b} \Sigma_{ab}^{(\pi\sigma)} \Sigma_{ba}^{(\sigma\pi)}$ 
we used the form in Eq.~(\ref{Yps:eom}), which was rewritten 
by using the equation of motion (\ref{EOM Npi B}).

Next 
$\sum_a \left( \Sigma \cdot \Sigma \right)_{aa}^{(\sigma\sigma)}$ is given by
\begin{eqnarray}
&&
\sum_a \left( \Sigma \cdot \Sigma \right)_{aa}^{(\sigma\sigma)}
=
\sum_{a,b} \left[ 
  \Sigma_{ab}^{(\sigma\pi)} \Sigma_{ba}^{(\pi\sigma)}
  + \Sigma_{ab}^{(\sigma\sigma)} \Sigma_{ba}^{(\sigma\sigma)}
  + \sum_{\alpha} 
    \Sigma_{ab}^{(\sigma V_\alpha)} \Sigma_{(V_\alpha ba}^{\sigma)}
\right]
\ ,
\end{eqnarray}
where
\begin{eqnarray}
&&
\sum_{a,b} \Sigma_{ab}^{(\sigma\pi)} \Sigma_{ba}^{(\pi\sigma)}
=
\sum_{a,b} \Sigma_{ab}^{(\pi\sigma)} \Sigma_{ba}^{(\sigma\pi)}
\ ,
\\
&&
\sum_{a,b} \Sigma_{ab}^{(\sigma\sigma)} \Sigma_{ba}^{(\sigma\sigma)}
\nonumber\\
&& \ 
=
\frac{a^2}{8} \, N_f
\, \mbox{tr} \left[ 
  \left(
    \overline{\cal A}_\mu \overline{\cal A}^\mu
  \right)^2
\right]
{} + \frac{1}{8} \, N_f
\, \mbox{tr} \left[
  \left(
    \left( \overline{\cal V}_\mu - \overline{V}_\mu \right)
    \left( \overline{\cal V}^\mu - \overline{V}^\mu \right)
  \right)^2
\right]
\nonumber\\
&& \quad
{} + \frac{a}{4} \, N_f
\, \mbox{tr} \left[
  \overline{\cal A}_\mu \overline{\cal A}^\mu
  \left( \overline{\cal V}_\nu - \overline{V}_\nu \right)
  \left( \overline{\cal V}^\nu - \overline{V}^\nu \right)
\right]
\nonumber\\
&& \quad
{}+ \frac{a^2}{8} 
\biggl(
  \mbox{tr} \left[ 
    \overline{\cal A}_\mu \overline{\cal A}^\mu
  \right]
\biggr)^2
{} + \frac{a^2}{4} 
\, \mbox{tr} \left[ 
  \overline{\cal A}_\mu \overline{\cal A}_\nu
\right]
\, \mbox{tr} \left[ 
  \overline{\cal A}^\mu \overline{\cal A}^\nu
\right]
\nonumber\\
&& \quad
{} + \frac{1}{8}
\biggl(
  \mbox{tr} \left[
    \left( \overline{\cal V}_\mu - \overline{V}_\mu \right)
    \left( \overline{\cal V}^\mu - \overline{V}^\mu \right)
  \right]
\biggr)^2
\nonumber\\
&& \quad
{} + \frac{1}{4}
\, \mbox{tr} \left[
  \left( \overline{\cal V}_\mu - \overline{V}_\mu \right)
  \left( \overline{\cal V}_\nu - \overline{V}_\nu \right)
\right]
\, \mbox{tr} \left[
  \left( \overline{\cal V}^\mu - \overline{V}^\mu \right)
  \left( \overline{\cal V}^\nu - \overline{V}^\nu \right)
\right]
\nonumber\\
&& \quad
{} + \frac{a}{4}
\, \mbox{tr} \left[
  \overline{\cal A}_\mu \overline{\cal A}^\mu
\right]
\, \mbox{tr} \left[
  \left( \overline{\cal V}_\nu - \overline{V}_\nu \right)
  \left( \overline{\cal V}^\nu - \overline{V}^\nu \right)
\right]
\nonumber\\
&& \quad
{} + \frac{a}{2}
\, \mbox{tr} \left[
  \overline{\cal A}_\mu
  \left( \overline{\cal V}_\nu - \overline{V}_\nu \right)
\right]
\, \mbox{tr} \left[
  \overline{\cal A}^\mu
  \left( \overline{\cal V}^\nu - \overline{V}^\nu \right)
\right]
\ ,
\\
&&
\sum_{a,b} \sum_{\alpha} 
   \Sigma_{ab}^{(\sigma V_\alpha)} \Sigma_{(V_\alpha ba}^{\sigma)}
=
- 2 M_\rho^2 \, N_f \,\mbox{tr}
\left[ 
  \left( \overline{\cal V}_\mu - \overline{V}_\mu \right)
  \left( \overline{\cal V}^\mu - \overline{V}^\mu \right)
\right]
\ .
\end{eqnarray}

Next $\sum_a \sum_\alpha 
\left( \Sigma \cdot \Sigma \right)_{(V_\alpha aa}^{V_\alpha)}$ is given by
\begin{eqnarray}
&&
\sum_a \sum_\alpha 
\left( \Sigma \cdot \Sigma \right)_{(V_\alpha aa}^{V_\alpha)}
=
\sum_{a,b} \sum_\alpha \left[ 
  \Sigma_{(V_\alpha ab}^{\pi)} \Sigma_{ba}^{(\pi V_\alpha)}
  + \Sigma_{(V_\alpha ab}^{\sigma)} \Sigma_{ba}^{(\sigma V_\alpha)}
  + \sum_{\beta} 
    \Sigma_{(V_\alpha V_\beta)ab} \Sigma_{ba}^{(V_\beta V_\alpha)}
\right]
\ ,
\end{eqnarray}
where
\begin{eqnarray}
&&
\sum_{a,b} \sum_\alpha
\Sigma_{(V_\alpha ab}^{\pi)} \Sigma_{ba}^{(\pi V_\alpha)}
=
\sum_{a,b} \sum_{\alpha} 
\Sigma_{ab}^{(\pi V_\alpha)} \Sigma_{(V_\alpha ba}^{\pi)}
=
- 2 a M_\rho^2 \, N_f 
\, \mbox{tr} \left[ 
  \overline{\cal A}_\mu \overline{\cal A}^\mu
\right]
\ ,
\\
&&
\sum_{a,b} \sum_\alpha
\Sigma_{(V_\alpha ab}^{\sigma)} \Sigma_{ba}^{(\sigma V_\alpha)}
=
\sum_{a,b} \sum_{\alpha} 
   \Sigma_{ab}^{(\sigma V_\alpha)} \Sigma_{(V_\alpha ba}^{\sigma)}
=
- 2 M_\rho^2 \, N_f \,\mbox{tr}
\left[ 
  \left( \overline{\cal V}_\mu - \overline{V}_\mu \right)
  \left( \overline{\cal V}^\mu - \overline{V}^\mu \right)
\right]
\ ,
\\
&&
\sum_{a,b} \sum_{\alpha,\beta}
\Sigma_{(V_\alpha V_\beta)ab} \Sigma_{ba}^{(V_\beta V_\alpha)}
=
8 \, N_f \, \mbox{tr}
\left[ \overline{V}_{\mu\nu} \overline{V}^{\mu\nu} \right]
\ .
\end{eqnarray}

Summing over the above
$\sum_a \left( \Sigma \cdot \Sigma \right)_{aa}^{(\pi\pi)}$,
$\sum_a \left( \Sigma \cdot \Sigma \right)_{aa}^{(\sigma\sigma)}$
and
$\sum_{a,b} \sum_\alpha
\Sigma_{(V_\alpha ab}^{\pi)} \Sigma_{ba}^{(\pi V_\alpha)}$,
we obtain
\begin{eqnarray}
&&
\frac{1}{2} \, \mbox{tr} 
\biggl( \widetilde{\Sigma} \cdot \widetilde{\Sigma} \biggr)
\nonumber\\
&& \ 
=
\frac{1}{2}
\sum_{a,b} \Sigma_{ab}^{(\pi\pi)} \Sigma_{ba}^{(\pi\pi)}
+ \frac{1}{2}
\sum_{a,b} \Sigma_{ab}^{(\sigma\sigma)} \Sigma_{ba}^{(\sigma\sigma)}
+ \frac{1}{2}
\sum_{a,b} \sum_{\alpha,\beta}
\Sigma_{(V_\alpha V_\beta)ab} \Sigma_{ba}^{(V_\beta V_\alpha)}
\nonumber\\
&& \quad
{}+
\sum_{a,b} \Sigma_{ab}^{(\pi\sigma)} \Sigma_{ba}^{(\sigma\pi)}
{}+
\sum_{a,b} \sum_{\alpha} 
\Sigma_{ab}^{(\pi V_\alpha)} \Sigma_{(V_\alpha ba}^{\pi)}
+
\sum_{a,b} \sum_{\alpha} 
   \Sigma_{ab}^{(\sigma V_\alpha)} \Sigma_{(V_\alpha ba}^{\sigma)}
\nonumber\\
&& \ 
=
4 \, N_f \, \mbox{tr}
\left[ \overline{V}_{\mu\nu} \overline{V}^{\mu\nu} \right]
{}- 2 a M_\rho^2 \, N_f 
\, \mbox{tr} \left[ 
  \overline{\cal A}_\mu \overline{\cal A}^\mu
\right]
{}- 2 M_\rho^2 \, N_f \,\mbox{tr}
\left[ 
  \left( \overline{\cal V}_\mu - \overline{V}_\mu \right)
  \left( \overline{\cal V}^\mu - \overline{V}^\mu \right)
\right]
\nonumber\\
&& \quad
{} + \frac{5a^2-12a+8}{8} \, N_f
\, \mbox{tr} \left[ 
  \left(
    \overline{\cal A}_\mu \overline{\cal A}^\mu
  \right)^2
\right]
{} + \frac{a^4+1}{16} \, N_f
\, \mbox{tr} \left[
  \left(
    \left( \overline{\cal V}_\mu - \overline{V}_\mu \right)
    \left( \overline{\cal V}^\mu - \overline{V}^\mu \right)
  \right)^2
\right]
\nonumber\\
&& \quad
{} + \frac{a\left(1+4a-3a^2\right)}{8} \, N_f
\, \mbox{tr} \left[
  \overline{\cal A}_\mu \overline{\cal A}^\mu
  \left( \overline{\cal V}_\nu - \overline{V}_\nu \right)
  \left( \overline{\cal V}^\nu - \overline{V}^\nu \right)
\right]
\nonumber\\
&& \quad
{} + \frac{a\left(5a^2-12a+9\right)}{8} \, N_f
\, \mbox{tr}
\left[
  \overline{\cal A}_\mu
  \overline{\cal A}_\nu
  \left( \overline{\cal V}^\nu - \overline{V}^\nu \right)
  \left( \overline{\cal V}^\mu - \overline{V}^\mu \right)
\right]
\nonumber\\
&& \quad
{} + \frac{a^2\left(3-2a\right)}{8} \, N_f
\Biggl(
  \mbox{tr}
  \left[
    \overline{\cal A}_\mu
    \left( \overline{\cal V}_\nu - \overline{V}_\nu \right)
    \overline{\cal A}^\nu
    \left( \overline{\cal V}^\mu - \overline{V}^\mu \right)
  \right]
\nonumber\\
&& \qquad\qquad\qquad\qquad\qquad
  {}+
  \, \mbox{tr}
  \left[
    \overline{\cal A}_\mu
    \left( \overline{\cal V}^\mu - \overline{V}^\mu \right)
    \overline{\cal A}_\nu
    \left( \overline{\cal V}^\nu - \overline{V}^\nu \right)
  \right]
\Biggr)
\nonumber\\
&& \quad
{}+ \frac{5a^2-12a+8}{8} 
\biggl(
  \mbox{tr} \left[ 
    \overline{\cal A}_\mu \overline{\cal A}^\mu
  \right]
\biggr)^2
{}+ \frac{5a^2-12a+8}{4}
\, \mbox{tr} \left[ 
  \overline{\cal A}_\mu \overline{\cal A}_\nu
\right]
\, \mbox{tr} \left[ 
  \overline{\cal A}^\mu \overline{\cal A}^\nu
\right]
\nonumber\\
&& \quad
{} + \frac{a^4+1}{16}
\biggl(
  \mbox{tr} \left[
    \left( \overline{\cal V}_\mu - \overline{V}_\mu \right)
    \left( \overline{\cal V}^\mu - \overline{V}^\mu \right)
  \right]
\biggr)^2
\nonumber\\
&& \quad
{} + \frac{a^4+1}{8}
\, \mbox{tr} \left[
  \left( \overline{\cal V}_\mu - \overline{V}_\mu \right)
  \left( \overline{\cal V}_\nu - \overline{V}_\nu \right)
\right]
\, \mbox{tr} \left[
  \left( \overline{\cal V}^\mu - \overline{V}^\mu \right)
  \left( \overline{\cal V}^\nu - \overline{V}^\nu \right)
\right]
\nonumber\\
&& \quad
{} + \frac{a\left(1+4a-3a^2\right)}{8}
\, \mbox{tr} \left[
  \overline{\cal A}_\mu \overline{\cal A}^\mu
\right]
\, \mbox{tr} \left[
  \left( \overline{\cal V}_\nu - \overline{V}_\nu \right)
  \left( \overline{\cal V}^\nu - \overline{V}^\nu \right)
\right]
\nonumber\\
&& \quad
{} + \frac{a(9-6a+a^2)}{8}
\, \mbox{tr}
\left[
  \overline{\cal A}_\mu
  \overline{\cal A}_\nu
\right]
\, \mbox{tr}
\left[
  \left( \overline{\cal V}^\mu - \overline{V}^\mu \right)
  \left( \overline{\cal V}^\nu - \overline{V}^\nu \right)
\right]
\nonumber\\
&& \quad
{} + \frac{a(9-6a+a^2)}{8}
\biggl(
  \mbox{tr}
  \left[
    \overline{\cal A}_\mu
    \left( \overline{\cal V}^\mu - \overline{V}^\mu \right)
  \right]
\biggr)^2
\nonumber\\
&& \quad
{} + \frac{a\left(1+4a-3a^2\right)}{4}
\, \mbox{tr} \left[
  \overline{\cal A}_\mu
  \left( \overline{\cal V}_\nu - \overline{V}_\nu \right)
\right]
\, \mbox{tr} \left[
  \overline{\cal A}^\mu
  \left( \overline{\cal V}^\nu - \overline{V}^\nu \right)
\right]
\nonumber\\
&& \quad
{} + \frac{a(9-6a+a^2)}{8}
\, \mbox{tr}
\left[
  \overline{\cal A}_\mu
  \left( \overline{\cal V}_\nu - \overline{V}_\nu \right)
\right]
\, \mbox{tr}
\left[
  \left( \overline{\cal V}^\mu - \overline{V}^\mu \right)
  \overline{\cal A}^\nu
\right]
\nonumber\\
&& \quad
{} + \frac{4-3a}{8} \, N_f \, \frac{F_\chi^2}{F_\pi^2}
\, \mbox{tr} \left[
  \overline{\cal A}_\mu \overline{\cal A}^\mu
  \left( \overline{\chi} + \overline{\chi}^\dag \right)
\right]
{} + \frac{4-3a}{8} \, \frac{F_\chi^2}{F_\pi^2}
\, \mbox{tr} \left[
  \overline{\cal A}_\mu \overline{\cal A}^\mu
\right]
\, \mbox{tr} \left[ \overline{\chi} + \overline{\chi}^\dag \right]
\nonumber\\
&& \quad
{} + \frac{a^2}{8} \, N_f \, \frac{F_\chi^2}{F_\pi^2}
\, \mbox{tr} \left[
  \left( \overline{\cal V}_\mu - \overline{V}_\mu \right)
  \left( \overline{\cal V}^\mu - \overline{V}^\mu \right)
  \left( \overline{\chi} + \overline{\chi}^\dag \right)
\right]
\nonumber\\
&& \quad
{} + \frac{a^2}{8} \, \frac{F_\chi^2}{F_\pi^2}
\, \mbox{tr} \left[
  \left( \overline{\cal V}_\mu - \overline{V}_\mu \right)
  \left( \overline{\cal V}^\mu - \overline{V}^\mu \right)
\right]
\, \mbox{tr} \left[ \overline{\chi} + \overline{\chi}^\dag \right]
\nonumber\\
&& \quad
{} + \frac{3a(a-1)}{16} \, N_f \, \frac{F_\chi^2}{F_\pi^2}
\, \mbox{tr}
\biggl[
  \left( \overline{\chi} - \overline{\chi}^\dag \right)
  \left[ 
    \overline{\cal A}_\mu \,,\,
    \overline{\cal V}^\mu - \overline{V}^\mu
  \right]
\biggr]
\nonumber\\
&& \quad
{} + \frac{ N_f^2-4}{16N_f} \left(\frac{F_\chi^2}{F_\pi^2}\right)^2
\, \mbox{tr} \left[
  \left( \overline{\chi} + \overline{\chi}^\dag \right)^2
\right]
{} + \frac{ N_f^2+2}{16N_f^2} \left(\frac{F_\chi^2}{F_\pi^2}\right)^2
\biggl(
  \mbox{tr} \left[ \overline{\chi} + \overline{\chi}^\dag \right]
\biggr)^2
\nonumber\\
&& \quad
{} - \frac{a}{32} \, N_f \, \left(\frac{F_\chi^2}{F_\pi^2}\right)^2
\left(
  \mbox{tr}
  \biggl[
    \left( \overline{\chi} - \overline{\chi}^\dag \right)^2
  \biggr]
  - \frac{1}{N_f}
  \biggl(
    \mbox{tr}
    \left[ \overline{\chi} - \overline{\chi}^\dag \right]
  \biggr)^2
\right)
\ ,
\end{eqnarray}

Next let us calculate
$\Gamma_{\mu\nu}$ defined in Eq.~(\ref{def gam}):
\begin{eqnarray}
{\Gamma_{\mu\nu}}_{ab}^{(\pi\pi)} 
&=&
i\, \mbox{tr}
\Biggl[
  \biggl(
    a\, \overline{V}_{\mu\nu} 
    + (2-a) \overline{\cal V}_{\mu\nu}
  \biggr)
  \left[ T_a \,,\, T_b \right]
\Biggr]
\nonumber\\
&&
{}- \mbox{tr}
\Biggl[
  \biggl(
    \frac{a(2-a)}{2} 
    \left[ 
      \overline{\cal V}_\mu - \overline{V}_\mu
      \,,\, 
      \overline{\cal V}_\nu - \overline{V}_\nu
    \right]
    + \frac{4-3a}{2} 
    \left[ 
      \overline{\cal A}_\mu \,,\, \overline{\cal A}_\nu
    \right]
  \biggr)
  \left[ T_a \,,\, T_b \right]
\Biggr]
\ ,
\\
{\Gamma_{\mu\nu}}_{ab}^{(\pi\sigma)} 
&=&
\sqrt{a} \, \mbox{tr}
\Biggl[
  \biggl(
    i\, \overline{\cal A}_{\mu\nu}
    - \frac{1}{2} 
    \left[ 
      \overline{\cal V}_\mu - \overline{V}_\mu
      \,,\, 
      \overline{\cal A}_\nu
    \right]
    - \frac{1}{2} 
    \left[ 
      \overline{\cal A}_\mu
      \,,\, 
      \overline{\cal V}_\nu - \overline{V}_\nu
    \right]
  \biggr)
  \left[ T_a \,,\, T_b \right]
\Biggr]
\nonumber\\
&& 
{} - \sqrt{a} \, \frac{a-1}{2} \, \mbox{tr}
\Biggl[
  \left[ \overline{\cal V}_\mu - \overline{V}_\mu \,,\, T_a \right]
  \left[ \overline{\cal A}_\nu \,,\, T_b \right]
\Biggr]
\nonumber\\
&& 
{} + \sqrt{a} \, \frac{a-1}{2} \, \mbox{tr}
\Biggl[
  \left[ \overline{\cal V}_\nu - \overline{V}_\nu \,,\, T_a \right]
  \left[ \overline{\cal A}_\mu \,,\, T_b \right]
\Biggr]
\ ,
\\
{\Gamma_{\mu\nu}}_{ab}^{(\sigma\pi)}
&=&
\sqrt{a} \, \mbox{tr}
\Biggl[
  \biggl(
    i\, \overline{\cal A}_{\mu\nu}
    - \frac{1}{2} 
    \left[ 
      \overline{\cal V}_\mu - \overline{V}_\mu \,,\, 
      \overline{\cal A}_\nu
    \right]
    - \frac{1}{2} 
    \left[ 
      \overline{\cal A}_\mu \,,\, 
      \overline{\cal V}_\nu - \overline{V}_\nu
    \right]
  \biggr)
  \left[ T_a \,,\, T_b \right]
\Biggr]
\nonumber\\
&& 
{} - \sqrt{a} \, \frac{a-1}{2} \, \mbox{tr}
\Biggl[
  \left[ \overline{\cal A}_\mu \,,\, T_a \right]
  \left[ \overline{\cal V}_\nu - \overline{V}_\nu \,,\, T_b \right]
\Biggr]
\nonumber\\
&& 
{} + \sqrt{a} \, \frac{a-1}{2} \, \mbox{tr}
\Biggl[
  \left[ \overline{\cal A}_\nu \,,\, T_a \right]
  \left[ \overline{\cal V}_\mu - \overline{V}_\mu \,,\, T_b \right]
\Biggr]
\ ,
\\
{\Gamma_{\mu\nu}}_{ab}^{(\sigma\sigma)}
&=&
\mbox{tr}
\Biggl[
  \biggl(
    i\, \overline{V}_{\mu\nu} 
    + i\, \overline{\cal V}_{\mu\nu}
  \biggr)
  \left[ T_a \,,\, T_b \right]
\Biggr]
\nonumber\\
&& 
{}- \frac{1}{2} 
\mbox{tr}
\Biggl[
  \biggl(
    \left[ 
      \overline{\cal V}_\mu - \overline{V}_\mu \,,\, 
      \overline{\cal V}_\nu - \overline{V}_\nu
    \right]
    + (2-a)
    \left[ 
      \overline{\cal A}_\mu \,,\, \overline{\cal A}_\nu
    \right]
  \biggr)
  \left[ T_a \,,\, T_b \right]
\Biggr]
\ ,
\\
{ {\Gamma_{\mu\nu}}_{(V_\alpha} }^{V_\beta)}_{ab}
&=&
2 i
\mbox{tr}
\Biggl[
  \overline{V}_{\mu\nu}
  \left[ T_a \,,\, T_b \right]
\Biggr]
\, g_\alpha^\beta
\ ,
\end{eqnarray}
where we used
the relations in Eqs.~(\ref{rel:perp}) and
(\ref{rel:parallel}).
Using this we obtain
\begin{eqnarray}
&&
\frac{1}{12} \sum_{a,b}
  {\Gamma_{\mu\nu}}_{ab}^{(\pi\pi)} {\Gamma^{\mu\nu}}_{ba}^{(\pi\pi)} 
=
- \frac{a^2}{24} \, N_f\,\mbox{tr}
\left[ \overline{V}_{\mu\nu} \overline{V}^{\mu\nu} \right]
- \frac{a(2-a)}{12} \, N_f\,\mbox{tr}
\left[ \overline{V}_{\mu\nu} \overline{\cal V}^{\mu\nu} \right]
\nonumber\\
&& \qquad
{}- \frac{(2-a)^2}{24} \, N_f\,\mbox{tr}
\left[ \overline{\cal V}_{\mu\nu} \overline{\cal V}^{\mu\nu} \right]
{}- i \frac{a(4-3a)}{12} \, N_f\,\mbox{tr}
\left[
  \overline{V}_{\mu\nu} 
  \overline{\cal A}^\mu
  \overline{\cal A}^\nu
\right]
\nonumber\\
&& \qquad
{}- i \, \frac{a^2(2-a)}{12} \, N_f\,\mbox{tr}
\left[
  \overline{V}_{\mu\nu} 
  \left( \overline{\cal V}^\mu - \overline{V}^\mu \right)
  \left( \overline{\cal V}^\nu - \overline{V}^\nu \right)
\right]
\nonumber\\
&& \qquad
{}- i \frac{(4-3a)(2-a)}{12} \, N_f\,\mbox{tr}
\left[ 
  \overline{\cal V}_{\mu\nu}
  \overline{\cal A}^\mu
  \overline{\cal A}^\nu
\right]
\nonumber\\
&& \qquad
{}- i \, \frac{a(2-a)^2}{12} \, N_f\,\mbox{tr}
\left[ 
  \overline{\cal V}_{\mu\nu}
  \left( \overline{\cal V}^\mu - \overline{V}^\mu \right)
  \left( \overline{\cal V}^\nu - \overline{V}^\nu \right)
\right]
\nonumber\\
&& \qquad
{} - \frac{(4-3a)^2}{48} \, N_f\,\mbox{tr}
\left[
  \left(
    \overline{\cal A}_\mu
    \overline{\cal A}^\mu
  \right)^2
\right]
{} + \frac{(4-3a)^2}{48} \, N_f\,\mbox{tr}
\left[ 
  \overline{\cal A}_\mu
  \overline{\cal A}_\nu
  \overline{\cal A}^\mu
  \overline{\cal A}^\nu
\right]
\nonumber\\
&& \qquad
{} - \frac{a^2(2-a)^2}{48} \, N_f\,\mbox{tr}
\left[
  \left( 
    \left( \overline{\cal V}_\mu - \overline{V}_\mu \right)
    \left( \overline{\cal V}^\mu - \overline{V}^\mu \right)
  \right)^2
\right]
\nonumber\\
&& \qquad
{} + \frac{a^2(2-a)^2}{48} \, N_f\,\mbox{tr}
\left[ 
  \left( \overline{\cal V}_\mu - \overline{V}_\mu \right)
  \left( \overline{\cal V}_\nu - \overline{V}_\nu \right)
  \left( \overline{\cal V}^\mu - \overline{V}^\mu \right)
  \left( \overline{\cal V}^\nu - \overline{V}^\nu \right)
\right]
\nonumber\\
&& \qquad
{} + \frac{a(2-a)(4-3a)}{24} \, N_f\,\mbox{tr}
\left[ 
  \overline{\cal A}_\mu
  \overline{\cal A}_\nu
  \left( \overline{\cal V}^\mu - \overline{V}^\mu \right)
  \left( \overline{\cal V}^\nu - \overline{V}^\nu \right)
\right]
\nonumber\\
&& \qquad
{} - \frac{a(2-a)(4-3a)}{24} \, N_f\,\mbox{tr}
\left[ 
  \overline{\cal A}_\mu
  \overline{\cal A}_\nu
  \left( \overline{\cal V}^\nu - \overline{V}^\nu \right)
  \left( \overline{\cal V}^\mu - \overline{V}^\mu \right)
\right]
\ ,
\\
&&
\frac{1}{12} \sum_{a,b}
{\Gamma_{\mu\nu}}_{ab}^{(\sigma\sigma)} 
{\Gamma^{\mu\nu}}_{ba}^{(\sigma\sigma)} 
=
- \frac{1}{24} \, N_f\,\mbox{tr}
\left[ \overline{V}_{\mu\nu} \overline{V}^{\mu\nu} \right]
- \frac{1}{12} \, N_f\,\mbox{tr}
\left[ \overline{V}_{\mu\nu} \overline{\cal V}^{\mu\nu} \right]
- \frac{1}{24} \, N_f\,\mbox{tr}
\left[ \overline{\cal V}_{\mu\nu} \overline{\cal V}^{\mu\nu} \right]
\nonumber\\
&& \qquad
{}- i \frac{2-a}{12} \, N_f\,\mbox{tr}
\left[
  \overline{V}_{\mu\nu} 
  \overline{\cal A}^\mu
  \overline{\cal A}^\nu
\right]
{}- i \, \frac{1}{12} \, N_f\,\mbox{tr}
\left[
  \overline{V}_{\mu\nu} 
  \left( \overline{\cal V}^\mu - \overline{V}^\mu \right)
  \left( \overline{\cal V}^\nu - \overline{V}^\nu \right)
\right]
\nonumber\\
&& \qquad
{}- i \frac{2-a}{12} \, N_f\,\mbox{tr}
\left[ 
  \overline{\cal V}_{\mu\nu}
  \overline{\cal A}^\mu
  \overline{\cal A}^\nu
\right]
{}- i \, \frac{1}{12} \, N_f\,\mbox{tr}
\left[ 
  \overline{\cal V}_{\mu\nu}
  \left( \overline{\cal V}^\mu - \overline{V}^\mu \right)
  \left( \overline{\cal V}^\nu - \overline{V}^\nu \right)
\right]
\nonumber\\
&& \qquad
{} - \frac{(2-a)^2}{48} \, N_f\,\mbox{tr}
\left[
  \left(
    \overline{\cal A}_\mu
    \overline{\cal A}^\mu
  \right)^2
\right]
{} + \frac{(2-a)^2}{48} \, N_f\,\mbox{tr}
\left[ 
  \overline{\cal A}_\mu
  \overline{\cal A}_\nu
  \overline{\cal A}^\mu
  \overline{\cal A}^\nu
\right]
\nonumber\\
&& \qquad
{} - \frac{1}{48} \, N_f\,\mbox{tr}
\left[
  \left( 
    \left( \overline{\cal V}_\mu - \overline{V}_\mu \right)
    \left( \overline{\cal V}^\mu - \overline{V}^\mu \right)
  \right)^2
\right]
\nonumber\\
&& \qquad
{} + \frac{1}{48} \, N_f\,\mbox{tr}
\left[ 
  \left( \overline{\cal V}_\mu - \overline{V}_\mu \right)
  \left( \overline{\cal V}_\nu - \overline{V}_\nu \right)
  \left( \overline{\cal V}^\mu - \overline{V}^\mu \right)
  \left( \overline{\cal V}^\nu - \overline{V}^\nu \right)
\right]
\nonumber\\
&& \qquad
{} + \frac{2-a}{24} \, N_f\,\mbox{tr}
\left[ 
  \overline{\cal A}_\mu
  \overline{\cal A}_\nu
  \left( \overline{\cal V}^\mu - \overline{V}^\mu \right)
  \left( \overline{\cal V}^\nu - \overline{V}^\nu \right)
\right]
\nonumber\\
&& \qquad
{} - \frac{2-a}{24} \, N_f\,\mbox{tr}
\left[ 
  \overline{\cal A}_\mu
  \overline{\cal A}_\nu
  \left( \overline{\cal V}^\nu - \overline{V}^\nu \right)
  \left( \overline{\cal V}^\mu - \overline{V}^\mu \right)
\right]
\ ,
\\
&&
\frac{1}{12} \sum_{a,b} \sum_{\alpha,\beta}
{ {\Gamma_{\mu\nu}}_{(V_\alpha} }^{V_\beta)}_{ab}
{ {\Gamma^{\mu\nu}}_{(V_\beta} }^{V_\alpha)}_{ba}
=
- \frac{2}{3} \, N_f \, \mbox{tr}
\left[ \overline{V}_{\mu\nu} \overline{V}^{\mu\nu} \right]
\ ,
\\
&&
\frac{1}{6} \sum_{a,b}
{\Gamma_{\mu\nu}}_{ab}^{(\pi\sigma)} 
{\Gamma^{\mu\nu}}_{ba}^{(\sigma\pi)} 
=
- \frac{a}{12} \,N_f\,\mbox{tr}
\left[ \overline{\cal A}_{\mu\nu} \overline{\cal A}^{\mu\nu} \right]
- i \, \frac{a(a+1)}{12} \, N_f \, \mbox{tr}
\biggl[
  \overline{\cal A}_{\mu\nu}
  \left[ 
    \overline{\cal A}^\mu \,,\, 
    \overline{\cal V}^\nu - \overline{V}^\nu
  \right]
\biggr]
\nonumber\\
&& \qquad
{} - \frac{a(a^2+1)}{24} \, N_f \, \mbox{tr}
\left[
  \overline{\cal A}_\mu \overline{\cal A}^\mu 
  \left( \overline{\cal V}_\nu - \overline{V}_\nu \right) 
  \left( \overline{\cal V}^\nu - \overline{V}^\nu \right)
\right]
\nonumber\\
&& \qquad
{} + \frac{a(a^2+1)}{24} \, N_f \, \mbox{tr} \left[
  \overline{\cal A}_\mu \overline{\cal A}_\nu
  \left( \overline{\cal V}^\mu - \overline{V}^\mu \right) 
  \left( \overline{\cal V}^\nu - \overline{V}^\nu \right)
\right]
\nonumber\\
&& \qquad
{} - \frac{a^2}{24} \, N_f 
\Biggl(
\mbox{tr} \left[ 
  \overline{\cal A}_\mu 
  \left( \overline{\cal V}^\mu - \overline{V}^\mu \right)
  \overline{\cal A}_\nu 
  \left( \overline{\cal V}^\nu - \overline{V}^\nu \right)
\right]
+ \mbox{tr} \left[
  \left( \overline{\cal V}_\mu - \overline{V}_\mu \right) 
  \overline{\cal A}^\mu
  \left( \overline{\cal V}_\nu - \overline{V}_\nu \right)
  \overline{\cal A}^\nu
\right] \Biggr)
\nonumber\\
&& \qquad
{} + \frac{a^2}{12} \, N_f 
\, \mbox{tr} \left[
  \overline{\cal A}_\mu
  \left( \overline{\cal V}_\nu - \overline{V}_\nu \right)
  \overline{\cal A}^\mu
  \left( \overline{\cal V}^\nu - \overline{V}^\nu \right)
\right]
\nonumber\\
&& \qquad
{} - \frac{a(a-1)^2}{24} \, N_f 
\, \mbox{tr} \left[
  \overline{\cal A}_\mu \overline{\cal A}^\mu
\right]
\mbox{tr} \left[
  \left( \overline{\cal V}_\nu - \overline{V}_\nu \right)
  \left( \overline{\cal V}^\nu - \overline{V}^\nu \right)
\right]
\nonumber\\
&& \qquad
{} + \frac{a(a-1)^2}{24} \, N_f 
\, \mbox{tr} \left[
  \overline{\cal A}_\mu \overline{\cal A}_\nu
\right]
\mbox{tr} \left[
  \left( \overline{\cal V}^\mu - \overline{V}^\mu \right) \left( \overline{\cal V}^\nu - \overline{V}^\nu \right)
\right]
\nonumber\\
&& \qquad
{} + \frac{a(a-1)^2}{24} \, N_f 
\left( \mbox{tr} \left[
  \overline{\cal A}_\mu \left( \overline{\cal V}^\mu - \overline{V}^\mu \right)
\right] \right)^2
\nonumber\\
&& \qquad
{} - \frac{a(a-1)^2}{12} \, N_f 
\, \mbox{tr} \left[
  \overline{\cal A}_\mu \left( \overline{\cal V}_\nu - \overline{V}_\nu \right)
\right]
\mbox{tr} \left[
  \overline{\cal A}^\mu \left( \overline{\cal V}^\nu - \overline{V}^\nu \right)
\right]
\nonumber\\
&& \qquad
{} + \frac{a(a-1)^2}{24} \, N_f 
\, \mbox{tr} \left[
  \overline{\cal A}_\mu \left( \overline{\cal V}_\nu - \overline{V}_\nu \right)
\right]
\mbox{tr} \left[
  \left( \overline{\cal V}^\mu - \overline{V}^\mu \right) \overline{\alpha}^\nu_\perp 
\right]
\ .
\end{eqnarray}
Thus, we obtain
\begin{eqnarray}
&&
\frac{1}{12} \, \mbox{tr} 
\biggl( \Gamma_{\mu\nu} \cdot \Gamma^{\mu\nu} \biggr)
\nonumber\\
&& \ 
=
\frac{1}{12} \sum_{a,b}
\left[
  {\Gamma_{\mu\nu}}_{ab}^{(\pi\pi)} {\Gamma^{\mu\nu}}_{ba}^{(\pi\pi)} 
  + {\Gamma_{\mu\nu}}_{ab}^{(\sigma\sigma)}
    {\Gamma^{\mu\nu}}_{ba}^{(\sigma\sigma)}
  + 2 \, {\Gamma_{\mu\nu}}_{ab}^{(\pi\sigma)}
    {\Gamma^{\mu\nu}}_{ba}^{(\sigma\pi)}
  + { {\Gamma_{\mu\nu}}_{(V_\alpha} }^{V_\beta)}_{ab}
    { {\Gamma^{\mu\nu}}_{(V_\beta} }^{V_\alpha)}_{ba}
\right]
\nonumber\\
&& \quad
=
- \frac{17+a^2}{24} \, N_f\,\mbox{tr}
\left[ \overline{V}_{\mu\nu} \overline{V}^{\mu\nu} \right]
{}- \frac{1+2a-a^2}{12} \, N_f\,\mbox{tr}
\left[ \overline{V}_{\mu\nu} \overline{\cal V}^{\mu\nu} \right]
\nonumber\\
&& \qquad
{}- \frac{5-4a+a^2}{24} \, N_f\,\mbox{tr}
\left[ \overline{\cal V}_{\mu\nu} \overline{\cal V}^{\mu\nu} \right]
- \frac{a}{12} \,N_f\,\mbox{tr}
\left[ \overline{\cal A}_{\mu\nu} \overline{\cal A}^{\mu\nu} \right]
\nonumber\\
&& \qquad
{}- i \, \frac{2+3a-3a^2}{12} \, N_f\,\mbox{tr}
\left[
  \overline{V}_{\mu\nu} 
  \overline{\cal A}^\mu
  \overline{\cal A}^\nu
\right]
\nonumber\\
&& \qquad
{}- i \, \frac{1+2a^2-a^3}{12} \, N_f\,\mbox{tr}
\left[
  \overline{V}_{\mu\nu} 
  \left( \overline{\cal V}^\mu - \overline{V}^\mu \right)
  \left( \overline{\cal V}^\nu - \overline{V}^\nu \right)
\right]
\nonumber\\
&& \qquad
{}- i \frac{(2-a)(5-3a)}{12} \, N_f\,\mbox{tr}
\left[ 
  \overline{\cal V}_{\mu\nu}
  \overline{\cal A}^\mu
  \overline{\cal A}^\nu
\right]
\nonumber\\
&& \qquad
{}- i \, \frac{1+4a-4a^2+a^3}{12} \, N_f\,\mbox{tr}
\left[ 
  \overline{\cal V}_{\mu\nu}
  \left( \overline{\cal V}^\mu - \overline{V}^\mu \right)
  \left( \overline{\cal V}^\nu - \overline{V}^\nu \right)
\right]
\nonumber\\
&& \qquad
{}- i \, \frac{a(a+1)}{12} \, N_f \, \mbox{tr}
\biggl[
  \overline{\cal A}_{\mu\nu}
  \left[ 
    \overline{\cal A}^\mu \,,\, 
    \overline{\cal V}^\nu - \overline{V}^\nu
  \right]
\biggr]
\nonumber\\
&& \qquad
{} - \frac{10-14a+5a^2}{24} \, N_f\,\mbox{tr}
\left[
  \left(
    \overline{\cal A}_\mu
    \overline{\cal A}^\mu
  \right)^2
\right]
{} + \frac{10-14+5a^2}{24} \, N_f\,\mbox{tr}
\left[ 
  \overline{\cal A}_\mu
  \overline{\cal A}_\nu
  \overline{\cal A}^\mu
  \overline{\cal A}^\nu
\right]
\nonumber\\
&& \qquad
{} - \frac{1+4a^2-4a^3+a^4}{48} \, N_f\,\mbox{tr}
\left[
  \left( 
    \left( \overline{\cal V}_\mu - \overline{V}_\mu \right)
    \left( \overline{\cal V}^\mu - \overline{V}^\mu \right)
  \right)^2
\right]
\nonumber\\
&& \qquad
{} + \frac{1+4a^2-4a^3+a^4}{48} \, N_f\,\mbox{tr}
\left[ 
  \left( \overline{\cal V}_\mu - \overline{V}_\mu \right)
  \left( \overline{\cal V}_\nu - \overline{V}_\nu \right)
  \left( \overline{\cal V}^\mu - \overline{V}^\mu \right)
  \left( \overline{\cal V}^\nu - \overline{V}^\nu \right)
\right]
\nonumber\\
&& \qquad
{} - \frac{a(a^2+1)}{24} \, N_f \, \mbox{tr}
\left[
  \overline{\cal A}_\mu \overline{\cal A}^\mu 
  \left( \overline{\cal V}_\nu - \overline{V}_\nu \right) 
  \left( \overline{\cal V}^\nu - \overline{V}^\nu \right)
\right]
\nonumber\\
&& \qquad
{} + \frac{1+4a-5a^2+2a^3}{12} \, N_f \, \mbox{tr} \left[
  \overline{\cal A}_\mu \overline{\cal A}_\nu
  \left( \overline{\cal V}^\mu - \overline{V}^\mu \right) \left( \overline{\cal V}^\nu - \overline{V}^\nu \right)
\right]
\nonumber\\
&& \qquad
{} - \frac{(2-a)(1+4a-3a^2)}{24} \, N_f\,\mbox{tr}
\left[ 
  \overline{\cal A}_\mu
  \overline{\cal A}_\nu
  \left( \overline{\cal V}^\nu - \overline{V}^\nu \right)
  \left( \overline{\cal V}^\mu - \overline{V}^\mu \right)
\right]
\nonumber\\
&& \qquad
{} - \frac{a^2}{24} \, N_f 
\Biggl(
\mbox{tr} \left[ 
  \overline{\cal A}_\mu \left( \overline{\cal V}^\mu - \overline{V}^\mu \right)
  \overline{\cal A}_\nu \left( \overline{\cal V}^\nu - \overline{V}^\nu \right)
\right]
+ \mbox{tr} \left[
  \left( \overline{\cal V}_\mu - \overline{V}_\mu \right) \overline{\cal A}^\mu
  \left( \overline{\cal V}_\nu - \overline{V}_\nu \right) \overline{\cal A}^\nu
\right] \Biggr)
\nonumber\\
&& \qquad
{} + \frac{a^2}{12} \, N_f 
\, \mbox{tr} \left[
  \overline{\cal A}_\mu \left( \overline{\cal V}_\nu - \overline{V}_\nu \right)
  \overline{\cal A}^\mu \left( \overline{\cal V}^\nu - \overline{V}^\nu \right)
\right]
\nonumber\\
&& \qquad
{} - \frac{a(a-1)^2}{24} \, N_f 
\, \mbox{tr} \left[
  \overline{\cal A}_\mu \overline{\cal A}^\mu
\right]
\mbox{tr} \left[
  \left( \overline{\cal V}_\nu - \overline{V}_\nu \right)
  \left( \overline{\cal V}^\nu - \overline{V}^\nu \right)
\right]
\nonumber\\
&& \qquad
{} + \frac{a(a-1)^2}{24} \, N_f 
\, \mbox{tr} \left[
  \overline{\cal A}_\mu \overline{\cal A}_\nu
\right]
\mbox{tr} \left[
  \left( \overline{\cal V}^\mu - \overline{V}^\mu \right)
  \left( \overline{\cal V}^\nu - \overline{V}^\nu \right)
\right]
\nonumber\\
&& \qquad
{} + \frac{a(a-1)^2}{24} \, N_f 
\left( \mbox{tr} \left[
  \overline{\cal A}_\mu \left( \overline{\cal V}^\mu - \overline{V}^\mu \right)
\right] \right)^2
\nonumber\\
&& \qquad
{} - \frac{a(a-1)^2}{12} \, N_f 
\, \mbox{tr} \left[
  \overline{\cal A}_\mu
  \left( \overline{\cal V}_\nu - \overline{V}_\nu \right)
\right]
\mbox{tr} \left[
  \overline{\cal A}^\mu
  \left( \overline{\cal V}^\nu - \overline{V}^\nu \right)
\right]
\nonumber\\
&& \qquad
{} + \frac{a(a-1)^2}{24} \, N_f 
\, \mbox{tr} \left[
  \overline{\cal A}_\mu
  \left( \overline{\cal V}_\nu - \overline{V}_\nu \right)
\right]
\mbox{tr} \left[
  \left( \overline{\cal V}^\mu - \overline{V}^\mu \right)
  \overline{\cal A}^\nu 
\right]
\ .
\end{eqnarray}
Finally,
\begin{eqnarray}
&&
\frac{1}{2} \, \mbox{tr} 
\biggl( \overline{\Sigma} \cdot \overline{\Sigma} \biggr)
+
\frac{1}{12} \, \mbox{tr} 
\biggl( \Gamma_{\mu\nu} \cdot \Gamma^{\mu\nu} \biggr)
\nonumber\\
&& \ 
=
- 2 a M_\rho^2 \, N_f 
\, \mbox{tr} \left[ 
  \overline{\cal A}_\mu \overline{\cal A}^\mu
\right]
{}- 2 M_\rho^2 \, N_f \,\mbox{tr}
\left[ 
  \left( \overline{\cal V}_\mu - \overline{V}_\mu \right) \left( \overline{\cal V}^\mu - \overline{V}^\mu \right)
\right]
\nonumber\\
&& \qquad
{}+ \frac{79-a^2}{24} \, N_f\,\mbox{tr}
\left[ \overline{V}_{\mu\nu} \overline{V}^{\mu\nu} \right]
{}- \frac{5-4a+a^2}{24} \, N_f\,\mbox{tr}
\left[ \overline{\cal V}_{\mu\nu} \overline{\cal V}^{\mu\nu} \right]
\nonumber\\
&& \qquad
{}- \frac{a}{12} \,N_f\,\mbox{tr}
\left[ \overline{\cal A}_{\mu\nu} \overline{\cal A}^{\mu\nu} \right]
{}- \frac{1+2a-a^2}{12} \, N_f\,\mbox{tr}
\left[ \overline{V}_{\mu\nu} \overline{\cal V}^{\mu\nu} \right]
\nonumber\\
&& \qquad
{}- i \, \frac{2+3a-3a^2}{12} \, N_f\,\mbox{tr}
\left[
  \overline{V}_{\mu\nu} 
  \overline{\cal A}^\mu
  \overline{\cal A}^\nu
\right]
\nonumber\\
&& \qquad
{}- i \, \frac{1+2a^2-a^3}{12} \, N_f\,\mbox{tr}
\left[
  \overline{V}_{\mu\nu} 
  \left( \overline{\cal V}^\mu - \overline{V}^\mu \right)
  \left( \overline{\cal V}^\nu - \overline{V}^\nu \right)
\right]
\nonumber\\
&& \qquad
{}- i \frac{(2-a)(5-3a)}{12} \, N_f\,\mbox{tr}
\left[ 
  \overline{\cal V}_{\mu\nu}
  \overline{\cal A}^\mu
  \overline{\cal A}^\nu
\right]
\nonumber\\
&& \qquad
{}- i \, \frac{1+4a-4a^2+a^3}{12} \, N_f\,\mbox{tr}
\left[ 
  \overline{\cal V}_{\mu\nu}
  \left( \overline{\cal V}^\mu - \overline{V}^\mu \right)
  \left( \overline{\cal V}^\nu - \overline{V}^\nu \right)
\right]
\nonumber\\
&& \qquad
{}- i \, \frac{a(a+1)}{12} \, N_f \, \mbox{tr}
\biggl[
  \overline{\cal A}_{\mu\nu}
  \left[ 
    \overline{\cal A}^\mu \,,\, 
    \overline{\cal V}^\nu - \overline{V}^\nu
  \right]
\biggr]
\nonumber\\
&& \qquad
{} + \frac{7-11a+5a^2}{12} \, N_f\,\mbox{tr}
\left[
  \left(
    \overline{\cal A}_\mu
    \overline{\cal A}^\mu
  \right)^2
\right]
{} + \frac{10-14+5a^2}{24} \, N_f\,\mbox{tr}
\left[ 
  \overline{\cal A}_\mu
  \overline{\cal A}_\nu
  \overline{\cal A}^\mu
  \overline{\cal A}^\nu
\right]
\nonumber\\
&& \qquad
{} + \frac{1-2a^2+2a^3+a^4}{24} \, N_f\,\mbox{tr}
\left[
  \left( 
    \left( \overline{\cal V}_\mu - \overline{V}_\mu \right)
    \left( \overline{\cal V}^\mu - \overline{V}^\mu \right)
  \right)^2
\right]
\nonumber\\
&& \qquad
{} + \frac{1+4a^2-4a^3+a^4}{48} \, N_f\,\mbox{tr}
\left[ 
  \left( \overline{\cal V}_\mu - \overline{V}_\mu \right)
  \left( \overline{\cal V}_\nu - \overline{V}_\nu \right)
  \left( \overline{\cal V}^\mu - \overline{V}^\mu \right)
  \left( \overline{\cal V}^\nu - \overline{V}^\nu \right)
\right]
\nonumber\\
&& \qquad
{} + \frac{a(1+6a-5a^2)}{12} \, N_f \, \mbox{tr}
\left[
  \overline{\cal A}_\mu \overline{\cal A}^\mu 
  \left( \overline{\cal V}_\nu - \overline{V}_\nu \right) 
  \left( \overline{\cal V}^\nu - \overline{V}^\nu \right)
\right]
\nonumber\\
&& \qquad
{} + \frac{1+4a-5a^2+2a^3}{12} \, N_f \, \mbox{tr} \left[
  \overline{\cal A}_\mu \overline{\cal A}_\nu
  \left( \overline{\cal V}^\mu - \overline{V}^\mu \right) \left( \overline{\cal V}^\nu - \overline{V}^\nu \right)
\right]
\nonumber\\
&& \qquad
{} - \frac{1-10a+13a^2-6a^3}{24} \, N_f\,\mbox{tr}
\left[ 
  \overline{\cal A}_\mu
  \overline{\cal A}_\nu
  \left( \overline{\cal V}^\nu - \overline{V}^\nu \right)
  \left( \overline{\cal V}^\mu - \overline{V}^\mu \right)
\right]
\nonumber\\
&& \qquad
{} + \frac{a^2(4-3a)}{12} \, N_f 
\Biggl(
\mbox{tr} \left[ 
  \overline{\cal A}_\mu
  \left( \overline{\cal V}^\mu - \overline{V}^\mu \right)
  \overline{\cal A}_\nu
  \left( \overline{\cal V}^\nu - \overline{V}^\nu \right)
\right]
\nonumber\\
&& \qquad\qquad\qquad\qquad\qquad
+ \mbox{tr} \left[
  \left( \overline{\cal V}_\mu - \overline{V}_\mu \right)
  \overline{\cal A}^\mu
  \left( \overline{\cal V}_\nu - \overline{V}_\nu \right)
  \overline{\cal A}^\nu
\right] \Biggr)
\nonumber\\
&& \qquad
{} + \frac{a^2}{12} \, N_f 
\, \mbox{tr} \left[
  \overline{\cal A}_\mu \left( \overline{\cal V}_\nu - \overline{V}_\nu \right)
  \overline{\cal A}^\mu \left( \overline{\cal V}^\nu - \overline{V}^\nu \right)
\right]
\nonumber\\
&& \qquad
{}+ \frac{8-12a+5a^2}{8} 
\biggl(
  \mbox{tr} \left[ 
    \overline{\cal A}_\mu \overline{\cal A}^\mu
  \right]
\biggr)^2
{}+ \frac{8-12a+5a^2}{4} 
\, \mbox{tr} \left[ 
  \overline{\cal A}_\mu \overline{\cal A}_\nu
\right]
\, \mbox{tr} \left[ 
  \overline{\cal A}^\mu \overline{\cal A}^\nu
\right]
\nonumber\\
&& \qquad
{} + \frac{1+a^4}{16}
\biggl(
  \mbox{tr} \left[
    \left( \overline{\cal V}_\mu - \overline{V}_\mu \right) \left( \overline{\cal V}^\mu - \overline{V}^\mu \right)
  \right]
\biggr)^2
\nonumber\\
&& \qquad
{} + \frac{1+a^4}{8}
\, \mbox{tr} \left[
  \left( \overline{\cal V}_\mu - \overline{V}_\mu \right) \left( \overline{\cal V}_\nu - \overline{V}_\nu \right)
\right]
\, \mbox{tr} \left[
  \left( \overline{\cal V}^\mu - \overline{V}^\mu \right) \left( \overline{\cal V}^\nu - \overline{V}^\nu \right)
\right]
\nonumber\\
&& \qquad
{} + \frac{a\left(1+7a-5a^2\right)}{12}
\, \mbox{tr} \left[
  \overline{\cal A}_\mu \overline{\cal A}^\mu
\right]
\, \mbox{tr} \left[
  \left( \overline{\cal V}_\nu - \overline{V}_\nu \right) \left( \overline{\cal V}^\nu - \overline{V}^\nu \right)
\right]
\nonumber\\
&& \qquad
{} + \frac{a(7-5a+a^2)}{6}
\, \mbox{tr}
\left[
  \overline{\cal A}_\mu
  \overline{\cal A}_\nu
\right]
\, \mbox{tr}
\left[
  \left( \overline{\cal V}^\mu - \overline{V}^\mu \right)
  \left( \overline{\cal V}^\nu - \overline{V}^\nu \right)
\right]
\nonumber\\
&& \qquad
{} + \frac{a(7-5a+a^2)}{6}
\biggl(
  \mbox{tr}
  \left[
    \overline{\cal A}_\mu
    \left( \overline{\cal V}^\mu - \overline{V}^\mu \right)
  \right]
\biggr)^2
\nonumber\\
&& \qquad
{} + \frac{a\left(1+7a-5a^2\right)}{6}
\, \mbox{tr} \left[
  \overline{\cal A}_\mu \left( \overline{\cal V}_\nu - \overline{V}_\nu \right)
\right]
\, \mbox{tr} \left[
  \overline{\cal A}^\mu \left( \overline{\cal V}^\nu - \overline{V}^\nu \right)
\right]
\nonumber\\
&& \qquad
{} + \frac{a(7-5a+a^2)}{8}
\, \mbox{tr}
\left[
  \overline{\cal A}_\mu
  \left( \overline{\cal V}_\nu - \overline{V}_\nu \right)
\right]
\, \mbox{tr}
\left[
  \left( \overline{\cal V}^\mu - \overline{V}^\mu \right)
  \overline{\cal A}^\nu
\right]
\nonumber\\
&& \qquad
{} + \frac{4-3a}{8} \, N_f \, \frac{F_\chi^2}{F_\pi^2}
\, \mbox{tr} \left[
  \overline{\cal A}_\mu \overline{\cal A}^\mu
  \left( \overline{\chi} + \overline{\chi}^\dag \right)
\right]
{} + \frac{4-3a}{8} \, \frac{F_\chi^2}{F_\pi^2}
\, \mbox{tr} \left[
  \overline{\cal A}_\mu \overline{\cal A}^\mu
\right]
\, \mbox{tr} \left[ \overline{\chi} + \overline{\chi}^\dag \right]
\nonumber\\
&& \qquad
{} + \frac{a^2}{8} \, N_f \, \frac{F_\chi^2}{F_\pi^2}
\, \mbox{tr} \left[
  \left( \overline{\cal V}_\mu - \overline{V}_\mu \right) \left( \overline{\cal V}_\nu - \overline{V}_\nu \right)
  \left( \overline{\chi} + \overline{\chi}^\dag \right)
\right]
\nonumber\\
&& \qquad
{} + \frac{a^2}{8} \, \frac{F_\chi^2}{F_\pi^2}
\, \mbox{tr} \left[
  \left( \overline{\cal V}_\mu - \overline{V}_\mu \right) \left( \overline{\cal V}_\nu - \overline{V}_\nu \right)
\right]
\, \mbox{tr} \left[ \overline{\chi} + \overline{\chi}^\dag \right]
\nonumber\\
&& \qquad
{} + \frac{3a(a-1)}{16} \, N_f \, \frac{F_\chi^2}{F_\pi^2}
\, \mbox{tr}
\biggl[
  \left( \overline{\chi} - \overline{\chi}^\dag \right)
  \left[ 
    \overline{\cal A}_\mu \,,\,
    \overline{\cal V}^\mu - \overline{V}^\mu
  \right]
\biggr]
\nonumber\\
&& \qquad
{} + \frac{ N_f^2-4}{16N_f} \left(\frac{F_\chi^2}{F_\pi^2}\right)^2
\, \mbox{tr} \left[
  \left( \overline{\chi} + \overline{\chi}^\dag \right)^2
\right]
{} + \frac{ N_f^2+2}{16N_f^2} \left(\frac{F_\chi^2}{F_\pi^2}\right)^2
\biggl(
  \mbox{tr} \left[ \overline{\chi} + \overline{\chi}^\dag \right]
\biggr)^2
\nonumber\\
&& \qquad
{} - \frac{a}{32} \, N_f \, \left(\frac{F_\chi^2}{F_\pi^2}\right)^2
\left(
  \mbox{tr}
  \biggl[
    \left( \overline{\chi} - \overline{\chi}^\dag \right)^2
  \biggr]
  - \frac{1}{N_f}
  \biggl(
    \mbox{tr}
    \left[ \overline{\chi} - \overline{\chi}^\dag \right]
  \biggr)^2
\right)
\ .
\end{eqnarray}

{}From the above lengthy equation, $\Gamma_{\rm FP}$ in 
Eq.~(\ref{gam fp}) and $\Gamma_{\rm PV}^{(1)}$ in Eq.~(\ref{gam pv1})
we can read the the divergent corrections to the parameters at 
${\cal O}(p^2)$.
Together with the bare parameteres they are given by
\begin{eqnarray}
&&
F_\pi^2 + 
\frac{N_f}{(4\pi)^2}
\biggl(
  - \frac{2-a}{2} \, \Lambda^2
  - \frac{3a}{4} \, 
     M_\rho^2 \ln \frac{\Lambda^2}{M_\rho^2}
\biggr)
\ ,
\\
&&
F_\sigma^2 +
\frac{N_f}{(4\pi)^2}
\biggl(
  - \frac{1+a^2}{4} \, \Lambda^2 
  - \frac{3}{4} \, M_\rho^2 
    \ln \frac{\Lambda^2}{M_\rho^2}
\biggr)
\ ,
\\
&&
- \frac{1}{2g^2} + 
\frac{N_f}{(4\pi)^2} \, \frac{87-a^2}{48}
\ln \frac{\Lambda^2}{M_\rho^2}
\ ,
\\
&&
\frac{F_\chi^2}{4} -
\frac{N_f^2-1}{2N_f} \, \frac{\Lambda^2}{(4\pi F_\pi)^2}
F_\chi^2
\ .
\end{eqnarray}

The logarithmically divergent corrections to the coefficients of 
${\cal O}(p^4)$ terms are listed in Table~\ref{tab:zwy div}.
Here the normalization is fixed by requiring that
$z_i + \left( \Gamma_{z_i} / \left(4(4\pi)^2\right) \right)
\ln \left( \Lambda^2 / M_\rho^2 \right)$
is finite.
[The normalizations for $\Gamma_{y_i}$ and $\Gamma_{w_i}$ are defined
in the same way.]
\begin{table}[hptb]
\begin{center}
\renewcommand{\arraystretch}{1.45}
\begin{tabular}{|c|l|}
\hline
\ $z_1$\  & \ $\displaystyle - N_f \, \frac{5-4a+a^2}{12}$ \\
$z_2$ & \ $\displaystyle - N_f \, \frac{a}{6}$ \\
$z_3$ & \ $\displaystyle - N_f \, \frac{1+2a-a^2}{6}$\\
$z_4$ & \ $\displaystyle - N_f \, \frac{2+3a-3a^2}{6}$\\
$z_5$ & \ $\displaystyle - N_f \, \frac{1+2a^2-a^3}{6}$\\
$z_6$ & \ $\displaystyle - N_f \, \frac{(2-a)(5-3a)}{6}$\\
$z_7$ & \ $\displaystyle - N_f \, \frac{1+4a-4a^2+a^3}{6}$\\
$z_8$ & \ $\displaystyle N_f \, \frac{a (a+1)}{6}$\\
\hline
\multicolumn{2}{c}{}\\
\hline
$w_1$ & \ $\displaystyle N_f\, \frac{4-3a}{4}$\\
$w_2$ & \ $\displaystyle \frac{4-3a}{4}$\\
$w_3$ & \ $\displaystyle N_f\, \frac{a^2}{4}$\\
$w_4$ & \ $\displaystyle \frac{a^2}{4}$\\
$w_5$ & \ $\displaystyle - N_f\, \frac{3a(a-1)}{8}$\\
$w_6$ & \ $\displaystyle \frac{N_f^2-4}{8N_f}$\\
$w_7$ & \ $\displaystyle \frac{N_f^2+2}{8N_f}$\\
$w_8$ & \ $\displaystyle - N_f \, \frac{a}{16}$\\
$w_9$ & \ $\displaystyle \frac{a}{16}$\\
\hline
\end{tabular}
\hspace{20pt}
\begin{tabular}{|c|l|}
\hline
\ $y_1$\ & \ $\displaystyle N_f \, \frac{7-11a+5a^2}{6}$\\
$y_2$ & \ $\displaystyle N_f \, \frac{10-14a+5a^2}{12}$\\
$y_3$ & \ $\displaystyle N_f \, \frac{1-2a^2+2a^3+a^4}{12}$\\
$y_4$ & \ $\displaystyle N_f \, \frac{1+4a^2-4a^3+a^4}{24}$\\
$y_5$ & \ $\displaystyle N_f \, \frac{a(1+6a-5a^2)}{6}$\\
$y_6$ & \ $\displaystyle N_f \, \frac{1+4a-5a^2+2a^3}{6}$\\
$y_7$ & \ $\displaystyle - N_f \, \frac{1-10a+13a^2-6a^3}{12}$\\
$y_8$ & \ $\displaystyle N_f \, \frac{a^2(4-3a)}{6}$\\
$y_9$ & \ $\displaystyle N_f \, \frac{a^2}{6}$\\
$y_{10}$ & \ $\displaystyle \frac{8-12a+5a^2}{4}$\\
$y_{11}$ & \ $\displaystyle \frac{8-12a+5a^2}{2}$\\
$y_{12}$ & \ $\displaystyle \frac{1+a^4}{8}$\\
$y_{13}$ & \ $\displaystyle \frac{1+a^4}{4}$\\
$y_{14}$ & \ $\displaystyle \frac{a(1+7a-5a^2)}{6}$\\
$y_{15}$ & \ $\displaystyle \frac{a(7-5a+a^2)}{3}$\\
$y_{16}$ & \ $\displaystyle \frac{a(7-5a+a^2)}{3}$\\
$y_{17}$ & \ $\displaystyle \frac{a(1+7a-5a^2)}{3}$\\
$y_{18}$ & \ $\displaystyle \frac{a(7-5a+a^2)}{3}$\\
\hline
\end{tabular}
\renewcommand{\arraystretch}{1}
\end{center}
\caption[Coefficients of the divergent corrections to ${\cal O}(p^4)$
parameters of the HLS]{%
Coefficients of the divergent corrections to 
$z_i$ in 
Eq.~(\ref{Lag: z terms}), $w_i$ in Eq.~(\ref{Lag: w terms}) and
$y_i$ in Eq.~(\ref{Lag: y terms}).
The normalization is fixed by requiring that
$z_i + \left( \Gamma_{z_i} / \left(4(4\pi)^2\right) \right)
\ln \left( \Lambda^2 / M_\rho^2 \right)$
is finite.
}
\label{tab:zwy div}
\end{table}

\newpage

\mmsection{References}

\end{document}